\documentclass[a4paper,12pt, epsfig]{article}
\pdfoutput=1 
\def\letter{0}\def\pr{0}
\usepackage{epsfig}
\usepackage{epstopdf}
\usepackage{textcomp}
\usepackage{graphicx}
\usepackage{ifthen}
\usepackage{ulem}

\usepackage{amsmath}
\usepackage[psamsfonts]{amssymb}
\usepackage{euscript}

\usepackage{latexsym}
\usepackage[arrow,matrix,curve]{xy}

\usepackage[hypertexnames=false]{hyperref}
\hypersetup{
    colorlinks=true,
    linkcolor=blue,
    filecolor=magenta,   
    citecolor=black,   
    urlcolor=cyan,
    }

\newskip\humongous \humongous=0pt plus 1000pt minus 1000pt

\newif\ifdtup

\def\,{\hspace{-.1cm}}
\def\hsp{,\hspace{.7cm}}

\def\ddp#1{d^3\vp_{#1}}
\def\dvx{\int d^3\vx}

\def\ddpp#1{d^3\vpp_{#1}}

\def\fc#1#2 {\frac{n}{q}#1\frac{n}{q}#2}

\def\kt{\kappa}
\def\kt{\mathfrak{K}}
\def\ks{|\kt\rangle}

\def\hp{H^{\prime(t)}}
\def\rv{{\rm{vac}}}

\newcommand{\vac}{\ensuremath{|0\rangle}}
\newcommand{\ovac}{\ensuremath{|\Omega\rangle}}
\newcommand{\ps}{\ensuremath{|\vp_0\rangle}}

\renewcommand{\sin}{\textrm{sin}}

\renewcommand{\sinh}{\textrm{sinh}}
\renewcommand{\cosh}{\textrm{cosh}}
\renewcommand{\tanh}{\textrm{tanh}}
\newcommand{\sech}{\textrm{sech}}
\newcommand{\csch}{\textrm{csch}}

\def\exp#1{\hbox{\rm exp}\left[#1\right]}

\def\sign#1{{\rm sign}\left(#1\right)}

\renewcommand{\theequation}{\arabic{section}.\arabic{equation}}
\renewcommand{\(}{\begin{equation}}
\renewcommand{\)}{end{equation} \vspace{-.05in}\linebreak}

\newcounter{saveeqn}
\newcounter{savealpheqn}

\newcommand{\alpheqn}{\setcounter{saveeqn}{\value{equation}}%
  \stepcounter{saveeqn}\setcounter{equation}{0}%
  \renewcommand{\theequation}{\mbox{\arabic{section}.\arabic{saveeqn}
\alph{equation}}}
  \renewcommand{\)}{\end{equation}}}
\def\part#1{\frac{\partial}{\partial{#1}}}%
\def\group#1{\refstepcounter{equation}\setcounter{saveeqn}
 {\value{equation}}%
  \label{#1}\setcounter{equation}{0}%
\renewcommand{\theequation}{\mbox{\arabic{section}.\arabic{saveeqn}
\alph{equation}}}
  \renewcommand{\)}{\end{equation}}}
\newcommand{\reseteqn}{\setcounter{equation}{\value{saveeqn}}%
  \renewcommand{\theequation}{\arabic{section}.\arabic{equation}}%
  \renewcommand{\)}{\end{equation}}}

\newcommand{\aalpheqn}{\setcounter{saveeqn}{\value{equation}}%
  \stepcounter{saveeqn}\setcounter{equation}{0}%
  \renewcommand{\theequation}{\mbox{
        \Alph{subsection}.\arabic{saveeqn}\alph{equation}}}
   \renewcommand{\)}{\end{equation}}}
\newcommand{\areseteqn}{\setcounter{equation}{\value{saveeqn}}%
  \renewcommand{\theequation}{\Alph{subsection}.\arabic{equation}}%
  \renewcommand{\)}{\end{equation}}}

\renewcommand{\thefootnote}{\alph{footnote}}
\renewcommand{\(}{\begin{equation}}
\renewcommand{\)}{\end{equation}}
\newcommand{\ba}{\begin{eqnarray}}
\newcommand{\ea}{\end{eqnarray}}

\renewcommand{\b}{\beta}

\renewcommand{\sl}{{\sqrt{\lambda}}}

\newcommand{\cbp}{\mathop{\vtop{\ialign{##\crcr
   $\hfil\displaystyle{}\hfil$\crcr\noalign{\kern-13pt\nointerlineskip}
   \BIG{)}\hskip0pt\crcr\noalign{\kern3pt}}}}}
\newcommand{\pa}{\mathop{\vtop{\ialign{##\crcr

$\hfil\displaystyle{\oplus}\hfil$\crcr\noalign{\kern+1pt\nointerlineskip
}
   \hspace{.08in}$^{\alpha=0}$\hskip6pt\crcr\noalign{\kern3pt}}}}}
\renewcommand{\hsp}{,\hspace{.3in}}
\newcommand{\p}{^\prime}
\newcommand{\pp}{^{\prime\prime}}

\catcode`\@=11
\def\vereq#1#2{\lower3pt\vbox{\baselineskip1.5pt \lineskip1.5pt
\ialign{$\m@th#1\hfill##\hfil$\crcr#2\crcr\sim\crcr}}}
\catcode`\@=12

\renewcommand{\(}{\begin{equation}}
\renewcommand{\)}{\end{equation}}

\def\vx{{\vec{x}}}
\def\vp{{\vec{p}}}
\def\vpp{{\vec{p}^{
\hspace{.05cm}\prime
}}}
\def\vppp{{\vec{p}^{
\hspace{.05cm}\prime\prime
}}}
\def\vk{{\vec{k}}}
\def\vkp{{\vec{k}\p}}

\def\pin#1{\int \frac{d#1}{2\pi}}
\def\ppin#1{\int\hspace{-17pt}\sum \frac{d#1}{2\pi}}
\def\ppink#1{\int\hspace{-17pt}\sum\frac{d^{#1}\vk}{(2\pi)^{#1}}}

\def\ppinkp#1{\int\hspace{-17pt}\sum\frac{d^{#1}k\p}{(2\pi)^{#1}}}
\def\dint{\int\hspace{-12pt}\sum }
\def\pink#1{\int \frac{d^{#1}k}{(2\pi)^{#1}}}

\def\pinkp#1{\int \frac{d^{#1}k\p}{(2\pi)^{#1}}}

\def\pinvp#1{\int \frac{d^{#1}\vp}{(2\pi)^{#1}}}
\def\pinvk#1{\int\hspace{-17pt}\sum \frac{d^{#1}\vk}{(2\pi)^{#1}}}
\def\kinv#1#2{\int\hspace{-17pt}\sum \frac{d^{#1}\vec{#2}}{(2\pi)^{#1}}}
\def\pinv#1#2{\int \frac{d^{#1}\vec{#2}}{(2\pi)^{#1}}}
\def\pinvx#1#2{\int d^{#1}\vec{#2}}
\def\pinvi#1#2#3{\int \frac{d^{#1}\vec{#2}_{#3}}{(2\pi)^{#1}}}
\def\ppinv#1#2{\int \frac{d^{#1}\vec{#2}^{
\hspace{.07cm}\prime
}}{(2\pi)^{#1}}}

\def\Bd#1{B^\ddag_{k_{#1}}}
\def\Ad#1{A^\ddag_{\vp_{#1}}}
\def\Btd#1{B^{(t)\ddag}_{k_{#1}}}
\def\Bt#1{B^{(t)}_{k_{#1}}}
\def\Bdp#1{B^\ddag_{k\p_{#1}}}

\def\cc{\mathcal{C}}
\def\df{\mathcal{D}_{f}}
\def\dv{\mathcal{D}_{v}}
\def\dft{\mathcal{D}_{f}^{(t)}}
\def\dfd{\mathcal{D}_{f}^{(t)\dag}}

\def\avkd{A^\ddag_{\vec{k}}}
\def\avpd{A^\ddag_{\vec{p}}}
\def\avkm{A_{-\vec{k}}}
\def\avpm{A_{-\vec{p}}}
\def\avk{A_{\vec{k}}}
\def\avp{A_{\vec{p}}}
\def\omvk{\omega_{\vec{k}}}
\def\omvp{\omega_{\vec{p}}}

\def\navkd#1{A^\ddag_{\vec{k}_{#1}}}
\def\navpd#1{A^\ddag_{\vec{p}_{#1}}}

\def\navpm#1{A_{-\vec{p}_{#1}}}
\def\navk#1{A_{\vec{k}_{#1}}}
\def\navp#1{A_{\vec{p}_{#1}}}

\def\nomvp#1{\omega_{\vec{p}_{#1}}}

\def\O#1#2{|\Omega\rangle_{#1}^{#2}}

\def\I{\mathcal{I}}

\def\os{\omega_S}

\def\gx{(\gamma(x-vt))}

\def\red#1{\textcolor{red}{Jarah: #1}}
\def\blu#1{\textcolor{blue}{Hui: #1}}
\usepackage[dvipsnames]{xcolor}
\def\gre#1{\textcolor{ForestGreen}{Hengyuan: #1}}
\def\hui#1{\textcolor{Mulberry}{Hui: #1}}

\newcommand{\beas}{\begin{eqnarray*}}
\newcommand{\eeas}{\end{eqnarray*}}

\newcommand{\bquo}{\begin{quote}}
\newcommand{\enqu}{\end{quote}}

\def\lim#1{\stackrel{\rm{lim}}{{}_{#1}}}

\newcommand{\R}{{\mathbb R}}

    \newcommand{\g}{\mathfrak g}

\def\ch{{\mathcal{H}}}
\def\co{{\mathcal{O}}}

\def\ok#1{\omega_{k_{#1}}}
\def\okp#1{\omega_{k\p_{#1}}}
\def\ovp#1{\omega_{\vp_{#1}}}
\def\ovpp#1{\omega_{\vpp_{#1}}}
\def\ovppp#1{\omega_{\vppp_{#1}}}
\def\okt#1{\omega_{\kt_{#1}}}
\def\V#1{V^{(#1)}(\sqrt{\lambda}f(x))}
\def\Vg#1{V^{(#1)}(\sqrt{\lambda}f(\gamma(x-vt))}
\def\ck{\csch\left(\frac{\pi k}{2\b}\right)}

\def\mb{\mathcal{B}}
\def\mc{\mathcal{C}}
\def\md{\mathcal{D}}
\def\me{\mathcal{E}}
\def\gt{\tilde{\g}}
\def\dim{2}
\def\dimtwo{4}
\def\dimthree{6}
\def\phip{\phi}
\def\pip{\pi}

\newcommand{\beq}{\begin{equation}}
\newcommand{\eeq}{\end{equation}}
\newcommand{\bea}{\begin{eqnarray}}
\newcommand{\eea}{\end{eqnarray}}

\newskip\humongous \humongous=0pt plus 1000pt minus 1000pt

\newif\ifdtup

\catcode`\@=11

\ifthenelse{\equal{\letter}{0}}{ 

\@addtoreset{equation}{section}
\def\theequation{\arabic{section}.\arabic{equation}}

\def\@normalsize{\@setsize\normalsize{15pt}\xiipt\@xiipt
\abovedisplayskip 14pt plus3pt minus3pt%
\belowdisplayskip \abovedisplayskip
\abovedisplayshortskip \z@ plus3pt%
\belowdisplayshortskip 7pt plus3.5pt minus0pt}

\def\small{\@setsize\small{13.6pt}\xipt\@xipt
\abovedisplayskip 13pt plus3pt minus3pt%
\belowdisplayskip \abovedisplayskip
\abovedisplayshortskip \z@ plus3pt%
\belowdisplayshortskip 7pt plus3.5pt minus0pt
\def\@listi{\parsep 4.5pt plus 2pt minus 1pt
      \itemsep \parsep
      \topsep 9pt plus 3pt minus 3pt}}

\relax

\def\section{\@startsection{section}{1}{\z@}{3.5ex plus 1ex minus  .2ex}{2.3ex plus .2ex}{\large\bf}}

\def\thesection{\arabic{section}}
\def\thesubsection{\arabic{section}.\arabic{subsection}}

\def\appendix{\setcounter{section}{0}
 \def\thesection{Appendix \Alph{section}}
 \def\thesubsection{\Alph{section}.\arabic{subsection}}
 \def\theequation{\Alph{section}.\arabic{equation}}}
\renewcommand{\theequation}{\arabic{section}.\arabic{equation}}

}{
\renewcommand{\theequation}{\arabic{equation}}

} 

\begin{document}
\def\thefootnote{\fnsymbol{footnote}}
\def\thetitle{The Domain Wall Soliton's Tension}
\def\autone{Jarah Evslin}
\def\autthree{Hui Liu}
\def\autfour{Baiyang Zhang}
\def\auttwo{Hengyuan Guo}
\def\affd{Yerevan Physics Institute, 2 Alikhanyan Brothers St., Yerevan, 0036, Armenia}
\def\affb{University of the Chinese Academy of Sciences, YuQuanLu 19A, Beijing 100049, China}
\def\affa{Institute of Modern Physics, NanChangLu 509, Lanzhou 730000, China}
\def\affc{School of Physics and Astronomy, Sun Yat-sen University, Zhuhai 519082, China}
\def\affe{Institute of Contemporary Mathematics, School of Mathematics and Statistics, Henan University, Kaifeng, Henan 475004, P. R. China}


\ifthenelse{\equal{\pr}{1}}{
\title{\thetitle}
\author{\autone}
\author{\auttwo}
\author{\autthree}
\affiliation {\affa}
\affiliation {\affb}
\affiliation {\affc}

}{

\begin{center}
{\large {\bf \thetitle}}

\bigskip

\bigskip


{\large 
\noindent  \autone{${}^{12}$}
\footnote{jarah@impcas.ac.cn}, 
\autthree{${}^{3}$} \footnote{hui.liu@yerphi.am} and  
\autfour{${}^{4}$\footnote{byzhang@henu.edu.cn}}
}


\vskip.7cm

1) \affa\\
2) \affb\\
3) \affd\\
4) \affe\\

\end{center}

}

\begin{abstract}
\noindent
We calculate the one-loop tension of the domain wall soliton in the $\phi^4$ double-well model.  Our result agrees with previous results from Dashen, Hasslacher and Neveu (1974) in 1+1d and Jaimunga, Semenoff and Zarembo (1999) in 2+1d.  After an additional 25 year interval, we have obtained a one-loop tension correction of $0.0410959m^3$ in 3+1d.  In this case, unlike lower-dimensional cases, even after normal ordering there are ultraviolet divergences that require both mass and also coupling constant renormalization.  We renormalized the coupling so that the three-point interaction in the effective potential is given by its tree level value at zero external momenta.
\end{abstract}


%
\setcounter{footnote}{0}
\renewcommand{\thefootnote}{\arabic{footnote}}

\ifthenelse{\equal{\pr}{1}}
{
\maketitle
}{}

\section{Introduction}

Many powerful methods \cite{dhn2,gs74,cl75,tom75,fk77,slava15} exist for treating quantum solitons in 1+1 dimensions, or in supersymmetric theories \cite{susy1,susy2}.  However, most of these become prohibitively difficult to apply in nonsupersymmetric settings in more dimensions.  One noteable exception are spectral methods, which have been applied in 3+1 dimensions to calculate one-loop mass corrections to domain walls \cite{muro1,muro2,muro3} and the Nielsen-Olesen vortex \cite{no} in the abelian Higgs model \cite{no1,no2,no3}.  There has even been progress towards the computation of the one-loop correction to the mass of the 't Hooft-Polyakov monopole \cite{mono1,mono2}.  These methods are able to tackle more complicated problems, however it is difficult to see how to extend them beyond one loop.  

Due to its role in confinement in supersymmetric gauge theories \cite{sw2} and its proposed role in confinement in real world QCD \cite{thooftconf,mandconf}, the ultimate goal of recent research in this direction \cite{weigel24} has been the understanding of the quantum 't Hooft-Polyakov monopole.  To say the least, due to the nonperturbative regime of interest, a one loop understanding will not be sufficient.

In Refs.~\cite{mekink,me2loop} we developed a new method, linearized soliton perturbation theory (LSPT).  In the spirit of Ref.~\cite{cahill76}, it is a Hamiltonian method, which begins by constructing the state corresponding to a soliton.  In 1+1 dimensions, it has long been appreciated that solitons in quantum field theory correspond to coherent states \cite{vinc72,cornwall74,taylor78}, up to perturbative corrections.  However, beyond 1+1 dimensions, coherent states lead to divergent energy densities \cite{erice}.  Recently, it has been shown \cite{noi4dlett} that, at one loop, this divergence can be cured by squeezing the coherent state\footnote{In fact, even in 1+1 dimensions the squeezing is necessary to obtain a Hamiltonian eigenstate, but there it only shifts the expected energy by a finite amount.}.  This opens the door to an application of LSPT beyond 1+1 dimensions.

In Ref.~\cite{noi3d} we used LSPT to compute the one-loop correction to the tension of the domain wall in 2+1 dimensions, recovering the old result of Ref.~\cite{zar98}.  This case, however, is rather trivial as the infinite energy density divergences only arise at two loops in 2+1 dimensions.  We provided a quick and dirty quantization of the domain wall soliton in 3+1 dimensions in Ref.~\cite{noi4dlett}.  This construction began with the ground state constructed by minimizing the bare potential, whose parameters are the bare parameters, which are infinite.  Needless to say, this state does not exist.  We then squeezed it and finally constructed a coherent state by acting with the displacement operator that shifts the fields by the classical field theory solution, again obtained from the bare potential.  Needless to say, this solution is a function of the bare parameters, and so it is infinite and this displacement operator does not exist.  We obtained an expression for the domain wall tension as the sum of two terms.  The first is a higher dimensional generalization of the kink mass formula of Ref.~\cite{cahill76}.  This contribution is negative and scheme independent.  The second is the contribution of the counterterms, which depends on our renormalization condition.  We found that while each contribution is separately divergent, their divergences cancel.  However we did not evaluate the finite parts.

In Ref.~\cite{noi4dlungo} we solved this problem properly, using renormalized perturbation theory from the start.  We found the vacuum for the renormalized potential and acted on it with the displacement operator constructed from the domain wall solution to the renormalized Hamiltonian.  Order by order, this led to the same state and the same expression for the tension found in the previous, much shorter, paper.  With this done, one need never return to this long method again, as the state and tension obtained using the the short method have been verified using renormalized perturbation theory.

The goal in the present paper is to apply this construction to obtain the one-loop correction to the tension of the domain wall soliton in the (3+1)-dimensional $\phi^4$ double-well model, now including the finite contributions.  This result is, in principle, implicit in the existing literature \cite{muro1,muro2,muro3}, and in the future we hope that a researcher knowledgeable in spectral techniques may use them to compute this same correction and so compare our results.  However LSPT has the advantage that it can be straightforwardly extended to any number of loops and also to form factors \cite{hengyuanff}, amplitudes \cite{mesonmult,elastic,hengyuanstokes} and decay rates \cite{alberto} which appear to be out of reach of spectral methods, and we feel that this new derivation is an important first step in its application to (3+1)-dimensional quantum solitons. 


\section{Renormalization using the Effective Potential}

\subsection{Generalities}

Faddeev and Korepin \cite{fk77} have stressed that, since the ultraviolet of a calculation with a soliton is like that without a soliton, any cancellation of ultraviolet divergences in one case will cancel that in the other.  As a result, in Ref.~\cite{noi4dlungo} we chose to renormalize using the Schrodinger picture, using calculations which can be carried over from the sector with no soliton, to the sector with a soliton, one at a time, while not affecting the high energy momenta.  In this way we were assured that we would cancel divergences in the soliton sector.  We checked that the divergences were indeed canceled.  

That processes was rather tedious, even though it was only applied to divergences.  In the present paper, we also wish to calculate finite contributions to the tension.  As any renormalization procedure that yields finite observables is expected to differ from any other by a finite amount, we will therefore use a much simpler procedure in the present paper, together with simpler renormalization conditions.  Namely, we will use the standard Heisenberg picture renormalization, where we fix counterterms by demanding that higher order corrections to the effective action vanish.

In this section, we consider a scalar theory in $d+1$ dimensions with a cubic and a quartic interaction with strengths $g$ and $\lambda$.  We will perform a perturbative expansion in $g$, assuming that $\lambda$ is of order $O(g^2)$, as it will be for a general polynomial potential.  Then we specialize to the double-well model in Subsec.~\ref{dwsez}.

We will consider the Hamiltonian
\beq
H=H_2+H_I\hsp H_2=\frac{1}{2}\pinvx{d}{x}\left[:\pi^2(\vx):+:\left(\partial_i\phi(\vx)\right)^2:+m^2:\phi^2(x):\right] 
\eeq
where $::$ is the usual Schrodinger picture normal ordering in terms of a plane wave decomposition, and $\phi(\vx)$ and $\pi(\vx)$ are a Schrodinger picture field and its conjugate momentum.  The interactions are
\beq
H_I=\pinvx{d}{x}\left[ (\Delta+g):\phi^3(\vx): +\frac{\lambda}{4} :\phi^4(\vx):-\frac{\delta m^2}{2}:\phi^2(\vx):\right] \label{hi}
\eeq
where $\Delta$ is a counterterm for the cubic interaction and $\delta m^2$ is the mass renormalization counterterm.  In our leading order approximation we will not need a counterterm for the quartic coupling, and also wave function renormalization is only necessary at two loops and beyond.  

It will be convenient to introduce the Heisenberg picture fields 
\beq
\phi(\vx,t)=e^{iHt}\phi(\vx)e^{-iHt}.
\eeq

\subsection{Coupling Constant Renormalization}

The three-point correlation function of three such fields is given by
\bea
\langle\Omega|T\{\phi(\vx_1,t_1)\phi(\vx_2,t_2)\phi(0,0)\}|\Omega\rangle&=&-i\pinv{2}{E}\pinv{2d}{p}\frac{e^{-iE_1t_1-iE_2t_2+i\vx_1\cdot \vp_1+i\vx_2\cdot \vp_2}}{(E_1^2-\ovp{1}^2+i\epsilon)(E_2^2-\ovp{2}^2+i\epsilon)}\nonumber\\
&&\times\frac{\Gamma(\vp_1,E_1,\vp_2,E_2)}{((E_1+E_2)^2-\omega_{\vp_1+\vp_2}^2+i\epsilon)}\label{teq}
\eea
where $T$ indicates time ordering.  In principle, there are also higher order corrections to the external legs.  For brevity, we will ignore these, calculating $\Gamma$ exactly order by order.  Of course they would need to be computed to obtain higher order corrections to the three point function.

We may expand the effective potential $\Gamma=\sum_i \Gamma_i$ so that $\Gamma_i$ is of order $O(g^i)$.  At leading order
\beq
\Gamma_1=-6gi
\eeq
leading, for example at $t_1>t_2>0$,  to a leading three-point function of
\bea
&&-\frac{3g}{4}\pinv{2d}{p}\frac{e^{i\vx_1\cdot \vp_1+i\vx_2\cdot \vp_2}}{\ovp 1\ovp 2\omega_{\vp_1+\vp_2}}\left[\frac{e^{-i\ovp 1t_1-i\ovp 2t_2}+e^{-i\ovp 2(t_1-t_2)-i\omega_{\vp_1+\vp_2}t_1}}{\ovp 1+\ovp 2+\omega_{\vp_1+\vp_2}}\right.\\
&&\ \ \ \ \left. + \frac{e^{-i\ovp 1t_1}\left(e^{-i(\omega_{\vp_1+\vp_2}-\ovp 1)t_2}-e^{-i\ovp 2t_2}\right)}{\ovp 1+\ovp 2-\omega_{\vp_1+\vp_2}}
+\frac{e^{-i\ovp 1(t_1-t_2)-i\omega_{\vp_1+\vp_2}t_2}-e^{-i\ovp 2(t_1-t_2)-i\omega_{\vp_1+\vp_2}t_1}}{\omega_{\vp_1+\vp_2}+\ovp 2-\ovp 1}
\right].\nonumber
\eea

The relation (\ref{teq}) is invertible
\bea
\Gamma(\vp_1,E_1,\vp_2,E_2)&=&i(E_1^2-\ovp{1}^2+i\epsilon)(E_2^2-\ovp{2}^2+i\epsilon)((E_1+E_2)^2-\omega_{\vp_1+\vp_2}^2+i\epsilon)\\
&&\times \int d^2 t\pinvx{2d}{x}\ e^{iE_1t_1+iE_2t_2-i\vx_1\cdot \vp_1-i\vx_2\cdot \vp_2}\langle\Omega|\phi(\vx_1,t_1)\phi(\vx_2,t_2)\phi(0,0)|\Omega\rangle.
\nonumber
\eea
The effective potential $\Gamma$ is particularly convenient as it may be computed using amputated Feynman diagrams.  Therefore, we choose the renormalization condition
\beq
\Gamma(\vec{0},0,\vec{0},0)=\Gamma_1(\vec{0},0,\vec{0},0).
\eeq

\begin{figure}[htbp]
\centering
\includegraphics[width = 0.85\textwidth]{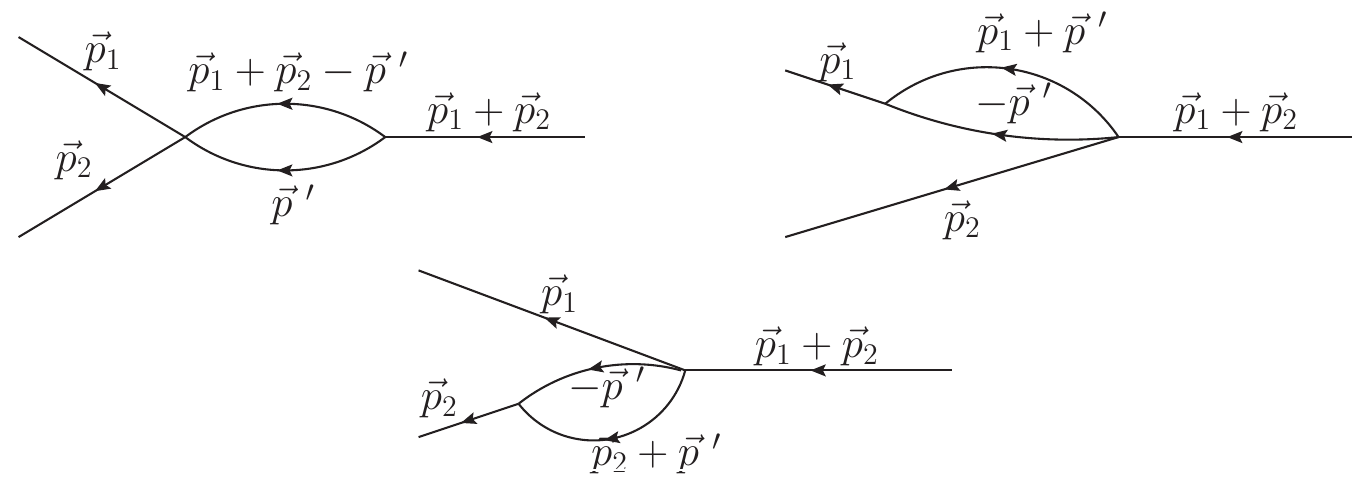}
\caption{These three amputated diagrams each yield a divergent contribution to the three point function.}\label{irredfig}
\end{figure}

At the first subleading order, there are three contributions to the effective potential
\beq
0=\Gamma_3=\Gamma_{3a}+\Gamma_{3b}+\Gamma_{3c}.
\eeq
The first arises from the counterterm and is simply
\beq
\Gamma_{3a}=-6i\Delta.
\eeq
The next contribution comes from three divergent, irreducible, amputated diagrams shown in Fig.~\ref{irredfig}
\beq
\Gamma_{3b}(\vp_1,E_1,\vp_2,E_2)=I(-\vp_1-\vp_2,-E_1-E_2)+I(\vp_1,E_1)+I(\vp_2,E_2)
\eeq
where
\bea
I(\vp,E)&=&18g\lambda\pinv{d}{p\p}\pin{E\p}\frac{1}{(E^{\prime 2}-\ovpp{}^2+i\epsilon)((E+E\p)^2-\omega_{\vp+\vpp}^2+i\epsilon)}\nonumber\\
&=&-\frac{9ig\lambda}{2}\pinv{d}{p\p}\frac{1}{\ovpp{}\omega_{\vp+\vpp}}\left(\frac{1}{E-\ovpp{}-\omega_{\vp+\vpp}}-\frac{1}{E+\ovpp{}+\omega_{\vp+\vpp}}\right).
\eea

\begin{figure}[htbp]
\centering
\includegraphics[width = 0.65\textwidth]{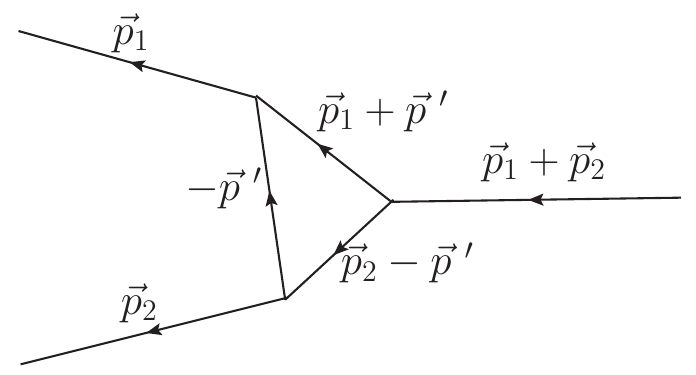}
\caption{This diagram provides a finite contribution to the three point function.}\label{trifig}
\end{figure}

Finally there is the contribution from the triangle diagram in Fig.~\ref{trifig}
\bea
\Gamma_{3c}(\vp_1,E_1,\vp_2,E_2)&=&216g^3\pinv{d}{p\p}\pin{E\p}\frac{1}{(E^{\prime 2}-\ovpp{}^2+i\epsilon)((E_1+E\p)^2-\omega_{\vp_1+\vpp}^2+i\epsilon)}\nonumber\\
&&\times\frac{1}{((E\p-E_2)^2-\omega_{\vpp-\vp_2}^2+i\epsilon)}\nonumber\\
&=&-27 i g^3\pinv{d}{p\p}\frac{B}{\ovpp{}\omega_{\vp_1+\vpp}\omega_{\vpp-\vp_2}}\hsp
B=B_1+B_2+B_3
\eea
where
\bea
B_1&=&\left(\frac{1}{E_1+\ovpp{}-\omega_{\vp_1+\vpp}}-\frac{1}{E_1+\ovpp{}+\omega_{\vp_1+\vpp}}\right)\left(\frac{1}{\ovpp{}-E_2-\omega_{\vpp-\vp_2}}-\frac{1}{\ovpp{}-E_2+\omega_{\vpp-\vp_2}}
\right)\nonumber\\
B_2&=&\left(\frac{-1}{E_1-\omega_{\vp_1+\vpp}+\ovpp{}}+\frac{1}{E_1-\omega_{\vp_1+\vpp}-\ovpp{}}\right)\nonumber\\
&&\times\left(\frac{-1}{E_1+E_2-\omega_{\vp_1+\vpp}+\omega_{\vpp-\vp_2}}+\frac{1}{E_1+E_2-\omega_{\vp_1+\vpp}-\omega_{\vpp-\vp_2}}\right)\nonumber\\
B_3&=&\left(\frac{1}{E_2+\omega_{\vpp-\vp_2}-\ovpp{}}-\frac{1}{E_2+\omega_{\vpp-\vp_2}+\ovpp{}}\right)\nonumber\\
&&\times\left(\frac{1}{E_1+E_2-\omega_{\vp_1+\vpp}+\omega_{\vpp-\vp_2}}+\frac{1}{E_1+E_2+\omega_{\vp_1+\vpp}+\omega_{\vpp-\vp_2}}\right).
\eea
We will impose our renormalization condition at $\vp_i=E_i=0$ where some of these denominators vanish.  Therefore, we will need to rewrite this expression so that these denominators do not appear
\bea
B&=&\frac{1}{\left(E_2-\ovpp{}-\omega_{\vpp-\vp_2}\right)\left(E_1+E_2-\omega_{\vp_1+\vpp}-\omega_{\vpp-\vp_2}\right)}\\
&&+\frac{1}{\left(E_2+\ovpp{}+\omega_{\vpp-\vp_2}\right)\left(E_1+E_2+\omega_{\vp_1+\vpp}+\omega_{\vpp-\vp_2}\right)}
\nonumber\\&&
+\frac{1}{\left(E_1+\ovpp{}+\omega_{\vp_1+\vpp}\right)\left(E_1+E_2+\omega_{\vp_1+\vpp}+\omega_{\vpp-\vp_2}\right)}\nonumber\\
&&+\frac{1}{\left(E_1-\ovpp{}-\omega_{\vp_1+\vpp}\right)\left(E_1+E_2-\omega_{\vp_1+\vpp}-\omega_{\vpp-\vp_2}\right)}\nonumber\\
&&-\frac{1}{\left(E_1-\ovpp{}-\omega_{\vp_1+\vpp}\right)\left(E_2+\ovpp{}+\omega_{\vpp-\vp_2}\right)}\nonumber\\
&&-\frac{1}{\left(E_1+\ovpp{}+\omega_{\vp_1+\vpp}\right)\left(E_2-\ovpp{}-\omega_{\vpp-\vp_2}\right)}.\nonumber
\eea
This can be simplified further, but it will not be useful here because we will impose the renormalization condition $\Gamma_3=0$ at $\vp_1=\vp_2=E_1=E_2=0$, where these expressions simplify to
\beq
I(\vec{0},0)=\frac{9ig\lambda}{2}\pinv{d}{p\p}\frac{1}{\ovpp{}^3}\hsp\Gamma_{3b}(\vec{0},0,\vec{0},0)=\frac{27ig\lambda}{2}\pinv{d}{p\p}\frac{1}{\ovpp{}^3}
\eeq
and
\beq
B=\frac{3}{2\ovpp{}^2}\hsp
\Gamma_{3c}(\vec{0},0,\vec{0},0)=-\frac{81i g^3}{2} \pinv{d}{p\p}\frac{1}{\ovpp{}^5}.
\eeq
Our renormalization condition then fixes the counterterm
\bea
\Delta=\frac{9g}{4}\pinv{d}{p\p}\left(\frac{\lambda}{\ovpp{}^3}-\frac{3g^2}{\ovpp{}^5}\right). \label{deltaeq}
\eea

\subsection{The Double-Well Model} \label{dwsez}

Now we restrict attention to the double-well model, which is characterized by the interaction
\beq
\frac{\lambda_0}{4}:\phi^4(\vx):.
\eeq
This is shifted to the tree level vacuum by the transformation
\beq
\phi(\vx)\rightarrow \phi(\vx)-\frac{m_0}{\sqrt{2\lambda_0}}
\eeq
leading to the total interaction
\beq
\frac{\lambda_0}{4}:\left(\phi(\vx)-\frac{m_0}{\sqrt{2\lambda_0}}\right)^4:
\eeq
and so the cubic interaction is
\beq
-\frac{m_0\sqrt{\lambda_0}}{\sqrt{2}}:\phi^3(\vx):.
\eeq
Using the definitions of the renormalized parameters
\beq
\sqrt{\lambda}=\sqrt{\lambda_0}+\delta\sqrt{\lambda}\hsp m^2=m_0^2+\delta m^2
\eeq
where the counterterms $\delta\sl$ and $\delta m^2$ are taken to be of order $O(\lambda^{3/2})$ and $O(\lambda)$, we expand the cubic interaction to order $O(\lambda^{3/2})$ as
\beq
\left(-\frac{m\sqrt{\lambda}}{\sqrt{2}}
+\frac{m\delta\sqrt{\lambda}}{\sqrt{2}}+\frac{\sqrt{\lambda}\delta m^2}{2\sqrt{2}m}\right):\phi^3(\vx):.
\eeq
Identifying this with our general interaction Hamiltonian (\ref{hi}) we find \cite{noi4dlungo}
\beq
\Delta= m\sqrt{\frac{\lambda}{2}}\left(\frac{\delta m^2}{2m^2}+\frac{\delta\sl}{\sl} \right)\hsp g=-m\sqrt{\frac{\lambda}{2}}. \label{mano}
\eeq
We have considered the $\phi^4$ model because it is the only renormalizable model in 3+1 dimensions that is bounded from below.  However, the nonrenormalizability is not visible at one loop, and so one may consider other potentials corresponding to low energy effective theories.  The generalization of this quick derivation is obvious, one simply replaces the coefficient of the cubic term with the third derivative of the potential evaluated at the relevant minimum of the potential.

One might worry that we have chosen a particular minimum of the potential, corresponding to one vacuum of the theory, but that the cubic coupling is different at the other minima and so our renormalization only applies on one side of the wall.  However, it has long been appreciated that ultraviolet divergences are independent of spontaneous symmetry breaking, and so in particular are independent of the choice of vacuum.

Substituting our choice of model (\ref{mano}) into our general result (\ref{deltaeq}) we find
\beq
 \frac{\delta m^2}{2m^2}+\frac{\delta\sl}{\sl} =\frac{9\lambda}{4}\pinv{d}{p\p}\left(-\frac{1}{\ovpp{}^3}+\frac{3m^2}{2\ovpp{}^5}\right).
\eeq
If we used a similar off-shell renormalization for $\delta m^2$ then we would obtain a very short expression for $\delta\sl$, but we use the conventional on-shell renormalization for the mass.  No on-shell renormalization is possible for the 3-point self-coupling.

Using the integrals
\beq
\pinv{d}{p\p}\frac{1}{\ovpp{}^3}=\left\{\begin{tabular}{lll}
$\frac{1}{\pi m^2}$&&1+1d\\
$\frac{1}{2\pi m}$&&2+1d\\
$\frac{1}{2\pi^2}\int_0^\infty dp \frac{p^2}{\omega_p^3}$&&3+1d
\end{tabular}
\right.
\eeq
and
\beq
\pinv{d}{p\p}\frac{1}{\ovpp{}^5}=\left\{\begin{tabular}{lll}
$\frac{2}{3\pi m^4}$&&1+1d\\
$\frac{1}{6\pi m^3}$&&2+1d\\
$\frac{1}{6\pi^2 m^2}$&&3+1d
\end{tabular}
\right.
\eeq
we conclude
\beq
\frac{\delta m^2}{2m^2}+\frac{\delta\sl}{\sl} =\lambda\left\{\begin{tabular}{lll}
$0$&&1+1d\\
$-\frac{9}{16\pi m}$&&2+1d\\
$\frac{9}{16\pi^2 }-\frac{9}{8\pi^2}\int_0^\infty dp \frac{p^2}{\omega_p^3}$&&3+1d.
\end{tabular}
\right.\label{combo}
\eeq

\subsection{Mass Renormalization}

\begin{figure}[htbp]
\centering
\includegraphics[width = 0.65\textwidth]{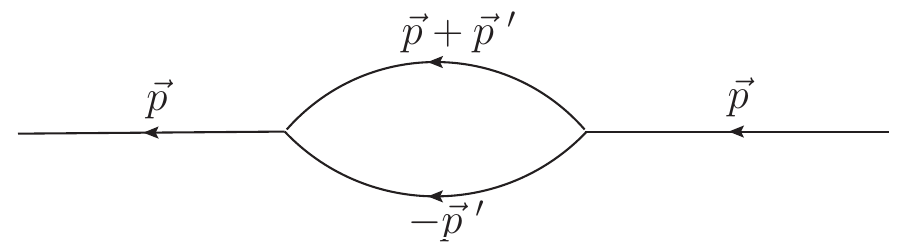}
\caption{This diagram provides a divergent contribution to the propagator.}\label{massfig}
\end{figure}

To obtain $\delta m^2$, we proceed similarly for the two point function.  We start with the relation between the two-point function and the meson self-energy $\Pi$
\beq
\langle\Omega|T\{\phi(\vx,t)\phi(0,0)\}|\Omega\rangle=i\pin{E}\pinv{d}{p}\frac{e^{-iEt+i\vx\cdot \vp}\Pi(E,\vp)}{(E^2-\ovp{}^2+i\epsilon)^2}.
\eeq
As before, for brevity we have ignored corrections to the external legs which will not be important for the calculation of $\Pi$, as we may obtain it from the two-point function by considering only one particle irreducible diagrams.  We expand $\Pi=\sum_i\Pi_i$ so that $\Pi_i$ is of order $O(g^i)$.  At leading order one obtains the inverse propagator 
\beq
\Pi_0(E,\vp)=E^2-\ovp{}^2+i\epsilon
\eeq
leading to the usual $1/(2\ovp{})$ in the two-point function.  

At order $O(\lambda)$ there are two contributions
\beq
\Pi_2(E,\vp)=\Pi_{2a}(E,\vp)+\Pi_{2b}(E,\vp).
\eeq
The first arises from the mass counterterm
\beq
\Pi_{2a}(E,\vp)=i\delta m^2
\eeq
while the second arises from a single loop in Fig.~\ref{massfig}
\bea
\Pi_{2b}(E,\vp)&=&18g^2\pinv{d}{p\p}\pin{E\p}\frac{1}{(E^{\prime 2}-\ovpp{}^2+i\epsilon)((E+E\p)^2-\omega_{\vp+\vpp}^2+i\epsilon)}\nonumber\\
&=&\frac{9ig^2}{2}\pinv{d}{p\p}\frac{1}{\ovpp{}\omega_{\vp+\vpp}}\left(\frac{1}{E+\ovpp{}+\omega_{\vp+\vpp}}-\frac{1}{E-\ovpp{}-\omega_{\vp+\vpp}}\right).
\eea
Fixing the renormalization condition $\Pi(E,\vp)=\Pi_0(E,\vp)$ at some $(E,\vp)$ we find
\beq
\delta m^2=-\frac{9g^2}{2}\pinv{d}{p\p}\frac{1}{\ovpp{}\omega_{\vp+\vpp}}\left(\frac{1}{E+\ovpp{}+\omega_{\vp+\vpp}}-\frac{1}{E-\ovpp{}-\omega_{\vp+\vpp}}\right).
\eeq

Unlike the coupling constant renormalization, here $(E,\vp)$ may be taken to be on-shell, so that $E=\ovp{}$.  In this case, using (\ref{mano}), one finds the usual result
\beq
\delta m^2=-\frac{9m^2\lambda}{4}\pinv{d}{p\p}\frac{1}{\ovpp{}\omega_{\vp+\vpp}}\left(\frac{1}{\ovpp{}+\omega_{\vp+\vpp}-\ovp{}}+\frac{1}{\ovp{}+\ovpp{}+\omega_{\vp+\vpp}}\right).
\eeq
As a result of Lorentz invariance, this is independent of $\vp$.  For simplicity, we will write it at $\vp=0$
\beq
\frac{\delta m^2}{m^2}=-9\lambda\pinv{d}{p\p}\frac{1}{\ovpp{}\left(3m^2+4p^{\prime 2}\right)}=\lambda\left\{\begin{tabular}{lll}
$-\frac{\sqrt{3}}{2m^2}$&&1+1d\\
$-\frac{9{\rm{ln}(3)}}{8\pi m}$&&2+1d\\
$-\frac{9}{2\pi^2}\int_0^\infty dp \frac{p^2}{\omega_p(3m^2+4p^2)}$&&3+1d.
\end{tabular}
\right.
\eeq
Combining this with (\ref{combo}) we find the coupling constant renormalization
\beq
\frac{\delta\sl}{\sl} =\lambda\left\{\begin{tabular}{lll}
$\frac{\sqrt{3}}{4m^2}$&&1+1d\\
$\frac{9}{16\pi m}\left({\rm{ln}}(3)-1\right)$&&2+1d\\
$\frac{9}{16\pi^2}+\frac{9}{8\pi^2}\int_0^\infty dp \ p^2\left(-\frac{1}{\omega_p^3}+\frac{2}{\omega_p(3m^2+4p^2)}
\right)$&&3+1d.
\end{tabular}
\right.
\eeq

\section{The Domain Wall Tension}

\subsection{Generalities}

At one-loop, the domain wall tension is
\beq
\rho=\rho_0+\rho_1
\eeq
where the contribution from the counterterms $\delta m^2$ and $\delta\sl$ is included in the bare tension
\beq
\rho_0=\frac{m_0^3}{3\lambda_0}=\frac{(m^2-\delta m^2)^{3/2}}{3(\sqrt{\lambda}-\delta\sqrt{\lambda})^2}=\frac{m^3}{3\lambda}+\frac{m^3}{\lambda}\left[-\frac{\delta m^2}{2m^2}+\frac{2\delta\sqrt\lambda}{3\sl}\right]. \label{ten}
\eeq
Inserting our $\delta m^2$ and $\delta \sl$ one finds
\beq
\rho_0=\frac{m^3}{3\lambda}+m^3
\pinv{d}{p}\left(-\frac{3}{2\ovp{}^3}+\frac{15}{2\ovp{}\left(3m^2+4p^{2}\right)}+\frac{9m^2}{4\ovp{}^5}\right).
\eeq
In 1+1 or 2+1 dimensions this is just
\beq
\rho_0=\frac{m^3}{3\lambda}+\left\{\begin{tabular}{lll}
$\frac{5}{4\sqrt{3}}m$&&1+1d\\
$\frac{3}{16\pi}\left({5\rm{ln}}(3)-2\right)m^2$&&2+1d .
\end{tabular}
\right. \label{c12}
\eeq

In 3+1 dimensions, it becomes
\bea
\rho_0&=&\frac{m^3}{3\lambda}+\frac{3m^3}{8\pi^2}+\pin{p_x}C(p_x)\label{C}\\
C(p_x)&=&m^3\int_0^\infty dp_r \left(-\frac{3p_r}{4\pi\left(m^2+p_x^2+p_r^2\right)^{3/2}}+\frac{15p_r}{4\pi \sqrt{m^2+p_x^2+p_r^2}\left(3m^2+4p_x^{2}+4p_r^2\right)}\right)\nonumber
\eea
Here we have adopted cylindrical coordinates, rather than spherical coordinates, as they respect the symmetries of the domain wall.  The $p_r$ integral $C(p_x)$ is finite but then the $p_x$ integral is logarithmically divergent.  One may regularize this by cutting off the $p_x$ integration at $\Lambda$.

\subsection{Calculating $\rho_1$}

The scheme-dependent correction due to one-loop mass and coupling renormalization, $\rho_0$, needs to be added to the scheme-independent, negative one-loop quantum correction that arises from the squeeze.  In Ref.~\cite{noi3d} we showed that the one-loop correction to the tension of a domain wall in any dimension is given by the formula for the kink mass found by Cahill, Comtet and Glauber in Ref.~\cite{cahill76}
\beq
\rho_1=-\frac{1}{4}\ppin{k_1}\pinv{3}{p} |\gt_{-k_1}(p_1)|^2 \frac{(\omega_{k_1p_2p_3}-\omega_{p_1p_2p_3})^2}{\omega_{p_1p_2p_3}}\hsp \omega_{pqr}=\sqrt{m^2+p^2+q^2+r^2} \label{c76}
\eeq
where the normal modes are
\bea
\tilde{\g}_{B}(p)&=&-\frac{\sqrt{6}\pi p}{m^{3/2}}  \csch\left(\frac{\pi p}{m}\right)
\hsp
\tilde{\g}_{S}(p)=-\frac{2i\sqrt{3}\pi p}{m^{3/2}}  \sech\left(\frac{\pi p}{m}\right)\\
\tilde{\g}_k(p)&=&\frac{2k^2-m^2}{\ok{}\sqrt{m^2+4k^2}}2\pi\delta(p+k)+\frac{6\pi p}{\ok{}\sqrt{m^2+4k^2}} \csch\left(\frac{\pi (p+k)}{m}\right).\nonumber
\eea

We decompose $\rho_1$ into its contributions from each normal mode $B$, $S$ and $k_x$
\bea
\rho_1&=&\rho_{1B}+\rho_{1S}+\rho_{1C}\hsp \rho_{1C}=\pin{p_x}\pin{k_x}\hat{\rho}_{1}(k_x,p_x)\hsp
p_r=\sqrt{p_y^2+p_z^2}\nonumber
\\
\rho_{1 B}&=&-\frac{1}{8\pi}\pin{p_x}\gt_{B}(p_x)\gt_{B}(-p_x)
\int_0^\infty dp_r p_r \frac{\left(p_r-\sqrt{m^2+p_x^2+p_r^2}\right)^2}{\sqrt{m^2+p_x^2+p_r^2}}\nonumber\\
\rho_{1 S}&=&-\frac{1}{8\pi}\pin{p_x}\gt_{S}(p_x)\gt_{S}(-p_x)
\int_0^\infty dp_r p_r\frac{\left(\sqrt{\frac{3m^2}{4}+p_r^2}-\sqrt{m^2+p_x^2+p_r^2}\right)^2}{\sqrt{m^2+p_x^2+p_r^2}}\nonumber\\
\hat{\rho}_{1}(k_x,p_x)&=&-\frac{1}{8\pi} \gt_{-k_x}(p_x)\gt_{k_x}(-p_x)
\int_0^\infty dp_r p_r\frac{\left(\sqrt{m^2+k_x^2+p_r^2}-\sqrt{m^2+p_x^2+p_r^2}\right)^2}{\sqrt{m^2+p_x^2+p_r^2}}.\nonumber
\eea
In 1+1 or 2+1 dimensions this leads to the old results \cite{dhn2,zar98,mekink,noi3d}
\bea
\rho_{1B}&=&\left\{\begin{tabular}{lll}
$-0.272m$&&1+1d\\
$-0.0477465 m^2$&&2+1d 
\end{tabular}
\right.
\hsp
\rho_{1S}=\left\{\begin{tabular}{lll}
$-0.020 m$&&1+1d\\
$-0.0072502 m^2$&&2+1d 
\end{tabular}
\right. \nonumber \\
\rho_{1C}&=&\left\{\begin{tabular}{lll}
$-0.041m$&&1+1d\\
$-0.03156 m^2$&&2+1d 
\end{tabular}
\right.\hsp
\rho_{1}=\left\{\begin{tabular}{lll}
$\left(\frac{1}{4\sqrt{3}}-\frac{3}{2\pi}\right) m$&&1+1d\\
$\left({\rm{ln}}(3)-4\right)\frac{3m^2}{32\pi}$&&2+1d .
\end{tabular}
\right. 
\eea
In other words, in 1+1 dimension we find
\beq
\rho=\frac{m^3_0}{3\lambda_0}+\left(\frac{1}{4\sqrt{3}}-\frac{3}{2\pi}\right) m=\frac{m^3}{3\lambda}+\left(\frac{3}{2\sqrt{3}}-\frac{3}{2\pi}\right) m
\eeq
while in 2+1 dimensions
\beq
\rho=\frac{m^3_0}{3\lambda_0}+\left({\rm{ln}}(3)-4\right)\frac{3m^2}{32\pi}=\frac{m^3}{3\lambda}+\left(\frac{33}{32\pi}{\rm{ln}}(3)-\frac{3}{4\pi}\right)m^2.
\eeq
Note that these $O(\lambda^0)$ mass corrections to $m^3_0/3\lambda_0$ are somewhat smaller than the scheme-dependent $O(\lambda^0)$ mass corrections to $m^3/3\lambda$ in Eq.~(\ref{c12}), which are chosen to enforce our renormalization conditions.  Intuitively, the scheme-independent mass corrections arise from vacuum loops, while those of (\ref{c12}) arise from loops connected to two or three external legs with a choice of momentum and energy.   With our choice of scheme we find that, as a result of the renormalization, the one-loop tension correction is now positive.

In 3+1 dimensions, we may analytically perform the $p_r$ integrations using the identity
\beq
\int_0^\infty dp_r p_r \frac{(\sqrt{a+p_r^2}-\sqrt{m^2+p_x^2+p_r^2})^2}{\sqrt{m^2+p_x^2+p_r^2}}=\frac{2 a^{3/2} - 3 a \sqrt{m^2+p_x^2} + (m^2+p_x^2)^{3/2}}{3}. \label{intid}
\eeq
Setting $a=0$ in the identity (\ref{intid}) one easily finds the zero-mode contribution to the tension
\bea
\rho_{1 B}&=&-\frac{1}{24\pi}\pin{p_x}\gt_{B}(p_x)\gt_{B}(-p_x)
(m^2+p_x^2)^{3/2}\\
&=&-\frac{\pi}{4m^3}\pin{p_x}{p_x^2}(m^2+p_x^2)^{3/2}{\rm{csch}}^2\left(\frac{\pi p_x}{m}\right)\sim -0.0178498m^3.\nonumber
\eea
Next, setting $a=3m^2/4$, one finds the shape mode contribution
\bea
\rho_{1 S}&=&-\frac{1}{24\pi}\pin{p_x}\gt_{S}(p_x)\gt_{S}(-p_x)\left[ 
\frac{3\sqrt 3 m^3}{4} - \frac{9m^2}{4} \sqrt{m^2+p_x^2} + (m^2+p_x^2)^{3/2}
\right]\\
&=&-\frac{\pi}{8m^3}\pin{p_x}{p_x^2}\left[ 
{3\sqrt 3 m^3} - {9m^2} \sqrt{m^2+p_x^2} + 4(m^2+p_x^2)^{3/2}
\right]\sech^2\left(\frac{\pi p_x}{m}\right)\nonumber\\
&\sim&-0.00423897m^3.
\eea
As in the case of the kink in 1+1d and the domain wall string in 2+1d, the shape mode contribution is smaller than the zero mode contribution.  Roughly, in 1+1d it was smaller by a factor of 14, in 2+1d by a factor of 7, now it is only smaller by a factor of 4.  This is to be expected, as in higher dimensions, the higher $p_r$ modes are more strongly weighted, and these have a similar frequency for the zero and shape modes.  

The partitioning of the finite part of the counterterm contribution $\rho_0$ among the modes is arbitrary, and here we will choose to include the third term of the first line of (\ref{C}) together with the continuum modes and we will then need to add the first two terms separately.  The continuum contribution $\rho_c$ to $\rho$ can then be decomposed in the plane-wave basis as
\beq
\rho_c=\pin{p_x}\left[C(p_x)+\pin{k_x}\hat{\rho}_1(k_x,p_x)\right]
\eeq
where
\bea
\hat{\rho}_{1}(k_x,p_x)&=&-\frac{1}{8\pi}\gt_{-k_x}(p_x)\gt_{k_x}(-p_x)
\int_0^\infty dp_r p_r\frac{\left(\sqrt{m^2+k_x^2+p_r^2}-\sqrt{m^2+p_x^2+p_r^2}\right)^2}{\sqrt{m^2+p_x^2+p_r^2}}\nonumber\\
&=&-\frac{3\pi}{2}\left[
\frac{p_x^2}{\ok{x}^2(m^2+4k_x^2)}\csch^2\left(\frac{\pi(p_x+k_x)}{m}\right)
\right]\nonumber\\
&&\times
\left[ 
2 (m^2+k_x^2)^{3/2} - 3 (m^2+k_x^2) \sqrt{m^2+p_x^2} + (m^2+p_x^2)^{3/2}
\right].
\eea

\begin{figure}[htbp]
\centering
\includegraphics[width = 0.6\textwidth]{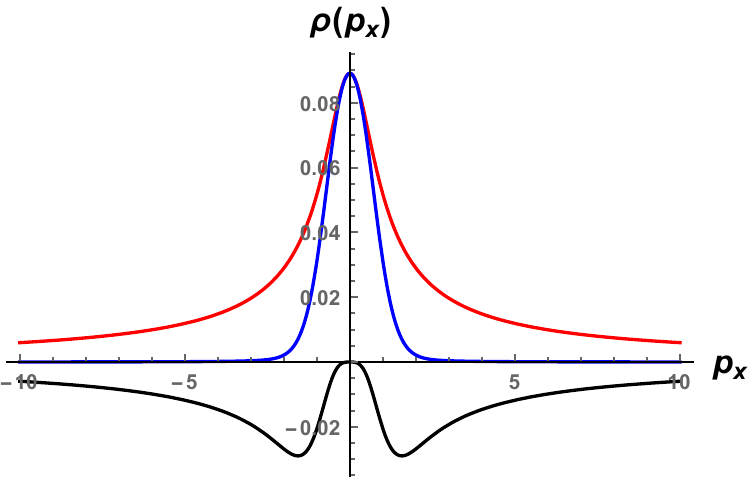}
\caption{The contribution to the tension from continuum modes arising at each $p_x$ from counterterms $C$ (red), one-loop corrections $\pin{k_x}\hat{\rho}_1(k_x,p_x)$ (black) and their total (blue).  The individual contributions asymptotically scale as $1/p_x$ leading to a logarithmic ultraviolet divergence in $\rho$, but their divergent contributions cancel.}\label{pdiff}
\end{figure}

Integrating we find the contribution
\beq
\rho_c\sim 0.0251892 m^3.
\eeq
As one can see in Fig.~\ref{pdiff}, the continuum contribution $\pin{k_x}\hat{\rho}_1(k_x,p_x)$ now dominates over the discrete contributions, unlike the case in lower dimensions.  Indeed, the continuum contribution is now infinite.  However, it is more than canceled by the scheme-dependent contribution from the counterterms, which overwhelmingly dominate the one-loop tension.

Finally, one must include the other contributions from $\rho_0$ in Eq.~(\ref{C}).  These are the classical tension $m^3/(3\lambda)$ and the other counterterm contribution from the triangle diagrams
\beq
\rho_C=\frac{3m^3}{8\pi^2}\sim 0.0379954m^3.
\eeq
In all we find that in 3+1 dimensions the tension is
\beq
\rho\sim \frac{m^3}{3\lambda}+0.0410959 m^3+O(m^3 \lambda).
\eeq
Note that the coefficient of the leading correction is positive.  However, this depended on our choice of renormalization conditions.

\section{Applications}

While we were motivated by the desire to extend LSPT to 3+1 dimensions, quantum domain walls themselves are of intrinsic interest \cite{blanmuri,lanmuri,tokmuri}.  Domain wall networks recently attracted attention \cite{gw1,gw2} as a potential source of the stochastic gravitational wave background observed by pulsar timing arrays \cite{pt1,pt2,pt3,pt4}, and even as a source of gravitational waves at lower frequencies \cite{gw3}.  In addition, the observation of supermassive black holes at larger and larger redshifts is in ever stronger tension with the usual supermassive black hole assembly paradigms, strongly suggesting the existence of black hole seeds \cite{bhrev}, which may be provided by domain walls \cite{bh1,bh2,bh3,bh4}.

Very strong bounds on domain wall abundances imply that domain walls must annihilate early.  The mechanisms behind this early annihilation are not yet established, and so it is important to understand both the interactions of domain walls with radiation and also their internal excitations \cite{albertomuro}.  Needless to say, both problems require an understanding of the underlying quantum theory, as some radiative processes have no classical analogue and also the internal excitations are quantized.  We have already applied LSPT to such problems in the case of kinks in 1+1 dimensions \cite{mesonmult,hengyuanstokes}, and so we now see no obstruction to extending this treatment to the more phenomenologically relevant setting of domain walls in 3+1 dimensions.

So far we have only treated one loop divergences, and so one may wonder whether our approach will fail at two loops.  Moreover, we have only treated divergences that are logarithmic in the ultraviolet cutoff.  We expect, based on the general arguments of Ref.~\cite{fk77}, that if we remove ultraviolet divergences in the vacuum sector then we will also remove it in the soliton sector.  However, since we use a cutoff which is not Lorentz invariant, one may expect that our counterterms will in general not be Lorentz invariant.  In the case of logarithmic divergences this problem is avoided because a change in the momentum scale of the renormalization leads to a shift which is inversely proportional to the cutoff, and so vanishes as the regulator is removed.  However, in the case of linear or quadratic divergences, one may expect that a renormalization condition at one momentum scale does not remove divergences at other scales.  Thus we expect our approach to apply as is to the renormalization of the (2+1)-dimensional domain wall string at any order.  

However, in the case of the (3+1)-dimensional domain wall, a quadratic divergence appears already at two loops.  Thus, if we wish to extend our results to 3+1 dimensions at two loops, we will need to introduce a regulator that respects Lorentz symmetry.  We do not yet know if dimensional regularization, or the equivalent zeta function regularization, may be applied as we do not know how the soliton solution should be modified in the regularized theory.  Similarly, Pauli-Villars regularization has not yet been incorporated in LSPT as we do not know how to treat the ghosts at a finite value of the regulator.  In gauge theories, excepting triangle anomaly, such divergences fortunately tend to cancel.  A notable exception would be provided by the elementary Higgs self-coupling, should this Standard Model coupling be confirmed in a precision setting \cite{higgs1,higgs2,higgs3} and should the Higgs be elementary.

\section*{Acknowledgement}

\noindent
JE is supported by NSFC MianShang grants 11875296 and 11675223.   HL is supported by the Ministry of Education, Science, Culture and Sport of the Republic of Armenia under the Postdoc-Armenia Program, grant number 24PostDoc/2‐1C009. JE and HL are supported by the Ministry of Education, Science, Culture and Sport of the Republic of Armenia under the Remote Laboratory Program, grant number 24RL-1C047. BYZ is supported by the Young Scientists Fund of the National Natural Science Foundation of China (Grant No. 12305079).

\end{document}

\red{After here comes most of the work, but I don't think it should go in the paper}

\section{Schrodinger Picture: Coupling Constant Renormalization}

\subsection{The Renormalization Condition}

It is well known that $n$-point functions can be factorized into amputated $n$-point functions contracted with propagators
\bea
\langle\Omega|\phi(\vx_1)\phi(\vx_2)\phi(\vx_3)|\Omega\rangle&=&\int d^{3d}\vec{y} \int d^3\vec{t} \langle\phi(\vec{y}_1,t_1)\phi(\vec{y}_2,t_2)\phi(\vec{y}_3,t_3)\rangle_{\rm{amp}}\label{fatt}\\
&&\hspace{-1cm}\times \langle\Omega|T\phi(\vec{y}_1,t_1)\phi(\vx_1)|\Omega\rangle\langle\Omega|T\phi(\vec{y}_2,t_2)\phi(\vx_2)|\Omega\rangle\langle\Omega|T\phi(\vec{y}_3,t_3)\phi(\vx_3)|\Omega\rangle\nonumber
\eea
where we have employed the vector notation for the triplets
\beq
\vec{t}=(t_1,t_2,t_3)\hsp
\vec{y}=(\vec{y}_1,\vec{y}_2,\vec{y}_3).
\eeq
We have also introduced the Heisenberg picture field
\beq
\phi(\vec{y}_i,t_i)=e^{iHt_i}\phi(\vec{y}_i)e^{-iHt_i}
\eeq
and the time ordering $T$ which in this case reverses the order when the corresponding $t_i$ is negative.

The amputated correlation function can be decomposed in powers of the coupling
\beq
\langle\phi(\vec{y}_1,t_1)\phi(\vec{y}_2,t_2)\phi(\vec{y}_3,t_3)\rangle_{\rm{amp}}=\sum_{i=0}^\infty \langle\phi(\vec{y}_1,t_1)\phi(\vec{y}_2,t_2)\phi(\vec{y}_3,t_3)\rangle_{\rm{amp}}^i
\eeq
where each term is of order $\lambda^i/2$.  We renormalize the coupling constant by imposing the renormalization condition
\beq
\langle\phi(\vx_1,t_1)\phi(\vx_2,t_2)\phi(\vx_3,t_3)\rangle_{\rm{amp}}=\langle\phi(\vx_1,t_1)\phi(\vx_2,t_2)\phi(\vx_3,t_3)\rangle_{\rm{amp}}^1.
\eeq
In other words, the counterterms must be chosen so as the cancel all contributions arising from loops in the amputated three point function, which corresponds to the one-particle irreducible part of the vertex.

While it is not possible to satisfy this condition at all $\vx$, we will impose it only at certain values $\vp$ of its Fourier transform, as is usual in the case of coupling constant renormalization.  Any such choice will lead to a finite tension, but different choices lead to different conventions for the renormalized $\lambda$ for the same theory and so the tension as a function of $\lambda$ depends on this choice. 

\subsection{Leading Order}

Let us begin with the propagator $\langle\Omega|T\phi(\vec{y}_i,t_i)\phi(\vx_i)\rangle$.  First consider $t_i>0$.  Then, at all orders, we find
\bea
\langle\Omega|T\phi(\vec{y}_i,t_i)\phi(\vx_i)|\Omega\rangle&=&\langle\Omega|\phi(\vec{y}_i,t_i)\phi(\vx_i)|\Omega\rangle=\langle\Omega|e^{iHt_i}\phi(\vec{y}_i)e^{-iHt_i}\phi(\vx_i)|\Omega\rangle\\
&=&\langle\Omega|\phi(\vec{y}_i)e^{-iHt_i}\phi(\vx_i)|\Omega\rangle.\nonumber
\eea
Let us decompose the propagator in powers of $\lambda$
\beq
\langle\Omega|T\phi(\vec{y}_i,t_i)\phi(\vx_i)|\Omega\rangle=\sum_{n=0}^\infty \langle\Omega|T\phi(\vec{y}_i,t_i)\phi(\vx_i)|\Omega\rangle^n
\eeq
where $\langle\Omega|T\phi(\vec{y}_i,t_i)\phi(\vx_i)|\Omega\rangle^n$ is of order $O(\lambda^{n/2})$.

At order unity this reduces to
\bea
\langle\Omega|T\phi(\vec{y}_i,t_i)\phi(\vx_i)|\Omega\rangle^0&=&{}_0\langle\Omega|\phi(\vec{y}_i)e^{-iH_2t_i}\phi(\vx_i)|\Omega\rangle_0\nonumber\\
&=&
\pinvi d p 1 e^{-i\vp_1\cdot \vec{y}_i}\pinvi d p 2  e^{-i\vp_2\cdot \vec{x}_i}
{}_0\langle -p_1|e^{-iH_2t_i}|p_2\rangle_0\nonumber\\
&=&\pinv d p \frac{e^{-i\vp\cdot (\vec{y}_i-\vec{x}_i)-i\ovp{} t_i}}{2\ovp{}}.
\eea
The case $t_i<0$ proceeds similarly, but now that $e^{iHt_i}$ term survives, and so at general $t_i$ one finds that the leading order propagator is
\beq
\langle\Omega|T\phi(\vec{y}_i,t_i)\phi(\vx_i)|\Omega\rangle^0=\pinv d p \frac{e^{-i\vp\cdot (\vec{y}_i-\vec{x}_i)-i\ovp{} |t_i|}}{2\ovp{}}. \label{propmin}
\eeq
The first correction to this formula occurs at order $O(\lambda)$, corresponding to $n=2$.  

The leading order 3-point function arises at order $O(\sl)$ where we find
\bea
\langle\Omega|\phi(\vx_1)\phi(\vx_2)\phi(\vx_3)|\Omega\rangle&=&2\ {}_0\langle\Omega|\phi(\vx_1)\phi(\vx_2)\phi(\vx_3)|\Omega\rangle_1\label{min}\\
&&\hspace{-4cm}=m\sqrt{2\lambda}  \pinv {2d} {p\p}  {}_0\langle\Omega|\phi(\vx_1)\phi(\vx_2)\phi(\vx_3)\frac{|\vpp_1,\vpp_2,-\vpp_1-\vpp_2\rangle_0}{\ovpp 1+\ovpp 2+\omega_{\vpp_1+\vpp_2}}
\nonumber\\
&&\hspace{-4cm}=m\sqrt{2\lambda} \pinv {3d} p e^{-i(\vp_1\cdot \vx_1+\vp_2\cdot \vx_2+\vp_3\cdot \vx_3)}  \pinv {2d} {p\p}  {}_0\langle-\vp_1,-\vp_2,-\vp_3|\frac{|\vpp_1,\vpp_2,-\vpp_1-\vpp_2\rangle_0}{\ovpp 1+\ovpp 2+\omega_{\vpp_1+\vpp_2}}
\nonumber\\
&&\hspace{-4cm}=\frac{3m\sl}{2\sqrt{2}} \pinv {2d} p  \frac{e^{-i\left[\vp_1\cdot (\vx_1-\vx_3)+\vp_2\cdot (\vx_2-\vx_3)\right]} }{\ovp 1\ovp 2\omega_{\vp_1+\vp_2}\left(\ovp 1+\ovp 2+\omega_{\vp_1+\vp_2}\right)}.
\nonumber
\eea

We will now show that
\beq
\langle\phi(\vx_1,t_1)\phi(\vx_2,t_2)\phi(\vx_3,t_3)\rangle_{\rm{amp}}^1=3\sqrt{2}i m\sl\delta^d(\vx_1-\vx_3)\delta^d(\vx_2-\vx_3)\delta(t_1-t_3)\delta(t_2-t_3)
\eeq
satisfies the factorization condition (\ref{fatt}).

Using these Dirac delta functions to integrate $\vec{y}_2$, $\vec{y}_3$, $t_2$ and $t_3$ the right hand side of $(\ref{fatt})$ simplifies to
\bea
&&3\sqrt{2}i m\sl \int d^{d}\vec{y} \int d t \langle\Omega|T\phi(\vec{y},t)\phi(\vx_1)|\Omega\rangle\langle\Omega|T\phi(\vec{y},t)\phi(\vx_2)|\Omega\rangle\langle\Omega|T\phi(\vec{y},t)\phi(\vx_3)|\Omega\rangle\nonumber\\
&&\hspace{2cm}=\frac{3i m\sl}{4\sqrt{2}} \int d^{d}\vec{y} \int d t \pinv {3d} p \frac{e^{-i\sum_{j=1}^3\left[\vp_j\cdot (\vec{y}-\vec{x}_j)+\ovp j |t|\right]}}{\ovp 1\ovp 2\ovp 3}.
\eea
When we perform the $t$ integration, we drop the rapidly oscillating boundary terms, leaving the contributions at $t=0$ from the integral over positive and negative $t$, which are equal and lead to a factor of two
\bea
\frac{3 m\sl}{2\sqrt{2}} \int d^{d}\vec{y}  \pinv {3d} p \frac{e^{-i\sum_{j=1}^3\vp_j\cdot (\vec{y}-\vec{x}_j)}}{\ovp 1\ovp 2\ovp 3\left(\ovp 1+\ovp 2+\ovp 3\right)}.
\eea
Integration over $\vec{y}$ leads to a $(2\pi)^d\delta(\vp_1+\vp_2+\vp_3)$ which can be used to perform the $\vp_3$ integral, yielding the three point function found in Eq.~(\ref{min}), which is the left hand side of Eq.~(\ref{fatt}) as claimed.

\subsection{Subleading Order}

\subsubsection{The Vacuum}

Recall that we will fix $\delta\sl$ by imposing
\beq
\langle\phi(\vx_1,t_1)\phi(\vx_2,t_2)\phi(\vx_3,t_3)\rangle_{\rm{amp}}^3=0.
\eeq
The left hand side can be determined by the factorization property (\ref{fatt}) at order $O(\lambda^{3/2})$ together with the propagator at order $O(1)$, which we already found in Eq.~(\ref{propmin}).  

We do not need to find the propagator at order $O(\lambda)$, so long as we only compute the one particle irreducible contributions
\beq
\langle\Omega|\phi(\vx_1)\phi(\vx_2)\phi(\vx_3)|\Omega\rangle_{\rm{1PI}}
\eeq
to the three-point function $\langle\Omega|\phi(\vx_1)\phi(\vx_2)\phi(\vx_3)|\Omega\rangle$.  This is because these contributions already determine the amputated three-point function $\langle\phi(\vx_1,t_1)\phi(\vx_2,t_2)\phi(\vx_3,t_3)\rangle_{\rm{amp}}^3$ via the identity
\bea
\langle\Omega|\phi(\vx_1)\phi(\vx_2)\phi(\vx_3)|\Omega\rangle^3_{\rm{1PI}}&=&\int d^{3d}\vec{y} \int d^3\vec{t} \langle\phi(\vec{y}_1,t_1)\phi(\vec{y}_2,t_2)\phi(\vec{y}_3,t_3)\rangle_{\rm{amp}}^3\\
&&\hspace{-3cm}\times \langle\Omega|T\phi(\vec{y}_1,t_1)\phi(\vx_1)|\Omega\rangle^0\langle\Omega|T\phi(\vec{y}_2,t_2)\phi(\vx_2)|\Omega\rangle^0\langle\Omega|T\phi(\vec{y}_3,t_3)\phi(\vx_3)|\Omega\rangle^0\nonumber
\eea

At order $O(\lambda)$, the only term in the vacuum state that will contribute to the 1PI correlation function is $|\Omega\rangle_2^4$ which is determined from
\beq
H_2\O 2 4=-H_3^1\O 1 3-H_4^4\O 0 0.
\eeq
These two contributions are respectively
\bea
H_3^1\O 1 3&=&\left(-\frac{3m\sqrt{\lambda}}{2\sqrt{2}} \pinv{2d}p A^\ddag_{\vp_1}A^\ddag_{\vp_2}\frac{A_{\vp_1+\vp_2}}{\omega_{\vp_1+\vp_2}}
\right)\left( m\sqrt{\frac{\lambda}{2}}\pinv {2d}{p\p}\frac{|\vpp_1,\vpp_2,-\vpp_1-\vpp_2\rangle_0}{\ovpp 1+\ovpp 2+\omega_{\vpp_1+\vpp_2}}
\right)\nonumber\\
&=&-\frac{9m^2\lambda}{4}\pinv{3d} p\frac{|\vp_1,\vp_2,\vp_3,-\vp_1-\vp_2-\vp_3\rangle_0}{\omega_{\vp_1+\vp_2}\left( \ovp 1+\ovp 2+\omega_{\vp_1+\vp_2}
\right)}\nonumber\\
H_4^4\O 0 0&=&\frac{\lambda}{4}\pinv{3d} p |\vp_1,\vp_2,\vp_3,-\vp_1-\vp_2-\vp_3\rangle_0
\eea
leading to
\bea
\O 2 4=\frac{\lambda}{4}\pinv{3d} p \left(
-1+\frac{9m^2}{\omega_{\vp_1+\vp_2}\left( \ovp 1+\ovp 2+\omega_{\vp_1+\vp_2}
\right)}
\right) \frac{|\vp_1,\vp_2,\vp_3,-\vp_1-\vp_2-\vp_3\rangle_0}{\ovp 1+\ovp 2+\ovp 3+\omega_{\vp_1+\vp_2+\vp_3}}.
\eea

At order $O(\lambda^{3/2})$ only $\O 3 3$ will contribute to the irreducible three-point function.  Dropping the one particle reducible contributions from $\O 2 2$ and $\O 2 6$, it is determined by
\beq
H_2\O 3 3=-H_3^{-1}\O 2 4-H_4^0\O 1 3-H_5^3\O 0 0.\label{o33eq}
\eeq

Some of these contributions are one particle reducible.  In the case of $H_3^{-1}\O 2 4$, the $H_3^{-1}$ operator contains two annihilation operators.  Any contraction of these annihilation operators will be irreducible in the case of the term that arose from $H_4^4\O 0 0$.  However, in the case of the term that arose from $H_3^1\O 1 1$, only contractions with one of $\{\vp_1,\vp_2\}$ and one of $\{\vp_3,-\vp_1-\vp_2-\vp_3\}$ lead to irreducible contributions.  We will use the $\supset$ symbol to indicate that reducible terms are set to zero.  Without loss of generality, we may contract with $\vp_2$ and $\vp_3$ in the two cases, multiplying by the correct combinatoric factor.  Then we find
\bea
H_3^{-1}\O 2 4&=&\left(-\frac{3m\sqrt{\lambda}}{4\sqrt{2}} \pinv{2d}{p\p} A^\ddag_{\vpp_1+\vpp_2}\frac{A_{\vpp_1}}{\ovpp 1}\frac{A_{\vpp_2}}{\ovpp 2}
\right)\O 2 4
\\
&&\hspace{-2cm}\supset \frac{3m\lambda^{3/2}}{16\sqrt{2}}\pinv{3d} p \left(
12-\frac{72m^2}{\omega_{\vp_1+\vp_2}\left( \ovp 1+\ovp 2+\omega_{\vp_1+\vp_2}
\right)}
\right) \frac{|\vp_1,\vp_2+\vp_3,-\vp_1-\vp_2-\vp_3\rangle_0}{\ovp 2\ovp 3\left(\ovp 1+\ovp 2+\ovp 3+\omega_{\vp_1+\vp_2+\vp_3}\right)}\nonumber\\
&&\hspace{-2cm}=\frac{9m\lambda^{3/2}}{4\sqrt{2}}\pinv{2d} p\left[\pinv{d} {p\p}
\frac{1}{\ovpp{}\omega_{\vp_1+\vpp}\left(\ovp 2+\ovpp{}+\omega_{\vp_1+\vpp}+\omega_{\vp_1+\vp_2}\right)}\right.\nonumber\\
&&\left.\times\left(
1-\frac{6m^2}{\omega_{\vp_2+\vpp}\left( \ovp 2+\ovpp{}+\omega_{\vp_2+\vpp}
\right)}
\right) \right]
|\vp_1,\vp_2,-\vp_1-\vp_2\rangle_0
\eea

In the case of $H_4^0\O 1 3$, only the term involving the quartic interaction leads to an irreducible contribution
\bea
H_4^0\O 1 3&\supset&\left(\frac{3\lambda}{8}\pinv{3d}{p\p} A^\ddag_{\vpp_1+\vpp_2+\vpp_3}A^\ddag_{-\vpp_3}\frac{A_{\vpp_1}}{\ovpp 1}\frac{A_{\vpp_2}}{\ovpp 2}\right)
\left( m\sqrt{\frac{\lambda}{2}}\pinv {2d}{p}\frac{|\vp_1,\vp_2,-\vp_1-\vp_2\rangle_0}{\ovp 1+\ovp 2+\omega_{\vp_1+\vp_2}}
\right)\nonumber\\
&=&\frac{9m\lambda^{3/2}}{4\sqrt{2}}\pinv{d}{p\p}\pinv {2d}{p}\frac{|-\vpp,\vp_1+\vp_2+\vpp,-\vp_1-\vp_2\rangle_0}{\ovp 1\ovp 2\left(\ovp 1+\ovp 2+\omega_{\vp_1+\vp_2} \right)}\nonumber\\
&=&\frac{9m\lambda^{3/2}}{4\sqrt{2}}\pinv {2d}{p}\left[ \pinv{d}{p\p} \frac{1}{\ovpp{}\omega_{\vpp+\vp_1}\left(\ovpp{}+\omega_{\vpp+\vp_1}+\ovp 1 \right)}
\right]
|\vp_1,\vp_2,-\vp_1-\vp_2\rangle_0.\nonumber
\eea

The last contribution to $\O 3 3$ arises from
\beq
H_5^3=\Delta\pinv{2d}{p} A^\ddag_{\vp_1}A^\ddag_{\vp_2}A^\ddag_{-\vp_1-\vp_2}\hsp
\Delta=\frac{m\sl}{\sqrt{2}}\left(\frac{\delta\sl}{\sl}+\frac{1}{2}\frac{\delta m^2}{m^2} 
\right).
\eeq
It consists of a single term
\beq
H_5^3\O 0 0=\Delta\pinv{2d}{p} |\vp_1,\vp_2,-\vp_1-\vp_2\rangle.
\eeq

Plugging these three contributions into (\ref{o33eq}) we find
\bea
\O 3 3&=&\pinv{2d} p \gamma_3^3(\vp_1,\vp_2) |\vp_1,\vp_2,-\vp_1-\vp_2\rangle\label{gamma}\\
\gamma_3^3(\vp_1,\vp_2)&\supset&-\frac{1}{\ovp 1+\ovp 2+\omega_{\vp_1+\vp_2}}\left[\Delta+ \frac{9m\lambda^{3/2}}{4\sqrt{2}}\pinv{d}{p\p}
\frac{1}{\ovpp{}\omega_{\vpp+\vp_1}}
\left(\frac{1}{\ovpp{}+\omega_{\vpp+\vp_1}+\ovp 1}
\right.\right.\nonumber\\
&&\left.\left.
+\frac{1}{\left(\ovp 2+\ovpp{}+\omega_{\vp_1+\vpp}+\omega_{\vp_1+\vp_2}\right)}
\left(
1-\frac{6m^2}{\omega_{\vp_2+\vpp}\left( \ovp 2+\ovpp{}+\omega_{\vp_2+\vpp}
\right)}
\right) 
\right)
\right].\nonumber
\eea

\subsubsection{The Three Point Function}

Finally we are ready to compute the three point function.  At order $O(\lambda^{3/2})$ it is
\beq
\langle\Omega|\phi(\vx_1)\phi(\vx_2)\phi(\vx_3)|\Omega\rangle=2\ {}_0\langle\Omega|\phi(\vx_1)\phi(\vx_2)\phi(\vx_3)|\Omega\rangle_3+2\ {}^3_1\langle\Omega|\phi(\vx_1)\phi(\vx_2)\phi(\vx_3)|\Omega\rangle_2
\eeq
where we recall that only $\O 2 4$ and $\O 3 3$ will contribute to the amputated correlation functions.  Our renormalization condition implies that the irreducible parts of the two terms on the right hand side must sum to zero, or more precisely a certain Fourier component should sum to zero.  Let us compute these two terms in turn. \red{Maybe these inner products are wrong ... need to symmetrize over momenta?}

The first is
\bea
{}_0\langle\Omega|\phi(\vx_1)\phi(\vx_2)\phi(\vx_3)|\Omega\rangle_3^3&=&\frac{1}{8}\pinv{3d}p e^{-i\sum_j^3\vx_j\cdot \vp_j} {}_0\langle\Omega|\frac{A_{-\vp_1}A_{-\vp_2}A_{-\vp_3}}{\ovp 1\ovp 2\ovp 3}\nonumber\\
&&\hspace{-2cm}\times \pinv{2d} {p\p} \gamma_3^3(\vpp_1,\vpp_2) |\vpp_1,\vpp_2,-\vpp_1-\vpp_2\rangle\nonumber\\
&&\hspace{-2cm}=\frac{3}{4}\pinv{2d} {p} \frac{e^{-i\left[\vp_1\cdot(\vx_1-\vx_3)+\vp_2\cdot(\vx_2-\vx_3)\right]}\gamma_3^3(\vp_1,\vp_2)}{\ovp 1\ovp 2\omega_{\vp_1+\vp_2}} \label{cor1}.
\eea

\red{Symmetrizing it is
\bea
\frac{1}{8}\pinv{2d} {p} \frac{e^{-i\left[\vp_1\cdot(\vx_1-\vx_3)+\vp_2\cdot(\vx_2-\vx_3)\right]}}{\ovp 1\ovp 2\omega_{\vp_1+\vp_2}} &&\hspace{-.5cm}\left( \gamma_3^3(\vp_1,\vp_2)+\gamma_3^3(\vp_2,\vp_1)+\gamma_3^3(\vp_1,-\vp_1-\vp_2)\right.\nonumber\\
&&\hspace{-3.5cm}\left.+\gamma_3^3(-\vp_1-\vp_2,\vp_1)+\gamma_3^3(\vp_2,-\vp_1-\vp_2)+\gamma_3^3(-\vp_1-\vp_2,\vp_2)\right).
\eea
It will turn out that only the choice of one external momentum matters, the one attached to the trivalent vertex.  So, for each choice, we get a factor of $1/3$.  We will consider this by simply dividing (\ref{cor1}) by 3.  Later we might want
to sum over the three choices of external vertex, but we ignore this sum for now.}

In the second contribution, again we must be careful to remove the one particle reducible terms, which will not contribute to the amputated correlator.  For example, we are not interested in the tadpole contribution arising from the commutators of the fields in the $\phi^3$ operator.  Thus, we may normal order it, to obtain
\bea
{}^3_1\langle\Omega|\phi(\vx_1)\phi(\vx_2)\phi(\vx_3)|\Omega\rangle_2^4&=&\left( m\sqrt{\frac{\lambda}{2}}\pinv {2d}{p\p}\frac{{}_0\langle\vpp_1,\vpp_2,-\vpp_1-\vpp_2|}{\ovpp 1+\ovpp 2+\omega_{\vpp_1+\vpp_2}}
\right)\\
&&\hspace{-4cm}\times\left(\frac{1}{4}\pinv{3d}p e^{-i\sum_j^3\vx_j\cdot \vp_j}A^\ddag_{\vp_1}\frac{A_{-\vp_2}A_{-\vp_3}}{\ovp 2\ovp 3}
\right)\nonumber\\
&&\hspace{-4cm}\times\frac{\lambda}{4}\pinv{3d} {p\pp} \left(
-1+\frac{9m^2}{\omega_{\vppp_1+\vppp_2}\left( \ovppp 1+\ovppp 2+\omega_{\vppp_1+\vppp_2}
\right)}
\right) \frac{|\vppp_1,\vppp_2,\vppp_3,-\vppp_1-\vppp_2-\vppp_3\rangle_0}{\ovppp 1+\ovppp 2+\ovppp 3+\omega_{\vppp_1+\vppp_2+\vppp_3}}\nonumber\\
&&\hspace{-4cm}=\frac{3m\lambda^{3/2}}{32\sqrt{2}}\pinv{3d}p\pinv{d}{p\p}\pinv{3d}{p\pp}
\frac{{}_0\langle \vpp,-\vp_1-\vpp|}{\ovp 1\left(\ovp 1+\ovpp{}+\omega_{\vp_1+\vpp}\right)}e^{-i\sum_j^3\vx_j\cdot \vp_j}\frac{A_{-\vp_2}A_{-\vp_3}}{\ovp 2\ovp 3}\nonumber\\
&&\hspace{-4cm}\times \left(
-1+\frac{9m^2}{\omega_{\vppp_1+\vppp_2}\left( \ovppp 1+\ovppp 2+\omega_{\vppp_1+\vppp_2}
\right)}
\right) \frac{|\vppp_1,\vppp_2,\vppp_3,-\vppp_1-\vppp_2-\vppp_3\rangle_0}{\ovppp 1+\ovppp 2+\ovppp 3+\omega_{\vppp_1+\vppp_2+\vppp_3}}.\nonumber
\eea
Again, not all contractions in this expression will lead to irreducible diagrams.  Consider the parenthesis on the last line.  In the case of the $-1$ term, all contractions of $A$ operators with $\vppp$ indeed lead to irreducible diagrams.  In the case of the $9m^2$ term, only contractions of one $A$ operator with one of $\{\vppp_1,\vppp_2\}$ 
and the other $A$ operator with one of $\{\vppp_3,-\vppp_1-\vppp_2-\vppp_3\}$ will be irreducible.  We will handle this by contracting $\vp_2$ with $\vppp_3$ and $\vp_3$ with $\vppp_3$ but associating a different combinatoric factor with each term, to count the number of irreducible possibilities
\bea
{}^3_1\langle\Omega|\phi(\vx_1)\phi(\vx_2)\phi(\vx_3)|\Omega\rangle_2^4&=&\frac{3m\lambda^{3/2}}{32\sqrt{2}}\pinv{3d}p\pinv{d}{p\p}\pinv{d}{p\pp}
\frac{{}_0\langle \vpp,-\vp_1-\vpp|}{\ovp 1\left(\ovp 1+\ovpp{}+\omega_{\vp_1+\vpp}\right)}\nonumber\\
&&\hspace{-4cm}\times\frac{e^{-i\sum_j^3\vx_j\cdot \vp_j}}{\ovp 2\ovp 3} \left(
-12+\frac{72m^2}{\omega_{\vppp+\vp_2}\left( \ovppp{}+\ovp 2+\omega_{\vppp+\vp_2}
\right)}
\right) \frac{|\vppp,-\vppp-\vp_2-\vp_3\rangle_0}{\ovppp{}+\ovp 2+\ovp 3+\omega_{\vppp+\vp_2+\vp_3}}\nonumber\\
&&\hspace{-4cm}=\frac{9m\lambda^{3/2}}{8\sqrt{2}}\pinv{3d}p\frac{e^{-i\sum_j^3\vx_j\cdot \vp_j}}{\ovp 1\ovp 2\ovp 3} \pinv{d}{p\p}
\frac{{}_0\langle -\vp_1-\vpp|}{\ovpp{}\left(\ovp 1+\ovpp{}+\omega_{\vp_1+\vpp}\right)}\nonumber\\
&&\hspace{-4cm}\times\left(
-1+\frac{6m^2}{\omega_{\vpp+\vp_2}\left( \ovpp{}+\ovp 2+\omega_{\vpp+\vp_2}
\right)}
\right) \frac{|-\vpp-\vp_2-\vp_3\rangle_0}{\ovpp{}+\ovp 2+\ovp 3+\omega_{\vpp+\vp_2+\vp_3}}\nonumber\\
&&\hspace{-4cm}=\frac{9m\lambda^{3/2}}{16\sqrt{2}}\pinv{2d}p\frac{e^{-i[\vp_1\cdot(\vx_1-\vx_3)+\vp_2\cdot(\vx_2-\vx_3)]}}{\ovp 1\ovp 2\omega_{\vp_1+\vp_2}} \pinv{d}{p\p}
\frac{1}{\ovpp{}\omega_{\vp_1+\vpp}\left(\ovp 1+\ovpp{}+\omega_{\vp_1+\vpp}\right)}\nonumber\\
&&\hspace{-4cm}\times\left(
-1+\frac{6m^2}{\omega_{\vpp+\vp_2}\left( \ovpp{}+\ovp 2+\omega_{\vpp+\vp_2}
\right)}
\right) \frac{1}{\ovpp{}+\ovp 2+\omega_{\vp_1+\vp_2}+\omega_{\vpp+\vp_1}}\label{cor2}.
\eea

\subsubsection{The Counterterm}

Demanding that the Fourier transform with respect to $\vp$ of the sum of (\ref{cor1}) and (\ref{cor2}) vanishes at some fixed $\vp$ one obtains the condition
\bea
\gamma_3^3(\vp_1,\vp_2)&=&\frac{3m\lambda^{3/2}}{4\sqrt{2}}\pinv{d}{p\p}
\frac{1}{\ovpp{}\omega_{\vp_1+\vpp}\left(\ovp 1+\ovpp{}+\omega_{\vp_1+\vpp}\right)}\\
&&\times \frac{1}{\left(\ovp 2+\ovpp{}+\omega_{\vp_1+\vpp}+\omega_{\vp_1+\vp_2}\right)}\left(
1-\frac{6m^2}{\omega_{\vpp+\vp_2}\left( \ovpp{}+\ovp 2+\omega_{\vpp+\vp_2}
\right)}
\right)\nonumber
\eea
where it is understood that we only use the irreducible part of $\gamma_3^3$, given in Eq.~(\ref{gamma}).  In other words
\bea
\Delta
&=& -\frac{9m\lambda^{3/2}}{4\sqrt{2}}\pinv{d}{p\p}
\frac{1}{\ovpp{}\omega_{\vpp+\vp_1}\left(\ovpp{}+\omega_{\vpp+\vp_1}+\ovp 1\right)}
\label{delta}\\
&&\hspace{-.7cm}
\times\left[1+\frac{ \left(
\ovp 1+\ovpp{}+\omega_{\vp_1+\vpp}\right)+
\left(\ovp 1+\ovp 2+\omega_{\vp_1+\vp_2}\right)
}{\left(\ovp 2+\ovpp{}+\omega_{\vp_1+\vpp}+\omega_{\vp_1+\vp_2}\right)}\left(
1-\frac{6m^2}{\omega_{\vp_2+\vpp}\left( \ovp 2+\ovpp{}+\omega_{\vp_2+\vpp}
\right)}
\right)\right].\nonumber
\eea
Note that at large $\vpp$, the integrand tends to $-9m\lambda^{3/2}/(4\sqrt{2}\vp^{\prime 3})$ as found in our previous paper.

In the case of the mass renormalization, $\delta m^2$ apparently depended on $\vp$ but in fact, upon performing the integral, one could see that it was independent.  In this case, $\Delta$ and so $\delta\sl$ really does depend on $\vp_1$, $\vp_2$ and $\vp_3$ although the large $\vpp$ asymptotics do not.  Therefore, one needs to fix the renormalization at some value of these external momenta.

Any choice will remove the ultraviolet divergences.  By analogy with Peskin and Schroeder's (10.40) we choose to impose the renormalization condition at
\beq
\vp_1=-\vp_2.
\eeq
\red{Check that the resulting $\Delta$ is independent of $\vp_1$ in all dimensions.  It probably isn't, maybe this only works in Peskin and Schroeder because the 0-momentum particle is also massless so it keeps a zero 4-momentum under boosts which change $\vp_1$ to an arbitrary value.  In the Yukawa coupling with massive particles Coleman uses a totally different convention, but with complex momenta, aack.  So this isn't a good choice of $\vp_1$ and $\vp_2$, maybe we should just set them all to zero?}

\end{document}

Solutions of the linearized equations of motion of a classical field theory, which are superpositions of plane waves, correspond to the perturbative Fock space of multimeson states in the corresponding quantum field theory.  In Ref.~\cite{skyrme}, Skyrme suggested that even intrinsically nonlinear solutions of a classical field theory, which we will loosely call solitons, correspond to states in a quantum field theory.  Skyrme's suggestion was eventually transformed into a series of concrete formalisms \cite{dhn2,gs74,cl75,tom75,fk77,cq1,cq2,gw22} for treating solitons in quantum field theories.  

It is generally believed \cite{vinc72,cornwall74} that solitons correspond to approximately coherent states in quantum field theories.  Finding the soliton states and their quantum corrections is important for various applications.  In the case of perturbative excitations, one is usually not interested in quantum corrections to states because the LSZ formula allows one to calculate the S-matrix using only the uncorrected states. However no LSZ formula is known for scattering in a soliton sector.  In the case of perturbative states the spectrum can be calculated using standard interaction picture perturbation theory.  But in the case of soliton states, this is complicated by the zero modes, which imply that normal ordered products of operators have nonvanishing and in general divergent expectation values on the soliton ground state.  

Systematic approaches to finding quantum corrections to these coherent states were first considered at order $O(1)$ in Ref.~\cite{taylor78}, at one loop in Ref.~\cite{mekink} and beyond in Ref.~\cite{me2loop}.  Recently it has been appreciated that these quantum corrections have important consequences.  For example they may drastically increase the lifetime of the oscillon \cite{noiosc} and they also eliminate some divergences \cite{cocorr23}.  The present work concerns this last point.

In Ref.~\cite{erice}, Coleman noted that the coherent state construction, although it works wonderfully in 1+1 dimensions, leads to an infinite energy density in higher dimensions.  He called upon the students at the Erice school to find the correct construction of soliton states in higher dimensions.  In the present paper we answer this call, solving the simplest manifestation of this problem.  

In the (3+1)-dimensional $\phi^4$ double-well theory, the coherent state domain wall tension is divergent already at one loop.  The leading quantum correction to the coherent states was already implicit in Ref.~\cite{cahill76}.  In the true vacuum $\ovac$ all perturbative excitations must be in their ground state, so that $A_p\ovac_0=0$ where $A_p$ is the meson annihilation operator which creates plane waves and $\ovac_0$ is the leading order approximation to the true vacuum.  However in the soliton ground state, the perturbative excitations are not plane waves but rather are normal modes.  Thus the ground state must be annihilated by the operators $B_k$ that annihilate normal modes, as well as various momentum operators for zero modes.  These are related to the $A_p$ by a Bogoliubov transformation \cite{wentzel}.  A Bogoliubov transformation is implemented by a squeeze operator.  And so it is known that soliton states are not ordinary coherent states, but rather are squeezed coherent states.  The squeeze is the leading order correction to the coherent state, and all other corrections are suppressed by powers of the coupling.

Our main result is that the squeeze itself is sufficient to remove the one-loop divergence in the tension of the domain wall in the $(3+1)-$dimensional double-well model.  We announced this result already in Ref.~\cite{noi4dlett}.  However, a full derivation was not presented.  The present paper handles several important issues not treated there.  For example, the original theory is normal ordered at the bare mass scale, but this needs to be transformed to a finite mass scale.  The change in renormalization condition leads to new divergences.  We show in the present paper that these divergences only appear beyond one loop.  Furthermore, in Ref.~\cite{noi4dlett} we did not define the renormalization conditions, but rather simply wrote down the divergent parts of the mass and coupling counterterms that we claimed would lead to finite quantities.  In the present paper, renormalization conditions are chosen and imposed order by order in the vacuum sector.  

This computation is performed in the Schrodinger picture, which makes it far more complicated.  However, as stressed in Ref.~\cite{fk77}, calculations in the vacuum sector that are equal to those in the soliton sector in the ultraviolet imply that the cancellation of ultraviolet divergences in the vacuum sector is equivalent to the cancellation of those divergences in the soliton sector.  Thus, the seemingly masochistic calculations below in fact bring the added value of ensuring the finiteness of the corresponding quantities in the domain wall sector.

We also show that infrared divergences cancel, which was not shown in Ref.~\cite{noi4dlett}.  This cancellation is particularly delicate as even a small mismatch, when integrated over all of space in the vacuum sector, would lead to an infinite energy.

We begin in Sec.~\ref{teorsez} by reviewing the double-well model and its spontaneous symmetry breaking.  In Sec.~\ref{vacsez} we renormalize the ultraviolet divergences that arise at one loop by introducing counterterms for the overall energy, the mass and the coupling constant.  Wave function renormalization, which makes this theory trivial, is only necessary to remove the divergences at two loops and so will not appear here.  Once the counterterms are fixed, the renormalized mass and coupling are fixed, and our freedom is more or less exhausted, except for a choice of displacement operator to construct the coherent state\footnote{Even here, subleading corrections to the displacement operator will be compensated by the perturbative corrections to the state, and so there is no true freedom.  In fact, in Ref.~\cite{noi4dlett} the displacement operator was constructed using the bare parameters and here it is constructed using the renormalized parameters and, nonetheless, our results are consistent.}.  We therefore proceed to calculate the one-loop correction to the mass of the domain wall soliton in Sec.~\ref{wallsez}.

\section{The Theory} \label{teorsez}

\subsection{Definitions}

We will be interested in a theory of a scalar field $\phi(\vx)$ and its conjugate momentum $\pi(\vx)$ in 3+1 dimensions.  It is described by the Hamiltonian
\beq
\hat H=\int d^3\vx:\hat\ch^0(\vx):_{m_0}\hsp
\hat\ch^0(\vx)=\frac{\pi^2(\vx)+\sum_{i=1}^3\left(\partial_i \phi(\vx)\right)^2}{2}+\frac{\lambda_0 \phi^4(\vx)-m_0^2 \phi^2(\vx)}{4}+A
\eeq
where the normal ordering $::_{m_0}$ is defined at the mass  $m_0$.  More precisely, one uses the Schrodinger picture decomposition
\bea
\phi(\vx)&=&\pinv{3}{p}e^{-i\vp\cdot\vx}\left(A^{(0)\ddag}_\vp+\frac{A^{(0)}_{-\vp}}{2\omega^0_\vp}\right)\hsp 
\pi(\vx)=i\pinv{3}{p}e^{-i\vp\cdot\vx}\left(\omega^0_\vp A^{(0)\ddag}_\vp-\frac{A^{(0)}_{-\vp}}{2}\right)\nonumber\\
\omega^0_{\vp}&=&\sqrt{m_0^2+p^2}\hsp A^{(0)\ddag}_\vp=\frac{A^{(0)\dag}_\vp}{2\omega^0_\vp}
\eea
and places all $A^{(0)\ddag}$ to the left of the $A^{(0)}$.  The $c$-number $A$ is a counterterm that will be fixed momentarily.

Now we will introduce renormalized quantities $m$ and $\lambda$ via the definitions
\beq
m^2=m_0^2+\delta m^2\hsp \sl=\sqrt{\lambda_0}+\delta\sl \label{ct}
\eeq
where the counterterms $\delta m^2$ and $\delta\sl$ are taken to be of order $O(\lambda)$ and $O(\lambda^{3/2})$\gre{why we can not said $\delta\sl$ is $0(\lambda)$ because $\sl$ is $0(\sl)$}. \red{The logic here is as follows.  I keep the minimum number of counterterms that I will need to satisfy the renormalization conditions.  So first I guess which counterterms I'll need.  Then I impose the renormalization conditions and check to see if they can be satisfied with these counterterms.  If they can't be satisfied, then I need to go back and add some more.  So here the claim is that you won't need an order $O(\lambda)$ contribution to satisfy the renormalization conditions.  This is shown later in the draft.  But here it is too early, because I haven't even written the renormalization conditions yet.}
\gre{So I think we should maybe normal order after the  shift to determine the contribution of the effect of normal order in quantum correction level of different order }In principle, these may contain higher order contributions.  However this order is sufficient for the calculations in the present note.  We will also expand the counterterm $A$ in powers of $\lambda$
\beq
A=\sum_{i=0}^\infty A_i
\eeq
where $A_i$ is of order $O(\lambda^{i/2-1})$.

It is known that this theory is trivial in the sense that if $\lambda_0$ and $m_0$ are fixed, as one takes the ultraviolet cutoff $\Lambda\rightarrow\infty$, wave function renormalization makes the theory free \cite{froh82,ai21}.  While we are of course interested in the behavior as $\Lambda$ increases, we will always hold $\lambda$ and $m$ fixed while varying $\Lambda$.  This leads to a divergence in $\lambda_0$ and $m_0$ in the strict $\Lambda\rightarrow\infty$ limit.  Be this as it may, we are only interested in the question of whether the domain wall tension is bounded as $\Lambda$ increases with $\lambda$ and $m$ fixed.

\subsection{Changing the Normal Ordering}

Instead of normal ordering at the scale $m_0$, it will be convenient to normal order at the scale $m$, which remains bounded as the ultraviolet cutoff $\Lambda$ is taken to infinity.  Now, one uses the Schrodinger picture decomposition
\bea
\phi(\vx)&=&\pinv{3}{p}e^{-i\vp\cdot\vx}\left(A^{\ddag}_\vp+\frac{A_{-\vp}}{2\omega_\vp}\right)\hsp 
\pi(\vx)=i\pinv{3}{p}e^{-i\vp\cdot\vx}\left(\omega_\vp A^{\ddag}_\vp-\frac{A^{}_{-\vp}}{2}\right)\nonumber\\
\omega_{\vp}&=&\sqrt{m^2+p^2}\hsp A^{\ddag}_\vp=\frac{A^{\dag}_\vp}{2\omega_\vp}
\eea
and places all $A^{\ddag}$ to the left of the $A^{}$.  The corresponding Hamiltonian density is defined by
\beq
\hat H=\int d^3\vx:\hat\ch(\vx):_{m}.
\eeq

To evaluate $\hat\ch(\vx)$, we use the identities
\bea
 :\pi^2(\vx):_{m_0}&=&:\pi^2(\vx):_{m}+\frac{1}{2}\pinv{3}{p}\left({}{\omega_p}-{}{\omega^0_p}\right)\\
 \sum_j:\partial_j\phi(\vx)\partial_j\phi(\vx):_{m_0}&=& \sum_j:\partial_j\phi(\vx)\partial_j\phi(\vx):_{m}+\frac{1}{2}\pinv{3}{p}\left(\frac{p^2}{\omega_p}-\frac{p^2}{\omega^0_p}\right)\nonumber\\
:\phi^4(\vx):_{m_0}&=&:\phi^4(\vx):_{m}+6I:\phi^2(\vx):_m+3I^2\hsp :\phi^2(\vx):_{m_0}=:\phi^2(\vx):_{m}+I\nonumber\\
I&=&\frac{1}{2}\pinv{3}{p}\left(\frac{1}{\omega_p}- \frac{1}{\omega^0_p}\right)
\nonumber
\eea
to derive
\beq
\hat\ch(\vx)=\hat\ch^0(\vx)+\frac{3\lambda_0}{2}I\phi^2(\vx)+\frac{3\lambda_0}{4}I^2-\frac{3m_0^2}{4}I
+\frac{1}{4}\pinv{3}{p}\frac{\left(\omega_p-\omega^0_p\right)^2}{\omega_p}. \label{hh}
\eeq

Note that the momentum integrals are divergent.  They require ultraviolet cutoffs at a scale~$\Lambda$.  We always consider the large $\Lambda$ limit after the small $\lambda$ limit, so that it does not affect our power counting.  More precisely, we consider the limit such that $\lambda\Lambda^N$ goes  to zero for all $N$.

Let us count the powers of $\lambda$ in the new terms.  First, the $\lambda_0$ consists of $\lambda$ plus higher order terms, and so it is of at least order $O(\lambda)$.  Next, $I$ is an integral of an expression that is proportional to $\delta m^2$ plus higher powers of $\delta m^2$, and all terms are at least of the order of $\delta m^2$, which is of order $O(\lambda)$.  Finally, the expression $\sqrt{m^2+p^2}-\sqrt{m_0^2+p^2}$ again consists of powers of $\delta m^2$, of which each is of order at least $O(\lambda)$.  We thus conclude that the fifth term on the right hand side of Eq.~(\ref{hh}) is of order at least $O(\lambda^2)$, while the third begins at order $O(\lambda^3)$.  In the one loop treatment in this paper, we will not need any terms of order greater than $O(\lambda^{3/2})$, corresponding to $\delta\sl$.  Therefore we may and will simply drop these terms in the rest of this note.   

What about the second term?  Its coefficient is of order $O(\lambda^2)$, however it depends on $\phi(\vx)$ whose order is $O(1/\sl)$ as a result of the spontaneous symmetry breaking.  Thus the second term is of order $O(\lambda)$ overall and we need to keep it.  We also need to keep the fourth term, as it is of order $O(\lambda).$

Inserting (\ref{ct}) into the Hamiltonian density, we may now write it in terms of renormalized quantities and counterterms
\bea
\hat\ch(\vx)&=&\frac{\pi^2(\vx)+\sum_{i=1}^3\left(\partial_i \phi(\vx)\right)^2}{2}+\frac{\lambda \phi^4(\vx)-m^2 \phi^2(\vx)}{4}-\frac{3m^2I}{4}\label{hhp}\\
&&\hspace{-1cm}+\frac{\left(-2\sl\delta\sl+(\delta\sl)^2 \right) \phi^4(\vx)+\left(\delta m^2+6\left(\sl-\delta\sl\right)^2 I\right) \phi^2(\vx)}{4}+A
\nonumber
\eea
where the second line contains the counterterms.
\subsection{Spontaneous Symmetry Breaking}

This theory has two vacua, which are not located at $\langle\phi\rangle=0$.  This is inconvenient for perturbation theory, which is most easily written as an expansion in moments of $\phi(\vx)$.  However, we can shift $\langle\phi\rangle$ using the displacement operator
\beq
\dv={\rm{exp}}\left(-iv\dvx\ \pi(\vx)\right)
\eeq
which satisfies the relation
\beq
\phi(\vx)\dv=\dv\left(\phi(\vx)+v\right).
\eeq
We can use the unitary operator $\dv$ to transform the Hilbert space.  

We do not wish to change our theory, merely to redefine the fields and states so that one vacuum, which corresponds to the classical solution $\phi(x,t)=-m_0/\sqrt{2\lambda_0}$, now lies at the origin.  

As usual, such a passive transformation requires one to transform all operators as well.   More precisely, our passive transformation is defined as follows:  We act on all states with the operator $\dv$ and we conjugate all operators by $\dv$.  Therefore, in the new frame, time evolution is generated by the Hamiltonian density
\beq
H[\phi(\vx),\pi(\vx)]=\dv^\dag \hat H[\phi(\vx),\pi(\vx)]\dv=\hat H[\phi(\vx)+v,\pi(\vx)].
\eeq

Note that conjugation with $\dv$ only shifts fields by $c$-numbers and so does not affect the normal ordering.  Therefore, one may transform $\hat\ch$ to $\ch$ by simply replacing $\phi(\vx)$ with $\phi(\vx)+v$
\bea
\ch(\vx)&=&\frac{\pi^2(\vx)+\sum_{i=1}^3\left(\partial_i \phi(\vx)\right)^2}{2}+\frac{\lambda (\phi(\vx)+v)^4-m^2 (\phi(\vx)+v)^2}{4}\label{chsb}\\
&&+\frac{\left(-2\sl\delta\sl+(\delta\sl)^2 \right) (\phi(\vx)+v)^4}{4}\nonumber\\
&&+\frac{\left(\delta m^2+6\left(\sl-\delta\sl\right)^2 I\right)  (\phi(\vx)+v)^2}{4}-\frac{3
m^2
I}{4}+A.\nonumber
\eea

Similarly to the counterterms, we will expand $v$ as

\beq
v=-\frac{m}{\sqrt{2\lambda}}+\delta v\hsp \delta v=\sum_{i=1}^\infty \delta v_i
\eeq
where $\delta v_i$ is of order $O(\lambda^{i/2})$.  Therefore our Hamiltonian density is
\bea
\ch(\vx)&=&\frac{\pi^2(\vx)+\sum_{i=1}^3\left(\partial_i \phi(\vx)\right)^2}{2}+\frac{\lambda \left(\phi(\vx)-\frac{m}{\sqrt{2\lambda}}+\delta v\right)^4-m^2 \left(\phi(\vx)-\frac{m}{\sqrt{2\lambda}}+\delta v\right)^2}{4}\nonumber\\
&&+\frac{\left(-2\sl\delta\sl+(\delta\sl)^2 \right) \left(\phi(\vx)-\frac{m}{\sqrt{2\lambda}}+\delta v\right)^4}{4}\nonumber\\
&&+\frac{\left(\delta m^2+6\left(\sl-\delta\sl\right)^2 I\right)  \left(\phi(\vx)-\frac{m}{\sqrt{2\lambda}}+\delta v\right)^2}{4}+\frac{-3
m^2
I}{4}+A.\label{hpad}
\eea
Notice that in the last line the $3m^2I/4$ in the first term cancels a $-3m^2I/4$ in the second term.  These terms arose from the fact that we shifted the normal ordering when expanding about a maximum in the potential, and would never have appeared had we first expanded about the true vacuum and then changed the normal ordering.

\subsection{Expanding the Hamiltonian}

We will expand the Hamiltonian and Hamiltonian density in powers of $\lambda$
\beq
H=\sum H_i\hsp H_i=\dvx :\ch_i(\vx):_m
\eeq
where $H_i$ and $\ch_i(\vx)$ are of order $O(\lambda^{i/2-1})$.  We will set $A_1$ to zero, and we will check that this choice is consistent with our renormalization conditions at each order.

At leading order we find
\beq
\ch_0(\vx)=-\frac{m^4}{16\lambda}+A_0. \label{ch0}
\eeq
Next, a potentially dangerous  tadpole arises at nonperturbative order $O(1/\sl)$.  However, with the choice $A_1=0$, 
$\ch_1(\vx)=0$.  The last order that is not perturbatively suppressed by powers of $\sl$ is
\beq
\ch_2(\vx)=\frac{\pi^2(\vx)+\sum_{i=1}^3\left(\partial_i \phi(\vx)\right)^2+m^2\phi^2(\vx)}{2}+\frac{m^4}{8\lambda}\left(-\frac{\delta\sl}{\sl}+\frac{\delta m^2}{m^2}\right)+A_2.
\eeq

The lowest order interaction term is
\beq
\ch_3(\vx)=-m\sqrt{\frac{\lambda}{2}}\phi^3(\vx)+\phi(\vx)\left[m^2\delta v_1+\frac{m^3}{\sqrt{2\lambda}}\left(\frac{\delta\sl}{\sqrt{\lambda}}-\frac{\delta m^2}{2m^2}\right)\right]+A_3.
\eeq
Note that the $\delta\sl$ and $\delta m^2$ diverge as the cutoff $\Lambda$ tends to infinity.  This divergence is not necessarily a problem, as the counterterms, like the bare terms, cannot be measured.  However, if $\delta v_1$ diverges, then the two vacuum sectors are infinitely separated and one may wonder whether the domain walls that we are studying exist.  It will be reassuring later that we will find that $\delta v_1$ in fact remains bounded as $\Lambda$ tends to infinity.

The next to leading order interaction is
\bea
\ch_4(\vx)&=&\frac{\lambda}{4}\phi^4(\vx)+\left[-\frac{3m\sl\delta v_1}{\sqrt{2}}-\frac{3m^2\delta\sl}{2\sl}+\frac{\delta m^2}{4} \right]\phi^2(\vx)\\
&&\hspace{-1cm}+{m^2\delta v_2}{}\phi(\vx)+\frac{m^4\left(\delta\sl\right)^2}{16\lambda^2}+\frac{m^3\delta v_1}{\sqrt{2\lambda}}\left[ 
\frac{\sl}{\sqrt{2}m}\delta v_1+\frac{\delta\sl}{\sl}-\frac{\delta m^2}{2m^2}\right]+A_4.\nonumber
\eea
The highest order interaction that we will need is
\bea
\ch_5(\vx)&=&\left[\lambda\delta v_1+\sqrt{2}m\delta\sl\right]\phi^3(\vx)+\left[-\frac{3m\sl\delta v_2}{\sqrt{2}} \right]\phi^2(\vx)\\
&&\hspace{-1.5cm}+\left[m^2\delta v_3-3m\sqrt{\frac{\lambda}{2}}\left(\delta v_1\right)^2-3\frac{m^2\delta\sl\delta v_1}{\sl} -\frac{m^3\left(\delta\sl\right)^2}{2\sqrt{2}\lambda^{3/2}}+\frac{\delta m^2\delta v_1}{2}-\frac{3m\sl I}{\sqrt{2}}\right]\phi(\vx)\nonumber\\
&&\hspace{-1.5cm}+\frac{m^3\delta v_2}{\sqrt{2\lambda}}\left[\frac{\sqrt{2\lambda}}{m}\delta v_1 +\frac{\delta\sl}{\sl}
-\frac{\delta m^2}{2m^2}\right]+A_5.\nonumber
\eea

\section{Renormalizing the Vacuum Sector} \label{vacsez}

\subsection{Renormalization Conditions} \label{rcsez}

The Hamiltonian $\hat{H}$ contained three counterterms: $A$, $\delta m^2$ and $\delta\sl$.  The shift to $H$ also depended on a free parameter $\delta v$.  To fix these four quantities, we need four renormalization conditions.  We choose the following
\begin{enumerate}
  \item \hypertarget{first}{$H\ovac=0$}
  \item \hypertarget{second}{$H|\vp\rangle=\ovp{}|\vp\rangle$} 
  \item \hypertarget{third}{${}_0\langle \vp_1 \vp_2|\vp_0\rangle_{i\neq 1}=0$}
  \item \hypertarget{fourth}{$\langle\Omega|\phi(\vx)|\Omega\rangle=0$}.
\end{enumerate}
Here $\ovac$ is the vacuum. $|\vp_1\cdots \vp_n\rangle$ is an $n$-meson state which is simultaneously a Hamiltonian eigenstate and also a momentum eigenstate with momentum $\sum \vp_i$.  Each state is expanded
\beq
|\Psi\rangle=\sum_{i=0}^\infty |\Psi\rangle_i
\eeq
where $|\Psi\rangle_i$ is of order $O(\lambda^{i/2})$.  We define $\ovac$ to be the Hamiltonian eigenstate such that
\beq
A_\vp\ovac_0=0.
\eeq
The states are normalized by the three conditions
\renewcommand{\labelenumi}{\Roman{enumi})}
\begin{enumerate}
\item ${}_0\langle \Omega|\Omega\rangle_0=1$ 
  \item $|\vp_1\cdots \vp_n\rangle_0=A^\ddag_{\vp_1}\cdots A^\ddag_{\vp_n}|\Omega\rangle_0$  
  \item 
   \hypertarget{terzo}{${}_0\langle \vp_1\cdots \vp_m|\vp_1\cdots \vp_m\rangle_{n>0}=0.$}
\end{enumerate}
However the renormalization conditions are independent of the normalization.

Note that the renormalization conditions are expressed in terms of states in the vacuum sector Fock space.  Intuitively, we demand a cancellation of divergences in the vacuum sector.  Once this is done, the four parameters are fixed.  The Hamiltonian eigenstates in the domain wall sector are then determined, and there is no more freedom that can be used to eliminate any domain wall sector divergences.

This Schrodinger picture renormalization is far less efficient then the usual interaction picture renormalization.  We use it because the Schrodinger picture calculations are identical in the ultraviolet to corresponding calculations in the domain wall sector.  The general argument of Ref.~\cite{fk77} then implies that ultraviolet finiteness in the vacuum sector of a given quantity will lead to finiteness in the domain wall sector of the same quantity.

\subsection{Order $O(1)$ and below}

At order $O(1/\lambda)$ we consider $\ch_0$ given in Eq.~(\ref{ch0}).  The \hyperlink{first}{first} renormalization condition implies
\beq
A_0=\frac{m^4}{16\lambda}\hsp H_0=0. \label{a0}
\eeq
As $\ch_1$ vanishes, the renormalization conditions are trivially satisfied at order $O(1/\sl)$.

To proceed to order $O(1)$, note that the free Hamiltonian may be written
\beq
H_2=\pinv{3}{p}\omega_{\vp}A^\ddag_\vp A_\vp+\dvx\left[ \frac{m^4}{8\lambda}\left(-\frac{\delta\sl}{\sl}+\frac{\delta m^2}{m^2}\right)+A_2\right].
\eeq
The \hyperlink{first}{first} renormalization condition at $O(1)$ then implies
\beq
A_2= \frac{m^4}{8\lambda}\left(\frac{\delta\sl}{\sl}-\frac{\delta m^2}{m^2}\right)\hsp H_2=\pinv{3}{p}\omega_{\vp}A^\ddag_\vp A_\vp. \label{a2}
\eeq
This automatically also satisfies the \hyperlink{second}{second} renormalization condition at order $O(1)$.  The \hyperlink{third}{third} renormalization condition is trivially satisfied at this order as
\beq
{}_0\langle \vp_1 \vp_2|\vp_3\rangle_0=4\ovp 1\ovp 2 {}_0\langle\Omega|A_{\vp_1}A_{\vp_2} A^{\ddag}_{\vp_3}\ovac_0=4\ovp 1\ovp 2 {}_0\langle\Omega|[A_{\vp_1}A_{\vp_2}, A^{\ddag}_{\vp_3}]\ovac_0=0
\eeq
which vanishes as
\beq
[A_{\vp_1}, A^{\ddag}_{\vp_2}]=(2\pi)^3\delta^3(\vp_1-\vp_2)
\eeq
and $A\ovac_0=0$.  Similarly the \hyperlink{fourth}{fourth} renormalization condition is satisfied at this order
\beq
{}_0\langle\Omega|\phi(\vx)\ovac_0=\pinv{3}{p}e^{-i\vp\cdot\vx}\left({}_0\langle\Omega|A^{\ddag}_\vp\ovac_0+\frac{{}_0\langle\Omega|A_{-\vp}\ovac_0}{2\omega_\vp}\right)=0.
\eeq

\subsection{Order $O(\sl)$}

The vacuum sector can be decomposed into $n$-meson Fock spaces, which are spanned respectively by the states
\beq
|\vp_1\cdots \vp_n\rangle_0=A^\ddag_{\vp_1}\cdots A^\ddag_{\vp_n}\ovac_0.
\eeq
The operator $H_2$ preserves $n$.  For any given state $|\psi\rangle$ in the vacuum sector, we will write this decomposition as
\beq
|\psi\rangle=\sum_n |\psi\rangle^n
\eeq
where $|\psi\rangle^n$ lies in the $n$-meson Fock space, and so is a linear combination of the states $|\vp_1\cdots \vp_n\rangle_0$.

At order $O(\sl)$ the \hyperlink{first}{first} renormalization condition is
\beq
H_2\ovac_1=-H_3\ovac_0. \label{slcond}
\eeq
Let us impose this condition in each $n$-meson Fock space.

First, note that the zero-meson Fock space consists of $\ovac_0$ which is annihilated by $H_2$.  Therefore, restricting to the zero-meson Fock space, the left hand side vanishes, and so the right hand side must also vanish.  The right hand side, restricted to the zero-meson Fock space, is $A_3\ovac_0$ and so we learn
\beq
A_3=0.
\eeq

\begin{figure}[htbp]
\centering
\includegraphics[width = 0.35\textwidth]{vac1.eps}
\caption{This graph represents the calculation of $\ovac_1^3$, which is $-H_2^{-1}H_3\ovac_0^0$.  The vertex represents the operator $-H_2^{-1}H_3$.    The order $i$ of $\ovac_i$ increases as one moves to the left.  Consider a vertical slice.  It intersects some number of lines $n$.  This represents the $n$-meson Fock space.  To the right of the vertex there are no lines, reflecting the fact that $\ovac_0=\ovac_0^0$ lives in the zero-meson Fock space.  To the left, there are three lines, as we are calculating a contribution $\ovac_1^3$ to $\ovac_1$ in the three-meson Fock space.} \label{v1fig}
\end{figure}

To shorten expressions, let us introduce the notation
\beq
\delta v\p_1=\delta v_1+\frac{m}{\sqrt{2\lambda}}\left(\frac{\delta\sl}{\sqrt{\lambda}}-\frac{\delta m^2}{2m^2}\right). 
\eeq
Then
\beq
H_3\ovac_0=-m\sqrt{\frac{\lambda}{2}}\pinv{3}{p_1}\pinv{3}{p_2}|\vp_1,\vp_2,-\vp_1-\vp_2\rangle_0+m^2\delta v\p_1 |\vp=\Vec{0}\rangle_0.
\eeq
\par 
Now we can solve (\ref{slcond}) by inverting $H_2$, where the inverse is unique once normalization condition (\hyperlink{terzo}{III}) in Subsec.~\ref{rcsez} is imposed
\beq
\ovac_1=m\sqrt{\frac{\lambda}{2}}\pinv{3}{p_1}\pinv{3}{p_2}\frac{|\vp_1,\vp_2,-\vp_1-\vp_2\rangle_0}{\ovp 1+\ovp 2+\omega_{\vp_1+\vp_2}}-m\delta v\p_1 |\vp=\Vec{0}\rangle_0. \label{vac1}
\eeq
With this perturbative correction to $\ovac$, the \hyperlink{first}{first} renormalization condition is solved.

We will introduce a graphical representation of such quantum corrections to states, which will be helpful later when we want to specify which contributions we are calculating.  In the case of Eq.~(\ref{vac1}), the graphical representation is in Fig.~\ref{v1fig}.  The perturbative corrections $|\psi\rangle_i$ are calculated with $i$ increasing as one moves along the arrows, which we will point to the left.  Each line represents a meson and each vertex represents $(E-H_2)^{-1}H_j$ for some $j$.  Therefore, if there are $n$ incoming mesons and $m$ outgoing mesons, the graph describes a contribution to $|\vp_1\cdots \vp_n\rangle^m$.

Let us skip to the fourth condition.  At order $O(\sl)$
\beq
\langle \Omega| \phi(\vx) \ovac=2\ {}_0\langle \Omega| \phi(\vx) \ovac_1=-\delta v\p_1
\eeq
and so we learn
\beq
\delta v\p_1=0\hsp
\delta v_1=\frac{m}{\sqrt{2\lambda}}\left(-\frac{\delta\sl}{\sqrt{\lambda}}+\frac{\delta m^2}{2m^2}\right)\hsp \ch_3(\vx)=-m\sqrt{\frac{\lambda}{2}}\phi^3(\vx). \label{v1}
\eeq

Now we are ready for the \hyperlink{second}{second} renormalization condition.  To make the calculation faster, we will decompose the vacuum sector states by meson number $n$
\beq
|\Psi\rangle=\sum_n |\Psi\rangle^n\hsp |\Psi\rangle_i=\sum_n |\Psi\rangle_i^n
\eeq
which is defined as the number of $A^\ddag$ operators that act on $\ovac_0$.  We will also decompose terms in the Hamiltonian by meson number, defined so that the total meson number of an operator plus a state is conserved when an operator acts on a state.  In the case of the leading interaction, this decomposition is
\bea
H_3&=&\sum_{n=0}^3 H_3^{3-2n}\hsp
H_3^{3}=-\frac{m\sl}{\sqrt{2}}\pinv{3}{p_1}\pinv{3}{p_2} \Ad 1\Ad 2 A^\ddag_{-\vp_1-\vp_2}\\
H_3^{1}&=&-\frac{3m\sl}{2\sqrt{2}}\pinv{3}{p_1}\pinv{3}{p_2} \frac{\Ad 1\Ad 2 A_{\vp_1+\vp_2}}{\omega_{\vp_1+\vp_2}}\nonumber\\
H_3^{-1}&=&-\frac{3m\sl}{4\sqrt{2}}\pinv{3}{p_1}\pinv{3}{p_2} \frac{\Ad 1 A_{-\vp_2} A_{\vp_1+\vp_2}}{\ovp 2\omega_{\vp_1+\vp_2}}\nonumber\\
H_3^{-3}&=&-\frac{m\sl}{8\sqrt{2}}\pinv{3}{p_1}\pinv{3}{p_2} \frac{A_{-\vp_1} A_{-\vp_2} A_{\vp_1+\vp_2}}{\ovp 1\ovp 2\omega_{\vp_1+\vp_2}}.
\eea

\begin{figure}[htbp]
\centering
\includegraphics[width = 0.6\textwidth]{mes1.eps}
\caption{This is a graphical representation of $|\vp\rangle_1^4$ (left) and $|\vp\rangle_1^2$ (right).  In each case, to the right of the vertex there is a single meson, as $|\vp\rangle_0=|\vp\rangle_0^1$ is contained in the one-meson Fock space.}\label{m1fig}
\end{figure}

Now we decompose the \hyperlink{second}{second} renormalization condition at meson number $n$
\beq
H_3^{n-1}|\vp\rangle^1_0=(\ovp{}-H_2)|\vp\rangle^n_1.
\eeq
As $|\vp\rangle_0=A^\ddag_\vp\ovac_0$, it contains a single meson and so will be annihilated by $H_3^{-1}$ as a result of the normal ordering.  Thus the left hand side is only nonvanishing at meson numbers $n=2$ and $n=4$.  Remembering that $H_3^{n-1}$ has $2-n/2$ annihilation operators and $|\vp\rangle_0$ contains one creation operator, the commutator yields a factor of the number of contractions, which in this case is unity for both $n=2$, which has one contraction, and $n=4$ which has none.

Altogether we find
\bea
H_3^3|\vp\rangle^1_0&=&-\frac{m\sl}{\sqrt{2}}\pinv{3}{p_1}\pinv{3}{p_2}|\vp,\vp_1,\vp_2,-\vp_1-\vp_2\rangle_0\\
H_3^1|\vp\rangle^1_0&=&-\frac{3m\sl}{2\sqrt{2}\ovp{}}\pinv{3}{p_1}|\vp_1,\vp-\vp_1\rangle_0.\nonumber
\eea
Inverting the $(\ovp{}-H_2)$ one finds
\bea
|\vp\rangle_1^4&=&\frac{m\sl}{\sqrt{2}}\pinv{3}{p_1}\pinv{3}{p_2}\frac{|\vp,\vp_1,\vp_2,-\vp_1-\vp_2\rangle_0}{\ovp 1 +\ovp 2 +\omega_{\vp_1+\vp_2}}\\
|\vp\rangle^2_1&=&\frac{3m\sl}{2\sqrt{2}\ovp{}}\pinv{3}{p_1}\frac{|\vp_1,\vp-\vp_1\rangle_0}{\ovp 1+\omega_{\vp-\vp_1}-\ovp{}}\nonumber
\eea
which is depicted in Fig.~\ref{m1fig}.

The \hyperlink{third}{third} renormalization condition at this order is the case $i=1$.  However the renormalization condition explicitly does not apply to $i=1$, and so it is trivially satisfied.

\subsection{Order $O(\lambda)$}

\subsubsection{The Vacuum State $\ovac$}

At order $O(\lambda)$ the \hyperlink{first}{first} renormalization condition restricted to the $n$-meson Fock space is
\beq
H_4^n\ovac^0_0+H_3^{n-3}\ovac_1^3=-H_2\ovac_2^n. \label{lam1}
\eeq

Let us define the shorthand,
\bea
A_4\p&=&\frac{m^4\left(\delta\sl\right)^2}{16\lambda^2}+\frac{m^3\delta v_1}{\sqrt{2\lambda}}\left[ 
\frac{\sl}{\sqrt{2}m}\delta v_1+\frac{\delta\sl}{\sl}-\frac{\delta m^2}{2m^2}\right]+A_4\nonumber\\
&=&\frac{m^4\left(\delta\sl\right)^2}{16\lambda^2}-\frac{m^2\delta v_1^2}{2}+A_4.
\eea
Then $\ch_4(\vx)$ reduces to
\bea
\ch_4(\vx)&=&\frac{\lambda}{4}\phi^4(\vx)-\frac{\delta m^2}{2} \phi^2(\vx)+m^2\delta v_2\phi(\vx)+A_4\p.
\eea

Let us begin with the one-meson sector $n=1$.  As $H_3^{-2}=0$, on the left hand side we need only act with $\ch_4^1$ yielding
\beq
H_4^1\ovac_0=m^2\delta v_2 |\vp=\vec{0}\rangle_0.
\eeq
Inverting $H_2$ one finds
\beq
\ovac_2^1=-m\delta v_2 |\vp=\vec{0}\rangle_0.
\eeq
This leads to a tadpole
\beq
\langle \Omega| \phi(\vx) \ovac=2\ {}_0\langle \Omega| \phi(\vx) \ovac_2=-\delta v_2.
\eeq
The \hyperlink{fourth}{fourth} renormalization condition then implies
\beq
\delta v_2=0\hsp
\ch_4(\vx)=\frac{\lambda}{4}\phi^4(\vx)-\frac{\delta m^2}{2} \phi^2(\vx)+A_4\p.
\eeq
The fact that $\delta v_2$ is finite implies that the distance between the vacua is finite, which is a consistency check of our calculation thus far.

Now we are ready to decompose $H_4$ by meson number
\bea
H_4&=&\sum_{n=0}^4 H_4^{4-2n}\hsp
H_4^{4}=\frac{\lambda}{4}\pinv{3}{p_1}\pinv{3}{p_2}\pinv{3}{p_3} \Ad 1\Ad 2\Ad 3 A^\ddag_{-\vp_1-\vp_2-\vp_3}\\
H_4^{2}&=&\frac{\lambda}{2}\pinv{3}{p_1}\pinv{3}{p_2}\pinv{3}{p_3} \frac{\Ad 1\Ad 2\Ad 3 A_{\vp_1+\vp_2+\vp_3}}{\omega_{\vp_1+\vp_2+\vp_3}}-\frac{\delta m^2}{2}\pinv{3}{p}\Ad {}A^\ddag_{-\vp}\nonumber\\
H_4^{0}&=&\frac{3\lambda}{8}\pinv{3}{p_1}\pinv{3}{p_2}\pinv{3}{p_3} \frac{\Ad 1\Ad 2A_{-\vp_3} A_{\vp_1+\vp_2+\vp_3}}{\ovp 3\omega_{\vp_1+\vp_2+\vp_3}}-\frac{\delta m^2}{2}\pinv{3}{p}\frac{\Ad {}A_{\vp}}{\ovp{}}+\dvx A\p_4\nonumber\\
H_4^{-2}&=&\frac{\lambda}{8}\pinv{3}{p_1}\pinv{3}{p_2}\pinv{3}{p_3} \frac{\Ad 1 A_{-\vp_2}A_{-\vp_3} A_{\vp_1+\vp_2+\vp_3}}{\ovp 2\ovp 3\omega_{\vp_1+\vp_2+\vp_3}}-\frac{\delta m^2}{8}\pinv{3}{p}\frac{A_{-\vp}A_{\vp}}{\ovp{}^2}\nonumber\\
H_4^{-4}&=&\frac{\lambda}{64}\pinv{3}{p_1}\pinv{3}{p_2}\pinv{3}{p_3} \frac{A_{-\vp_1} A_{-\vp_2}A_{-\vp_3} A_{\vp_1+\vp_2+\vp_3}}{\ovp 1\ovp 2\ovp 3\omega_{\vp_1+\vp_2+\vp_3}}.\nonumber
\eea

We have now assembled all of the ingredients to solve the renormalization condition (\ref{lam1}) one meson number $n$ at a time, with the contributions drawn in Fig.~\ref{v2fig}.  At $n=6$, the only contribution to the left hand side arises from
\beq
H_3^3\ovac^3_1=-\frac{m^2\lambda}{2}\pinv{3}{p_1}\pinv{3}{p_2}\pinv{3}{p_3} \pinv{3}{p_4}\frac{|\vp_1,\vp_2,-\vp_1-\vp_2,\vp_3,\vp_4,-\vp_3-\vp_4\rangle_0}{\ovp 1+\ovp 2+\omega_{\vp_1+\vp_2}}.
\eeq
Inverting $H_2$ one then finds
\beq
\ovac_2^6=\frac{m^2\lambda}{2}\pinv{3}{p_1}\cdots \frac{d^3\vp_4}{(2\pi)^3}\frac{|\vp_1,\vp_2,-\vp_1-\vp_2,\vp_3,\vp_4,-\vp_3-\vp_4\rangle_0}{\left(\ovp 1+\ovp 2+\omega_{\vp_1+\vp_2}\right)\left(\omega_{\vp_1+\vp_2}+\omega_{\vp_3+\vp_4}+\sum_{i=1}^4 \ovp i
\right)}.
\eeq

\begin{figure}[htbp]
\centering
\includegraphics[width = 0.85\textwidth]{vac2.eps}
\caption{We represent the contributions to $\ovac_2^6$ in panel (1), $\ovac_2^4$ in panels $(2)$ and $(3)$, $\ovac_2^2$ in panels $(4)$ and $(5)$ and $\ovac_2^0$ in panels $(6)$ and $(7)$.}\label{v2fig}
\end{figure}

At the lower meson numbers, both $H_3$ and $H_4$ contribute.  At meson number $n=4$ these contributions are
\bea
H_4^4\ovac_0&=&\frac{\lambda}{4}\pinv{3}{p_1}\pinv{3}{p_2}\pinv{3}{p_3}|\vp_1,\vp_2,\vp_3,-\vp_1-\vp_2-\vp_3\rangle_0\\
H_3^1\ovac_1^3&=&-\frac{9m^2\lambda}{4}\pinv{3}{p_1}\pinv{3}{p_2}\pinv{3}{p_3}\frac{|\vp_1,\vp_2,\vp_3,-\vp_1-\vp_2-\vp_3\rangle_0}{\omega_{\vp_1+\vp_2}\left(\ovp 1+\ovp 2 +\omega_{\vp_1+\vp_2}\right)}\nonumber
\eea








where we have included a factor of three in $H_3^1\ovac_1^3$ arising from the three possible contractions of the $A$ in $H_3^1$ with the three $A^\ddag$ operators in $\ovac_1^3$.  Inverting $H_2$ we find 
\beq
\ovac_2^4=\frac{\lambda}{4}\pinv{3}{p_1}\cdots \frac{d^3\vp_3}{(2\pi)^3}\left[\frac{9m^2}{\omega_{\vp_1+\vp_2}\left(\ovp 1+\ovp 2 +\omega_{\vp_1+\vp_2}\right)}-1 
\right]\frac{|\vp_1,\vp_2,\vp_3,-\vp_1-\vp_2-\vp_3\rangle_0}{\ovp 1+\ovp 2+\ovp 3+\omega_{\vp_1+\vp_2+\vp_3}}.
\eeq
The calculation at meson number $n=2$ is similar, with contributions
\bea
H_4^2\ovac_0&=&-\frac{\delta m^2}{2}\pinv{3}{p}|\vp,-\vp\rangle_0\\
H_3^{-1}\ovac_1^3&=&-\frac{9m^2\lambda}{4}\pinv{3}{p}\ppinv{3}{p}\frac{|\vp,-\vp\rangle_0}{\ovpp{}\omega_{\vp+\vpp}\left(\ovp{}+\ovpp{}+\omega_{\vp+\vpp} \right)}\nonumber
\eea
where we have included a factor of six from the possible contractions of the two $A$ operators in $H_3^{-1}$ with the three $A^\ddag$ operators in $\ovac_1^3$.  Note that the $\vpp$ integral is logarithmically divergent.  Any logarithmic divergence in a state could take us out of the vacuum sector Fock space.  The corresponding component of the state is 
\beq
\ovac_2^2=\pinv{3}{p}\left[\frac{\delta m^2}{4}+\frac{9m^2\lambda}{8}\ppinv{3}{p}\frac{1}{\ovpp{}\omega_{\vp+\vpp}\left(\ovp{}+\ovpp{}+\omega_{\vp+\vpp} \right)}
\right]\frac{|\vp,-\vp\rangle_0}{\ovp{}}.
\eeq


We therefore need the expression in the square brackets to be finite.  This constrains $\delta m^2$ to be, up to a finite quantity
\beq
\delta m^2\sim -\frac{9m^2\lambda}{4}\pinv{3}{p}\frac{1}{p^3}+{\rm{finite}}\hsp p=|\vp|. \label{dma}
\eeq
Momentarily we will calculate $\delta m^2$ and check this condition.

Finally we turn to the zero-meson Fock space.  Here the relevant term in $H_4^0$ contains a $\dvx$ of an $\vx$-independent quantity and so is infrared divergent.  Similarly, working as above one would find that $H_3^{-3}\ovac_1^3$ contains a squared Dirac delta, and so it is also divergent.  The right hand side of (\ref{lam1}) vanishes as we have set $\ovac_2^0=0$.

The problem is that we may only perform the $\vx$ integration after summing these terms.  Before the $\vx$ integration, the two contributions are
\bea
\ch^0_4(\vx)\ovac_0&=&A\p_4\ovac_0\\
\ch_3^{-3}(\vx)\ovac_1^3&=&-\frac{3m^2\lambda}{8}\pinv{3}{p_1}\pinv{3}{p_2}\frac{\ovac_0}{\ovp 1\ovp 2\omega_{\vp_1+\vp_2}\left(\ovp1+\ovp2+\omega_{\vp_1+\vp_2}\right)}\nonumber
\eea
where we have included a factor of six in the second term for the $3!$ contractions of the three $A$ operators with the three $A^\ddag$ operators.  

The sum of these two terms needs to be exactly zero, because otherwise the $\vx$ integration will lead to a divergence.  Thus we conclude
\beq
A_4\p=\frac{3m^2\lambda}{8}\pinv{3}{p_1}\pinv{3}{p_2}\frac{1}{\ovp 1\ovp 2\omega_{\vp_1+\vp_2}\left(\ovp1+\ovp2+\omega_{\vp_1+\vp_2}\right)}. \label{a4}
\eeq
This exhibits a quadratic ultraviolet divergence.

\subsubsection{The One-Meson State $|\vp\rangle$}

The \hyperlink{second}{second} renormalization condition
\beq
H_4^{n-1}|\vp\rangle^1_0+H_3^{n-4}|\vp\rangle^4_1+H_3^{n-2}|\vp\rangle^2_1=(\ovp{}-H_2)|\vp\rangle_2^n \label{lamp}
\eeq
yields the subleading correction $|\vp\rangle_2$ to the one-meson state $|\vp\rangle$.

The simplest Fock sector is the seven-meson sector, as the only contribution arises from
\beq
H_3^3|\vp\rangle_1^4=-\frac{m^2\lambda}{2}\pinv{3}{p_1}\cdots \frac{d^3\vp_4}{(2\pi)^3}\frac{|\vp,\vp_1,\vp_2,-\vp_1-\vp_2,\vp_3,\vp_4,-\vp_3-\vp_4\rangle_0}{\ovp 1+\ovp 2+\omega_{\vp_1+\vp_2}}
\eeq
and so
\beq
|\vp\rangle_2^7=\frac{m^2\lambda}{2}\pinv{3}{p_1}\cdots \frac{d^3\vp_4}{(2\pi)^3}\frac{|\vp,\vp_1,\vp_2,-\vp_1-\vp_2,\vp_3,\vp_4,-\vp_3-\vp_4\rangle_0}{\left(\ovp 1+\ovp 2+\omega_{\vp_1+\vp_2}\right)\left(\omega_{\vp_1+\vp_2}+\omega_{\vp_3+\vp_4}+\sum_{i=1}^4 \ovp i
\right)}.
\eeq
This contribution is shown in Fig.~\ref{m257fig}.

\begin{figure}[htbp]
\centering
\includegraphics[width = 0.85\textwidth]{mes2-57.eps}
\caption{The only contribution to $|\vp\rangle_2^7$ is shown in panel $(1)$.  The other panels show the contributions to $|\vp\rangle_2^5$.}\label{m257fig}
\end{figure}

There are four contributions to $|\vp\rangle_2^5$
\bea
H_4^4|\vp\rangle_0^1&=&\frac{\lambda}{4}\pinv{3}{p_1}\pinv{3}{p_2}\pinv{3}{p_3}|\vp,\vp_1,\vp_2,\vp_3,-\vp_1-\vp_2-\vp_3\rangle_0\\
H_3^1|\vp\rangle_1^4&=&-\frac{9m^2\lambda}{4}\pinv{3}{p_1}\pinv{3}{p_2}\pinv{3}{p_3}\frac{|\vp,\vp_1,\vp_2,\vp_3,-\vp_1-\vp_2-\vp_3\rangle_0}{\omega_{\vp_1+\vp_2}\left(\ovp 1+\ovp 2 +\omega_{\vp_1+\vp_2}\right)}\nonumber\\
&&-\frac{3m^2\lambda}{4}\pinv{3}{p_1}\pinv{3}{p_2}\pinv{3}{p_3}\frac{|\vp_3,\vp-\vp_3,\vp_1,\vp_2,-\vp_1-\vp_2\rangle_0}{\ovp{}\left(\ovp 1+\ovp 2 +\omega_{\vp_1+\vp_2}\right)}\nonumber\\
H_3^3|\vp\rangle_1^2&=&-\frac{3m^2\lambda}{4}\pinv{3}{p_1}\pinv{3}{p_2}\pinv{3}{p_3}\frac{|-\vp_3,\vp+\vp_3,\vp_1,\vp_2,-\vp_1-\vp_2\rangle_0}{\ovp{}\left(\ovp 3 +\omega_{\vp+\vp_3}-\ovp{}\right)}.\nonumber
\eea

Altogether these lead to
\bea
|\vp\rangle_2^5&=&\frac{\lambda}{4}\pinv{3}{p_1}\cdots \frac{d^3\vp_3}{(2\pi)^3}\left[
\left(\frac{9m^2}{\omega_{\vp_1+\vp_2}\left(\ovp 1+\ovp 2 +\omega_{\vp_1+\vp_2}\right)}-1
\right)\frac{|\vp,\vp_1,\vp_2,\vp_3,-\vp_1-\vp_2-\vp_3\rangle_0}{\ovp 1+\ovp 2+\ovp 3+\omega_{\vp_1+\vp_2+\vp_3}}
\right.\nonumber\\
&&\hspace{-1.5cm}\left.+\frac{3m^2}{\ovp{}}\left(
\frac{1}{\left(\ovp 1+\ovp 2 +\omega_{\vp_1+\vp_2}\right)}
+\frac{1}{\left(\ovp 3 +\omega_{\vp+\vp_3}-\ovp{}\right)}
\right)\frac{|-\vp_3,\vp+\vp_3,\vp_1,\vp_2,-\vp_1-\vp_2\rangle_0}{\omega_{\vp+\vp_3}+\omega_{\vp_1+\vp_2}-\ovp{}
+\sum_{i=1}^3\ovp i}\right]\nonumber\\
&=&\frac{\lambda}{4}\pinv{3}{p_1}\cdots \frac{d^3\vp_3}{(2\pi)^3}\left[
\left(\frac{9m^2}{\omega_{\vp_1+\vp_2}\left(\ovp 1+\ovp 2 +\omega_{\vp_1+\vp_2}\right)}-1
\right)\frac{|\vp,\vp_1,\vp_2,\vp_3,-\vp_1-\vp_2-\vp_3\rangle_0}{\ovp 1+\ovp 2+\ovp 3+\omega_{\vp_1+\vp_2+\vp_3}}
\right.\nonumber\\
&&\hspace{-0.0cm}\left.+\frac{3m^2}{\ovp{}}\frac{|-\vp_3,\vp+\vp_3,\vp_1,\vp_2,-\vp_1-\vp_2\rangle_0}{\left(\ovp 1+\ovp 2 +\omega_{\vp_1+\vp_2}\right)\left(\ovp 3 +\omega_{\vp+\vp_3}-\ovp{}\right)}\right]
.
\eea


There are five contributions to $|\vp\rangle_2^3$.  The first two arise from $H_4$
\beq
H_4^2|\vp\rangle_0^1=\frac{\lambda}{2\ovp{}}\pinv{3}{p_1}\pinv{3}{p_2}|\vp_1,\vp_2,\vp-\vp_1-\vp_2\rangle_0
-\frac{\delta m^2}{2}\pinv{3}{p_1}|\vp,\vp_1,-\vp_1\rangle_0.
\eeq
The others arise from $H_3$
\bea
H_3^{-1}|\vp\rangle_1^4&=&-\frac{9m^2\lambda}{4}\pinv{3}{p_1}\ppinv{3}{p}\frac{|\vp,\vp_1,-\vp_1\rangle_0}{\ovpp{}\omega_{\vp_1+\vpp}\left(\ovp 1+\ovpp{}+\omega_{\vp_1+\vpp}\right)}
\\
&&-\frac{9m^2\lambda}{4\ovp{}}\pinv{3}{p_1}\pinv{3}{p_2}\frac{|-\vp_1,-\vp_2,\vp+\vp_1+\vp_2\rangle_0}{\omega_{\vp_1+\vp_2}\left(\ovp 1+\ovp 2+\omega_{\vp_1+\vp_2} 
\right)}
\nonumber\\
H_3^{1}|\vp\rangle_1^2&=&-\frac{9m^2\lambda}{4\ovp{}}\pinv{3}{p_1}\pinv{3}{p_2}\frac{|-\vp_1,-\vp_2,\vp+\vp_1+\vp_2\rangle_0}{\omega_{\vp_1+\vp_2}\left(\omega_{\vp_1+\vp_2}+\omega_{\vp+\vp_1+\vp_2}-\ovp{} 
\right)}.\nonumber
\eea

\begin{figure}[htbp]
\centering
\includegraphics[width = 0.85\textwidth]{mes2-3.eps}
\caption{These are the five contributions to $|\vp\rangle_2^3$, the part of the order $O(\lambda)$ correction to the Hamiltonian eigenstate $|\vp\rangle$ which lies in the three-meson Fock space.}\label{m23fig}
\end{figure}

Inverting $\ovp{}-H_2$ we find the leading three-meson contribution to the one-meson Hamiltonian eigenstate
\bea
|\vp\rangle_2^3&=&\pinv{3}{p_1}
\left(\frac{\delta m^2}{4} 
+\frac{9m^2\lambda}{8}\ppinv{3}{p}\frac{1}{\ovpp{}\omega_{\vp_1+\vpp}\left(\ovp 1+\ovpp{}+\omega_{\vp_1+\vpp}\right)}\right)
\frac{|\vp,\vp_1,-\vp_1\rangle_0}{\ovp{1}}\nonumber
\\
&&\hspace{-1cm}+\frac{\lambda}{2\ovp{}}\pinv{3}{p_1}\pinv{3}{p_2}\left[ \frac{9m^2}{2\omega_{\vp_1+\vp_2}}\left(\frac{1}{\ovp 1+\ovp 2+\omega_{\vp_1+\vp_2}}+\frac{1}{\omega_{\vp_1+\vp_2}+\omega_{\vp+\vp_1+\vp_2}-\ovp{}} 
\right)
-1
\right]\nonumber\\
&&\times\frac{|-\vp_1,-\vp_2,\vp+\vp_1+\vp_2\rangle_0}{\left(\ovp 1+\ovp 2+\omega_{\vp+\vp_1+\vp_2}-\ovp{}\right)}
\eea
which is shown in Fig.~\ref{m23fig}.  Note that the ultraviolet divergence on the first line is canceled if Eq.~(\ref{dma}) is satisfied.

There are also five contributions to $|\vp\rangle_2^1$, shown in Fig.~\ref{m21fig}.  Two of these, shown in panels (4) and (5), suffer from infrared divergences, and so we will write them before $\dvx$ integration as
\bea
\ch_4^0(\vx)|\vp\rangle_0^1&\supset& A\p_4|\vp\rangle_0^1\label{dis}\\
\ch_3^{-3}|\vp\rangle_1^4&\supset&-\frac{3m^2\lambda}{8}\ppinv{3}{p_1}\ppinv{3}{p_2} \frac{1}{\ovpp 1\ovpp 2 \omega_{\vpp_1+\vpp_2}\left(\ovpp 1+\ovpp 2+\omega_{\vpp_1+\vpp_2}
\right)}
|\vp\rangle_0^1\nonumber
\eea
where the $\supset$ notation indicates that we have only considered the $c$-number term in $\ch_4$ and only certain contractions with $\ch_3$.  We will consider the other terms momentarily. 

The sum of the two terms in Eq.~(\ref{dis}) is exactly zero as a result of Eq.~(\ref{a4}).  Of course this is no surprise, the role of the counterterm $A_4$ is to cancel disconnected diagrams in the vacuum sector.  Note that the cancellation needs to be and is exact, as any mismatch would diverge when integrated over $\vx$.  In the soliton sector the cancellation will not be exact, even in the case of the domain wall ground state, but it will be $x_1$-dependent and so the integral over $x_1$ will converge.

The other three contributions, on the other hand, diverge in the ultraviolet
\bea
H_4^0|\vp\rangle_0^1&\supset&-\frac{\delta m^2}{2\ovp{}}|\vp\rangle_0
\\
H_3^{-3}|\vp\rangle_1^4&\supset&-\frac{9m^2\lambda}{8\ovp{}}\ppinv{3}{p}\frac{|\vp\rangle_0}{\ovpp{}\omega_{\vp+\vpp}\left(\ovp{}+\ovpp{}+\omega_{\vp+\vpp} \right)}
\nonumber\\
H_3^{-1}|\vp\rangle_1^2&=&-\frac{9m^2\lambda}{8\ovp{}}\ppinv{3}{p}\frac{|\vp\rangle_0}{\ovpp{}\omega_{\vp+\vpp}\left(\ovpp{}+\omega_{\vp+\vpp}-\ovp{} \right)}
\nonumber
\eea
where the $\supset$ symbol indicates that we are considering the terms corresponding to panels $(1-3)$ in Fig.~\ref{m21fig}, which we ignored above.  

\begin{figure}[htbp]
\centering
\includegraphics[width = 0.85\textwidth]{mes2-1.eps}
\caption{These are the five contributions to $|\vp\rangle_2^1$.  The bottom two enjoy infrared divergences that cancel one another.  Our \hyperlink{terzo}{normalization condition} implies that the total of all five terms vanishes, which allows us to calculate $\delta m^2$.}\label{m21fig}
\end{figure}

These three terms also need to cancel precisely, as the right hand side of Eq.~(\ref{lamp}) vanishes in the one-meson sector because $(\ovp{}-H_2)|\vp\rangle_0=0$, or alternately because of our \hyperlink{terzo}{convention (III)} $|\vp\rangle_2^1=0$.  Adding the three contributions and setting the sum to zero we find
\beq
\delta m^2=-\frac{9m^2\lambda}{2}\ppinv{3}{p}\frac{\ovpp{}+\omega_{\vp+\vpp}}{\ovpp{}\omega_{\vp+\vpp}\left(\left(\ovpp{}+\omega_{\vp+\vpp}\right)^2-\ovp{}^2 \right)}=\left(-\frac{9}{8\pi^2}{\rm{ln}}\left(\frac{2\Lambda}{m}\right)+\frac{3\sqrt{3}}{16\pi}\right)m^2\lambda \label{dm}
\eeq



where $\vpp$ is integrated over a sphere of radius $\Lambda$.  This satisfies Eq.~(\ref{dma}) and so the coefficients are indeed finite.  Note also that it is independent of $\vp$, so that if the \hyperlink{second}{second} renormalization condition is applied at one value of $\vp_0$, it holds at all values.  This follows from the Lorentz covariance of the renormalization condition.
\red{This depends on $\vp$.  The \hyperlink{second}{second} renormalization condition only holds at $\vp_0$, so we should impose $\vp=\vp_0$ here where $\vp_0$ can be chosen arbitrarily.  However, Lorentz invariance seems to say that if the second condition holds at one value of $\vp$, it should hold at all values.  But that would imply that the integral is independent of $\vp$.  Is it?  Anyway, even if it is in some sense, the cutoff will ruin the Lorentz invariance.  The cutoff chooses a frame, and $\delta m^2$ depends on that frame.  But the \hyperlink{second}{second} renormalization condition should still hold for all $\vp_0$ if it holds for one ... does it?}

Finally we turn to the \hyperlink{third}{third} renormalization condition.  The inner product is proportional to $|\vp_0\rangle_2^2$.  However, only odd meson numbers have appeared at this order.  Therefore, this renormalization condition is trivially satisfied.

\subsection{Order $O(\lambda^{3/2})$}

\subsubsection{Setup}

We have checked that all renormalization conditions can be satisfied at order $O(\lambda)$.  A similar calculation at order $O(\lambda^{3/2})$ would be somewhat lengthy and so we will perform it elsewhere.  For the purpose of testing the finiteness of the domain wall tension, we only need the asymptoic behavior of the divergent counterterm $\delta\sl$, which falls from the \hyperlink{third}{third} renormalization condition at order $O(\lambda^{3/2})$.  This renormalization condition concerns $|\vp\rangle_3^2$, to which we now turn.

At meson number two, the \hyperlink{second}{second} renormalization condition at order $O(\lambda^{3/2})$ is
\beq
H_5^1|\vp\rangle_0^1+H_4^{-2}|\vp\rangle_1^4+H_4^{0}|\vp\rangle_1^2+H_3^{-3}|\vp\rangle_2^5+H_3^{-1}|\vp\rangle_2^3+H_3^{1}|\vp\rangle_2^1=(\ovp{}-H_2)|\vp\rangle_3^2. \label{lam3}
\eeq

\subsubsection{Coupling Renormalization}

We are free to choose $\vp_0$, $\vp_1$ and $\vp_2$ in the \hyperlink{third}{third} renormalization condition.  Let us choose values such that neither $\vp_1$ nor $\vp_2$ is equal to $\vp_0$.  There will be contributions from components of $|\vp\rangle_3^2$ consisting of two mesons, one of which is $\vp$.  This choice means that, for the purpose of imposing the \hyperlink{third}{third} renormalization condition, we can drop all such contributions in Eq.~(\ref{lam3}).  In particular, we are not interested in the tadpole terms in $H_5$, which yield contributions of this form\footnote{Presumably the tadpole contributions would anyway cancel as a result of the \hyperlink{fourth}{fourth} renormalization condition, which fixes $\delta v_3$, but we will not impose this.}.  

Therefore we are only interested in the cubic term in $\ch_5(\vx)$
\beq
\ch_5(\vx)\supset m\sqrt{\frac{\lambda}{2}}\left(\frac{\delta m^2}{2m^2}+\frac{\delta\sl}{\sl} \right)\phi^3(\vx)
\eeq
which leads to
\beq
H_5^1\supset  \frac{3m}{4}\sqrt{\frac{\lambda}{2}}\left(\frac{\delta m^2}{m^2}+2\frac{\delta\sl}{\sl} \right)\pinv{3}{p_1}\pinv{3}{p_2} \frac{\Ad 1\Ad 2 A_{\vp_1+\vp_2}}{\omega_{\vp_1+\vp_2}}.
\eeq
One then finds the first contribution
\beq
H_5^1|\vp\rangle_0^1=\frac{3m}{4\ovp{}}\sqrt{\frac{\lambda}{2}}\left(\frac{\delta m^2}{m^2}+2\frac{\delta\sl}{\sl} \right)\pinv{3}{p_1}|-\vp_1,\vp+\vp_1\rangle_0^2. \label{dlcon}
\eeq
Our strategy to evaluate the ultraviolet divergent piece of $\delta\sl$ will be to ensure that this contribution cancels the ultraviolet divergent contributions from the other terms on the left hand side of Eq.~(\ref{lam3}).

\subsubsection{Reducible Ultraviolet Divergent Contributions}

First we turn to reducible divergences, involving insertions and loops on external legs of the three-point function.  These may be removed by amputating external legs.

In our application of the renormalization conditions at order $O(\lambda)$ we have already seen that two ultraviolet divergences cancel in the construction of $|\vp\rangle_2^1$.  After this cancellation, no more ultraviolet divergences remain in $H_3^1|\vp\rangle_2^1$ and so we will not consider this contribution further.

\begin{figure}[htbp]
\centering
\includegraphics[width = 0.85\textwidth]{red.eps}
\caption{These three contributions to $|\vp\rangle_3^2$ contain divergences that are reducible, in the sense that they can be amputated by cutting an external leg.  In these cases the external leg is outgoing.}\label{redfig}
\end{figure}

However, a related cancellation occurs between the following three contributions, drawn in Fig.~\ref{redfig}
\bea
H_4^{0}|\vp\rangle_1^2&\supset&-\frac{3m\delta m^2\sl}{2\sqrt{2}\ovp{}}\pinv{3}{p_1}\frac{|-\vp_1,\vp+\vp_1\rangle_0}{\ovp 1\left(\ovp 1+\omega_{\vp+\vp_1}-\ovp{}\right)}
\\
H_3^{-1}|\vp\rangle_2^3&\supset&-\frac{27m^3\lambda^{3/2}}{8\sqrt{2}\ovp{}}
\pinv{3}{p_1}\left[\ppinv{3}{p}\frac{1}{\omega_{\vp_1+\vpp}\ovpp{}\left(\omega_{\vp_1+\vpp}+\ovpp{}+\omega_{\vp+\vp_1}-\ovp{}\right)}\right]\nonumber\\
&&\times\frac{|-\vp_1,\vp+\vp_1\rangle_0}{\ovp 1\left(\omega_{\vp_1}+\omega_{\vp+\vp_1}-\ovp{}\right)} 
\nonumber\\
H_3^{-3}|\vp\rangle_2^5&\supset&-\frac{27m^3\lambda^{3/2}}{8\sqrt{2}\ovp{}}\pinv{3}{p_1}\left[\ppinv{3}{p}\frac{1}{\omega_{\vpp+\vp_1}\ovpp{}\left(\omega_{\vp+\vp_1}+2\omega_{\vp_1}+\ovpp{}+\omega_{\vp_1+\vpp}-\ovp{}\right)}\right]\nonumber\\
&&\times\frac{|-\vp_1,\vp+\vp_1\rangle_0}{\ovp 1\left(\omega_{\vp_1}+\omega_{\vp+\vp_1}-\ovp{}\right)} .
\nonumber
\eea
At high $\vpp=|\vpp|$, the quantities in square brackets diverge as $\ppinv{3}{p} 1/(2p^{\prime 3})$ and so, at each value of $\vp_1$, they cancel the divergence in the first term where $\delta m^2$ was given in Eq.~(\ref{dm}).  Note that the terms outside of the square brackets have the same $\vp$ and $\vp_1$ dependence, and so the cancellation applies at all values of $\vp$ and $\vp_1$.

A similar triplet of divergences corresponds to reducible loops that arise at a lower order than the initial $\vp$ interaction.  These are all terms in $H_3^{-1}|\vp\rangle_2^3$.  Their sum is finite because, as we argued in the $O(\lambda)$ calculation, $|\vp\rangle_2^3$ is finite as a result of Eq.~(\ref{dma}).  There are also four related diagrams with a reducible loop, shown in Fig.~\ref{finfig}.  They contain four powers of $p\p$ in the denominator and so are finite.

\begin{figure}[htbp]
\centering
\includegraphics[width = 0.85\textwidth]{fin.eps}
\caption{These four contributions to $|\vp\rangle_3^2$ contain a loop, but are power counting finite because the $(E-H_2)^{-1}$ at a vertex which does not lie on the loop contributes a factor of $1/\vpp$.}\label{finfig}
\end{figure}

Finally, three terms shown in Fig.~\ref{red2fig} contain reducible loops or counterterm insertions on the momentum $\vp$ leg
\bea
H_4^{-2}|\vp\rangle_1^4&\supset&-\frac{3m\delta m^2\sl}{4\sqrt{2}\ovp{}^2}\pinv{3}{p_1}\frac{|-\vp_1,\vp+\vp_1\rangle_0}{\left(\omega_{\vp_1}+\omega_{\vp+\vp_1}+\ovp{}\right)} .
\nonumber\\
H_3^{-3}|\vp\rangle_2^5&\supset&-\frac{27m^3\lambda^{3/2}}{16\sqrt{2}\ovp{}^2}\pinv{3}{p_1}\left[\ppinv{3}{p}\frac{1}{\ovpp{}\omega_{\vp+\vpp}}\left(\frac{1}{\ovpp{}+\omega_{\vp+\vpp}+\ovp 1+\omega_{\vp+\vp_1}} \right.\right.\nonumber\\
&&\left.\left.
+\frac{1}{\ovp 1+\omega_{\vp+\vp_1}+\ovpp{}+\omega_{\vp+\vpp}}
\right)
\right]
\frac{|-\vp_1,\vp+\vp_1\rangle_0}{\left(\ovp{}+\ovp{1}+\omega_{\vp+\vp_1}\right)}.
\nonumber
\eea
The last two terms are equal.  Adding them, the term in the square bracket diverges as $\ppinv{3}{p} 1/(2p^{\prime 3})$.  Therefore again, at each value of $\vp_1$, they cancel the divergence in the first term as a result of Eq.~(\ref{dm}).  As expected, reducible loops do not contribute to the divergence in $\delta\sl$.
\begin{figure}[htbp]
\centering
\includegraphics[width = 0.85\textwidth]{red2.eps}
\caption{As in Fig.~\ref{redfig}, but these divergences lie on the incoming leg.}\label{red2fig}
\end{figure}

\subsubsection{Irreducible Ultraviolet Divergent Contributions}

\begin{figure}[htbp]
\centering
\includegraphics[width = 0.65\textwidth]{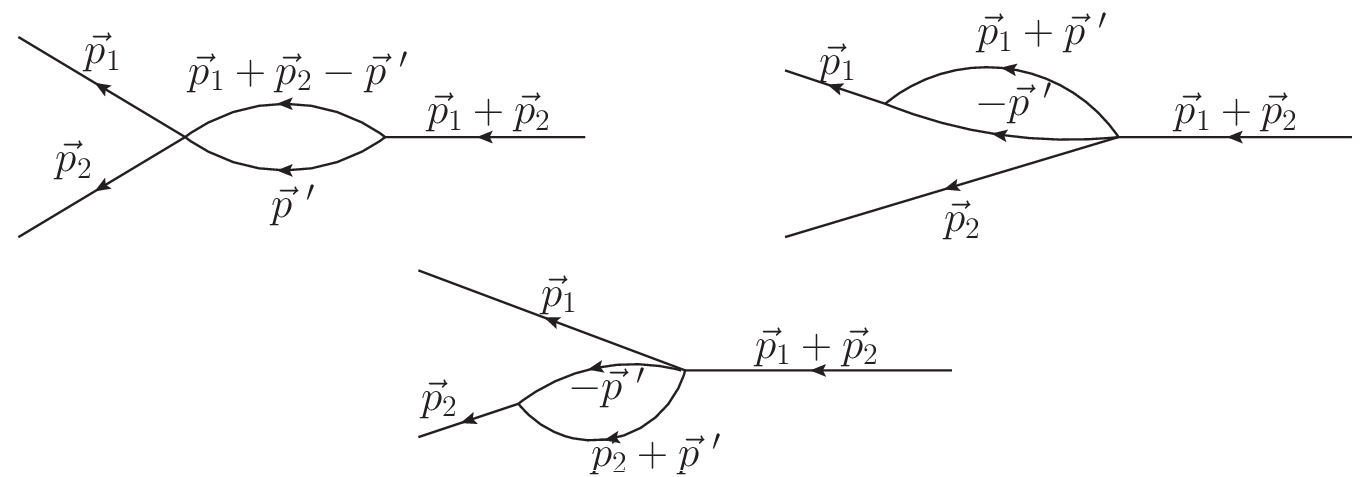}
\caption{These are the divergent contributions of interest to $|\vp\rangle_3^2$.  The \hyperlink{third}{third} renormalization condition implies that they must cancel the counterterm in Eq.~(\ref{dlcon}), which corresponds to a diagram of a single, degree three vertex.  This allows us to fix $\delta\sl$.}\label{irredfig}
\end{figure}

There are four contributions with irreducible loops shown in Fig.~\ref{irredfig}
\bea
H_4^0|\vp\rangle_1^2&\supset&\frac{9m\lambda^{3/2}}{8\sqrt{2}\ovp{}}\pinv{3}{p_1}\left[ \ppinv{3}{p}\frac{1}{\ovpp{}\omega_{\vpp+\vp}\left(\ovpp{}+\omega_{\vp+\vpp}-\ovp{} 
\right)}\right]|-\vp_1,\vp+\vp_1\rangle_0
\\
H_3^{-3}|\vp\rangle_2^5&\supset&\frac{9m\lambda^{3/2}}{8\sqrt{2}\ovp{}}\pinv{3}{p_1}\left[ \ppinv{3}{p}\frac{1}{\ovpp{}\omega_{\vpp+\vp}\left(\ovp 1+\ovpp{}+\omega_{\vp+\vp_1}+\omega_{\vp+\vpp} 
\right)}\right]|-\vp_1,\vp+\vp_1\rangle_0
\nonumber\\
H_3^{-1}|\vp\rangle_2^3&\supset&\frac{9m\lambda^{3/2}}{4\sqrt{2}\ovp{}}\pinv{3}{p_1}\left[ \ppinv{3}{p}\frac{1}{\ovpp{}\omega_{\vpp+\vp_1}\left(\ovpp{}+\omega_{\vp_1+\vpp}+\omega_{\vp+\vp_1}-\ovp{} 
\right)}\right]|-\vp_1,\vp+\vp_1\rangle_0
\nonumber\\
H_4^{-2}|\vp\rangle_1^4&\supset&\frac{9m\lambda^{3/2}}{4\sqrt{2}\ovp{}}\pinv{3}{p_1}\left[ \ppinv{3}{p}\frac{1}{\ovpp{}\omega_{\vpp+\vp_1}\left(\ovp 1+\ovpp{}+\omega_{\vp_1+\vpp} 
\right)}\right]|-\vp_1,\vp+\vp_1\rangle_0.
\nonumber
\eea
The terms in square brackets each diverge in the ultraviolet as $\ppinv{3}{p}1/(2p^{\prime 3})$ and so, up to finite terms, these four contributions sum to
\beq
\frac{27m\lambda^{3/2}}{8\sqrt{2}\ovp{}}\pinv{3}{p_1} \left[ \ppinv{3}{p}\frac{1}{p^{\p 3}}\right]|-\vp_1,\vp+\vp_1\rangle_0=
\frac{27m\lambda^{3/2}}{16\sqrt{2}\pi^2\ovp{}}{\rm{ln}}(\Lambda)\pinv{3}{p_1}|-\vp_1,\vp+\vp_1\rangle_0.
\eeq

This must cancel the contribution (\ref{dlcon}) from the counterterms.  Therefore, up to finite corrections
\beq
\left(\frac{\delta m^2}{m^2}+2\frac{\delta\sl}{\sl}\right)=-\frac{9\lambda}{2}\ppinv{3}{p}\frac{1}{p^{\p 3}}=-\frac{9\lambda}{4\pi^2}{\rm{ln}}(\Lambda).
\eeq
Using the value of $\delta m^2$ reported in Eq.~(\ref{dm}) we find, up to a finite term, the coupling constant renormalization
\beq
\delta\sl=-\frac{9\lambda^{3/2}}{16\pi^2}{\rm{ln}}(\Lambda).
\eeq
\gre{dim problem for $ln \lambda$,we can review (3.50)}
Inserting this and Eq.~(\ref{dm}) into Eq.~(\ref{v1}), we see that $\delta v_1$ is finite.  This is a necessary condition for the operator $\dv$ to exist.

\section{The Domain Wall Sector} \label{wallsez}

\subsection{Definitions}

Consider a solution
\beq
\phi(\vx,t)=F(\vx)=f(x_1)
\eeq
to the classical field equations with the renormalized parameters, corresponding to the first line of Eq.~(\ref{hhp}).  In other words
\beq
\partial_1^2 f(x_1)=-\frac{m^2}{2}f(x_1)+f^3(x_1). \label{eom}
\eeq
We will be interested in the domain wall solution
\beq
f(x_1)=\frac{m}{\sqrt{2\lambda}}\tanh{\left(\frac{mx_1}{2}\right)}.
\eeq
We may also consider a vacuum solution $f=\pm m/\sqrt{2\lambda}$ but in that case the following reasoning would lead us to the results of the previous sections.

Now construct the displacement operator
\beq
\df=\exp{-i\dvx F(\vx) \pi(\vx)}
\eeq
which satisfies
\beq
\phi(\vx)\df=\df\left(\phi(\vx)+F(\vx)\right)\hsp [\pi(\vx),\df]=0.
\eeq
Acting as in the vacuum sector, we may use the displacement operator to construct a domain wall sector Hamiltonian
\beq
H\p[\phi,\pi]=\df^\dag \hat H[\phi,\pi]\df=\hat H[\phi+F,\pi]\hsp H\p=\dvx :\ch\p(\vx):_m.
\eeq

The result looks rather like Eq.~(\ref{chsb})
\bea
\ch(\vx)&=&\frac{\pi^2(\vx)+\left(\partial_1(\phi(\vx)+f(x_1))\right)^2+\sum_{i=2}^3\left(\partial_i \phi(\vx)\right)^2}{2}\label{hp}\\
&&+\frac{\lambda (\phi(\vx)+f(x_1))^4-m^2 (\phi(\vx)+f(x_1))^2}{4}+A
\nonumber\\
&&\hspace{-2cm}+\frac{\left(-2\sl\delta\sl+(\delta\sl)^2 \right) (\phi(\vx)+f(x_1))^4+\left(\delta m^2+6\left(\sl-\delta\sl\right)^2 I\right)  (\phi(\vx)+f(x_1))^2}{4}.\nonumber
\eea

Why have we constructed a new Hamiltonian?  Because if any state $|\psi\rangle$ is in the vacuum sector, then $\df\dv^\dag|\psi\rangle$ will be in the domain wall sector.  If
\beq
H\df\dv^\dag|\psi\rangle=E\df\dv^\dag|\psi\rangle \label{nonp}
\eeq
then
\beq
H\p|\psi\rangle=E|\psi\rangle. \label{pert}
\eeq
However the Fock space of nonperturbative states $\df\dv^\dag|\psi\rangle$ consisting of the domain wall plus a finite number of mesons is in one to one correspondence with the ordinary perturbative Fock space of states $|\psi\rangle$.  Thus by using $H\p$ instead of $H$ we can find the domain wall states and their energies using ordinary perturbation theory.  

Said differently, the domain wall states $\df\dv^\dag|\psi\rangle$ are intrinsically nonperturbative, because $\df$ and $\dv$ contain an exponentiated $1/\sl$, and they satisfy the difficult equation (\ref{nonp}).  However, to find the spectrum of $H$ in the domain wall sector, it is sufficient to solve the equation (\ref{pert}) for the perturbative states $|\psi\rangle$.  In other words, the domain wall sector consists of perturbative eigenstates of $H\p$.

\subsection{Decomposing the Domain Wall Hamiltonian}

Since we have already calculated the counterterms, now to calculate the domain wall tension to order $O(\lambda^0)$ we need only consider the terms in the Hamiltonian up to order $O(\lambda^0)$.  Let us now decompose $H\p$
\beq
H\p=\sum_{i=0}^\infty H\p_i\hsp H\p_i=\dvx :\ch\p_i(\vx):_m
\eeq
where $H\p_i$ is of order $O(\lambda^{i/2-1})$.

At leading order (\ref{hp}) reduces to
\beq
\ch\p_0(\vx)=\frac{f^{\prime 2}(x_1)}{2}+\frac{\lambda f^4(x_1)-m^2f^2(x_1)}{4}+\frac{m^4}{16\lambda}
\eeq
where we have used the value of $A_0$ found in Eq.~(\ref{a0}).  Of course this is just the energy density of the $\phi^4$ kink in a theory with the bare parameters replaced by the renormalized parameters, and so it integrates to the classical kink mass
\beq
\rho=\sum_{i=0}^\infty\rho_i\hsp \rho_i=\int dx_1 :\ch\p_{2i}(\vx):_m\hsp
\rho_0=\frac{m^3}{3\lambda}.
\eeq
In 3+1 dimensions, $\rho$ is not the kink mass but rather the tension of the domain wall, which extends in the $x_2$ and $x_3$ directions.  We conclude that at leading order the tension $\rho_0$ of the domain wall agrees with the classical result, which is not a surprise.

At order $O(1/\sl)$, after integrating by parts, Eq.~(\ref{hp}) yields
\beq
\ch\p_1(\vx)= \phi(vx)\left[-\partial_1^2 f(x_1)
+\lambda f^3(x_1)-\frac{m^2}{2} f(x_1)
\right]=0
\eeq
\gre{typo for $\phi(\vx)$}
where in the last step we used the truncated classical equation of motion (\ref{eom}).  Of course this is not entirely trivial, as the full classical equation of motion contains the bare parameters.  However, the difference will not arise until order $O(\sl)$ and so will not affect the one-loop tension.  We may therefore expect an unpleasant surprise in the calculation of the two-loop tension, but this will have to await further work.

Finally we turn to the domain wall Hamiltonian density at order $O(\lambda^0)$
\bea
\ch\p_2(\vx)&=&\frac{\pi^2(\vx)+\sum_{i=1}^3\left(\partial_i \phi(\vx)\right)^2+(3\lambda f^2(x_1)-m^2/2)\phi^2(\vx)}{2}\\
&&+\frac{f^2(x_1)\delta m^2-2f^4(x_1)\sl\delta\sl
}{4}+\frac{m^4}{8\lambda}\left(\frac{\delta\sl}{\sl}-\frac{\delta m^2}{m^2}\right)\nonumber
\eea
where we have used the value of $A_2$ reported in Eq.~(\ref{a2}).

The first line is an operator, let us call it $\ch^{\prime A}_2(\vx)$ while the second is a $c$-number that we will call $\ch^{\prime B}_2(\vx)$.  We will find that both contain ultraviolet divergences. 

\subsection{Divergences from the $c$-number Term}

To obtain the next correction $\rho_1$ to the tension, we need to integrate both terms over $x_1$
\beq
\rho_1=\rho_1^A+\rho_1^B.
\eeq
Let us begin with the $c$-number term
\bea
\rho_1^B&=&\frac{\delta m^2}{m^2}\int dx_1 \left(\frac{m^2 f^2(x_1)}{4}-\frac{m^4}{8\lambda}\right)+\frac{\delta\sl}{\sl}\int dx_1\left(-\frac{\lambda f^4(x_1)}{2}+ \frac{m^4}{8\lambda}\right)\nonumber\\
&=&\frac{m^3}{2\lambda}\frac{\delta m^2}{m^2}-\frac{2m^3}{3\lambda}\frac{\delta\sl}{\sl}=\frac{3m^3}{16\pi^2}{\rm{ln}}(\Lambda)+{\rm{finite}}.\label{b}
\eea
We will drop the finite term in the last step because we have not calculated the finite contributions to $\delta\sl$.

We believe that in Ref.~\cite{erice}, when Coleman stated that the energy density of a coherent state corresponding to a soliton is infinite beyond 1+1 dimensions, he was referring to this divergence.  Indeed, the tension of the coherent domain wall state $\df\dv^\dag\ovac_0$ is $\rho_0+\rho^B_1$, which we have just shown is divergent.  In the next section we will construct what we believe is the correct domain wall state, which is a squeezed coherent state where the squeeze exactly cancels this divergence.

\subsection{Constructing the Domain Wall State}

We are interested in the lowest energy eigenstate of the Hamiltonian $H\p_2$, and the corresponding eigenvalue of the tension operator
\beq
\rho_1^A=\frac{1}{2}\int dx_1 \left( :\pi^2(\vx):_m+\sum_{i=1}^3:\left(\partial_i \phi(\vx)\right)^2:_m+\V2:\phi^2(\vx):_m
\right)
\eeq
where have introduced the shorthand
\beq
\V2=3\lambda f^2(x_1)-m^2/2.
\eeq
\gre{all the notation $\V2$ Better to be $ V^{2}[\sqrt{\lambda}f(\vx)]$ to keep consistent with ref[23]?}
This problem was solved in Ref.~\cite{me2d}.  Let us review the solution here.

The classical equation of motion corresponding to $H\p_2$ is
\beq
\left(-\partial_t^2+\partial_i^2\right)\phi(\vx,t)=\V2 \phi(\vx,t).
\eeq
A basis of constant, negative frequency solutions is
\beq
\phi(\vx,t)=\g_{k_1k_2k_3}(\vx)e^{-i\omega_{k_1k_2k_3}t}\hsp
\g_{k_1k_2k_3}(\vx)=\g_{k_1}(x_1)e^{-i(k_2x_2+k_3x_3)}. \label{phis}
\eeq
Here the functions $\g_{k}(x)$ are solutions to the P\"oschl-Teller potential
\bea
\g_k(x)&=&\frac{e^{-ikx}}{\ok{} \sqrt{m^2+4k^2}}\left[2k^2-m^2+(3/2)m^2\sech^2(m x/2)-3im k\tanh(m x/2)\right]\nonumber\\
\g_S(x)&=&\frac{\sqrt{3 m}}{2}\tanh(m x/2)\sech(m x/2)\hsp
\g_B(x)=-\sqrt{\frac{{3m}}{8}}\sech^2(m x/2).\label{nmode}
\eea
Note that in (\ref{phis}) the abstract index $k_1$ runs over the real numbers and also the two discrete values $B$ and $S$ whereas in the first line of (\ref{nmode}), the same letter $k$ only runs over the real numbers.
The indices $B$ and $S$ represent the zero mode and shape mode of the (1+1)-dimensional $\phi^4$ kink.  The frequencies $\omega$ are defined to be
\beq
\omega_{B k_2 k_3}=\sqrt{k_2^2+k_3^2}\hsp \omega_{S k_2 k_3}=\sqrt{\frac{3m^2}{4}+k_2^2+k_3^2}\hsp \omega_{k_1k_2 k_3}=\sqrt{m^2+k_1^2+k_2^2+k_3^2} \label{omk}
\eeq
where $k_1$, like $k_2$ and $k_3$, runs over the real numbers.

The equation of motion for $H\p_2$ implies that the functions $\g_k(x)$ satisfy the Sturm-Liouville equations
\beq
\V2\g_k(x)=\left(\partial_x^2+\omega^2_{k 0 0}\right)\g_k(x)
\eeq
or equivalently
\beq
\V2\g_{k_1 k_2 k_3}(\vx)=\left(\nabla^2+\omega^2_{k_1 k_2 k_3}\right)\g_{k_1k_2k_3}(\vx). \label{sl}
\eeq

Following Ref.~\cite{cahill76} we decompose the field $\phi(\vx)$ and its momentum $\pi(\vx)$ in normal modes
\bea
\phi(\vx)&=&\ppink{3}\g_{\vk}(\vx)\phi_\vk\nonumber\\
\pi(\vx)&=&\ppink{3}\g_{\vk}(\vx)\pi_\vk
\eea
where $\dint$ sums over real triplets $(k_1,k_2,k_3)$ and also $(B,k_1,k_2)$ and $(S,k_2,k_3)$ where $B$ and $S$ are the abstract indices representing the zero mode and shape mode.  

Integrating by parts, $\ch^{\prime A}_2(\vx)$ can be simplified using the Sturm-Liouville equations (\ref{sl})
\bea
\ch^{\prime A}_2(\vx)&=&\frac{\pi^2(\vx)+\phi(\vx)\left[\V2-\nabla^2\right]\phi(\vx)}{2} \label{chpa}\\
&=&\frac{1}{2}\ppink{3}\ppinkp{3}\g_{\vk}(\vx)\g_{\vkp}(\vx)\left[\pi_{\vk}\pi_{\vkp}+\omega^2_{\vk_2}\phi_{\vk}\phi_{\vkp}\right].\nonumber
\eea
\gre{we may change the $k\p$ short-hand def of $\ppinkp{3}$ to be $\vec{k}\p$ inside it in the begining  }
The Sturm-Liouville completeness relation can be used to fix the normalizations
\beq
\int dx \g_{k}(x)\g_{k\p}(x)=2\pi \delta(k+k\p)\hsp
\int dx \g_{S}(x)\g_{S}(x)=\int dx \g_B(x)\g_B(x)=1
\eeq
where other combinations yield zero by the orthogonality of the solutions.  Inserting this into (\ref{chpa}) we find
\bea
\int dx_1 \ch^{\prime A}_2(\vx)&=&\frac{1}{2}\ppin{k_1}\int\frac{dk_2dk_3dk\p_2dk\p_3}{(2\pi)^4}
e^{-i(k_2+k\p_2)x_2-i(k_3+k\p_3)x_3}\label{dena}\\
&&\times\left[\pi_{k_1k_2k_3}\pi_{-k_1k\p_2k\p_3}+\omega^2_{k_1k\p_2k\p_3}\phi_{k_1k_2k_3}\phi_{-k_1k\p_2k\p_3}\right].\nonumber
\eea

What is the eigenvector of $H^{\prime A}_2$ with minimum eigenvalue?  Integrating (\ref{dena}) over $x_2$ and $x_3$, we see that $k_2=-k\p_2$ and $k_3=-k\p_3$ and so this reduces to a sum of quantum harmonic oscillators.  The ground state $\vac_0$ is therefore the state annihilated by $\pi_\vk- i\omega_{\vk}\phi_\vk$.  The normal ordering of an operator quadratic in the fields only shifts the operator by a $c$-number, and so does not affect the eigenvectors or the ordering of their eigenvalues, and so does not alter this conclusion.


\subsection{The Tension In General}

The $A$ contribution to the tension is the eigenvalue of
\beq
\rho^{A}_1=\int dx_1 :\ch_2^{\prime A}(\vx):_m
\eeq
acting on $\vac_0$.  We will now turn to the evaluation of this contribution.

Let us define the operators
\beq
B_{\vk}=\frac{\phi_\vk}{2}+i\frac{\pi_\vk}{2\omega_{\vk}}
\eeq
that annihilate $\vac_0$ and as usual we will define the combination
\beq
B^\ddag_{\vk}=\frac{B^\dag_{\vk}}{2\omega_{\vk}}.
\eeq
Note that these are not defined in the case $\vk=(B00)$ and so in that case we will instead adopt the notation
\beq
\phi_{\vk=(B00)}=\phi_0\hsp\pi_{\vk=(B00)}=\pi_0
\eeq
and state that $\pi_0\vac_0=0$.  We will see that the following expressions are infrared finite and so the contributions from the small $k$ limit vanish, allowing us to safely ignore these operators.  This is in contrast with the 1+1 dimensional kink, in which the majority of the kink mass arises from the zero mode.

If the Hamiltonian $H\p_2$ were normal ordered with respect to the $B$ operators, so that each $B^\ddag$ appears on the left, then $H\p_2$ would annihilate $\vac_0$.  In that case the tension would be given by $\rho^B$ which diverges, and in fact is equal to that of a coherent state.  However it is normal ordered with respect to the $A$ operators, and it is this difference which will remove the divergence.

As normal ordering bilinears in fields yields a $c$-number, the difference between these normal orderings contributes a constant to the energy, which in fact is our tension $\rho_1^A$.

Combining the decompositions of the fields in normal modes and plane waves, one finds
\bea
\phi_{k_1k_2k_3}&=&\dvx \g_{-k_1}(x_1) e^{i(k_2x_2+k_3x_3)}\phi(\vx)\\
&=&\dvx \g_{-k_1}(x_1) e^{i(k_2x_2+k_3x_3)}\pinv{3}{p} e^{-i\vp\cdot\vx}\left(A^\ddag_{\vp}+\frac{A_{-\vp}}{2\ovp{}}\right)
\nonumber\\
&=&\pin{p_1}\tilde{\g}_{-k_1}(p_1)\left(A^\ddag_{p_1k_2k_3}+\frac{A_{-p_1-k_2-k_3}}{2\omega_{p_1k_2k_3}}\right)\nonumber\\
\pi_{k_1k_2k_3}&=&i\pin{p_1}\tilde{\g}_{-k_1}(p_1)\left(\omega_{p_1k_2k_3}A^\ddag_{p_1k_2k_3}-\frac{A_{-p_1-k_2-k_3}}{2}\right)\nonumber
\eea
where $\tilde{\g}$ is the Fourier transform of $\g$.

This implies
\bea
\omega^2_{k_1k\p_2k\p_3}:\phi_{k_1k_2k_3}\phi_{-k_1k\p_2k\p_3}:_m&=&\pin{p_1}\pin{p\p_1}\tilde{\g}_{-k_1}(p_1)\tilde{\g}_{k_1}(p\p_1)\omega^2_{k_1k\p_2k\p_3}\nonumber\\
&&\hspace{-1cm}\times\left(A^\ddag_{p_1k_2k_3}A^\ddag_{p\p_1k\p_2k\p_3}+A^\ddag_{p_1k_2k_3}\frac{A_{-p\p_1-k\p_2-k\p_3}}{2\omega_{p\p_1k\p_2k\p_3}}\right.\nonumber\\
&&\hspace{-1cm}\left.+A^\ddag_{p\p_1k\p_2k\p_3}\frac{A_{-p_1-k_2-k_3}}{2\omega_{p_1k_2k_3}}+\frac{A_{-p_1-k_2-k_3}}{2\omega_{p_1k_2k_3}}\frac{A_{-p\p_1-k\p_2-k\p_3}}{2\omega_{p\p_1k\p_2k\p_3}}\right)
\nonumber\\
:\pi_{k_1k_2k_3}\pi_{-k_1k\p_2k\p_3}:_m&=&
\pin{p_1}\pin{p\p_1}\tilde{\g}_{-k_1}(p_1)\tilde{\g}_{k_1}(p\p_1)\omega_{p_1k_2k_3}\omega_{p\p_1k\p_2k\p_3}\nonumber\\
&&\hspace{-1cm}\times\left(-A^\ddag_{p_1k_2k_3}A^\ddag_{p\p_1k\p_2k\p_3}+A^\ddag_{p_1k_2k_3}\frac{A_{-p\p_1-k\p_2-k\p_3}}{2\omega_{p\p_1k\p_2k\p_3}}\right.\nonumber\\
&&\hspace{-1cm}\left.+A^\ddag_{p\p_1k\p_2k\p_3}\frac{A_{-p_1-k_2-k_3}}{2\omega_{p_1k_2k_3}}-\frac{A_{-p_1-k_2-k_3}}{2\omega_{p_1k_2k_3}}\frac{A_{-p\p_1-k\p_2-k\p_3}}{2\omega_{p\p_1k\p_2k\p_3}}\right). \label{aa}
\eea

Now the normal ordering symbol has disappeared, because we have enforced the normal ordering.  All that remains is to act these operators on $\vac_0$.  We know that this is annihilated by the $B$ operators, and so we need to use a Bogoliubov transformation to change the $A$ operators into $B$ operators
\bea
A^\ddag_{p_1k_2k_3}&=&\frac{1}{2}\dvx e^{i(p_1x_1+k_2x_2+k_3x_3)}\left(\phi(\vx)-i\frac{\pi(\vx)}{\omega_{p_1k_2k_3}}\right)\\
&=&\frac{1}{2}\dvx e^{i(p_1x_1+k_2x_2+k_3x_3)}\ppinkp{3}\g_{k\p_1}(x_1)e^{-i(k\p_2x_2+k\p_3x_3)}\nonumber\\
&&\times\left(B^\ddag_{k\p_1k\p_2k\p_3}+\frac{B_{-k\p_1k\p_2k\p_3}}{2\omega_{k\p_1k\p_2k\p_3}} 
+\frac{\omega_{k\p_1k\p_2k\p_3}}{\omega_{p_1k_2k_3}}B^\ddag_{k\p_1k\p_2k\p_3}-\frac{B_{-k\p_1k\p_2k\p_3}}{2\omega_{p_1k_2k_3}}
\right)\nonumber\\
&=&\frac{1}{2}\ppin{k\p_1}\tilde\g_{k\p_1}(-p_1)\left[\left(1+\frac{\omega_{k\p_1k_2k_3}}{\omega_{p_1k_2k_3}}\right)B^\ddag_{k\p_1k_2k_3}+\left(1-\frac{\omega_{k\p_1k_2k_3}}{\omega_{p_1k_2k_3}}\right)\frac{B_{-k\p_1k_2k_3}}{2\omega_{k\p_1k_2k_3}} 
\right]\nonumber\\
\frac{A_{-p_1-k_2-k_3}}{2\omega_{p_1k_2k_3}}&=&\frac{1}{2}\ppin{k\p_1}\tilde\g_{k\p_1}(-p_1)\left[\left(1-\frac{\omega_{k\p_1k_2k_3}}{\omega_{p_1k_2k_3}}\right)B^\ddag_{k\p_1k_2k_3}+\left(1+\frac{\omega_{k\p_1k_2k_3}}{\omega_{p_1k_2k_3}}\right)\frac{B_{-k\p_1k_2k_3}}{2\omega_{k\p_1k_2k_3}} 
\right].\nonumber
\eea

Consider a term such as $AA$ in Eq.~(\ref{aa}).  The contribution to the tension comes from the difference between the plane wave normal ordering $::_m$ and the normal mode normal ordering $::_b$, which puts all $B$ to the right of $B^\ddag$.  Therefore it comes from the $B$ term in the first $A$ and the $B^\ddag$ term in the second $A$.  Acting on $\vac_0$, it is therefore equal to the commutator of the two
\bea
A^\ddag_{p_1k_2k_3}A^\ddag_{p\p_1k\p_2k\p_3}-:A^\ddag_{p_1k_2k_3}A^\ddag_{p\p_1k\p_2k\p_3}:_b&=&
\frac{1}{4}\ppin{k_1}\tilde\g_{k_1}(-p_1)\ppin{k\p_1}\tilde\g_{k\p_1}(-p\p_1)\\
&&\hspace{-5cm}\ \ \times\left(1-\frac{\omega_{k_1k_2k_3}}{\omega_{p_1k_2k_3}}\right) \left(1+\frac{\omega_{k_1k\p_2k\p_3}}{\omega_{p\p_1k\p_2k\p_3}}\right) 
\left[\frac{B_{-k_1 k_2k_3}}{2\omega_{k_1k_2k_3}},B^\ddag_{k\p_1k\p_2k\p_3} 
\right]\nonumber\\
&&\hspace{-5cm}=\frac{1}{8}\ppin{k_1}\frac{\tilde \g_{k_1}(-p_1)\tilde \g_{-k_1}(-p\p_1)}{\omega_{\vk}}\left(1-\frac{\omega_{k_1k_2k_3}}{\omega_{p_1k_2k_3}}\right) \left(1+\frac{\omega_{k_1k_2k_3}}{\omega_{p\p_1k_2k_3}}\right) \nonumber\\
&&\hspace{-5cm}\ \ \times(2\pi)^2\delta(k\p_2-k_2)\delta(k\p_3-k_3)\nonumber
\eea
and similarly
\bea
A^\ddag_{p_1k_2k_3}\frac{A_{-p\p_1-k\p_2-k\p_3}}{2\omega_{p\p_1k\p_2k\p_3}}-:A^\ddag_{p_1k_2k_3}\frac{A_{-p\p_1-k\p_2-k\p_3}}{2\omega_{p\p_1k\p_2k\p_3}}:_b&=&\frac{1}{8}\ppin{k_1}\frac{\tilde \g_{k_1}(-p_1)\tilde \g_{-k_1}(-p\p_1)}{\omega_{\vk}}\\
&&\hspace{-5cm}\times\left(1-\frac{\omega_{k_1k_2k_3}}{\omega_{p_1k_2k_3}}\right) \left(1-\frac{\omega_{k_1k_2k_3}}{\omega_{p\p_1k_2k_3}}\right)(2\pi)^2\delta(k\p_2-k_2)\delta(k\p_3-k_3)\nonumber\\
A^{\ddag}_{p\p_1k\p_2k\p_3}\frac{A_{-p_1-k_2-k_3}}{2\omega_{p_1k_2k_3}}-:A^\ddag_{p\p_1k\p_2k\p_3}\frac{A_{-p_1-k_2-k_3}}{2\omega_{p_1k_2k_3}}:_b&=&\frac{1}{8}\ppin{k_1}\frac{\tilde \g_{k_1}(-p_1)\tilde \g_{-k_1}(-p\p_1)}{\omega_{\vk}}\nonumber\\
&&\hspace{-5cm}\times\left(1-\frac{\omega_{k_1k_2k_3}}{\omega_{p_1k_2k_3}}\right) \left(1-\frac{\omega_{k_1k_2k_3}}{\omega_{p\p_1k_2k_3}}\right)(2\pi)^2\delta(k\p_2-k_2)\delta(k\p_3-k_3)\nonumber\\
\frac{A_{-p_1-k_2-k_3}}{2\omega_{p_1k_2k_3}}\frac{A_{-p\p_1-k\p_2-k\p_3}}{2\omega_{p\p_1k\p_2k\p_3}}-:\frac{A_{-p_1-k_2-k_3}}{2\omega_{p_1k_2k_3}}\frac{A_{-p\p_1-k\p_2-k\p_3}}{2\omega_{p\p_1k\p_2k\p_3}}:_b&=&\frac{1}{8}\ppin{k_1}\frac{\tilde \g_{k_1}(-p_1)\tilde \g_{-k_1}(-p\p_1)}{\omega_{\vk}}\nonumber\\
&&\hspace{-5cm}\times\left(1+\frac{\omega_{k_1k_2k_3}}{\omega_{p_1k_2k_3}}\right) \left(1-\frac{\omega_{k_1k_2k_3}}{\omega_{p\p_1k_2k_3}}\right)(2\pi)^2\delta(k\p_2-k_2)\delta(k\p_3-k_3).\nonumber
\eea

\hui{The position of the equation number of the above equation. }Inserting these into Eq.~(\ref{aa}) we find that the contributions to the tension are
\bea
\omega^2_{k_1k\p_2k\p_3}\left(:\phi_{k_1k_2k_3}\phi_{-k_1k\p_2k\p_3}:_m-:\phi_{k_1k_2k_3}\phi_{-k_1k\p_2k\p_3}:_b\right)
&=&\frac{1}{4}\pin{p_1}\pin{p\p_1}\tilde{\g}_{-k_1}(p_1)\tilde{\g}_{k_1}(p\p_1)\nonumber\\
&&\hspace{-6cm}\ \ \times \omega^2_{\vk}
\ppin{k\p_1}{\tilde \g_{k\p_1}(-p_1)\tilde \g_{-k\p_1}(-p\p_1)}\left(\frac{2}{\omega_{k\p_1k_2k_3}}-\frac{1}{\omega_{p_1k_2k_3}}-\frac{1}{\omega_{p\p_1k_2k_3}}\right)
\nonumber\\
&&\hspace{-6cm}\ \ \times(2\pi)^2\delta(k\p_2-k_2)\delta(k\p_3-k_3)
\nonumber\\
:\pi_{k_1k_2k_3}\pi_{-k_1k\p_2k\p_3}:_m-:\pi_{k_1k_2k_3}\pi_{-k_1k\p_2k\p_3}:_b&=&\frac{1}{4}\pin{p_1}\pin{p\p_1}\tilde{\g}_{-k_1}(p_1)\tilde{\g}_{k_1}(p\p_1)\nonumber\\&&\hspace{-6cm}\ \ \times 
\ppin{k\p_1}{\tilde \g_{k\p_1}(-p_1)\tilde \g_{-k\p_1}(-p\p_1)}\left(2\omega_{k\p_1k_2k_3}-{\omega_{p_1k_2k_3}-\omega_{p\p_1k_2k_3}}{}\right)\nonumber\\
&&\hspace{-6cm}\ \ \times {}(2\pi)^2\delta(k\p_2-k_2)\delta(k\p_3-k_3).
\nonumber
\eea

This in turn must be inserted into the tension (\ref{dena})
\bea
\int dx_1 \left(:\ch^{\prime A}_2(\vx):_m-:\ch^{\prime A}_2(\vx):_b\right)&=&\frac{1}{8}\ppink{3}\pin{p_1}\pin{p\p_1}\tilde{\g}_{-k_1}(p_1)\tilde{\g}_{k_1}(p\p_1)\\
&&\hspace{-6cm}\times
\ppin{k\p_1}{\tilde \g_{k\p_1}(-p_1)\tilde \g_{-k\p_1}(-p\p_1)}\left[2\omega_{k\p_1k_2k_3}-\omega_{p_1k_2k_3}-\omega_{p\p_1k_2k_3}+\frac{2\omega_\vk^2}{\omega_{k\p_1k_2k_3}}-\frac{\omega^2_\vk}{\omega_{p_1k_2k_3}}-\frac{\omega^2_\vk}{\omega_{p\p_1k_2k_3}} 
\right]\nonumber
\eea
where we remind the reader that the $::_b$ term annihilates $\vac_0$ as it consists of terms of the form $B^\ddag B$.

Notice that no summand in the square bracket has both $p_1$ and $p_1\p$.  Thus, whichever is missing may be integrated, using the completeness relation for the $\tilde\g$ to yield a $2\pi\delta(k_1+k\p_1)$, or a Kronecker $\delta$ when $k$ is a discrete mode.  This can be used to perform the $k\p_1$ integration, and so we find
\bea
\int dx_1 :\ch^{\prime A}_2(\vx):_m\vac_0&=&
\frac{1}{4}\ppink{3}\pin{p_1}|\tilde{\g}_{-k_1}(p_1)|^2
\left[2\omega_{k\p_1k_2k_3}-\omega_{p_1k_2k_3}-\frac{\omega^2_\vk}{\omega_{p_1k_2k_3}} 
\right]\vac_0\nonumber\\
&=&-\frac{1}{4}\ppink{3}\pin{p_1}|\tilde{\g}_{-k_1}(p_1)|^2\frac{\left(\omega_\vk-\omega_{p_1k_2k_3}\right)^2}{\omega_{p_1k_2k_3}}\vac_0.
\eea

We conclude that the corresponding contribution to the tension is
\beq
\rho_1^A=-\frac{1}{4}\ppink{3}\pin{p_1}|\tilde{\g}_{-k_1}(p_1)|^2\frac{\left(\omega_\vk-\omega_{p_1k_2k_3}\right)^2}{\omega_{p_1k_2k_3}}.
\eeq
\gre{this result could be directly the higher-dim generalization of  be same like the our 2+1 dim  in previous paper ref[23].but a  little difference with exchange of sign $\vk$ and $p$,but why we need to derive again and  get a slight different result?  }\red{In our old paper we first gave a derivation for point-like solitons, which gave an infinite energy.  Then we described how to modify that derivation for domain walls without actually redoing the derivation.  Here, instead, we derive the tension directly for domain walls.}
This is a higher-dimensional generalization of the formula for a kink mass presented in Ref.~\cite{cahill76}.

\subsection{The Tension in this Model}

The logarithmic ultraviolet divergence in $\rho_1^A$ arises from the continuum modes, for which $k_1$ may be arbitrarily large.  Thus we will restrict our attention to those modes.  In Ref.~\cite{mekink} the Fourier transform
\beq
\tilde{\g}_k(p)=\frac{2k^2-m^2}{\ok{}\sqrt{m^2+4k^2}}2\pi\delta(p+k)+\frac{6\pi p}{\ok{}\sqrt{m^2+4k^2}} \csch\left(\frac{\pi (p+k)}{m}\right)
\eeq
was computed. The Dirac $\delta$ term does not contribute to the tension, because each occurrence is multiplied by a zero.  

The continuum contribution is therefore
\bea
\rho_1^A&=&-\pin{k_1}\pin{p_1}\frac{9\pi^2 p_1^2}{(m^2+k_1^2)(m^2+4k_1^2)}\csch^2\left(\frac{\pi (p_1-k_1)}{m}\right)\int\frac{dk_2dk_3}{(2\pi)^2}\frac{\left(\omega_\vk-\omega_{p_1k_2k_3}\right)^2}{\omega_{p_1k_2k_3}}\nonumber\\
&=&-\frac{3\pi}{2}\pin{k_1}\pin{p_1}\frac{p_1^2\sqrt{m^2+k_1^2}}{(m^2+4k_1^2)}\csch^2\left(\frac{\pi (p_1-k_1)}{m}\right)\nonumber\\
&&\times\left(2-3\sqrt{\frac{m^2+p_1^2}{m^2+k_1^2}}+\left(\frac{m^2+p_1^2}{m^2+k_1^2}\right)^{3/2}\right).
\eea
As a result of the $\csch$, at large $k_1$ or $p_1$, the support of the tension is at 
\beq
\epsilon=p_1-k_1
\eeq
of order $O(m)$.  In particular, this is much smaller than $p_1$ or $k_1$.  We thus expand to second order in $\epsilon/k_1$ to obtain the limiting behavior at large $k_1$
\beq
2-3\sqrt{\frac{m^2+p_1^2}{m^2+k_1^2}}+\left(\frac{m^2+p_1^2}{m^2+k_1^2}\right)^{3/2}\sim\frac{3\epsilon^2k_1^2}{(m^2+k_1^2)^2}\sim\frac{3\epsilon^2}{k_1^2}
\eeq
and so
\bea
\rho^A_1&\sim& -\frac{9\pi}{8}\pin{k_1}\frac{1}{|k_1|}\pin{\epsilon}\epsilon^2\csch^2\left(\frac{\pi\epsilon}{m}\right)\nonumber\\
&=&-\frac{3m^3}{16\pi}\pin{k_1}\frac{1}{|k_1|}=-\frac{3m^3}{16\pi^2}{\rm{ln}}(\Lambda) \label{a}
\eea
where we have included a factor of two from the fact that $|k_1|$ is large but the sign must be summed over.
\gre{considering  the dimension like you do for the $\delta m^2$ in(3.50), here all the $ln(\lambda)$ should be $ln\frac{\lambda}{m}$}
This divergent negative energy cancels, up to finite contributions, the positive energy divergence from $\rho_1^B$ reported in Eq.~(\ref{b}).  This cancellation is our main result.

\section{Remarks}

49 years ago, Coleman noted that solitons above 1+1 dimensions cannot be described by coherent states, because their energy densities would be infinite.  We have reproduced this ultraviolet divergence in Eq.~(\ref{b}).  

However, solitons are not described by coherent states even in 1+1 dimensions.  At order $e^{-1/\sl}$ they are described by coherent states, but there is a correction of order unity.  This correction changes the true perturbative vacuum $\ovac_0$, which is annihilated by the $A_\vp$ operators, to the squeezed state $\vac_0$ which is annihilated by the $B_\vk$ operators.  Physically this is because the normal modes are turned off in the ground state, instead of the plane waves.  These states are related by a Bogoliubov transformation, and so $\vac_0$ is a squeezed state.  The soliton ground state is not the coherent state $\df\ovac_0$, rather it is the squeezed coherent state $\df\vac_0$, plus perturbative corrections that are suppressed by powers of $\sl$.

This squeeze contributes a divergent negative energy density, calculated in Eq.~(\ref{a}), which exactly cancels the divergence resulting from the displacement operator in the coherent state, leaving a finite energy density and even a finite tension.

Thus, at one loop, we feel that we have succeeded in constructing the quantum state corresponding to a domain wall in the 3+1 dimensional double-well theory. 

At this point the generalization to multiple loops is still not obvious.  In principle, it may depend on our choices of operator $\df$, although it may well be that any change in $\df$ could be absorbed into a change in $\vac$.  Also, our somewhat unconventional Schrodinger picture renormalization conditions may lead to divergences.

Our derivation is not the same as that of Ref.~\cite{noi4dlett}.  One crucial difference is that, in that reference, the operator $\df$ was constructed using the bare mass and coupling.  This made the construction very fast, but it is not clear that such an operator exists.  In the present note the operator $\df$ was constructed using the renormalized mass and coupling.  This operator exists and leads to the same results.  In fact, the two operators only differ by terms which are suppressed by powers of $\sl$, and so changing the construction of $\df$ is compensated by the different perturbative corrections that arise.

Domain walls have a number of applications, featuring in many cosmological and phenomenological models \cite{okun}, and displaying a rich dynamics \cite{bp21,bp23}.  In 2+1 dimensions, the one-loop quantum correction to the domain wall tension has been calculated in Ref.~\cite{zar98}.  Keeping the finite terms in the calculation above, one would arrive at the domain wall tension in 3+1 dimensions.  The two-loop quantum correction in 2+1 dimensions has a similar divergence structure to the one-loop correction in 3+1 dimensions, and so the same approach may be used to calculate this two-loop correction.  To move to two loops in 3+1 dimensions, one needs to understand the role of quadratic divergences and field renormalization.  

The Skyrme model \cite{skyrmion} is nonrenormalizable, but at one loop is a sensible theory.  It describes nuclei at large $N$ \cite{wittskyrme}, but is often thought to be a poor description of nuclei at $N=3$ because the predicted binding energies are an order of magnitude too large.  There has recently been a surge of interest in one-loop quantum corrections to Skyrmions \cite{qs1,qs2,qs3,qs4,qs5,qs6}.  Recent calculations \cite{bjarkequant} of some contributions to the one loop energy suggest that these quantum corrections largely solve the binding energy problem, making the model phenomenologically viable.  Thus, it would be of interest to apply our formalism to the Skyrme model, where a full calculation of one-loop corrections is in principle possible, albeit numerically challenging beyond the smallest nuclei.

Our motivation for studying ultraviolet divergences is not only to allow the methods of Refs.~\cite{mekink} and \cite{me2loop} to be applied to more than one spatial dimension, but also to allow a more general field content.  For example, recently there has been a renaissance in studies of kinks in interesting models with fermions \cite{ferm1,ferm2,ferm3,ferm4,ferm5,ferm6} to which we will turn before our ultimate goal of constructing the 't Hooft-Polyakov monopole state.  Needless to say, we are also interested in solitons in theories with yet more challenging field content \cite{zhong1,zhong2,zhong3}.

\section* {Acknowledgement}

\noindent
JE is supported by NSFC MianShang grants 11875296 and 11675223.

\end{document}

If a classical field theory exhibits a soliton solution, then generically the Hilbert space of the quantum field theory will exhibit a solitonic sector.  In other words, in addition to the usual Fock space of perturbative excitations of the vacuum, there will be an additional Fock space of perturbative excitations of the soliton.  In principle, ultraviolet divergences appear in both the vacuum and solitons sector.  However, both are described by the same Hamiltonian which has the same set of counterterms.  Thus the renormalization conditions may be applied only once, and the same counterterms must eliminate ultraviolet divergences in both sectors.

There are many convincing arguments in the literature that this is indeed the case.  It was noted at one loop in 1+1 dimensions in Ref.~\cite{gjs75}, but postulated to hold for all renormalizable theories in Ref~\cite{polyakov75}.  A diagramatic argument that divergences in the S-matrix are eliminated was presented in Ref.~\cite{fk77}.  However, Coleman has argued that this is not the case if one constructs the soliton sector using the usual coherent state construction of Refs.~\cite{vinc72,cornwall74,taylor78}.  This conventional construction in fact leads to an infinite energy density for states in the soliton sector for field theories in more than 1+1 dimensions.  This leads one to ask just how the solitonic sector is to be constructed beyond 1+1 dimensions for this divergence to be avoided.

As has been recently emphasized in Ref.~\cite{cocorr23}, the coherent state corresponding to a soliton is somewhat deformed.  The leading deformation is a squeeze that puts all bound and continuum normal modes in their ground state and averages over zero modes.  In the present note, we show that this squeeze is already sufficient to eliminate the simplest manifestation of the divergence discussed by Coleman, that appearing at one loop in the domain wall soliton of the 3+1 dimensional $\phi^4$ double-well model.

\section{The Domain Wall Soliton Tension}

Consider the 3+1 dimensional theory of a scalar $\phi(\vx)$ and its conjugate momentum $\pi(\vx)$ described by the Hamiltonian
\beq
H=\int d^3\vx:\ch(\vx):\hsp
\ch(\vx)=\frac{\pi^2(\vx)+\sum_{i=1}^3\left(\partial_i \phi(\vx)\right)^2}{2}+\frac{\lambda_0 \phi^4(\vx)-m_0^2 \phi^2(\vx)}{4}+A.
\eeq
Here the normal ordering $::$ is defined at the mass scale $m_0$, and the $c$-number $A$ is a counterterm.  

Classically, there are two minimum energy configurations given by $\phi(\vx)=\pm m_0/\sqrt{2\lambda_0}$.  There is also a domain wall soliton solution 
\beq
\phi(\vx)=\frac{m_0}{\sqrt{2\lambda_0}}{\rm{tanh}}\left(\frac{m x_1}{2}\right)
\eeq
with tension
\beq
\rho_0=\frac{m_0^3}{3\lambda_0}.
\eeq
The profile in the $x_1$ direction is just that of the $\phi^4$ kink, whose mass is $\rho_0$, while it is independent of the other directions.  Such higher-dimensional extensions of the kink have been of interest lately, even finding applications in cosmology \cite{bp21,bp23}.

A simple formula for the one-loop correction to the kink mass was provided in Ref.~\cite{cahill76}.   It was shown in Ref.~\cite{me3d} that essentially this same formula provides the one-loop correction to the tension of the domain wall soliton
\beq
\rho_1=-\frac{1}{4}\ppin{k_1}\pinv{3}{p} |\gt_{-k_1}(p_1)|^2 \frac{(\omega_{k_1p_2p_3}-\omega_{p_1p_2p_3})^2}{\omega_{p_1p_2p_3}}\hsp \omega_{pqr}=\sqrt{m_0^2+p^2+q^2+r^2}. \label{c76}
\eeq
This derivation explicitly used the construction of the kink as a deformed, coherent state.

Here $\dint$ sums over the bound normal modes of the kink and integrates over the continuum modes.  $\gt_k(p)$ is the Fourier transform of the normal mode indexed by $k$.

More precisely, Ref.~\cite{me3d} considered the two dimensional case, where this expression is finite.  In fact, the tension of the 2+1 dimensional domain wall was found explicitly in Ref.~\cite{zar98}.  However, in 3+1 dimensions, $\phi_1$ enjoys an ultraviolet divergence arise from the continuum normal modes
\beq
\tilde{\g}_k(p)=\frac{2k^2-m_0^2}{\ok{}\sqrt{m_0^2+4k^2}}2\pi\delta(p+k)+\frac{6\pi p}{\ok{}\sqrt{m_0^2+4k^2}} \csch\left(\frac{\pi (p+k)}{m_0}\right).
\eeq
The Dirac $\delta$ term does not contribute, as it is multiplied by a double zero.  However the other term leads to a contribution to $\rho_1$ of 
\beq
-\frac{3\pi}{2}\pin{k_1}\frac{\omega_{k_1}}{m_0^2+4k_1^2}\pin{p_1}p_1^2\csch^2\left[\frac{\pi(p_1-k_1)}{m_0}\right]\left( 2-3\frac{\omega_{p_1}}{\ok1}+\frac{\omega^2_{p_1}}{\ok{1}^2}\right).
\eeq
At large $|k_1|$ this contribution to $\rho_1$ tends to
\beq
-\frac{3m_0^3}{16\pi}\pin{k_1} \frac{1}{k_1}.
\eeq
An ultraviolet cutoff of $|k_1|\leq \Lambda$ then leads to a leading divergence of
\beq
\rho_1=-\frac{3m_0^3}{16\pi^2}{\rm{ln}}(\Lambda).
\eeq
The total tension is $\rho_0+\rho_1$.  Is this divergent at $\Lambda\rightarrow\infty$?  Not necessarily, as $m_0$ and $\lambda_0$ also diverge.  To answer this question, let us expand in powers of $\lambda$
\beq
m_0^2=m^2-\delta m^2+O(\lambda^2)\hsp
\sqrt{\lambda_0}=\sqrt{\lambda}-\delta\sqrt\lambda+O(\lambda^{5/2})
\eeq
where $\delta m^2$ and $\delta\sqrt\lambda$ are of order $O(\lambda)$ and $O(\lambda^{3/2})$ respectively.  At this stage we have not fixed a renormalization condition, however we assume that a renormalization condition is chosen such that $m$ and $\lambda$ are finite.

Then at leading order $O(\lambda^0)$, dropping finite terms, the tension is
\beq
\rho_0+\rho_1=\frac{(m^2-\delta m^2)^{3/2}}{3(\sqrt{\lambda}-\delta\sqrt{\lambda})^2}-\frac{3m^3}{16\pi^2}{\rm{ln}}(\Lambda)=\frac{m^3}{3\lambda}-\frac{m\delta m^2}{2\lambda}+\frac{2m^3\delta\sqrt\lambda}{3\lambda^{3/2}}-\frac{3m^3}{16\pi^2}{\rm{ln}}(\Lambda). \label{ten}
\eeq
The first term in the rightmost expression is of order $O(1/\lambda)$ and is manifestly finite.  The question is therefore whether the sum of the last three terms is finite.

\section{Renormalizing a Vacuum Sector}
We have now reduced the problem of finding the domain wall tension to the problem of finding the counterterms $\delta m^2$ and $\delta\sqrt\lambda$.  These counterterms must not only eliminate divergences, for example in $\rho_0+\rho_1$, in the soliton sector, but also in the vacuum sector.  Therefore they may be fixed uniquely through a calculation in the vacuum sector.

Needless to say, the renormalization of the $\phi^4$ theory in the vacuum sector is a subject that was solved long ago.  There are many different renormalization prescriptions that lead to answers whose finite parts disagree but whose infinite parts agree.  Furthermore, it has been appreciated for over 50 years that the spontaneous symmetry breaking of this theory does not affect the infinite parts of the counterterms.  We will therefore not dwell on this problem, but simply review the main steps.

First of all, at each order $A$ is chosen to make the vacuum energy finite.  With this choice, contributions arising from diagrams with disconnected components with no external legs will be, up to finite terms, canceled by contributions from $A$.

Next, the divergence in the propagator can be fixed using a renormalization condition.  This invariably involves the ultraviolet divergent diagram in Fig.~\ref{m2fig} which, dropping finite terms, fixes
\beq
\frac{\delta m^2}{m^2}-\frac{6\delta\sqrt\lambda}{\sqrt\lambda}=\frac{9\lambda}{2}\int\frac{d^3\vp}{(2\pi)^3}\frac{1}{|\vp|^3}.
\eeq
The same ultraviolet cutoff yields
\beq
\frac{\delta m^2}{m^2}-\frac{6\delta\sqrt\lambda}{\sqrt\lambda}=\frac{9\lambda}{4\pi^2}\rm{ln}(\Lambda). \label{m2eq}
\eeq
\begin{figure}[htbp]
\centering
\includegraphics[width = 0.75\textwidth]{m2.eps}
\caption{This divergent diagram contributes to $\delta m^2/m^2-6\delta\sqrt\lambda/\sqrt\lambda$}\label{m2fig}
\end{figure}

One next uses third and final renormalization condition to fix the three point interaction, which at one loop contains the divergent contribution in Fig.~\ref{dlfig} as well as reducible divergences that are canceled by (\ref{m2eq}) and disconnected contributions that are canceled by $A$.  Again applying an ultraviolet cutoff, this diagram leads to
\beq
{\delta\sqrt\lambda}=-\frac{9\lambda^{3/2}}{8}\int\frac{d^3\vp}{(2\pi)^3}\frac{1}{|\vp|^3}=-\frac{9\lambda^{3/2}}{16\pi^2}\rm{ln}(\Lambda).
\eeq
Inserting this into (\ref{m2eq}) one finds
\beq
\delta m^2=-\frac{9m^2\lambda}{8\pi^2}\rm{ln}(\Lambda).
\eeq

\begin{figure}[htbp]
\centering
\includegraphics[width = 0.75\textwidth]{dl.eps}
\caption{This divergent diagram contributes to $\delta\sqrt\lambda$}\label{dlfig}
\end{figure}

Inserting this and (\ref{m2eq}) into (\ref{ten}), one sees that the divergent parts of the three last terms indeed cancel, and so the domain wall tension $\rho_0+\rho_1$ is finite at one loop.

\section{Conclusion}

Recall that Ref.~\cite{me3d} derived the one-loop tension (\ref{c76}) from the construction $\df\vac_0$ where $\df$ was the displacement operator that created a coherent state.  Here $\vac_0$ was the tree level vacuum squeezed so that, instead of being annihilated by the operators that create plane waves, it is annihilated by the operators that create normal modes except for the zero mode.  In the case of the zero mode, it was annihilated by the momentum operator for the center of mass.  In other words, it is a Bogoliubov transformation of the true vacuum $|\Omega\rangle$, and so is described by a squeeze, which was explicitly constructed in Ref.~\cite{mestate}.

Therefore we may summarize our results as follows.  Coleman's statement that the coherent state $\df|\Omega\rangle$ has infinite energy density is true.  The corresponding tension is $\rho_0$, which indeed is infinite.  However the correct ground state is, at $O(\lambda^0)$, given by $\df\vac_0$ which is a squeezed coherent state.  This state has finite energy density and leads to a finite tension for the domain wall soliton.  This resolves the puzzle raised by Coleman at one loop, in a scalar theory.  It remains to test it at higher loops and with more general matter content.

\section* {Acknowledgement}

\noindent
JE is supported by NSFC MianShang grants 11875296 and 11675223.

\end{document}

In the this section, we will be concerned with the long-ago solved problem of renormalizing the $\phi^4$ scalar field theory in a vacuum sector.  Somewhat unusually, we will work in the Schrodinger picture.

\subsection{Counterterms in the Hamiltonian}
Coleman (14.62)
\beq
\hat{H}=\int d^{\dim}\vx :\hat{\mathcal{H}}(\vx):_a\hsp
\hat{\mathcal{H}}(\vx)=\frac{\pi^2(\vx)+\partial_i \phi(\vx)\partial_i \phi(\vx)}{2}-\frac{m^2_0\phi^2(\vx)}{4}+\frac{\lambda\phi^4(\vx)}{4}+A
\eeq
 
Coleman (14.64)
\bea
\hat{\mathcal{H}}(\vx)&=&\frac{\pi^{2}(\vx)+\partial_i \phip(\vx)\partial_i \phip(\vx)}{2}-\frac{m^2\phi^{2}(\vx)}{4}+\frac{\lambda \phi^{4}(\vx)}{4}+\delta m^2\frac{\phi^{2}(\vx)}{4}+A\label{chhat}\\
\delta m^2&=&m^2- m^2_0\nonumber
\eea

Let ${\ovac}^s$ be a vacuum and $|p\rangle^s$ be the one-meson states in the corresponding vacuum sector.  Then we impose two renormalization conditions
\bea
\hat H|\hat \Omega\rangle^s&=&0\nonumber\\
\hat H|\vp_0\rangle^s&=&\omega_{\vp_0}|\vp_0\rangle^s\hsp \omega_\vp=\sqrt{m^2+\vp^2}. \label{rc3h}
\eea
Here $\vp_0$ plays the role of a renormalization scale and can be chosen at will.  We will generally choose $\vp_0=\vec{0}.$  Note that if the potential is Lorentz-invariant, for example if there are no impurities, then the different renormalization scales are related by a Lorentz transformation and so, if the renormalization condition is satisfied at $\vp_0$, it will be satisfied at all momenta.


\subsection{Counterterm in the Displacement Operator}
Of course, there are two vacua, each with the expectation value ${}^s\langle \Omega|\phi(x){\ovac}^s$ shifted by some vacuum expectation value.  We study these by shifting $\phi(\vx)$, which can be achieved using the displacement operator
\beq
\dv={\rm{Exp}}\left[-iv\int d^{\dim}\vx \pi(\vx) 
\right]\hsp v=\sum_{i=0}^\infty v_i
\eeq
where $v_i$ is of order $O(\lambda^{(i-1)/2})$.  Let us choose one vacuum, for example the right vacuum in which 
\beq
v_0=\frac{m_0}{\sqrt{2\lambda}}.
\eeq
In this vacuum, we may define the right vacuum Hamiltonian $H$ by
\beq
H[\phi,\pi]=\dv^\dag \hat{H}[\phi,\pi]\dv=\hat{H}[\phi+v,\pi]. \label{hr}
\eeq
If $|\psi\rangle^s$ is an eigenstate of $\hat H$ then
\beq
|\psi\rangle=\dv^\dag|\psi\rangle^s
\eeq
is an eigenstate of $H$ with the same eigenvalue.  Therefore we can forget the $s$ superscripts, and work in terms of the states $|\psi\rangle$, rewriting our renormalization conditions (\ref{rc3h}) in terms of $H$. 
\bea
H|\hat \Omega\rangle&=&0\nonumber\\
H|\vp_0\rangle&=&\omega_{\vp_0}|\vp_0\rangle\hsp \omega_\vp=\sqrt{m^2+\vp^2}. \label{rc3}
\eea

To fix $v$, we will impose a \hyperlink{third}{third} renormalization condition \red{Coleman (14.41)}
\beq
\langle \Omega|\phi(x){\ovac}=0. \label{tad}
\eeq

Now that all the definitions have been made, we can calculate the counterterms to any order in $\lambda$ and the renormalized mass $m$.


\subsection{Expanding the Hamiltonian}

\subsubsection{The Expansion}

The first step is to calculate $H[\phip,\pip]$ using Eq.~(\ref{hr}) to obtain
\beq
H[\phip,\pip]=\hat{H}[\phip+v,\pip]=\int d^{\dim}\vx \ch(\vx)
\eeq
where, using (\ref{chhat})
\beq
{\mathcal{H}}(\vx)=\frac{\pi^{2}(\vx)+\partial_i \phip(\vx)\partial_i \phip(\vx)}{2}-\frac{m^2(\phip(\vx)+v)^{2}}{4}+\frac{\lambda  (\phip(\vx)+v)^{4}}{4}
+\delta m^2\frac{(\phip(\vx)+v)^{2}}{4}+A.
\eeq

We will expand the Hamiltonian order by order
\beq
H=\sum_{i=0}^\infty H_i\hsp H_i=\int d^{\dim}\vx \ch_i(\vx)
\eeq
where $H_i$ is of order $O(\lambda^{i/2-1})$.  We also expand

\beq
\delta m^2=m^2-
m_0^2=\sum_{i=1}^\infty \delta m^2_i\hsp A=\sum_{i=0}^\infty A_i
\eeq
where $\delta m^2_i$ and $A_i$ are of order $O(\lambda^{i/2})$ or order $O(\lambda^{i/2-1})$ respectively. \par 
\gre{how about if we add expansion with $\phi^n$ firstly see appendix} 

One easily finds
\beq
\ch_0=-\frac{m^2v_0^2}{4}+\frac{\lambda v_0^4}{4}+A_0=-\frac{m^4}{16\lambda}+A_0.
\eeq

This is a $c$-number and so the \hyperlink{first}{first} renormalization condition in (\ref{rc3}) implies $\ch_0=0$.  We conclude
\beq
A_0=\frac{m^4}{16\lambda}.
\eeq
\subsubsection{Order $O(1/\sl)$}

At the next order one finds
\bea
\ch_1&=&-\frac{m^2v_0(\phip(\vx)+v_1)}{2}+\lambda  v_0^3(\phip(\vx)+v_1)+\delta m^2_1 \frac{v_0^2}{4}+A_1
=A_1+\delta m^2_1 \frac{m^2}{8\lambda}.
\eea
The \hyperlink{first}{first} renormalization condition implies that the $c$-number term in $\ch_1$ vanishes.  This is consistent with the choice of scheme
\beq
\delta m^2_1=A_1=0.
\eeq
This choice appears to be consistent with our renormalization conditions.  The correction $v_1$ did not appear in the final expression, and, since we want quantum corrections to be suppressed by powers of $\lambda$, we will also set $v_1=0$.

\subsubsection{Order $O(\lambda^0)$}


The renormalization conditions will impose that $v_2$ have a finite, nonzero value.  Of course, since it is finite in 2+1 dimensions, unlike 3+1, we could instead choose $v_2=0$ which would simplify the calculations below and resemble our treatment of the 1+1 dimensional case.  However, so as to always follow our renormalization conditions, we will not set $v_2=0$.

Then the $O(\lambda^0)$ terms in the Hamiltonian density are
\bea
{\mathcal{H}}_2(\vx)&=&\frac{\pi^{2}(\vx)+\partial_i \phip(\vx)\partial_i \phip(\vx)}{2}-\frac{m^2\phi^{2}(\vx)}{4}-\frac{m^2 v_2 v_0}{2}+\frac{3\lambda v_0^2 \phi^{2}(\vx)}{2}+\lambda v_0^3 v_2+A_2+\delta m^2_2 \frac{v_0^2}{4}\nonumber\\
&=&\frac{\pi^{2}(\vx)+\partial_i \phip(\vx)\partial_i \phip(\vx)}{2}+\frac{m^2\phi^{2}(\vx)}{2}+A_2+\delta m^2_2 \frac{m^2}{8\lambda}.
\eea

Again the \hyperlink{first}{first} renormalization condition implies that the scalar part vanishes
\beq
A_2=-\delta m^2_2 \frac{m^2}{8\lambda}.
\eeq
This leaves
\beq
{\mathcal{H}}_2(\vx)=\frac{\pi^{2}(\vx)+\partial_i \phip(\vx)\partial_i \phip(\vx)}{2}+\frac{m^2\phi^{2}(\vx)}{2}. \label{freemass}
\eeq
In other words, the leading order Hamiltonian describes a free scalar theory with mass $m$.  We impose that the normal ordering $::_a$ is defined at this mass scale $m$.

To proceed, we will decompose the Schrodinger picture field $\phi(\vx)$ and its dual momentum $\pi(\vx)$ into plane waves
\beq
\phi(\vx)=\pinvp{\dim} e^{-i\vx\cdot\vp}\left(A^\ddag_p+\frac{A_{-p}}{2\omega_p}\right)\hsp \,\,
\pi(\vx)=i\pinvp{\dim} e^{-i\vx\cdot\vp}\left(\omega_p A^\ddag_p-\frac{A_{-p}}{2}\right)\hsp\,\, A^\ddag_p=\frac{A^\dag_p}{2\omega_p}.
\eeq
The canonical commutation relations of $\phi(\vx)$ and $\pi(\vx)$ imply that $A^\ddag$ and $A$ satisfy an oscillator algebra
\beq
[A_{\vp_1},A^\ddag_{\vp_2}]=(2\pi)^\dim \delta^{\dim}(\vp_1-\vp_2).
\eeq
Then the order $O(\lambda^0)$ Hamiltonian is
\beq
H_2=
\pinvp{\dim} \omega_{\vp} A^\ddag_p A_{-p}. \label{h2}
\eeq


\subsubsection{Order $O(\sl)$}

We will also need the next order Hamiltonian density
\bea
{\mathcal{H}}_3(\vx)&=&-\frac{m^2v_2\phi(\vx)}{2}-\frac{m^2v_0 v_3}{2}
+3\lambda v_0^2 v_2 \phi(\vx)+\lambda v_0 \phi^{3}(\vx)+\lambda v_0^3 v_3\nonumber\\
&&+\delta m^2_3 \frac{v_0^2}{4} 
+A_3
+\delta m^2_2 v_0 \frac{\phip(\vx)}{2}\nonumber\\
&=&\frac{m}{2}\left(2mv_2+\frac{\delta m^2_2}{\sqrt{2\lambda}} 
\right)\phi(\vx)
+\sqrt\frac{\lambda}{2} m \phi^{3}(\vx)+A_3+\delta m^2_3 \frac{m^2}{8\lambda}\nonumber
\eea

Again the \hyperlink{first}{first} renormalization condition implies that the $c$-number term vanishes
\beq
A_3=-\delta m^2_3 \frac{m^2}{8\lambda}
.
\eeq
The \hyperlink{third}{third} renormalization condition (\ref{tad}) implies that the tadpoles vanish, which at leading order implies that the $\phip$ terms in $H$ vanish
\beq
v_2=-\frac{\delta m^2_2}{2m\sqrt{2\lambda}}.
\eeq
Intuitively this is the statement that if the meson mass increases, corresponding to a more tachyonic mass at the false vacuum, then the vacuum expectation value of $\phi(\vx)$ will decrease.  In fact, it is simply the second term in an expansion of $v=\sqrt{m^2-\delta m^2}/\sqrt{2\lambda}$.  At this point it is consistent to set $v_2=\delta m^2_2=0$.  This choice however will fall afoul of the \hyperlink{second}{second} renormalization condition in Eq.~(\ref{rc3}).

Once the renormalization conditions are imposed, the leading order interaction Hamiltonian is
\beq
{\mathcal{H}}_3(\vx)=\lambda v_0 \phi^{3}(\vx)=m\sqrt\frac{\lambda}{2}\phi^{3}(\vx).
\eeq
The corresponding term in the Hamiltonian is
\bea
H_3&=&m\sqrt\frac{\lambda}{2}
\int d^{\dim}\vx \int\frac{d^{\dim}\vp_1 d^{\dim}\vp_2 d^{\dim}\vp_3}{(2\pi)^{\dimthree}} e^{-i\vx\cdot (\vp_1+\vp_2+\vp_3)}\left(A^\ddag_{\vp_1}+\frac{A_{-\vp_1}}{2\omega_{\vp_1}}\right)\\
&&\times \left(A^\ddag_{\vp_2}+\frac{A_{-\vp_2}}{2\omega_{\vp_2}}\right)\left(A^\ddag_{\vp_3}+\frac{A_{-\vp_3}}{2\omega_{\vp_3}}\right)\nonumber\\
&=&m\sqrt\frac{\lambda}{2}   \int\frac{d^{\dim}\vp_1 d^{\dim}\vp_2}{(2\pi)^{\dimtwo}}\left(A^\ddag_{\vp_1}+\frac{A_{-\vp_1}}{2\omega_{\vp_1}}\right)\left(A^\ddag_{\vp_2}+\frac{A_{-\vp_2}}{2\omega_{\vp_2}}\right)\left(A^\ddag_{-\vp_1-\vp_2}+\frac{A_{-\vp_1-\vp_2}}{2\omega_{\vp_1+\vp_2}}\right).\nonumber
\eea

\subsubsection{Order $O(\lambda)$}

Finally, to compute the leading order counterterms, we write the $O(\lambda)$ Hamiltonian density
\bea
{\mathcal{H}}_4(\vx)&=&-\frac{v_3 m^2}{2}\phi(x)-\frac{m^2 v_2^2}{4}-\frac{m^2 v_0 v_4}{2}+3\lambda v_0v_2\phi^2(\vx)+\frac{\lambda}{4}\phi^4(\vx)+\frac{3\lambda v_0^2 v_2^2}{2}\\&&+{\lambda v_0^3 v_4}+3\lambda v_0^2 v_3 \phi(\vx)+A_4+\frac{\delta m^2_2}{4}\phi^2(\vx)+\frac{\delta m^2_2 v_0v_2}{2}+\frac{\delta m^2_3v_0}{2}\phi(\vx)+\frac{\delta m^2_4 v_0^2}{4}\nonumber\\
&=&\frac{\lambda}{4}\phi^4(\vx)-\frac{\delta m^2_2}{2}\phi^2(\vx)+\frac{m}{2}\left(\frac{\delta m^2_3}{\sqrt{2\lambda}}+\frac{v_3 m}{2}\right)\phi(x)+A_4-\frac{\left(\delta m^2_2\right)^2}{16\lambda}+
\frac{m^2\delta m^2_4 }{8\lambda}.\nonumber
\eea

Now the renormalization condition does not imply that the $c$-number term vanishes, as there will be a contribution to the energy at $O(\lambda)$ from a loop diagram.  However, at this order it does still imply that the linear term vanishes
\beq
v_3=-\frac{\delta m^2_3}{2m\sqrt{2\lambda}}.
\eeq
In all, we conclude that the $O(\lambda)$ Hamiltonian density is
\beq
{\mathcal{H}}_4(\vx)=\frac{\lambda}{4}\phi^4(\vx)-\frac{\delta m^2_2}{2}\phi^2(\vx)+\hat{A}_4\hsp \hat{A}_4=A_4-\frac{\left(\delta m^2_2\right)^2}{16\lambda}+
\frac{m^2\delta m^2_4 }{8\lambda}.
\eeq

\section{Fixing the Counterterms at order $O(\lambda)$}
We will fix the counterterms by finding the vacuum and one-meson states while imposing the renormalization conditions.

\subsection{Finding the Vacuum}

We expand all states $|\Psi\rangle$ in powers of the coupling
\beq
|\Psi\rangle=\sum_{i=0}^\infty |\Psi\rangle_i
\eeq
where $|\Psi\rangle_i$ is suppressed by a factor of $\lambda^{i/2}$.

Let ${\ovac}$ be the vacuum state, defined by the Hamiltonian eigenstate whose leading contribution satisfies
\beq
A_p{\ovac}_0=0\hsp {}_0\langle \Omega{\ovac}_0=1.
\eeq
Note that the \hyperlink{first}{first} renormalization condition can only be satisfied if ${\ovac}$ is a Hamiltonian eigenstate, and so we may apply the renormalization condition below with no need to impose the eigenstate condition separately.

\subsubsection{Order $O(\lambda^0)$}

At order $O(\lambda^0)$ this indeed is a Hamiltonian eigenstate, as ${\ovac}$ reduces to ${\ovac}_0$ and
\beq
H_2{\ovac}_0=0.
\eeq
This also implies that the \hyperlink{first}{first} renormalization condition is satisfied at order $O(\lambda^0)$.

\subsubsection{Order $O(\sl)$}

Let us define the state
\beq
|\vp\rangle_0=A^\ddag_\vp{\ovac}_0
\eeq
whose adjoint is
\beq
{}_0\langle \vp|={}_0\langle \Omega| \frac{A_\vp}{2\omega_\vp}.
\eeq
Note that 
\beq
{}_0\langle \vp_1|\vp_2\rangle_0=\frac{(2\pi)^\dim\delta^{\dim}(\vp_1-\vp_2)}{2\omega_{\vp_1}}.
\eeq
This is readily generalized to states
\beq
|\vp_1\cdots \vp_n\rangle_0=A^\ddag_{\vp_1}\cdots A^\ddag_{\vp_n}{\ovac}_0.
\eeq

Now the renormalization condition
\beq
H{\ovac}=0
\eeq
at order $O(\sl)$ implies
\beq
H_2{\ovac}_1=-H_3{\ovac}_0.
\eeq
As a result of the normal ordering, the only term in $H_3$ that contributes to the last expression is that with only factors of $B^\ddag$ and no $B$
\beq
H_3{\ovac}_0=m
\sqrt{\frac{\lambda}{2}}\int\frac{d^{\dim}\vp_1 d^{\dim}\vp_2}{(2\pi)^{\dimtwo}}|\vp_1,\vp_2,-\vp_1-\vp_2\rangle_0.
\eeq
We then find
\beq
{\ovac}_1=-m\sqrt{\frac{\lambda}{2
}}\int\frac{d^{\dim}\vp_1 d^{\dim}\vp_2}{(2\pi)^{\dimtwo}}\frac{|\vp_1,\vp_2,-\vp_1-\vp_2\rangle_0}{\omega_{\vp_1}+\omega_{\vp_2}+\omega_{\vp_1+\vp_2}}. \label{om1}
\eeq

\subsubsection{Order $O(\lambda)$} 

At next order, the \hyperlink{first}{first} renormalization condition is
\beq
H_4{\ovac}_0+H_3{\ovac}_1+H_2{\ovac}_2=0. \label{omtwo}
\eeq

The states $|\vp_1\cdots \vp_n\rangle_0$ span the multimeson Fock space.  Now let us decompose each coefficient in this basis
\beq
|\Psi\rangle_j=\sum_{n=0}^\infty |\Psi\rangle^n_j
\eeq
where $|\Psi\rangle^n_j$ is the projection  of $|\Psi\rangle_j$ onto the $n$-meson Fock space.

The only contribution to the leading order counterterms will arise from the component  ${\ovac}^0_2$ of ${\ovac}_2$ parallel to ${\ovac}_0$.  The term $H_4$ does not contribute to this component except for the $\hat A_4$ term, as a result of the normal ordering.  Therefore we find
\beq
\int d^{\dim}\vx \hat A_4{\ovac}_0+H_2{\ovac}_2^0=-m
\sqrt{\frac{\lambda}{2}}\int d^{\dim}\vx\int\frac{d^{\dim}\vp_1 d^{\dim}\vp_2 d^{\dim}\vp_3}{(2\pi)^{\dimthree}}e^{-i\vx\cdot(\vp_1+\vp_2+\vp_3)}\frac{B_{-\vp_1}B_{-\vp_2}B_{\vp_3}}{8\omega_{\vp_1}\omega_{\vp_2}\omega_{\vp_3}}{\ovac}_1. \label{id}
\eeq
We cannot perform the $\vx$ integral yet, because the term on the right has a constant energy density and so the total energy would be divergent.  We need to combine the integrands on the left and right hand sides of the equation before the integral may be performed.

Note that
\beq
H_2{\ovac}_n^0=0
\eeq
as these states contain no mesons.  So the second term on the left hand side vanishes and ${\ovac}_2^0$ is arbitrary.  The freedom to choose ${\ovac}_2^0$ in fact is just the freedom to choose the normalization of ${\ovac}.$  We will fix the convention
\beq
{\ovac}_{i>1}^0=0.
\eeq

The right hand side may be evaluated using (\ref{om1})
\beq
\int d^{\dim}\vx \hat A_4=\frac{3m^2\lambda}{8
}\int d^{\dim}\vx \int\frac{d^{\dim}\vp_1 d^{\dim}\vp_2}{(2\pi)^{\dimtwo}}\frac{1}{\omega_{\vp_1}\omega_{\vp_2}\omega_{\vp_1+\vp_2}({\omega_{\vp_1}+\omega_{\vp_2}+\omega_{\vp_1+\vp_2}})}. 
\eeq
As the integrands are homogeneous, we may identify them
\beq
\hat A_4=\frac{3m^2\lambda}{8
} \int\frac{d^{\dim}\vp_1 d^{\dim}\vp_2}{(2\pi)^{\dimtwo}}\frac{1}{\omega_{\vp_1}\omega_{\vp_2}\omega_{\vp_1+\vp_2}({\omega_{\vp_1}+\omega_{\vp_2}+\omega_{\vp_1+\vp_2}})}. \label{a4}
\eeq
In one spatial dimension this integral would converge, but in more dimensions we need to regularize, for example by choosing a momentum cutoff $\Lambda$
\beq
\hat A_4^\Lambda=\frac{3m^2\lambda}{8} \int_{|\vp_1|,|\vp_2|,|\vp_1+\vp_2|\leq \Lambda}\frac{d^{\dim}\vp_1 d^{\dim}\vp_2}{(2\pi)^{\dimtwo}}\frac{1}{\omega_{\vp_1}\omega_{\vp_2}\omega_{\vp_1+\vp_2}({\omega_{\vp_1}+\omega_{\vp_2}+\omega_{\vp_1+\vp_2}})}.
\eeq

\subsubsection{The Norm and Infrared Regulator}
With these coefficients, we can compute the norm of ${\ovac}$ up to order $O(\lambda)$
\beq
\langle \Omega{\ovac}={}_0\langle \Omega{\ovac}_0+{}_1\langle \Omega{\ovac}_1
=1+{}_1\langle \Omega{\ovac}_1.
\eeq

The last term is divergent.  Intuitively, the problem is as follows.  The vacuum ${\ovac}$ is a Hamiltonian eigenstate which is an infinite superposition of meson eigenstates.  A generic member of this superposition contains a roughly constant density $\rho$ of virtual mesons.  $\rho$ is suppressed by some power of $\lambda$.  However, the number of mesons in a volume $V$ is of order $\rho V$ in such a generic state.   Thus if one considers a volume $V$ much bigger than $1/\rho$ then the number of mesons is large, and the higher order terms dominate the perturbative expansion of the norm.

The solution to this problem is that an infrared cutoff $V$ needs to be fixed such that $V\ll 1/\rho$.  In practice, when the infrared regularization is removed, it is sufficient to take $V m^{\dim}\rightarrow\infty$ with $V\lambda m^3\rightarrow 0$.  This will guarantee that lower order terms in the perturbative $\lambda$ expansion dominate the norm.

Including this regulator, the interaction becomes
\bea
H_3&=&m\sqrt\frac{\lambda}{2}
\int_V d^{\dim}\vx \int\frac{d^{\dim}\vp_1 d^{\dim}\vp_2 d^{\dim}\vp_3}{(2\pi)^{\dimthree}} e^{-i\vx\cdot (\vp_1+\vp_2+\vp_3)}\left(A^\ddag_{\vp_1}+\frac{A_{-\vp_1}}{2\omega_{\vp_1}}\right)\\
&&\times \left(A^\ddag_{\vp_2}+\frac{A_{-\vp_2}}{2\omega_{\vp_2}}\right)\left(A^\ddag_{\vp_3}+\frac{A_{-\vp_3}}{2\omega_{\vp_3}}\right)\nonumber
\eea
so that
\beq
H_3{\ovac}_0=m
\sqrt{\frac{\lambda}{2}}\int_V d^{\dim}\vx\int\frac{d^{\dim}\vp_1 d^{\dim}\vp_2 d^{\dim}\vp_3}{(2\pi)^{\dimthree}} e^{-i\vx\cdot (\vp_1+\vp_2+\vp_3)}|\vp_1,\vp_2,\vp_3\rangle_0.
\eeq
We then find that infrared regularized $O(\sl)$ correction to the vacuum state
\beq
{\ovac}_1=-m\sqrt{\frac{\lambda}{2
}}\int_V d^{\dim}\vx\int\frac{d^{\dim}\vp_1 d^{\dim}\vp_2 d^{\dim}\vp_3}{(2\pi)^{\dimthree}} e^{-i\vx\cdot (\vp_1+\vp_2+\vp_3)}\frac{|\vp_1,\vp_2,\vp_3\rangle_0}{\omega_{\vp_1}+\omega_{\vp_2}+\omega_{\vp_3}}.
\eeq
The norm squared is then
\beq
{}_1\langle \Omega{\ovac}_1=\frac{3\lambda m^2}{16
}\int_V d^{\dim}\vx_1 \int_V d^{\dim}\vx_2\int\frac{d^{\dim}\vp_1 d^{\dim}\vp_2 d^{\dim}\vp_3}{(2\pi)^{\dimthree}} \frac{e^{-i(\vx_1-\vx_2)\cdot (\vp_1+\vp_2+\vp_3)}}{\omega_{\vp_1}\omega_{\vp_2}\omega_{\vp_3}({\omega_{\vp_1}+\omega_{\vp_2}+\omega_{\vp_3}})^2}.
\eeq

Now, dropping a term suppressed by $V m^{\dim}$, we may perform the $\vx_2$ integration with an infinite regulator, yielding $(2\pi)^3\delta^{\dim}(\vp_1+\vp_2+\vp_3)$ and so
\bea
{}_1\langle \Omega{\ovac}_1&=&m^2\frac{3\lambda}{16
}\int_V d^{\dim}\vx \int\frac{d^{\dim}\vp_1 d^{\dim}\vp_2 }{(2\pi)^{2}} \frac{1}{\omega_{\vp_1}\omega_{\vp_2}\omega_{\vp_1+\vp_2}({\omega_{\vp_1}+\omega_{\vp_2}+\omega_{\vp_1+\vp_2}})^2}\\
&=&m^2\frac{3V\lambda}{16
} \int\frac{d^{\dim}\vp_1 d^{\dim}\vp_2 }{(2\pi)^{2}} \frac{1}{\omega_{\vp_1}\omega_{\vp_2}\omega_{\vp_1+\vp_2}({\omega_{\vp_1}+\omega_{\vp_2}+\omega_{\vp_1+\vp_2}})^2}\nonumber
\eea
Now this term has a coefficient $V\lambda$, which we recall tends to zero, in units of $m$, as the infrared regulator is removed.  In three or more spatial dimensions, this expression will also enjoy ultraviolet divergences which we may resolve using
a cutoff when any $|p_i|$ equals $\Lambda$, as in Ref.~\cite{andy}.

In summary, we conclude that once the infrared regulator is removed with the limits as specified above
\beq
\langle \Omega{\ovac}=1.
\eeq


\subsection{Finding the One Meson States at One Loop}

To fix all of the counterterms, we do not need to construct a complete basis of states.  It suffices to construct the vacuum state ${\ovac}$ and the one-meson states $|\vp\rangle$.  Extending these constructions to higher order allows one to compute the counterterms to higher orders.

We define the one meson state $|\vp\rangle$ to be the Hamiltonian eigenstate such that
\beq
|\vp\rangle_0=A^\ddag_\vp {\ovac}_0
\eeq
and also the one meson terms at higher order $i>0$ vanish
\beq
|\vp\rangle_i^1=0. \label{max}
\eeq

Our strategy will be as follows.  First, we will construct the state $|\vp\rangle$ order by order using the \hyperlink{second}{second} renormalization condition in Eq.~(\ref{rc3}).  

At leading order the second condition is
\beq
H_2|\vp_0\rangle_0=\omega_{\vp_0}|\vp_0\rangle_0.
\eeq
We may compute the left hand side, using Eq.~(\ref{h2}) and we indeed produce the right hand side.  Thus the renormalization condition is authomatically satisfied at leading order. 

At order $O(\sl)$ the renormalization condition is
\bea
(H_2-\omega_{\vp_0})|\vp_0\rangle_1&=&-H_3|\vp_0\rangle_0\\
&=&-3m\sqrt\frac{\lambda}{2}
\int_V d^{\dim}\vx \int\frac{d^{\dim}\vp_1 d^{\dim}\vp_2}{(2\pi)^{\dimtwo}} \frac{e^{-i\vx\cdot (\vp_1+\vp_2-\vp_0)}}{2\omega_{\vp_0}}|\vp_1\vp_2\rangle_0\nonumber\\
&&-m\sqrt\frac{\lambda}{2}
\int_V d^{\dim}\vx \int\frac{d^{\dim}\vp_1 d^{\dim}\vp_2 d^{\dim}\vp_3}{(2\pi)^{\dimthree}} e^{-i\vx\cdot (\vp_1+\vp_2+\vp_3)}|\vp_0\vp_1\vp_2\vp_3\rangle_0.\nonumber
\eea
The leading correction to the one-meson state is therefore
\bea
|\vp_0\rangle_1&=&-\frac{3}{2}m\sqrt\frac{\lambda}{2}
\int_V d^{\dim}\vx \int\frac{d^{\dim}\vp_1 d^{\dim}\vp_2}{(2\pi)^{\dimtwo}} \frac{e^{-i\vx\cdot (\vp_1+\vp_2-\vp_0)}}{\omega_{\vp_0}(\omega_{\vp_1}+\omega_{\vp_2}-\omega_{\vp_0})}|\vp_1\vp_2\rangle_0\\
&&-m\sqrt\frac{\lambda}{2}
\int_V d^{\dim}\vx \int\frac{d^{\dim}\vp_1 d^{\dim}\vp_2 d^{\dim}\vp_3}{(2\pi)^{\dimthree}} \frac{e^{-i\vx\cdot (\vp_1+\vp_2+\vp_3)}}{(\omega_{\vp_1}+\omega_{\vp_2}+\omega_{\vp_3})}|\vp_0\vp_1\vp_2\vp_3\rangle_0.\nonumber
\eea
When the IR regulator is taken to infinity this reduces to
\bea
|\vp_0\rangle_1&=&-\frac{3}{2}m\sqrt\frac{\lambda}{2}
\int\frac{d^{\dim}\vp_1}{(2\pi)^{\dim}} \frac{1}{\omega_{\vp_0}(\omega_{\vp_1}+\omega_{\vp_0-\vp_1}-\omega_{\vp_0})}|\vp_1,\vp_0-\vp_1\rangle_0\\
&&-m\sqrt\frac{\lambda}{2}
\int\frac{d^{\dim}\vp_1 d^{\dim}\vp_2}{(2\pi)^{\dimtwo}} \frac{1}{(\omega_{\vp_1}+\omega_{\vp_2}+\omega_{\vp_1+\vp_2})}|\vp_0,\vp_1,\vp_2,-\vp_1-\vp_2\rangle_0.\nonumber
\eea

At order $O(\lambda)$ we again impose the \hyperlink{second}{second} renormalization condition, restricting to the one-meson Fock space
\bea
H_4|\vp_0\rangle_0{\big{\vert}}_{\rm{one-meson}}&=&-H_3|\vp_0\rangle_1{\big{\vert}}_{\rm{one-meson}}. \label{renm3}
\eea
More precisely, restricting our attention to the one-meson terms
\bea
H_4|\vp_0\rangle_0{\big{\vert}}_{\rm{one-meson}}&=&\int_V d^{\dim}\vx \left( \hat A_4-\delta m^2_2 \frac{:\phi^{2}(\vx):_a}{2}\right)|\vp_0\rangle_0{\big{\vert}}_{\rm{one-meson}}\label{h41}\\
&=&\int_V d^{\dim}\vx \left( \hat A_4-{\delta m^2_2}\int\frac{d\vp_1 d\vp_2}{(2\pi)^{\dimtwo}}e^{-i\vx\cdot(\vp_1+\vp_2)}\frac{B^\ddag_{\vp_1} B_{-\vp_2}}{2\omega_{\vp_2}}\right) |\vp_0\rangle_0{\big{\vert}}_{\rm{one-meson}}\nonumber\\
&=&\left(
-\frac{\delta m^2_2}{2\omega_{\vp_0}}+\int_V d^{\dim}\vx  \hat A_4\right) |\vp_0\rangle_0\nonumber
\eea
and
\bea
H_3|\vp_0\rangle_1{\big{\vert}}_{\rm{one-meson}}&=&-\frac{9m^2\lambda}{8
\omega_{\vp_0}}\left(\int\frac{d\vp_1}{(2\pi)^{\dim}}   \frac{1}{\omega_{\vp_1}\omega_{\vp_0-\vp_1}(\omega_{\vp_1}+\omega_{\vp_0-\vp_1}-\omega_{\vp_0})}\right)|\vp_0\rangle_0\label{h31}\\
&&-\frac{9m^2\lambda}{8
\omega_{\vp_0}}\left(\int\frac{d\vp_1}{(2\pi)^{\dim}}   \frac{1}{\omega_{\vp_1}\omega_{\vp_0+\vp_1}(\omega_{\vp_1}+\omega_{\vp_0+\vp_1}+\omega_{\vp_0})}\right)|\vp_0\rangle_0\nonumber\\
&&-\frac{3m^2\lambda}{8
}\int d^{\dim}\vx\left(\int\frac{d\vp_1d\vp_2}{(2\pi)^{\dimtwo}}  
\frac{1}{\omega_{\vp_1}\omega_{\vp_2}\omega_{\vp_1+\vp_2}(\omega_{\vp_1}+\omega_{\vp_2}+\omega_{\vp_1+\vp_2})}\right)|\vp_0\rangle_0.\nonumber
\eea
We recognize the coefficient in the last line as the spatial integral over $\hat A_4$, as given in Eq.~(\ref{a4}).  The last line therefore cancels the last term on the last line of Eq.~(\ref{h41}).

Overall, inserting Eqs.~(\ref{h41}) and (\ref{h31}) into the renormalization condition (\ref{renm3}) and identifying the coefficients of $|\vp_0\rangle_0$, we find
\beq
\delta m^2_2=
-\frac{9m^2\lambda}{2}
\int\frac{d^{\dim}\vp_1}{(2\pi)^{\dim}} \frac{\omega_{\vp_1}+\omega_{\vp_0-\vp_1}}{\omega_{\vp_1}\omega_{\vp_0-\vp_1}((\omega_{\vp_1}+\omega_{\vp_0-\vp_1})^2-\omega^2_{\vp_0})}.
\eeq
Again this would diverge in three spatial dimensions, but not in less.  In one spatial dimension, the integral would equal $-2/(3\sqrt{3} m^3)$ and so the right hand side is $-12\sqrt 3 \lambda
$.  One then would find 
$\delta m^2_2= -12\sqrt 3 \lambda$.

We conclude that, here at one loop, we did not need mass renormalization.  However, the counterterm at two loops will be infinite, and so mass renormalization will eventually be necessary.  For simplicity, we could set $\delta m^2_2=0$, and we would not find any divergences.  However we will chose to impose our renormalization conditions consistently at all loops, whether they are necessary or not.


How should $\vp_0$ be chosen?  We could impose the renormalization condition at all values of $\vp_0$.  To do this, we would need to reparametrize the names of the states $|\vp\rangle_0$.  Indeed, it is not hard to show that with such a reparametrization, the state $|\vp\rangle$ would have momentum $\vp$.  

In this note, we will opt for a simpler alternative.  We only apply the \hyperlink{third}{third} renormalization condition to the case $\vp_0=0$.

In either case, whether one choses to reparameterize $\vp$ or else only apply the renormalization condition at $\vp=0$, one only fixes the counterterm at $\vp=0$ since this extremum is parameterization invariant.  Therefore one obtains
\bea
\delta m^2_2&=&-
{9m^2\lambda}
\int\frac{d^{\dim}\vp_1}{(2\pi)^{\dim}} \frac{1}{\omega_{\vp_1}(3m^2+4\vp_1^2)}=-
{9m^2\lambda}
\int_0^\infty\frac{d p_r}{2\pi} \frac{p_r}{\omega_{p_r}(3m^2+4p_r^2)}\nonumber\\
&=&-\frac{9\ {\rm{ln}}(3)}{8\pi} \lambda m
. \label{dm2}
\eea
In one spatial dimension this would become
\beq
\delta m^2_2=-\frac{\sqrt 3}{2} \lambda.
\eeq

\section{Fixing the Counterterms at Order $O(\lambda^2)$}

We are interested in the leading divergence to the mass renormalization.  As the goal of the present note is to see whether it will pose any problems in our consideration of the kink sector.  This divergence arises at order $O(\lambda^2)$.  And so we will need to extend the considerations above to this order.

\subsection{The Hamiltonian at Higher Orders}

\subsubsection{Order $O(\lambda^{3/2})$}

At this order the Hamiltonian density is
\bea
\ch_5&=&-\frac{m^2(v_4\phi(\vx)+v_5v_0+v_3 v_2)}{2}+\lambda v_2 \phi^3(\vx)+{3}\lambda v_0v_3\phi^2(\vx)\\
&&+3\lambda\left(v_0^2 v_4+v_0 v_2^2 \right)\phi(x)+\lambda\left(v_0^3 v_5+3v_0^2 v_2 v_3\right)+\delta m^2_2\left(\frac{v_2\phi(\vx)+v_0 v_3}{2} \right)\nonumber\\
&&+\delta m^2_3 \left(\frac{\phi^2(\vx)}{4}+\frac{v_0v_2}{2} \right) +\delta m^2_4 \left(\frac{v_0 \phi(\vx)}{2} \right) + \delta m^2_5 \frac{v_0^2}{4}
+A_5\nonumber\\
&=&-m^2\left(\frac{v_4}{2}\phi(\vx)+v_5 \frac{m}{2\sqrt{2\lambda}}+\frac{\delta m^2_2\delta m^2_3}{16m^2\lambda}\right)-\sqrt\frac{\lambda}{2} \frac{\delta m^2_2}{2m} \phi^3(\vx)-\frac{3\delta m^2_3}{4}\phi^2(\vx)\nonumber\\
&&+3 \left(\frac{m^2}{2} v_4+\frac{\left(\delta m^2_2\right)^2}{8m\sqrt{2\lambda}} \right)\phi(x)
+\left(\frac{m^3}{2\sqrt{2\lambda}} v_5+\frac{3}{16\lambda} \delta m^2_2\delta m^2_3\right)\nonumber\\
&&
-\delta m^2_2\left({\frac{\delta m^2_2}{4 m\sqrt{2\lambda}}\phi(\vx)+\frac{\delta m^2_3}{8\lambda}} \right)\nonumber\\
&&+\delta m^2_3 \left(\frac{\phi^2(\vx)}{4}-\frac{\delta m^2_2}{8\lambda} \right) +\delta m^2_4 \left(\frac{m \phi(\vx)}{2\sqrt{2\lambda}} \right) + \delta m^2_5 \frac{m^2}{8\lambda}
+A_5\nonumber\\
&=&-\sqrt\frac{\lambda}{2} \frac{\delta m^2_2}{2m} \phi^3(\vx)-\frac{\delta m^2_3}{2}\phi^2(\vx)+T_5 \phi(\vx)+\hat{A}_5
\eea
where we have defined
\bea
T_5
&=&m^2 v_4+\frac{4m^2\delta m^2_4+\left(\delta m^2_2\right)^2}{8m\sqrt{2\lambda}}\\
\hat A_5&=&A_5+\frac{m^2\delta m^2_5}{8\lambda}-\frac{\delta m^2_2\delta m^2_3}{8\lambda}
.\nonumber
\eea

The \hyperlink{first}{first} renormalization condition implies $\hat A_5=0$.  However the same is not true of the tadpole $T_5$, as it may cancel a two-loop diagram containing a three-point and a four-point vertex.  Nonetheless, we conclude
\beq
v_4=\frac{T_5}{m^2}-\frac{\left(\delta m^2_2\right)^2}{8m^3\sqrt{2\lambda}}-\frac{\delta m^2_4}{2m\sqrt{2\lambda}}. \label{v4}
\eeq

\subsubsection{Order $O(\lambda^2)$}

Finally we need the Hamiltonian density at order $O(\lambda^2)$
\bea
\ch_6&=&-\frac{m^2(2v_5\phi(\vx)+2v_6v_0+2v_4 v_2+v_3^2)}{4}+\lambda v_3 \phi^3(\vx)+{3}\lambda \left(v_0v_4+\frac{v_2^2}{2}\right)\phi^2(\vx)\\
&&+3\lambda\left(v_0^2 v_5+2v_0 v_2 v_3 \right)\phi(x)+\lambda\left(v_0^3 v_6+3v_0^2 v_2 v_4+\frac{3}{2}v_0^2v_3^2+v_0 v_2^3\right)\nonumber\\
&&+\delta m^2_2\left(\frac{v_3\phi(\vx)+v_0 v_4}{2} +\frac{v_2^2}{4}\right)
+\delta m^2_3\left(\frac{v_2\phi(\vx)+v_0 v_3}{2} \right)\nonumber\\
&&+\delta m^2_4 \left(\frac{\phi^2(\vx)}{4}+\frac{v_0v_2}{2} \right) +\delta m^2_5 \left(\frac{v_0 \phi(\vx)}{2} \right) + \delta m^2_6 \frac{v_0^2}{4}
+A_6\nonumber\\
&=&-\sl \frac{\delta m^2_3}{2m\sqrt{2}} \phi^3(\vx) + \left(
\sqrt\frac{\lambda}{2}\frac{3 T_5}{m}-\frac{\delta m^2_4}{2}
\right)\phi^2(\vx)+T_6 \phi(\vx)+\hat A_6\nonumber
\eea
\gre{ we can simplify the $\phi^2$ term as $-\frac{\delta m_4^2}{2}\phi^2$ }
\gre{ we can ignore it if the $T_5\neq 0$ for general case}
where we will not need the expressions
\bea
T_6&=&m^2 v_5+\frac{\delta m^2_2\delta m^2_3}{4m \sqrt{2\lambda}}
+\frac{m\delta m^2_5}{2\sqrt{2\lambda}} \label{a6}
\\
\hat A_6&=& A_6
-\frac{\left(\delta m^2_3\right)^2}{16\lambda}
- \frac{\delta m^2_2\delta m^2_4}{8\lambda}  + \frac{m^2}{8\lambda}\delta m^2_6 
\nonumber
.\nonumber
\eea

\subsection{The Vacuum State}

\subsubsection{Order $O(\lambda)$}
The \hyperlink{first}{first} renormalization condition at order $O(\lambda)$ is Eq.~(\ref{omtwo}).  We have already imposed this equation in the zero meson sector.  Now we will impose it in the two, four and six meson sectors.

Projecting onto the two meson sector
\bea
H_4{\ovac}_0&=&-\frac{\delta m^2_2}{2}\int d\vx :\phi^2(\vx):_a{\ovac}_0
=-\frac{\delta m^2_2}{2}\int d\vx\int\frac{\ddp 1\ddp 2}{(2\pi)^4}e^{-i\vx\cdot(\vp_1+\vp_2)}|\vp_1\vp_2\rangle_0\nonumber\\
&=&-\frac{\delta m^2_2}{2}\int\frac{\ddp {}}{(2\pi)^2}|\vp,-\vp\rangle_0
\eea
and
\bea
H_3{\ovac}_1&=&\frac{3m}{4}\sqrt\frac{\lambda}{2}   \int\frac{d^{\dim}\vpp_1 d^{\dim}\vpp_2}{(2\pi)^{\dimtwo}}\frac{A^\ddag_{\vpp_1+\vpp_2}A_{\vpp_1}A_{\vpp_2}}{\omega_{\vpp_1}\omega_{\vpp_2}}
\left(-m\sqrt{\frac{\lambda}{2
}}\int\frac{d^{\dim}\vp_1 d^{\dim}\vp_2}{(2\pi)^{\dimtwo}}\frac{|\vp_1,\vp_2,-\vp_1-\vp_2\rangle_0}{\omega_{\vp_1}+\omega_{\vp_2}+\omega_{\vp_1+\vp_2}}
\right)\nonumber\\
&=&-\frac{9m^2\lambda}{4}\int\frac{\ddp{}}{(2\pi)^2}\left[\int\frac{\ddpp{}}{(2\pi)^2} \frac{1}{\omega_{\vpp}\omega_{\vp-\vpp}\left(\omega_\vpp+\omega_{\vp-\vpp}+\omega_\vp\right)}
\right]|\vp,-\vp\rangle_0
\eea
leading to
\beq
{\ovac}_2^2=\int\frac{\ddp{}}{(2\pi)^2}\left[\frac{\delta m^2_2}{2}+\frac{9m^2\lambda}{4}\int\frac{\ddpp{}}{(2\pi)^2} \frac{1}{\omega_{\vpp}\omega_{\vp-\vpp}\left(\omega_\vpp+\omega_{\vp-\vpp}+\omega_\vp\right)}
\right]\frac{|\vp,-\vp\rangle_0}{\omega_\vp}.
\eeq
One can see from the form of (\ref{dm2}) that although the two terms in square brackets are each finite in 2+1 dimensions, in 3+1 dimensions they would both logarithmically diverge but these logarithmic divergences would cancel.

Similarly, projecting onto the four meson sector
\bea
H_4\ovac_0&=&\frac{\lambda}{4}\int d^2\vx \int \frac{d^2\vp_1 \cdots d^2 \vp_4}{(2\pi)^8}e^{-i\vx\cdot(\vp_1+\vp_2+\vp_3+\vp_4)}|\vp_1\vp_2\vp_3\vp_4\rangle_0\nonumber\\
&=&\frac{\lambda}{4} \int \frac{d^2\vp_1 \cdots d^2 \vp_3}{(2\pi)^6}|\vp_1,\vp_2,\vp_3,-\vp_1-\vp_2-\vp_3\rangle_0
\eea
and
\bea
H_3{\ovac}_1&=&\frac{3m}{2}\sqrt\frac{\lambda}{2}   \int\frac{d^{\dim}\vpp_1 d^{\dim}\vpp_2}{(2\pi)^{\dimtwo}}\frac{A^\ddag_{\vpp_1}A^\ddag_{\vpp_2}A_{\vpp_1+\vpp_2}}{\omega_{\vpp_1+\vpp_2}}
\left(-m\sqrt{\frac{\lambda}{2
}}\int\frac{d^{\dim}\vp_1 d^{\dim}\vp_2}{(2\pi)^{\dimtwo}}\frac{|\vp_1,\vp_2,-\vp_1-\vp_2\rangle_0}{\omega_{\vp_1}+\omega_{\vp_2}+\omega_{\vp_1+\vp_2}}
\right)\nonumber\\
&=&-\frac{9m^2\lambda}{2}\int\frac{\ddp 1\ddp 2\ddp 3}{(2\pi)^6} \frac{|\vp_1,\vp_2,\vp_3,-\vp_1-\vp_2-\vp_3\rangle_0}{\omega_{\vp_1+\vp_2}\left(\omega_{\vp_1}+\omega_{\vp_2}+\omega_{\vp_1+\vp_2}\right)}
\eea
leading to
\beq
{\ovac}_2^4=\lambda\int\frac{\ddp 1\ddp 2\ddp 3}{(2\pi)^6} \left[\frac{9m^2}{2\omega_{\vp_1+\vp_2}\left(\omega_{\vp_1}+\omega_{\vp_2}+\omega_{\vp_1+\vp_2}\right)}-\frac{1}{4}\right] \frac{|\vp_1,\vp_2,\vp_3,-\vp_1-\vp_2-\vp_3\rangle_0}{(\omega_{\vp_1}+\omega_{\vp_2}+\omega_{\vp_3}+\omega_{\vp_1+\vp_2+\vp_3})}.
\eeq

Finally, projecting on to the six meson sector, one obtains
\bea
H_3{\ovac}_1&=&m\sqrt\frac{\lambda}{2}   \int\frac{d^{\dim}\vpp_1 d^{\dim}\vpp_2}{(2\pi)^{\dimtwo}}A^\ddag_{\vpp_1}A^\ddag_{\vpp_2}A^\ddag_{-\vpp_1-\vpp_2}
\left(-m\sqrt{\frac{\lambda}{2
}}\int\frac{d^{\dim}\vp_1 d^{\dim}\vp_2}{(2\pi)^{\dimtwo}}\frac{|\vp_1,\vp_2,-\vp_1-\vp_2\rangle_0}{\omega_{\vp_1}+\omega_{\vp_2}+\omega_{\vp_1+\vp_2}}
\right)\nonumber\\
&=&-\frac{m^2\lambda}{2}\int\frac{\ddp 1\ddp 2\ddp 3\ddp 4}{(2\pi)^8} \frac{|\vp_1,\vp_2,\vp_3,\vp_4,-\vp_1-\vp_2,-\vp_3-\vp_4\rangle_0}{\left(\omega_{\vp_1}+\omega_{\vp_2}+\omega_{\vp_1+\vp_2}\right)}
\eea
and so the last component in the $O(\lambda)$ state $\ovac_2$ is
\beq
\ovac_2^6=\frac{m^2\lambda}{2}\int\frac{\ddp 1\ddp 2\ddp 3\ddp 4}{(2\pi)^8} \frac{|\vp_1,\vp_2,\vp_3,\vp_4,-\vp_1-\vp_2,-\vp_3-\vp_4\rangle_0}{\left(\omega_{\vp_1}+\omega_{\vp_2}+\omega_{\vp_1+\vp_2}\right)\left( \omega_{\vp_1}+\omega_{\vp_2}+\omega_{\vp_1+\vp_2}+\omega_{\vp_3}+\omega_{\vp_4}+\omega_{\vp_3+\vp_4}
\right)}.
\eeq

\subsubsection{Order $O(\lambda^{3/2})$: The Tadpole}

The vacuum $\ovac$ enters in two renormalization conditions.  The \hyperlink{first}{first} renormalization condition will allow us to use it to obtain $A_6$.  On the other hand, the tadpole condition (\ref{tad}), which at this order is $\ovac_3^1=0$, will allow us to obtain $T_5$ and so $v_4$ as a function of $\delta m^2_4$.  The first calculation will depend on $\ovac_3^3$ while the second will depend on $\ovac_3^1$.  Therefore we must calculate these two contributions.


First let us evaluate the tadpole.  Projecting on to the one meson sector, the tree-level tadpole is
\beq
H_5\ovac_0=\dvx T_5 \phi(\vx)\ovac_0=T_5\dvx\int\ddp{}e^{-i\vx\cdot\vp}A^\ddag_\vp\ovac_0=T_5|\vp=\vec{0}\rangle_0
\eeq
where $|\vp=\vec{0}\rangle_0$ is $|\vp\rangle_0$ evaluated at $\vp=0$.  To this, we must add various perturbative corrections.  The $O(\lambda)$ interactions $H_4$ lead to two contributions
\bea
H_4\ovac_1^3&=&-m\sqrt{\frac{\lambda}{2
}}\int d^2\vx :\left[\frac{\lambda}{4}\phi^4(\vx)-\frac{\delta m^2_2}{2}\phi^2(\vx)+\hat{A}_4 \right]:_a\int\frac{d^{\dim}\vp_1 d^{\dim}\vp_2}{(2\pi)^{\dimtwo}}\frac{|\vp_1,\vp_2,-\vp_1-\vp_2\rangle_0}{\omega_{\vp_1}+\omega_{\vp_2}+\omega_{\vp_1+\vp_2}}\nonumber\\
&=&-m\sqrt{\frac{\lambda}{2
}} \left[\frac{\lambda}{8}\int\frac{\ddpp 1\ddpp 2\ddpp 3}{(2\pi)^6}\frac{A^\ddag_{\vpp_1}A_{\vpp_2}A_{\vpp_3}A_{\vpp_1-\vpp_2-\vpp_3}}{\omega_{\vpp_2}\omega_{\vpp_3}\omega_{\vpp_1+\vpp_2-\vpp_3}}
-\frac{\delta m^2_2}{8}\int\frac{\ddpp{}}{(2\pi)^2}\frac{A_{\vpp}A_{-\vpp}}{\omega_{\vpp}^2}
\right]\nonumber\\
&&\ \ \times\int\frac{d^{\dim}\vp_1 d^{\dim}\vp_2}{(2\pi)^{\dimtwo}}\frac{|\vp_1,\vp_2,-\vp_1-\vp_2\rangle_0}{\omega_{\vp_1}+\omega_{\vp_2}+\omega_{\vp_1+\vp_2}}
\nonumber\\
&=&m\sqrt{\frac{\lambda}{2
}} \left[-\frac{3\lambda}{4}\int\frac{\ddpp 1\ddpp 2}{(2\pi)^4}\frac{1}{\omega_{\vpp_1}\omega_{\vpp_2}\omega_{\vpp_1+\vpp_2}\left( \omega_{\vpp_1}+\omega_{\vpp_2}+\omega_{\vpp_1+\vpp_2}\right)}\right.\nonumber\\
&&\ \ \ \left.
+\frac{3\delta m^2_2}{4}\int\frac{\ddpp{}}{(2\pi)^2}\frac{1}{\omega_{\vpp}^2\left(2\omega_{\vpp}+m\right)}
\right] |\vp=\vec{0}\rangle_0.
\eea
Finally there are five contributions arising from the $O(\sl)$ interactions $H_3$
\bea
H_3\ovac_2^4&=&\frac{m}{8}\sqrt\frac{\lambda}{2}   \int\frac{d^{\dim}\vpp_1 d^{\dim}\vpp_2}{(2\pi)^{\dimtwo}}\frac{A_{\vpp_1}A_{\vpp_2}A_{-\vpp_1-\vpp_2}}{\omega_{\vpp_1}\omega_{\vpp_2}\omega_{\vpp_1+\vpp_2}}\\
&&\hspace{-2cm}\times \lambda\int\frac{\ddp 1\ddp 2\ddp 3}{(2\pi)^6} \left[\frac{9m^2}{2\omega_{\vp_1+\vp_2}\left(\omega_{\vp_1}+\omega_{\vp_2}+\omega_{\vp_1+\vp_2}\right)}-\frac{1}{4}\right] \frac{|\vp_1,\vp_2,\vp_3,-\vp_1-\vp_2-\vp_3\rangle_0}{(\omega_{\vp_1}+\omega_{\vp_2}+\omega_{\vp_3}+\omega_{\vp_1+\vp_2+\vp_3})}\nonumber\\
&&\hspace{-1cm}=\frac{m\lambda^{3/2}}{8\sqrt{2}}\int\frac{d^{\dim}\vpp_1 d^{\dim}\vpp_2}{(2\pi)^4}\left[ \frac{27m^2}{\omega_{\vpp_1}\omega_{\vpp_2}\omega_{\vpp_1+\vpp_2}^2\left(\omega_{\vpp_1}+\omega_{\vpp_2}+\omega_{\vpp_1+\vpp_2}\right)\left(\omega_{\vpp_1}+\omega_{\vpp_2}+m+\omega_{\vpp_1+\vpp_2}\right)}\right.\nonumber\\
&&+\frac{81m^2}{\omega_{\vpp_1}\omega_{\vpp_2}\omega_{\vpp_1+\vpp_2}^2\left(\omega_{\vpp_1}+\omega_{\vpp_2}+\omega_{\vpp_1+\vpp_2}\right)\left(\omega_{\vpp_1}+\omega_{\vpp_2}+\omega_{\vpp_1+\vpp_2}+m\right)}
\nonumber\\
&&\left.
-\frac{6}{\omega_{\vpp_1}\omega_{\vpp_2}\omega_{\vpp_1+\vpp_2}\left(\omega_{\vpp_1}+\omega_{\vpp_2}+\omega_{\vpp_1+\vpp_2}+m\right)}
\right]  |\vp=\vec{0}\rangle_0\nonumber\\
&&\hspace{-1cm}=\frac{3m\lambda^{3/2}}{4\sqrt{2}}\int\frac{d^{\dim}\vpp_1 d^{\dim}\vpp_2}{(2\pi)^4} \frac{\left(18m^2+\omega_{\vpp_1+\vpp_2}\left(\omega_{\vpp_1}+\omega_{\vpp_2}+\omega_{\vpp_1+\vpp_2}\right)\right) |\vp=\vec{0}\rangle_0}{\omega_{\vpp_1}\omega_{\vpp_2}\omega_{\vpp_1+\vpp_2}^2\left(\omega_{\vpp_1}+\omega_{\vpp_2}+\omega_{\vpp_1+\vpp_2}\right)\left(\omega_{\vpp_1}+\omega_{\vpp_2}+\omega_{\vpp_1+\vpp_2}+m\right)}\nonumber
\eea
and
\bea
H_3\ovac_2^2&=&\frac{3m}{4}\sqrt\frac{\lambda}{2}   \int\frac{d^{\dim}\vpp_1 d^{\dim}\vpp_2}{(2\pi)^{\dimtwo}}\frac{A^\ddag_{\vpp_1}A_{\vpp_2}A_{\vpp_1-\vpp_2}}{\omega_{\vpp_2}\omega_{\vpp_1+\vpp_2}}\\
&&\times \int\frac{\ddp{}}{(2\pi)^2}\left[\frac{\delta m^2_2}{2}+\frac{9m^2\lambda}{4}\int\frac{\ddpp{}}{(2\pi)^2} \frac{1}{\omega_{\vpp}\omega_{\vp-\vpp}\left(\omega_\vpp+\omega_{\vp-\vpp}+\omega_\vp\right)}
\right]\frac{|\vp,-\vp\rangle_0}{\omega_\vp}\nonumber\\
&&\hspace{-1cm}=\frac{3m}{16}\sqrt\frac{\lambda}{2} 
\int\frac{\ddpp{1}}{(2\pi)^2}\left[2\delta m^2_2+\int\frac{\ddpp{2}}{(2\pi)^2} \frac{9m^2\lambda}{\omega_{\vpp_2}\omega_{\vpp_1+\vpp_2}\left(\omega_{\vpp_2}+\omega_{\vpp_1+\vpp_2}+\omega_{\vpp_1}\right)}
\right]\frac{|\vp=0\rangle_0}{\omega_{\vpp_1}^3}.\nonumber
\eea

\begin{figure}[htbp]
\centering
\includegraphics[width = 0.75\textwidth]{tadpole.eps}
\caption{The tadpole contributions in the order in which they appear in the text, arising from (1) $H_5\ovac_0$, (2-3) $H_4\ovac_1^3$, (4-6) $H_3\ovac_2^4$ and (7-8) $H_3\ovac_2^2$ }\label{tadfig}
\end{figure}

The tadpole condition (\ref{tad}) implies that these eight contributions must all sum to zero, and so
\bea
T_5&=&-\frac{3m}{4}\sqrt{\frac{\lambda}{2}}\delta m^2_2 \int\frac{\ddp{1}}{(2\pi)^2}\frac{1}{\omega_\vp^2}\left[
\frac{1}{(2\omega_\vp +m)}+\frac{1}{2\omega_\vp}\right]
\\
&&+\frac{3m\lambda^{3/2}}{4\sqrt{2}}\int\frac{d^{\dim}\vp_1 d^{\dim}\vp_2}{(2\pi)^4}\frac{1}{\omega_{\vp_1}\omega_{\vp_2}\omega_{\vp_1+\vp_2}\left( \omega_{\vp_1}+\omega_{\vp_2}+\omega_{\vp_1+\vp_2}\right)}\nonumber\\
&& 
\times\left[1-\frac{\left(18m^2+\omega_{\vp_1+\vp_2}\left(\omega_{\vp_1}+\omega_{\vp_2}+\omega_{\vp_1+\vp_2}\right)\right)}{\omega_{\vp_1+\vp_2}\left(\omega_{\vp_1}+\omega_{\vp_2}+\omega_{\vpp_1+\vpp_2}+m\right)}-\frac{9m^2}{4\omega_{\vp_1}^2}
\right]\nonumber\\
&=&-\frac{3m}{4}\sqrt{\frac{\lambda}{2}}\delta m^2_2 \int\frac{\ddp{1}}{(2\pi)^2}\frac{1}{\omega_\vp^2}\left[
\frac{1}{(2\omega_\vp +m)}+\frac{1}{2\omega_\vp}\right]
\\
&&+\frac{3m^2\lambda^{3/2}}{4\sqrt{2}}\int\frac{d^{\dim}\vp_1 d^{\dim}\vp_2}{(2\pi)^4}\frac{1}{\omega_{\vp_1}\omega_{\vp_2}\omega_{\vp_1+\vp_2}\left( \omega_{\vp_1}+\omega_{\vp_2}+\omega_{\vp_1+\vp_2}\right)}\nonumber\\
&& 
\times\left[\frac{\omega_{\vp_1+\vp_2}-18m}{\omega_{\vp_1+\vp_2}\left(\omega_{\vp_1}+\omega_{\vp_2}+\omega_{\vpp_1+\vpp_2}+m\right)}-\frac{9m}{4\omega_{\vp_1}^2}
\right].\nonumber
\eea
Recall that $T_5$ is determined in terms of $v_4$ and $\delta m^2_4$, which at this point are unknown.  Therefore by fixing $T_5$, following Eq.~(\ref{v4}), we have fixed $v_4$ as a function of $\delta m^2_4$.  In turn we will fix $\delta m^2_4$ when we calculate the order $O(\lambda^2)$ correction to the one-meson state.

\subsubsection{Order $O(\lambda^{3/2})$: The Vacuum Energy}

Projecting to the three meson sector, the \hyperlink{first}{first} renormalization condition fixes $\ovac_3^3$ in terms of
\beq
H_5\ovac_0=-\sqrt\frac{\lambda}{2} \frac{\delta m^2_2}{2m} \int d^2 \vx :\phi^3(\vx):_a\ovac_0=-\sqrt\frac{\lambda}{2} \frac{\delta m^2_2}{2m} \int\frac{\ddp 1\ddp 2}{(2\pi)^4}|\vp_1,\vp_2,-\vp_1-\vp_2\rangle_0
\eeq
and
\bea
H_4\ovac_1&=&-m\sqrt{\frac{\lambda}{2
}}\int d^2\vx :\left[\frac{\lambda}{4}\phi^4(\vx)-\frac{\delta m^2_2}{2}\phi^2(\vx)+\int d^2\vx\hat{A}_4 \right]:_a\int\frac{d^{\dim}\vp_1 d^{\dim}\vp_2}{(2\pi)^{\dimtwo}}\frac{|\vp_1,\vp_2,-\vp_1-\vp_2\rangle_0}{\omega_{\vp_1}+\omega_{\vp_2}+\omega_{\vp_1+\vp_2}}\nonumber\\
&=&-m\sqrt{\frac{\lambda}{2
}} \left[\frac{3\lambda}{8}\int\frac{\ddpp 1\ddpp 2\ddpp 3}{(2\pi)^6}\frac{A^\ddag_{\vpp_1}A^\ddag_{\vpp_2}A_{\vpp_3}A_{\vpp_1+\vpp_2-\vpp_3}}{\omega_{\vpp_3}\omega_{\vpp_1+\vpp_2-\vpp_3}}
-\frac{\delta m^2_2}{2}\int\frac{\ddpp{}}{(2\pi)^2}\frac{A^\ddag_{\vpp}A_{\vpp}}{\omega_{\vpp}}+\int d^2\vx\hat{A}_4
\right]\nonumber\\
&&\ \ \times\int\frac{d^{\dim}\vp_1 d^{\dim}\vp_2}{(2\pi)^{\dimtwo}}\frac{|\vp_1,\vp_2,-\vp_1-\vp_2\rangle_0}{\omega_{\vp_1}+\omega_{\vp_2}+\omega_{\vp_1+\vp_2}}
\nonumber\\
&=&m\sqrt{\frac{\lambda}{2
}} \int\frac{d^{\dim}\vp_1 d^{\dim}\vp_2}{(2\pi)^{\dimtwo}} \left[
-\int\frac{\ddpp{}}{(2\pi)^2}\frac{9\lambda}{4\omega_{\vpp}\omega_{\vp_1+\vp_2-\vpp}\left(\omega_\vpp+ \omega_{\vp_1+\vp_2-\vpp}+\omega_{\vp_1+\vp_2}\right)}\right.\nonumber\\
&&\left. +\frac{\delta m^2_2}{2(\omega_{\vp_1}+\omega_{\vp_2}+\omega_{\vp_1+\vp_2})}\left(\frac{1}{\omega_{\vp_1}}+\frac{1}{\omega_{\vp_2}}+\frac{1}{\omega_{\vp_1+\vp_2}} 
\right)\right.\nonumber\\
&&\left.-\frac{1}{(\omega_{\vp_1}+\omega_{\vp_2}+\omega_{\vp_1+\vp_2})}\int d^2\vx\hat{A}_4
\right] |\vp_1,\vp_2,-\vp_1-\vp_2\rangle_0 \label{h41}.
\eea
Note that the range of integration is invariant under permutations of $\vp_1$, $\vp_2$ and $\vp_1+\vp_2$ as are other factors, and so the three terms multiplying $\delta m^2_2$ each yield the same contribution to the state.  

In addition to the contribution from $H_5$ and the three contributions from $H_4$, there are eleven contributions from $H_3$.  As usual, leaving the projection onto the three-meson Fock space implicit, four of these are
\bea
H_3\ovac^6_2&=&\frac{m}{8}\sqrt\frac{\lambda}{2}   \int\frac{\ddpp 1\ddpp 2}{(2\pi)^{\dimtwo}}\frac{A_{\vpp_1}A_{\vpp_2}A_{-\vpp_1-\vpp_2}}{\omega_{\vpp_1}\omega_{\vpp_2}\omega_{\vpp_1+\vpp_2}}\\
&&\hspace{-2cm}\times \frac{m^2\lambda}{2}\int\frac{\ddp 1\ddp 2\ddp 3\ddp 4}{(2\pi)^8} \frac{|\vp_1,\vp_2,\vp_3,\vp_4,-\vp_1-\vp_2,-\vp_3-\vp_4\rangle_0}{\left(\omega_{\vp_1}+\omega_{\vp_2}+\omega_{\vp_1+\vp_2}\right)\left( \omega_{\vp_1}+\omega_{\vp_2}+\omega_{\vp_1+\vp_2}+\omega_{\vp_3}+\omega_{\vp_4}+\omega_{\vp_3+\vp_4}
\right)}\nonumber\\
&=&\frac{m^3\lambda^{3/2}}{16\sqrt{2}} \int\frac{d^{\dim}\vp_1 d^{\dim}\vp_2}{(2\pi)^{\dimtwo}} \left[
 \int d^2\vx\int\frac{\ddpp 1\ddpp 2}{(2\pi)^{\dimtwo}}
 \frac{6}{\left( \omega_{\vp_1}+\omega_{\vp_2}+\omega_{\vp_1+\vp_2}+\omega_{\vpp_1}+\omega_{\vpp_2}+\omega_{\vpp_1+\vpp_2}
\right)}
 \right.\nonumber\\
 &&
 \times\frac{1}{{\omega_{\vpp_1}\omega_{\vpp_2}\omega_{\vpp_1+\vpp_2}}}\left( 
 \frac{1}{\omega_{\vpp_1}+\omega_{\vpp_2}+\omega_{\vpp_1+\vpp_2}}+\frac{1}{\omega_{\vp_1}+\omega_{\vp_2}+\omega_{\vp_1+\vp_2}}
 \right)\nonumber\\
 &&\left. + \int\frac{\ddpp{}}{(2\pi)^{2}}\frac{18}{\omega_{\vp_1}\omega_\vpp\omega_{\vp_1+\vpp}\left( 2\omega_{\vp_1}+\omega_{\vpp}+\omega_{\vp_1+\vpp}+\omega_{\vp_2}+\omega_{\vp_1+\vp_2}
\right)}\right.\nonumber\\
&&\left.\times\left(\frac{1}{\left(\omega_{\vp_1}+\omega_{\vpp}+\omega_{\vp_1+\vpp}\right)}+\frac{1}{\left(\omega_{\vp_1}+\omega_{\vp_2}+\omega_{\vp_1+\vp_2}\right)}
 \right)
\right] |\vp_1,\vp_2,-\vp_1-\vp_2\rangle_0.\nonumber
\\
&&\hspace{-1cm}=\frac{3m^3\lambda^{3/2}}{8\sqrt{2}} \int\frac{d^{\dim}\vp_1 d^{\dim}\vp_2}{(2\pi)^{\dimtwo}} \frac{1}{\left( \omega_{\vp_1}+\omega_{\vp_2}+\omega_{\vp_1+\vp_2}\right)}\left[
\int\frac{\ddpp{}}{(2\pi)^{2}}\frac{3}{\omega_{\vp_1}\omega_\vpp\omega_{\vp_1+\vpp}\left( \omega_{\vp_1}+\omega_{\vpp}+\omega_{\vp_1+\vpp}\right)}
 \right.\nonumber\\
&&\hspace{-1cm}\ \ \left. + 
 \int d^2\vx\int\frac{\ddpp 1\ddpp 2}{(2\pi)^{\dimtwo}}
 \frac{1}{{\omega_{\vpp_1}\omega_{\vpp_2}\omega_{\vpp_1+\vpp_2}}\left(\omega_{\vpp_1}+\omega_{\vpp_2}+\omega_{\vpp_1+\vpp_2}
\right)}\right] |\vp_1,\vp_2,-\vp_1-\vp_2\rangle_0.\nonumber
\eea
Here we have implicitly handled the infrared divergence as in Eq.~(\ref{id}), by not yet performing the divergent $\vx$ integration in $H_3$.  This term cancels the $\hat{A}_4$ term in Eq.~(\ref{h41}).

\begin{figure}[htbp]
\centering
\includegraphics[width = 0.75\textwidth]{Vac3.eps}
\caption{Contributions to $\ovac_3^3$ arising from (1) $H_5\ovac_0$, (2-4) $H_4\ovac_1$ and (5-8) $H_3\ovac_2^6$}\label{vac3fig}
\end{figure}

\begin{figure}[htbp]
\centering
\includegraphics[width = 0.75\textwidth]{Vac3b.eps}
\caption{Contributions to $\ovac_3^3$ arising from (1-4) $H_3\ovac_2^4$ and (5-7) $H_3\ovac_2^2$}\label{vac3bfig}
\end{figure}

Four terms arise from $H_3\ovac^4_2$
\bea
H_3\ovac^4_2&=&\frac{3m}{4}\sqrt\frac{\lambda}{2}   \int\frac{\ddpp 1\ddpp 2}{(2\pi)^{\dimtwo}}\frac{A^\ddag_{\vpp_1}A_{\vpp_2}A_{\vpp_1-\vpp_2}}{\omega_{\vpp_2}\omega_{\vpp_1-\vpp_2}}\\
&&\hspace{-2cm}\times \lambda\int\frac{\ddp 1\ddp 2\ddp 3}{(2\pi)^6} \left[\frac{9m^2}{2\omega_{\vp_1+\vp_2}\left(\omega_{\vp_1}+\omega_{\vp_2}+\omega_{\vp_1+\vp_2}\right)}-\frac{1}{4}\right] \frac{|\vp_1,\vp_2,\vp_3,-\vp_1-\vp_2-\vp_3\rangle_0}{(\omega_{\vp_1}+\omega_{\vp_2}+\omega_{\vp_3}+\omega_{\vp_1+\vp_2+\vp_3})}\nonumber\\
&=&\frac{9m\lambda^{3/2}}{4\sqrt{2}}\int\frac{\ddp 1\ddp 2}{(2\pi)^4}\int\frac{\ddpp {}}{(2\pi)^2}\frac{1}{\omega_\vpp\omega_{\vp_1+\vp_2+\vpp}\left(\omega_{\vp_1}+\omega_{\vp_2}+\omega_{\vpp}+\omega_{\vp_1+\vp_2+\vpp}\right)}\nonumber\\
&&\times\left[ 
\frac{3m^2}{\omega_{\vp_1+\vp_2}\left(\omega_{\vp_1}+\omega_{\vp_2}+\omega_{\vp_1+\vp_2}\right)}+\frac{6m^2}{\omega_{\vp_1+\vpp}\left(\omega_{\vp_1}+\omega_{\vpp}+\omega_{\vp_1+\vpp}\right)}\right.\nonumber\\
&&\left.
+\frac{3m^2}{\omega_{\vp_1+\vp_2}\left(\omega_{\vp_1+\vp_2+\vpp}+\omega_{\vpp}+\omega_{\vp_1+\vp_2}\right)}
-1\right] |\vp_1,\vp_2,-\vp_1-\vp_2\rangle_0\nonumber
\eea
and three from  $H_3\ovac^2_2$
\bea
H_3\ovac^2_2&=&\frac{3m}{2}\sqrt\frac{\lambda}{2}   \int\frac{\ddpp 1\ddpp 2}{(2\pi)^{\dimtwo}}\frac{A^\ddag_{\vpp_1}A^\ddag_{\vpp_2}A_{-\vpp_1-\vpp_2}}{\omega_{\vpp_1+\vpp_2}}\\
&&\hspace{-1cm}\times \int\frac{\ddp{}}{(2\pi)^2}\left[\frac{\delta m^2_2}{2}+\frac{9m^2\lambda}{4}\int\frac{\ddpp{3}}{(2\pi)^2} \frac{1}{\omega_{\vpp_3}\omega_{\vp-\vpp_3}\left(\omega_{\vpp_3}+\omega_{\vp-\vpp_3}+\omega_\vp\right)}
\right]\frac{|\vp,-\vp\rangle_0}{\omega_\vp}\nonumber\\
&=&\frac{3m}{4}\sqrt\frac{\lambda}{2}   \int\frac{\ddp 1\ddp 2}{(2\pi)^{\dimtwo}}\frac{1}{\omega_{\vp_1}^2}\left[2\delta m^2_2\right.\nonumber\\
&&\left.+{9m^2\lambda}{}\int\frac{\ddpp{}}{(2\pi)^2} \frac{1}{\omega_{\vpp}\omega_{\vp_1+\vpp}\left(\omega_{\vpp}+\omega_{\vp_1}+\omega_{\vp_1+\vpp}\right)}
\right] |\vp_1,\vp_2,-\vp_1-\vp_2\rangle_0\nonumber
\eea
Here we note that the term with $\ddpp{}$ integration arises from two diagrams, which are equal, representing the case in which the meson resulting from the fusion of two initial mesons is the same as the same that splits, and the case in which the other meson splits.

Summing these contributions together, we find that, projecting again to the three-meson sector
\beq
H_5\ovac_0+H_4\ovac_1+H_3\ovac_2=\int\frac{\ddp 1\ddp 2}{(2\pi)^4} B(\vp_1,\vp_2) |\vp_1,\vp_2,-\vp_1-\vp_2\rangle_0=-H_2|\Omega\rangle_3^3
\eeq
where
\bea
B(\vp_1,\vp_2)&=&\frac{1}{2m}\sqrt{\frac{\lambda}{2}}\delta m^2_2\left[ 
-1+\frac{3m^2}{\omega_{\vp_1}\left(\omega_{\vp_1}+\omega_{\vp_2}+\omega_{\vp_1+\vp_2}\right)}+\frac{3m^2}{\omega_{\vp_1}^2}\right]\\
&&+3m\lambda^{3/2}\int\frac{\ddp{}}{(2\pi)^{2}}\frac{1}{\omega_{\vpp}}\left[ 
-\frac{3}{4\omega_{\vp_1+\vp_2+\vpp}\left(\omega_{\vpp}+\omega_{\vp_1+\vp_2}+\omega_{\vp_1+\vp_2+\vpp}\right)}\right.\nonumber\\
&&\left.+\frac{3m^2}{8\omega_{\vp_1}\omega_{\vp_1+\vpp}\left(\omega_{\vp_1}+\omega_{\vp_2}+\omega_{\vp_1+\vp_2}\right)\left(\omega_{\vpp}+\omega_{\vp_1}+\omega_{\vp_1+\vpp}\right)}\right.\nonumber\\
&&\left.
+\frac{3}{4\omega_{\vp_1+\vp_2+\vpp}\left(\omega_{\vp_1}+\omega_{\vp_2}+\omega_{\vpp}+\omega_{\vp_1+\vp_2+\vpp}\right)}\left(
\frac{3m^2}{\omega_{\vp_1+\vp_2}\left(\omega_{\vp_1}+\omega_{\vp_2}+\omega_{\vp_1+\vp_2}\right)}\right.\right.\nonumber\\
&&\left.\left.+\frac{6m^2}{\omega_{\vp_1+\vpp}\left(\omega_{\vp_1}+\omega_{\vpp}+\omega_{\vp_1+\vpp}\right)}
+\frac{3m^2}{\omega_{\vp_1+\vp_2}\left(\omega_{\vp_1+\vp_2+\vpp}+\omega_{\vpp}+\omega_{\vp_1+\vp_2}\right)}
-1
\right)\right.\nonumber\\
&&\left.+\frac{9m^2}{4\omega_{\vp_1}^2\omega_{\vp_1+\vpp}\left(\omega_{\vp_1}+\omega_{\vpp}+\omega_{\vp_1+\vpp}\right)}\right].\nonumber
\eea
Therefore we conclude
\beq
|\Omega\rangle_3^3=-\int\frac{\ddp 1\ddp 2}{(2\pi)^4} \frac{B(\vp_1,\vp_2)}{\left(\omega_{\vp_1}+\omega_{\vp_2}+\omega_{\vp_1+\vp_2}\right)} |\vp_1,\vp_2,-\vp_1-\vp_2\rangle_0.
\eeq

\subsubsection{Order $O(\lambda^2)$}

Finally we are ready to impose the \hyperlink{first}{first} renormalization condition at order $O(\lambda)^2$.  We are interested in the components in the zero-meson Fock space.  Projecting onto this Fock space, the first contribution to the renormalization condition is
\bea
H_6\ovac_0&=&\dvx \hat{A}_6\ovac_0.
\eea
In this subsection we will determine $\hat{A}_6$ and so $A_6$ via Eq.~(\ref{a6}).  

The first contribution to the renormalization condition $H\ovac=0$ is
\bea
H_5\ovac_1^3&=&\sqrt\frac{\lambda}{2} \frac{\delta m^2_2}{16 m}
\int\frac{d^{\dim}\vpp_1 d^{\dim}\vpp_2}{(2\pi)^{\dimtwo}}\frac{A_{\vpp_1}A_{\vpp_2}A_{-\vpp_1-\vpp_2}}{\omega_{\vpp_1}\omega_{\vpp_2}\omega_{\vpp_1+\vpp_2}}
\left(m\sqrt{\frac{\lambda}{2}}\right)\int\frac{d^{\dim}\vp_1 d^{\dim}\vp_2}{(2\pi)^{\dimtwo}}\frac{|\vp_1,\vp_2,-\vp_1-\vp_2\rangle_0}{\omega_{\vp_1}+\omega_{\vp_2}+\omega_{\vp_1+\vp_2}}\nonumber\\
&&\hspace{-1cm}=\dvx \frac{3\lambda \delta m^2_2}{16}
\int\frac{d^{\dim}\vpp_1 d^{\dim}\vpp_2}{(2\pi)^{\dimtwo}}\frac{1}{\omega_{\vpp_1}\omega_{\vpp_2}\omega_{\vpp_1+\vpp_2}\left( \omega_{\vpp_1}+\omega_{\vpp_2}+\omega_{\vpp_1+\vp_2}
\right)}
\ovac_0\nonumber\\
&=&\dvx \frac{\delta m^2_2}{2} \hat{A}_4 \ovac_0
\eea
followed by
\bea
H_4\ovac_2^4&=&\frac{\lambda}{64}\int\frac{d^{\dim}\vpp_1 d^{\dim}\vpp_2d^{\dim}\vpp_3}{(2\pi)^{6}}\frac{A_{\vpp_1}A_{\vpp_2}A_{\vpp_3}A_{-\vpp_1-\vpp_2-\vpp_3}}{\omega_{\vpp_1}\omega_{\vpp_2}\omega_{\vpp_3}\omega_{\vpp_1+\vpp_2+\vpp_3}}\\
&&\hspace{-2cm}\times
\lambda\int\frac{\ddp 1\ddp 2\ddp 3}{(2\pi)^6} \left[\frac{9m^2}{2\omega_{\vp_1+\vp_2}\left(\omega_{\vp_1}+\omega_{\vp_2}+\omega_{\vp_1+\vp_2}\right)}-\frac{1}{4}\right] \frac{|\vp_1,\vp_2,\vp_3,-\vp_1-\vp_2-\vp_3\rangle_0}{(\omega_{\vp_1}+\omega_{\vp_2}+\omega_{\vp_3}+\omega_{\vp_1+\vp_2+\vp_3})}
\nonumber\\
&=&\dvx\frac{3\lambda ^2}{8}\int\frac{d^{\dim}\vpp_1 d^{\dim}\vpp_2d^{\dim}\vpp_3}{(2\pi)^{6}}\frac{1}{\omega_{\vpp_1}\omega_{\vpp_2}\omega_{\vpp_3}\omega_{\vpp_1+\vpp_2+\vpp_3}(\omega_{\vpp_1}+\omega_{\vpp_2}+\omega_{\vpp_3}+\omega_{\vpp_1+\vpp_2+\vpp_3})}\nonumber\\
&&\times
 \left[\frac{9m^2}{2\omega_{\vpp_1+\vpp_2}\left(\omega_{\vpp_1}+\omega_{\vpp_2}+\omega_{\vpp_1+\vpp_2}\right)}-\frac{1}{4}\right] {
 \ovac_0}{}\nonumber
\eea
and
\bea
H_4\ovac_2^2&=&-\frac{\delta m^2_2}{8}\int\frac{d^{\dim}\vpp_1 }{(2\pi)^{2}}\frac{A_{\vpp_1}A_{-\vpp_1}}{\omega^2_{\vpp_1}}
\\
&&\times\int\frac{\ddp{}}{(2\pi)^2}\left[\frac{\delta m^2_2}{2}+\frac{9m^2\lambda}{4}\int\frac{\ddpp{2}}{(2\pi)^2} \frac{1}{\omega_{\vpp_2}\omega_{\vp-\vpp_2}\left(\omega_{\vpp_2}+\omega_{\vp-\vpp_2}+\omega_\vp\right)}
\right]\frac{|\vp,-\vp\rangle_0}{\omega_\vp}\nonumber\\
&=&\dvx\left[ -\frac{\left(\delta m^2_2\right)^2}{8}\int\frac{\ddpp{}}{(2\pi)^2}\frac{1}{\omega_{\vpp}^3}
 \right.\nonumber\\
 &&\left.-\frac{9m^2\lambda\delta m^2_2}{16}\int\frac{\ddpp{1}\ddpp 2}{(2\pi)^4}\frac{1}{\omega_{\vpp_1}^3\omega_{\vpp_2}\omega_{\vpp_1-\vpp_2}\left(\omega_{\vpp_2}+\omega_{\vpp_1-\vpp_2}+\omega_{\vpp_1}\right)}
\right]\vac_0.\nonumber
\eea

\begin{figure}[htbp]
\centering
\includegraphics[width = 0.75\textwidth]{Vac4.eps}
\caption{Contributions to the three-loop vacuum energy arising from (1) $H_6\ovac_0$, (2) $H_5\ovac_1^3$, (3-4) $H_4\ovac_2^4$ and (5-6) $H_4\ovac_2^2$.  The contributions from $H_3\ovac_3^3$, which are most of the contributions, are obtained by contracting the three external lines on the left sides of the diagrams in Figs.~\ref{vac3fig} and \ref{vac3bfig}.}\label{vac4fig}
\end{figure}

Finally, all of the contributions to $\ovac^3_3$ from the previous section are included, with the three final mesons contracted
\bea
H_3\ovac^3_3&=&-\frac{m}{8}\sqrt\frac{\lambda}{2}   \int\frac{d^{\dim}\vpp_1 d^{\dim}\vpp_2}{(2\pi)^{\dimtwo}}\frac{A_{\vpp_1}A_{\vpp_2}A_{-\vpp_1-\vpp_2}}{\omega_{\vpp_1}\omega_{\vpp_2}\omega_{\vpp_1+\vpp_2}}\\
&&\times \int\frac{\ddp 1\ddp 2}{(2\pi)^4} \frac{B(\vp_1,\vp_2)}{\left(\omega_{\vp_1}+\omega_{\vp_2}+\omega_{\vp_1+\vp_2}\right)} |\vp_1,\vp_2,-\vp_1-\vp_2\rangle_0\nonumber\\
&=&-\dvx \frac{3m}{4}\sqrt\frac{\lambda}{2}   \int\frac{d^{\dim}\vpp_1 d^{\dim}\vpp_2}{(2\pi)^{\dimtwo}}\frac{B(\vpp_1,\vpp_2)}{\omega_{\vpp_1}\omega_{\vpp_2}\omega_{\vpp_1+\vpp_2}\left(\omega_{\vpp_1}+\omega_{\vpp_2}+\omega_{\vpp_1+\vpp_2}\right)}\ovac_0.\nonumber
\eea

Summing these contributions, we find that the three-loop contribution to the vacuum energy density is
\bea
\hat A_6&=&-\frac{\delta m^2_2}{2} \hat{A}_4 -\frac{3\lambda ^2}{8}\int\frac{d^{\dim}\vpp_1 d^{\dim}\vpp_2d^{\dim}\vpp_3}{(2\pi)^{6}}\frac{1}{\omega_{\vpp_1}\omega_{\vpp_2}\omega_{\vpp_3}\omega_{\vpp_1+\vpp_2+\vpp_3}(\omega_{\vpp_1}+\omega_{\vpp_2}+\omega_{\vpp_3}+\omega_{\vpp_1+\vpp_2+\vpp_3})}\nonumber\\
&&\times
 \left[\frac{9m^2}{2\omega_{\vpp_1+\vpp_2}\left(\omega_{\vpp_1}+\omega_{\vpp_2}+\omega_{\vpp_1+\vpp_2}\right)}-\frac{1}{4}\right]+\frac{\left(\delta m^2_2\right)^2}{8}\int\frac{\ddpp{}}{(2\pi)^2}\frac{1}{\omega_{\vpp}^3}
 \nonumber\\
 &&
 +\frac{3m}{16}\sqrt\frac{\lambda}{2}   \int\frac{d^{\dim}\vpp_1 d^{\dim}\vpp_2}{(2\pi)^{\dimtwo}}\frac{4\omega_{\vpp_1}^2B(\vpp_1,\vpp_2)+3m\delta m^2_2\sqrt{2\lambda}}{\omega_{\vpp_1}^3\omega_{\vpp_2}\omega_{\vpp_1+\vpp_2}\left(\omega_{\vpp_1}+\omega_{\vpp_2}+\omega_{\vpp_1+\vpp_2}\right)}
\eea
where we remind the reader that Eq.~(\ref{a6}) yields $A_6$ in terms of $\hat A_6$.  This relation depends on several mass counterterms that we have not fixed, some of which we will not fix.  However, the $c$-number term in $H_6$ is $\hat A_6$, not $A_6$, and so only $\hat A_6$ is necessary for computing quantities such as quantum corrections to soliton masses.

\subsection{Finding the One Meson States at Two Loops}

In the previous section, we calculated the $O(\lambda)$ correction to the one mesons states arising from two applications of the 3-point interaction $H_3$.  It resulted in a finite, one-loop correction to the mass counterterm $\delta m^2$.  In the present subsection we will similarly compute the corrections arising at order $O(\lambda^2)$.  We will find that they leads to a divergent, two-loop correction to $\delta m^2$.

\subsubsection{Order $O(\lambda)$ with three mesons}
As a result of our normalization convention (\ref{max}), there is no one-meson contribution to $\ps_1$.  

\begin{figure}[htbp]
\centering
\includegraphics[width = 0.75\textwidth]{p2.eps}
\caption{Contributions to $\ps_2^3$, the three meson part of the order $\lambda$ correction to the one-meson state, arising from (1-2) $H_4\ps_0$, (3) $H_3\ps_1^2$ and (4-5) $H_3\ps_1^4$.}\label{p2fig}
\end{figure}

We next turn to $\ps_2^3$.  Projecting, implicitly as usual, onto the three-meson sector, the first two contributions arise from
\bea
H_4\ps_0&=&\dvx\left[\frac{\lambda}{4}\phi^4(\vx)-\frac{\delta m^2_2}{2}\phi^2(\vx)
\right]\ps_0\\
&=&\left[\frac{\lambda}{2}\int\frac{\ddpp 1\ddpp 2\ddpp 3}{(2\pi)^{6}}\frac{A^\ddag_{\vpp_1}A^\ddag_{\vpp_2}A^\ddag_{\vpp_3}A_{\vpp_1+\vpp_2+\vpp_3}}{\omega_{\vpp_1+\vpp_2+\vpp_3}}
-\frac{\delta m^2_2}{2}\int\frac{\ddpp {}}{(2\pi)^{2}}A^\ddag_{\vpp}A^\ddag_{-\vpp}
\right]\ps_0\nonumber\\
&=&\frac{\lambda}{2\ovp{0}}\int\frac{\ddp 1\ddp 2}{(2\pi)^{4}}|\vp_1,\vp_2,\vp_0-\vp_1-\vp_2\rangle_0-\frac{\delta m^2_2}{2}\int\frac{\ddp {}}{(2\pi)^{2}}|\vp_0,\vp,-\vp\rangle_0.
\eea
Only the first of these contributes to the ultraviolet divergence in $\delta m^2_4$.

There is also one contribution from
\bea
H_3\ps_1^2&=&\frac{3m}{2}\sqrt\frac{\lambda}{2}   \int\frac{\ddpp 1\ddpp 2}{(2\pi)^{\dimtwo}}\frac{A^\ddag_{\vpp_1}A^\ddag_{\vpp_2}A_{\vpp_1+\vpp_2}}{\omega_{\vpp_1+\vpp_2}}\\
&&\times\left[ -\frac{3}{2}m\sqrt\frac{\lambda}{2}
\int\frac{d^{\dim}\vp_1}{(2\pi)^{\dim}} \frac{1}{\ovp 0(\ovp 1+\omega_{\vp_0-\vp_1}-\ovp 0)}
|\vp_1,\vp_0-\vp_1\rangle_0
\right]\nonumber\\
&=&-\frac{9m^2\lambda}{4\ovp 0}\int\frac{\ddp 1\ddp 2}{(2\pi)^{4}}\frac{|\vp_1,\vp_2,\vp_0-\vp_1-\vp_2\rangle_0}{\omega_{\vp_1+\vp_2}(\omega_{\vp_1+\vp_2}+\omega_{\vp_1+\vp_2-\vp_0}-\ovp 0)}
\eea
and two from
\bea
H_3\ps_1^4&=&\frac{3m}{4}\sqrt\frac{\lambda}{2}   \int\frac{\ddpp 1\ddpp 2}{(2\pi)^{\dimtwo}}\frac{A^\ddag_{\vpp_1}A_{\vpp_2}A_{\vpp_1-\vpp_2}}{\ovp 2\omega_{\vpp_1-\vpp_2}}\\
&&\times\left( -m\sqrt\frac{\lambda}{2}\right)
\int\frac{d^{\dim}\vp_1 d^{\dim}\vp_2}{(2\pi)^{\dimtwo}} \frac{1}{(\omega_{\vp_1}+\omega_{\vp_2}+\omega_{\vp_1+\vp_2})}|\vp_0,\vp_1,\vp_2,-\vp_1-\vp_2\rangle_0
\nonumber\\
&=&-\frac{9m^2\lambda}{4}\left[\int\frac{\ddp{}\ddpp {}}{(2\pi)^{\dimtwo}}\frac{|\vp_0,\vp,-\vp\rangle_0}{\ovpp{} \omega_{\vp-\vpp}(\omega_{\vpp}+\omega_{\vp-\vpp}+\omega_{\vp})}\right.
\nonumber\\
&&\left.
+\int\frac{\ddp{1}\ddp {2}}{(2\pi)^{\dimtwo}}\frac{|\vp_1,\vp_2,\vp_0-\vp_1-\vp_2\rangle_0}{\omega_{\vp_1+\vp_2} \ovp 0(\ovp 1+\ovp 2+\omega_{\vp_1+\vp_2})}\right]
\nonumber
\eea

Combining these contributions we find
\bea
\ps_2^3&=&\int\frac{\ddp{}}{(2\pi)^{\dimtwo}}\left[\frac{\delta m^2_2}{4\ovp{}}+\frac{9m^2\lambda}{8}\int\frac{\ddpp {}}{(2\pi)^{2}}\frac{1}{\ovp{}\ovpp{} \omega_{\vp-\vpp}(\omega_{\vpp}+\omega_{\vp-\vpp}+\omega_{\vp})}  
\right]{|\vp_0,\vp,-\vp\rangle_0}\nonumber\\
&&
+\frac{\lambda}{\ovp 0}\int\frac{\ddp{1}\ddp {2}}{(2\pi)^{\dimtwo}}\left[  
-\frac{1}{2}
+\frac{9m^2}{4\omega_{\vp_1+\vp_2}(\omega_{\vp_1+\vp_2}+\omega_{\vp_1+\vp_2-\vp_0}-\ovp 0)}\right.\nonumber\\
&&\left.+\frac{9m^2}{4\omega_{\vp_1+\vp_2} (\ovp 1+\ovp 2+\omega_{\vp_1+\vp_2})}
\right]\frac{|\vp_1,\vp_2,\vp_0-\vp_1-\vp_2\rangle_0}{\ovp 1+\ovp 2+\omega_{\vp_0-\vp_1-\vp_2}-\ovp 0}.
\eea

\subsubsection{Order $O(\lambda)$ with five mesons}

There are four contributions to $\ps_2^5$.  The only one which will contribute to the divergent part of $\delta m^2_4$ is, projecting implicitly onto the five-meson sector
\bea
H_4\ps_0&=&\frac{\lambda}{4}\int\frac{\ddp 1\ddp 2\ddp 3}{(2\pi)^{\dimtwo}}|\vp_0,\vp_1,\vp_2,\vp_3,-\vp_1-\vp_2-\vp_3\rangle_0.
\eea

Two contributions arise from 
\bea
H_3\ps_1^4&=&\frac{3m}{2}\sqrt\frac{\lambda}{2}   \int\frac{\ddpp 1\ddpp 2}{(2\pi)^{\dimtwo}}\frac{A^\ddag_{\vpp_1}A^\ddag_{\vpp_2}A_{\vpp_1+\vpp_2}}{\omega_{\vpp_1+\vpp_2}}\\
&&\times \left( -m\sqrt\frac{\lambda}{2}\right)
\int\frac{d^{\dim}\vp_1 d^{\dim}\vp_2}{(2\pi)^{\dimtwo}} \frac{1}{(\omega_{\vp_1}+\omega_{\vp_2}+\omega_{\vp_1+\vp_2})}|\vp_0,\vp_1,\vp_2,-\vp_1-\vp_2\rangle_0
\nonumber\\
&=&-\frac{3m^2\lambda}{8}\int\frac{\ddp 1\ddp 2\ddp 3}{(2\pi)^{6}}\left[\frac{3|\vp_0,\vp_1,\vp_2,\vp_3,-\vp_1-\vp_2-\vp_3\rangle_0}{\omega_{\vp_1+\vp_2}\left(\ovp1+\ovp2+\omega_{\vp_1+\vp_2}\right)}\right.\nonumber\\
&&\left.
+\frac{|\vp_1,\vp_0-\vp_1,\vp_2,\vp_3,-\vp_2-\vp_3\rangle_0}{\ovp 0 (\omega_{\vp_2}+\omega_{\vp_3}+\omega_{\vp_2+\vp_3})}\right]\nonumber
\eea
and one from
\bea
H_3\ps_1^2&=&m\sqrt\frac{\lambda}{2}   \int\frac{\ddpp 1\ddpp 2}{(2\pi)^{\dimtwo}}{A^\ddag_{\vpp_1}A^\ddag_{\vpp_2}A^\ddag_{-\vpp_1-\vpp_2}}\\
&&\times\left[ -\frac{3}{2}m\sqrt\frac{\lambda}{2}
\int\frac{d^{\dim}\vp_1}{(2\pi)^{\dim}} \frac{1}{\ovp 0(\ovp 1+\omega_{\vp_0-\vp_1}-\ovp 0)}
|\vp_1,\vp_0-\vp_1\rangle_0
\right]\nonumber\\
&=&-\frac{3m^2\lambda}{8} \int\frac{\ddp 1\ddp 2\ddp 3}{(2\pi)^{6}} \frac{|\vp_1,\vp_0-\vp_1,\vp_2,\vp_3,-\vp_2-\vp_3\rangle_0}{\ovp 0(\ovp 1+\omega_{\vp_0-\vp_1}-\ovp 0)}.\nonumber
\eea

Combining these four contributions we find
\bea
\ps_2^5&=&
\eea

\subsubsection{Order $O(\lambda)$ with seven mesons}

\subsubsection{Order $O(\lambda^{3/2})$ with two mesons}

\subsubsection{Order $O(\lambda^{3/2})$ with three mesons and $\delta m^2_3$}
The only contribution to $|\vp_0\rangle_3^3$ arises from the mass term in $H_5$, which is proportional to $\delta m^2_3$.  As result of our convention (\ref{max}), $|\vp_0\rangle_3^3=0$.  Therefore we conclude
\beq
\delta m^2_3=0.
\eeq

\subsubsection{Order $O(\lambda^{3/2})$ with four mesons}

\subsubsection{Fixing $\delta m^2_4$}

\section{The Domain Wall Sector}

\red{The main result is that $\ch\p_4$ contains a term like $-(1/2)\delta m^2_4 f^2(x)$ which has a logarithmic divergence coming from a meson that turns into three mesons that travel forward or backward in time and become one meson, calculated in the previous section.  The other terms in $\delta m^2_4$ are finite.  This divergence cancels the logarithmic divergence in the $|V_{kkk}|^2$ term in our old two-loop formula for the kink mass.  Both terms are of order $O(\lambda)$.   Note that $V^{(3)}(\sl f(x))$ is proportional to $f(x)$ in the $\phi^4$ theory.  However the $x$ coordinates in the two $V_{kkk}$'s are different, which hopefully doesn't cause problems.  Maybe the high $k$ limit helps here, picking out terms where the $x$'s are the same?  Or maybe there's no elegant solution and you just need to choose the quantum corrections to $f(x)$ order by order to make it work?  But you need to deal with energies and tadpoles altogether in the same corrections somehow.  I guess that with no tadpole you get the minimum energy, so that should be a finite one if there is any.}






\appendix
\section{Hengyuan calculation: Hamiltonian  expansion}
Expansion with $\phi^n$ firstly 
\beq
\begin{aligned}
\ch_0\p=&-\frac{m^2}{4}v^2+\frac{\lambda}{4}v^4+\frac{1}{4}\delta m^2v^2+A\\
\ch_1\p=&-\frac{m^2}{2}\phi v+\lambda\phi v^3+\frac{1}{2}\delta m^2\phi v\\
\ch_2\p=&\frac{\pi^2+(\partial_i\phi)^2}{2}-\frac{m^2}{4}\phi^2+\frac{3\lambda}{2}\phi^2v^2+\frac{1}{4}\delta m^2\phi^2\\
\ch_3\p=&\lambda \phi^3v\\\nonumber
\ch_4\p=&\frac{\lambda}{4}\phi^4\\\nonumber
\end{aligned}
\eeq
Where to make comparison with expansion about $0(\lambda^{i/2})$, we expand the $v$ and $\delta m^2$ too
\beq
\begin{aligned}
 v^2=&\bigg[v_0^2+2v_0v_1+(2v_0v_2+v_1^2)+(2v_0v_3+2v_1v_2)+(2v_0v_4+2v_1v_3+v_2^2)\\
&+(2v_0v_5+2v_1v_4+2v_2v_3)+(2v_0v_6+2v_1v_5+2v_2v_4)+\cdots)\bigg]   
\end{aligned}
\eeq
which is of order $0(\lambda^{-1}),0(\lambda^{-1/2}),0(\lambda^{0}),0(\lambda^{1/2}),0(\lambda^{1}),\cdots$ Because the quantum correction are supressed by $\lambda$ order. so we have set $v_1=0$ firstly.then we get:
\beq
v^2=\bigg[v_0^2+2v_0v_2+2v_0v_3+(2v_0v_4+v_2^2)+(2v_2v_3+2v_0v_5)+(2v_2v_4+2v_0v_6)\cdots \bigg]
\eeq
which is of order  $0(\lambda^{-1}),0(\lambda^{0}),0(\lambda^{1/2}),0(\lambda^{1}),0(\lambda^{3/2}),0(\lambda^{2})\cdots$ 
 And similiar we have:
\beq
v^3=\bigg[v_0^3+3v_0^2v_2+3v_0^2v_3+(3v_0v_2^2+3v_0^2v_4),(6v_0v_2v_3+3v_0^2v_5)+\cdots\bigg]
\eeq
which is of order $0(\lambda^{-3/2}),0(\lambda^{-1}),0(\lambda^{-1/2}),0(\lambda^{0}),0(\lambda^{1/2}),0(\lambda^{1})\cdots$ 
\beq
v^4=\bigg[v_0^4+4v_0^3v_2+4v_0^3v_3+(6v_0^2v_2^2+4v_0^3v_34)+(12v_0^2v_2v_3+4v_0^3v_5)+(4v_0v_2^3+6v_0^2v_3^2+12v_0^2v_2v_4+4v_0^3v_6)+\cdots \bigg] 
\eeq
which is of order $0(\lambda^{-2}),0(\lambda^{-3/2}),0(\lambda^{-1}),0(\lambda^{-1/2}),0(\lambda^0),\cdots$. 
\beq
\delta m^2=\delta m_1^2+\delta m_2^2+\delta m_3^2+\delta m_4^2+\delta m_5^2+\delta m_6^2\cdots
\eeq
which is of order $0(\lambda^{1/2}),0(\lambda^{1}),0(\lambda^{3/2}),0(\lambda^{2}),0(\lambda^{5/2}),0(\lambda^{3})\cdots$. 
And according to the renormalization condition the, we know that $\delta m_1^2=0$ \par 
Then for $H_0$ we can do the expansion by $\lambda$ as
\beq
\begin{aligned}
\ch_0\p=&-\frac{m^2}{4}v_0^2+\frac{\lambda}{4}v_0^4+A_0+A_1
         +\bigg(-\frac{m^2}{4}2v_0v_2+\frac{\lambda}{4}4v_0^3v_2+\frac{1}{4}\delta m_2^2v_0^2+A_2\bigg)\\
        &+\bigg(-\frac{m^2}{2}v_0v_3+\lambda v_0^3v_3+\frac{\delta m_3^2 v_0^2}{4}+A_3\bigg)\\
        &+\bigg(-\frac{m^2}{4}(2v_0v_4+v_2^2)+\frac{\lambda}{4}(6v_0^2v_2^2+4v_0^3v_4)
         +\frac{1}{4}(\delta m_2^22v_0v_2+\delta m_4^2v_0^2)+A_4\bigg)\\
        &+\bigg(-\frac{m^2}{4}(2v_2v_3+2v_0v_5)+\frac{\lambda}{4}(12v_0^2v_2v_3+4v_0^3v_5)
         +\frac{1}{4}(\delta m_2^22v_0v_3+\delta m_3^2 2v_0v_2+\delta m_5^2v_0^2)+A_5\bigg)\\
        &+\bigg(-\frac{m^2}{4}(2v_2v_4+2v_3^2+2v_0v_6)+\frac{\lambda}{4}(4v_0v_2^3+6v_0^2v_3^2+12v_0^2v_2v_4+4v_0^3v_6)\\
         &+\frac{1}{4}(\delta m_2^2(v_2^2+2v_0v_4)+\delta m_3^2 2v_0v_3+\delta m_4^2 2v_0v_2+\delta m_6^2 v_0^2)+A_6\bigg)
         \cdots\nonumber
\end{aligned}
\eeq
 which is order of $0(\lambda^{-1}),0(\lambda^{-1/2}),0(\lambda^{0}),0(\lambda^{1/2}),0(\lambda^{1}),0(\lambda^{1/2}),0(\lambda^{1}),0(\lambda^{3/2}),0(\lambda^{2})\cdots$.
\par 
Then for $H_1$ we can do the expansion by $\lambda$ as
\beq
\begin{aligned}
\ch_1\p=&(-\frac{m^2}{2}v_0+\lambda v_0^3)\phi+(-\frac{m^2}{2}v_2+3\lambda v_0^2v_2+\frac{1}{2}\delta m_2^2v_0)\phi
         +\bigg(-\frac{m^2}{2}v_3+3\lambda v_0^2v_3+\frac{1}{2}\delta m_3^2v_0\bigg)\phi\\
        &+\bigg(-\frac{m^2}{2}v_4+\lambda(v_0^2v_4+3v_0v_2^2)+\frac{1}{2}(\delta m_2^2v_2+\delta m_4^2v_0)\bigg)\phi\\
        &+\bigg(-\frac{m^2}{2}v_5+\lambda(6v_0v_2v_3+3v_0^2v_5)+\frac{1}{2}(\delta m_2^2v_3+\delta m_3^2v_2+\delta m_5^2v_0)\bigg)\phi
        +\cdots\nonumber
\end{aligned}
\eeq
 which is order of $0(\lambda^{-1/2}),0(\lambda^{1/2}),0(\lambda^1),0(\lambda^{3/2}),0(\lambda^{2}),\cdots$.
Then for $H_2$ we can do the expansion by $\lambda$ as
\beq
\begin{aligned}
\ch_2\p=&(\frac{\pi^2+(\partial_i\phi)^2}{2}-\frac{m^2}{4}\phi^2+\frac{3\lambda}{2}\phi^2v_0^2)+(3\lambda v_0v_2+\frac{1}{4}\delta m_2^2)\phi^2
        +(3\lambda v_0v_3+\frac{1}{4}\delta m_3^2)\phi^2\\
        &+\bigg(\frac{3\lambda}{2}(v_2^2+2v_0v_4)+\frac{1}{4}\delta m_4^2\bigg)\phi^2+\cdots
\end{aligned}
\eeq
which is order of $0(\lambda^{0}),0(\lambda^{1}),0(\lambda^{3/2}),0(\lambda^2),\cdots $.  
For $H_3$ we can do the expansion by $\lambda$ as
\beq
\begin{aligned}
\ch_3\p=\lambda\phi^3v_0+\lambda\phi^3v_2+\lambda\phi^3v_3 
\end{aligned}
\eeq
 which is order of $0(\lambda^{1/2}),0(\lambda^{3/2}),0(\lambda^{2})\cdots$
For $H_4$ we can do the expansion by $\lambda$ as
\beq
\begin{aligned}
\ch_4\p=\frac{\lambda}{4}\phi^4\nonumber
\end{aligned}
\eeq
It is of order $0(\lambda^{1})$
Then based on the above expansion,we get the Hamiltonian expansion by order of $\lambda^{i/2}$ with denotation $H_{i+2}$.

\subsection{$o(\lambda^{-1}): H_0$}
\beq
\begin{aligned}
\ch_0=&-\frac{m^2}{4}v_0^2+\frac{\lambda}{4}v_0^4+A_0=-\frac{m^4}{16}+A_0
\end{aligned}
\eeq
With renormalization condition, we have $H_0=0$ we have 
\beq 
A_0=\frac{m^4}{16}
\eeq 
\subsection{$o(\lambda^{-1/2}): H_1$}
\beq
\begin{aligned}
\ch_1=A_1+(-\frac{m^2}{2}v_0+\lambda v_0^3)\phi=A_1
\end{aligned}
\eeq
So with renormalization condition: we need $H_1=0$, Then we have:
\beq
A_1=0
\eeq
\subsection{$o(\lambda^{0}): H_2$}
\beq
\begin{aligned}
\ch_2=&\bigg(-\frac{m^2}{4}2v_0v_2+\frac{\lambda}{4}4v_0^3v_2+\frac{1}{4}\delta m_2^2v_0^2+A_2\bigg)
      +(\frac{\pi^2+(\partial_i\phi)^2}{2}-\frac{m^2}{4}\phi^2+\frac{3\lambda}{2}\phi^2v_0^2)\\
     =&\frac{\pi^2+(\partial_i\phi)^2}{2}+\frac{m^2}{2}\phi^2+\frac{\delta m_2^2}{8\lambda}m^2+(\frac{\delta m_2^2}{8\lambda}m^2+A_2)
\end{aligned}
\eeq
The renormalization condition need the c-number term vanish, which indicate:
\beq
A_2=-\frac{\delta m_2^2}{8\lambda}m^2
\eeq
so we have:
\beq
\ch_2=\frac{\pi^2+(\partial_i\phi)^2}{2}+\frac{m^2}{2}\phi^2
\eeq
In indicate the free particle scalar theory with mass $m^2$. And we do the plane wave expansion:
\beq
\phi(\vx)=\pinvp{2}\frac{1}{\sqrt{2\omvp}} (a_{\vp}^{\dag}e^{-i\vp\vx}+a_{\vp}e^{i\vp\vx})
        =\pinvp{2}(\avpd+\frac{\avpm}{2\omvp})e^{-i\vp\vx}
\eeq
\beq
\pi(\vx)=-i\pinvp{2}\frac{\sqrt{\omega_{\vp}}}{\sqrt{2}}(-a_{\vp}^{\dag}e^{-i\vp\vx}+a_{\vk}e^{i\vp\vx})
        =i\pinvp{2}(\omvp{}A_{\vp}^{\ddag}-\frac{A_{-\vp}}{2})e^{-i\vp\vx}
\eeq
 Where we have the operator transformation for simplication of computation:
 \beq
 \avpd=\frac{1}{\sqrt{2\omvp}}a^{\dag}_{\vp},\hsp
  \avp=\sqrt{2\omvp}a_{\vp}
\eeq
The commutation still hold:
\beq
[\navp1,\navpd2]=(2\pi)^2 \delta(\vp_1-\vp_2)
\eeq
after normal order like we do in normal proccedure in QFT. we have:
\beq
\ch_2=\pinvp{2}\omvp\avpd\avpm
\eeq

aft
\subsection{$o(\lambda^{1/2}): H_3$}
\beq
\begin{aligned}
\ch_3=&-\frac{m^2}{2}v_0v_3+\lambda v_0^3v_3+\frac{\delta m_3^2}{4}v_0^2+A_3
       +(-\frac{m^2}{2}v_2+3\lambda v_0^2v_2+\frac{1}{2}\delta m_2^2v_0)\phi+\lambda\phi^3v_0\\
    =&\frac{\delta m_3^2}{8\lambda}m^2+A_3
       +(m^2v_2+\frac{1}{2}\delta m_2^2v_0)\phi+\lambda\phi^3v_0\\ 
\end{aligned}
\eeq
The renormalization condition need the c-number term vanish, which indicate:
\beq
A_3=-\frac{\delta m_3^2}{8\lambda}m^2
\eeq
And tadpole term vanish indicate:
\beq
v_2=-\frac{\delta m_2^2v_0}{2m^2}=-\frac{\delta m_2^2}{2m\sqrt{2\lambda}}
\eeq
Then we have:
\beq
\ch_3=\lambda\phi^3v_0=m\sqrt{\frac{\lambda}{2}}\phi^3(\vx)
\eeq
With mode expansion, we have:
\bea
H_3&=&m\sqrt\frac{\lambda}{2}
\int d^{\dim}\vx \int\frac{d^{\dim}\vp_1 d^{\dim}\vp_2 d^{\dim}\vp_3}{(2\pi)^{\dimthree}} e^{-i\vx\cdot (\vp_1+\vp_2+\vp_3)}\left(A^\ddag_{\vp_1}+\frac{A_{-\vp_1}}{2\omega_{\vp_1}}\right)\\
&&\times \left(A^\ddag_{\vp_2}+\frac{A_{-\vp_2}}{2\omega_{\vp_2}}\right)\left(A^\ddag_{\vp_3}+\frac{A_{-\vp_3}}{2\omega_{\vp_3}}\right)\nonumber\\
&=&m\sqrt\frac{\lambda}{2}   \int\frac{d^{\dim}\vp_1 d^{\dim}\vp_2}{(2\pi)^{\dimtwo}}\left(A^\ddag_{\vp_1}+\frac{A_{-\vp_1}}{2\omega_{\vp_1}}\right)\left(A^\ddag_{\vp_2}+\frac{A_{-\vp_2}}{2\omega_{\vp_2}}\right)\left(A^\ddag_{-\vp_1-\vp_2}+\frac{A_{\vp_1+\vp_2}}{2\omega_{\vp_1+\vp_2}}\right).\nonumber
\eea
\subsection{$o(\lambda^{1}): H_4$}
\beq
\begin{aligned}
\ch_4=&\bigg(-\frac{m^2}{4}(2v_0v_4+v_2^2)+\frac{\lambda}{4}(6v_0^2v_2^2+4v_0^3v_4)
         +\frac{1}{4}(\delta m_2^22v_0v_2+\delta m_4^2v_0^2)+A_4\bigg)\\
      &\bigg(-\frac{m^2}{2}v_3+3\lambda v_0^2v_3+\frac{1}{2}\delta m_3^2v_0\bigg)\phi
       +(3\lambda v_0v_2+\frac{1}{4}\delta m_2^2)\phi^2
       +\frac{\lambda}{4}\phi^4\\
    =&\frac{m^2v_2^2}{2}-\frac{(\delta m_2^2)^2}{8\lambda}+\frac{m^2\delta m_4^2}{8\lambda}+A_4
       +\bigg(m^2v_3+\frac{1}{2}\delta m_3^2v_0\bigg)\phi-\frac{1}{2}\delta m_2^2\phi^2
       +\frac{\lambda}{4}\phi^4\\
    =&\frac{\lambda}{4}\phi^4-\frac{\delta m_2^2}{2}\phi^2+\bigg(m^2v_3+\frac{1}{2}\delta m_3^2v_0\bigg)\phi
         -\frac{(\delta m_2^2)^2}{16\lambda}+\frac{m^2\delta m_4^2}{8\lambda}+A_4
\end{aligned}
\eeq
The renormalization condition indicate odd term vanish, then  we have:
\beq
v_3=-\frac{\delta m_3^2v_0}{2m^2}=-\frac{\delta m_3^2}{2m\sqrt{2\lambda}}
\eeq
then we have
\bea
\ch_4&=&\frac{\lambda}{4}\phi^4-\frac{\delta m_2^2}{2}\phi^2+\hat A_4\\
\hat A_4&=&-\frac{(\delta m_2^2)^2}{16\lambda}+\frac{m^2\delta m_4^2}{8\lambda}+A_4 \label{a4}
\eea
While we can expand:
\bea
\phi^2&=&\int d^{\dim}\vx \int\frac{d^{\dim}\vp_1 d^{\dim}\vp_2}{(2\pi)^{\dimtwo}} e^{-i\vx\cdot(\vp_1+\vp_2)}
(A^\ddag_{\vp_1}+\frac{A_{-\vp_1}}{2\omega_{\vp_1}})\left(A^\ddag_{\vp_2}+\frac{A_{-\vp_2}}{2\omega_{\vp_2}}\right)\nonumber\\
&=&\int\frac{d^{\dim}\vp_1 }{(2\pi)^{\dim}}\left(A^\ddag_{\vp_1}+\frac{A_{-\vp_1}}{2\omega_{\vp_1}}\right)
    \left(A^\ddag_{-\vp_1}+\frac{A_{\vp_1}}{2\omega_{\vp_1}}\right).\nonumber
\eea
\bea
\phi^4&=&\int d^{\dim}\vx \int\frac{d^{\dim}\vp_1 d^{\dim}\vp_2}{(2\pi)^{\dimtwo}} e^{-i\vx\cdot(\vp_1+\vp_2+\vp_3+\vp_4)}
(\avpd1+\frac{\avpm1}{2\omega_{\vp_1}})\left(A^\ddag_{\vp_2}+\frac{A_{-\vp_2}}{2\omega_{\vp_2}}\right)\nonumber\\
&=&\int\frac{d^{\dim}\vp_1 }{(2\pi)^{\dim}}\left(A^\ddag_{\vp_1}+\frac{A_{-\vp_1}}{2\omega_{\vp_1}}\right)
    \left(A^\ddag_{\vp_2}+\frac{A_{-\vp_2}}{2\omega_{\vp_2}}\right)\left(A^\ddag_{\vp_3}+\frac{A_{-\vp_3}}{2\omega_{\vp_3}}\right)
    \left(A^\ddag_{-\vp_1-\vp_2-\vp_3}+\frac{A_{\vp_1+\vp_2+\vp_3}}{2\omega_{\vp_1+\vp_2+\vp_3}}\right)\nonumber
\eea

 \subsection{$o(\lambda^{3/2}): H_5$}
\beq
\begin{aligned}
\ch_5=&\bigg(-\frac{m^2}{4}(2v_2v_3+2v_0v_5)+\frac{\lambda}{4}(12v_0^2v_2v_3+4v_0^3v_5)
         +\frac{1}{4}(\delta m_2^22v_0v_3+\delta m_3^2 2v_0v_2+\delta m_5^2v_0^2)+A_5\bigg)\\
      &+\bigg(-\frac{m^2}{2}v_4+3\lambda(v_0^2v_4+v_0v_2^2)+\frac{1}{2}(\delta m_2^2v_2+\delta m_4^2v_0)\bigg)\phi
      +(3\lambda v_0v_3+\frac{1}{4}\delta m_3^2)\phi^2+\lambda\phi^3v_2\\
    =&-\frac{\delta m_2^2\delta m_3^2}{8\lambda}+\frac{\delta m_5^2}{8\lambda}m^2+A_5
       +\bigg(m^2v_4+\frac{4m^2\delta m_4^2+(\delta m_2^2)^2}{8m\sqrt{2\lambda}}\bigg)\phi
       -\frac{\delta m_3^2}{2}\phi^2
       -\sqrt{\frac{\lambda}{2}}\frac{\delta m_2^2}{2m}\phi^3
 \end{aligned}
\eeq
we set 
\beq
T_5=\bigg(m^2v_4+\frac{4m^2\delta m_4^2+(\delta m_2^2)^2}{8m\sqrt{2\lambda}}\bigg)
\eeq
Then we have:
\beq
\ch_5=-\sqrt{\frac{\lambda}{2}}\frac{\delta m_2^2}{2m}\phi^3 -\frac{\delta m_3^2}{2}\phi^2+T_5\phi+\hat{A_5}\hsp
\hat{A_5}=-\frac{\delta m_2^2\delta m_3^2}{8\lambda}+\frac{\delta m_5^2}{8\lambda}m^2+A_5
\eeq
For renormalization condition we have $A_5=0$, such that:

 \subsection{$o(\lambda^{2}): H_6$}
\beq
\begin{aligned}
\ch_6=&\bigg(-\frac{m^2}{4}(2v_2v_4+v_3^2+2v_0v_6)+\frac{\lambda}{4}(4v_0v_2^3+6v_0^2v_3^2+12v_0^2v_2v_4+4v_0^3v_6)\\
         &+\frac{1}{4}(\delta m_2^2(v_2^2+2v_0v_4)+\delta m_3^2 2v_0v_3+\delta m_4^2 2v_0v_2+\delta m_6^2 v_0^2)+A_6\bigg)\\
         &\bigg(-\frac{m^2}{2}v_5+\lambda(6v_0v_2v_3+3v_0^2v_5)+\frac{1}{2}(\delta m_2^2v_3+\delta m_3^2v_2+\delta m_5^2v_0)\bigg)\phi\\
         &+\bigg(\frac{3\lambda}{2}(v_2^2+2v_0v_4)+\frac{1}{4}\delta m_4^2\bigg)\phi^2+\lambda\phi^3v_3\\
      =&-\sqrt{\frac{\lambda}{2}}\frac{\delta m_3^2}{2m}\phi^3-\frac{\delta m_4^2}{2}\phi^2+\bigg(m^2v_5+\frac{\delta m_2^2\delta m_3^2}{4m\sqrt{2\lambda}}
      +\frac{m\delta m_5^2}{2\sqrt{2\lambda}}\bigg)\phi
      +\bigg(-\frac{(\delta m_3^2)^2}{16\lambda}-\frac{\delta m_2^2\delta m_4^2}{8\lambda}+\frac{\delta m_6^2m^2}{8\lambda}+A_6\bigg)
\end{aligned}
\eeq
We set 
\beq
T_6=m^2v_5+\frac{\delta m_2^2\delta m_3^2}{4m^3\sqrt{2\lambda}}+\frac{\delta m_5^2}{2m\sqrt{2\lambda}}
    =m^5v_5+\frac{\delta m_2^2\delta m_3^2+2m^2\delta m_5^2}{4m^3\sqrt{2\lambda}}
\eeq
Then we have:
\beq
\ch_6=-\sqrt{\frac{\lambda}{2}}\frac{\delta m_3^2}{2m}\phi^3-\frac{\delta m_4^2}{2}\phi^2+T_6\phi+\hat{A_6}\hsp
\hat{A_6}=-\frac{(\delta m_3^2)^2}{16\lambda}-\frac{\delta m_2^2\delta m_4^2}{8\lambda}+\frac{\delta m_6^2m^2}{8\lambda}+A_6
\eeq

\section{Hengyuan calculation: vacuum state expansion}
The schronger equation is 
\beq
H{\ovac}=0.
\eeq
While:
\beq
{\ovac}=\sum_{i-0}{\ovac}_i
\eeq
while${\ovac}_i$  is order of $\lambda^{i}$. then the equation give renormalization order by order 
\beq
\begin{aligned}
\lambda^{0}\hsp     &H_2{\ovac}_0=0.\\
\lambda^{1/2}\hsp   &H_2{\ovac}_1+H_3{\ovac}_0=0.\\
\lambda^{1}\hsp     &H_2{\ovac}_2+H_3{\ovac}_1+H_4{\ovac}_0=0.\\
\lambda^{3/2}\hsp   &H_2{\ovac}_3+H_3{\ovac}_2+H_4{\ovac}_1+H_5{\ovac}_0=0.\\
\lambda^{2}\hsp     &H_2{\ovac}_4+H_3{\ovac}_3+H_4{\ovac}_2+H_5{\ovac}_1+H_6{\ovac}_0=0.\\
\cdots
\end{aligned}
\eeq
It is noted that we do the normal order to all the $H_i $above\par 
What is more: we do the expansion of single ${\ovac}_i=\sum\Omega\rangle_i^n$, while$\Omega\rangle_i^n$ is the n-meson state component  of the i-th order quantum correction vacuum sate.

\subsection{$\lambda^0$ }
The leadinig order schrodinger equation is 
\beq
H_2{\ovac}_0=0.
\eeq
While we have got that:
\beq
H_2=\pink{2}\omvk\avkd\avkm
\eeq
Then the leading order renormalization condition generate 
\beq
\avk{\ovac}_o=0
\eeq
and it is obvious that we need the higher order correction ${\ovac}_i^0 $ for$bi>=1 $. which in physical means, we don't have the component which is lienar dependwnt with the ${\ovac}_o=$. it indicate:
\beq
H_2{\ovac}_i^0=0
\eeq
this will contribute a constrain to the state in later section.
\subsection{$\lambda^{1/2}$ }
The leadinig order schrodinger equation is 
\beq
H_2{\ovac}_1+H_3{\ovac}_0=0.
\eeq
While we have get
\bea
H_3&=&m\sqrt\frac{\lambda}{2}
\int d^{\dim}\vx \int\frac{d^{\dim}\vp_1 d^{\dim}\vp_2 d^{\dim}\vp_3}{(2\pi)^{\dimthree}} e^{-i\vx\cdot (\vp_1+\vp_2+\vp_3)}\left(A^\ddag_{\vp_1}+\frac{A_{-\vp_1}}{2\omega_{\vp_1}}\right)\\
&&\times \left(A^\ddag_{\vp_2}+\frac{A_{-\vp_2}}{2\omega_{\vp_2}}\right)\left(A^\ddag_{\vp_3}+\frac{A_{-\vp_3}}{2\omega_{\vp_3}}\right)\nonumber\\
&=&m\sqrt\frac{\lambda}{2}   \int\frac{d^{\dim}\vp_1 d^{\dim}\vp_2}{(2\pi)^{\dimtwo}}\left(A^\ddag_{\vp_1}+\frac{A_{-\vp_1}}{2\omega_{\vp_1}}\right)\left(A^\ddag_{\vp_2}+\frac{A_{-\vp_2}}{2\omega_{\vp_2}}\right)\left(A^\ddag_{-\vp_1-\vp_2}+\frac{A_{\vp_1+\vp_2}}{2\omega_{\vp_1+\vp_2}}\right).\nonumber
\eea

The only term that contribute nonzero part of schrondinger equation is  
\bea
H_3{\ovac}_0 \supset &&  m\sqrt\frac{\lambda}{2} \int\frac{d^{\dim}\vp_1 d^{\dim}\vp_2}{(2\pi)^{\dimtwo}}\left(A^\ddag_{\vp_1}A^\ddag_{\vp_2}A^\ddag_{-\vp_1-\vp_2}\right){\ovac}_0.\\\nonumber 
 &&=m\sqrt\frac{\lambda}{2} \int\frac{d^{\dim}\vp_1 d^{\dim}\vp_2}{(2\pi)^{\dimtwo}}|\vp_1,\vp_2,-\vp_1-\vp_2\rangle_0.
\eea
Then with leading order schrodinger equation and comutation $[\navk1,\navkd2]=(2\pi)^2\delta(k_1-k_2)$. we get:

\beq
{\ovac}_1=-m\sqrt\frac{\lambda}{2} \int\frac{d^{\dim}\vp_1 d^{\dim}\vp_2}{(2\pi)^{\dimtwo}}
                  \frac{|\vp_1,\vp_2,-\vp_1-\vp_2\rangle_0}{\nomvp1+\nomvp2+\omega_{\vp_1+\vp_2}}
\eeq

\subsection{$\lambda^{1}$ }
The second order schrodinger equation is 
\beq
H_2{\ovac}_2+H_3{\ovac}_1+H_4{\ovac}_0=0.
\eeq
While we have 
\bea
\ch_4&=&:\frac{\lambda}{4}\phi^4-\frac{\delta m_2^2}{2}\phi^2+\hat A_4:\\\nonumber
\ch_3 &=&:\lambda\phi^3v_0=:m\sqrt{\frac{\lambda}{2}}\phi^3(\vx):
\eea
The only term that contribute  ${\ovac}_2^0$ part of ${\ovac}_2$ in schrodinger equation is  
\beq
\int d^2\vx\hat A_4{\ovac}_0
+m\sqrt\frac{\lambda}{2}\int d^2\vx\int\frac{d^{\dim}\vp_1 d^{\dim}\vp_2 d^\dim\vp_3}{(2\pi)^{\dimthree}}e^{-i\vx(\vp_1+\vp_2+\vp_3)}
A_{\vp_1}A_{\vp_2}\frac{A_{-\vp_3}}{2\nomvp3}{\ovac}_1
+H_2{\ovac}_2^0=0
\eeq
Because$ H_2{\ovac}_2^0=0$ as we point before, so we have:
\bea
\int d^2\vx\hat A_4{\ovac}_0
&=&-m\sqrt\frac{\lambda}{2}\int d^2\vx\int\frac{d^{\dim}\vp_1 d^{\dim}\vp_2 d^\dim\vp_3}{(2\pi)^{\dimthree}}e^{-i\vx(\vp_1+\vp_2+\vp_3)}
        \frac{A_{\vp_1}A_{\vp_2}A_{-\vp_3}}{8\nomvp1\nomvp2\nomvp3}{\ovac}_1\\\nonumber
&=&m^2\frac{\lambda}{2}\int d^2\vx\int\frac{d^{\dim}\vp_1 d^{\dim}\vp_2 d^\dim\vp_3}{(2\pi)^{\dimthree}}e^{-i\vx(\vp_1+\vp_2-\vp_3)}
       \frac{A_{\vp_1}A_{\vp_2}A_{-\vp_3}}{8\nomvp1\nomvp2\nomvp3}\\\nonumber
&&\times\bigg(\int\frac{d^{\dim}\vp_1 d^{\dim}\vp_2}{(2\pi)^{\dimtwo}}
                  \frac{\navpd1\navpd2 A^\ddag_{-\vp_1-\vp_2}}{\nomvp1+\nomvp2+\omega_{\vp_1+\vp_2}}\bigg){\ovac}_0\\\nonumber 
&=&m^2\frac{3\lambda}{8}\int d^2\vx\int\frac{d^{\dim}\vp_1 d^{\dim}\vp_2}{(2\pi)^{\dimtwo}}
       \frac{1}{\nomvp1\nomvp2\nomvp3(\nomvp1+\nomvp2+\omega_{\vp_1+\vp_2}}){\ovac}_0\\\nonumber   
\eea
So we get:
\beq
\hat A_4=\frac{3\lambda}{8}m^2\int d^2\vx\int\frac{d^{\dim}\vp_1 d^{\dim}\vp_2}{(2\pi)^{\dimtwo}}
       \frac{1}{\nomvp1\nomvp2\nomvp3(\nomvp1+\nomvp2+\omega_{\vp_1+\vp_2}})
\eeq
The left term in ${\ovac}_2$ left: ${\ovac}_2^2, {\ovac}_2^4, {\ovac}_2^6$m, Where\par 
${\ovac}_2^2 $ comes from $\navpd1\navpm2\navpm3{\ovac}_1 $ terms in $H_3{\ovac}_1 $ and $\navpd1\navpd2{\ovac}_0 $ terms in  $H_2{\ovac}_0$ \par  
\beq
\begin{aligned}
H_3{\ovac}_1 \supset &m\sqrt\frac{\lambda}{2} \int d^2\vx\int\frac{d^{\dim}\vp_1 d^{\dim}\vp_2d^{\dim}\vp_3}{(2\pi)^{\dimthree}}
      \left(A^\ddag_{\vp_1}\frac{A_{-\vp_2}A_{-\vp_3}}{4\nomvp2\nomvp3}\right) e^{-\vx(\vp_1+\vp_2+\vp_3)}{\ovac}_1.\\
 =&-m\frac{\lambda}{2}\int d^2\vx\int\frac{d^{\dim}\vp_1 d^{\dim}\vp_2d^{\dim}\vp_3}{(2\pi)^{\dimtwo}}A^\ddag_{\vp_1}\frac{A_{-\vp_2}A_{-\vp_3}}{4\nomvp2\nomvp3}
      e^{-\vx(\vp_1+\vp_2+\vp_3)}
      \bigg(\int\frac{d^{\dim}\vp_1 d^{\dim}\vp_2}{(2\pi)^{\dimtwo}}
      \frac{\navpd1\navpd2 A^\ddag_{-\vp_1-\vp_2}}{\nomvp1+\nomvp2+\omega_{\vp_1+\vp_2}}\bigg){\ovac}_0
\end{aligned}
\eeq
\bea
H_4{\ovac}_0 \supset && -\int d^2\vx\frac{\delta m_2^2}{2}\int\frac{d^{\dim}\vp_1 d^{\dim}\vp_2}{(2\pi)^{\dimtwo}}
      \left(A^\ddag_{\vp_1}A^\ddag_{\vp_2}\right)e^{-i\vx(\vp_1+\vp_2)}{\ovac}_0.\\\nonumber 
 &&=-\frac{\delta m_2^2}{2}\int\frac{d^2\vp_1 }{(2\pi)^{2}}
      \left(A^\ddag_{\vp_1}A^\ddag_{-\vp_1}\right){\ovac}_0.\\\nonumber    
\eea
${\ovac}_2^4$ comes from $\navpd1\navpd2\navpm3{\ovac}_1$ terms in $H_3{\ovac}_1$ and $\navpd1\navpd2\navpd3\navpd4{\ovac}_0$ terms in  $H_2{\ovac}_0$, indivisually:
\bea
H_3{\ovac}_1 \supset &&  m\sqrt\frac{\lambda}{2} \int\frac{d^{\dim}\vp_1 d^{\dim}\vp_2d^{\dim}\vp_3}{(2\pi)^{\dimthree}}
      \left(A^\ddag_{\vp_1}A^\ddag_{\vp_2}\frac{A_{-\vp_3}}{2\nomvp3}\right)e^{-i\vx(\vp_1+\vp_2+\vp_3)}{\ovac}_1.\\\nonumber  
 &&= -m\frac{\lambda}{2} \int\frac{d^{\dim}\vp_1 d^{\dim}\vp_2d^{\dim}\vp_3}{(2\pi)^{\dimtwo}}A^\ddag_{\vp_1}A^\ddag_{\vp_2}\frac{A_{-\vp_3}}{2\nomvp3}
      \bigg(\int\frac{d^{\dim}\vp_1 d^{\dim}\vp_2}{(2\pi)^{\dimtwo}}
      \frac{\navpd1\navpd2 A^\ddag_{-\vp_1-\vp_2}}{\nomvp1+\nomvp2+\omega_{\vp_1+\vp_2}}\bigg){\ovac}_0\\\nonumber   
\eea

\beq
\begin{aligned}
H_4{\ovac}_0 \supset && \frac{\lambda}{4}\int d^2\vx\int d^2\vx\int\frac{d^{\dim}\vp_1 d^{\dim}\vp_2d^{\dim}\vp_3d^{\dim}\vp_4}{(2\pi)^{8}}
      \left(A^\ddag_{\vp_1}A^\ddag_{\vp_2}A^\ddag_{\vp_3}A^\ddag_{\vp_4}\right)e^{-i\vx(\vp_1+\vp_2+\vp_3+\vp_4)}{\ovac}_0.\\\nonumber 
 &&=-\frac{\delta m_2^2}{2}\int\frac{d^{\dim}\vp_1 d^{\dim}\vp_2d^{\dim}\vp_3}{(2\pi)^6}
      A^\ddag_{\vp_1}A^\ddag_{\vp_2}A^\ddag_{\vp_3}A^\ddag_{-\vp_1-\vp_2-\vp_3}{\ovac}_0\\\nonumber   
\end{aligned}
\eeq
${\ovac}_2^6$ comes from $\navpd1\navpd2\navpd3{\ovac}_1$ terms in $H_3{\ovac}_1$ \par 
\bea
H_3{\ovac}_1 \supset &&  m\sqrt\frac{\lambda}{2}\int d^2\vx\int\frac{d^{\dim}\vp_1 d^{\dim}\vp_2d^{\dim}\vp_3}{(2\pi)^{\dimthree}}
      \left(A^\ddag_{\vp_1}A^\ddag_{\vp_2}A^\ddag_{\vp_3}\right)e^{-i\vx(\vp_1+\vp_2+\vp_3)}{\ovac}_1.\\\nonumber  
 &&= -m\frac{\lambda}{2} \int\frac{d^{\dim}\vp_1 d^{\dim}\vp_2d^{\dim}\vp_3}{(2\pi)^{\dimtwo}}A^\ddag_{\vp_1}A^\ddag_{\vp_2}A^\ddag_{-\vp_1-\vp_2}
      \bigg(\int\frac{d^{\dim}\vp_1 d^{\dim}\vp_2}{(2\pi)^{\dimtwo}}
      \frac{\navpd1\navpd2 A^\ddag_{-\vp_1-\vp_2}}{\nomvp1+\nomvp2+\omega_{\vp_1+\vp_2}}\bigg){\ovac}_0\\\nonumber   
\eea
Then we can have the full expresion of ${\ovac}_2={\ovac}_2^2+{\ovac}_2^4+{\ovac}_2^6$

\subsection{$\lambda^{3/2}$ }
The  order third schrodinger equation is 
\beq
H_2{\ovac}_3+H_3{\ovac}_2+H_4{\ovac}_1+H_5{\ovac}_0=0.
\eeq
While we have 
\bea
\ch_5&=&-\sqrt{\frac{\lambda}{2}}\frac{\delta m_2^2}{2m}\phi^3 -\frac{\delta m_3^2}{2}\phi^2+T_5\phi+\hat{A_5}\\\nonumber
\ch_4&=&:\frac{\lambda}{4}\phi^4-\frac{\delta m_2^2}{2}\phi^2+\hat A_4:\\\nonumber
\ch_3&=&:\lambda\phi^3v_0=:m\sqrt{\frac{\lambda}{2}}\phi^3(\vx):
\eea
 There is no term which relate ${\ovac}_3^0$ part of ${\ovac}_3$ in schrodinger equation.  it only have
 \beq
 {\ovac}_3={\ovac}_3^1+{\ovac}_3^3+{\ovac}_3^5+{\ovac}_3^7+{\ovac}_3^9
 \eeq
Where ${\ovac}_3^0$ contribution comes from\par 
\beq
\begin{aligned}
H_5{\ovac}_1 \supset & \hat A_5{\ovac}_0.\\
\end{aligned}
\eeq
Where ${\ovac}_3^1$ contribution comes from\par 
\beq
\begin{aligned}
H_3{\ovac}_2 \supset &m\sqrt\frac{\lambda}{2} \int d^2\vx\int\frac{d^{\dim}\vp_1 d^{\dim}\vp_2d^{\dim}\vp_3}{(2\pi)^{\dimthree}}
      \left(A^\ddag_{\vp_1}\frac{A_{-\vp_2}A_{-\vp_3}}{4\nomvp2\nomvp3}\right) e^{-\vx(\vp_1+\vp_2+\vp_3)}{\ovac}_2^2.\\
H_3{\ovac}_2 \supset &m\sqrt\frac{\lambda}{2} \int d^2\vx\int\frac{d^{\dim}\vp_1 d^{\dim}\vp_2d^{\dim}\vp_3}{(2\pi)^{\dimthree}}
      \left(\frac{A_{-\vp_1}A_{-\vp_2}A_{-\vp_3}}{8\nomvp1\nomvp2\nomvp3}\right) e^{-\vx(\vp_1+\vp_2+\vp_3)}{\ovac}_2^4.\\
H_4{\ovac}_1 \supset & -\int d^2\vx\frac{\delta m_2^2}{2}\int\frac{d^{\dim}\vp_1 d^{\dim}\vp_2}{(2\pi)^{\dimtwo}}
      \left(\frac{A_{-\vp_1}A_{-\vp_2}}{4\nomvp1\nomvp2}\right)e^{-i\vx(\vp_1+\vp_2)}{\ovac}_1.\\\nonumber 
H_4{\ovac}_1 \supset & -\frac{\lambda}{4}\int d^2\vx\int\frac{d^{\dim}\vp_1 d^{\dim}\vp_2d^{\dim}\vp_3 d^{\dim}\vp_4}{(2\pi)^{8}}
      \left(\navpd1\frac{\navpm2\navpm3\navpm4}{8\nomvp2\nomvp3\nomvp4}\right)e^{-i\vx(\vp_1+\vp_2+\vp_3+\vp_4)}{\ovac}_1\\\nonumber 
H_5{\ovac}_0 \supset & T_5\int d^2\vx\int\frac{d^{\dim}\vp_1}{(2\pi)^{2}}
      \left(A^\ddag_{\vp_1}\right) e^{-\vx\cdot \vp_1}{\ovac}_0.\\
\end{aligned}
\eeq
${\ovac}_3^3$ comes from:
\beq
\begin{aligned}
H_3{\ovac}_2 \supset &m\sqrt\frac{\lambda}{2} \int d^2\vx\int\frac{d^{\dim}\vp_1 d^{\dim}\vp_2d^{\dim}\vp_3}{(2\pi)^{\dimthree}}
      \left(A^\ddag_{\vp_1}A^\ddag_{-\vp_2}\frac{A_{-\vp_3}}{2\nomvp3}\right) e^{-\vx(\vp_1+\vp_2+\vp_3)}{\ovac}_2^2.\\
H_3{\ovac}_2 \supset &m\sqrt\frac{\lambda}{2} \int d^2\vx\int\frac{d^{\dim}\vp_1 d^{\dim}\vp_2d^{\dim}\vp_3}{(2\pi)^{\dimthree}}
      \left(A^\ddag_{\vp_1}\frac{A_{-\vp_2}A_{-\vp_2}}{4\nomvp2\nomvp3}\right) e^{-\vx(\vp_1+\vp_2+\vp_3)}{\ovac}_2^4.\\
H_3{\ovac}_2 \supset &m\sqrt\frac{\lambda}{2} \int d^2\vx\int\frac{d^{\dim}\vp_1 d^{\dim}\vp_2d^{\dim}\vp_3}{(2\pi)^{\dimthree}}
      \left(\frac{A_{-\vp_1}A_{-\vp_2}A_{-\vp_2}}{8\nomvp1\nomvp2\nomvp3}\right) e^{-\vx(\vp_1+\vp_2+\vp_3)}{\ovac}_2^6.\\
H_4{\ovac}_1 \supset & -\int d^2\vx\frac{\delta m_2^2}{2}\int\frac{d^{\dim}\vp_1 d^{\dim}\vp_2}{(2\pi)^{\dimtwo}}
      \left(A^\ddag_{\vp_1}\frac{A_{-\vp_2}}{2\nomvp2}\right)e^{-i\vx(\vp_1+\vp_2)}{\ovac}_1.\\\nonumber 
H_4{\ovac}_1 \supset & \frac{\lambda}{4}\int d^2\vx\int\frac{d^{\dim}\vp_1 d^{\dim}\vp_2d^{\dim}\vp_3 d^{\dim}\vp_4}{(2\pi)^{8}}
      \left(\navpd1\navpd2\frac{\navpm3\navpm4}{4\nomvp3\nomvp4}\right)e^{-i\vx(\vp_1+\vp_2+\vp_3+\vp_4)}{\ovac}_1\\\nonumber 
H_4{\ovac}_1 \supset & \int d^2\hat A_4\vx{\ovac}_1\\\nonumber 
H_5{\ovac}_0 \supset &\frac{\delta m m_3^2}{2}\int d^2\vx\int\frac{d^{\dim}\vp_1 d^{\dim}\vp_2}{(2\pi)^{\dimtwo}}
      \left(A^\ddag_{\vp_1}A^\ddag_{\vp_2}\right) e^{-\vx(\vp_1+\vp_2)}{\ovac}_0.\\
\end{aligned}
\eeq
${\ovac}_3^5$ comes from:
\beq
\begin{aligned}
H_3{\ovac}_1 \supset &m\sqrt\frac{\lambda}{2} \int d^2\vx\int\frac{d^{\dim}\vp_1 d^{\dim}\vp_2d^{\dim}\vp_3}{(2\pi)^{\dimthree}}
      \left(A^\ddag_{\vp_1}A^\ddag_{-\vp_2}A^\ddag_{-\vp_3}\right) e^{-\vx(\vp_1+\vp_2+\vp_3)}{\ovac}_2^2.\\
H_3{\ovac}_1 \supset &m\sqrt\frac{\lambda}{2} \int d^2\vx\int\frac{d^{\dim}\vp_1 d^{\dim}\vp_2d^{\dim}\vp_3}{(2\pi)^{\dimthree}}
      \left(A^\ddag_{\vp_1}A^\ddag_{\vp_2}\frac{A_{-\vp_3}}{2\nomvp3}\right) e^{-\vx(\vp_1+\vp_2+\vp_3)}{\ovac}_2^4.\\
H_3{\ovac}_1 \supset &m\sqrt\frac{\lambda}{2} \int d^2\vx\int\frac{d^{\dim}\vp_1 d^{\dim}\vp_2d^{\dim}\vp_3}{(2\pi)^{\dimthree}}
      \left(A^\ddag_{\vp_1}\frac{A_{-\vp_2}A_{-\vp_2}}{4\nomvp2\nomvp3}\right) e^{-\vx(\vp_1+\vp_2+\vp_3)}{\ovac}_2^6.\\
H_4{\ovac}_0 \supset & -\int d^2\vx\frac{\delta m_2^2}{2}\int\frac{d^{\dim}\vp_1 d^{\dim}\vp_2}{(2\pi)^{\dimtwo}}
      \left(A^\ddag_{\vp_1}A^\ddag_{\vp_2}\right)e^{-i\vx(\vp_1+\vp_2)}{\ovac}_1.\\\nonumber 
H_4{\ovac}_0 \supset & -\frac{\lambda}{4}\int d^2\vx\int\frac{d^{\dim}\vp_1 d^{\dim}\vp_2d^{\dim}\vp_3 d^{\dim}\vp_4}{(2\pi)^{8}}
      \left(\navpd1\navpd2\navpm3\frac{\navpm4}{2\nomvp4}\right)e^{-i\vx(\vp_1+\vp_2+\vp_3+\vp_4)}{\ovac}_1\\\nonumber 
H_5{\ovac}_0 \supset &\sqrt{\frac{\lambda}{2}}\frac{\delta m m_2^2}{2m}\int d^2\vx\int\frac{d^{\dim}\vp_1d^{\dim}\vp_2d^{\dim}\vp_3}{(2\pi)^{6}}
      \left(A^\ddag_{\vp_1}A^\ddag_{\vp_2}A^\ddag_{\vp_3}\right) e^{-\vx(\vp_1+\vp_2+\vp_3)}{\ovac}_0.\\
\end{aligned}
\eeq
${\ovac}_3^7$ comes from:
\beq
\begin{aligned}
H_3{\ovac}_1 \supset &m\sqrt\frac{\lambda}{2} \int d^2\vx\int\frac{d^{\dim}\vp_1 d^{\dim}\vp_2d^{\dim}\vp_3}{(2\pi)^{\dimthree}}
      \left(A^\ddag_{\vp_1}A^\ddag_{\vp_2}A^\ddag_{\vp_3}\right) e^{-\vx(\vp_1+\vp_2+\vp_3)}{\ovac}_2^4.\\
H_3{\ovac}_1 \supset &m\sqrt\frac{\lambda}{2} \int d^2\vx\int\frac{d^{\dim}\vp_1 d^{\dim}\vp_2d^{\dim}\vp_3}{(2\pi)^{\dimthree}}
      \left(A^\ddag_{\vp_1}A^\ddag_{\vp_2}\frac{A_{-\vp_3}}{2\nomvp3}\right) e^{-\vx(\vp_1+\vp_2+\vp_3)}{\ovac}_2^6.\\
\end{aligned}
\eeq
${\ovac}_3^0$ comes from:
\beq
\begin{aligned}
H_3{\ovac}_1 \supset &m\sqrt\frac{\lambda}{2} \int d^2\vx\int\frac{d^{\dim}\vp_1 d^{\dim}\vp_2d^{\dim}\vp_3}{(2\pi)^{\dimthree}}
      \left(A^\ddag_{\vp_1}A^\ddag_{\vp_2}A^\ddag_{\vp_3}\right) e^{-\vx(\vp_1+\vp_2+\vp_3)}{\ovac}_2^6.\\
\end{aligned}
\eeq
Then we can have the full expression of 
\beq
{\ovac}_3={\ovac}_3^1+{\ovac}_3^3+{\ovac}_3^5+{\ovac}_3^7+{\ovac}_3^9  \label{v3}
\eeq

\subsection{$\lambda^{2}$ }
The  order third schrodinger equation is 
\beq
H_2{\ovac}_4+H_3{\ovac}_3+H_4{\ovac}_2+H_5{\ovac}_1+H_6{\ovac}_0=0.
\eeq
While we have 
\bea
\ch_6&=&-\sqrt{\frac{\lambda}{2}}\frac{\delta m_3^2}{2m}\phi^3-\frac{\delta m_4^2}{2}\phi^2+T_6\phi+\hat{A_6}\\\nonumber
\ch_5&=&-\sqrt{\frac{\lambda}{2}}\frac{\delta m_2^2}{2m}\phi^3 -\frac{\delta m_3^2}{2}\phi^2+T_5\phi+\hat{A_5}\\\nonumber
\ch_4&=&:\frac{\lambda}{4}\phi^4-\frac{\delta m_2^2}{2}\phi^2+\hat A_4:\\\nonumber
\ch_3&=&:\lambda\phi^3v_0=:m\sqrt{\frac{\lambda}{2}}\phi^3(\vx):
\eea
where  
 \beq
 {\ovac}_4={\vac}_0^0+{\ovac}_4^1+{\ovac}_4^2+{\ovac}_4^3+{\ovac}_4^4+{\ovac}_4^5+{\ovac}_4^6+{\ovac}_4^8+{\ovac}_4^{10}+{\ovac}_4^{12}
 \eeq
the terms which contribute ${\ovac}_4^0$ comes from:
\beq
\begin{aligned}
H_3{\ovac}_3 \supset &m\sqrt\frac{\lambda}{2} \int d^2\vx\int\frac{d^{\dim}\vp_1 d^{\dim}\vp_2d^{\dim}\vp_3}{(2\pi)^{\dimthree}}
      \left(\frac{A_{-\vp_1}A_{-\vp_2}A_{-\vp_3}}{8\nomvp1\nomvp2\nomvp3}\right) e^{-\vx(\vp_1+\vp_2+\vp_3)}{\ovac}_3^3.\\
H_4{\ovac}_2 \supset & -\int d^2\vx\frac{\delta m_2^2}{2}\int\frac{d^{\dim}\vp_1 d^{\dim}\vp_2}{(2\pi)^{\dimtwo}}
      \left(\frac{A_{-\vp_1}A_{-\vp_2}}{4\nomvp1\nomvp2}\right)e^{-i\vx(\vp_1+\vp_2)}{\ovac}_2^2.\\\nonumber 
H_4{\ovac}_2 \supset & \frac{\lambda}{4}\int d^2\vx\int\frac{d^{\dim}\vp_1 d^{\dim}\vp_2d^{\dim}\vp_3 d^{\dim}\vp_4}{(2\pi)^{8}}
      \left(\frac{\navpm1\navpm2\navpm3\navpm4}{16\nomvp1\nomvp2\nomvp3\nomvp4}\right)e^{-i\vx(\vp_1+\vp_2+\vp_3+\vp_4)}{\ovac}_2^4\\\nonumber 
H_5{\ovac}_1 \supset & \sqrt{\frac{\lambda}{2}}\frac{\delta m_2^2}{2m}\int d^2\vx\int\frac{d^{\dim}\vp_1 d^{\dim}\vp_2d^{\dim}\vp_3} {(2\pi)^{6}}
      \left(\frac{\navpm1\navpm2\navpm3}{8\nomvp1\nomvp2\nomvp3}\right)e^{-i\vx(\vp_1+\vp_2+\vp_3)}{\ovac}_1^3\\\nonumber 
H_6{\ovac}_0 \supset & \hat A_6{\ovac}_0\\\nonumber 
\end{aligned}
\eeq

Where ${\ovac}_4^1$ contribution comes from\par 
\beq
\begin{aligned}
H_5{\ovac}_1 \supset & -\frac{\delta m_3^2}{2}\int d^2\vx\int\frac{d^{\dim}\vp_1 d^{\dim}\vp_2}{(2\pi)^{4}}
      \left(\frac{\navpm1\navpm2}{4\nomvp1\nomvp2}\right)e^{-i\vx(\vp_1+\vp_2)}{\ovac}_1^3\\\nonumber 
H_6{\ovac}_0 \supset & T_6\int d^2\vx\int\frac{d^{\dim}\vp_1 }{(2\pi)^{2}}
      \left(\navpd1\right)e^{-i\vx\cdot \vp_1}{\ovac}_0\\\nonumber 
\end{aligned}
\eeq

${\ovac}_4^2$ comes from:
\beq
\begin{aligned}
H_3{\ovac}_3 \supset &-m\sqrt\frac{\lambda}{2} \int d^2\vx\int\frac{d^{\dim}\vp_1 d^{\dim}\vp_2d^{\dim}\vp_3}{(2\pi)^{\dimthree}}
      \left(A^\ddag_{\vp_1}A^\ddag_{\vp_2}\frac{A_{-\vp_3}}{2\nomvp3}\right) e^{-\vx(\vp_1+\vp_2+\vp_3)}{\ovac}_3^1.\\
H_3{\ovac}_3 \supset &-m\sqrt\frac{\lambda}{2} \int d^2\vx\int\frac{d^{\dim}\vp_1 d^{\dim}\vp_2d^{\dim}\vp_3}{(2\pi)^{\dimthree}}
      \left(A^\ddag_{\vp_1}\frac{A_{-\vp_2}A_{-\vp_3}}{4\nomvp2\nomvp3}\right) e^{-\vx(\vp_1+\vp_2+\vp_3)}{\ovac}_3^3.\\
H_3{\ovac}_1 \supset &-m\sqrt\frac{\lambda}{2} \int d^2\vx\int\frac{d^{\dim}\vp_1 d^{\dim}\vp_2d^{\dim}\vp_3}{(2\pi)^{\dimthree}}
      \left(\frac{A_{-\vp_1}A_{-\vp_2}A_{-\vp_2}}{8\nomvp1\nomvp2\nomvp3}\right) e^{-\vx(\vp_1+\vp_2+\vp_3)}{\ovac}_3^5.\\
H_4{\ovac}_2 \supset & -\int d^2\vx\frac{\delta m_2^2}{2}\int\frac{d^{\dim}\vp_1 d^{\dim}\vp_2}{(2\pi)^{\dimtwo}}
      \left(A^\ddag_{\vp_1}\frac{A_{-\vp_2}}{2\nomvp2}\right)e^{-i\vx(\vp_1+\vp_2)}{\ovac}_2^2.\\\nonumber 
H_4{\ovac}_2 \supset & -\int d^2\vx\frac{\delta m_2^2}{2}\int\frac{d^{\dim}\vp_1 d^{\dim}\vp_2}{(2\pi)^{\dimtwo}}
      \left(\frac{A_{-\vp_1}A_{-\vp_2}}{4\nomvp1\nomvp2}\right)e^{-i\vx(\vp_1+\vp_2)}{\ovac}_2^4.\\\nonumber 
H_4{\ovac}_2 \supset & \frac{\lambda}{4}\int d^2\vx\int\frac{d^{\dim}\vp_1 d^{\dim}\vp_2d^{\dim}\vp_3 d^{\dim}\vp_4}{(2\pi)^{8}}
      \left(\navpd1\navpd2\frac{\navpm3\navpm4}{4\nomvp3\nomvp4}\right)e^{-i\vx(\vp_1+\vp_2+\vp_3+\vp_4)}{\ovac}_2^2\\\nonumber 
H_4{\ovac}_2 \supset & \frac{\lambda}{4}\int d^2\vx\int\frac{d^{\dim}\vp_1 d^{\dim}\vp_2d^{\dim}\vp_3 d^{\dim}\vp_4}{(2\pi)^{8}}
      \left(\navpd1\frac{\navpm2\navpm3\navpm4}{8\nomvp2\nomvp3\nomvp4}\right)e^{-i\vx(\vp_1+\vp_2+\vp_3+\vp_4)}{\ovac}_2^4\\\nonumber 
H_4{\ovac}_2 \supset & \frac{\lambda}{4}\int d^2\vx\int\frac{d^{\dim}\vp_1 d^{\dim}\vp_2d^{\dim}\vp_3 d^{\dim}\vp_4}{(2\pi)^{8}}
      \left(\frac{\navpm1\navpm2\navpm3\navpm4}{16\nomvp1\nomvp2\nomvp3\nomvp4}\right)e^{-i\vx(\vp_1+\vp_2+\vp_3+\vp_4)}{\ovac}_2^6\\\nonumber 
H_4{\ovac}_2 \supset & \hat A_4{\ovac}_2^2\\\nonumber 
H_5{\ovac}_1 \supset &-\sqrt{\frac{\lambda}2}\frac{\delta m_2^2}{2m}\int d^2\vx\int\frac{d^{\dim}\vp_1 d^{\dim}\vp_2d^{\dim}\vp_3} {(2\pi)^{6}}
      \left(\navpd1\frac{\navpm2\navpm3}{4\nomvp2\nomvp3}\right)e^{-i\vx(\vp_1+\vp_2+\vp_3)}{\ovac}_1\\\nonumber 
H_5{\ovac}_1 \supset &\frac{\delta m_2^2}{2m}\int d^2\vx T_5\int\frac{d^{\dim}\vp_1 }{(2\pi)^{2}}
      \left(\frac{\navpm1}{2\nomvp1}\right)e^{-i\vx \cdot \vp_1}{\ovac}_1\\\nonumber 
H_6{\ovac}_0 \supset &-\frac{\delta m_4^2}{2}\int d^2\vx\int\frac{d^{\dim}\vp_1d^{\dim}\vp_2 }{(2\pi)^{4}}
      \left(\navpd1\navpd2\right)e^{-i\vx(\vp_1+\vp_2)}{\ovac}_0\\\nonumber 
\end{aligned}
\eeq

${\ovac}_4^3$ comes from:
\beq
\begin{aligned}
H_5{\ovac}_1 \supset &-\frac{\delta m_2^2}{2}\int d^2\vx\int\frac{d^{\dim}\vp_1 d^{\dim}\vp_2} {(2\pi)^{4}}
      \left(\navpd1\frac{\navpd2}{2\nomvp2}\right)e^{-i\vx(\vp_1+\vp_2)}{\ovac}_1\\\nonumber 
H_5{\ovac}_1 \supset &\int d^2\vx\hat A_5{\ovac}_1^3\\\nonumber 
H_6{\ovac}_0 \supset &-\sqrt{\frac{\lambda}{2}}\frac{\delta m_3^2}{2m}\int d^2\vx\int\frac{d^{\dim}\vp_1d^{\dim}\vp_2 d^{\dim}\vp_3}{(2\pi)^{6}}
      \left(\navpd1\navpd2\navpd3\right)e^{-i\vx(\vp_1+\vp_2+\vp_3)}{\ovac}_0\\\nonumber 
\end{aligned}
\eeq

${\ovac}_4^4$ comes from:
\beq
\begin{aligned}
H_3{\ovac}_3 \supset &-m\sqrt\frac{\lambda}{2} \int d^2\vx\int\frac{d^{\dim}\vp_1 d^{\dim}\vp_2d^{\dim}\vp_3}{(2\pi)^{\dimthree}}
      \left(A^\ddag_{\vp_1}A^\ddag_{\vp_2}\navpd3\right) e^{-\vx(\vp_1+\vp_2+\vp_3)}{\ovac}_3^1.\\
H_3{\ovac}_3 \supset &-m\sqrt\frac{\lambda}{2} \int d^2\vx\int\frac{d^{\dim}\vp_1 d^{\dim}\vp_2d^{\dim}\vp_3}{(2\pi)^{\dimthree}}
      \left(A^\ddag_{\vp_1}\avpd2\frac{A_{-\vp_3}}{2\nomvp3}\right) e^{-\vx(\vp_1+\vp_2+\vp_3)}{\ovac}_3^3.\\
H_3{\ovac}_3 \supset &-m\sqrt\frac{\lambda}{2} \int d^2\vx\int\frac{d^{\dim}\vp_1 d^{\dim}\vp_2d^{\dim}\vp_3}{(2\pi)^{\dimthree}}
      \left(\navpd1\frac{A_{-\vp_2}A_{-\vp_3}}{4\nomvp2\nomvp3}\right) e^{-\vx(\vp_1+\vp_2+\vp_3)}{\ovac}_3^5.\\
H_3{\ovac}_3 \supset &-m\sqrt\frac{\lambda}{2} \int d^2\vx\int\frac{d^{\dim}\vp_1 d^{\dim}\vp_2d^{\dim}\vp_3}{(2\pi)^{\dimthree}}
      \left(\navpd1\frac{\navpm1A_{-\vp_2}A_{-\vp_3}}{8\nomvp1\nomvp2\nomvp3}\right) e^{-\vx(\vp_1+\vp_2+\vp_3)}{\ovac}_3^7.\\      
H_4{\ovac}_2 \supset & -\int d^2\vx\frac{\delta m_2^2}{2}\int\frac{d^{\dim}\vp_1 d^{\dim}\vp_2}{(2\pi)^{\dimtwo}}
      \left(A^\ddag_{\vp_1}\navpd2\right)e^{-i\vx(\vp_1+\vp_2)}{\ovac}_2^2.\\\nonumber 
H_4{\ovac}_2 \supset & -\int d^2\vx\frac{\delta m_2^2}{2}\int\frac{d^{\dim}\vp_1 d^{\dim}\vp_2}{(2\pi)^{\dimtwo}}
      \left(A^\ddag_{\vp_1}\frac{\navpm2}{2\nomvp2}\right)e^{-i\vx(\vp_1+\vp_2)}{\ovac}_2^4.\\\nonumber 
H_4{\ovac}_2 \supset & -\int d^2\vx\frac{\delta m_2^2}{2}\int\frac{d^{\dim}\vp_1 d^{\dim}\vp_2}{(2\pi)^{\dimtwo}}
      \left(\frac{\navpm1\navpm2}{4\nomvp1\nomvp2}\right)e^{-i\vx(\vp_1+\vp_2)}{\ovac}_2^6.\\\nonumber       
H_4{\ovac}_2 \supset & \frac{\lambda}{4}\int d^2\vx\int\frac{d^{\dim}\vp_1 d^{\dim}\vp_2d^{\dim}\vp_3 d^{\dim}\vp_4}{(2\pi)^{8}}
      \left(\navpd1\navpd2\navpd3\frac{\navpm4}{2\nomvp4}\right)e^{-i\vx(\vp_1+\vp_2+\vp_3+\vp_4)}{\ovac}_2^2\\\nonumber 
H_4{\ovac}_2 \supset & \frac{\lambda}{4}\int d^2\vx\int\frac{d^{\dim}\vp_1 d^{\dim}\vp_2d^{\dim}\vp_3 d^{\dim}\vp_4}{(2\pi)^{8}}
      \left(\navpd1\navpd2\frac{\navpm3\navpm4}{4\nomvp3\nomvp4}\right)e^{-i\vx(\vp_1+\vp_2+\vp_3+\vp_4)}{\ovac}_2^4\\\nonumber 
H_4{\ovac}_2 \supset & \frac{\lambda}{4}\int d^2\vx\int\frac{d^{\dim}\vp_1 d^{\dim}\vp_2d^{\dim}\vp_3 d^{\dim}\vp_4}{(2\pi)^{8}}
      \left(\navpd1\frac{\navpm2\navpm3\navpm4}{8\nomvp2\nomvp3\nomvp4}\right)e^{-i\vx(\vp_1+\vp_2+\vp_3+\vp_4)}{\ovac}_2^6 \\\nonumber       
H_4{\ovac}_2 \supset & \int d^2\vx\hat A_4{\ovac}_2^4\\\nonumber 
H_5{\ovac}_1 \supset &-\sqrt{\frac{\lambda}2}\frac{\delta m_2^2}{2m}\int d^2\vx\int\frac{d^{\dim}\vp_1 d^{\dim}\vp_2d^{\dim}\vp_3} {(2\pi)^{6}}
      \left(\navpd1\navpd2\frac{\navpm3}{2\nomvp3}\right)e^{-i\vx(\vp_1+\vp_2+\vp_3)}{\ovac}_1\\\nonumber 
H_5{\ovac}_1 \supset &T_5\frac{\delta m_2^2}{2m}\int d^2\vx\int\frac{d^{\dim}\vp_1 }{(2\pi)^{2}}
      \left(\navpd1\right)e^{-i\vx \cdot \vp_1}{\ovac}_1\\\nonumber 
\end{aligned}
\eeq
${\ovac}_3^5$ comes from:
\beq
\begin{aligned}
H_5{\ovac}_1 \supset &-\sqrt{\frac{\lambda}{2}}\frac{\delta m_2^2}{2m}\int d^2\vx\int\frac{d^{\dim}\vp_1 d^{\dim}\vp_2} {(2\pi)^{4}}
      \left(\navpd1\navpd2\right)e^{-i\vx(\vp_1+\vp_2)}{\ovac}_1\\\nonumber 
\end{aligned}
\eeq

${\ovac}_4^6$ comes from:
\beq
\begin{aligned}
H_3{\ovac}_3 \supset &-m\sqrt\frac{\lambda}{2} \int d^2\vx\int\frac{d^{\dim}\vp_1 d^{\dim}\vp_2d^{\dim}\vp_3}{(2\pi)^{\dimthree}}
      \left(\navpd1\navpd2\navpd3\right) e^{-\vx(\vp_1+\vp_2+\vp_3)}{\ovac}_3^3.\\
H_3{\ovac}_3 \supset &-m\sqrt\frac{\lambda}{2} \int d^2\vx\int\frac{d^{\dim}\vp_1 d^{\dim}\vp_2d^{\dim}\vp_3}{(2\pi)^{\dimthree}}
      \left(\navpd1\navpd2\frac{A_{-\vp_3}}{2\nomvp3}\right) e^{-\vx(\vp_1+\vp_2+\vp_3)}{\ovac}_3^5.\\
H_3{\ovac}_3 \supset &-m\sqrt\frac{\lambda}{2} \int d^2\vx\int\frac{d^{\dim}\vp_1 d^{\dim}\vp_2d^{\dim}\vp_3}{(2\pi)^{\dimthree}}
      \left(\navpd1\frac{A_{-\vp_2}A_{-\vp_3}}{4\nomvp2\nomvp3}\right) e^{-\vx(\vp_1+\vp_2+\vp_3)}{\ovac}_3^7.\\ 
H_3{\ovac}_3 \supset &-m\sqrt\frac{\lambda}{2} \int d^2\vx\int\frac{d^{\dim}\vp_1 d^{\dim}\vp_2d^{\dim}\vp_3}{(2\pi)^{\dimthree}}
      \left(\frac{\navpm1 A_{-\vp_2}A_{-\vp_3}}{8\nomvp1\nomvp2\nomvp3}\right) e^{-\vx(\vp_1+\vp_2+\vp_3)}{\ovac}_3^9.\\  
H_4{\ovac}_2 \supset & -\int d^2\vx\frac{\delta m_2^2}{2}\int\frac{d^{\dim}\vp_1 d^{\dim}\vp_2}{(2\pi)^{\dimtwo}}
      \left(A^\ddag_{\vp_1}\navpd2\right)e^{-i\vx(\vp_1+\vp_2)}{\ovac}_2^4.\\\nonumber 
H_4{\ovac}_2 \supset & -\int d^2\vx\frac{\delta m_2^2}{2}\int\frac{d^{\dim}\vp_1 d^{\dim}\vp_2}{(2\pi)^{\dimtwo}}
      \left(\navpd1\frac{\navpm2}{2\nomvp2}\right)e^{-i\vx(\vp_1+\vp_2)}{\ovac}_2^6.\\\nonumber           
H_4{\ovac}_2 \supset & \frac{\lambda}{4}\int d^2\vx\int\frac{d^{\dim}\vp_1 d^{\dim}\vp_2d^{\dim}\vp_3 d^{\dim}\vp_4}{(2\pi)^{8}}
      \left(\navpd1\navpd2\navpd3\navpd4\right)e^{-i\vx(\vp_1+\vp_2+\vp_3+\vp_4)}{\ovac}_2^2\\\nonumber 
H_4{\ovac}_2 \supset & \frac{\lambda}{4}\int d^2\vx\int\frac{d^{\dim}\vp_1 d^{\dim}\vp_2d^{\dim}\vp_3 d^{\dim}\vp_4}{(2\pi)^{8}}
      \left(\navpd1\navpd2\navpd3\frac{\navpm4}{2\nomvp4}\right)e^{-i\vx(\vp_1+\vp_2+\vp_3+\vp_4)}{\ovac}_2^4\\\nonumber 
H_4{\ovac}_2 \supset & \frac{\lambda}{4}\int d^2\vx\int\frac{d^{\dim}\vp_1 d^{\dim}\vp_2d^{\dim}\vp_3 d^{\dim}\vp_4}{(2\pi)^{8}}
      \left(\navpd1\navpd2\frac{\navpm3\navpm4}{4\nomvp3\nomvp4}\right)e^{-i\vx(\vp_1+\vp_2+\vp_3+\vp_4)}{\ovac}_2^6 \\\nonumber      
H_4{\ovac}_2 \supset & \hat A_4{\ovac}_2^6\\\nonumber 
H_5{\ovac}_1 \supset &-\sqrt{\frac{\lambda}2}\frac{\delta m_2^2}{2m}\int d^2\vx\int\frac{d^{\dim}\vp_1 d^{\dim}\vp_2d^{\dim}\vp_3} {(2\pi)^{6}}
      \left(\navpd1\navpd2\navpd3\right)e^{-i\vx(\vp_1+\vp_2+\vp_3)}{\ovac}_1\\\nonumber 
\end{aligned}
\eeq
 ${\ovac}_4^8$ comes from:
\beq
\begin{aligned}
H_3{\ovac}_3 \supset &-m\sqrt\frac{\lambda}{2} \int d^2\vx\int\frac{d^{\dim}\vp_1 d^{\dim}\vp_2d^{\dim}\vp_3}{(2\pi)^{\dimthree}}
      \left(\navpd1\navpd2\navpd3\right) e^{-\vx(\vp_1+\vp_2+\vp_3)}{\ovac}_3^5.\\
H_3{\ovac}_3 \supset &-m\sqrt\frac{\lambda}{2} \int d^2\vx\int\frac{d^{\dim}\vp_1 d^{\dim}\vp_2d^{\dim}\vp_3}{(2\pi)^{\dimthree}}
      \left(\navpd1\navpd2\frac{A_{-\vp_3}}{2\nomvp3}\right) e^{-\vx(\vp_1+\vp_2+\vp_3)}{\ovac}_3^7.\\ 
H_3{\ovac}_3 \supset &-m\sqrt\frac{\lambda}{2} \int d^2\vx\int\frac{d^{\dim}\vp_1 d^{\dim}\vp_2d^{\dim}\vp_3}{(2\pi)^{\dimthree}}
      \left(\navpd1\frac{A_{-\vp_2}A_{-\vp_3}}{4\nomvp2\nomvp3}\right) e^{-\vx(\vp_1+\vp_2+\vp_3)}{\ovac}_3^9.\\  
H_4{\ovac}_2 \supset & -\int d^2\vx\frac{\delta m_2^2}{2}\int\frac{d^{\dim}\vp_1 d^{\dim}\vp_2}{(2\pi)^{\dimtwo}}
      \left(\navpd1\navpd2\right)e^{-i\vx(\vp_1+\vp_2)}{\ovac}_2^6.\\\nonumber           
H_4{\ovac}_2 \supset & \frac{\lambda}{4}\int d^2\vx\int\frac{d^{\dim}\vp_1 d^{\dim}\vp_2d^{\dim}\vp_3 d^{\dim}\vp_4}{(2\pi)^{8}}
      \left(\navpd1\navpd2\navpd3\navpd4\right)e^{-i\vx(\vp_1+\vp_2+\vp_3+\vp_4)}{\ovac}_2^4\\\nonumber 
H_4{\ovac}_2 \supset & \frac{\lambda}{4}\int d^2\vx\int\frac{d^{\dim}\vp_1 d^{\dim}\vp_2d^{\dim}\vp_3 d^{\dim}\vp_4}{(2\pi)^{8}}
      \left(\navpd1\navpd2\navpd3\frac{\navpm4}{2\nomvp4}\right)e^{-i\vx(\vp_1+\vp_2+\vp_3+\vp_4)}{\ovac}_2^6 \\\nonumber       
\end{aligned}
\eeq
${\ovac}_4^{10}$ comes from:
\beq
\begin{aligned}
H_3{\ovac}_3 \supset &-m\sqrt\frac{\lambda}{2} \int d^2\vx\int\frac{d^{\dim}\vp_1 d^{\dim}\vp_2d^{\dim}\vp_3}{(2\pi)^{\dimthree}}
      \left(\navpd1\navpd2\navpd3\right) e^{-\vx(\vp_1+\vp_2+\vp_3)}{\ovac}_3^7.\\ 
H_3{\ovac}_3 \supset &-m\sqrt\frac{\lambda}{2} \int d^2\vx\int\frac{d^{\dim}\vp_1 d^{\dim}\vp_2d^{\dim}\vp_3}{(2\pi)^{\dimthree}}
      \left(\navpd1\navpd2\frac{A_{-\vp_3}}{2\nomvp3}\right) e^{-\vx(\vp_1+\vp_2+\vp_3)}{\ovac}_3^9.\\  
H_4{\ovac}_2 \supset & \frac{\lambda}{4}\int d^2\vx\int\frac{d^{\dim}\vp_1 d^{\dim}\vp_2d^{\dim}\vp_3 d^{\dim}\vp_4}{(2\pi)^{8}}
      \left(\navpd1\navpd2\navpd3\navpd4\right)e^{-i\vx(\vp_1+\vp_2+\vp_3+\vp_4)}{\ovac}_2^6 \\\nonumber       
\end{aligned}
\eeq
 ${\ovac}_4^{12}$ comes from:
\beq
\begin{aligned}
H_3{\ovac}_3 \supset &-m\sqrt\frac{\lambda}{2} \int d^2\vx\int\frac{d^{\dim}\vp_1 d^{\dim}\vp_2d^{\dim}\vp_3}{(2\pi)^{\dimthree}}
      \left(\navpd1\navpd2\navpd3\right) e^{-\vx(\vp_1+\vp_2+\vp_3)}{\ovac}_3^9.\\       
\end{aligned}
\eeq
Then we can have the full expresion of ${\ovac}_4={\ovac}_4^1+{\ovac}_4^2+{\ovac}_4^3+{\ovac}_4^4+{\ovac}_4^5+{\ovac}_4^6+{\ovac}_4^2+{\ovac}_4^8+{\ovac}_4^{10}+{\ovac}_4^{12}$

\section{Hengyuan calculation: one-meson state}
We define the meson state with momenutum $ p_0 $as the first order excitation state on the vacuum
\beq
|\vp_0\rangle_0=\avpd {\ovac}_0
\eeq
 while with quantum correction, like the vacuum state we have $|\vp_0\rangle=\sum_{i=0}|\vp_0\rangle_i$,where it is expressed by order $\lambda^{i/2} $ and $|\vp_0\rangle_0$ is order $\lambda^0$. Then meson state also satisfy the renormalization  condition:
 \beq
 H|\vp_0\rangle=\nomvp0|\vp_0\rangle\hsp \nomvp0=\sqrt{m^2+\vp_0^2}
 \eeq
Where the quantum correction of energy arise from the mass term correction order by order.  and one meson state can be expand as
\beq
|\vp_0\rangle=\sum_i|\vp_0\rangle_i
\eeq
\par  
We can calculate order by order :
\beq
\begin{aligned}
\lambda^{0}\hsp     &H_2|\vp_0\rangle_0=\nomvp0|\vp_0\rangle_0\\
\lambda^{1/2}\hsp   &H_2|\vp_0\rangle_1+H_3|\vp_0\rangle_0=\nomvp0|\vp_0\rangle_1\\
\lambda^{1}\hsp     &H_2|\vp_0\rangle_2+H_3|\vp_0\rangle_1+H_4|\vp_0\rangle_0=\nomvp0|\vp_0\rangle_2\\
\lambda^{3/2}\hsp   &H_2|\vp_0\rangle_3+H_3|\vp_0\rangle_2+H_4|\vp_0\rangle_1+H_5|\vp_0\rangle_0=\nomvp0|\vp_0\rangle_3\\
\lambda^{2}\hsp     &H_2|\vp_0\rangle_4+H_3|\vp_0\rangle_3+H_4|\vp_0\rangle_2+H_5|\vp_0\rangle_2+H_6|\vp_0\rangle_0=\nomvp0|\vp_0\rangle_4\\
\cdots
\end{aligned}
\eeq
For each correction$|\vp_0\rangle_i$  in have component $|\vp_0\rangle_i^n$ which can be given by renormalization condition or physical symmetry.\par 
Then we can calculate order by order to fix the quantum correction of the state and fix the counterterm.

\subsection{$\lambda^{1/2}$}
\beq
\begin{aligned}
H_2|\vp_0\rangle_1+H_3|\vp_0\rangle_0=\nomvp0|\vp_0\rangle_1\\
\end{aligned}
\eeq
The can be expressed exactly as:
\beq
\begin{aligned}
&(\int\frac{d^2\vk}{2\pi}\omvk\avpd\avpm-\nomvp0)|\vp_0\rangle_1=-H_3|\vp_0\rangle_0\\
=&-m\sqrt\frac{\lambda}{2}\int d^2\vx\int\frac{d^{\dim}\vp_1 d^{\dim}\vp_2d^{\dim}\vp_3}{(2\pi)^{\dimthree}}
      \left(\navpd1\navpd2\navpd3\right) e^{-i\vx(\vp_1+\vp_2+\vp_3)}|\vp_0\rangle_0\\  
 &-m\sqrt\frac{\lambda}{2}\int d^2\vx\int\frac{d^{\dim}\vp_1 d^{\dim}\vp_2d^{\dim}\vp_3}{(2\pi)^{\dimthree}}
      \left(3\navpd1\navpd2\frac{\navpm3}{2\nomvp3}\right) e^{-i\vx(\vp_1+\vp_2+\vp_3)}|\vp_0\rangle_0\\ 
=&-m\sqrt\frac{\lambda}{2}\int d^2\vx\int\frac{d^{\dim}\vp_1 d^{\dim}\vp_2d^{\dim}\vp_3}{(2\pi)^{\dimthree}}
     e^{-i\vx(\vp_1+\vp_2+\vp_3)}|\vp_0\vp_1\vp_2\vp_3\rangle_0\\  
  &-\frac{3}{2}m\sqrt\frac{\lambda}{2}\int d^2\vx\int\frac{d^{\dim}\vp_1 d^{\dim}\vp_2}{(2\pi)^{\dimtwo}}
      e^{-i\vx(\vp_1+\vp_2-\vp_0)}\frac{1}{2\nomvp0}|\vp_1\vp_2\rangle_0\\
\end{aligned}
\eeq
Then we can get:
\beq
\begin{aligned}
|\vp_0\rangle_1=&|\vp_0\rangle_1^2+|\vp_0\rangle_1^4\\
|\vp_0\rangle_1^4=&-m\sqrt\frac{\lambda}{2}\int d^2\vx\int\frac{d^{\dim}\vp_1 d^{\dim}\vp_2d^{\dim}\vp_3}{(2\pi)^{\dimthree}}
     \frac{e^{-i\vx(\vp_1+\vp_2+\vp_3)}}{\nomvp1+\nomvp2+\nomvp3} |\vp_0\vp_1\vp_2\vp_3\rangle_0\\  
|\vp_0\rangle_1^2=&-\frac{3}{2}m\sqrt\frac{\lambda}{2}\int d^2\vx\int\frac{d^{\dim}\vp_1 d^{\dim}\vp_2}{(2\pi)^{\dimtwo}}
      \frac{e^{-i\vx(\vp_1+\vp_2-\vp_0)}}{\nomvp0(\nomvp1+\nomvp2-\nomvp0)}|\vp_1\vp_2\rangle_0\\ 
\end{aligned}
\eeq

\subsection{$\lambda^{1}$}
\beq
\begin{aligned}
H_2|\vp_0\rangle_2+H_3|\vp_0\rangle_1+H_4|\vp_0\rangle_0=\nomvp0|\vp_0\rangle_2\\\label{2ordermeson}
\end{aligned}
\eeq
Recall that $\ch_4=\frac{\lambda}{4}\phi^4-\frac{\delta m_2^2}{2}\phi^2+\hat A_4\hsp \hat A_4=-\frac{(\delta m_2^2)^2}{16\lambda}+\frac{m^2\delta m_4^2}{8\lambda}+A_4$, \par 
From the form of $H_4$ ,and $H_3$ we know the $|\vp_0\rangle_2$ have the component :
\beq
|\vp_0\rangle_2=|\vp_0\rangle_2^1+|\vp_0\rangle_2^3+|\vp_0\rangle_2^5+|\vp_0\rangle_2^7
\eeq
we see  the $H_4|\vp_0\rangle_0$ contribute  $|\vp_0\rangle_2$ with terms $|\vp_0\rangle_2^1$ from $\hat A_4|\vp_0\rangle_0, \phi^2|\vp_0\rangle_0$, $|\vp_0\rangle_2^3$ from $\hat \phi^2|\vp_0\rangle_0,\phi^4|\vp_0\rangle_0$, $|\vp_0\rangle_2^5$ from $\frac{\lambda}{4}\phi^4|\vp_0\rangle_0$. \par 
And recall the form of $|\vp_0\rangle_1=|\vp_0\rangle_1^2+|\vp_0\rangle_1^4$, we can see $H_3|\vp_0\rangle_1$ contribute  $|\vp_0\rangle_2$ with terms $|\vp_0\rangle_2^1, |\vp_0\rangle_2^3,|\vp_0\rangle_2^5,|\vp_0\rangle_2^7$.\par 
It is reasonable we impose renormalization condition:
\beq
|\vp_0\rangle_2^1=0
\eeq
which is Just similiar as we impose for vacuum sector ${\ovac}_i^0=0\hsp i>=1$.\par 
So We can have 
\beq
|\vp_0\rangle_2^1 \supset \bigg(H_3|\vp_1\rangle_1+H_4|\vp_0\rangle_0\bigg)|_{one-meson}=0
\eeq
Where we 
\beq
\begin{aligned}
&H_4|\vp_0\rangle_0|_{one-meson}
=\int d^2\vx\hat A_4|\vp_0\rangle_0-\frac{\delta m_2^2}{2}\int d^2\vx\int\frac{d^{\dim}\vp_1 d^{\dim}\vp_2}{(2\pi)^{\dimtwo}}\frac{2\navpd1\navpm2}{2\nomvp2}e^{-i\vx(\vp_1+\vp_2)}|\vp_0\rangle_0 \\
=&\int d^2\vx\hat A_4|\vp_0\rangle_0-\frac{\delta m_2^2}{2}\int\frac{d^{\dim}\vp_1}{(2\pi)^2}\frac{\navpd1}{\nomvp0}e^{-i\vx(\vp_1-\vp_0)}|\Omega\rangle_0 \\
=&\int d^2\vx\hat A_4|\vp_0\rangle_0-\frac{\delta m_2^2}{2\nomvp0}|\vp_0\rangle_0\\
=&(\int d^2\vx\hat A_4-\frac{\delta m_2^2}{2\nomvp0})|\vp_0\rangle_0\label{h4vo1}
\end{aligned}
\eeq

\red{I think that in the first line, the $\delta m^2_2/2$ should be $\delta m^2/2$, because you get a factor of two by choosing which $\phi$ in $\phi^2$ has the $A^\ddag$ and the $A$.  As a result, the $\delta m^2_2/4$ becomes $\delta m^2/2$.}
\gre{ the $\delta m^2$ include many , but here  we just need $\lambda$ order. which is indeed the order. and  I make a typo there should be a factor 2 because of symmetry of $\vp_1,\vp_2$}
\beq
\begin{aligned}
&H_3|\vp_1\rangle_1|_{one-meson}\\
=&m\sqrt\frac{\lambda}{2} \int d^2\vx\int\frac{d^{\dim}\vp_1 d^{\dim}\vp_2d^{\dim}\vp_3}{(2\pi)^{\dimthree}}
      \left(\frac{\navpm1\navpm2\navpm3}{8\nomvp1\nomvp2\nomvp3}\right) e^{-i\vx(\vp_1+\vp_2+\vp_3)}\\
 &\times\bigg(-m\sqrt\frac{\lambda}{2}\int d^2\vx\int\frac{d^{\dim}\vp_1 d^{\dim}\vp_2d^{\dim}\vp_3}{(2\pi)^{\dimthree}}
     \frac{e^{-i\vx(\vp_1+\vp_2+\vp_3)}}{\nomvp1+\nomvp2+\nomvp3} |\vp_0\vp_1\vp_2\vp_3\rangle_0\bigg)\\
&m\sqrt\frac{\lambda}{2} \int d^2\vx\int\frac{d^{\dim}\vp_1 d^{\dim}\vp_2d^{\dim}\vp_3}{(2\pi)^{\dimthree}}
      \left(3\navpd1\frac{\navpm2\navpm3}{4\nomvp2\nomvp3}\right) e^{-i\vx(\vp_1+\vp_2+\vp_3)}\\
 &\times\bigg(-\frac{3}{2}m\sqrt\frac{\lambda}{2}\int d^2\vx\int\frac{d^{\dim}\vp_1 d^{\dim}\vp_2}{(2\pi)^{\dimtwo}}
      \frac{e^{-i\vx(\vp_1+\vp_2-\vp_0)}}{\nomvp0(\nomvp1+\nomvp2-\nomvp0)}|\vp_1\vp_2\rangle_0\bigg)\\
=&-\frac{\lambda}{2}m^2\int d^2\vx\int d^2\vx\p
     \bigg(\int\frac{d^{\dim}\vp_1 d^{\dim}\vp_2d^{\dim}\vp_3}{(2\pi)^{\dimthree}}
     \left(\frac{6e^{-i\vx\p(-\vp_1-\vp_2-\vp_3)}e^{-i\vx(\vp_1+\vp_2+\vp_3)}}{8\nomvp1\nomvp2\nomvp3(\nomvp1+\nomvp2+\nomvp3)}\right)|\vp_0\rangle_0\\
 &-\frac{\lambda}{2}m^2\int d^2\vx\int d^2\vx\p\int\frac{d^{\dim}\vp_1 d^{\dim}\vp_2d^{\dim}\vp_3}{(2\pi)^{\dimthree}}\left(\frac{18e^{-i\vx\p(-\vp_1-\vp_2-\vp_0)}e^{-i\vx(\vp_1+\vp_2+\vp_3)}}{8\nomvp1\nomvp2\nomvp0(\nomvp1+\nomvp2+\nomvp3)}\right)\bigg)|\vp_3\rangle_0\\
 &-\frac{3\lambda}{4}m^2\int d^2\vx \int d^2\vx\p\int\frac{d^{\dim}\vp_1\p d^{\dim}\vp_1 d^{\dim}\vp_2}{(2\pi)^{\dimthree}}
     \frac{6e^{-i\vx\p(\vp_1\p-\vp_1-\vp_2)}e^{-i\vx(\vp_1+\vp_2-\vp_0)}}{4\nomvp1\nomvp2(\nomvp1+\nomvp2-\nomvp0)} |\vec{p_1}\p\rangle_0\\
=&-\frac{3\lambda}{8}m^2\int d^2\vx\int\frac{d^{\dim}\vp_1d^{\dim}\vp_2}{(2\pi)^{\dimtwo}}
     \frac{1}{\nomvp1\nomvp2\omega_{\vp_1+\vp_2}(\nomvp1+\nomvp2+\omega_{\vp_1+\vp_2})}|\vp_0\rangle_0\\
 &-\frac{9\lambda}{8}m^2\int\frac{d^{\dim}\vp_1}{(2\pi)^{2}}
    \frac{1}{\nomvp1\nomvp0\omega_{\vp_0+\vp_1}(\nomvp0+\nomvp1+\omega_{\vp_0+\vp_1})}|\vp_0\rangle_0\\
 &-\frac{9\lambda}{8}m^2\int \frac{d^{\dim}\vp_1 }{(2\pi)^{2}}
     \frac{1}{\nomvp0\nomvp1\omega_{\vp_0-\vp_1}(\nomvp1+\omega_{\vp_0-\vp_1}-\nomvp0)} |\vp_0\rangle_0\label{h3v11}
\end{aligned} 
\eeq
\red{It looks like a factor of 3 disappears here, in the second equal sign, in the term where $H_3$ acts on the four meson state without forming a disconnected diagram.  You need to choose 3 mesons out of 4 to annihilate, with the condition that they can't be the same three that just got created in the first step.  So there's a factor of 3 for which $A$ annihilates the original meson, and then a factor of 6 for which $A$ annihilates which of the two new mesons ... so altogether there should be a factor of 18 (not 6) after the second equal sign in the first term (and you still divide by 8 and divide by 2).   There was also an overall factor mistake in my $\delta m^2_2$ which I just fixed. }\gre{ yes i have just correct it}
Where with symmetry  we used the  formula:
\beq
\begin{aligned}
&\navpm1\p\navpm2\p\navpm3\p\navpd0\navpd1\navpd2\navpd3\\
=&6(6\pi)^6\delta(\vp_1\p-\vp_1)(\delta\vp_2\p-\vp_2)(\delta\vp_3\p-\vp_3)\navpd0
+18(6\pi)^6\delta(\vp_1\p-\vp_0)(\delta\vp_2\p-\vp_1)(\delta\vp_3\p-\vp_2)\navpd3 \\
\end{aligned}
\eeq
\red{Where do the $6\pi$ factors come from?  Anyway, I think the problem is in the second term on the right hand side.  The coefficient should be 18.  Overall there are 24 ways to contract.  6 are in the first term.  So 24-6=18 should be in the second.}\gre{Yes you are right the 18 should be in the second term.}
Then compare the (\ref{h4vo1}) and  (\ref{h3v11}), we know the renormalization condition $|\vp_0\rangle_2^1=0$ generate:
\beq
\begin{aligned}
\hat A_4=&-\frac{(\delta m_2^2)^2}{16\lambda}+\frac{m^2\delta m_4^2}{8\lambda}+A_4\\
=&-\frac{3\lambda}{8}m^2\int\frac{d^{\dim}\vp_1d^{\dim}\vp_2}{(2\pi)^{\dimtwo}}
     \frac{1}{\nomvp1\nomvp2\omega_{\vp_1+\vp_2}(\nomvp1+\nomvp2+\omega_{\vp_1+\vp_2})}
\end{aligned}
\eeq
\beq
\begin{aligned}
\delta m_2^2=&-2\nomvp0\times\bigg(\frac{9\lambda}{8}m^2\int\frac{d^{\dim}\vp_1}{(2\pi)^{2}}
    \frac{1}{\nomvp1\nomvp0\omega_{\vp_0+\vp_1}(\nomvp0+\nomvp1+\omega_{\vp_0+\vp_1})}\\
 &+\frac{9\lambda}{8}m^2\int \frac{d^{\dim}\vp_1 }{(2\pi)^{2}}
     \frac{1}{\nomvp0\nomvp1\omega_{\vp_0-\vp_1}(\nomvp1+\omega_{\vp_0-\vp_1}-\nomvp0)}\bigg)\\
  =&-\frac{9\lambda}{2}m^2\nomvp0
\int \frac{d^{\dim}\vp_1 }{(2\pi)^{2}}
     \frac{(\nomvp1+\omega_{\vp_0+\vp_1})}{\nomvp0\nomvp1\omega_{\vp_0+\vp_1}((\nomvp1+\omega_{\vp_0+\vp_1})^2-\nomvp0^2)}\\
\end{aligned}
\eeq
 We choose the renormalization conditon that:$ \vp_0=0$ as the begin point in the renormalziation grounp.
 then we get:
 \beq
\begin{aligned}
\delta m_2^2=&-\frac{9\lambda}{2}m^3\int \frac{d^{\dim}\vp_1 }{(2\pi)^{2}}
             \frac{(\nomvp1+\omega_{\vp_1})}{m\nomvp1^2((\nomvp1+\omega_{\vp_1})^2-m^2)}\\
     =&-\frac{9\lambda}{2}m^3\int\int_{0}^{\infty}  \frac{2\pi |\vp_1| d|\vp_1| }{(2\pi)^{2}}\frac{(\nomvp1+\omega_{\vp_1})}       {m\nomvp1^2((\nomvp1+\omega_{\vp_1})^2-m^2)}\\
     =&-\frac{9\lambda m}{2}\int_{0}^{\infty/m} \frac{xdx }{2\pi} \frac{2\sqrt{1+x^2}}{(1+x^2)(4(1+x^2)-1)}\\
     =&-\frac{9\ln{3}} {8\pi}\lambda m
\end{aligned}
\eeq
\red{I integrate this with Mathematica and get a different result (on the last line) ... I get the above result if I change $xdx\rightarrow  dx$.   It would be good to check this.}
\gre{Yes you are right, I am sorry that I make some typo input,   checked it again to get this new one?  Is it same with yours?}
This is one of the important  result we want to get.\par 
To get the $2-loop$ mass correction $\delta m_4^2$, we know we just need exact form of $|\vp_0\rangle_2^3,|\vp_0\rangle_2^5,|\vp_0\rangle_2^7$ which generate $|\vp_0\rangle_3^2,|\vp_0\rangle_3^4$,and with these we get all term contribute $|\vp_0\rangle_4^2$, with is and renormalizaiton condition $|\vp_0\rangle_4^2=0$, we can get exact form of $\delta m_4^2$. so we focus on the exact form of $|\vp_0\rangle_2^3,|\vp_0\rangle_2^5,|\vp_0\rangle_2^7 $ her.\par 
Where $|\vp_0\rangle_2^3$  come from $H_3|\vp_0\rangle_1^2,H_3|\vp_0\rangle_1^4,H_4|\vp_0\rangle_0$, which are exactly.\gre{update at 4:00 pm in 4.19}:
\beq
\begin{aligned}
H_3|\vp_0\rangle_1  \supset 
&m\sqrt\frac{\lambda}{2}\int d^2\vx\int\frac{d^{\dim}\vp_1 d^{\dim}\vp_2d^{\dim}\vp_3}{(2\pi)^{\dimthree}}
   \left(3\navpd1\navpd2\frac{A_{-\vp_3}}{2\nomvp3}\right) 
   e^{-i\vx(\vp_1+\vp_2+\vp_3)}|\vp_0\rangle_1^2.\\
   &+m\sqrt\frac{\lambda}{2}\int d^2\vx\int\frac{d^{\dim}\vp_1 d^{\dim}\vp_2d^{\dim}\vp_3}{(2\pi)^{\dimthree}}\left(3\navpd1\frac{A_{-\vp_2}A_{-\vp_3}}{4\nomvp2\nomvp3}\right) 
   e^{-i\vx(\vp_1+\vp_2+\vp_3)}|\vp_0\rangle_1^4.\\  
=&m\sqrt\frac{\lambda}{2}\int d^2\vx\int\frac{d^{\dim}\vp_1 d^{\dim}\vp_2d^{\dim}\vp_3}{(2\pi)^{\dimthree}}
   \left(3\navpd1\navpd2\frac{A_{-\vp_3}}{2\nomvp3}\right) 
   e^{-i\vx(\vp_1+\vp_2+\vp_3)}.\\
   &\times\bigg(-\frac{3}{2}m\sqrt\frac{\lambda}{2}\int d^2\vx\int\frac{d^{\dim}\vp_1 d^{\dim}\vp_2}{(2\pi)^{\dimtwo}}
      \frac{e^{-i\vx(\vp_1+\vp_2-\vp_0)}}{\nomvp0(\nomvp1+\nomvp2-\nomvp0)}|\vp_1\vp_2\rangle_0\bigg)\\
   &+m\sqrt\frac{\lambda}{2}\int d^2\vx\int\frac{d^{\dim}\vp_1 d^{\dim}\vp_2d^{\dim}\vp_3}{(2\pi)^{\dimthree}}\left(3\navpd1\frac{A_{-\vp_2}A_{-\vp_3}}{4\nomvp2\nomvp3}\right) 
   e^{-i\vx(\vp_1+\vp_2+\vp_3)}.\\ 
   &\times\bigg(-m\sqrt\frac{\lambda}{2}\int d^2\vx\int\frac{d^{\dim}\vp_1 d^{\dim}\vp_2d^{\dim}\vp_3}{(2\pi)^{\dimthree}}
     \frac{e^{-i\vx(\vp_1+\vp_2+\vp_3)}}{\nomvp1+\nomvp2+\nomvp3} |\vp_0\vp_1\vp_2\vp_3\rangle_0 \bigg)\\
=&-\frac{9m^2\lambda}{4}\int\frac{d^{\dim}\vp_1 d^{\dim}\vp_2}{(2\pi)^{\dimtwo}}
\frac{|\vp_1,\vp_2,\vp_0-\vp_1-\vp_2\rangle_0}{\nomvp0\omega_{\vp_0-\vp_1}(\nomvp1-\vp_0+\omega_{\vp_0-\vp_1})}\\
&-\frac{9m^2\lambda}{4}\int\frac{d^{\dim}\vp_1 d^{\dim}\vp_2}{(2\pi)^{4}}
\bigg(\frac{|\vp_0,\vp_1,-\vp_1\rangle_0}{\nomvp1\omega_{\vp_1+\vp_2}(\nomvp1+\nomvp2+\omega_{\vp_1+\vp_2})}+
      \frac{|\vp_2,-\vp_1-\vp_2,\vp_0+\vp_1\rangle_0}{\nomvp0\nomvp1(\nomvp1+\nomvp2+\omega_{\vp_1+\vp_2})}\bigg) \\ 
H_4|\vp_0\rangle_0 \supset 
 &\frac{\lambda}{4}\int d^2\vx\int\frac{d^{\dim}\vp_1 d^{\dim}\vp_2d^{\dim}\vp_3 d^{\dim}\vp_4}{(2\pi)^{8}}
      \left(4\navpd1\navpd2\navpd3\frac{A_{-\vp_4}}{2\nomvp4}\right)e^{-i\vx(\vp_1+\vp_2+\vp_3+\vp_4)}|\vp_0\rangle_0.\\ 
 &-\frac{\delta m_2^2}{2}\int d^2\vx\int\frac{d^{\dim}\vp_1 d^{\dim}\vp_2}{(2\pi)^{\dimtwo}}
      \left(\navpd1\navpd2\right)e^{-i\vx(\vp_1+\vp_2)}|\vp_0\rangle_0\\
=&\frac{\lambda}{2}\int\frac{d^{\dim}\vp_1 d^{\dim}\vp_2}{(2\pi)^{4}}
      \frac{|\vp_1,\vp_2,\vp_0-\vp_1-\vp_2\rangle_0}{2\nomvp0}
 -\frac{\delta m_2^2}{2}\int\frac{d^{2}\vp_1 }{(2\pi)^{2}}
     |\vp_0,\vp_1,-\vp_1\rangle_0\\
\end{aligned}
\eeq
Then based on equation of motion (\ref{2ordermeson}) in 3-meson case we get:
\beq
\begin{aligned}
|\vp_0\rangle_2^3
=&\frac{9m^2\lambda}{4}\int\frac{d^{\dim}\vp_1 d^{\dim}\vp_2}{(2\pi)^{\dimtwo}}
\frac{|\vp_1,\vp_2,\vp_0-\vp_1-\vp_2\rangle_0}{\nomvp0\omega_{\vp_0-\vp_1}(\nomvp1-\vp_0+\omega_{\vp_0-\vp_1})(\nomvp1+\nomvp1+\omega_{\vp_0-\vp_1-\vp_2}-3\nomvp0)}\\
&+\frac{9m^2\lambda}{4}\int\frac{d^{\dim}\vp_1 d^{\dim}\vp_2}{(2\pi)^{4}}
\bigg(\frac{|\vp_0,\vp_1,-\vp_1 \rangle_0}{\nomvp1\omega_{\vp_1+\vp_2}(\nomvp1+\nomvp2+\omega_{\vp_1+\vp_2})(\nomvp0+2\nomvp1-3\nomvp0)}\\
&+\frac{|\vp_2,-\vp_1-\vp_2,\vp_0+\vp_1\rangle_0}{\nomvp0\nomvp1(\nomvp1+\nomvp2+\omega_{\vp_1+\vp_2})(\nomvp2+\omega_{\vp_1+\vp_2}+\omega_{\vp_0+\vp_1}-3\nomvp0)}\bigg) \\ 
&-\frac{\lambda}{2}\int\frac{d^{\dim}\vp_1 d^{\dim}\vp_2}{(2\pi)^{4}}
      \frac{|\vp_1,\vp_2,\vp_0-\vp_1-\vp_2\rangle_0}{2\nomvp0(\nomvp1+\nomvp2+\omega_{\vp_0-\vp_1-\vp_2}-3\nomvp0)}
+\frac{\delta m_2^2}{2}\int\frac{d^{2}\vp_1 }{(2\pi)^{2}}
     \frac{|\vp_0,\vp_1,-\vp_1\rangle_0}{\nomvp0+2\nomvp1-3\nomvp0}\\
\end{aligned}
\eeq

$|\vp_0\rangle_2^5$  come from $H_3|\vp_0\rangle_1^2,H_3|\vp_0\rangle_1^4,H_4|\vp_0\rangle_0$, which are exactly:
\beq
\begin{aligned}
H_3|\vp_0\rangle_1 \supset 
&m\sqrt\frac{\lambda}{2}\int d^2\vx\int\frac{d^{\dim}\vp_1 d^{\dim}\vp_2d^{\dim}\vp_3}{(2\pi)^{\dimthree}}
   \left(\navpd1\navpd2\navpd3\right) 
   e^{-i\vx(\vp_1+\vp_2+\vp_3)}.\\
   &\times\bigg(-\frac{3}{2}m\sqrt\frac{\lambda}{2}\int d^2\vx\int\frac{d^{\dim}\vp_1 d^{\dim}\vp_2}{(2\pi)^{\dimtwo}}
      \frac{e^{-i\vx(\vp_1+\vp_2-\vp_0)}}{\nomvp0(\nomvp1+\nomvp2-\nomvp0)}|\vp_1\vp_2\rangle_0\bigg)\\
   &+m\sqrt\frac{\lambda}{2}\int d^2\vx\int\frac{d^{\dim}\vp_1 d^{\dim}\vp_2d^{\dim}\vp_3}{(2\pi)^{\dimthree}}\left(3\navpd1\navpd2\frac{A_{-\vp_3}}{2\nomvp3}\right) 
   e^{-i\vx(\vp_1+\vp_2+\vp_3)}\\
   &\times\bigg(-m\sqrt\frac{\lambda}{2}\int d^2\vx\int\frac{d^{\dim}\vp_1 d^{\dim}\vp_2d^{\dim}\vp_3}{(2\pi)^{\dimthree}}
     \frac{e^{-i\vx(\vp_1+\vp_2+\vp_3)}}{\nomvp1+\nomvp2+\nomvp3} |\vp_0\vp_1\vp_2\vp_3\rangle_0 \bigg)\\
=&-\frac{3\lambda m^2}{4}\int
\frac{d^{\dim}\vp_1 d^{\dim}\vp_2d^{\dim}\vp_3d}{(2\pi)^{6}}
  \frac{|\vp_1,\vp_2,\vp_3,\vp_0-\vp_1,-\vp_2-\vp_3\rangle_0.}{\nomvp0(\nomvp1+\omega_{\vp_0-\vp_1}-\nomvp0)}\\
   &-\frac{9\lambda m^2}{8}\int\frac{d^{\dim}\vp_1 d^{\dim}\vp_2d^{\dim}\vp_3}{(2\pi)^{\dimthree}}
   \frac{|\vp_0,\vp_1,\vp_2,\vp_3,-\vp_1-\vp_2-\vp_3\rangle_0}{\omega_{\vp_1+\vp_2}(\nomvp1+\nomvp2+\omega_{\vp_1+\vp_2})}.\\ 
   &-\frac{9\lambda m^2}{8}\int\frac{d^{\dim}\vp_1 d^{\dim}\vp_2d^{\dim}\vp_3}{(2\pi)^{\dimthree}}
   \frac{|\vp_1,\vp_2,\vp_3,\vp_0-\vp_1,-\vp_1-\vp_2\rangle_0}{\nomvp0(\nomvp1+\nomvp2+\omega_{\vp_1+\vp_2})}.\\
H_4|\vp_0\rangle_0 \supset 
 &\frac{\lambda}{4}\int d^2\vx\int\frac{d^{\dim}\vp_1 d^{\dim}\vp_2d^{\dim}\vp_3 d^{\dim}\vp_4}{(2\pi)^{8}}
      \left(\navpd1\navpd2\navpd3\navpd4\right)e^{-i\vx(\vp_1+\vp_2+\vp_3+\vp_4)}|\vp_0\rangle_0.\\ 
=&\frac{\lambda}{4}\int\frac{d^{\dim}\vp_1 d^{\dim}\vp_2d^{\dim}\vp_3}{(2\pi)^{6}}
     |\vp_0,\vp_1,\vp_2,\vp_3,-\vp_1-\vp_2-\vp_3\rangle_0.\\ 
\end{aligned}
\eeq
Then we get:
\beq
\begin{aligned}
|\vp_0\rangle_2^5
=&+\frac{3\lambda m^2}{4}\int
\frac{d^{\dim}\vp_1 d^{\dim}\vp_2d^{\dim}\vp_3d}{(2\pi)^{6}}
  \frac{|\vp_1,\vp_2,\vp_3,\vp_0-\vp_1,-\vp_2-\vp_3\rangle_0.}{\nomvp0(\nomvp1+\omega_{\vp_0-\vp_1}-\nomvp0)(\nomvp1+\nomvp2+\nomvp3+\omega_{\vp_0-\vp_1}+\omega_{\vp_2+\vp_3}-5\nomvp0)}\\
   &+\frac{9\lambda m^2}{8}\int\frac{d^{\dim}\vp_1 d^{\dim}\vp_2d^{\dim}\vp_3}{(2\pi)^{\dimthree}}
   \frac{|\vp_0,\vp_1,\vp_2,\vp_3,-\vp_1-\vp_2-\vp_3\rangle_0}{\omega_{\vp_1+\vp_2}(\nomvp1+\nomvp2+\omega_{\vp_1+\vp_2})(\nomvp0+\nomvp1+\nomvp2+\nomvp3+\omega_{\vp_1+\vp_2+\vp_3}-5\nomvp0)}.\\ 
   &+\frac{9\lambda m^2}{8}\int\frac{d^{\dim}\vp_1 d^{\dim}\vp_2d^{\dim}\vp_3}{(2\pi)^{\dimthree}}
   \frac{|\vp_1,\vp_2,\vp_3,\vp_0-\vp_1,-\vp_1-\vp_2\rangle_0}{\nomvp0(\nomvp1+\nomvp2+\omega_{\vp_1+\vp_2})(\nomvp1+\nomvp2+\nomvp3+\omega_{\vp_0-\vp_1}+\omega_{\vp_1+\vp_2}-5\nomvp0)}.
&-\frac{\lambda}{4}\int\frac{d^{\dim}\vp_1 d^{\dim}\vp_2d^{\dim}\vp_3}{(2\pi)^{6}}
     \frac{|\vp_0,\vp_1,\vp_2,\vp_3,-\vp_1-\vp_2-\vp_3\rangle_0}{\nomvp0+\nomvp1+\nomvp2+\nomvp3+\omega_{-\vp_1-\vp_2-\vp_3}-5\nomvp0}.\\ 
\end{aligned}
\eeq

$|\vp_0\rangle_2^7$  come from$H_3|\vp_0\rangle_1^4$ which are exactly:
\beq
\begin{aligned}
H_3|\vp_0\rangle_1^4  \supset &
 m\sqrt\frac{\lambda}{2}\int d^2\vx\int\frac{d^{\dim}\vp_1 d^{\dim}\vp_2d^{\dim}\vp_3}{(2\pi)^{\dimthree}}
   \left(\navpd1\navpd2\navpd3\right) 
   e^{-i\vx(\vp_1+\vp_2+\vp_3)}\\
   &\bigg(-m\sqrt\frac{\lambda}{2}\int d^2\vx\int\frac{d^{\dim}\vp_1 d^{\dim}\vp_2d^{\dim}\vp_3}{(2\pi)^{\dimthree}}
     \frac{e^{-i\vx(\vp_1+\vp_2+\vp_3)}}{\nomvp1+\nomvp2+\nomvp3} |\vp_0\vp_1\vp_2\vp_3\rangle_0 \bigg)\\\\
=&-\frac{\lambda m^2}{2}\int\frac{d^{\dim}\vp_1 d^{\dim}\vp_2d^{\dim}\vp_3d^{\dim}\vp_4}{(2\pi)^{8}}
|\vp_0,\vp_1,\vp_2,\vp_3,\vp_4,-\vp_1-\vp_2,-\vp_3-\vp_4\rangle_0\\
\end{aligned}
\eeq
Then we get:
\beq
\begin{aligned}
|\vp_0\rangle_2^7
=\frac{\lambda m^2}{2}\int\frac{d^{\dim}\vp_1 d^{\dim}\vp_2d^{\dim}\vp_3d^{\dim}\vp_4}{(2\pi)^{8}}
\frac{|\vp_0,\vp_1,\vp_2,\vp_3,\vp_4,-\vp_1-\vp_2,-\vp_3-\vp_4\rangle_0}{\nomvp0+\nomvp1+\nomvp2+\nomvp3+\nomvp4+\omega_{\vp_1+\vp_2}+
\omega_{\vp_3+\vp_4}-7\nomvp0}\\
\end{aligned}
\eeq
Now we get all the material we need for capture the $\delta m_4^2$
\gre{above correction update at 11:50 pm in Apr.19}:

\subsection{$\lambda^{3/2}$}
\beq
\begin{aligned}
H_2|\vp_0\rangle_3+H_3|\vp_0\rangle_2+H_4|\vp_0\rangle_1+H_5|\vp_0\rangle_0=\nomvp0|\vp_0\rangle_3\label{h2m3}
\end{aligned}
\eeq
Where \beq
\ch_5=-\sqrt{\frac{\lambda}{2}}\frac{\delta m_2^2}{2m}\phi^3 -\frac{\delta m_3^2}{2}\phi^2+T_5\phi+\hat{A_5}\hsp
\hat{A_5}=-\frac{\delta m_2^2\delta m_3^2}{8\lambda}+\frac{\delta m_5^2}{8\lambda}m^2+A_5
\eeq
 So we can get:
 \beq
|\vp_0\rangle_3= |\vp_0\rangle_3^0+|\vp_0\rangle_3^2+|\vp_0\rangle_3^6+|\vp_0\rangle_3^6+|\vp_0\rangle_3^8+|\vp_0\rangle_3^{10}\\
\eeq
Compared with the vacuum sector: $|{\Omega}\rangle_3$ in (\ref{v3}), We think the two term just totally cancel the disconnected diagram. \par  
To determine the $\delta m_4^2$ or in other words the terms which contribute the $|\vp_0\rangle_4^1$ in$ \lambda^2$ order, we see we only need the term $|\vp_0\rangle_3^2,|\vp_0\rangle_3^4$ from the (\ref{h6p0}). \par 
where $|\vp_0\rangle_3^2$  come from: 
\beq
\begin{aligned}
H_3|\vp_0\rangle_2  \supset &
 m\sqrt\frac{\lambda}{2} \int d^2\vx\int\frac{d^{\dim}\vp_1 d^{\dim}\vp_2d^{\dim}\vp_3}{(2\pi)^{\dimthree}}
   \left(3\navpd1\frac{A_{-\vp_2}A_{-\vp_3}}{4\nomvp2\nomvp3}\right) 
   e^{-i\vx(\vp_1+\vp_2+\vp_3)}|\vp_0\rangle_2^3.\\
   &+m\sqrt\frac{\lambda}{2}\int d^2\vx\int\frac{d^{\dim}\vp_1 d^{\dim}\vp_2d^{\dim}\vp_3}{(2\pi)^{\dimthree}}\left(\frac{A_{-\vp_1}A_{-\vp_2}A_{-\vp_3}}{8\nomvp1\nomvp2\nomvp3}\right) 
   e^{-i\vx(\vp_1+\vp_2+\vp_3)}|\vp_0\rangle_2^5.\\   
H_4|\vp_0\rangle_1 \supset 
 &\frac{\lambda}{4}\int d^2\vx\int\frac{d^{\dim}\vp_1 d^{\dim}\vp_2d^{\dim}\vp_3 d^{\dim}\vp_4}{(2\pi)^{8}}
      \left(12\navpd1\navpd2\frac{A_{-\vp_3}A_{-\vp_4}}{4\nomvp3\nomvp4}\right)e^{-i\vx(\vp_1+\vp_2+\vp_3+\vp_4)}|\vp_0\rangle_1^2.\\ 
 &-\frac{\delta m_2^2}{2}\int d^2\vx\int\frac{d^{\dim}\vp_1 d^{\dim}\vp_2}{(2\pi)^{\dimtwo}}
      \left(2\navpd1\frac{A_{-\vp_2}}{2\nomvp2}\right)e^{-i\vx(\vp_1+\vp_2)}|\vp_0\rangle_1^2\\
 &+\frac{\lambda}{4}\int d^2\vx\int\frac{d^{\dim}\vp_1 d^{\dim}\vp_2d^{\dim}\vp_3 d^{\dim}\vp_4}{(2\pi)^{8}}
      \left(4\navpd1\frac{\navpm2 A_{-\vp_3}A_{-\vp_4}}{8\nomvp2\nomvp3\nomvp4}\right)e^{-i\vx(\vp_1+\vp_2+\vp_3+\vp_4)}|\vp_0\rangle_1^4.\\ 
 &-\frac{\delta m_2^2}{2}\int d^2\vx\int\frac{d^{\dim}\vp_1 d^{\dim}\vp_2}{(2\pi)^{\dimtwo}}
      \left(\frac{\navpm1A_{-\vp_2}}{4\nomvp1\nomvp2}\right)e^{-i\vx(\vp_1+\vp_2)}|\vp_0\rangle_1^4\\
 &+\hat A_4|\vp_0\rangle_1^2\\
H_5|\vp_0\rangle_0 \supset 
&-\sqrt{\frac{\lambda}{2}}\frac{\delta m_2^2}{2m}\int d^2\vx\int
   \frac{d^{\dim}\vp_1d^{\dim}\vp_2d^{\dim}\vp_3}{(2\pi)^{6}}
      \left(3\navpd1\navpd2\frac{\navpm3}{2\nomvp3}\right) e^{-i\vx(\vp_1+\vp_2+\vp_3)}|\vp_0\rangle_0.\\
&+T_5\int d^2\vx\int\frac{d^{\dim}\vp_1}{(2\pi)^{2}}
      \left(\navpd1\right) e^{-i\vx\cdot \vp_1}|\vp_0\rangle_0.\\
\end{aligned}
\eeq

$|\vp_0\rangle_3^4$  come from: 
\beq
\begin{aligned}
H_3|\vp_0\rangle_2  \supset &
 m\sqrt\frac{\lambda}{2} \int d^2\vx\int\frac{d^{\dim}\vp_1 d^{\dim}\vp_2d^{\dim}\vp_3}{(2\pi)^{\dimthree}}
   \left(3\navpd1\navpd2\frac{A_{-\vp_3}}{\nomvp3}\right) 
   e^{-i\vx(\vp_1+\vp_2+\vp_3)}|\vp_0\rangle_2^3.\\
   &+m\sqrt\frac{\lambda}{2}\int d^2\vx\int\frac{d^{\dim}\vp_1 d^{\dim}\vp_2d^{\dim}\vp_3}{(2\pi)^{\dimthree}}\left(3\navpd1\frac{A_{-\vp_2}A_{-\vp_3}}{4\nomvp2\nomvp3}\right) 
   e^{-i\vx(\vp_1+\vp_2+\vp_3)}|\vp_0\rangle_2^5.\\   
   &+m\sqrt\frac{\lambda}{2}\int d^2\vx\int\frac{d^{\dim}\vp_1 d^{\dim}\vp_2d^{\dim}\vp_3}{(2\pi)^{\dimthree}}\left(\frac{\navpm1A_{-\vp_2}A_{-\vp_3}}{8\nomvp1\nomvp2\nomvp3}\right) 
   e^{-i\vx(\vp_1+\vp_2+\vp_3)}|\vp_0\rangle_2^7.\\   
H_4|\vp_0\rangle_1 \supset 
 &\frac{\lambda}{4}\int d^2\vx\int\frac{d^{\dim}\vp_1 d^{\dim}\vp_2d^{\dim}\vp_3 d^{\dim}\vp_4}{(2\pi)^{8}}
      \left(12\navpd1\navpd2\navpd3\frac{A_{-\vp_4}}{2\nomvp4}\right)e^{-i\vx(\vp_1+\vp_2+\vp_3+\vp_4)}|\vp_0\rangle_1^2.\\ 
 &-\frac{\delta m_2^2}{2}\int d^2\vx\int\frac{d^{\dim}\vp_1 d^{\dim}\vp_2}{(2\pi)^{\dimtwo}}
      \left(\navpd1\navpd2\right)e^{-i\vx(\vp_1+\vp_2)}|\vp_0\rangle_1^2\\
 &+\frac{\lambda}{4}\int d^2\vx\int\frac{d^{\dim}\vp_1 d^{\dim}\vp_2d^{\dim}\vp_3 d^{\dim}\vp_4}{(2\pi)^{8}}
      \left(6\navpd1\navpd2\frac{A_{-\vp_3}A_{-\vp_4}}{4\nomvp3\nomvp4}\right)e^{-i\vx(\vp_1+\vp_2+\vp_3+\vp_4)}|\vp_0\rangle_1^4.\\ 
 &-\frac{\delta m_2^2}{2}\int d^2\vx\int\frac{d^{\dim}\vp_1 d^{\dim}\vp_2}{(2\pi)^{\dimtwo}}
      \left(\navpd1\frac{\navpd2}{2\nomvp2}\right)e^{-i\vx(\vp_1+\vp_2)}|\vp_0\rangle_1^4\\
 &+\hat A_4|\vp_0\rangle_1^4\\
H_5|\vp_0\rangle_0 \supset 
&-\sqrt{\frac{\lambda}{2}}\frac{\delta m_2^2}{2m}\int d^2\vx\int
   \frac{d^{\dim}\vp_1d^{\dim}\vp_2d^{\dim}\vp_3}{(2\pi)^{6}}
      \left(\navpd1\navpd2\navpd3\right) e^{-i\vx(\vp_1+\vp_2+\vp_3)}|\vp_0\rangle_0.\label{h5p0}
\end{aligned}
\eeq
And we see we need $|\vp_0\rangle_2^3,|\vp_0\rangle_2^5,|\vp_0\rangle_2^7$ to generate $|\vp_0\rangle_3^2,|\vp_0\rangle_3^4$

\subsection{$\lambda^{2}$}
\beq
\begin{aligned}
H_2|\vp_0\rangle_4+H_3|\vp_0\rangle_3+H_4|\vp_0\rangle_2+H_5|\vp_0\rangle_1+H_6|\vp_0\rangle_0=\nomvp0|\vp_0\rangle_4\\
\end{aligned}
\eeq
Recall that
\beq
\ch_6=-\sqrt{\frac{\lambda}{2}}\frac{\delta m_3^2}{2m}\phi^3-\frac{\delta m_4^2}{2}\phi^2+T_6\phi+\hat{A_6}\hsp
\hat{A_6}=-\frac{(\delta m_3^2)^2}{16\lambda}-\frac{\delta m_2^2\delta m_4^2}{8\lambda}+\frac{\delta m_6^2m^2}{8\lambda}+A_6
\eeq
we know the $|\vp_0\rangle_4$ have the component :
\beq
|\vp_0\rangle_4=
|\vp_0\rangle_4^1+|\vp_0\rangle_4^3+|\vp_0\rangle_4^5+|\vp_0\rangle_4^7+|\vp_0\rangle_4^9+|\vp_0\rangle_4^11
+|\vp_0\rangle_4^{13}
\eeq
Then we can impose renormalization condition:
\beq
|\vp_0\rangle_4^1=0
\eeq
\beq
|\vp_0\rangle_2^1 \supset \bigg(H_3|\vp_1\rangle_1+H_4|\vp_0\rangle_0\bigg)|_{one-meson}=0
\eeq
Where we see the term which contribute $|\vp_0\rangle_4^1=0$ come from:
\beq
\begin{aligned}
H_3|\vp_0\rangle_3 
\supset 
&m\sqrt\frac{\lambda}{2} \int d^2\vx\int\frac{d^{\dim}\vp_1 d^{\dim}\vp_2d^{\dim}\vp_3}{(2\pi)^{\dimthree}}
      \left(\frac{6\navpd1\navpm2\navpm3}{4\nomvp2\nomvp3}\right) e^{-i\vx(\vp_1+\vp_2+\vp_3)}|\vp_0\rangle_3^2\\
&+m\sqrt\frac{\lambda}{2} \int d^2\vx\int\frac{d^{\dim}\vp_1 d^{\dim}\vp_2d^{\dim}\vp_3}{(2\pi)^{\dimthree}}
      \left(\frac{\navpm1\navpm2\navpm3}{8\nomvp1\nomvp2\nomvp3}\right) e^{-i\vx(\vp_1+\vp_2+\vp_3)}|\vp_0\rangle_3^4\\
H_4|\vp_0\rangle_2 \supset 
  &\frac{\lambda}{4}\int d^2\vx\int\frac{d^{\dim}\vp_1 d^{\dim}\vp_2d^{\dim}\vp_3 d^{\dim}\vp_4}{(2\pi)^{8}}
      \left(\navpd1\frac{\navpm2\navpm3\navpm4}{8\nomvp2\nomvp3\nomvp4}\right)e^{-i\vx(\vp_1+\vp_2+\vp_3+\vp_4)}
      |\vp_0\rangle_2^3 \\\nonumber
 &+\frac{\delta m_2^2}{2}\int d^2\vx\int\frac{d^{\dim}\vp_1 d^{\dim}\vp_2d^{\dim}}{(2\pi)^{4}}
      \left(\frac{\navpm1\navpm1}{4\nomvp1\nomvp1}\right)e^{-i\vx(\vp_1+\vp_2)}
      |\vp_0\rangle_2^3 \\\nonumber
 H_5|\vp_0\rangle_1 \supset    
 &\sqrt{\frac{\lambda}{2}}\frac{\delta m_3^2}{2m}\int d^2\vx\int\frac{d^{\dim}\vp_1 d^{\dim}\vp_2d^{\dim}\vp_3}{(2\pi)^{6}}
      \left(\navpd1\frac{\navpm2\navpm3}{4\nomvp2\nomvp3}\right)e^{-i\vx(\vp_1+\vp_2+\vp_3)}
      |\vp_0\rangle_1^2 \\\nonumber 
  &+T_5\int d^2\vx\int\frac{d^{\dim}\vp_1}{(2\pi)^{2}}
      \left(\frac{\navpm1}{2\nomvp2}\right)e^{-i\vx \vp_1}
      |\vp_0\rangle_1^2 \\\nonumber 
H_6|\vp_0\rangle_0 \supset    
  &A_6|\vp_0\rangle_0 \\\nonumber \label {h6p0}
\end{aligned}
\eeq
We see we only need the exactly $|\vp_0\rangle_3^2,|\vp_0\rangle_3^4$ and $|\vp_0\rangle_2^3,|\vp_0\rangle_1^2$.
 
\end{document}
\section{}

In his Erice lectures \cite{erice}, Coleman suggested an open problem.  It was already known that in 1+1 dimensional scalar theories, quantum states with solitons correspond to coherent states \cite{vinc72}, albeit with perturbative corrections \cite{taylor78,cocorr23}.  The coherent state construction works in these theories essentially because their ultraviolet divergences can be removed by normal ordering.  Moving beyond this narrow class of theories on the other hand, the coherent state constructon leads to various pathologies, for example, Coleman claims that the expectation value of the Hamiltonian density is infinite.  The open question, is how to construct the states corresponding to solitons in this larger class of theories.  Coleman writes, ``A good place to begin exploring would be a super-renormalizable theory in two spatial dimensions."

In this paper, we follow Coleman's suggestion.  We try to construct solitons in a scalar theory in 2+1 dimensions.  We do not yet complete the problem, rather we push it as far as we can before we run into the troublesome ultraviolet divergences.  More precisely, we work to linear order in perturbations about the soliton, corresponding to one loop in the original formulation of the theory.  We explicitly construct a perturbative expansion, and use it to calculate the $O(\lambda^0)$ quantum correction to the tension of the domain wall present in this theory.  

The next step in this program requires a choice for the construction of the subleading correction to the state.  Then several consistency checks will be necessary.  One must be sure that the tadpole cancellation present in 1+1 dimensions is not ruined by the renormalization.  Also, one must check that a choice of counterterms which cancels the ultraviolet divergences in the vacuum sector, automatically also does so in the soliton sector.  The results of the present paper are independent of this choice, and so we believe can serve as a springboard for this next, critical step in Coleman's program.

We begin in Sec.~\ref{classsez} with a review of classical solitons.  Our main construction appears in Sec.~\ref{quantsez}, where we apply canonical quantization.  The soliton states are written as a nonperturbative displacement operator, which creates the coherent states, acting on a state.  We show that this later state can be constructed and evolved in perturbation theory using an operator called the soliton Hamiltonian, which we construct.  This procedure is a straightforward generalization of Ref.~\cite{cahill76,mekink} to more dimensions.   Unfortunately Derrick's theorem tells us that higher-dimensional scalar theories, as we have constructed, do not have localized soliton solutions.  In Sec.~\ref{exsez} we apply to construction of the previous section to a domain wall solution in the (2+1)-dimensional $\phi^4$ double-well theory.  The solution is just the kink of the (1+1)-dimensional theory lifted up a dimension.

\section{Classical Solitons} \label{classsez}

Let us consider a theory in $d+1$ dimensions consisting of a scalar field $\phi(\vx)$ with conjugate momentum $\pi(\vx)$ and governed by a Hamiltonian which in the Schrodinger picture is
\beq
H=\int d^{\dim}\vx :\ch:_a\hsp
\ch=\frac{\pi^2(\vx)+\nabla\phi(\vx)\cdot \nabla\phi(\vx)}{2}+\frac{V(\sl\phi(\vx))}{\lambda}.
\eeq
The normal ordering $::_a$ is the usual plane-wave normal ordering, defined at the mass scale corresponding to the mass $m$ of the perturbative meson far from the soliton.

The corresponding classical theory, defined by ignoring the normal ordering and allowing $\phi(\vx,t)$ and $\pi(\vx,t)$ to depend on time, is characterized by the classical equation of motions
\beq
\ddot\phi(\vx,t)=\nabla^2\phi(\vx,t)-\frac{V^{(1)}(\sl\phi(\vx,t))}{\sl} \label{eom}
\eeq
where $V^{(n)}$ is the $n$-th derivative of $V$ with respect to its argument ${\sqrt{\lambda}\phi}$.

We will be interested in two solutions.  First, consider a time-independent soliton
\beq
\phi(\vx,t)=f(\vx).
\eeq
In this case, the classical equation of motion (\ref{eom}) is
\beq
\nabla^2 f(\vx)=\frac{V^{(1)}(\sl f(\vx))}{\sl}.
\eeq

Second we are interested in small perturbations about this solution
\beq
\phi(\vx,t)=f(\vx)+\g(\vx,t).
\eeq
In this case, to linear order in $\g$, the equation of motion is
\beq
V^{(2)}(\sl f(\vx))\g(\vx,t)+\ddot \g(\vx,t)=\nabla^2 \g(\vx,t).
\eeq

We will decompose $\g(\vx,t)$ into components $\g(\vx)$ with fixed frequencies, which can be taken to be real for a stable soliton
\beq
\g_\vk(\vx,t)=\g_\vk(\vx)e^{-i\ok{} t}.
\eeq
For each component, labeled by the abstract index $\vk$, the equation of motion becomes
\beq
V^{(2)}(\sl f(\vx))\g_\vk(\vx)=\left(\omega_{\vk}^2+\nabla^2\right) \g_\vk(\vx).   \label{sl}
\eeq
We define the functions $\g_\vk(\vx,t)$ to be the solutions of this equation.  The index $\vk$ will in general run over discrete and also continuous values.  The continuous values include a vector space $\R^d$, which is the reason for the vector symbol on the $\vk$, defined up to signs by $\omega_\vk=\sqrt{m^2+\vk^2}$.  In the case of the continuous values, we will normalize the $\g_k(\vx)$ via
\beq
\int d^{\dim}\vx \g_{\vk_1}(\vx)\g_{\vk_2}(\vx)=(2\pi)^\dim\delta^{\dim}(\vk_1+\vk_2)\hsp \g_{\vk}^*(\vx)=\g_{-\vk}(\vx). \label{dirac}
\eeq
In the case of discrete indices, $\g_\vk(\vx)$ will be taken to be real and
\beq
\int d^{\dim}\vx \g_{\vk_1}(\vx)\g_{\vk_2}(\vx)=\delta_{\vk_1,\vk_2}.
\eeq
More generally, some values of $\vk$ inhabit lower dimensions submanifolds, and we will use the obvious hybrids in which continuous directions are normalized with Dirac delta functions and discrete labels of manifolds are normalized with Kronecker $\delta$.  Often we will use a shorthand in which the normalization condition in (\ref{dirac}) is written, but it is implied that if some component of $\vk$ is discrete, then the corresponding $2\pi\delta$ should be replaced with a Kronecker $\delta$.

\section{Quantum Solitons} \label{quantsez}

\subsection{Soliton Hamiltonian}
Define the displacement operator
\beq
\df=\exp{-i\int d^{\dim}\vx f(\vx) \pi(\vx)}
\eeq
and the soliton Hamiltonian
\beq
H\p=\df^\dag H \df.
\eeq

Explicitly, the soliton Hamiltonian is
\beq
H\p[\phi(\vx),\pi(\vx)]=H[\phi(\vx)+f(\vx),\pi(\vx)].
\eeq
One can check that this identity holds despite the normal ordering.  We will expand $H\p$ in powers of the coupling $\sl$
\beq
H\p=\sum_{j=0}^\infty H\p_j
\eeq
where $H\p_j$ is a functional of the fields times of a coefficient of order $\lambda^{j/2-1}$.  It is defined to consist of terms which, when normal ordered using $::_a$, are $n$-linear in $\phi(x)$ and $\pi(x)$.  One easily finds
\beq
H\p_0=Q_0\hsp H\p_1=0
\eeq
where $Q_0$ is the energy of the classical solution $\phi(\vx,t)=f(\vx)$.

The most important step in perturbation theory is order $O(\lambda^0)$, as any failure to diagonalize the Hamiltonian exactly at this order will not be suppressed at small $\lambda$.  The contribution to the Hamiltonian at this order is
\bea
H\p_2&=&A+B+C\hsp
A=\frac{1}{2} \int d^{\dim}\vx :\pi^2(\vx):_a\\
B&=&\frac{1}{2} \int d^{\dim}\vx :(\nabla \phi)^2(\vx):_a\hsp
C=\frac{1}{2} \int d^{\dim}\vx V^{(2)}(\sl f(\vx)):\phi^2(\vx):_a.\nonumber
\eea

\subsection{Decompositions}
We will consider two decompositions of the fields
\bea
\phi(\vx)&=&\pinvp{\dim} e^{-i\vx\cdot\vp}\phi_{\vp}=\pinvk{d} \g_{\vk}(\vx)\phi_{\vk}\label{dec}\\
\pi(\vx)&=&\pinvp{\dim} e^{-i\vx\cdot\vp}\pi_{\vp}=\pinvk{d} \g_{\vk}(\vx)\pi_{\vk}.\nonumber
\eea
To avoid a proliferation of hats and tildes, we use the same notation for $\phi_{\vp}$ and $\phi_{\vk}$ although they represent distinct bases of the space of operators, they will be distinguished only by the letter used for the index.  The $\dint$ symbol is an integration over continuous indices $\vk$, dividing by $2\pi$ for each dimension, plus a sum over discrete indices.  In general, the space of $\vk$ has components of various dimensions and these are each integrated over and the integrals are summed.

The completeness relations (\ref{dirac}) allow these decompositions to be inverted.  The canonical commutation relations
\beq
[\phi(\vx_1),\pi(\vx_2)]=i \delta^{\dim} (\vx_1-\vx_2)
\eeq
then lead to the usual commutation relations in the plane wave and normal mode bases
\beq
[\phi_{\vp_1},\pi_{\vp_2}]=i(2\pi)^\dim\delta^{\dim}(\vp_1+\vp_2)\hsp [\phi_{\vk_1},\pi_{\vk_2}]=i(2\pi)^\dim\delta^{\dim}(\vk_1+\vk_2)
\eeq
where again it is implicit that in the case of lower dimensional submanifolds in the $\vk$ space, the transverse $2\pi\delta$ should be replaced with  Kronecker deltas.

\subsection{Harmonic Oscillators}
Using the $\vk$ decompositions in Eq.~(\ref{dec}) one finds
\beq
A=\frac{1}{2}\kinv{d}{k_1}\kinv{d}{k_2}\int d^{\dim}\vx \g_{\vk_1}(x)\g_{\vk_2}(x):\pi_{\vk_1}\pi_{\vk_2}:_a=\frac{1}{2}\kinv{d}{k}:\pi_\vk\pi_{-\vk}:_a
\eeq
where we have used the completeness relation (\ref{dirac}).  Similarly, integrating by parts and dropping a rapidly oscillating boundary term
\beq
B=-\frac{1}{2}\kinv{d}{k_1}\kinv{d}{k_2}\int d^{\dim}\vx \g_{\vk_1}(x)\nabla^2\g_{\vk_2}(x):\phi_{\vk_1}\phi_{\vk_2}:_a \label{beq}
\eeq
while the defining equation (\ref{sl}) leads to
\beq
C=\frac{1}{2}\kinv{d}{k_1}\kinv{d}{k_2}\int d^{\dim}\vx \g_{\vk_1}(x)\left(\nabla^2+\omega_{\vk_2}^2\right)\g_{\vk_2}(x):\phi_{\vk_1}\phi_{\vk_2}:_a. \label{ceq} 
\eeq
Adding (\ref{beq}) and (\ref{ceq}) and again using the completeness (\ref{dirac}) one obtains
\beq
B+C=\frac{1}{2}\kinv{d}{k_1}\kinv{d}{k_2} \omega_{\vk_2}^2:\phi_{\vk_1}\phi_{\vk_2}:_a\int d^{\dim}\vx\g_{\vk_2}(x)\g_{\vk_1}(x)=\frac{1}{2}\kinv{d}{k}\omega_{\vk}^2:\phi_\vk\phi_{-\vk}:_a. \label{bceq}
\eeq

Adding all of these contributions we find the soliton Hamiltonian at order $O(\lambda^0)$
\beq
H\p_2=\frac{1}{2}\kinv{d}{k}\left(:\pi_\vk\pi_{-\vk}:_a+\omega_{\vk}^2:\phi_\vk\phi_{-\vk}:_a\right). \label{h2}
\eeq
If it were not for the normal ordering, this would be a sum of harmonic oscillators, one at each $\vk$.  The ground state at leading order in perturbation theory would be the ground state of each harmonic oscillator, while the excited states would be created by the corresponding creation operators $B_\vk^{\ddag}$.  There in general will be zero modes, for example if the Hamiltonian is translation invariant or has some similar internal symmetry.  For these, $\omega_\vk=0$ and so only the corresponding $\pi_\vk^2$ term is present.  This describes the quantum mechanics of a free particle describing the position with respect to that symmetry, and one must impose that the ground state is annihilated by each such $\pi_\vk$, while excited states correspond to exponentials in $i\phi_{\vk}$.

What is the effect of the normal ordering?  Since these operators are linear, it can only add a constant.  We refer to this constant as $Q_1$ when $H\p_2$ is ordered in the form $B^\ddag B$.  It is the one-loop correction to the soliton mass \cite{cahill76,mekink}.  We will now compute it. 

\section{One-Loop Mass Correction}

\subsection{Plane-Wave Decomposition}

The normal ordering is defined in terms of the usual plane wave decomposition of the fields, corresponding to the middle expressions in (\ref{dec}).  Using this decomposition, one easily finds
\beq
A=\frac{1}{2}\pinv{d}{p_1}\pinv{d}{p_2}\int d^{\dim}\vx e^{-ix(\vp_1+\vp_2)}:\pi_{\vp_1}\pi_{\vp_2}:_a=\frac{1}{2}\pinv{d}{p}:\pi_\vp\pi_{-\vp}:_a.
\eeq

As the decompositions are both in complete bases, one may map from one to the other via
\beq
\phi_\vk=\pinv{d}{p}\gt_{-\vk}(\vp)\phi_{\vp}\hsp \pi_\vk=\pinv{d}{p}\gt_{-\vk}(\vp)\pi_{\vp} \label{phid}
\eeq
where we have defined the Fourier transform
\beq
\gt_\vk(\vp)=\int d^{\dim}\vx \g_\vk(\vx)e^{-i\vp\cdot\vx}.
\eeq
This allows us to rewrite $B+C$, given in Eq.~(\ref{bceq}), in the plane-wave basis
\beq
B+C=\frac{1}{2}\kinv{d}{k}\omega_{\vk}^2\pinv{d}{p_1}\pinv{d}{p_2}\gt_{-\vk}(\vp_1)\gt_{\vk}(\vp_2):\phi_{\vp_1}\phi_{\vp_2}:_a. 
\eeq

Now we use the standard Schrodinger picture decomposition into creation and annihilation operators
\beq
\phi_\vp=A^\ddag_\vp+\frac{A_{-\vp}}{2\omega_{\vp}}\hsp
\pi_\vp=i\omega_{\vp}A^\ddag_\vp-\frac{iA_{-\vp}}{2}\hsp
A^\ddag_\vp=\frac{A^\dag_\vp}{2\omega_\vp}. \label{pd}
\eeq

The normal ordering $::_a$ is defined to be the operation that places all $A^\ddag$ to the left of all $A$.  And so we may finally evaluate the normal ordering
\bea
A&=&\frac{1}{2}\pinv{d}{p}\left[-\omega_{\vp}^2A^\ddag_\vp A^\ddag_{-\vp}+\omega_\vp A^\ddag_\vp A_\vp -\frac{A_\vp A_{-\vp}}{4}  
\right]\label{abc}\\
B+C&=&\frac{1}{2}\kinv{d}{k}\omega_{\vk}^2\pinv{d}{p_1}\pinv{d}{p_2}\gt_{-\vk}(\vp_1)\gt_{\vk}(\vp_2)\nonumber\\
&&\times\left[ 
A^\ddag_{\vp_1}A^\ddag_{\vp_2}+\frac{A^\ddag_{\vp_1}A_{-\vp_2}}{2\omega_{\vp_2}}+\frac{A^\ddag_{\vp_2}A_{-\vp_1}}{2\omega_{\vp_1}}+\frac{A_{-\vp_1}A_{-\vp_2}}{4\omega_{\vp_1}\omega_{\vp_2}}
\right]
.\nonumber
\eea

\subsection{Back to the Normal Mode Basis}
Now that the normal ordering symbol has disappeared, we can freely move between bases with Bogoliubov transforms.  We will now need to move back to the normal mode basis.  We will decompose the index $\vk$ into zero modes, for which $\omega_\vk=0$, and nonzero modes, for which it is taken to be positive.  We will now consider nonzero modes.  With a page of calculations, following the example worked out in Ref.~\cite{mekink}, the argument below can be easily modified to the case of zero modes, and one can derive that the final results below will hold for zero modes just by setting $\omega_\vk=0$, although this substitution cannot be used at intermediate steps.

In the case of nonzero modes, we will make the decomposition into creation and annihilation operators
\beq
\phi_\vk=B^\ddag_\vk+\frac{B_{-\vk}}{2\omega_{\vk}}\hsp
\pi_\vk=i\omega_{\vk}B^\ddag_\vk-\frac{iB_{-\vk}}{2}\hsp
B^\ddag_\vk=\frac{B^\dag_\vk}{2\omega_\vk}. \label{bd}
\eeq
In the case of discrete modes, it is understood that $B_{-\vk}$ is defined to be $B_{\vk}$ as the corresponding $\g_{\vk}$ were taken to be real.  Define the state $\vac_0$ by
\beq
B_{\vk}\vac_0=0. \label{bz}
\eeq
We will also impose that it is annihilated by $\pi_\vk$ for each zero mode $\vk$, but that will not be relevant now.  

The one-loop correction to the soliton mass, $Q_1$, is the eigenvalue of $H\p_2$
\beq
H\p_2\vac_0=Q_1\vac_0.
\eeq
Therefore we are only interested in those terms in $H\p_2$ which do not annihilate $\vac_0$.  We have already seen in Eq.~(\ref{h2}) that any such term must be a scalar, and so there cannot be any $B^\ddag B^\ddag$ terms.  This leaves terms of the form $B B^\ddag$.  We can simplify these using (\ref{bz}) which implies the identity
\beq
B_{\vk_1} B^\ddag_{\vk_2}\vac_0=[B_{\vk_1}, B^\ddag_{\vk_2}]\vac_0=(2\pi)^\dim\delta^{\dim}(\vk_1-\vk_2)\vac_0. \label{id}
\eeq

Our strategy will therefore be to calculate $Q_1$ by isolating all $B^\ddag B\vac_0$ terms $H\p_2\vac_0$ and applying the identity (\ref{id}) to simplify them.  We will obtain these terms by plugging the Bogoliubov transform\footnote{This is derived from Eqs.~(\ref{phid}), (\ref{pd}) and (\ref{bd}).}
\bea
A^\ddag_\vp&=&\frac{1}{2}\kinv{d}{k}\frac{\gt_{\vk}(-\vp)}{\omega_\vp}\left[ (\omega_\vp+\omega_\vk)B^\ddag_\vk +(\omega_\vp-\omega_\vk)\frac{B_{-\vk}}{2\omega_\vk}
\right]\label{bog}\\
\frac{A_{-\vp}}{2\omega_{\vp}}&=&\frac{1}{2}\kinv{d}{k}\frac{\gt_{\vk}(-\vp)}{\omega_\vp}\left[(\omega_\vp-\omega_\vk) B^\ddag_\vk +(\omega_\vp+\omega_\vk)\frac{B_{-\vk}}{2\omega_\vk}
\right]\nonumber
\eea
into Eq.~(\ref{abc}).

Only two combinations of ladder operators do not annihilate the kink ground state, namely $B^\ddag B^\ddag$ and $B B^\ddagger$.  The $B^\ddag B^\ddag$ terms in $A$ and $B+C$ are
\beq
A \supset -\frac{1}{2} \kinv{d}{k} \omega _ {\vk} ^{2} B^{\ddagger}_ {\vk}B^{\ddagger}_ {-\vk} \hsp
B+C \supset \frac{1}{2} \kinv{d}{k} \omega _ {\vk} ^{2} B^{\ddagger}_ {\vk}B^{\ddagger}_ {-\vk} .
\eeq
Therefore $H\p_2=A+B+C$  contains no $B^\ddag B^\ddag$ terms, and only $B B^\ddag$ terms remain.

Restricting our attention to terms proportional to $B B^\ddag$, in the case of the $\pi^2$ term, one finds
\bea
A\vac_0&=&\frac{1}{8}\pinv{d}{p}\kinv{d}{k_1}\kinv{d}{k_2}\frac{\gt_{\vk_1}(-\vp)}{\omega_\vp}\frac{\gt_{\vk_2}(\vp)}{\omega_\vp}\frac{\omega_\vp^2}{2\omega_{\vk_1}} \left[
-(\omega_{\vp}+\omega_{\vk_2})(\omega_{\vp}-\omega_{\vk_1})\right.\nonumber\\
&&\left.+2(\omega_{\vp}-\omega_{\vk_2})(\omega_{\vp}-\omega_{\vk_1})-(\omega_{\vp}-\omega_{\vk_2})(\omega_{\vp}+\omega_{\vk_1})
\right]
B_{-\vk_1}B^\ddag_{k_2}\vac_0.
\eea
The identity (\ref{id}) then yields
\bea
A\vac_0&=&\frac{1}{4}\pinv{d}{p}\kinv{d}{k}\gt_{\vk}(-\vp)\gt_{-\vk}(\vp)(\omega_\vk-\omega_\vp)\vac_0.
\eea
Similarly
\bea
(B+C)\vac_0&=&\frac{1}{8}\kinv{d}{k}\pinv{d}{p_1}\pinv{d}{p_2}\kinv{d}{k\p}\omega_{\vk}^2\gt_{-\vk}(\vp_1)\gt_\vk(\vp_2)\gt_{\vk\p}(-\vp_1)\gt_{-\vk\p}(-\vp_2)\nonumber\\
&&\times\left[ 
\frac{2}{\omega_{\vk\p}}-\frac{\omega_{\vp_1}+\omega_{\vp_2}}{\omega_{\vp_1}\omega_{\vp_2}}
\right]\vac_0=(D+E)\vac_0
\eea
where $D$ and $E$ correspond to the first and second terms in the square bracket.  In the case of the term $D$, we perform the $\vp_2$ integral using the completeness relation, which yields a $(2\pi)^\dim\delta^{\dim}(\vk-\vk\p)$, which is used to perform the $k\p$ integration.  We thus find
\beq
D=\frac{1}{4}\kinv{d}{k}\pinv{d}{p}\omega_\vk \gt_{-\vk}(\vp)\gt_\vk (-\vp).
\eeq
At this point the $\vp$ integral could be performed, yielding an infinite answer.  This is to be expected, only the sum of all terms in $H\p_2$ is finite and the integral should not be performed until the sum is taken.

The term $E$ may be evaluated by performing the $\vk\p$ integral, which yields a $(2\pi)^\dim\delta^{\dim}(\vp_1-\vp_2)$, which in turn is used to perform the $\vp_2$ integral.  One finds
\beq
E=-\frac{1}{4}\kinv{d}{k}\pinv{d}{p} \gt_{-\vk}(\vp)\gt_\vk(-\vp)\frac{\omega_\vk^2}{\omega_\vp}.
\eeq
Adding the scalars $D$ and $E$ to the eigenvalue of $A$, one finds our main result for the one-loop mass correction 
\beq
Q_1=-\frac{1}{4}\kinv{d}{k}\pinv{d}{p} \gt_{-\vk}(\vp)\gt_\vk(-\vp)\frac{(\omega_\vk-\omega_\vp)^2}{\omega_\vp}. \label{padrone}
\eeq
This generalizes the famous formula from Ref.~\cite{cahill76} to solitons in arbitrary dimensions.

One can now write the leading order soliton Hamiltonian as
\beq
H\p_2=Q_1+\frac{\pi_0^2}{2}+\kinv{d}{k}\omega_{\vk}B^\ddag_{\vk}B_{\vk} \label{h2f}
\eeq
where $\pi_0$ is $\pi_\vk$ in the case in which $\omega_{\vk}=0$.  If there are multiple such values of $\vk$, corresponding to zero modes of various classically broken symmetries, then the corresponding $\pi_0$ terms should be summed.

\subsection{Limitations}

Our master formula (\ref{padrone}) appears very general.  We have not even assumed that the theory is renormalizable.  However some caution is in order.  First of all, one needs to check that the expression for $Q_1$ is convergent.  As we describe below, this is not the case for infinitely extended solitons, but that is not a problem as one instead is interested in densities or tensions in such cases.  There may also be ultraviolet divergences if, for example, $\gt_{-\vk}(\vp)$ does not fall faster than $\vp^{(3-d)/2}$ when $\vk-\vp$ is held fixed.  

Another problem is that Derrick's theorem tells us that there are no localized solitons in the scalar theories that we have considered beyond the 1+1 dimensional case, which was handled already in Ref.~\cite{cahill76}.  In practice this exercise has been useful for settings which are somewhat different.  First, one may stabilize the solutions with additional fields, such as gauge fields.  In this case the one-loop correction $Q_1$ will arise, but one must add the corrections arising from the gauge fields.  We intend to study such theories in future work.  Second, one may consider time-dependent solutions such as oscillons and Q-balls.  We have recently shown that the generalization to such cases is feasible.  Finally, one may consider extended solutions.  For these $Q_1$ will be infinite, but the mass per volume will be finite, and can be calculated via a straightforward generalization of the argument above.  This case will be considered in the next section.  

The more serious problem is mass renormalization.  We have used the bare mass in the normal ordered Hamiltonian to construct $\df$ and so $H\p$.  In a theory that requires mass renormalization, this bare mass is infinite.  As a result, the factors of $\omega$ in (\ref{padrone}) are all infinite and the argument makes no sense.  Of course, we are working at order $O(\lambda^0)$ and such divergences appear at order $O(\lambda)$, so formally in the sense of an asymptotic expansion this is not a problem.  Whether it nonetheless leads to physical results in such theories is unclear.  We will investigate this problem in future work, in which we will explore higher order corrections with the necessary counterterms introduced, following Ref.~\cite{andy}.  We will need to determine, in particular, how $\df$ is to be renormalized.  This is in fact the open problem posed by Coleman in Ref.~\cite{erice}.  Our hope is that the correct handling of the renormalization of $\df$ will imply that eliminating divergences in the vacuum sector eliminates them in the soliton secctor, and that (\ref{padrone}) proves to be the correct one-loop mass correction in all renormalizable theories, as the naive asymptotic expansion suggests.

For now, we note that there are a few theories that are not finite yet do not require mass normalization, to which the above treatment may be applied immediately.  We will provide an example in the following section.

We note that at one loop, spectral methods are also available to calculate mass corrections \cite{specrev} and even form factors \cite{gw24}.  However, these do not generalize in any obvious way to higher loops, whereas in future work we intend to demonstrate that our approach can be extended to higher-loop corrections.

\section{Example: $\phi^4$ Domain Wall} \label{exsez}

Needless to say, $Q_1$ is divergent for a soliton that extends along an infinite direction.  In this case one should calculate not the infinite mass correction, but rather the correction to the tension.  Let us first see this divergence in the case of the $\phi^4$ double-well model in $2+1$ dimensions.  Renormalizability tells us that more generally one may consider a potential which is at most sextic in $\phi(\vx)$, and the generalization to this case will be obvious.

\subsection{The Classical Domain Wall}

Let the spatial directions be $x$ and $y$.  Consider the potential
\beq
V(\sl\phi)=\frac{\lambda \phi^2}{4}\left(\sl\phi-\sqrt{2}m\right)^2
\eeq
and the classical solution 
\beq
f(x,y)=\frac{m}{\sqrt{2\lambda}}\left(1+\tanh\left({\frac{mx}{2}}\right)\right).
\eeq
The solution is identical to that of the kink in 1+1 dimensions, but now it is infinitely extended in the $y$ direction and we will call it a domain wall \cite{vach84}.

The classical domain wall tension is \cite{morris18,muro22}
\beq
\rho_0=\frac{m^3}{3\lambda}. 
\eeq
This is the same formula as the mass $Q_0$ of the $\phi^4$ kink in 1+1 dimensions.  However, one should recall that in $2+1$ dimensions $\lambda$ has dimensions of mass whereas in $1+1$ dimensions it has dimensions of mass${}^2$.

\subsection{Normal Modes}

The normal modes $\g_\vk(x,y)$ can be factorized
\beq
\g_{k_xk_y}(x,y)=\g_{k_x}(x) e^{-i k_y y}
\eeq
where the normal modes in the $x$ direction are those of the $\phi^4$ kink, described by the exact solutions of the Poschl-Teller potential
\bea
\g_k(x)&=&\frac{e^{-ikx}}{\ok{} \sqrt{m^2+4k^2}}\left[2k^2-m^2+(3/2)m^2\sech^2(m x/2)-3im k\tanh(m x/2)\right]\nonumber\\
\g_S(x)&=&\frac{\sqrt{3 m}}{2}\tanh(m x/2)\sech(m x/2)\hsp
\g_B(x)=-\sqrt{\frac{{3m}}{8}}\sech^2(m x/2).\label{nmode}
\eea
Here the indices $B$ and $S$ represent the zero mode and shape mode of the kink in 1+1 dimensions.  The corresponding frequencies of the $\g_{k_xk_y}(x,y)$ are
\beq
\omega_{B k_y}=|k_y|\hsp \omega_{S k_y}=\sqrt{\frac{3m^2}{4}+k_y^2}\hsp \omega_{k_xk_y}=\sqrt{m^2+k_x^2+k_y^2}. \label{omk}
\eeq
There is a single zero mode, corresponding to the case $k_y=0$ of $\g_{B k_y}$.  

Unlike the 1+1 dimensional case, there is no mass gap, as $k_y$ can be arbitrarily small but positive leading to an arbitrarily small $\omega_{B ky}$.  These correspond to long wavelength vibrations of the domain wall $x$ coordinate.  At every step it is essential to check that they do not lead to infrared divergences.

As in previous sections, we use the abstract vector notation $\vk$ to represent pairs $(k_x,k_y)$ of continuum modes as well as pairs $(B,k_y)$ and $(S,k_y)$.

The Fourier transform is
\beq
\gt_{k_xk_y}(p_x,p_y)=\int dx\int dy \g_{k_x}(x)e^{-i(p_x x+(p_y+k_y) y)}=2\pi\delta(p_y+k_y)\gt_{k_x}(p_x). \label{gte}
\eeq
This can be easily substituted into our master formula for the one-loop mass correction (\ref{padrone}).  The mass correction is quadratic in $\gt$, yielding two factors of $\delta(p_y+k_y)$.  The first may be used to perform the $k_y$ or $p_y$ integration.  The second, leaves an infinity. This, of course, is the expected answer for the mass correction to a domain wall of infinite length.

\subsection{The One-Loop Tension}
Can one modify the derivation to obtain the one-loop correction not to the mass, but rather the tension?  Recall that in a local quantum field theory, the $\vx$ integration in the Hamiltonian should be performed last.  Therefore, ideally, one could write the Hamiltonian as an integral $\int dy$, perform the entire derivation to arrive at the density as a function of $y$ and declare that the answer is the tension.  Unfortunately, this method of handling the infrared divergence from the infinite $y$ integration is not compatible with our normal ordering, which is defined in terms of momenta and not positions.

The normal ordering was implemented in Eq.~(\ref{abc}).  And so any modification must be made after that point in the derivation.  The divergent Dirac $\delta$ in $\gt$ entered our derivation through the Bogoliubov transformation in Eq.~(\ref{bog}), which in turn obtained $\gt$ from the field transformation (\ref{phid}).  This was derived by inverting decompositions (\ref{dec}).  The inversion required an integral of the field over all $y$ coordinates.  This integral is not defined in the present case, leading to the above infrared divergence.  Moreover, as a result of the locality of the theory, this integral should be performed after the momentum integrals.  We are justified in reordering the integrals only when they satisfy the usual Fubini's Theorem conditions, which is not the case here.

Therefore, each $\gt_\vk(\vp)$ should in general be expanded following the derivation of Eq.~(\ref{gte}) as
\beq
\gt_{k_xk_y}(p_x,p_y)=\int dx\int dy \g_{k_x}(x)e^{-i(p_x x+(p_y+k_y) y)}=\gt_{k_x}(p_x)\int dy e^{-i(p_y+k_y) y}. 
\eeq
Clearly, this oscillates rapidly when $k_y\neq -p_y$ and so the $k_y$ integration will have support at $-p_y$ and we may safely use one of the Dirac $\delta$ functions to perform the $k_y$ integral. 

Therefore we conclude that (\ref{padrone}) becomes
\bea
Q_1&=&-\frac{1}{4}\kinv{2}{k}\pinv{2}{p} \gt_{-k_x}(p_x)2\pi\delta(k_y-p_y)\gt_{k_x}(-p_x)\int dy  e^{-i(-p_y+k_y) y}
\frac{(\omega_\vk-\omega_\vp)^2}{\omega_\vp}\nonumber\\
&=&\int dy\left[-\frac{1}{4}\ppin{k_x}\pinv{2}{p} \gt_{-k_x}(p_x)\gt_{k_x}(-p_x)
\frac{(\omega_{k_xp_y}-\omega_{p_xp_y})^2}{\omega_{p_xp_y}}\right].
\eea
Identifying the term in square brackets with the one-loop correction to the domain wall tension $\rho_1(y)$ one obtains
\beq
Q_1=\int dy \rho_1(y)\hsp
\rho_1(y)=-\frac{1}{4}\ppin{k_x}\pinv{2}{p} \gt_{-k_x}(p_x)\gt_{k_x}(-p_x)
\frac{(\omega_{k_xp_y}-\omega_{p_xp_y})^2}{\omega_{p_xp_y}}. \label{teq}
\eeq
This formula is valid for 2+1 dimensional scalar models with quartic or sextic potentials.  We remind the reader that
\beq
\omega_{p_xp_y}=\sqrt{m^2+p_x^2+p_y^2}
\eeq
while $\omega_{k_x p_y}$ is given in Eq.~(\ref{omk}).  Clearly $\rho_1(y)$ is independent of $y$, due to the flatness of the wall.

\subsection{Numerical One-Loop Tension Correction}
In this subsection we will numerically evaluate our expression (\ref{teq}) for the tension $\rho_1$ in the case of the $\phi^4$ double-well model.  In this case $k_x$ needs to be integrated over all real values, corresponding to continuum modes, plus it should be summed over the discrete zero mode and shape mode.  The relevant Fourier transformations have been computed in Ref.~\cite{mekink}  
\bea
\tilde{\g}_{B}(p)&=&-\frac{\sqrt{6}\pi p}{m^{3/2}}  \csch\left(\frac{\pi p}{m}\right)
\hsp
\tilde{\g}_{S}(p)=-\frac{2i\sqrt{3}\pi p}{m^{3/2}}  \sech\left(\frac{\pi p}{m}\right)\\
\tilde{\g}_k(p)&=&\frac{2k^2-m^2}{\ok{}\sqrt{m^2+4k^2}}2\pi\delta(p+k)+\frac{6\pi p}{\ok{}\sqrt{m^2+4k^2}} \csch\left(\frac{\pi (p+k)}{m}\right).\nonumber
\eea





We will decompose the tension into contributions from distinct normal modes $k_x$
\bea
\rho_1&=&\rho_{1B}+\rho_{1S}+\pin{k_x}\rho_{1k_x}\\
\rho_{1 B}&=&-\frac{1}{4}\pinv{2}{p} \gt_{B}(p_x)\gt_{B}(-p_x)
\frac{\left(|p_y|-\sqrt{m^2+p_x^2+p_y^2}\right)^2}{\sqrt{m^2+p_x^2+p_y^2}}\nonumber\\
\rho_{1 S}&=&-\frac{1}{4}\pinv{2}{p} \gt_{S}(p_x)\gt_{S}(-p_x)
\frac{\left(\sqrt{\frac{3m^2}{4}+p_y^2}-\sqrt{m^2+p_x^2+p_y^2}\right)^2}{\sqrt{m^2+p_x^2+p_y^2}}\nonumber\\
\rho_{1k_x}&=&-\frac{1}{4}\pinv{2}{p} \gt_{-k_x}(p_x)\gt_{k_x}(-p_x)
\frac{\left(\sqrt{m^2+k_x^2+p_y^2}-\sqrt{m^2+p_x^2+p_y^2}\right)^2}{\sqrt{m^2+p_x^2+p_y^2}}.\nonumber
\eea
The $p_y$ integrations may be performed analytically using the identity
\beq
\pin{p_y}\frac{(\sqrt{a+p_y^2}-\sqrt{m^2+p_x^2+p_y^2})^2}{\sqrt{m^2+p_x^2+p_y^2}}=\frac{m^2+p_x^2+a\left({\rm{ln}}\left(\frac{a}{m^2+p_x^2}\right)-1\right)}{2\pi}. \label{intid}
\eeq

In the case of the zero mode, which will turn out to be the dominant contribution, $a=0$ and one easily evaluates all integrals analytically
\bea
\rho_{1 B}&=&-\frac{1}{8\pi}\pin{p_x} \left(m^2+p_x^2\right) \gt_{B}(p_x)\gt_{B}(-p_x)\\
&=&-\frac{3\pi}{4m^3}\pin{p_x} \left(m^2+p_x^2\right) p_x^2\csch^2\left(\frac{\pi p_x}{m}\right)\nonumber\\
&=&-\frac{3}{20\pi}m^2.\nonumber
\eea
In the case of the shape mode, $a=3m^2/4$ and so
\bea
\rho_{1 S}&=&-\frac{1}{8\pi}\pin{p_x} \left(m^2+p_x^2+\frac{3m^2}{4}\left[{\rm{ln}}\left(\frac{m^2}{m^2+p_x^2}\right)+{\rm{ln}}\left(\frac{3}{4}\right)-1\right]\right) \gt_{S}(p_x)\gt_{S}(-p_x)\nonumber\\
&=&-\frac{3\pi}{2m^3}\pin{p_x} \left(m^2+p_x^2+\frac{3m^2}{4}\left[{\rm{ln}}\left(\frac{m^2}{m^2+p_x^2}\right)+{\rm{ln}}\left(\frac{3}{4}\right)-1\right]\right)p_x^2\sech^2\left(\frac{\pi p_x}{m}\right)\nonumber\\
&=&-\frac{27}{160\pi}m^2+\frac{9\pi}{8m}\pin{p_x} p_x^2\ {\rm{ln}}\left(\frac{4e(m^2+p_x^2)}{3m^2}\right)\sech^2\left(\frac{\pi p_x}{m}\right).
\eea

\begin{figure}[htbp]
\centering
\includegraphics[width = 0.65\textwidth]{rhokmarNew.pdf}
\caption{The contribution $\rho_{1k_x}$ to the one loop tension arising from each continuum normal mode $k_x$. The global minima are obtained at $k_x/m\approx \pm 0.87$ with $\rho_{1k_x}\approx -0.03$.}\label{rhokfig}
\end{figure}

In the case of the continuum modes, our identity (\ref{intid}) becomes
\beq
\pin{p_y}\frac{(\sqrt{m^2+k_x^2+p_y^2}-\sqrt{m^2+p_x^2+p_y^2})^2}{\sqrt{m^2+p_x^2+p_y^2}}=\frac{p_x^2-k_x^2-(m^2+k_x^2){\rm{ln}}\left(\frac{m^2+p_x^2}{m^2+k_x^2}\right)}{2\pi}. 
\eeq
This leaves
\bea
\rho_{1k_x}&=&-\frac{1}{8\pi}\pin{p_x} \gt_{-k_x}(p_x)\gt_{k_x}(-p_x)
\left[ p_x^2-k_x^2-(m^2+k_x^2){\rm{ln}}\left(\frac{m^2+p_x^2}{m^2+k_x^2}\right)
\right].\nonumber\\
&=&-\frac{9\pi}{2(m^2+k_x^2)(m^2+4k_x^2)}\pin{p_x} p_x^2
\csch^2\left(\frac{\pi (k_x+p_x)}{m}\right)\nonumber\\
&&\times \left[ p_x^2-k_x^2-(m^2+k_x^2){\rm{ln}}\left(\frac{m^2+p_x^2}{m^2+k_x^2}\right)
\right]
\eea
which is plotted in Fig.~\ref{rhokfig}.

Naively this looks logarithmically divergent.  At large $k_x$, the csch in $\gt_{k_x}(p_x)$ has support at $p_x=-k_x+C$ where $C$ is fixed in this limit.  Then the argument of the logarithm  in $\rho_{1k_x}$ becomes
\beq
\frac{p_x^2}{k_x^2}=1-\frac{2C}{k_x}
\eeq
whose logarithm contributes a $-2C$ to the term in square brackets, canceling the term from $p_x^2-k_x^2$.  Therefore the term in square brackets remains finite in the ultraviolet, while the $\gt^2$ prefactor scales as $1/k^2$.  As the $p_x$ integration has fixed support in the support of the csch term, the scaling is that of $\int dk 1/k^2$  which is convergent.  Thus we have not yet encountered the ultraviolet divergence in the Hamiltonian density predicted in Ref.~\cite{erice}, it will need to wait for the next order.

We note that, as in the case of the kink, the Dirac $\delta$ term in $\gt_k$ does not contribute, as it is multiplied by a double zero in the squared frequency difference.  Each factor of the Dirac $\delta$ vanishes when folded in to the corresponding zero.

Numerically we find
\beq
\rho_{1B}=-0.0477465 m^2\hsp
\rho_{1S}=-0.0072502 m^2\hsp
\pin{k_x}\rho_{1k_x}=-0.03156 m^2.
\eeq
In all $\rho_1=-0.08656m^2$.  Similarly to the case of the kink mass in 1+1 dimensions \cite{mekink}, the largest contribution arises from the zero mode, followed by the continuum modes, and last the shape mode.  However, in the case of the domain wall, the contribution from the continuum modes and zero mode differ by less than a factor of two, in contrast with the factor of eight in the case of the kink.

This tension has been computed previously in Refs.~\cite{zar98,kon89} using spectral methods.  Our result agrees with that of Ref.~\cite{zar98}, obtained using spectral methods, considering that our definitions of $m$ and $\lambda$ are twice the definitions there.  It does not agree with that of Ref.~\cite{kon89}, obtained using the zeta function regularization methods of Ref.~\cite{kon87}.  As noted in Ref.~\cite{zar98}, this discrepancy arises because the bound modes have not been included in the zeta function approach.  This is potentially dangerous, as Ref.~\cite{kon87} has enjoyed a resurgence in popularity in the last decade as it has been applied repeatedly to false vacuum decay in an interesting series of papers by the Munich \cite{mun18} and Ljubljana \cite{lj20} groups.

\subsection{Excited States}
The spectrum of excited states of the domain wall is now obvious.  Begin with the ground state $\vac_0$ which is annihilated by all $B$ operators, and the zero mode $\pi_\vk$ with $k_x=k_y=0$.  At order $O(\lambda^0)$ excited states are created by acting with $B^\ddag$.  Each $B^\ddag_\vk$ increases the energy by $\omega_\vk$.  

Unlike the case of the kink, some of these have degenerate energy while not being related by any symmetry.  For example, if $k_y\geq m$ then $B^\ddag_{B k_y}\vac_0$, which describes physical vibrations of the domain wall in the $x$ direction, has the same energy as a continuum state with the same frequency.  However the former has more $y$-momentum, which is separately conserved, and so this state does not unbind from the wall and escape into the bulk.  The same argument applies to states $B^\ddag_{S k_y}\vac_0$.  It does not apply to multiple excitations of bound states, as in some cases these have both degenerate energy and also degenerate momentum with bulk states.  However, they will mix only at order $O(\lambda)$, and so their decay to bulk modes will be slow.  This is similar to the case of the doubly-excited shape mode of the kink~\cite{alberto}.

\section{Remarks}

A word of caution is in order.  As there is no mass gap, in general one expects various infrared divergences.  In particular, states of interest will generally have infinite numbers of excitations of $B^\ddag_{Bk_y}$ with small values of $k_y$.  While these do not appear to lead to any divergences in the $O(\lambda^0)$ study presented here, infrared divergences may be expected at higher orders.  Indeed, the domain wall worldsheet theory is a 1+1 dimensional field theory with a massless scalar, which leads to various infinite matrix elements \cite{coleman}.  These will need to be treated as they arise in the computations of observables.

This is not to say that the state eliminated by all $B$ operators is not a Hamiltonian eigenstate, it is the ground state of the leading order soliton Hamiltonian $H\p_2$.  However, consider the following argument.  Eq.~(\ref{h2}) shows that each mode $k$ is described by a Harmonic oscillator with frequency squared equal to $\omega^2$, which, except in the ultrarelativistic case, is of order $O(m^2)$ in the case of the kink.  Now, consider an incoming meson.  The leading interaction $H\p_3$, in the presence of an incoming kink that is not ultrarelativistic, leads to another quadratic term in the field with coefficient of order $O(\sl m)$.  In the semiclassical approximation this is much smaller than $O(m^2)$ and so the amplitude for an interaction is small and perturbation theory is valid.  

What about the case of the domain wall?  Now, Eq.~(\ref{h2}) tells us that the frequency squared coefficient of $\phi^2_{Bk_y}$ is equal to $k_y^2$.  When $|k_y|$ is small enough, this is of the same order as the $O(\sl m)$ interaction contribution.  Therefore, the probability of the oscillator at each such $\vec{k}$ being excited is of order unity.  In general, one then expects infinitely many excitations, taking the state out of the Fock space of finite meson-number excitations.  Of course this is the usual situation in the presence of massless particles \cite{bloch37,kf}, in recent times, at least in four or more dimensions, being attributed to a memory effect \cite{scalar17,wald19,ir22}.  It is not a pathology, but rather an interesting part of the physics to which we hope to turn in the near future when we extend the present study to include interactions $H\p_3$. 

Our next step will be to proceed to the next order, where a loop diagram leads to a divergence in the meson propagator and a two-loop diagram leads to a divergence in the tension of the domain wall, just the divergence noted by Coleman.  In the vacuum sector, these divergences are well-known \cite{miro20,anand20} and they can treated with standard counterterms.  We will need to formulate the appropriate renormalization conditions in the Schrodinger picture, choose a displacement operator, and check that $H\p_1$ continues to vanish and also that the ultraviolet divergences are removed in the soliton sector.  The key step of course will be finding a displacement operator with these two properties, if it exists.

If this step is successful, then one can proceed to calculate quantum corrections to systems of phenomenological interest.  The urgent need for such corrections in the case of models of nuclei has recently been highlighted in Refs.~\cite{nakayama23,chris23,nochris23}.  Recently, there has even been progress in understanding quantum corrections to Q-balls \cite{q24}.  A successful treatment of ultraviolet divergences will also allow us to treat fermions, allowing us to study fermion-soliton scattering \cite{loginov22,gani22,loginov24}.  Besides allowing for more dimensions and richer field content, once we are able to tame ultraviolet divergences in the soliton sector, we may also approach models with nontrivial kinetic terms, such as the modified dilaton, allowing an application to the kinks in Refs.~\cite{gravkink95,zhong23,zhong24}.

\section* {Acknowledgement}

\noindent
JE is supported by NSFC MianShang grants 11875296 and 11675223.

\end{document}

\bibitem{Evslin:2022xdz}
J.~Evslin,
``Kink form factors,''
JHEP \textbf{07} (2022), 033
doi:10.1007/JHEP07(2022)033
[arXiv:2203.15445 [hep-th]].

\bibitem{Guo:2022ord}
H.~Guo,
``Leading quantum correction to the \ensuremath{\Phi}4 kink form factor,''
Phys. Rev. D \textbf{106} (2022) no.9, 096001
doi:10.1103/PhysRevD.106.096001
[arXiv:2209.03650 [hep-th]].

\bibitem{Evslin:2022wyx}
J.~Evslin and A.~Garc\'\i{}a Mart\'\i{}n-Caro,
``Spontaneous emission from excited quantum kinks,''
JHEP \textbf{12} (2022), 111
doi:10.1007/JHEP12(2022)111
[arXiv:2210.13791 [hep-th]].

\bibitem{Liu:2023dqt}
H.~Liu, J.~Evslin and B.~Zhang,
``Meson production from kink-meson scattering,''
Phys. Rev. D \textbf{107} (2023) no.2, 025012
doi:10.1103/PhysRevD.107.025012

\bibitem{Evslin:2023ypw}
J.~Evslin and H.~Liu,
``(Anti-)Stokes scattering on kinks,''
JHEP \textbf{03} (2023), 095
doi:10.1007/JHEP03(2023)095
[arXiv:2301.04099 [hep-th]].

\subsection{Motivation}
The present work is motivated by three questions.  The first concerns oscillons.  Oscillons are time-dependent classical solutions whose existence depends on the sign of a leading nonlinearity \cite{sk}.  Therefore, in a sense, they are present in half of all theories and these include many of phenomenological interest.   Oscillons are generally \cite{osc1,osc2,osc3} created by any violent event, such as a first order phase transition or soliton collision.  They are unstable, yet in general they are the longest lived unstable excitations.  As a result, at intermediate timescales after a violent event they dominate the dynamics.  

However, it is widely believed that the situation in the quantum theory is very different.  Refs.~\cite{hertzberg,tanmay} have argued that in the quantum theory oscillons experience a rapid decay into pairs of elementary quanta.  Therefore, it is believed that once quantum corrections are considered, oscillons do not dominate the dynamics at any time.  In Ref.~\cite{noiosc}, it was argued that one-loop corrections to the states might radically alter these conclusions.  Yet such corrections have never been calculated.  We aim to present a formalism which will allow the calculation of such corrections.

Second, it has long been known \cite{sug,csw,frac91} that kink-antikink collisions lead to an interesting fractal pattern of escape windows.  Needless to say, fractal patterns can be affected qualitatively by even arbitrarily small perturbations \cite{pert0,pert1,pert2,pert3}.  Thus various authors have wondered throughout the ages just what effect the leading quantum corrections have on this resonance structure, although the first serious study of this question appeared only quite recently~\cite{qpert23}.

These first two questions are rather straightforward applications.  The third is more speculative.  The Unruh effect in many ways resembles the two-particle decay claimed for oscillons in Refs.~\cite{hertzberg,tanmay}.  It is therefore interesting, in our opinion to calculate the leading quantum corrections to Rindler space states and to see how they affect the radiation spectrum.  Indeed, an intriguing paper \cite{akh21} has recently suggested that the usual choice of states, on the future wedge, leads to a secular evolution, similarly to the choice of oscillon states in the above works, suggesting that such quantum corrections indeed arise automatically in the natural course of evolution. 

\subsection{Goals}

To attack these problems, one first needs to choose a formalism for treating quantum states corresponding to classically nontrivial configurations.  Several such formalisms are available.  At the linear level there are powerful spectral methods \cite{gw22} and also the classical-quantum correspondence of Refs.~\cite{cq1,cq2}.  However, we are interested in not only linear but also the higher order nonlinear effects.  The usual tool of choice in this regime is the collective coordinate method of Ref.~\cite{gs74}.  However, after separating the collective coordinates, one needs a complicated canonical transformation to return to canonical coordinates, leading to an infinite number of terms in the Hamiltonian.  An additional infinite number of terms is required \cite{gj76} at the nonlinear level so that this transformation will leave the path integral measure invariant.  In the end, all of these terms lead to a formalism which is powerful but also cumbersome.

We therefore plan to address all of these questions using the linearized soliton perturbation theory introduced in Refs.~\cite{mekink,me2loop}.  This is a variant of the strategy in Ref.~\cite{cahill76} which avoids a notorious problem \cite{rebhan} arising from separately regularizing the vacuum and nontrivial sectors by regularizing only once, and then using a unitary transformation that is equivalent to the coherent state construction of solitons \cite{taylor78}, with the necessary \cite{cocorr23} corrections, to relate the sectors.

So far this formalism has largely been applied to states that are obtained by quantizing time-independent solutions of the classical equations of motion.  That is not to say that dynamics has not been studied, indeed kink-meson scattering amplitudes have been calculated exhaustively at the leading order \cite{memult,mestokes}, and also the elastic kink-meson scattering amplitude has recently been calculated \cite{elastico}.  However, the soliton is heavier than the perturbative degrees of freedom by a factor of the inverse coupling $1/\lambda$, and so the effect of these perturbative processes on the the soliton  is suppressed by the coupling $\lambda$.  As $\lambda\hbar$ is dimensionless, the perturbative degrees of freedom vanish in the classical limit $\hbar\rightarrow 0$.  Thus classically there was no dynamics.

For the three motivating questions, we are interested in a different regime.  We are interested in classical solutions that are themselves time-dependent, corresponding to a nonperturbative time dependence in the quantum field theory.

How can we treat a nonperturbative time-dependence using perturbation theory?  We do it in the same way as we treat a nonperturbative soliton in perturbation theory:  We use a nonperturbative unitary transformation to separate out the nonperturbative part.  We refer to the transformed states and operators as the comoving frame.  The introduction and study of this frame are the main results of this paper.

\subsection{Summary}

There is one case in which linearized soliton perturbation theory has already been applied to a classically time-dependent soliton.  This is the case of the moving kink, treated in Ref.~\cite{memove} and used to calculate form factors in Refs.~\cite{meff,hengyuan}.  This was possible, without the introduction of the comoving frame, because it is related by a Lorentz transformation to a solution with no time-dependence.   Therefore our strategy will be to use this example to learn how to construct the comoving frame in general, for solutions that may not be kinks at all.  

We begin in Sec.~\ref{classsez} with a quick review of the classical kink.  We review the quantization of the stationary kink in Sec.~\ref{quantsez}.  Then we quantize the moving kink in Sec.~\ref{movesez}.  We see the obstructions to a perturbative approach.  Therefore, in Sec.~\ref{cosez} we introduce the comoving frame, in the case of the moving kink.  Once we have understood the comoving frame in this example, the generalization to an arbitrary time-dependent solution of the classical equations of motion is clear, and so we present our general construction in Sec.~\ref{gensez}.

\section{Classical Kink} \label{classsez}

We believe that the results of this paper can be readily generalized to more phenomenologically interesting theories.  However, we will work in a 1+1 dimensional theory of a scalar $\phi(x)$ and its conjugate $\pi(x)$, described by the Hamiltonian 
\beq
H=\int d x: \mathcal{H}(x):_a, \quad \mathcal{H}(x)=\frac{\pi^2(x)}{2}+\frac{\left(\partial_x \phi(x)\right)^2}{2}+\frac{V(\sqrt{\lambda} \phi(x))}{\lambda}. \label{heq}
\eeq
Here $\lambda$ is the coupling constant, $V$ is a degenerate potential and we have included the usual normal ordering $::_a$ which acts trivially in the classical theory, but eliminates all ultraviolet divergences in the quantum theory, which we turn to in Sec.~\ref{quantsez}.  If one wishes to add more dimensions or fermions, then counterterms will be necessary to treat the corresponding divergences.  

The classical equations of motion are
\beq
\dot\phi(x,t)=\frac{\delta H}{\delta \pi(x)}=\pi(x,t)
\hsp
\dot\pi(x,t)=-\frac{\delta H}{\delta \phi(x)}=\partial_x^2\phi(x,t)-\frac{V^{(1)}(\sl\phi(x,t))}{\sl}
\eeq
where we have defined
\beq
V^{(n)}(\sqrt{\lambda} \phi(x))=\frac{\partial^n V(\sqrt{\lambda} \phi(x))}{(\partial \sqrt{\lambda} \phi(x))^n}.
\eeq
Putting these together yields the classical equation of motion
\beq
\ddot\phi(x,t)-\partial_x^2\phi(x,t)+\frac{V^{(1)}(\sl\phi(x,t))}{\sl} =0.\label{eom}
\eeq

\subsection{Stationary Classical Kink}

Consider a static kink solution
\beq
\phi(x,t)=f(x)\hsp f\pp(x)=\frac{\V1}{\sl}. \label{bps}
\eeq
The corresponding classical energy is
\beq
Q_0=\int dx \left[ \frac{f^{2}(x)}{2}+\frac{V(\sqrt{\lambda} f(x))}{\lambda}
\right]. \label{q0}
\eeq

Let us multiply the rightmost expression in Eq.~(\ref{bps}) by $f\p(x)$
\beq
\frac{1}{2}\partial_x f^{\p 2}(x)=f\pp(x)f\p(x)=\frac{\V1 f\p(x)}{\sl}=\frac{\partial_x V(\sl f(x))}{\lambda}
\eeq
and integrate, yielding
\beq
\frac{f^{2}(x)}{2}=\frac{V(\sl f(x))}{\lambda}+C
\eeq
where $C$ is a constant of integration.

Let
\beq
V(\pm\infty)=0
\eeq
then
\beq
C=0\hsp \frac{f^{2}(x)}{2}=\frac{V(\sl f(x))}{\lambda}.
\eeq
So we learn that the two terms in (\ref{q0}) are equal
\beq
\frac{Q_0}{2}=\int dx  \frac{f^{2}(x)}{2} =\int dx  \frac{V(\sqrt{\lambda} f(x))}{\lambda}.
\eeq

Consider a small perturbation
\beq
\phi(x,t)=f(x)+\g(x) e^{\pm i\omega t}.
\eeq
The linearized equation of motion is the Sturm-Liouville equation
\bea
0&=&\ddot\phi(x,t)-\partial_x^2\phi(x,t)+\frac{V^{(1)}(\sl\phi(x,t))}{\sl} \label{sl} \\
&=&\left[-\omega^2 \g(x)-\g\pp(x)+\V2 \g(x)
\right]e^{\pm i\omega t}.\nonumber
\eea
There are three kinds of solutions, classified by their frequencies $\omega$.  There is always a unique, real zero mode $\g_B(x)$ with $\omega_B=0$.  For every real $k$ there is a continuum normal mode $\g_k(x)$ with $\ok{}=\sqrt{m^2+k^2}$.  Finally there may be discrete, real shape modes $\g_S(x)$ with $0<\omega_S<m$.  We will define the meson mass $m$ momentarily.  We will fix the normalizations of the normal modes using the completeness relation
\beq
\int dx {\g}_{B}(x)^2=1,\
\int dx {\g}_{k_1} (x) {\g}^*_{k_2}(x)=2\pi \delta(k_1-k_2),\ 
\int dx {\g}_{S_1}(x){\g}_{S_2}(x)=\delta_{S_1S_2}. \label{comp}
\eeq
The sign of $\g_B(x)$ is fixed using
\beq
\g_B(x)=-\frac{f\p(x)}{\sqrt{Q_0}}. \label{gb}
\eeq

\subsection{Moving Classical Kink}
The moving kink
\beq
\phi(x,t)=f\gx\hsp \pi(x,t)=\partial_t\phi(x,t)=-v\gamma f\p\gx
\eeq
satisfies the equations of motion
\bea
\ddot\phi(x,t)-\partial_x^2\phi(x,t)+\frac{V^{(1)}(\sl\phi(x,t))}{\sl}&=&(v^2\gamma^2-\gamma^2)f\pp\gx+\frac{V^{(1)}(\sl f\gx)}{\sl}\nonumber\\
&=&-f\pp\gx+\frac{V^{(1)}(\sl f\gx}{\sl}=0.\nonumber
\eea

Its energy is
\bea
E&=&\int dx \left[ 
\frac{\pi^2(x)}{2}+\frac{\left(\partial_x \phi(x)\right)^2}{2}+\frac{V(\sqrt{\lambda} \phi(x))}{\lambda}
\right]\label{en}\\
&=&\int dx \left[ 
\frac{v^2\gamma^2 f^{2}\gx}{2}+\frac{\gamma^2 f^{2}\gx}{2}+\frac{V(\sqrt{\lambda} f\gx)}{\lambda}
\right]\nonumber\\
&=&\int \frac{dy}{\gamma} \left[ 
\frac{1+v^2}{1-v^2}\frac{f^{2}(y)}{2}+\frac{V(\sqrt{\lambda} f(y)}{\lambda}
\right]
=
\int \frac{dy}{\gamma}\frac{f^{2}(y)}{1-v^2}=Q_0\gamma.
\nonumber
\eea


\section{Quantum Kink at Rest} \label{quantsez}

The normal ordering in Eq.~(\ref{heq}) is defined at the mass of the field $\phi(x)$ near one of the minima of the potential
\beq
m^2=V^{(2)}(\sqrt{\lambda} f(\pm \infty))
.
\eeq
The values obtained at the minima on the two sides of the kink must be equal, or else the kink will begin to accelerate \cite{wstabile}.

\subsection{Ground State Kink}
The defining frame is a choice of identification of the vectors in the Hilbert space, which we denote with kets, with the states in the quantum theory.  In the defining frame, let $|K\rangle$ be vector corresponding to the ground state kink at rest, in other words it is defined to be the Hamiltonian eigenstate with the lowest energy in the kink sector, which consists of states with a single kink and a finite number of mesons.

We may rewrite this state using the unitary displacement operator $\df$
\beq
|K\rangle=\df\vac\hsp 
\df=\exp{-i\int dx f(x) \pi(x)}
\eeq
where $\vac$ is defined to be $\df^\dag|K\rangle$.  In the defining frame, $\vac$ represents a state in the vacuum sector, which consists of states with finite numbers of mesons and no kinks.

In the defining frame, the energy of a state is the corresponding eigenvalue of the Hamiltonian $H$, while time evolution is generated by the action of $H$.  The state $|K\rangle$ is defined to be a Hamiltonian eigenstate
\beq
H|K\rangle=H\df\vac=Q\df\vac=Q|K\rangle. \label{hvac}
\eeq

We define the kink frame to be a different identification of the vectors in the Hilbert space with quantum states.  In particular, a state called $|\Psi\rangle$ in the defining frame is called $\df^\dag|\Psi\rangle$ in the kink frame.  For example, the kink ground state is denoted by $|K\rangle$ in the defining frame but in the kink frame we call it $\vac$.  In summary, the vector $\vac$ represents a vacuum sector state in the defining frame but it represents a kink sector state in the kink frame. 

The transition between the frames is a passive transformation, and so it transforms not only the coordinates on the Hilbert space, but also the operators acting on those coordinates.  For example, it transforms the Hamiltonian into the kink Hamiltonian $H\p$
\beq
H\p=\df^\dag H\df. \label{hpdef}
\eeq
The ket $\vac$ is a kink Hamiltonian eigenstate
\beq
H\p\vac=Q\vac.
\eeq
Note that this follows from Eqs.~(\ref{hvac}) and (\ref{hpdef}), it is a purely algebraic identity independent of the frame which is used to identify $\vac$ with a physical state.

More generally, for any operator $\co$ acting in the defining frame, we may define an operator $\co\p$ acting in the kink frame
\beq
\co\df|\Psi\rangle=\df\co\p|\Psi\rangle\hsp \co\p=\df^\dag\co\df.
\eeq
For example, the ground state kink is translation invariant, which in the defining frame means
\beq
P\df\vac=0\hsp P=\int dx \phi(x)\partial_x \pi(x)
\eeq
and, upon multiplying by $\df^\dag$ becomes the kink frame statement
\beq
P\p\vac=0\hsp P\p=\df^\dag P\df.
\eeq
To evaluate this further, let us decompose the ground state kink rest mass $Q$ in orders of $\lambda$
\beq
Q=\sum_{i=0}^\infty Q_i
\eeq
and, following \cite{cahill76}, decompose the Schrodinger picture fields in terms of the normal modes $\g_i(x)$ that solve the Sturm-Liouville equation (\ref{sl}) with frequency $\omega_i$
\bea
\phi(x) &=&\phi_0 \mathfrak{g}_B(x)+\ppin{k} \left(B_k^{\ddag}+\frac{B_{-k}}{2 \omega_k}\right) \mathfrak{g}_k(x) \label{dec}\\
\pi(x) &=&\pi_0 \mathfrak{g}_B(x)+i \ppin{k}\left(\omega_k B_k^{\ddag}-\frac{B_{-k}}{2}\right) \mathfrak{g}_k(x) \nonumber
\eea
where we adopt the conventions
\beq
B_k^\ddag=\frac{B_k^{\dagger}}{2 \omega_k}\hsp
B_{-S}=B_S.
\eeq
The symbol $\dint$ represents the sum of an integral over continuum modes $k$ with a sum $\sum_S$ over shape modes $S$.

Then, using Eq.~(\ref{gb}), one may calculate
\beq
P\p=\sqrt{Q_0}\pi_0+P
\eeq
where the first term translates the kink center of mass and the second translates the mesons.

There is a perturbative expansion in powers of $\sl$
\beq
\vac=\sum_{i=0}^\infty\vac_i\hsp H\p=\sum_{i=0}^\infty H\p_i 
\eeq
such that
\beq
H\p_0=Q_0\hsp H\p_1=0\hsp H\p_2\vac_0=Q_1\vac_0\hsp \pi_0\vac_0=B_k\vac_0=0.
\eeq
Each successive term in $\vac$ and $H\p$ contains an additional power of $\sl$ while each term in $Q$ contains a power of $\lambda$.   In particular, $Q_0$ is of order $O(1/\lambda)$ and so
\beq
0=P\p\vac=\sqrt{Q_0}\pi_0\vac+P\vac
\eeq
implies
\beq
\pi_0\vac_1=-\frac{1}{\sqrt{Q_0}}P\vac_0.
\eeq
The left hand side is the momentum of the center of mass of the kink while the right hand side is minus the meson momentum.  Of course, these two contributions to the momentum must cancel because we are working in the center of mass frame as $P\p\vac=0$. 

\subsection{Excited Kink}

A kink may be excited by adding a meson or shape mode, which we will refer to formally using the index $k$.  In the kink frame this excited state corresponds to the ket vector $|k\rangle$ and in the defining frame to the vector $\df|k\rangle$.  We demand that it is translation invariant
\beq
P\df|k\rangle=0\hsp P\p|k\rangle=0.
\eeq
It is also required to be a Hamiltonian eigenstate
\beq
H\df|k\rangle=(Q+\ok{}+O(\lambda))\df|k\rangle\hsp H\p|k\rangle=(Q+\ok{}+O(\lambda))|k\rangle.
\eeq
At leading order, in the kink frame, the corresponding ket vector is
\beq
|k\rangle_0=\Bd{}\vac_0\hsp
|k\rangle=\sum_{i=0}^\infty|k\rangle_i
\eeq
and so in the defining frame at leading order it is $\df|k\rangle_0$.  The translation invariance condition fixes the leading correction, up to a term in the kernel of $\pi_0$, to be
\bea
\pi_0|k\rangle_1&=&-\frac{1}{\sqrt{Q_0}}P|k\rangle_0=-\frac{1}{\sqrt{Q_0}}P\Bd{}\vac_0=-\frac{1}{\sqrt{Q_0}}\left([P,\Bd{}]+\Bd{}P\right)\vac_0\\
&=&-\frac{1}{\sqrt{Q_0}}[P,\Bd{}]\vac_0+\Bd{}\pi_0\vac_1.
\eea
In other words, up to corrections of order $O(\sl)$ in the kernel of $\pi_0$ and arbitrary corrections of $O(\lambda)$
\beq
|k\rangle=\left(\Bd{}-\frac{[P,\Bd{}]}{\sqrt{Q_0}}\right)\vac.
\eeq
In particular, this approximation to $|k\rangle$ is sufficient to ensure translation invariance $P\p|k\rangle=0$ at leading order. 

To evaluate the commutator term, let us expand
\beq
\phi(x)=\ppin{k}\phi_k \g_k(x)\hsp \pi(x)=\ppin{k}\pi_k \g_k(x).
\eeq
Here, breaking with our usual notation, the $k$ index runs over zero modes as well as shape modes and continuum modes.  However, in the case of continuum and shape modes
\beq
\phi_k=\Bd{}+\frac{B_{-k}}{2\ok{}}
\hsp
\pi_k=i\ok{}\Bd{}-i\frac{B_{-k}}{2}.
\eeq
Therefore
\beq
[\phi(x),\Bd{}]=\frac{\g_{-k}(x)}{2\ok{}}\hsp
[\pi(x),\Bd{}]=-i\frac{\g_{-k}(x)}{2}.
\eeq
Assembling the pieces
\bea
[P,\Bd{}]&=&\int dx\left( [\phi(x),\Bd{}]\partial_x\pi(x)- \partial_x\phi(x) [\pi(x),\Bd{}]\right)\\
&=&\frac{1}{2}\ppin{k\p}\Delta_{-k k\p}\left(\frac{\pi_{k\p}}{\ok{}}+i\phi_{k\p}
\right)\nonumber\\
&=&\frac{1}{2}\Delta_{-k B}\left(\frac{\pi_{0}}{\ok{}}+i\phi_{0}\right)
+\frac{i}{2}\ppin{k\p}\Delta_{-k k\p}\left(\frac{2\okp{}\Bdp{}-B_{-k\p}}{2\ok{}}+\Bdp{}+\frac{B_{-k\p}}{2\okp{}}
\right)
\nonumber\\
&=&\frac{\Delta_{-k B}}{2\ok{}}\left(\pi_{0}+i\ok{}\phi_{0}\right)
+\frac{i}{4\ok{}}\ppin{k\p}\Delta_{-k k\p}\left({2(\okp{}+\ok{})\Bdp{}+\left(\frac{\ok{}}{\okp{}}-1\right)B_{-k\p}}
\right).
\nonumber
\eea

As a result of the time independence of $f(x)$, one has
\beq
 \partial_t \df =\df \partial_t.
\eeq
Therefore, time evolution of any kink frame vector $|\Psi\rangle$ can be easily derived from the defining frame time evolution equation
\beq
\partial_t\df|\Psi\rangle=-iH\df|\Psi\rangle.
\eeq
One finds
\beq
\partial_t|\Psi\rangle=\partial_t\df^\dag\df|\Psi\rangle=\df^\dag \partial_t \df|\Psi\rangle=-i\df^\dag H \df|\Psi\rangle=-iH\p|\Psi\rangle.
\eeq
So we learn that the kink Hamiltonian $H\p$ generates time evolution in the kink frame.

\section{Moving Quantum Kink} \label{movesez}

\subsection{Boost Operator}
In the defining frame, we may boost any state using the boost operator
\beq
\Lambda=-tP[\pi(x),\phi(x)]+\int dx x \ch(\pi(x),\phi(x)).
\eeq
Using
\bea
[P,\phi(x)]&=&-\int dy [\pi(y),\phi(x)]\partial_y\phi(y)=i\partial_x\phi(x)\\
\left[P,\pi(x)\right]&=&\int dy [\phi(y),\pi(x)]\partial_y\pi(y)=i\partial_x\pi(x)\nonumber
\eea
at time $t=0$ one finds
\beq
[P,\Lambda]=i\int dx x\partial_x\ch(\pi(x),\phi(x))=-iH.
\eeq
Also the Poincar\'e algebra implies
\beq
[H,\Lambda]=-iP\hsp [H,P]=0.
\eeq

In the kink frame this becomes
\beq
\Lambda\p=\df^\dag\Lambda \df=\int dx x \ch(\pi(x),\phi(x)+f(x))=\int dx x\ch\p(\pi(x),\phi(x))
\eeq
where $\ch\p$ is the density of the kink Hamiltonian.

\subsection{Moving Ground State Kink}

Consider an eigenstate of the both the Hamiltonian $H$  and also the momentum $P$.  As they mix under boosts
\beq
e^{-i\alpha\Lambda}He^{i\alpha\Lambda}=H\cosh{\alpha}+P\sinh{\alpha}\hsp
e^{-i\alpha\Lambda}Pe^{i\alpha\Lambda}=P\cosh{\alpha}+H\sinh{\alpha}
\eeq
a boost of this state will also be an eigenstate of both $H$ and $P$, but the eigenvalues will change.

\subsubsection{The Defining Frame}

Define the state 
\beq
|K\rangle^v=e^{i\alpha\Lambda}|K\rangle
\eeq
to be the kink ground state boosted to a velocity $v$ or equivalently a rapidity $\alpha=$arctanh$(v)$.  Now the eigenvalues are
\beq
H|K\rangle^v=Q\gamma |K\rangle^v\hsp P|K\rangle^v=Qv\gamma |K\rangle^v.
\eeq
As $|K\rangle^v$ is written in the defining frame, the time evolution is just
\beq
\partial_t|K\rangle^v=-iH|K\rangle^v.
\eeq

\subsubsection{An Inertial Frame}

The inertial frame is defined by the passive transformation $\df$
\beq
|K\rangle^v=\df \vac^v.
\eeq
In the inertial frame, the moving kink state is represented by an eigenvector of the kink Hamiltonian and the kink momentum
\bea
H\p\vac^v&=&H\p\df^\dag|K\rangle^v=\df^\dag H|K\rangle^v=Q\gamma\df^\dag|K\rangle^v=Q\gamma\vac^v\label{hp}\\
P\p\vac^v&=&P\p\df^\dag|K\rangle^v=\df^\dag P|K\rangle^v=Qv\gamma\df^\dag|K\rangle^v=Qv\gamma\vac^v.\nonumber
\eea

Note that $\vac^v$ is not annihilated by $P\p$, indeed it has an eigenvalue proportional to $Q$ which is of order $O(1/\lambda)$, potentially mixing various orders in perturbation theory when this condition is imposed.  

To see that this is reasonable, let us try to understand the state $\vac^v$ more explicitly.  It is
\beq
\vac^v=\df^\dag|K\rangle^v=\df^\dag e^{i\alpha\Lambda}|K\rangle=e^{i\alpha\Lambda\p}\df^\dag|K\rangle=e^{i\alpha\Lambda\p}\vac.
\eeq
We may expand $\Lambda\p$ in powers of the coupling
\bea
\Lambda\p&=&\sum_{i=0}^\infty \Lambda\p_i\hsp
\Lambda\p_0=0\\
\Lambda\p_1&=&\int dx x \left[\partial_x\phi(x)\partial_x f(x) +\V1 \phi(x) \right]\nonumber\\
&=&\int dx \phi(x)\left( x \left[-\partial^2_x f(x) +\V1  \right]-\partial_x f(x)\right)\nonumber=\sqrt{Q_0}\phi_0\nonumber\\
\Lambda\p_2&=&\frac{1}{2}\int dx x \left[\pi^2(x)+\left(\partial_x\phi(x)\right)^2+\V2 \phi^2(x)\right].\nonumber
\eea

Recall that the semiclassical limit requires $\alpha\ll 1$.  Let us therefore take the limit $\alpha\rightarrow 0$ but we do not impose that $\alpha/\sqrt{\lambda}\rightarrow 0$.  In this limit
\bea
e^{i\alpha\Lambda\p}&=&e^{i\alpha\Lambda\p_1}+i\int_0^\alpha d\beta e^{i(\alpha-\beta)\Lambda\p_1}\Lambda\p_2 e^{i\beta\Lambda\p_1}\\
&=&e^{i\alpha\sqrt{Q_0}\phi_0}+i\int_0^\alpha d\beta e^{i(\alpha-\beta)\sqrt{Q_0}\phi_0}\Lambda\p_2 e^{i\beta\sqrt{Q_0}\phi_0}.\nonumber
\eea
Using
\beq
[\pi_0,e^{i\theta\phi_0}]=\theta e^{i\theta\phi_0}\hsp
[\pi(x),e^{i\theta\phi_0}]=\theta\g_B(x) e^{i\theta\phi_0}
\eeq
one finds
\beq
e^{i(\alpha-\beta)\sqrt{Q_0}\phi_0}\pi^2(x) e^{i\beta\sqrt{Q_0}\phi_0}=e^{i\alpha\sqrt{Q_0}\phi_0}\left(\pi(x)+\beta\sqrt{Q_0}\g_B(x)\right)^2.
\eeq
The $\beta^2$ term leads to an $x$ integral proportional to $x\g_B^2(x)$ which is odd, and so it vanishes.  The linear term, when integrated over $\beta$, is proportional to $\alpha^2$, which we drop as we are working at linear order in $\alpha$.  In all, the Lorentz transformation is
\beq
e^{i\alpha\Lambda\p}=e^{-i\alpha\sqrt{Q_0}\phi_0}\left(1+i\alpha\Lambda\p_2\right). \label{lt}
\eeq

Our inertial frame ground state kink is then
\beq
\vac^v=e^{i\alpha\sqrt{Q_0}\phi_0}\left(1+i\alpha\Lambda\p_2\right)\vac.
\eeq
Now 
\bea
P\p\vac^v&=&\left(\sqrt{Q_0}\pi_0+P\right)e^{i\alpha\sqrt{Q_0}\phi_0}\left(1+i\alpha\Lambda\p_2\right)\vac\\
&=&(Q_0\alpha+P)\vac^v +\sqrt{Q_0}e^{i\alpha\sqrt{Q_0}\phi_0}\pi_0\left(1+i\alpha\Lambda\p_2\right)\vac.\nonumber
\eea
The $Q_0\alpha$ term is of order $O(1/\lambda)$ and it exactly reproduces the eigenvalue in Eq.~(\ref{hp}).  Therefore this nonvanishing eigenvalue is simply a result of the $e^{i\alpha\sqrt{Q_0}\phi_0}$ coefficient in the inertial frame state.  It can be factored out, and then the usual perturbation theory derivation of the states follows with $P\p$ annihilating the rest of the state.  This is done automatically in the comoving frame.

\section{The Comoving Frame} \label{cosez}

\subsection{Definition}

The comoving frame is defined using the passive transformation
\bea
\dft&=&\exp{-i\int dx \left(f\gx\pi(x)-\partial_t f\gx \phi(x)\right)}\\
&=&\exp{-i\int dx \left( f\gx\pi(x)+\gamma v f\p\gx \phi(x)\right)}\nonumber
\eea
which shifts $\phi(x)$ and $\pi(x)$ by the classical moving kink solution.  It is sometimes convenient to regard $t$ as a parameter, so that one comoving frame is introduced for each value of $t$.  One can use the comoving frame in which $t$ is equal to the time, but that choice is not necessary.

Consider a state that is represented by the ket $|\Psi\rangle$ in the defining frame.  In the comoving frame, this state is represented by the ket $|\Psi\rangle^{(t)}$
\beq
|\Psi\rangle^{(t)}=\dfd|\Psi\rangle.
\eeq

\subsection{The Ground State Kink}

In particular, in the comoving frame, we write the moving ground state kink as
\beq
\vac^{(t)v}=\dfd|K\rangle^v=\dfd\df\vac^v=\dfd\df e^{i\alpha\Lambda\p}\vac.
\eeq
Using the Baker-Campbell-Hausdorff formula we may evaluate
\bea
\dfd\df&=&\exp{i\int dx \left((f\gx-f(x))\pi(x)-\dot f\gx \phi(x)\right)}\nonumber\\
&&\times\exp{-i\int dx \frac{\dot f\gx f(x)}{2}}. \label{dfdf}
\eea
Note that the last factor is an overall constant phase.  This may be simplified using
\beq
-i\int dx \dot f\gx = iv\gamma\int dx f\p\gx=-iv\gamma \sqrt{Q_0}\phi_0=-i\sinh{\alpha}\sqrt{Q_0}\phi_0.
\eeq
Thus the $\phi(x)$ term in (\ref{dfdf}) cancels the $e^{i\alpha\Lambda\p_1}=e^{-i\alpha\sqrt{Q_0}\phi_0}$ term in $e^{i\alpha\Lambda\p}$, reported in Eq.~(\ref{lt}), up to corrections of order $\alpha^3\sqrt{Q_0}$, which we drop as higher powers of $\alpha$ arise at higher orders of perturbation theory.

Assembling these results, we find
\beq
\dfd\df e^{i\alpha\Lambda\p}=\exp{i\int dx \left(\left[f\gx-f(x)\right]\pi(x)-\frac{\dot f\gx f(x)}{2}\right)}
\left(1+i\alpha\Lambda\p_2\right) \label{otrans}
\eeq
and so in the comoving frame, the kink ground state corresponds to the ket
\beq
\vac^{(t)v}=\exp{i\int dx \left(\left[f\gx-f(x)\right]\pi(x)-\frac{\dot f\gx f(x)}{2}\right)}
\left(1+i\alpha\Lambda\p_2\right)\vac.
\eeq
We see that all that remains at order $O(1/\lambda)$ is a constant phase, which will be inconsequential and we will drop it from now on.  

At order $O(1/\sl)$ one finds a $\pi(x)$ tadpole term that Lorentz contracts and moves the boosted kink.  At time $t=0$ it does not move the boosted kink.  We are free to choose any time to be $t=0$, as this choice simply corresponds to the choice of kink position in $\df$ which does not affect the state.  The contraction appears only at order $O(v^2/\sl)$.

\subsection{Evolution}

Evolution in the defining frame is generated by $H$.  We will now use this fact to learn how to evolve kets in the comoving frame. 

Evolution in the comoving frame is described by
\beq
\partial_t\vac^{(t)v}=\partial_t\dfd|K\rangle^v=[\partial_t,\dfd]|K\rangle^v+\dfd\partial_t|K\rangle^v. \label{com}
\eeq

The second term is easily evaluated, as it corresponds to evolution in the defining frame
\beq
\dfd\partial_t|K\rangle^v=-i\dfd H|K\rangle^v=-i \dfd H\dft\vac^{(t)v}.
\eeq
To evaluate the first term, use 
\beq
[\partial_t,f\pi-\dot f\phi]=\dot f\pi-\ddot f\phi
\eeq
to find
\bea
[\partial_t,e^{i\int dx \left(f\pi-\dot f\phi\right)}]&=&\sum_{n=0}^\infty \frac{(-i)^n}{n!}\left[
\partial_t,
\left(\int dx f\pi-\dot f\phi\right)^n
\right]\\
&&\hspace{-2cm}=\sum_{n=1}^\infty \frac{i^n}{n!}\sum_{m=1}^n 
\left(\int dx f\pi-\dot f\phi\right)^{m-1}
\left(\int dx \dot f\pi-\ddot f\phi\right)
\left(\int dx f\pi-\dot f\phi\right)^{n-m}\nonumber\\
&&\hspace{-2cm}=\left(\int dx \dot f\pi-\ddot f\phi\right)
\sum_{n=1}^\infty \frac{i^n}{n!}\sum_{m=1}^n 
\left(\int dx f\pi-\dot f\phi\right)^{n-1}
\nonumber\\
&&\hspace{-2cm}\ \ +\sum_{n=1}^\infty \frac{i^n}{n!}\sum_{m=1}^n 
\left[\left(\int dx f\pi-\dot f\phi\right)^{m-1}
,\left(\int dx \dot f\pi-\ddot f\phi\right)\right]
\left(\int dx f\pi-\dot f\phi\right)^{n-m}\nonumber\\
&&\hspace{-2cm}=\left(\int dx \dot f\pi-\ddot f\phi\right)
\sum_{n=1}^\infty \frac{i^n}{(n-1)!}
\left(\int dx f\pi-\dot f\phi\right)^{n-1}
\nonumber\\
&&\hspace{-2cm}\ \ +\sum_{n=1}^\infty \frac{i^n}{n!}\sum_{m=1}^n 
(m-1)\int dx(if\ddot f-i\dot f^2)
\left(\int dx f\pi-\dot f\phi\right)^{m-2}\left(\int dx f\pi-\dot f\phi\right)^{n-m}
\nonumber\\
&&\hspace{-2cm}=i\int dx (\dot f\pi-\ddot f\phi)e^{-i\int dx f\pi-\dot f \phi}-\sum_{n=2}^\infty \frac{i^{n-1}}{(n-2)!}
\left(\dot f\phi-\int dx f\pi\right)^{n-2}
\int dx\frac{f\ddot f-\dot f^2}{2}
\nonumber\\
&&\hspace{-2cm}=-i \int dx\left(\dot f\pi-\ddot f\phi +\frac{f\ddot f-\dot f^2}{2}\right)\dft.\nonumber
\eea

Combining these terms one finds
\beq
\partial_t\vac^{(t)v}=-i\hp\vac^{(t)v}
\eeq
where the comoving Hamiltonian is defined to be
\bea
\hp&=& \dfd H\dft+\int dx\left(\ddot f\gx\phi(x)-\dot f\gx\pi(x)\right.\\
&&\left.+\frac{f\gx\ddot f\gx-\dot f^2\gx}{2}\right).\nonumber
\eea
This comoving Hamiltonian evolves the state while changing the comoving frame parameter $t$ so that it equals the time.  

Again we decompose it in powers $\hp_i$ of the coupling and evaluate it
\bea
\hp[\phi(x),\pi(x)]&=&H[\phi(x)+f\gx,\pi(x)+\dot f\gx]+\int dx\left(
\ddot f\gx\phi(x)
\right.\nonumber\\
&&\left.-\dot f\gx\pi(x)+\frac{f\gx\ddot f\gx-\dot f^2\gx}{2}\right)\nonumber\\
&=&\sum_{i=0}^\infty \hp_i.
\eea
The first term is
\beq
\hp_0=\int dx\left[\frac{f\gx\ddot f\gx+\left(\partial_x f\gx\right)^2}{2}+V(\sl f\gx)\right].
\eeq
Unlike Eq.~(\ref{en}), the result is not $Q_0\gamma$, as a result of the $f\ddot f-\dot f^2$ term which negates the kinetic energy.  While the commutator term in (\ref{com}) makes $\hp_0$ more complicated, it also exactly cancels the tadpole in $\dfd H\dft$
\beq
\hp_1=0.
\eeq
It does not contain terms at higher orders, so these are given by $\dfd H\dft$.  For example, leaving the normal ordering implicit,
\beq
\hp_2=\frac{1}{2}\int dx\left[ 
\pi^2(x)+(\partial_x\phi(x))^2+V^{(2)}(\sl f\gx)\phi^2(x)\right].
\eeq

\subsection{Operators}

Refs.~\cite{cahill76,me2loop} construct the spectrum of a stationary kink in the kink frame using kink frame operators.  In this subsection, we will see how to map from these operators to the comoving frame of a moving kink, so that the old construction may be imported to the case of a moving kink.

Consider an operator
\beq
\co\p=\df^\dag\co\df
\eeq
acting on a stationary kink state $|\Psi\rangle$\ in the kink frame.  How does it act on the corresponding boosted kink in the comoving frame?  

After the action of the operator, in the kink frame the state is represented by the vector
\beq
\co\p|\Psi\rangle=\df^\dag\co\df|\Psi\rangle
\eeq
which, in the defining frame, corresponds to
\beq
\co\df|\Psi\rangle.
\eeq
Now we may boost it, yielding a new state which in the defining frame is
\beq
e^{i\alpha\Lambda}\co\df|\Psi\rangle. \label{act}
\eeq
In the comoving frame, before acting with the operator the boosted state was
\beq
|\Psi\rangle^{v(t)}=\dfd e^{i\alpha\Lambda}\df|\Psi\rangle=\dfd\df e^{i\alpha\Lambda\p}|\Psi\rangle.
\eeq
Acting with the operator it becomes the state (\ref{act}) which in the comoving frame is represented by
\beq
\dfd e^{i\alpha\Lambda}\co\df|\Psi\rangle=
\dfd\df e^{i\alpha\Lambda\p}\co\p|\Psi\rangle=\co^{\prime v (t)}|\Psi\rangle^{v(t)}
\eeq
where
\beq
\co^{\prime v (t)}=\dfd\df e^{i\alpha\Lambda\p}\co\p e^{-i\alpha\Lambda\p} \df^\dag\dft.
\eeq
Using Eq.~(\ref{otrans}) this yields our master formula
\beq
\co^{\prime v (t)}=e^{i\int dx \left(f\gx-f(x)\right)\pi(x)}
\left(\co\p+i\alpha[\Lambda\p_2,\co\p]\right)e^{-i\int dx \left(f\gx-f(x)\right)\pi(x)}. \label{padrone}
\eeq
For any operator $\co\p$ acting on a stationary kink in the kink frame, we have found the corresponding operator $\co^{\prime v (t)}$ that acts on the moving kink in the comoving frame.  This formula can be used to migrate all results concerning a stationary kink to a moving kink.  For example, we know how to build the spectrum of a stationary kink using the operators $B^\ddag$, $B$, $\phi_0$ and $\pi_0$ in the kink frame, and so this map yields the corresponding construction for a moving kink in the comoving frame.

Recall that, choosing $t=0$, the Lorentz-contracting $\pi(x)$ terms are of order $O(v^2/\sl)$ and so may be ignored at lower orders.

\subsection{Fields}
Let us apply the transformation (\ref{padrone}) to the scalar field $\co\p=\phi(x)$ acting in the kink frame of the stationary kink.  Dropping the contraction terms, the action of the field $\phi(x)$ in the comoving frame of the moving kink corresponds to that of
\beq
\hat\phi(x)=\phi(x)-[i\alpha\Lambda\p_2,\phi(x)]=\phi(x)-\alpha x\pi(x)
\eeq
in the kink frame.  In other words, the operator $\hat\phi(x)$ acting on a state of a stationary kink in the kink frame changes it to the same state, up to a boost, as one would get by acting on the moving kink with $\phi(x)$ in the comoving frame.  Note that $\phi(x)=\hat\phip(x)$.

Similarly, acting with the operator $\pi(x)$ in the comoving frame has the same effect on the state as acting on the stationary kink with
\bea
\hat\pi(x)&=&\pi(x)-[i\alpha\Lambda\p_2,\pi(x)]=\pi(x)+\alpha \left(-\partial_x^2+V^{(2)}(\sl f\gx) 
\right)(x\phi(x))\nonumber\\
&=&\pi(x)+\alpha \ppin{k}\phi_k\left(-\partial_x^2+V^{(2)}(\sl f\gx) \right)(x\g_k(x))
\eea
in the kink frame.  Again dropping terms of order $O(v^2/\sl)$, the Lorentz contraction on the potential term can be dropped and at time $t=0$ we find the Sturm-Liouville operator acting on the normal modes, and so
\beq
\hat\pi(x)=\pi(x)+\alpha
\left(-\partial_x\phi(x)+x\ppin{k}\ok{}^2\g_k(x)\phi_k\right).
\eeq

These equations took a long time to derive, but in the Lagrangian formulation they resemble the usual Lorentz transformation where $\pi(x,t)=\dot\phi(x,t)$ and the Heisenberg equations of motion are satisfied for the comoving kink Hamiltonian.

Recall that in the kink frame of the stationary kink, the operators $B^\ddag$ and $B$ are creation and annihilation operators for mesons.  The operators $B^{\ddag\prime}$ and $B\p$ that play the same role in the comoving frame of the moving kink are more complicated to write explicitly, as one needs to add a commutator with respect to $\Lambda\p_2$.   

Nonetheless, we may decompose our operators $\phi(x)$ and $\pi(x)$ into $B^{\ddag\p}$ and $B\p$ to learn how the fields are related to the operators that create and destroy mesons in the comoving frame.  To do this, one recalls that $\phi(x)$ and $\pi(x)$ in the comoving frame act identically to $\hat\phi(x)$ and $\hat\pi(x)$ in the kink frame, which we may write in terms of $\phi(x)$ and $\pi(x)$ using the results above, and then decompose into $B^\ddag$ and $B$ as usual.  Then, by definition, the action in the comoving frame is given by replacing these with $B^{\ddag\p}$ and $B\p$.  In other words
\bea
\phi(x)&=&\phip(x)-\alpha x \pip(x)
\label{deco}\\
&=&(\phip_0-\alpha x\pip_0)\g_B(x)+\ppin{k}\g_k(x)\left[ 
\left(1-i\alpha x\ok{} \right)B^{\ddag\prime}{}
+\left(1+i\alpha x\ok{}  \right) \frac{B\p_{-k}}{2\ok{}}
\right]\nonumber\\
\pi(x)&=&\pip(x)+
\alpha\left(-\partial_x\phip(x)+x\ppin{k}\ok{}^2\g_k(x)\phip_k\right)\nonumber\\
&=&\pip_0\g_B(x)-\alpha\phip_0\partial_x\g_B(x)+\ppin{k}\left[-\alpha\partial_x\g_k(x)\left(B^{\ddag\prime}_k+ \frac{B\p_{-k}}{2\ok{}}
\right)\right.\nonumber\\
&&\left.
+\ok{}\g_k(x)\left[ 
\left(i+\alpha x\ok{}\right)B^{\ddag\prime}_k{}
+\left(-i+\alpha x \ok{}\right) \frac{B\p_{-k}}{2\ok{}}
\right]\right].
\nonumber
\eea
Notice that the expansion in the rapidity $\alpha$ is in fact an expansion in $\alpha x\ok{}$.  Let $\ok{}$ be of order $O(m)$, corresponding to mesons that are relativistic but not ultrarelativistic.  Physical effects such as loop corrections to the kink mass are indeed dominated by this range of $k$.  Then this approximation is only valid at $|x|\ll 1/(\alpha m)$.  The width of the classical kink is $1/m$, and so the approximation is valid over a range of $1/\alpha$ classical kink widths.  Recall that we are interested in nonrelativistic kinks, as perturbation theory is only expected to apply in this case, and so $\alpha\rightarrow 0$.  Therefore this maximum distance can be larger than many other characteristic scales in the problem. 

\subsection{The Main Lesson from this Example}

We are interested in generalizing these results to other time-dependent solutions.  The key to this generalization will be the following observation.  The decomposition (\ref{deco}) can be derived as follows.  One starts with the classical field theory solution for a perturbation of a stationary kink in the inertial frame $\phi(x,t)\rightarrow \phi(x,t)-f(x)$
\beq
\phi(x,t)=\phi_0 \g_B(x) +\pi_0 t\g_B(x)+ \ppin{k}\g_k(x)\left(\Bd{}e^{i\ok{} t}+\frac{B_{-k}}{2\ok{}}e^{-i\ok{} t}\right).
\eeq
Here $\phi_0,\ \pi_0,\ B^\ddag$\ and $B$ are $c$-number parameters defining the solution.  

Then Lorentz boost, yielding a new solution to the classical equations of motion (\ref{eom}), where we separate the perturbations from the boosted kink using the transformation $\phi(x,t)\rightarrow \phi(x,t)-f(\gamma(x-vt))$
\bea
\phi(x,t)&=&\phi_0 \g_B\gx +\pi_0 (\gamma(t-vx))\g_B\gx\label{clas}\\
&&+ \ppin{k}\g_k\gx\left(\Bd{}e^{i\ok{} (\gamma(t-vx))}+\frac{B_{-k}}{2\ok{}}e^{-i\ok{} (\gamma(t-vx))}\right).\nonumber
\eea
The decompositions of the quantum fields $\phi(x)$ and $\pi(x)$ follow from the rule
\beq
\phi(x)=\phi(x,0)\hsp \pi(x)=\dot\phi(x,0)
\eeq
applied to the classical solution (\ref{clas}).  In particular, one finds Eq.~(\ref{deco}) expanding to linear order in $v$.

\subsection{The Ground State Revisited}

In the kink frame of a ground state kink, the leading approximation $\vac_0$ to the ground state $\vac$ is annihilated by $\pi_0$ and the operators $B_k$.  In the comoving frame of a moving kink, these operators correspond to 
\beq
\pip_0=\int dx \pip(x)\g_B(x)=\pi_0-\alpha\int dx \partial_x\phi(x)\g_B(x)=\pi_0-\alpha\ppin{k}\Delta_{Bk}\phi_k
\eeq
and
\bea
B\p_{-k}&=&\int dx\g^*_{-k}(x)\left(\ok{}\phip(x)+i\pip(x)\right)\\
&=&B_{-k}+\alpha\int dx \g^*_{-k}(x) \left(x (\ok{}\pi(x)+i\ppin{k\p}\okp{}^2\g_{k\p}(x)\phi_{k\p})-i\partial_x\phi(x)\right)\nonumber\\
&=&B_{-k}+\alpha\int dx \g^*_{-k}(x) \ppin{k\p} \left[\pi_{k\p}x\ok{}+i\phi_{k\p}(x\okp{}^2-\partial_x)\right] \g_{k\p}(x)\nonumber\\
&=&B_{-k}+i\alpha\int dx \g^*_{-k}(x) \ppin{k\p} \left[\left(\okp{}\Bdp{}-\frac{B_{-k\p}}{2}\right)x\ok{}+\left(\Bdp{}+\frac{B_{-k\p}}{2\okp{}}
\right)(x\okp{}^2-\partial_x)\right] \g_{k\p}(x)\nonumber\\
&=&B_{-k}+i\alpha\int dx \g^*_{-k}(x) \ppin{k\p} \left[\Bdp{}\left(x\okp{}(\okp{}+\ok{})-\partial_x \right)+\frac{B_{-k\p}}{2}\left(x(\okp{}-\ok{})-\frac{\partial_x}{\okp{}}
\right)
\right] \g_{k\p}(x)\nonumber\\
&=&B_{-k}+i\alpha \ppin{k\p} \left[\Bdp{}\left(\hat\Delta_{k k\p}\okp{}(\okp{}+\ok{})-\Delta_{k k\p} \right)+\frac{B_{-k\p}}{2}\left(\hat\Delta_{k k\p}(\okp{}-\ok{})-\frac{\Delta_{k k\p}}{\okp{}}
\right)
\right]\nonumber
\eea
where
\beq
\hat\Delta_{kk\p}=\int dx x\g_k(x)\g_{k\p}(x).
\eeq
One can check that $\pip_0$ and $B\p_{-k}$ indeed annihilate the leading order comoving frame ground state $\left(1+i\alpha\Lambda\p_2\right)\vac_0$.

\subsection{General States}

There are two kinds of questions that one may ask.  The first is the initial value problem, in which one starts with an initial state at time $t=0$ and asks what state arises at time $t$.  This can, in principle, be solved in the comoving frame as time evolution is generated by $\hp$ which has been constructed above.

One may also ask the spectrum.  There are several distinct questions here.  The eigenstates of $\hp$ in the comoving frame are time-independent in the comoving frame, in the sense that the ket that describes the state in the comoving frame is the same at all times.  However, although the ket itself is time-independent, the state to which it corresponds depends on $t$.  Correspondingly, the defining frame kets are time-dependent.  The operator $\hp$, except for scalar terms, begins at order $O(\lambda^0)$ and so in principle this problem can be treated perturbatively.

On the other hand, one may ask about states that are truly time-independent.  These are states that are annihilated by the Hamiltonian $H$ in the defining frame, and so by $\dfd H\dft$ in the comoving frame.  The operator $\dfd H\dft$ has tadpole terms of order $O(1/\sl)$
\beq
\dfd H\dft=\int dx \left( \dot f\gx \pi(x) -\ddot f\gx \phi(x)\right)+O(\lambda^0).
\eeq
In the case of the moving kink, the $\ddot f$ term is of order $O(v^2/\sl)$ and so, if $v^2\ll O(\sl)$, it can be treated perturbatively.  

Dropping it, we find
\beq
\dfd H\dft=v \gamma \sqrt{Q_0 } \pi_0 +O(\lambda^0).
\eeq
One may recognize this tadpole as the part of the momentum operator $P\p$ corresponding to the kink center of mass, which was mixed with the Hamiltonian by the boost.  In particular, {\it{if the state was translation-invariant before the boost}}, so that it was annihilated by $P\p$ up to terms of order unity, then after the boost this term will be of order $O(1)$ because $\sqrt{Q_0}\pi_0$ on such a state is $P$  on the state, and $P$ is of order $O(1)$.  Therefore, for such states, there are no terms in $\dfd H\dft$ of order the inverse coupling and we conclude that the spectrum of translation-invariant states can be found perturbatively.

Physically this is reasonable.  Before the boost, translation-invariance means that one starts in a flat superposition of states with different kink positions $x_0$.  After the boost, evolving in time, the kink moves, and so $x_0$ shifts.  However, if the initial wave function of $x_0$ was constant, so that all kink positions were weighted equally, then this shift leaves the state invariant, and so the boosted state remains time-independent.  Of course, these are the kinds of states that we focused upon and often constructed in previous papers.
  
More generally, in the case of time-dependent solitons that are not simply moving but also have some internal dynamics, this tadpole term always appears, reflecting the fact that the quantum state changes as a result of the classical evolution.  But also in this general case, we expect that this evolution is trivial if the quantum state is a flat superposition over the entire classical trajectory.  This will be manifested by the fact that the tadpole term will annihilate the state, and so one need only solve the eigenvalue equation for the perturbative part of $\dfd H\dft$.  For example, in the case of breathers and oscillons, we expect to be able to obtain the spectra perturbatively if we restrict our attention to states that are flat quantum superpositions over the coherent states corresponding to each classical configurations that appears over a period of their evolution.

\section{A General Time-Dependent Solution} \label{gensez}

\subsection{Perturbations}

Consider a time-dependent classical solution $\phi(x,t)=f(x,t)$ of Eq.~(\ref{eom}).  Now let us consider small perturbations $\g(x,t)$
\beq
\phi(x,t)=f(x,t)+\g(x,t).
\eeq
The classical equations of motion are
\beq
0=\ddot\g(x,t)-\partial_x^2\g(x,t)+\frac{V^{(1)}(\sl(f(x,t)+\g(x,t)))-V^{(1)}(\sl(f(x,t)))}{\sl}.
\eeq
At linear order in $\g$ this becomes
\beq
0=\ddot\g(x,t)-\partial_x^2\g(x,t)+V^{(2)}(\sl(f(x,t)))\g(x,t). \label{geq}
\eeq

We are not always interested in Lorentz-invariant theories.  For example, kink-impurity scattering is phenomenologically rich and tractable with our methods, and yet the impurity necessarily breaks spatial translation symmetry and may even break time translation symmetry.  

However, in the case of a Lorentz-invariant theory, two solutions $\g$ are always present.  These are the spatial and temporal zero modes
\beq
\g_B(x,t)=c_B \partial_x f(x,t)\hsp \g_T(x,t)=c_T \partial_t f(x,t)
\eeq
corresponding to a spatial translation of the solution or a displacement of the solution along its trajectory.  Here $c_B$ and $c_T$ are normalization constants which may, at this point, be chosen freely.

The semiclassical treatment of solitons only is a reasonable approximation for heavy solitons, whose mass is much greater than any momentum scale.  They are necessarily nonrelativistic, but in a Lorentz-invariant theory there will be solutions related by Galilean invariance.  These correspond to the solutions
\beq
\g_V(x,t)=c_V t \g_B(x,t).
\eeq
Formally one may worry that the linear expansion may not be trusted at large $|t|$.  However, for the purpose of quantizing a soliton, $t$ can be chosen freely, as a shift in $t$ corresponds to a shift in the choice of $f(x,t)$ in the moduli space of solutions.

In addition to these three solutions, consider a set of solutions $\g_k(x,t)$ where the index $k$ will in general contain a continuum corresponding to real values of $k$ and also discrete shape modes $S$.  This set of solutions must be linearly independent, but also must span the possible initial perturbations.   

More precisely, consider the pairs
\beq
(\g_T(x,0),\dot\g_T(x,0)),\ (\g_B(x,0),\dot\g_B(x,0)),\ (\g_V(x,0),\dot\g_V(x,0)),\ (\g_k(x,0),\dot\g_k(x,0)).
\eeq
We demand that these pairs are linearly independent and also span the space of pairs of functions $(G(x,0),\dot G(x,0))$. Note that $\g_V(x,0)=0$ but $\dot\g_V(x,0)$ is nonzero and so is not an obstruction to this condition.  In the case of a time-independent solution, $\g_T$ vanishes, but we are not interested in that case as it has been covered in the literature.

In practice, one needs to simply search for solutions to (\ref{geq}) until these span the space of functions, throwing away solutions with linearly dependent initial data.

\subsection{The Comoving Frame}

The displacement operator is defined to be
\beq
\dft=\exp{-i\int dx \left(f(x,t)\pi(x)-\partial_t f(x,t) \phi(x)\right)}.
\eeq
The state $|\Psi\rangle^{(t)}$ in the comoving frame is defined to be the same state as $\dfd|\Psi\rangle$ in the defining frame.

Now the arguments in Sec.~\ref{cosez} may be imported, as the derivations are generally identical.  In particular, time evolution is generated by the comoving Hamiltonian
\bea
\hp&=& \dfd H\dft+\int dx\left(\ddot f(x,t)\phi(x)-\dot f(x,t)\pi(x)
+\frac{f(x,t)\ddot f(x,t)-\dot f^2(x,t)}{2}\right)\nonumber
\eea
while time-independent states are eigenstates of the inertial Hamiltonian $\dfd H\dft$.  In particular
\beq
\hp_1=0\hsp \hp_2=\frac{1}{2}\int dx\left(\pi^2(x)+(\partial_x\phi(x))^2+V^{(2)}(\sl f(x,t))\phi^2(x)\right). \label{hpe}
\eeq

Note that $\g_B,\ \g_T$\ and $\g_V$ are real, while the space of $\g_k$ is invariant under complex conjugation.  

\subsection{The Decomposition}

Following the example of the moving kink, for each value of the parameter $t$ we introduce the decomposition
\bea
\phi(x)&=&\g_B(x,0)\phi_0+\g_T(x,0)\phi_T+\ppin{k}\left(\g_k(x,0)\Bd{}+\g^*_k(x,0)\frac{B_{k}}{2\ok{}}\right)\nonumber\\
\pi(x)&=&\dot\g_B(x,0)\phi_0+\dot\g_T(x,0)\phi_T+\g_B(x,0)\pi_0\nonumber
+\ppin{k}\left(\dot\g_k(x,0)\Bd{}+\dot\g^*_k(x,0)\frac{B_{k}}{2\ok{}}\right).\nonumber
\eea

Using Eq.~(\ref{hpe}) one finds
\beq
[\hp_2,\phi(x)]=-i\pi(x)\hsp
[\hp_2,\pi(x)]=i\left(-\partial_x^2+V^{(2)}(\sl f(x,t))\right)\phi(x).
\eeq
Using the $\g$ equation of motion (\ref{geq}) the latter can be simplified
\beq
[\hp_2,\pi(x)]=-i\left( \ddot\g_B(x,0)\phi_0+\ddot\g_T(x,0)\phi_T+\ppin{k}\left(\ddot\g_k(x,0)\Bd{}+\ddot\g^*_k(x,0)\frac{B_{k}}{2\ok{}}\right)
\right).
\eeq
Iterating, it is not hard to see that
\bea
e^{i\hp_2 t}\phi(x) e^{-i\hp_2 t}&=&t\g_B(x,t)\pi_0+\g_B(x,t)\phi_0+\g_T(x,t)\phi_T+t\g_V(x,t)\pi_0\\
&&+\ppin{k}\left(\g_k(x,t)\Bd{}+\g^*_k(x,t)\frac{B_{k}}{2\ok{}}\right)\nonumber\\
e^{i\hp_2 t}\pi(x) e^{-i\hp_2 t}&=&\left(\g_B(x,t)+t\dot\g_B(x,t)\right)\pi_0+\dot\g_B(x,t)\phi_0+\dot\g_T(x,t)\phi_T+t\g_V(x,t)\pi_0\nonumber\\
&&+\ppin{k}\left(\dot\g_k(x,t)\Bd{}+\dot\g^*_k(x,t)\frac{B_{k}}{2\ok{}}\right).\nonumber
\eea
Or equivalently one can introduce a family of decompositions
\bea
\phi(x)&=&t\g_B(x,-t)\pi^{(t)}_0+\g_B(x,-t)\phi^{(t)}_0+\g_T(x,-t)\phi^{(t)}_T+t\g_V(x,-t)\pi^{(t)}_0\label{dect}\\
&&+\ppin{k}\left(\g_k(x,-t)\Btd{}+\g^*_k(x,-t)\frac{\Bt{}}{2\ok{}}\right)\nonumber\\
\pi(x)&=&\left(\g_B(x,-t)+t\dot\g_B(x,-t)\right)\pi^{(t)}_0+\dot\g_B(x,-t)\phi^{(t)}_0+\dot\g_T(x,-t)\phi^{(t)}_T+t\g_V(x,-t)\pi^{(t)}_0\nonumber\\
&&+\ppin{k}\left(\dot\g_k(x,-t)\Btd{}+\dot\g^*_k(x,-t)\frac{\Bt{}}{2\ok{}}\right)\nonumber
\eea
parametrized by $t$, where
\beq
\co^{(t)}=e^{i\hp_2 t}\co e^{-i\hp_2 t}. \label{cot}
\eeq
Note that Eq.~(\ref{cot}) implies that the algebra satisfied by these operators is independent of $t$
\beq
\left[\co^{(t_1)}_1,\co^{(t_1)}_2\right]=\
\left[\co^{(t_2)}_1,\co^{(t_2)}_2\right]=\
\left[\co_1,\co_2\right].
\eeq

Now clearly acting on a state with the operator $\co$ at time $t=0$ is equivalent to acting on it with $\co^{(t)}$ at time $t$, up to corrections of order $O(\sl)$.  Therefore we may use the algebra of the operators $\co$, that is to say the operators $\pi_0,\ \phi_0,\ \phi_t,\ B^\ddag_k$\ and $B_k$, to construct our comoving frame, one-soliton sector states at time $t=0$.  For example, we may impose that a state of interest is annihilated by some $\co_1$ and the excited state is created by acting with $\co_2$.   Then, at time $t$, our state will be annihilated by $\co^{(t)}_1$ and the excited state will be created by acting with $\co^{(t)}_2$.  This much is obvious, but then we may use the invertible expansions (\ref{dect}) to go between the $\co^{(t)}$ operators and the fields $\phi(x)$ and $\pi(x)$ at any time $t$, which allows us to use perturbation theory to study the static and dynamic properties of these states at any time.

\section{Remarks}

If the solution $f(x,t)$ is periodic then the parameter $t$ labeling the comoving frames is also periodic.  Therefore, after one period, one returns to the original frame.  If the original state was defined in the comoving frame by being annihilated by some set of operators $\co_i$, then after one period it will still be annihilated by those operators and so will still be in the same state.  In particular, it will not have decayed.  Thus, if such a state is sensible, for example if it does not poses classically unstable modes which would lead its perturbative expansion to be ill-defined, then it is stable at order $O(\lambda^0)$, in other words over timescales of order $O(1/m)$.  In this case one would have disproved the claim that oscillons decay already at the linear level in Refs.~\cite{hertzberg,tanmay}.  Indeed, this suggests more generally that the classical stability of a periodic classical solution generically implies an absence of decays via the emission of pairs of mesons.

So far we have only worked out one rather trivial example, the moving kink.  However, with this formalism developed, we hope to turn to more interesting applications.  We intend to approach quantum oscillons, estimating their lifetime, and also Rindler space, to see how quantum corrections affect the incoming radiation, trying to understand the remarkable yet perplexing results of Refs.~\cite{akh15,akh21,akh22}.

\section* {Acknowledgement}

\noindent
JE is supported by NSFC MianShang grants 11875296 and 11675223.

\end{document}

\section{Quantum Kink}

\subsection{The Kink Hamiltonian}

\bea
\df^{(t)}&=&\exp{-i\int dx \left(f\gx\pi(x)-\partial_t f\gx \phi(x)\right)}\\
&=&\exp{-i\int dx \left( f\gx\pi(x)+\gamma v f\p\gx \phi(x)\right)}
\eea

In the defining frame
\beq
i\partial_t|\psi\rangle=H|\psi\rangle= H \dft \dfd |\psi\rangle
\eeq
multiply by $\dfd$
\beq
\left(\dfd H\dft \right)\dfd |\psi\rangle=i\dfd \partial_t\dft\dfd|\psi\rangle=i\left(\partial_t+\dfd[\partial_t,\dft]
\right)\dfd|\psi\rangle
\eeq
Therefore the evolution of the kink state $\df^{(t)\dag}|\psi\rangle$ is given by
\beq
\partial_t \df^{(t)\dag}|\psi\rangle -i\left(\df^{(t)\dag} H\df -i\df^{(t)\dag}[\partial_t,\df^{(t)}]\right) \df^{(t)\dag}|\psi\rangle 
\eeq
We have shown that time evolution of kink states is generated by the operator
\beq
H^{(t)\prime}=\df^{(t)\dag} (-i\partial_t+H)\df^{(t)}=\sum_{i=0}^\infty H^{(t)\prime}_i \hsp H^{(t)\prime}_i=\int dx :\ch^{(t)\prime}_i:_a
\eeq
It is easily evaluated
\beq
-i\df^{(t)\dag} \partial_t\df^{(t)}=\int dx\left[ v\gamma f\p\gx \pi(x) +v^2\gamma^2 f\pp\gx\phi(x)\right]
\eeq

\bea
\df^{(t)\dag}\mathcal{H}(x)\df^{(t)}&=&\frac{\left(\pi(x)-\gamma vf\p\gx \right)^2}{2}+\frac{\left(\partial_x \left(\phi(x)+f\gx\right)\right)^2}{2}\nonumber\\
&+&\frac{V(\sqrt{\lambda} \left(\phi(x)+f\gx\right))}{\lambda}
\eea

\beq
H^{(t)\prime}_0=\int dx \left[ 
\frac{v^2\gamma^2 f^{2}\gx}{2}+\frac{\gamma^2 f^{2}\gx}{2}+\frac{V(\sqrt{\lambda} f\gx)}{\lambda}
\right]=Q_0\gamma
\eeq

\bea
H^{(t)\prime}_1&=&\int dx\phi(x)\left[-\partial_x^2 f\gx+v^2\gamma^2f\pp\gx+\frac{\Vg1}{\sl}f\gx
\right]\nonumber\\
&=&\int dx\phi(x)\left[-f\pp\gx+\frac{\Vg1}{\sl}f\gx
\right]
=0
\eea

\beq
\ch^{(t)\prime}_2=\frac{\pi^2(x)}{2}+\frac{(\partial_x\phi(x))^2}{2}+\frac{\Vg2}{2}\phi^2(x)
\eeq

We quantize via the canonical commutation relation
\beq
[\phi(x),\pi(y)]=i\delta(x-y).
\eeq
and so
\beq
[H^{(t)\prime}_2,\phi(x)]=-i\pi(x)\hsp
[H^{(t)\prime}_2,\pi(x)]=-i\partial_x^2\phi(x)+i\Vg2\phi(x) \label{heis}
\eeq

\subsection{The Decomposition}

Let us 
introduce a decomposition for every $t$
\bea
\phi(x)&=&\ppin{k}G_k\gx\left(\Btd{}+\frac{\Bt{}}{2\ok{}}\right)\\
\pi(x)&=&i\ppin{k}G_k\gx\left(\ok{}\Btd{}-\frac{\Bt{}}{2}\right)\nonumber
\eea
Impose the equal time commutation relations
\beq
[\Bt{1},\Btd{2}]=2\pi\delta(k_1+k_2)
\eeq
with other equal time commutators vanishing. \blu{Actually we need to derive these from the canonical commutation relations.}

Let us show that the $G_k\gx$ are orthogonal, by showing that the $G_k(y)$ are eigenvectors of a Hermitian operator
\beq
L=\partial^2_y -2i\gamma v\ok{}\partial_y-V^{(2)}(\sl f(y))
\eeq
with eigenvalue $-\ok{}$.  Consider $G_{k_1}(y)$ and $G_{k_2}(y)$ with $k_1\neq \pm k_2$, as those pairs need to be diagonalized separately.  These satisfy
\beq
L G_i(y)=-\ok{i}^2 G_i(y).
\eeq
Then, integrating by parts which flips the sign of the $\partial_y$ term and so complex conjugates $L$, one sees
\beq
\int G^*_{k_1}(y) L G_{k_2}(y)=-\ok{2}^2 \int G^*_{k_1}(y) G_{k_2}(y) = \int G_{k_2}(y) L^* G^*_{k_1}(y) = \ok{1}^2 \int G_{k_2}(y) G^*_{k_1}(y)
\eeq
and so
\beq
\int G_{k_2}(y) G^*_{k_1}(y)=0.
\eeq
We thus see that $G_{k_1}$ and $G_{k_2}$ are orthogonal unless $k_1=\pm k_2$.  Now we will normalize them so that
\beq
\int dx G^*_{k_1}\gx G_{k_2}\gx=2\pi\delta(k_1-k_2). \label{ortho}
\eeq

\blu{We need to change the text below so that $\tilde G$ becomes $G^*$.  Also, we need some kind of orthogonality relation for $\dot G$ since we will use that below for the decomposition of $\pi(x)$.}

We want to show that
\beq
e^{-iH^{(t)\prime}_2\Delta}\Btd{}e^{iH^{(t)\prime}_2\Delta}=e^{-i\ok{}\Delta}B^{(t+\Delta)\ddag}_k\hsp
e^{-iH^{(t)\prime}_2\Delta}\Bt{}e^{iH^{(t)\prime}_2\Delta}=e^{i\ok{}\Delta}B^{(t+\Delta)}_{-k}
\eeq
because if this is true, then the oscillator states are time-independent at leading order and we can quantize the moving kink.  At linear order in $\Delta$ this is
\beq
-i[H^{(t)\prime}_2,\Btd{}]=-i\ok{}\Btd{}+\frac{\partial \Btd{}}{\partial t}\hsp
-i[H^{(t)\prime}_2,\Bt{}]=i\ok{}\Bt{}+\frac{\partial \Bt{}}{\partial t}\label{want}
.
\eeq

Assume that the $G_k\gx$ are a complete basis of functions of $x$, at each $t$.  Then, at each $t$, there is a dual basis $\tilde \g_k\gx$ such that
\beq
\int dx \tilde \g_{k_1}\gx G_{k_2}\gx = 2\pi\delta(k_1-k_2).
\eeq
The time derivative of this relation yields
\beq
\int dx \tilde \g_{k_1}\gx G_{k_2}\gx \partial_t\tilde \g_{k_1}\gx x =-\int dx \tilde \g_{k_1}\gx \partial_t G_{k_2}\gx 
\eeq
We can use the dual basis to find $\Btd{}$ and $\Bt{}$
\bea
\Btd{}&=&\int dx \tilde \g_k\gx \left[\frac{\phi(x)}{2}-i\frac{\pi(x)}{2\ok{}}\right]\\
\Bt{}&=&\int dx \tilde \g_k\gx \left[\ok{}\phi(x)+i\pi(x)\right].\nonumber
\eea
The time derivative is
\bea
\partial_t\Btd{1}&=&\int dx \left[\frac{\phi(x)}{2}-i\frac{\pi(x)}{2\ok{1}}\right]\partial_t\tilde \g_{k_1}\gx\\
&=&-\int dx \tilde \g_{k_1}\gx\ppin{k_2}\partial_t G_{k_2}\gx\left[\frac{\Btd{2}}{2}+\frac{\Bt{2}}{4\ok{2}}+\frac{\ok{2}\Btd{2}}{2\ok{1}}-\frac{\Bt{2}}{4\ok{1}}\right]\nonumber\\
&&\hspace{-1.2cm}=-\int dx \tilde \g_{k_1}\gx\ppin{k_2}\partial_t G_{k_2}\gx\left[\Btd{2}+\left(\frac{\ok{2}-\ok{1}}{2\ok 1}\right)\left({\Btd{2}}{}-\frac{\Bt{2}}{2
\ok{2}}
\right)
\right]\nonumber\\
\partial_t\Bt{1}&=&-\int dx \tilde \g_{k_1}\gx\ppin{k_2}\partial_t G_{k_2}\gx\left[ \Bt{2}+
\right]\nonumber
\eea

The free time evolution of the ladder operators is
\bea
[H^{(t)\prime}_2,\Btd{1}]&=&\int dx \tilde \g_{k_1}\gx \left[\frac{-i\pi(x)}{2}+\frac{-\partial_x^2\phi(x)+\Vg2\phi(x)}{2\ok{1}}\right]\nonumber\\
&=&\int dx \tilde \g_{k_1}\gx \ppin{k_2}\left[\frac{\ok{2}\Btd{2}}{2}-\frac{\Bt{2}}{4}-\left(\frac{\Btd{2}}{2\ok 1}+\frac{\Bt{2}}{4\ok 1\ok{2}}\right)\partial_x^2 \right.\nonumber\\
&&\left.+\Vg2\left(\frac{\Btd{2}}{2\ok 1}+\frac{\Bt{2}}{4\ok 1\ok{2}}\right)\right]G_{k_2}\gx\nonumber\\
&=&\int dx \tilde \g_{k_1}\gx \ppin{k_2}\left[\frac{\ok{2}\Btd{2}}{2}-\frac{\Bt{2}}{4}
\right.\nonumber\\
&&-\left(\frac{\Btd{2}}{2\ok 1}+\frac{\Bt{2}}{4\ok 1\ok{2}}\right)\left[ -\ok{2}^2 +2i v\ok{2} \partial_x+\Vg 2\right] \nonumber\\
&&\left.
+\Vg2\left(\frac{\Btd{2}}{2\ok 1}+\frac{\Bt{2}}{4\ok 1\ok{2}}\right)
\right]G_{k_2}\gx\nonumber\\
&=&\int dx \tilde \g_{k_1}\gx \ppin{k_2}\left[\frac{\ok{2}\Btd{2}}{2}-\frac{\Bt{2}}{4}
\right.\nonumber\\
&&\left.-\left(\frac{\Btd{2}}{2\ok 1}+\frac{\Bt{2}}{4\ok 1\ok{2}}\right)\left[ -\ok{2}^2 +2i v\ok{2} \partial_x\right]\right]G_{k_2}\gx\nonumber\\
&=&\ok{1}\Btd{1}+i\int dx \tilde \g_{k_1}\gx\ppin{k_2}   \left(\frac{\ok 2\Btd{2}}{\ok 1}+\frac{\Bt{2}}{2\ok 1}\right)\partial_t G_{k_2}\gx\nonumber
\eea

It isn't quite (\ref{want}) because of the $\Bt{}$ term ...

\subsection{Take Two}

That didn't seem to work.  So instead let us decompose $\pi(x)$ not using $\omega G$.  Instead use $\partial_t(G\gx e^{\pm i\omega t})$ for the two coefficients in $\pi(x)$.  The coefficients are then
\beq
e^{\mp i\omega t} \partial_t(G\gx e^{\pm i\omega t})=\pm i\omega  G\gx-\gamma v G\p\gx
\eeq
and we decompose
\beq
\pi(x)=\ppin{k}\left[iG_k\gx\left(\ok{}\Btd{}-\frac{\Bt{}}{2}\right)-v\partial_x G_k\gx\left(\Btd{}+\frac{\Bt{}}{2\ok{}}\right)
\right] \label{dec2}
\eeq

\blu{Now we should calculate $[H,\phi]$ and $[H,\pi]$.  Assume (\ref{want}) and try to derive (\ref{heis}).  Of course the logic is the opposite, (\ref{heis}) is true and we want to show (\ref{want}), but since the decomposition is a basis if you show the equality in one direction you get the other for free.
}

\subsection{Finding $\Bt{}$}

Note that, dropping quickly oscillating boundary terms
\beq
\int dx \tilde \g_{k_1}\gx \partial_x G_{k_2}\gx = -\int dx  G_{k_2}\gx \partial_x \tilde \g_{k_1}\gx
\eeq
and so
\bea
\int dx \pi(x)\tilde \g_{k_1}\gx&=&i\left(\ok{1}\Btd{1}-\frac{\Bt{1}}{2}\right)\\
&&\hspace{-2cm}+v\int dx\ppin{k_2}G_{k_2}\gx \partial_x \tilde \g_{k_1}\gx\left(\Btd{2}+\frac{\Bt{2}}{2\ok{2}}\right)\nonumber
\eea

\bea
\int dx \pi(x)\partial_x\tilde \g_{k_1}\gx&=&i\int dx \ppin{k_2}G_{k_2}\gx\partial_x\tilde \g_{k_1}\gx\nonumber\\
&&\times\left(\ok{2}\Btd{2}-\frac{\Bt{2}}{2}\right)+O(v)
\eea

We can only isolate $\Bt{}$ and $\Btd{}$ perturbatively in $v$
\bea
\int dx \pi(x)\left(1\pm\frac{iv}{\ok 1}\partial_x\right)\tilde \g_{k_1}\gx&=&i\left(\ok{1}\Btd{1}-\frac{\Bt{1}}{2}\right)\\
&&\hspace{-5cm}+v\int dx \ppin{k_2}G_{k_2}\gx\partial_x\tilde \g_{k_1}\gx\nonumber\\
&&\hspace{-5cm}\times
\left[\left(1\mp\frac{\ok 2}{\ok 1}
\right)\Btd{2}+\left(1\pm\frac{\ok 2}{\ok 1}
\right)\frac{\Bt{2}}{2\ok{2}}\right]
+O(v^2)\nonumber
\eea

If we drop terms of order $v(\ok 2-\ok 1)$ then this becomes
\bea
\int dx \pi(x)\left(1+\frac{iv}{\ok 1}\partial_x\right)\tilde \g_{k_1}\gx&=&i\left(\ok{1}\Btd{1}-\frac{\Bt{1}}{2}\right)\\
&&\hspace{-5cm}+v\int dx \ppin{k_2}G_{k_2}\gx\partial_x\tilde \g_{k_1}\gx \frac{\Bt{2}}{\ok{2}}
+O(v^2)\nonumber\\
\int dx \pi(x)\left(1-\frac{iv}{\ok 1}\partial_x\right)\tilde \g_{k_1}\gx&=&i\left(\ok{1}\Btd{1}-\frac{\Bt{1}}{2}\right)\nonumber\\
&&\hspace{-5cm}+2v\int dx \ppin{k_2}G_{k_2}\gx\partial_x\tilde \g_{k_1}\gx \Btd{2}
+O(v^2)\nonumber
\eea

\beq
\int dx \phi(x)\tilde \g_{k_1}\gx=\Btd{1}+\frac{\Bt{1}}{2\ok{1}}
\eeq

\bea
\int dx \left[\ok 1\phi(x)+\pi(x)\left(-i-\frac{v}{\ok 1}\partial_x\right)
\right]\tilde \g_{k_1}\gx&=&2\ok{1}\Btd{1}\nonumber\\
&&\hspace{-7cm}-2iv\int dx \ppin{k_2}G_{k_2}\gx\partial_x\tilde \g_{k_1}\gx \Btd{2}
+O(v^2)\nonumber
\eea

Define the shorthand
\beq
\Delta_{k_2k_1}=\int dx G_{k_2}\gx\partial_x\tilde \g_{k_1}\gx
\eeq
Then
\bea
&&\int dx \left[\ok 1\phi(x)+\pi(x)\left(-i-\frac{v}{\ok 1}\partial_x\right)
\right]\tilde \g_{k_1}\gx\\
&&\hspace{3cm}=2\ok{1}\ppin{k_2}\left(2\pi\delta(k_2-k_1)-\frac{iv}{\ok{1}}\Delta_{k_2k_1}
\right)\Btd{1}\nonumber
\eea
Up to corrections of order $O(v^2)$ we may invert the term in the parenthesis
\bea
&&\ppin{k_2}\left(2\pi\delta(k_2-k_1)+\frac{iv}{\ok{1}}\Delta_{k_2k_1}
\right)\int dx \left[\ok 2\phi(x)+\pi(x)\left(-i-\frac{v}{\ok 2}\partial_x\right)
\right]\tilde \g_{k_2}\gx\nonumber\\
&&\hspace{3cm}=2\ok{1}\Btd{1}\nonumber
\eea

\blu{I need to go.  Do you want to try to repeat the steps in Subsec 3.2 with this new decomposition?  }

\section{Lorentz Transform Approach}

\subsection{Transforming the Fields}

Let the original coordinates be $(x\p,t\p)$, where the kink is at rest.  A Lorentz transformation changes these to
\beq
x=\gamma(x\p + v t\p)\hsp
t=\gamma(t\p+v x\p).
\eeq
On the original coordinates lived a field $\phip(x\p)$ and its conjugate $\pip(x\p)$ that satisfy
\beq
[\phip(x\p),\pip(y\p)]=i\delta(x\p-y\p).
\eeq
We decompose them as 
\bea
\phip(x\p)&=&\phi_0\g_B(x\p)+\ppin{k}\g_k(x\p)\left(\Bd{}+\frac{B_{-k}}{2\ok{}}\right)\label{dec3}\\
\pip(x\p)&=&\pi_0\g_B(x\p)+i\ppin{k}\g_k(x\p)\left(\ok{}\Bd{}-\frac{B_{-k}}{2}\right)\nonumber
\eea
where
\beq
[B_{-k_1},B^\ddag_{k_2}]=2\pi\delta(k_1+k_2)\hsp [\phi_0,\pi_0]=i.
\eeq
Note that the $B$, $\phi_0$ and $\pi_0$ are not primed, these will be the same operators in both frames. 

These operators live on a timeslice with constant $t\p$, although we are working in the Schrodinger picture so they are independent of the timeslice $t\p$ that we choose.  Now we want to defined operators $\phi(x)$ and $\pi(x)$ on a timeslice with constant $t$.

For simplicity, let us set $t\p=0$ for $\phip(x\p)$ and $t=0$ for $\phi(x)$, although this choice cannot affect our final result.  Then $t=\gamma vx\p$.  We define $\phi(x)$ by evolving $\phip(x\p)$ to $t=0$, in other words we need to evolve for a time $-\gamma v x\p$
\beq
\phi(x)=\phi(\gamma x\p)=e^{iH\p vx\p}\phip(x\p)e^{-iH\p vx\p}.
\eeq
Here $H\p$ is the kink Hamiltonian in the $(x\p,t\p)$ frame.  It generates time translations in $t\p$, leaving $x\p$ constant.

Truncating to $O(1)$ in the coupling constant
\beq
\phi(x)=\phi(\gamma x\p)=e^{iH\p_2 \gamma vx\p}\phip(x\p)e^{-iH\p_2 \gamma vx\p}=e^{iH\p_2  vx}\phip(x\p)e^{-iH\p_2  vx}.
\eeq
This is given by
\beq
H\p_2=\frac{\pi_0^2}{2}+\ppin{k}\ok{}\Bd{}B_{k}.
\eeq
One can easily derive the transformations
\bea
e^{iH\p_2 vx}\Bd{}e^{-iH\p_2 vx}&=&e^{ivx\ok{}}\Bd{}\hsp
e^{iH\p_2 vx}B_{-k}e^{-iH\p_2 vx}=e^{-ivx\ok{}}B_{-k}\\
e^{iH\p_2 vx}\phi_0 e^{-iH\p_2 vx}&=&\phi_0+v x \pi_0
\hsp
e^{iH\p_2 vx}\pi_0 e^{-iH\p_2 vx}=\pi_0.\nonumber
\eea
Plugging this into the decomposition (\ref{dec3}) we find
\beq
\phi(x)=\g_B(\gamma x)\left( \phi_0+v x \pi_0
\right)+\ppin{k}\g_k(\gamma x)\left(e^{iv\ok{} x}\Bd{}+e^{-iv\ok{} x}\frac{B_{-k}}{2\ok{}}\right).
\eeq

We define the conjugate momentum to be any operator such that $\pi(x)$
\beq
[\phi(x),\pi(y)]=i\delta(x-y).
\eeq
This has at least one solution
\beq
\pi(x)=\gamma\g_B(\gamma x)\pi_0+i\gamma\ppin{k}\g_k(\gamma x)\left(e^{iv\ok{} x}\ok{}\Bd{}+e^{-iv\ok{} x}\frac{B_{-k}}{2}\right).
\eeq

\subsection{The Comoving Hamiltonian}

At time $t=0$, the free part of the comoving Hamiltonian is
\beq
H\p_2=\frac{1}{2}\int d x: \left[  \pi^2(x)+\left(\partial_x \phi(x)\right)^2+V^{(2)}(\sl f(\gamma x))\phi^2(x)
\right]:_a.
\eeq
For simplicity, let us drop the normal ordering, although we need it to compute the correct one-loop mass correction $Q_1$.  Using
\beq
\int dx \g_B^2(\gamma x)=\frac{1}{\gamma}
\eeq
we may manipulate ... the problem is that the coefficients of the $B$ are not orthogonal, also the $B$ do not diagonalize the comoving Hamiltonian
\bea
H\p_2
&=&\frac{1}{2}\left[\gamma\pi_0^2+\int d x \left[  \pi^2(x)+\left(\partial_x \phi(x)\right)^2+V^{(2)}(\sl f(\gamma x))\phi^2(x)\right]\\
\eea

\subsection{The Kink Hamiltonian and Momentum}

The kink Hamiltonian generates time translations
\beq
H\p |\psi(t\p)\rangle=i\partial_{t\p}|\psi(t\p)\rangle
\eeq
in the $t\p$ direction, keeping $x\p$ fixed.  The kink momentum operator 
\beq
P\p=\sqrt{Q_0}\pi_0+P\hsp
P=\int dx\p \phip(x\p)\partial_{x\p}\pip(x\p)
\eeq
generates spatial translations in the $x\p$ direction, keeping $t\p$ fixed
\bea
[P\p,\pip(x\p)]&=&\int dy\p [\phip(y\p),\pip(x\p)]\partial_{y\p}\pip(y\p)=i\partial_{x\p}\pip(x\p)\nonumber
\\
\left[P\p,\phip(x\p)+f(x\p)\right]&=&\sqrt{Q_0}[\pi_0,\phip(x\p)]
-\int dy\p [\pip(y\p),\phip(x\p)]\partial_{y\p}\phip(y\p)=i\partial_{x\p}\left(\phip(x\p)+f\p(x\p)\right)\nonumber
\eea

Translations in the direction $t$ are therefore generated by 
\beq
\tilde{H}=\gamma(H\p-v P\p).
\eeq
This Hamiltonian is a disaster, because it has the $\sqrt{Q_0}\pi_0$ tadpole term from $P\p$.  It is order $O(\lambda^{-1/2})$ and so makes perturbation theory complicated by mixing the orders.  However it is not quite our comoving Hamiltonian.

\subsection{Lorentz Transformation Operator}

To Lorentz transform a nonlocal operator such as $\Bd{}$ one needs to use the full Lorentz transformation operator, which at $t\p=0$ is
\beq 
U=\exp{-i\alpha\int dx\p x\p :\ch\p(x\p):_a }
\eeq
where $\alpha=$arctanh$(v)$ is the rapidity.  At leading order this is
\bea
U_1&=&\exp{-i\alpha\int dx\p x\p \ch\p_1(x\p)}\\
&=&\exp{-i\alpha\int dx\p x\p \left[\left(\partial_{x\p}\phip(x\p)\right)\left(\partial_{x\p}f(x\p)\right)+V^{(1)}(\sl f(x\p))\phi^{\p}(x\p)
\right]}\nonumber\\
&=&\exp{-i\alpha\int dx\p \left(x\p \phip(x\p)\left[-\partial^2_{x\p}f(x\p)+V^{(1)}(\sl f(x\p))\right]
-\phip(x)\partial_{x\p}f(x\p)\right)}\nonumber\\
&=&\exp{-i\alpha\sqrt{Q_0}\int dx\p 
\phip(x)\g_B(x\p)}=e^{-i\alpha\sqrt{Q_0}\phi_0}.\nonumber
\eea
At the next order
\bea
U_2&=&\exp{-i\alpha\int dx\p x\p (\ch\p_1(x\p)+\ch\p_2(x\p)}\\
&=&\exp{-i\alpha\left(\sqrt{Q_0}\phi_0+\int dx\p x\p \left[\frac{\pi^{2}(x\p)+\left(\partial_{x\p}\phip(x\p)\right)^2+V^{(2)}(\sl f(x\p))\phi^{\p 2}(x\p)}{2} 
\right]\right)}\nonumber
\eea
To avoid clutter we do not write the normal ordering, which will be irrelevant below.

\end{document}
-\omega^2 G\gx+2i\gamma v\omega G\p\gx+\Vg 2 G\gx=

Just 40 years ago, the scattering of quantum solitons with fundamental quanta was a popular topic \cite{weigel89}.  The strategy was as follows.  First, one would consider kinks in (1+1)-dimensional models, using the collective coordinate approach of Refs.~\cite{gs74,tom75}.  For example, the elastic meson-kink scattering considered in the present work was first treated using collective coordinates in Ref.~\cite{uehara91}.  Once the lower-dimensional case was understood, the results would be imported into higher dimensional models \cite{hayashi92,erase}.  Suitable approximations were made \cite{adiabatic84} in this formal work and it was then applied to phenomenology \cite{diakonov97,ellis04}.

The house of cards reached its peak with the discovery of a predicted exotic hadron in Ref.~\cite{nakano03}.  Within a few years it became clear \cite{notheta} that this resonance, whose prediction was reasonably independent of the details of the model \cite{petrov16}, did not exist.  In fact, many of the predictions made over the previous twenty years were falsified one at a time.  Of course the problem could be with assumptions built into the models, or, as advocated in Ref.~\cite{weigel18}, it could be with the approximations used to treat calculations involving quantum solitons\footnote{Indeed, the importance of including the full continuum quantum degrees of freedom has recently by highlighted in Refs.~\cite{chris23,bjarke23}.}.

This second possibility motivates a lighter formalism, so that less severe approximations are necessary and as a result more reliable conclusions may be expected.  Such a formalism, linearized soliton perturbation theory, has been formulated in Refs.~\cite{mekink,me2loop}.  

The purpose of the present paper is to re-examine elastic meson-kink scattering using this new, simpler formalism.  Our answer, summarized in Eqs.~(\ref{abcd}) and (\ref{peq}), will disagree with the old result, Eq. (3.19) of Ref.~\cite{uehara91}.  At leading order, we find, for example, that there are contributions from states with two virtual mesons.  As we will show, such contributions in fact are essential to eliminate soliton-meson elastic scattering in the Sine-Gordon model, which, as their masses differ, would be in contradiction with the integrability of the model.  That contribution is not present in Ref.~\cite{uehara91} as loop contributions were intentionally dropped.  However that calculation, like ours, kept contributions to the amplitude that are quadratic in the coupling constant, which are of the same order as these loop corrections.  Indeed, this is evidenced by the fact that the loop corrections cancel the other contributions in the Sine-Gordon case.

Besides the fact that they are essential for consistency, these intermediate two-meson states are interesting for another reason.  In models in which the kink has bound normal mode excitations, called shape modes, there will be an intermediate state consisting of a twice-excited shape mode.  In models such as the $\phi^4$ model,  a single shape mode excitation is stable while a double excitation is unstable.  We therefore expect that our scattering probability will have a narrow peak at twice the energy of the shape mode, with a width equal to the shape mode's inverse lifetime.  More generally, we hope that kink-meson scattering can teach us about the unstable excited spectrum of the kink itself.  We will test this general expectation in future work, but the aforementioned peak will already be visible below.

We begin in Sec.~\ref{revsez} with a review of linearized soliton perturbation theory.  Our main calculation is presented in Sec.~\ref{calcsez}.  Finally, we turn to the Sine-Gordon model in Sec.~\ref{exsez}.

\section{Linearized Soliton Perturbation Theory} \label{revsez}

Consider a  (1+1)-dimensional quantum field theory with a scalar field $\phi(x)$ and its conjugate field $\pi(x)$.  We will work in the Schrodinger picture, where the Hamiltonian may be written as
\begin{equation}
H=\int d x: \mathcal{H}(x):_a\hsp \mathcal{H}(x)=\frac{\pi^2(x)}{2}+\frac{\left(\partial_x \phi(x)\right)^2}{2}+\frac{V(\sqrt{\lambda} \phi(x))}{\lambda}
\end{equation}
in terms of a degenerate potential $V$ and an expansion parameter $\sqrt{\lambda}$.  The corresponding classical equations of motion enjoy a stationary kink solution $\phi(x,t)=f(x)$ that interpolates between the minima of the potential.  

The usual normal ordering $::_a$ removes ultraviolet divergences arising from loops connected to a single vertex, which are the only ultraviolet divergences in such theories.  The normal ordering is defined at the mass scale $m$ where
\beq
m^2=V^{(2)}(\sqrt{\lambda} f(\pm \infty))\hsp
V^{(n)}(\sqrt{\lambda} \phi(x))=\frac{\partial^n V(\sqrt{\lambda} \phi(x))}{(\partial \sqrt{\lambda} \phi(x))^n}.
\eeq
If the masses corresponding to the two sign choices are different, then at one loop the kink will accelerate \cite{wstabile}.  We will not be interested in such cases.

We refer to the quanta of the $\phi(x)$ field as mesons.  The collection of states with no kinks, but a finite number of mesons, will be called the vacuum sector.  States with a single kink, together with a finite number of mesons, are referred to as the kink sector.  Any kink sector state may be created by acting the displacement operator
\beq
\df={{\rm Exp}}\left[-i\int dx f(x)\pi(x)\right]\hsp
\df^\dag \phi(x) \df = \phi(x)+f(x)  \label{dfd}
\eeq
on a vacuum sector state.  Intuitively this is clear, as $\df$ shifts the field $\phi(x)$ by the classical kink solution.

It may seem that the problem of constructing kink sector states is nonperturbative, as the operator $\df$ contains an exponential of $f(x)$ which is proportional to $1/\sl$.  This apparent problem can be resolved using a passive transformation to remove the $\df$ from each state.  More specifically, first we choose names $|\psi\rangle$ for all of the states.  This choice of names for kets is called the defining frame.  We will work in a different frame, called the kink frame, defined as follows.  In the kink frame, the state $|\psi\rangle$ is defined to be the state $\df^\dag|\psi\rangle$ in the defining frame.  One can easily show that if this state is time-independent, so that $\df^\dag|\psi\rangle$ is an eigenstate of $H$, then $|\psi\rangle$ is an eigenstate of the kink Hamiltonian
\beq
H\p=\df^\dag H\df
. 
\label{df}
\eeq
This is a manifestation of the familiar fact that, when making a passive transformation of coordinates, in this case coordinates $|\psi\rangle$ on the Hilbert space, one must remember to also transform all of the functions that act on those coordinates, in this case the operators.

What have we gained with this passive transformation?  Now all of the $\df$ operators have been removed from the names of our kink sector states, thus we can treat them using ordinary perturbation theory, with the caveat that the Hamiltonian in this frame is $H\p$.

In the vacuum sector, perturbation theory describes small perturbations about the vacuum.  Classically the constant frequency perturbations are plane waves.  In the kink sector, kink frame perturbation theory describes small perturbations about the kink.  Classically, small constant frequency $\omega$ perturbations are normal modes
\beq
\phi(x,t)=e^{-i\omega t}\g(x)
\eeq
which satisfy the Sturm-Liouville equation
\beq
\V{2}{\g}(x)=\omega^2{\g}(x)+{\g}^{\prime\prime}(x).  \label{sl}
\eeq
The solutions are classified by their frequencies $\omega$.  There is a single zero-mode $\g_B(x)$ with $\omega_B=0$.   For each real number $k$ there is a continuum mode $\g_k(x)$ with frequency
\beq
\ok{}=\sqrt{m^2+k^2}.
\eeq
There can also be discrete, real shape modes $\g_S(x)$ with frequencies $0<\omega_S<m$.   All modes are chosen to satisfy $\g^*_k=\g_{-k}$ and the completeness relations
\beq
\int dx |{\g}_{B}(x)|^2=1,\
\int dx {\g}_{k_1} (x) {\g}^*_{k_2}(x)=2\pi \delta(k_1-k_2),\ 
\int dx {\g}_{S_1}(x){\g}^*_{S_2}(x)=\delta_{S_1S_2}. \label{comp}
\eeq
The sign of $\g_B$ is fixed by the convention
\beq
\g_B(x)=-\frac{f\p(x)}{\sqrt{Q_0}}. \label{gb}
\eeq
Here $Q_i$ is the $O(\lambda^{i-1})$ term in the mass of the ground state kink.

Following Refs.~\cite{cahill76,mekink} we decompose the field and its conjugate as
\bea
\phi(x) &=&\phi_0 \mathfrak{g}_B(x)+\ppin{k} \left(B_k^{\ddag}+\frac{B_{-k}}{2 \omega_k}\right) \mathfrak{g}_k(x)\hsp
B^\ddag_k=\frac{B^\dag_k}{2\ok{}}\hsp
B^\ddag_S=\frac{B^\dag_S}{2\omega_S} \label{dec}\\
\pi(x) &=&\pi_0 \mathfrak{g}_B(x)+i \ppin{k}\left(\omega_k B_k^{\ddag}-\frac{B_{-k}}{2}\right) \mathfrak{g}_k(x)\hsp
B_S=B_{-S}\hsp \ppin{k}=\pin{k}+\sum_S. \nonumber
\eea
Here $\phi_0$ and $\pi_0$ represent the position and momentum of the kink center of mass.  The operators $B^\ddag_S$ and $B_S$ create and annihilate shape modes, while $B^\ddag_k$ and $B_k$ create and annihilate continuum modes.  The canonical commutation relations satisfied by $\phi(x)$ and $\pi(x)$ yield the commutators of $\pi_0,\ \phi_0, B^\ddag$ and $B$
\beq
\left[\phi_0, \pi_0\right]=i, \quad\left[B_{S_1}, B_{S_2}^{\ddagger}\right]=\delta_{S_1 S_2}, \quad\left[B_{k_1}, B_{k_2}^{\ddagger}\right]=2 \pi \delta\left(k_1-k_2\right).
\eeq

The perturbative expansion begins with the eigenstates of the free part of $H\p$.  These can be constructed as follows \cite{cahill76,mekink}.  The kink ground state $\vac_0$ is defined to be the state satisfying
\beq
\pi_0\vac_0=B_k\vac_0=B_S\vac_0=0. \label{v0}
\eeq
Similarly, an $n$ meson state $|k_1\cdots k_n\rangle_0$ is 
\beq
|k_1\cdots k_n\rangle_0=\Bd1\cdots\Bd n\vac_0.
\eeq

A basis of the kink sector is provided by states $\phi_0^m|k_1\cdots k_n\rangle_0$.  Any state in the kink sector may be decomposed in this basis.  In particular, if $|\psi\rangle$ is a Hamiltonian eigenstate in the kink sector, then we name the corresponding coefficients $\gamma$
\beq
|\psi\rangle=\sum_{m,n=0}^\infty |\psi\rangle^{mn}\hsp
|\psi\rangle^{mn}=\phi_0^m\ppink{n}\gamma_\psi^{mn}(k_1\cdots k_n)|k_1\cdots k_n\rangle_0.
 \label{gameqa}
\eeq
These coefficients can be found in a perturbative expansion.  Recalling that $Q_0^{-1/2}$ is of order $O(\sl)$, where $Q_0$ is the classical kink mass and $\sl$ is our perturbative parameter, this expansion can be written
\beq
\gamma^{mn}=\sum_i Q_0^{-i/2}\gamma_i^{mn}
\eeq
where $i$ is the order of the expansion.

In the present work, we are interested in Hamiltonian eigenstates $|k_1\rangle$ consisting of a single kink and a single meson of momentum $k_1$.  At leading order, kink-meson elastic scattering will be completely determined by the order $i=2$ perturbative correction to this state, which is determined by the coefficients $\gamma_{2k_1}^{mn}$.  In fact, the scattering amplitude only depends on a single coefficient, $\gamma_{2k_1}^{01}$. 

This coefficient was evaluated in Ref.~\cite{menorm}.  Let us recall its general form here.  First one separates out a $\delta$ function piece which depends on the choice of normalization
\beq
\gamma_{2k_1}^{21}(k_2)=\hat\gamma_{2k_1}^{21}(k_2)+2\pi\delta(k_2-k_1)Q_0\left(\hat\sigma_{ k_1}-\sigma_{ k_1}
\right) 
\eeq
where $\hat\gamma_{ k_1}(k_2)$ is continuous at $k_2= k_1$.  Then it can be written in terms of the functions $\rho$ and $\hat\gamma$, defined below, as
\beq
\gamma_{2 k_1}^{01}(k_2)=\frac{\rho_{ k_1}(k_2)-\hat \gamma_{2 k_1}^{21}(k_2)}{\ok 1-\ok{2}}. \label{g2}
\eeq

We have not yet defined the coefficients at $\ok 1=\ok 2$, corresponding to the location of the pole in Eq.~(\ref{g2}).   There are two such cases, corresponding to $k_1=\pm k_2$.  In the case $k_1=k_2$, the choice is physically irrelevant as it corresponds to a choice of normalization of the state.  In the case $k_1=-k_2$ the choice is physically relevant.  The states $|k_1\rangle$ and $|-k_1\rangle$ are degenerate Hamiltonian eigenstates, and so the choice of convention for treating this pole is equivalent to a choice of vector in this degenerate eigenspace.  Any choice will yield a Hamiltonian eigenstate $|k_1\rangle$, and so the choice must be made appropriately for the physics of a given problem.  This will be done in Sec.~\ref{calcsez}.

Finally, for completeness, let us recall the functions
\bea
\rho_{ k_1}(k_2)&=&
\frac{\lambda Q_0}{4\omega_{k_1}}V_{\I k_2 - k_1}
+\frac{\lambda Q_0}{8}\left(\frac{V_{\I  - k_1}}{\omega^2_{ k_1}}-\frac{\Delta_{- k_1 B}}{\omega_{ k_1}\sqrt{\lambda Q_0}}\right)V_{\I k_2}\label{rho}\\
&&+\sqrt{\lambda Q_0}\left[\ppinkp{2}\frac{\sqrt{\lambda Q_0}V_{- k_1 k\p_1 k\p_2}V_{-k\p_1-k\p_2k_2}}{16\omega_{ k_1}\okp1\okp2\left(\omega_{ k_1}-\okp1-\okp2\right)}\right.\nonumber\\
&&\left.+\ppin{k\p}\left(\frac{\left(-\okp{}\Delta_{k\p B}-\sqrt{\lambda Q_0}V_{\I  k\p}\right)
V_{-k\p- k_1 k_2}}{8\okp{}^2\omega_{ k_1}}+\frac{\sqrt{\lambda Q_0}V_{- k_1 k\p k_2}V_{\I -k\p}}{8\omega_{ k_1}\okp{}\left(\omega_{ k_1}-\okp{}-\ok2\right)}\right)\right.\nonumber\\
&&+\left.
\frac{ \left(-\ok2\Delta_{k_2 B}-\sqrt{\lambda Q_0}V_{\I  k_2}\right)V_{\I - k_1}}{8\omega_{ k_1}\ok2}\right]
-\frac{\lambda Q_0}{16}\ppinkp{2}\frac{V_{k_2k\p_1k\p_2}V_{- k_1-k\p_1-k\p_2}}{\omega_{ k_1}\okp1\okp2\left(\ok2+\okp1+\okp2\right)}
\nonumber
\eea
and
\bea
\hat\gamma_{2 k_1}^{21}(k_2)&=&\frac{3}{8}\left(-1+\frac{\ok2}{\omega_{ k_1}}\right)\Delta_{k_2 B}\Delta_{- k_1 B}-\frac{1}{4}\ppin{k\p}\left(\frac{\ok2}{\okp{}}+\frac{\okp{}}{\omega_{ k_1}}
\right)\Delta_{- k_1,-k\p}\Delta_{k_2k\p}\label{hg}\\
&&-\frac{\sqrt{\lambda Q_0}}{8\omega_{ k_1}}\left(\omega_{k_2}\Delta_{k_2 B}\frac{V_{\I - k_1}}{\omega_{ k_1}}+\omega_{ k_1}\Delta_{- k_1 B}\frac{V_{\I  k_2}}{\ok2}\right)+\frac{1}{8}\ppin{k\p}
\frac{\sqrt{\lambda Q_0}\Delta_{-k\p B}V_{- k_1 k_2 k\p}}{\omega_{ k_1}\left(\omega_{ k_1}-\ok2-\okp{}\right)}.\nonumber
\eea
Here we have defined the shorthand notation
\bea
\Delta_{k_1k_2}&=&\int dx \g_{k_1}(x) \g\p_{k_2}(x)\\
\I(x)&=&\pin{k}\frac{\left|{\g}_{k}(x)\right|^2-1}{2\omega_k}+\sum_S \frac{\left|{\g}_{S}(x)\right|^2}{2\omega_k}\nonumber\\
V_{k_1 k_2 k_3}&=&\int d x V^{(3)}(\sqrt{\lambda} f(x)) \mathfrak{g}_{k_1}(x) \mathfrak{g}_{k_2}(x) \mathfrak{g}_{k_3}(x)\nonumber\\
V_{\I k}&=&\int d x V^{(3)}(\sqrt{\lambda} f(x)) \I(x) \mathfrak{g}_{k}(x)\nonumber\\
V_{\I k_1 k_2}&=&\int d x V^{(4)}(\sqrt{\lambda} f(x)) \I(x) \mathfrak{g}_{k_1}(x) \mathfrak{g}_{k_2}(x).\nonumber
\eea
The indices $k_i$ here run over the zero mode $B$, all shape modes $S$ as well as the continuum modes $k$.  

\section{The Calculation} \label{calcsez}

In one-dimensional nonrelativistic quantum mechanics, there are two ways to calculate the reflection coefficient for a particle scattering off of a localized feature in the potential.  The first method is as follows.  One first solves the time-independent Schrodinger equation, imposing the boundary condition that there are no particles incoming from the right.  One next takes the inner product of the Hamiltonian eigenstate with an outgoing wave packet far to the left.  To get a nonzero answer, of course one must choose a Hamiltonian eigenstate whose energy is in the continuum.  Such states are not normalizable, but one may nonetheless define a sensible, finite inner product.

In the other method, one begins with an incoming wave packet on the left.  This is evolved in time using $e^{-iHt}$ until a late time, after it is far from the feature.  The part on each side will be a superposition of outgoing plane waves.  The coefficients of this decomposition into plane waves are the scattering amplitude as a function of momentum.

In the present note, we will consider the quantum field theory analogue of the first method.  In Ref.~\cite{ellong} we will redo the calculation using the second method, to check that the answers agree.

\subsection{Defining the Wave Packet}

Following the prescription above, we consider Hamiltonian eigenstates $|k_1\rangle$, consisting of a kink and a meson of momentum $k_1$.  How do we impose the boundary condition that there are no incoming mesons from the right?  Of course the Hamiltonian eigenstate itself has no dynamics, so the question itself only makes sense if we construct a wave packet of these Hamiltonian eigenstates.   A wave packet beginning largely near $x_0\ll -1/m$ with average momentum $k_0\gg 1/\sigma$ can be chosen to be
\beq
|t=0\rangle=\pin{k_1} e^{-\sigma^2(k_1-k_0)^2-i(k_1-k_0)x_0}|k_1\rangle.
\eeq
Recall that $|k_1\rangle$ is an eigenstate of the full kink Hamiltonian $H\p$, not just the free part, and so it is a sum of many different free eigenstates with various numbers of mesons.  It even includes the reflected part $|-k_1\rangle_0$ with some small coefficient of order $O(\lambda)$.  This small coefficient will be responsible for the scattering amplitude calculated here.

The wave packet is not a Hamiltonian eigenstate, so it evolves.  At time $t$ it is equal to
\beq
|t\rangle=\pin{k_1} e^{-\sigma^2(k_1-k_0)^2-i(k_1-k_0)x_0-iE_{k_1}t}|k_1\rangle. \label{t}
\eeq
Let us make the approximation $E_{k_1}=\ok 1$, which holds at leading order.  Now recall $\sigma\gg 1/m$.  This allows us to expand the frequency $\omega$
\beq
\ok{1}=\ok{0}+\frac{\partial \ok{0}}{\partial k_0}(k_1-k_0)+O\left((k_1-k_0)^2\right)=\ok{0}+\frac{k_0}{\ok 0}(k_1-k_0)+O\left((k_1-k_0)^2\right).
\eeq
The higher orders capture physics such as the spreading of the wave packet, which we will ignore from now on as they are small at large enough $\sigma$.  

Defining the position
\beq
x_t=x_0+\frac{k_0}{\ok 0} t
\eeq
one can rewrite (\ref{t}) as
\beq
|t\rangle=e^{-i\ok 0 t}\pin{k_1} e^{-\sigma^2(k_1-k_0)^2-i(k_1-k_0)x_t}|k_1\rangle. \label{t2}
\eeq
Apparently the main part of the wave packet is at position $x_t$ at time $t$.

Consider a pole in $|k_1\rangle$ at $k_1=-k_2$
\beq
|k_1\rangle=\pin{k_2}\left(F(k_1,k_2)+\frac{R(k_1)}{k_1+k_2}\right)|k_2\rangle_0 \label{ls}
\eeq
where $F(k_1,k_2)$ is analytic near $k_1=-k_2$.  At this point, we have not yet specified the function at the pole.  This choice, we have seen in earlier papers, corresponds to a choice of eigenstate $|k_1\rangle$ inside of a degenerate eigenspace.  Eq.~(\ref{ls}) is a Lippmann-Schwinger equation, and the ambiguity at the pole corresponds to the usual freedom to choose a solution in that context.

Inserting (\ref{ls}) into (\ref{t2}) one finds
\beq
|t\rangle=e^{-i\ok 0 t}\pin{k_2} I(k_2)|k_2\rangle_0\hsp
I(k_2)=\pin{k_1} e^{-\sigma^2(k_1-k_0)^2-i(k_1-k_0)x_t} \left(F(k_1,k_2)+\frac{R(k_1)}{k_1+k_2}\right).
\eeq
We see that the definition of the pole affects the integral $I(k_2)$.

\subsection{Choosing a Hamiltonian Eigenstate}

Consider a time $t$ such that $x_t\ll0$ and $\sigma\ll|x_t|$.  Physically the first condition means that the meson is not yet near the kink, while the second means that the meson wave packet is much smaller than the kink-meson separation.  This is not in contradiction with our standard assumptions that $\sigma\gg1/m$ and $\sigma\gg1/k_0$.   Choose an $r$ such that $\sigma\ll 1/r \ll |x_t|$.

Now let us evaluate $I(k_2)$ by integrating around a semicircle of radius $r$ that closes in the $+i$ side of the complex $k_1$ plane.  In principle, there may be contributions from nonanalytic parts of $F(k_1,k_2)$ far from $k_2=-k_1$.  From the general form of $|k\rangle$ found in other papers, this only occurs at real $k_2$ where it will contribute to other asymptotic final states.  We will simply ignore these, as our interest lies in elastic scattering. In Ref.~\cite{ellong} we will let $k_2$ be arbitrary and so will consider such processes.

Thus elastic scattering can only arise from the pole contribution
\beq
I(k_2)\Big|_{\rm{pole}}=\pin{k_1} e^{-\sigma^2(k_1-k_0)^2-i(k_1-k_0)x_t}\frac{R(k_1)}{k_1+k_2}.
\eeq
Recall that $\sigma r\ll 1$ and so the Gaussian factor is approximately unity.  Physically this is a consequence of the fact that the wave packet is so broad that there is negligible momentum smearing, although it is not so broad that the meson yet overlaps with the kink.

As $x_t\ll0$, the meson has not yet had time to scatter.  Thus, we wish to choose our initial condition such that no elastic scattering has occurred, so that $I(k_2)\big|_{\rm{pole}}$ vanishes.  As the residue $R(k_1)$ may in principle be nonzero, we achieve this by choosing the pole to be outside of our integration contour.  In other words, we wish to shift the pole in the $-i$ direction.  This can be accomplished with the choice
\beq
|k_1\rangle=\pin{k_2}\left(F(k_1,k_2)+\frac{R(k_1)}{k_1+k_2+i\epsilon}\right)|k_2\rangle_0.
\eeq
This $+i\epsilon$ is the same one that arises in the Lippmann-Schwinger equation for the ``in" state.

\subsection{Calculating the Scattering Probability}

Now that our initial eigenstate is determined, we may proceed to evolve it through the scattering, to $x_t\gg0$.  The appropriate contour for evaluating $I(k_2)$ closes in the $-i$ direction, and so picks up the pole at $k_1=-k_2-i\epsilon$.  The residue theorem then yields
\beq
I(k_2)\Big|_{\rm{pole}}=-iR(-k_2) e^{-\sigma^2(k_2+k_0)^2+i(k_2+k_0)x_t}
\eeq
and so the reflected part of the state is
\beq
|t\rangle_{\rm{reflected}}=-i e^{-i\ok 0 t}\pin{k_2} R(-k_2) e^{-\sigma^2(k_2+k_0)^2+i(k_2+k_0)x_t}|k_2\rangle_0.
\eeq
Here we have used the fact that, by energy conservation, the reflected part of the state is necessarily at $k_2=-k_1$ and so corresponds to the pole contribution.

As $\sigma |k_0|\gg1$ and $m\sigma\gg1$, one may approximate $R(-k_2)$ by its value at the peak of the Gaussian
\beq
|t\rangle_{\rm{reflected}}=-i e^{-i\ok 0 t} R(k_0)\pin{k_2} e^{-\sigma^2(k_2+k_0)^2+i(k_2+k_0)x_t}|k_2\rangle_0.
\eeq

The scattering probability is simply
\beq
P(k_0)=\frac{{}_{\rm{reflected}}\langle t|t\rangle_{\rm{reflected}}}{\langle t=0|t=0\rangle}=|R(k_0)|^2.
\eeq
The inner products were technically divergent as the states are nonrenormalizable.  Therefore they were computed using the reduced inner product of Ref.~\cite{menorm}.  This norm contains additional, subdominant terms, which change the meson number by one.  However it is clear that these vanish here, as all mesons, long after an interaction, are far from the kink but the subdominant terms contain factors of $\Delta_{kB}$ which have support close to the kink.  The only exception to this argument is Stokes scattering, in which the final state contains an excited shape mode, but then energy conservation would demand that the final state meson does not have momentum $-k_0$, which does not describe elastic scattering.  We will treat this term in Ref.~\cite{ellong} where we will see that it exactly cancels a final state correction.

\subsection{Calculating the Elastic Scattering Amplitude}

We have now seen that $R(k)$ is the elastic scattering amplitude for an incoming meson with momentum $k$.  The function $R(k)$ was, in turn, defined to be the residue of the pole in the Lippmann-Schwinger equation~(\ref{ls}).

Comparing the definition (\ref{ls}) with the result (\ref{g2}), one finds
\beq
R(k_0)=\frac{1}{Q_0}\frac{\partial k_ 0}{\partial \ok 0}\left( \rho_{ k_0}(-k_0) -\hat \gamma_{2 k_0}^{21}(-k_0)\right)=\frac{\ok 0}{Q_0 k_0}\left( \rho_{ k_0}(-k_0) -\hat \gamma_{2 k_0}^{21}(-k_0)\right).
\eeq
The expressions for $\rho$ and $\hat\gamma$ in Eqs.~(\ref{rho},\ref{hg}) simplify slightly to
\bea
\rho_{ k_0}(-k_0)&=&
\frac{\lambda Q_0}{4\ok{0}}V_{\I -k_0 - k_0}
-\frac{\sqrt{\lambda Q_0}}{4\ok 0}\Delta_{- k_0 B}V_{\I -k_0}\\
&&\hspace{-1.4cm}+\frac{\lambda Q_0}{16\ok 0}\ppinkp{2}\frac{V_{- k_0 k\p_1 k\p_2}V_{-k\p_1-k\p_2-k_0}}{\okp1\okp2}\left(\frac{1}{\ok 0-\okp1-\okp2+i\epsilon}-\frac{1}{\ok 0+\okp1+\okp2}\right)\nonumber\\
&&\hspace{-1.4cm}-\frac{\sqrt{\lambda Q_0}}{8\ok 0}\ppin{k\p}\frac{\left(\okp{}\Delta_{k\p B}+2\sqrt{\lambda Q_0}V_{\I  k\p}\right)
V_{-k\p- k_0 -k_0}}{\okp{}^2}\nonumber
\eea
and
\bea
-\hat\gamma_{2 k_0}^{21}(-k_0)&=&\frac{1}{4}\ppin{k\p}\left(\frac{\ok0}{\okp{}}+\frac{\okp{}}{\omega_{ k_0}}
\right)\Delta_{- k_0,-k\p}\Delta_{-k_0k\p}\\
&&+\frac{\sqrt{\lambda Q_0}}{4\omega_{ k_0}}\Delta_{-k_0 B}V_{\I - k_0}+\frac{\sqrt{\lambda Q_0}}{8\ok 0}\ppin{k\p}
\frac{\Delta_{-k\p B}V_{- k_0 -k_0 k\p}}{\okp{}}.\nonumber
\eea
Again we have chosen the sign in the pole to correspond the ``in" state from the Lippmann-Schwinger equation.  Note that the terms quadratic in $\Delta_{kB}$ have cancelled.  They would have led to elastic scattering with no intermediate mesons, corresponding to the Yukawa terms reported in Ref. \cite{uehara91}.

\begin{figure}[htbp]
\centering
\includegraphics[width = 0.45\textwidth]{figa.pdf}
\includegraphics[width = 0.45\textwidth]{figb.pdf}
\includegraphics[width = 0.45\textwidth]{figc.pdf}
\includegraphics[width = 0.45\textwidth]{figd.pdf}
\caption{Six diagrams represent the contributions $A(k_0)$, $B(k_0)$, $C(k_0)$ and $D(k_0)$ to the elastic scattering amplitude $R(k_0)$.  Here time runs to the left.}\label{abcdfig}
\end{figure}

Assembling these ingredients, one finds that the amplitude is the sum of the contributions from four kinds of processes
\beq
R(k_0)=\lambda(A(k_0)+B(k_0)+C(k_0)+D(k_0))
\eeq
where
\bea
A(k_0)&=&
\frac{1}{4\lambda Q_0 k_0}\ppin{k\p}\left(\frac{\ok0^2+\okp{}^2}{\okp{}}\right)\Delta_{- k_0,-k\p}\Delta_{-k_0k\p} \label{abcd}\\
B(k_0)&=&
\frac{V_{\I -k_0 - k_0}}{4 k_0}
\nonumber\\
C(k_0)&=&
-\frac{1}{4k_0}\ppin{k\p}\frac{V_{\I  k\p}
V_{-k\p- k_0 -k_0}}{\okp{}^2}
\nonumber\\
D(k_0)&=&
\frac{1}{8k_0}\ppinkp{2}\frac{\left(\okp1+\okp2\right)V_{- k_0 k\p_1 k\p_2}V_{-k_0 -k\p_1-k\p_2}}{\okp1\okp2\left(\ok 0^2-(\okp1+\okp2)^2+i\epsilon\right)}.
\nonumber
\eea
Correspondingly, the probability is the sum squared of these four contributions
\beq
P(k_0)=\lambda^2|A(k_0)+B(k_0)+C(k_0)+D(k_0)|^2. \label{peq}
\eeq

\subsection{The Four Processes}

The four kinds of processes are depicted schematically in Fig.~\ref{abcdfig}.  Only the meson lines are shown, although the kink is treated fully dynamically and one should remember that the internal lines $k\p$ have values which run over not only the continuum modes, but also any shape modes that the kink may possess.   $\I(x)$ is a loop factor that arises when a meson loop contains a single vertex.  The terms $V_{k_1\cdots k_n}$ are the coupling constants responsible for $n$-meson interactions.  On the other hand, $V_{\I k_1\cdots k_n}$ is the coupling constant for the interaction of $n$ mesons plus a meson loop.  The matrix $\Delta_{k_1 k_2}$ describes the interaction of a meson with the momentum of the kink center of mass, changing the meson momentum from $k_1$ to $k_2$.

In process $A$, an incoming meson scatters off the kink, giving a kick to the momentum of the kink's center of mass.  To see this, recall that $\phi_0$ and $\pi_0$ are the usual $x$ and $p$ in the quantum mechanical description of the kink center of mass, and the first collision multiplies the corresponding wave function by $x$.  Next the meson interacts again, yielding a $\phi_0^2$, while it bounces off the kink.  The free evolution of the kink contains a nonrelativistic kinetic term $\pi_0^2/2$ which eliminates the $\phi_0^2$ and returns the kink to its ground state after the meson has left.

The process $B$ corresponds to the meson reflecting off the kink, but the interaction is in fact a four-point interaction involving a virtual meson-antimeson pair that propagates briefly inside the kink.  

The process $C$ is similar, but when the meson reflects it leaves a virtual meson, which is annihilated by such a meson-antimeson pair.  In Ref.~\cite{metad} we showed that such tadpoles can be removed by a quantum correction to the classical kink solution appearing in the displacement function $\df$, chosen so as to include a linear term in the Hamiltonian which exactly cancels the tadpole diagram.  We saw that such a change of prescription does not affect the quantum corrections to the kink mass, it simply reorganizes them.  We suspect that also in the present case, such a choice would lead the tadpole diagrams to disappear, but the same contribution would arise from elsewhere. 

Process $D$ is the most interesting.  Here the meson reflects via the creation of two virtual mesons, one or both of which may be shape modes.  In particular, if they are both shape modes, we see that the denominator of $D(k_0)$ possesses a pole at which the incoming meson energy is the energy of the twice excited shape mode.  We expect that, including higher order terms, this pole will assume the usual Breit-Wigner shape of an unstable resonance, with a width equal to the inverse lifetime computed in Ref.~\cite{alberto}.  Note that there is no such contribution in Eq.~(3.19) of Ref.~\cite{uehara91}, in which there is at most one intermediate meson.

The denominator has an imaginary part, corresponding again to same $i\epsilon$ as in the Lippmann-Schwinger equation.  In this context, it was found in Ref.~\cite{memult} and an alternate derivation appeared in Ref.~\cite{menorm}.  The pole corresponds to the case in which the two-meson intermediate state is on-shell.  In the case of the Sine-Gordon model, the pole will be exactly canceled by a term in the $V_{- k_0 k\p_1 k\p_2}$ in the numerator, which is a consequence of the fact that meson multiplication is forbidden by integrability \cite{memult}.

\section{Example: The Sine-Gordon Model} \label{exsez}
The Sine-Gordon model is an integrable quantum field theory described by the potential
\beq
V(\sqrt{\lambda}\phi(x))=m^2\left(1-{\rm{cos}}(\sqrt{\lambda}\phi(x)\right).
\eeq
Its kink has profile
\beq
f(x)=\frac{4}{\sl}{\rm{arctan}}\left(e^{mx}\right)\hsp Q_0\lambda=8
\eeq
and is called the Sine-Gordon soliton, because it is a soliton in the original sense, it scatters without deformation.

In particular, as in all integrable field theories, the S-matrix only allows for elastic scattering of particles with the same mass.  The soliton and the meson do not have the same mass, and so their elastic scattering is not allowed.  It is therefore a critical test of our result that $R(k_0)=0$ in this case.

Let us now calculate $R(k_0)$.  From the potential and the soliton solution, one can easily find
\beq
V^{(3)}[\sqrt{\lambda} f(x)]=2 m^2  \tanh (m x)\sech (m x)\hsp
V^{(4)}[\sl f(x)]= m^2 \left(-1+2\sech^2(mx)\right).
\eeq
The normal modes $\g_k(x)$ of the Sine-Gordon kink, and the loop factor $\I(x)$ are given by
\beq
\g_k(x)=\frac{e^{-ikx}}{\ok{}}(k-im \tanh(mx))\hsp
\I(x)=-\frac{\sech^2(mx)}{2\pi}.
\eeq
From these, one can easily compute the other relevant functions
\bea
\Delta_{k_1k_2}&=&-i(k_1-k_2)\pi\delta(k_1+k_2)+\frac{i\pi}{2}\frac{(k_2^2-k_1^2)}{\omega_{k_1}\omega_{k_2}}{\rm{csch}}\left(\frac{\pi\left(k_1+k_2\right)}{2m}\right)\\
V_{\I-k-k}&=&\frac{k(4k^2+m^2)}{15m^2}\csch\left( \frac{k\pi}{m}\right)\hsp
V_{\I k}=\frac{i}{8m^2}\ok{}^3 \sech\left( \frac{k\pi}{2m}\right)\nonumber\\
V_{k_1k_2k_3}&=&\frac{\pi i(\ok 1+\ok 2+\ok 3)}{4\ok 1\ok 2\ok 3}(\ok 1+\ok 2-\ok 3)(\ok 1-\ok 2+\ok 3)\nonumber\\
&&\times(-\ok 1+\ok 2+\ok 3)
\sech\left( \frac{(k_1+k_2+k_3)\pi}{2m}\right).\nonumber
\eea

\begin{figure}[htbp]
\centering
\includegraphics[width = 0.65\textwidth]{SG.pdf}
\caption{Contributions $A$ (red), $B$ (black), $C$ (blue) and $D$ (brown) to the elastic scattering amplitude in the Sine-Gordon model}\label{sgfig}
\end{figure}

Substituting these into our general result (\ref{abcd}) one can find the individual contributions to soliton-meson scattering in the Sine-Gordon model
\bea
A(k_0)&=&\frac{\pi^2}{128m k_0\ok {0}^2}\pin{k\p}\frac{(\ok 0^4-\okp{}^4)(\okp{}^2-\ok 0^2)}{\okp{}^3}\csch\left( \frac{(k_0+k\p)\pi}{2m}\right)\csch\left( \frac{(k_0-k\p)\pi}{2m}\right)\nonumber
\\
B(k_0)&=&\frac{(4k_0^2+m^2)}{60m^2}\csch\left( \frac{k_0\pi}{m}\right)
\\
C(k_0)&=&\frac{\pi}{128m^2k_0\ok 0^2}\pin{k\p}\left(4\ok 0^2\okp {}^2-\okp{}^4\right)\sech\left( \frac{k\p\pi}{2m}\right)\sech\left( \frac{(2k_0+k\p)\pi}{2m}\right)
\nonumber\\
D(k_0)&=&-\frac{\pi^2 }{128 k_0\ok 0^2}\pinkp{2}\frac{(\okp 1+\okp 2)(\ok 0+\okp 1+\okp 2)^2}{\okp 1^3\okp 2^3
\left[(\okp 1+\okp2)^2-\ok 0^2 
\right]}(\ok 0+\okp 1-\okp2)^2\nonumber\\
&&\hspace{-1.5cm}\times(\ok 0-\okp 1+\okp2)^2 (-\ok 0+\okp 1+\okp2)^2\sech\left( \frac{(k\p_1+k\p_2+k_0)\pi}{2m}\right)\sech\left( \frac{(k\p_1+k\p_2-k_0)\pi}{2m}\right).\nonumber
\eea
These are each nonzero.  However, we have checked numerically that, up to at least one part in $10^{4}$, they cancel for all regimes of $k_0>0$.  The relative contributions can be seen in Fig.~\ref{sgfig}.

\section{Remarks}
This paper presented a quick and dirty calculation of the amplitude and probability of elastic kink-meson scattering.  It identified contributions from Lippmann-Schwinger pole terms, but it did not show that there are no other contributions.  Nonetheless, in the case of the Sine-Gordon model, it led to a seemingly miraculous cancellation of the amplitude which was demanded by integrability, and so provided a consistency check of our results.  In the future a more thorough investigation would nonetheless be warranted, perhaps by starting with an initial asymptotic meson state, evolving it, and calculating the probability that the final meson is moving backwards \cite{ellong}.

In the case of models whose kinks have shape modes, we have seen that our result contains a contribution $D(k_0)$ in which the intermediate state contains two excited shape modes.  These lead to poles in the scattering amplitude.  Near the poles, our perturbative expansion fails as the pole contribution is greater than $1/\sl$.  Here we should include bubble diagrams to fix this problem, and we suspect that after summing these bubble diagrams the pole will shift in the complex plane and assume the usual Breit-Wigner form for a resonance.

Is kink-meson scattering important?  Recently the interactions of kinks with radiation has entered the spotlight~\cite{mech22,multi22,dorey23,nav23,hahne23} as it has been discovered the interactions with bulk degrees of freedom play at least as important of a role in kink dynamics as shape modes~\cite{doreyprl}.  Yet, with some notable exceptions~\cite{kinkquantscat,alonso14,vach23}, so far this has largely been at the classical level, where reflectionless kinks are indeed reflectionless.  Here we see that at the quantum level, these interactions change qualitatively.

\section* {Acknowledgement}

\noindent
JE is supported by NSFC MianShang grants 11875296 and 11675223.

\end{document}

\subsection{Motivation}

The collective coordinate method of Ref.~\cite{gjscc} allows for arbitrary calculations involving quantum kinks\footnote{At one loop, there are many robust and efficient methods for treating solitons, beginning with Ref.~\cite{dhn2}.  Ref.~\cite{wrev} provides a recent review.}.  The position of the kink itself is quantized, and the fields are expanded about this time-dependent position.  However, the interplay of the kink position and the field expansion is complicated.  To bring the operators into a canonical form, to allow for quantization, one requires a nonlinear canonical transformation already in the classical theory.  This transformation does not leave the quantum path integral invariant, and so in the quantum theory, an infinite series of terms needs to be added to the Hamiltonian \cite{gj76}.  These complications have made all but the simplest problems impractical.  For example, two-loop corrections to kink masses have only been computed when they are already known as a result of integrabilility \cite{vega,verwaest} or supersymmetry \cite{shifmananomalia}.  Also, kink-meson scattering has been restricted to calculating the leading contribution to an effective Yukawa coupling \cite{hayashi1,hayashi2}.  However, recently it is led to promising developments in the calculation of form factors \cite{accel,andyprl,raggio}.

A new, simpler method, linearized kink perturbation theory, has been formulated in Refs.~\cite{mekink,me2loop}.  Here the kink fields are expanded as if the kink were at a fixed base point.   As a result, the fields are canonical from the beginning.  The price is that the distance of the kink from the base point is treated perturbatively, as a semiclassical expansion in the coupling.  Thus it is not reliable if the kink wave packet extends beyond the radius of convergence of the expansion.  The radius of convergence, in the sense of an asymptotic series, is more than the de Broglie wavelength of the kink, but less than its classical diameter.  This leaves the method applicable to localized kink wave packets\footnote{One should draw a distinction between a wave packet for the center of mass of the kink-meson system, whose size is treated perturbatively and thus is bounded, and a wave packet describing the relative position of a meson with respect to the kink, which is treated exactly and whose monochromatic limit poses no complications.}, or more generally localized soliton wave packets, as arise in many applications such as solitonic dark matter \cite{memono,fuzzy14}, pinned Abrikosov vortices and kink-impurity interactions \cite{impure,chris}.

However, sometimes one is interested in the opposite regime, in which the kink is in a translation-invariant state, such as its ground state or the ground state of a system of a kink and a finite number of mesons.  A translation-invariant state is a quantum superposition, summing over all possible simultaneous and equal translations of the kink and mesons, which necessarily keep the relative distances fixed.  This is relevant \cite{hayashirep}, for example, to treating proton-meson scattering using the Skyrme \cite{skyrme,smorg} model.  Here one must simultaneously consider kinks at positions arbitrarily far from the base point.  The perturbative approach above naively fails miserably.

The solution to this problem is to use translation-invariance\footnote{An alternative approach, which does not require translation-invariance, was presented in Ref.~\cite{point22}.  However so far zero-modes have not been included.}.  All of the information regarding a translation-invariant kink state is contained in the configuration involving the kink at the base point, and so a study of that case, together with translation-invariance, yields all quantities.  In the case of computations of kink masses, this is achieved \cite{me2loop} by projecting the state, perturbatively, onto the kernel of the momentum operator and then solving the Schrodinger equation for the power series expansion in the kink position about the base point.  If it is solved for all coefficients in the expansion about the base point, it is solved everywhere by translation invariance.  And indeed, the method has been shown to agree with previous calculations of form factors and two-loop mass corrections where available and, due to its simplicity, it has provided novel calculations of the mass of non-integrable, non-supersymmetric kinks \cite{phi42loop} and even excited kinks \cite{menormal}, as well as kink form factors in non-integrable models \cite{hengyuan}.

However, this problem becomes more severe when one tries to compute dynamical quantities.  Here one needs to calculate inner products of states.  These translation-invariant states are non-normalizable, and so their inner products do not exist.  Usually one can evade this problem by regularizing the state in the form of a wave packet and taking the limit in which the wave packet becomes large.  Unfortunately, in the case of linearized perturbation theory, this cannot be done as there is no way to treat a finite wave packet which is larger than the radius of convergence.  One may try to avoid the problem by compactifying space.  However, such a compactification requires particular boundary conditions and there is no guarantee that finite contributions do not remain when the compactification radius is taken to infinity.  Thus, if compactification can be avoided, it is better in our opinion to avoid it.

So far in dynamical problems we have side-stepped this complication.  In the case of excited kink decay \cite{alberto} and meson multiplication \cite{memult} the inner products that appeared were always in fractions where the same inner product appeared in both the numerator and the denominator of probabilities, and so we canceled them.  However, at higher orders, different states will appear in the numerator and denominator.  

\subsection{Reduced Inner Product}

In this note we suggest a new strategy for dealing with norms of translation-invariant states without regularizing the infinities.  As the inner products of interest always appear in both the numerator and denominator of an expression for an observable, we quotient both by the infinite volume translation group, being careful to keep the relevant Jacobian factor.  This strategy resembles gauging by the global translation symmetry, which simultaneously shifts the kink and also the mesons.  Intuitively this is also related to a compactification of radius zero, except that the distance between the kink and mesons is preserved and so they are effectively in an infinite space.

We restrict our attention to the kink sector.  This is the Fock space of any finite number of mesons in the presence of a quantum kink.  However a generalization to other topological sectors is obvious.

Our main result, a formula for the reduced inner product of any two translation-invariant states, is presented in Eq.~(\ref{padqft}).   Intuitively, the ordinary inner product can be written as an integral over the collective coordinate $x$ and the reduced inner product is defined by inserting $\delta(x)$.  The collective coordinate transforms under translations via the usual rigid shift.  In linearized perturbation theory one works using not the collective coordinate $x$, but rather the linearized coordinate $y$, whose transformation under translations is rather complicated.  Our formula (\ref{padqft}) replaces the $\delta(x)$ with $y\p(x)\delta(y)$, where the Jacobian factor $y\p(x)$ is an operator. Amazingly, as a result of the $\delta(y)$, this formula is simpler than the usual formula for the inner product, as it does not use the dependence of the state on the zero-modes, which have eigenvalue $y$.  This is not obviously inconsistent,  as, in the case of translation-invariant states, the zero-mode dependence is entirely fixed by the translation invariance \cite{me2loop}.  The price for this simplification is the addition of $y\p(x)$, which contains two finite quantum corrections above the naive inner product, which mix sectors whose meson numbers differ by one unit.  These corrections reflect the fact that the kink zero-mode mixes with the normal modes upon translation.


Three applications are presented.  First, this allows us to place formal manipulations, in which these norms were canceled in Refs.~\cite{alberto,memult}, on more solid footing.  Also, it allows us to treat cases where more complicated inner products arise, such as matrix elements of zero-modes.  In fact, this already happens in the order $O(\sqrt{\lambda})$ contribution to the meson multiplication amplitude, arising from an $O(\sqrt{\lambda})$ correction to the initial or final state and no interaction.  We thus apply our formalism to calculate these corrections, and to show that they vanish at this order.  

Finally, we use this to calculate the leading order correction to the one-meson state consisting of two-meson states.  It was already calculated in Ref.~\cite{menormal} but the result involved a pole at a location where the Hamiltonian is degenerate, and so distinct prescriptions for treating the pole yield legitimate, yet inequivalent, Hamiltonian eigenstates.  We find that one particular prescription for the pole yields the physically-motivated initial conditions for meson-kink scattering, in which the initial state never contains two mesons.

In Sec.~\ref{revsez} we review the linearized kink perturbation theory of Refs.~\cite{mekink} and \cite{me2loop}.  Next in Sec.~\ref{qmsez} we present our construction of the reduced inner product in quantum mechanics.  This construction is adapted to kink sectors of quantum field theories in Sec.~\ref{qftsez}.  In Sec.~\ref{exsez} we provide some examples of reduced inner products.  Finally in Sec.~\ref{multsez} we apply this formalism to evaluate the meson multiplication amplitude using translation-invariant states, finding the same result as Ref.~\cite{memult} where a basis of eigenstates of the free Hamiltonians were used and divergences in the numerator and denominator of various expressions were cancelled naively.  In addition, we find the quantum corrections to the initial state which are relevant to this experiment, corresponding to a prescription for treating the pole in Ref.~\cite{menormal}.

\section{Review} \label{revsez}

While we suspect that it may be generalized to theories of greater phenomenological interest, so far linearized kink perturbation theory has only been formulated for (1+1)-dimensional quantum field theories with a Schrodinger picture scalar field $\phi(x)$, conjugate to $\pi(x)$, and a Hamiltonian
\begin{equation}
H=\int d x: \mathcal{H}(x):_a\hsp \mathcal{H}(x)=\frac{\pi^2(x)}{2}+\frac{\left(\partial_x \phi(x)\right)^2}{2}+\frac{V(\sqrt{\lambda} \phi(x))}{\lambda}.
\end{equation}
The potential $V$ has degenerate minima and a classical kink solution $f(x)$ interpolates from one to another.  The normal ordering $::_a$ renders such theories UV-finite.  It is defined at the mass scale $m$, defined by
\beq
m^2=V^{(2)}(\sqrt{\lambda} f(\pm \infty))\hsp
V^{(n)}(\sqrt{\lambda} \phi(x))=\frac{\partial^n V(\sqrt{\lambda} \phi(x))}{(\partial \sqrt{\lambda} \phi(x))^n}.
\eeq
This in fact defines two values of the mass, one at the vacuum on each side of the kink.  If the masses are different, then one-loop corrections to the vacuum energy imply that one vacuum is a false vacuum, and the kink will accelerate towards it \cite{wstabile}.  Such a kink does not correspond to a Hamiltonian eigenstate in the quantum theory and we will not consider it further.  We will treat the theory using a semiclassical expansion in the coupling $\sqrt{\lambda}$.

We will consider several sectors of the Hilbert space.  The vacuum sector consists of configurations with no kinks, and a finite number of perturbative excitations of $\phi(x)$, which we will call mesons.  Here, one meson is a plane wave, which is created by a creation operator defined using the usual plane wave decomposition of $\phi(x)$ and $\pi(x)$.  More precisely, there is a vacuum sector for each minimum of the classical potential $V$, and sometimes one needs to distinguish between the vacuum sectors to the left and to the right of the kink.  

The kink sector consists of a single kink and a finite number of excitations.  We will refer to the excitations which are unbound again as mesons, and those which are bound as shape modes.  These are also created by creation operators, $B^\ddag_S$ and $B^\ddag_k$ respectively, defined by the decomposition of $\phi(x)$ and $\pi(x)$ in terms of normal modes $\g(x)$ \cite{cahill76}
\bea
\phi(x) &=&\phi_0 \mathfrak{g}_B(x)+\ppin{k} \left(B_k^{\ddag}+\frac{B_{-k}}{2 \omega_k}\right) \mathfrak{g}_k(x)\hsp
B^\ddag_k=\frac{B^\dag_k}{2\ok{}}\hsp
B^\ddag_S=\frac{B^\dag_S}{2\omega_S} \label{dec}\\
\pi(x) &=&\pi_0 \mathfrak{g}_B(x)+i \ppin{k}\left(\omega_k B_k^{\ddag}-\frac{B_{-k}}{2}\right) \mathfrak{g}_k(x)\hsp
B_S=B_{-S}\hsp \ppin{k}=\pin{k}+\sum_S \nonumber
\eea
where $\phi_0$ is the zero mode.

The normal modes $\g(x)$ are constant frequency $\omega$ solutions of the Sturm-Liouville equation for infinitesimal perturbations about a kink
\beq
\V{2}{\g}(x)=\omega^2{\g}(x)+{\g}^{\prime\prime}(x)\hsp \phi(x,t)=e^{-i\omega t}\g(x). \label{sl}
\eeq
There will always be one solution, $\g_B(x)$, with $\omega_B=0$ corresponding to a zero mode.  The shape modes are those with $0<\omega_S<m$.  Continuum modes have frequencies 
\beq
\ok{}=\sqrt{m^2+k^2}.
\eeq
All modes are assembled and normalized to satisfy $\g^*_k=\g_{-k}$ and the completeness relations
\beq
\int dx |{\g}_{B}(x)|^2=1,\
\int dx {\g}_{k_1} (x) {\g}^*_{k_2}(x)=2\pi \delta(k_1-k_2),\ 
\int dx {\g}_{S_1}(x){\g}^*_{S_2}(x)=\delta_{S_1S_2}. \label{comp}
\eeq
We fix the sign of $\g_B$ via
\beq
\g_B(x)=-\frac{f\p(x)}{\sqrt{Q_0}} \label{gb}
\eeq
where $Q_i$ is the $i$-loop correction to the energy of the ground state kink.  Note that the sign is not the same as in previous papers.  $Q_0$ is just the energy of the classical field configuration.

Any operator can be expanded in terms of $\phi(x)$ and $\pi(x)$ or alternatively in terms of $\phi_0,\ \pi_0,\ B^\ddag_k,\ B_k,\ B^\ddag_S$\ and\ $B_S$.  From the canonical commutation relations for $\phi(x)$ and $\pi(x)$, we find the algebra satisfied by this second basis
\beq
\left[\phi_0, \pi_0\right]=i, \quad\left[B_{S_1}, B_{S_2}^{\ddagger}\right]=\delta_{S_1 S_2}, \quad\left[B_{k_1}, B_{k_2}^{\ddagger}\right]=2 \pi \delta\left(k_1-k_2\right).
\eeq

We would like to perform calculations involving states in the one-kink sector.  The problem is that these states are nonperturbative.  This is easy to understand in classical field theory, where small perturbations about the kink correspond to field configurations $\phi(x,t)$ close to $f(x)$, which is far from zero, and so higher moments of $\phi(x,t)$ are not small.  The solution in classical field theory is to decompose $\phi(x,t)=f(x)+\eta(x,t)$ and work with $\eta(x,t)$, which is small and so can be treated perturbatively.  We would like an analogous procedure in the quantum theory, where $\phi(x)$ is a Schrodinger picture quantum field.

The replacement $\phi(x,t)\rightarrow\eta(x,t)$ in the classical theory can be achieved, in the quantum theory, by conjugating with the displacement operator $\df$
\beq
\df={{\rm Exp}}\left[-i\int dx f(x)\pi(x)\right]\hsp
\df^\dag \phi(x) \df = \phi(x)-f(x).  \label{dfd}
\eeq
This displacement operator is unitary and commutes with normal ordering.  It also acts on the states, mapping a vacuum sector state to a one-kink sector state.

So far we have worked in the defining frame of the Hilbert space.  This is the usual representation in which the Hamiltonian $H$ generates time translations and the momentum $P$ generates spatial translations.  Energies are eigenvalues of $H$ while $e^{-iHt}$ yields finite time evolution.  All states in all sectors can be written in the defining frame, using Dirac kets.

Now we want to write the same Hilbert space in a new frame, called the kink frame.  We define the state $|\psi\rangle$ in the kink frame to be the state $\df|\psi\rangle$ in the defining frame.  In this definition, $\df$ plays the role of a passive transformation, changing the coordinate system used to describe the Hilbert space without changing the state.  This passive transformation transforms not the states but rather the operators that act on these states.  For example, time and space translations in the kink frame are generated by the kink Hamiltonian $H\p$ and kink momentum $P\p$
\beq
H\p=\df^\dag H\df\hsp
P\p=\df^\dag P\df=P+\sqrt{Q_0}\pi_0. \label{df}
\eeq
As a consistency check, note that the energy of $|\psi\rangle$ in the kink frame, as measured by $H\p$, is equal to the energy of $\df|\psi\rangle$ in the vacuum frame, as measured by $H$
\beq
H\df|\psi\rangle=E\df|\psi\rangle\Rightarrow H\p|\psi\rangle=\df^\dag H\df|\psi\rangle=E|\psi\rangle.
\eeq
This is a trivial manipulation, but for kink states the eigenvalue equation for $H\p$ is perturbative while for $H$ it is nonperturbative.  Thus in the kink frame, kink states are within the range of perturbation theory, just like $\eta(x,t)$ in classical field theory.  Thus one can find kink states perturbatively in the kink frame, and if desired they can be transformed back to the defining frame using $\df$.

We expand the kink Hamiltonian
\beq
H\p=\sum_{i=0}^\infty H\p_i \label{semi}
\eeq
where $H\p_i$ is of order $O(\lambda^{i/2-1})$.  The terms $H\p_i$ were found in Ref.~\cite{mekink}
\bea
H\p_0&=&Q_0\hsp H\p_1=0\hsp
H\p_2=Q_1+H\p_{\text {free }}, \quad H\p_{\text {free }}=\frac{\pi_0^2}{2}+\omega_S B_S^{\ddag} B_S+\int \frac{d k}{2 \pi} \omega_k B_k^{\ddag} B_k
\nonumber\\
H\p_{n>2}&=&\lambda^{\frac{n}{2}-1}\int dx \frac{V^{(n)}(\sqrt{\lambda} f(x))}{n !}: \phi^n(x):_a.
\eea
Note that the terms in $H\p_{\rm{free}}$ correspond to solved systems in quantum mechanics.  The $\pi_0^2$ term is the kinetic energy for a free particle of mass $Q_0$, and so we see that $\phi_0/\sqrt{Q_0}$ and $\sqrt{Q_0}\pi_0$ are the position and momentum of the center of mass of the kink.  The factor of $\sqrt{Q_0}$ relating the kink position to the eigenvalue of $\phi_0$ will appear again later, as the leading term in the reduced norm. The other terms in $H\p_{\text {free }}$ are harmonic oscillators, one for each shape mode and continuum mode.

All Hamiltonian eigenstates $|\psi\rangle$ will be decomposed in a semiclassical expansion
\beq
|\psi\rangle=\sum_{i=0}^\infty|\psi\rangle_i  \label{semi}
\eeq
where $|\psi\rangle_i$ is of order $O(\lambda^{i/2})$.  The leading components $|\psi\rangle_0$ of all states $|\psi\rangle$ solve the leading order eigenvalue equation for $H\p$, and so are defined to be the eigenstates of $H\p_2$.   

In the kink frame, the ground state of the kink sector is $\vac$.  The leading component $\vac_0$ is the ground state of the system defined by each term in $H\p_{\rm{free}}$ and so satisfies
\beq
\pi_0\vac_0=B_k\vac_0=B_S\vac_0=0. \label{v0}
\eeq
Similarly one may define, at leading order, states with one kink and one or two mesons
\beq
|k\rangle_0=B^\ddag_k\vac_0\hsp
|kk\p\rangle_0=B^\ddag_k B^\ddag_{k\p}\vac_0. \label{2m}
\eeq

\section{Reduced Inner Products in Quantum Mechanics} \label{qmsez}

Our main result will be a finite, reduced inner product for translation-invariant states in the one kink sector.  It is derived by quotienting the ordinary inner product by the translation group, keeping careful track of the Jacobian term.  In this section, we will motivate our result by defining a similar reduced inner product in quantum mechanics.

The Jacobian term is nontrivial because the translation operator $P\p$ acts nonlinearly on the linearized coordinates $y$, defined to be the eigenvalues of $\phi_0$.  However, it acts linearly, as a simple shift, on the collective coordinate $x$.  We will derive our result by first computing the, rather trivial, reduced inner product for a state expressed in collective coordinates $x$, and later will derive a matching condition between collective and linearized coordinates $y$ which allows us to define the reduced inner product on a state expressed in terms of linearized coordinates.

\subsection{Collective Coordinates: Definitions}

We begin by defining the collective coordinate description of states in our quantum mechanical Hilbert space.

Let $|e^n\rangle$ be an orthogonal basis of states which are invariant under the translation operator $P\p$.  Each can be decomposed into eigenstates of the collective coordinate operator $\hat x$
\beq
|e^n\rangle=\int dx |nx\rangle_x\hsp
\hat x |n x\rangle_x=x|n x\rangle_x\hsp
[\hat x,P\p]=i
\eeq
where $n$ is an integer quantum number and
\beq
P\p\int dx F(x) |nx\rangle_x=-i\int dx F\p(x) |nx\rangle_x\hsp
{}_x\langle n_1 x_1|n_2x_2\rangle_x=\delta_{n_1 n_2}\delta(x_1-x_2). \label{xdef}
\eeq
Note that the first relation in (\ref{xdef}) implies that $P\p$ acts on this basis like a momentum operator in quantum mechanics
\beq
[\hat x,e^{-ix_2 P\p}]=x_2e^{-ix_2 P\p}\Rightarrow
e^{-ix_2 P\p}|n x_1\rangle_x=|n,x_1+x_2\rangle_x.
\eeq

While the $|e^n\rangle$ are not normalizable, we define the reduced inner product by
\beq\label{redee}
\langle e^m|e^n\rangle_{\rm{red}}=\delta_{mn}.
\eeq
Any translation-invariant $|\psi\rangle$ can be expanded
\beq
|\psi\rangle=\sum_n \psi_n|e^n\rangle.
\eeq
Therefore any reduced inner product is
\beq
\langle \phi|\psi\rangle_{\rm{red}}=\sum_n \phi^*_n \psi_n.
\eeq

\subsection{Linearized Coordinates: Inner Product}

Consider another basis of states $|ny\rangle_y$ where $\phi_0$ and $\pi_0$ are Hermitian operators such that
\beq
\phi_0|ny\rangle_y=y|ny\rangle_y\hsp \pi_0\int dy F(y) |ny\rangle_y=-i\int dy F\p(y)|ny\rangle_y.
\eeq
Define the integral quantum number $n$ such that 
\beq
{}_y\langle n_1 y_1|n_2 y_2\rangle_y=\delta_{n_1n_2} G_{n_1}(y_1,y_2) 
\eeq
for some functions $G_n$.

As $\phi_0$ is Hermitian, 
\beq
0={}_y\langle n y_1|(\phi_0-\phi_0)|n y_2\rangle_y=(y_1-y_2){}_y\langle n y_1|n y_2\rangle_y=(y_1-y_2)G_n(y_1,y_2)
\eeq
and so $G_n(y_1,y_2)$ is only nonvanishing if $y_1=y_2.$  Therefore we will write it simply as  $\delta(y_1-y_2)G_n(y_1)$ and
\beq
{}_y\langle n_1 y_1|n_2 y_2\rangle_y=\delta_{n_1n_2} \delta(y_1-y_2)G_{n_1}(y_1).
\eeq
As $\pi_0$ is Hermitian, for any functions $F_i(y)$ with compact support
\bea
0&=&\int dy_1\int dy_2\ {}_y\langle n y_1| F_1^*(y_1) (\pi_0-\pi_0) F_2(y_2)|n y_2\rangle_y\\
&=&\int dy_1\int dy_2\ {}_y\langle n y_1| \left[i F_1^{\prime *}(y_1)  F_2(y_2)+iF_1^{*}(y_1)  F\p_2(y_2)\right]|n y_2\rangle_y\nonumber\\
&=&i\int dy_1\int dy_2\left[F_1^{\prime *}(y_1)  F_2(y_2)+F_1^{*}(y_1)  F\p_2(y_2) \right]G_n(y_1)\delta(y_1-y_2)\nonumber\\
&=&i\int dy \partial_y(F_1^*(y)F_2(y))G_n(y)=-i\int dy F_1^*(y)F_2(y)\partial_y G_n(y).
\eea
As this is true for arbitrary functions with compact support, we find
\beq
\partial_y G_n(y)=0
\eeq
and so we replace $G_n(y)$ with $G_n$ and write
\beq
{}_y\langle n_1 y_1|n_2 y_2\rangle_y=\delta_{n_1n_2} \delta(y_1-y_2)G_{n_1}.
\eeq
Finally, we may renormalize the $|n y\rangle_y$ states by a factor of $1/\sqrt{G_n}$ so that 
\beq
{}_y\langle n_1 y_1|n_2 y_2\rangle_y=\delta_{n_1n_2} \delta(y_1-y_2). \label{yort}
\eeq

\subsection{Linearized Coordinates: Translations}

Consider the translation-invariant state
\beq\label{transinvar}
|\psi\rangle=\sum_n \int dy\hat\psi_n(y)|ny\rangle_y
\eeq
and assume that the translation operator is of the form
\beq
P\p=A\pi_0+B+C\phi_0 \label{pp}
\eeq
where $A$, $B$ and $C$ are matrices that commute with $\pi_0$ and $\phi_0$.  Note that the hat notation does not mean that $\hat \psi$ is an operator, but merely that it is the coefficient in the $y$ basis.

Acting the translation operator on the invariant state, one finds
\beq
P\p|\psi\rangle=\sum_{mn}\int dy \left[-iA_{mn}\hat \psi_n\p(y)+ B_{mn}\hat \psi_n(y)+ C_{mn}y\hat \psi_n(y)
\right]|my\rangle_y
\eeq
so that for invariant states
\beq
A\hat \psi\p(y)+i(B+yC)\hat \psi(y)=0.
\eeq
In particular, for small $\epsilon$
\beq
\hat\psi(\epsilon)=\hat\psi(0)+\epsilon\hat\psi\p(0)=\hat\psi(0)-i\epsilon A^{-1}B\hat\psi(0).
\eeq

Let $\hat{v}^j$ be an eigenvector of $A_{mn}$ such that
\beq
\sum_n A_{mn}\hat v^j_n=\lambda_j \hat v^j_m.
\eeq
Consider the translation-invariant state $|v^j\rangle$ defined by
\beq
|v^j\rangle=\sum_n \int dy \hat{v}^j_n(y) |n y\rangle_y\hsp \hat{v}^j_n(0)=\hat v^j_n.
\eeq
Then the translation generator yields
\bea
P\p|v^j\rangle=\sum_{mn}\int dy \left[-iA_{mn}\hat v_n^{j\prime}(y)+ B_{mn}\hat v^j_n(y)+ C_{mn}y\hat v^j_n(y)
\right]|my\rangle_y=0
\eea
so
\beq
\sum_n\left[-iA_{mn}\hat v_n^{j\prime}(y)+ B_{mn}\hat v^j_n(y)+ C_{mn}y\hat v^j_n(y)
\right]=0.
\eeq
In particular, for small $\epsilon$,
\beq
\hat v^j(\epsilon)=(1-i\epsilon A^{-1}B)\hat v^j. \label{dv}
\eeq

Now let us consider just the component at fixed $y$
\beq
|v^j,y\rangle_y=\sum_n \hat{v}^j_n(y) |n y\rangle_y\hsp |v^j\rangle=\int dy |v^j,y\rangle_y.
\eeq
This is not translation-invariant
\bea
P\p|v^j,0\rangle_y&=&P\p\sum_n \hat{v}^j_n |n 0\rangle_y=\sum_n \hat{v}^j_n P\p\int dy \delta(y) |n y\rangle_y=\sum_n \hat{v}^j_n \lim{\sigma\rightarrow 0} \frac{1}{\sigma\sqrt{2\pi}} P\p\int dy e^{-\frac{y^2}{2\sigma^2}} |n y\rangle_y\nonumber\\
&=&\sum_{mn} \hat{v}^j_n \lim{\sigma\rightarrow 0} \frac{1}{\sigma\sqrt{2\pi}} \int dy e^{-\frac{y^2}{2\sigma^2}}\left[ 
iA_{mn}\frac{y}{\sigma^2}+ B_{mn}+ C_{mn}y
\right] |m y\rangle_y\nonumber\\
&=&\sum_{mn} \hat{v}^j_n \lim{\sigma\rightarrow 0} \frac{1}{\sigma\sqrt{2\pi}} \int dy e^{-\frac{y^2}{2\sigma^2}}\left[ 
iA_{mn}\frac{y}{\sigma^2}+ B_{mn}
\right] |m y\rangle_y.
\eea
In the last equality we used the fact that, as $\sigma\rightarrow 0$, also $y\rightarrow 0$ with $y/\sigma$ fixed.  The coefficient of the $C$ term is $y$, which therefore goes to zero. 

Now let us consider a transformation by a finite distance $\epsilon$.  We will approximate it by the first order transformation inside of the limit, which is legitimate if, when we take $\sigma\rightarrow 0$, we also take $\epsilon/\sigma\rightarrow 0$.   The transformation is
\bea
e^{-i\epsilon P\p}|v^j,0\rangle_y&=&\sum_n \hat{v}^j_n \lim{\sigma\rightarrow 0} \frac{1}{\sigma\sqrt{2\pi}} e^{-i\epsilon P\p}\int dy e^{-\frac{y^2}{2\sigma^2}} |n y\rangle_y\\
&=&\sum_n \hat{v}^j_n \lim{\sigma\rightarrow 0} \frac{1}{\sigma\sqrt{2\pi}} (1-i\epsilon P\p)\int dy e^{-\frac{y^2}{2\sigma^2}} |n y\rangle_y\nonumber\\
&=&\sum_{mn} \hat{v}^j_n \lim{\sigma\rightarrow 0} \frac{1}{\sigma\sqrt{2\pi}} \int dy e^{-\frac{y^2}{2\sigma^2}}\left[\delta_{mn}
+\epsilon A_{mn}\frac{y}{\sigma^2}-i\epsilon B_{mn}
\right] |m y\rangle_y\nonumber\\
&=&\sum_{mn} \hat{v}^j_n \lim{\sigma\rightarrow 0} \frac{1}{\sigma\sqrt{2\pi}} \int dy e^{-\frac{y^2}{2\sigma^2}}\left[\delta_{mn}
+\epsilon \delta_{mn} \lambda_{j}\frac{y}{\sigma^2}-i\epsilon \lambda_j(A^{-1}B)_{mn}
\right] |m y\rangle_y.\nonumber
\eea
Using the expansion (\ref{dv})
\beq
e^{-\frac{(y-\lambda_j\epsilon)^2}{2\sigma^2}}\hat v^j(\lambda_j\epsilon)=e^{-\frac{y^2}{2\sigma^2}}\left[1+\frac{\lambda_j\epsilon y}{\sigma^2} -i\lambda_j\epsilon A^{-1}B
\right]\hat v^j
\eeq
we then conclude
\bea
e^{-i\epsilon P\p}|v^j,0\rangle_y&=&
\sum_{n} \hat v_n^j(\lambda_j\epsilon)  \lim{\sigma\rightarrow 0} \frac{1}{\sigma\sqrt{2\pi}} \int dy e^{-\frac{(y-\lambda_j\epsilon)^2}{2\sigma^2}}  |n y\rangle_y\nonumber\\
&=&\sum_n  \hat v_n^j(\lambda_j\epsilon)|n,\lambda_j\epsilon\rangle_y=|v^j,\lambda_j\epsilon\rangle_y.
\eea
We see that the linearized $y$ coordinates are like the collective $x$ coordinates, except that a translation by $\epsilon$ increases $x$ by $\epsilon$, while it increases $y$ by $\lambda_j\epsilon$.  In particular, this rate depends on the index $j$ on the state being transformed.

\subsection{Linearized Coordinates: Norm}\label{normsec}

To calculate the reduced norm in the linearized $y$ basis, we will need to tie the $y$ and $x$ bases together.  To do this, we will need to match their ordinary normalizations, which we will do by matching their norms.  Both $|v^j\rangle$ and $|v^j,0\rangle$ have infinite norms.  This motivates us to define
\beq
|v^j;\epsilon\rangle_y=\int_0^\epsilon dz e^{-i z P\p}|v^j,0\rangle_y.
\eeq
For small $\epsilon$, its norm is easily calculated
\bea\label{epnorm}
\left||v^j;\epsilon\rangle_y\right|^2
&=&\int_0^\epsilon dz_1\int_0^\epsilon dz_2\ {}_y\langle v^j,0|e^{-i(z_2-z_1) P\p}|v^j,0\rangle_y\\
&=&\sum_{n_1n_2}\int_0^\epsilon dz_1\int_0^\epsilon dz_2  \hat v_{n_1}^{j*}(\lambda_j z_1)\hat v_{n_2}^{j}(\lambda_jz_2){}_y\langle n_1,\lambda_j z_1|n_2,\lambda_jz_2\rangle_y\nonumber\\
&=&\sum_{n_1n_2}\int_0^\epsilon dz_1\int_0^\epsilon dz_2  \hat v_{n_1}^{j*}(\lambda_j z_1)\hat v_{n_2}^{j}(\lambda_jz_2)\delta_{n_1n_2}\delta(\lambda_j (z_1-z_2))
\nonumber\\
&=&\frac{1}{\lambda_j}\sum_{n}\int_0^\epsilon dz \hat v_{n}^{j*}(\lambda_jz)\hat v_{n}^{j}(\lambda_j z).\nonumber
\eea
Now, up to corrections of order $O(\epsilon)$ we can approximate $\hat v(\lambda_jz)=\hat v$.  
Then we find
\beq
\left||v^j;\epsilon\rangle_y\right|^2=\frac{1}{\lambda_j}\sum_{n}\hat v_{n}^{j*}\hat v_{n}^{j}\int_0^\epsilon dz =\frac{\epsilon\sum_{n}\hat v_{n}^{j*}\hat v_{n}^{j}}{\lambda_j}=\frac{\epsilon|\hat v^j|^2}{\lambda_j}.
\eeq

\subsection{Collective Coordinates: Norm}
 
 Let us write the same state $|v^j\rangle$ in the collective coordinate basis
\beq
|v^j\rangle=\sum_n v^j_n|e^n\rangle=\sum_n v^j_n\int dx |n x\rangle_x. \label{vdef}
\eeq
Again we can partition this state by the collective coordinate
\beq\label{collcoor}
|v^j,x\rangle_x=\sum_n v^j_n|n x\rangle_x
\eeq
where a translation by $\epsilon$ acts as
\beq
e^{-i\epsilon P\p}|v^j,0\rangle_x=|v^j,\epsilon\rangle_x.
\eeq
To obtain a quantity with a finite norm, we again define
\beq
|v^j;\epsilon\rangle_x=\int_0^\epsilon dx |v^j,x\rangle_x.
\eeq
Calculating as above, its norm is
\beq
\left||v^j;\epsilon\rangle_x\right|^2=\epsilon\sum_n v_n^{j*}v_n^j=\epsilon|v^j|^2.
\eeq

\subsection{Identifying Collective and Linearized Coordinates}

Now we want to identify the $x$ and $y$ bases of the Hilbert space.  Clearly this identification must preserve the norm, and so we choose
\beq
|v^j;\epsilon\rangle_y=\frac{1}{\sqrt{\lambda_j}}\frac{|\hat v^j|}{|v^j|}|v^j;\epsilon\rangle_x.\label{colla}
\eeq
Dividing by $\epsilon$ and taking the limit $\epsilon\rightarrow 0$ this becomes
\beq
\sum_n \hat v^j_n |n 0\rangle_y=|v^j,0\rangle_y=\frac{1}{\sqrt{\lambda_j}}\frac{|\hat v^j|}{|v^j|}|v^j,0\rangle_x=\frac{1}{\sqrt{\lambda_j}}\frac{|\hat v^j|}{|v^j|}\sum_n v_n^j |n 0\rangle_x
\eeq
and so
\beq
\sum_n \frac{\hat v^j_n}{|\hat v^j|} |n 0\rangle_y=\frac{1}{\sqrt{\lambda_j}}\sum_n \frac{v^j_n}{|v^j|} |n 0\rangle_x. 
\eeq
At leading order in $\epsilon$, a translation yields
\beq
\sum_n \frac{\hat v^j_n}{|\hat v^j|} |n, \lambda_j\epsilon\rangle_y=\frac{1}{\sqrt{\lambda_j}}\sum_n \frac{v^j_n}{|v^j|} |n \epsilon\rangle_x.
\eeq

\subsection{Orthogonality}

Recall from Eq.~(\ref{xdef}) that the $|n0\rangle_x$ basis is orthonormal, and from Eq.~(\ref{yort}) that the $|n0\rangle_y$ basis is orthonormal.  As $\hat v^j$ are eigenvectors of a matrix $A$, which we assume to be Hermitian, they will also be orthogonal.  

What about the $v^j$?  Up to a rescaling by $\lambda_j$, these are just $\hat v^j$ written in the $|n0\rangle_x$ basis instead of the $|n0\rangle_y$ basis.  As both bases are orthogonal one expects these to remain orthogonal.  Let us check that this is indeed the case.

Let us take the inner product of Eq.~(\ref{colla}) divided by $\sqrt{\epsilon}$ with itself, at two distinct eigenvalues $j_1\neq j_2$. We denote $\min\{\lambda_{j_1},\lambda_{j_2}\}$ as $\lambda_{\min}$ and $\max\{\lambda_{j_1},\lambda_{j_2}\}$ as $\lambda_{\max}$. The calculation here is similar as in Subsec.~\ref{normsec}. The left hand side yields
\bea
\frac{1}{\epsilon}{}_y\langle v^{j_1};\epsilon|v^{j_2};\epsilon\rangle_y
&=&\frac{1}{\epsilon}\int_0^\epsilon dz_1\int_0^\epsilon dz_2\ {}_y\langle v^{j_1},0|e^{-i(z_2-z_1) P\p}|v^{j_2},0\rangle_y\\
&=&\frac{1}{\epsilon}\sum_{n_1n_2}\int_0^\epsilon dz_1\int_0^\epsilon dz_2  \hat v_{n_1}^{j_1*}(\lambda_j z_1)\hat v_{n_2}^{j_2}(\lambda_jz_2){}_y\langle n_1,\lambda_{j_1} z_1|n_2,\lambda_{j_2} z_2\rangle_y\nonumber\\
&=&\frac{1}{\epsilon}\sum_{n_1n_2}\int_0^\epsilon dz_1\int_0^\epsilon dz_2  \hat v_{n_1}^{j_1*}(\lambda_j z_1)\hat v_{n_2}^{j_2}(\lambda_jz_2)\delta_{n_1n_2}\delta(\lambda_{j_1}z_1 - \lambda_{j_2}z_2)
\nonumber\\
&=&\frac{1}{\epsilon\lambda_{j_1}\lambda_{j_2}}\sum_{n_1n_2}\int_0^\epsilon d(\lambda_{j_1}z_1) \int_0^\epsilon d(\lambda_{j_2}z_2) \hat v_{n_1}^{j_1*}(\lambda_{j_1}z_1)\hat v_{n_2}^{j_2}(\lambda_{j_2} z_2)\delta_{n_1n_2}\delta(\lambda_{j_1}z_1 - \lambda_{j_2}z_2)\nonumber\\
&=&\frac{1}{\epsilon\lambda_{j_1}\lambda_{j_2}}\sum_{n}\int_0^{\lambda _{j_1}\epsilon} d \tilde{z}_1 \int_0^{\lambda_{j_2}\epsilon} d\tilde{z}_2 \hat v_{n}^{j_1*}(\tilde{z}_1)\hat v_{n}^{j_2}(\tilde{z}_2)\delta(\tilde{z}_1-\tilde{z}_2)\nonumber\\
&=&\frac{1}{\epsilon\lambda_{\min}\lambda_{\max}}\sum_{n}\int_0^{\lambda _{\min}\epsilon} d \tilde{z}  \hat v_{n}^{j_1*}(\tilde{z})\hat v_{n}^{j_2}(\tilde{z}).\nonumber
\eea
Again as in Subsec.~\ref{normsec}, up to corrections of order $O(\epsilon)$ we can approximate $\hat v(\tilde{z})=\hat v$. Then we find
\beq
\frac{1}{\epsilon}{}_y\langle v^{j_1};\epsilon|v^{j_2};\epsilon\rangle_y=\frac{1}{\epsilon\lambda_{\min}\lambda_{\max}}\sum_{n}\hat v_{n}^{j_1*}\hat v_{n}^{j_2}\int_0^{\lambda _{\min}\epsilon} d \tilde{z}=\frac{\sum_{n}\hat v_{n}^{j_1*}\hat v_{n}^{j_2}}{\lambda_{\max}}=0
\eeq
where the last equality used the orthogonality of the $\hat v$. 

 Here we assumed that all $\lambda_j>0$.  In the case of interest of quantum kinks, $A$ will be a positive scalar plus a correction suppressed by a power of the coupling, so this is the case at small coupling.

The calculation of the right hand side is similar
\bea
0&=&\frac{1}{\epsilon\sqrt{\lambda_{j_1} \lambda_{j_2}}}\frac{|\hat v^{j_1}||\hat v^{j_2}|}{|v^{j_1}||v^{j_2}|}{}_x\langle v^{j_1};\epsilon|v^{j_2};\epsilon\rangle_x\\
&=&\frac{1}{\epsilon\lambda_{j_1} \lambda_{j_2}}\int_0^\epsilon dx_1\int_0^\epsilon dx_2\ {}_x\langle v^{j_1},x_1|v^{j_2},x_2\rangle_x\nonumber\\
&=&\frac{1}{\epsilon\lambda_{j_1} \lambda_{j_2}} \sum_{n_1n_2}v_{n_1}^{j_1*} v_{n_2}^{j_2}\int_0^\epsilon dx_1\int_0^\epsilon dx_2\ {}_x\langle n_1,x_1|n_2,x_2\rangle_x\nonumber\\
&=&\frac{1}{\epsilon\lambda_{j_1} \lambda_{j_2}} \sum_{n_1n_2}v_{n_1}^{j_1*} v_{n_2}^{j_2}\int_0^\epsilon dx_1\int_0^\epsilon dx_2\ \delta_{n_1n_2}\delta(x_1-x_2)\nonumber\\
&=&\frac{1}{\epsilon\lambda_{j_1} \lambda_{j_2}} \sum_{n}v_{n}^{j_1*} v_{n}^{j_2}\int_0^\epsilon dx=\frac{ \sum_{n}v_{n}^{j_1*} v_{n}^{j_2}}{\lambda_{j_1} \lambda_{j_2}} \nonumber
\eea
where from the 2nd line to the 3rd line we used Eq.~(\ref{collcoor}).  In passing from the first line to the second, we used the identity~(\ref{vhatv}) which will be proved momentarily. So the $v$ are also orthogonal
\beq
\sum_n v^{j_1*}_n v^{j_2}_n=0.
\eeq


\subsection{Linearized Coordinates: Reduced Norm}

Finally we are ready to compute the reduced norm of $|v^j\rangle$.  The reduced norm squared is defined to be $|v^j|^2$.  The following manipulations are valid at small $y$, where the $y$-coordinate perturbation theory is valid
\bea
|v^j\rangle&=&\sum_n\int dy \hat v^j_n(y)|n y\rangle_y=\frac{1}{\sqrt{\lambda_j}}|\hat v^j|\sum_n \frac{v^j_n}{|v^j|}\int dy|n, y/\lambda_j\rangle_x\\
&=&{\sqrt{\lambda_j}}|\hat v^j|\sum_n \frac{v^j_n}{|v^j|}\int dx|n, x\rangle_x={\sqrt{\lambda_j}}|\hat v^j|\sum_n \frac{v^j_n}{|v^j|}|e^n\rangle.
\nonumber
\eea
Matching to Eq.~(\ref{vdef}), we find
\beq\label{vhatv}
\sqrt{\lambda_j}|\hat v^j|=|v^j|.
\eeq

The reduced norm is therefore
\bea\label{rednorm}
{}_{\rm{}}\langle v^j|v^j\rangle_{\rm{red}}&=&\lambda_j |\hat v^j|^2 \sum_{n_1n_2}\frac{v^{j*}_{n_1}v^{j}_{n_2}}{|v^j|^2}{}_{\rm{}}\langle e^{n_1}|e^{n_2}\rangle_{\rm{red}}\\
&=&\lambda_j |\hat v^j|^2 \sum_{n_1n_2}\frac{v^{j*}_{n_1}v^{j}_{n_2}}{|v^j|^2}\delta_{n_1n_2}=\lambda_j|\hat{v}^j|^2=\sum_{mn}\hat v^{j*}_m A_{mn}\hat v^j_n.
\nonumber
\eea
Similarly, if $j\neq k$ then the reduced inner product is 
\bea
{}_{\rm{}}\langle v^j|v^k\rangle_{\rm{red}}&=&\sqrt{\lambda_j\lambda_k}|\hat v^j||\hat v^k| \sum_{n_1n_2}\frac{v^{j*}_{n_1}v^{k}_{n_2}}{|v^j||v^k|}\delta_{n_1n_2}\\
&=&\sqrt{\lambda_j\lambda_k}|\hat v^j||\hat v^k| \sum_{n}\frac{v^{j*}_{n}v^{k}_{n}}{|v^j||v^k|}=0=\lambda_k \sum_{n}\hat v^{j*}_n\hat v^k_n=\sum_{mn}\hat v^{j*}_m A_{mn}\hat v^k_n.\nonumber
\eea

Now consider any two translation-invariant states $|\phi\rangle$ and $|\psi\rangle$.  Assume that $A$ is Hermitian, so that its eigenvectors are a basis of the vector space generated by the $|e^n\rangle$.  Then
\bea
|\psi\rangle&=&\sum_n \int dy\hat\psi_n(y)|ny\rangle_y=\sum_n \int dy\left[\hat\psi_n+O(y)\right]|ny\rangle_y=\sum_n \hat\psi_n|n\rangle_y=\sum_{jn}\hat\psi_n\left(\hat v^{-1}\right)^j_n |v^j\rangle\nonumber\\
|\phi\rangle&=&\sum_{jn}\hat\phi_n\left(\hat v^{-1}\right)^j_n |v^j\rangle\label{2trans}
\eea
where we have defined
\beq
|n\rangle_y=\int dy |ny\rangle_y. \label{ndef}
\eeq
Here we have dropped the term of order $O(y)$, as translation-invariance implies that the matching of the $y$ and $x$ kets can be applied at any value of $y$, and we apply it at $y=0$ where the $O(y)$ correction vanishes.

Their reduced inner product is
\bea
\langle \phi|\psi\rangle_{\rm{red}}&=&\sum_{n_1n_2j_1j_2}\hat\phi_{n_1}^*\left(\hat v^{*-1}\right)^{j_1}_{n_1}\hat\psi_{n_2}\left(\hat v^{-1}\right)^{j_2}_{n_2}
\langle v^{j_1}|v^{j_2}\rangle_{\rm{red}}\label{qmpadr}\\
&=&\hat\phi^* \left(\hat v^*\right)^{-1}\hat v^* A \hat v \hat v^{-1}\hat \psi=\hat\phi^* A \hat\psi.\nonumber
\eea

\subsection{Interpretation}

Let us pause to interpret our result (\ref{qmpadr}).  The inner product of
\beq
|\psi\rangle=\sum_n \int dy \hat\psi_n(y)|ny\rangle_y\hsp
|\phi\rangle=\sum_n \int dy \hat\phi_n(y)|ny\rangle_y
\eeq
is infrared divergent, due to the $y$ integral. However these inner products only appear in ratios, so it is sufficient to consider the inner product per unit of translation, dividing through by the volume of the translation group. 

Translation symmetry acts transitively on the $y$ coordinate, leaving the states invariant.  Therefore we can calculate this inner product density in a neighborhood of any fixed $y$.  Consider $y=0$.  Close to this point, we can approximate
\beq
|\psi\rangle=\sum_n \hat\psi_n \int dy |ny\rangle_y\hsp
|\phi\rangle=\sum_n \hat\phi_n \int dy |ny\rangle_y.
\eeq
Now let us define
\beq
|\psi y\rangle_y=\sum_n \hat\psi_n|ny\rangle_y\hsp
|\phi y\rangle_y=\sum_n \hat\phi_n|ny
\rangle_y.
\eeq

The inner product is still divergent, but combining (\ref{yort}) and (\ref{qmpadr}) we see that close to the base point $y=0$ it factorizes
\beq
{}_y\langle \phi y_1|A|\psi y_2\rangle_y=\langle \phi|\psi\rangle_{\rm{red}}\delta(y_1-y_2). \label{amp}
\eeq
We learn that the reduced inner product $\langle \phi|\psi\rangle_{\rm{red}}$ is given by the vector inner product of $\hat\psi$ and $\hat\phi$ with a Jacobian factor, $A$, resulting from the difference between the translation operator $P\p$ and a rigid shift in $y$.  Intuitively (\ref{amp}) may be written $\delta(x_1-x_2)=A\delta(y_1-y_2)$.

One may calculate the reduced inner product using (\ref{amp}).  To do this one first evaluates the left hand side at small $y$ and then amputates the $\delta(y_1-y_2)$.   This will be our strategy in quantum field theory.






\section{Reduced Inner Products for Quantum Kinks} \label{qftsez}

\subsection{Notation}

To pass from quantum mechanics to the case of a quantum field theory admitting quantum kinks, we make the following replacements.  First, the discrete quantum number $n$ is replaced by symmetrized $n$-tuples of continuum and shape normal mode labels $k$.  Here $k$ is a real number for continuum modes, and a discrete index for shape modes.  Now $n\geq 0$ and these states represent the $n$-meson Fock space in the kink sector.

We let $\phi_0$ be the operator whose eigenvalue is $y$.  Its dual momentum we recall is $\pi_0$.  We introduce the shorthand
\beq
\Delta_{ij}=\int dx \g_i(x) \g\p_j(x)
\eeq
where $i$ and $j$ run over the zero mode $B$, as well as continuum modes $k$ and shape modes $S$.

The translation operator $P\p$ is now given by
\bea
P\p&=&P+\sqrt{Q_0}\pi_0 \label{ppk}\\
P&=&\ppin{k}\Delta_{kB}\left[i\phi_0\left(-\ok{}B^\ddag_k+\frac{B_{-k}}{2}\right)+\pi_0\left(B^\ddag_k+\frac{B_{-k}}{2\ok{}}\right)\right]\nonumber\\
&&+i\ppink{2}\Delta_{k_1k_2}\left[\frac{\ok{2}-\ok{1}}{2}B^\ddag_{k_1}B^\ddag_{k_2}-\frac{1}{2}\left(1+\frac{\ok{1}}{\ok{2}}\right)B^\ddag_{k_1}B_{-k_2}+\frac{\ok{1}-\ok{2}}{8\ok{1}\ok{2}}B_{-k_1}B_{-k_2}
\right].\nonumber
\eea
Intuitively $P$ is the momentum operator for the mesons while $\sqrt{Q_0}\pi_0$ is the momentum operator for the kink.  Only $P\p$ is conserved.  Recalling our old decomposition (\ref{pp})
\beq
P\p=A\pi_0+B+C\phi_0
\eeq
we can match the $\pi_0$ coefficient in (\ref{ppk}) to obtain
\beq
A=\sqrt{Q_0}+ \ppin{k}\Delta_{kB}\left( B^\ddag_k+\frac{B_{-k}}{2\ok{}}\right).
\eeq

We decompose states as
\bea
|\psi\rangle&=&\sum_{m,n=0}^\infty |\psi\rangle^{mn}\hsp
|k_1\cdots k_n\rangle_0=\Bd1\cdots\Bd n\vac_0
\nonumber\\
|\psi\rangle^{mn}&=&\phi_0^m\ppink{n}\gamma_\psi^{mn}(k_1\cdots k_n)|k_1\cdots k_n\rangle_0\hsp \vac_0=\int dy |y\rangle_y.
 \label{gameqa}
\eea

To make contact with the decomposition in Sec.~\ref{qmsez}, note that
\bea\label{yk0}
|\psi\rangle^{mn}&=&\int dy |y,\psi\rangle_y^{mn}\hsp
|y,\psi\rangle^{mn}_y
=y^m\ppink{n}\gamma_\psi^{mn}(k_1\cdots k_n)|y,k_1\cdots k_n\rangle_y\nonumber\\
|y,k_1\cdots k_n\rangle_y&=&\Bd1\cdots\Bd n|y\rangle_y.
\eea
We see that here the role which was played by $\hat{\psi}_n(y)$ in quantum mechanics is now played by
\beq
\hat{\psi}_n(y) \rightarrow \sum_m \gamma_\psi^{mn}(k_1\cdots k_n) y^m.
\eeq
The discrete $n$ quantum number is replaced by an $n$-tuple of shape and continuum mode indices $k$
\beq
|n y\rangle_y \rightarrow |y,k_1\cdots k_n\rangle_y\hsp
\sum_n \rightarrow \sum_n \ppink{n}. \label{nymap}
\eeq

Recall that, in quantum mechanics, the coefficient at the origin was $\hat\psi_n=\hat\psi_n(0)$.  Similarly, setting $y=0$ here we obtain $\gamma^{0n}$
\beq
\hat{\psi}_n \rightarrow  \gamma_\psi^{0n}(k_1\cdots k_n) .
\eeq

\subsection{The Reduced Inner Product}

With these substitutions, we can derive the reduced inner product in quantum field theory by running through the same arguments as in Sec.~\ref{qmsez}.  In other words, one can construct invariant states in the collective coordinate basis and in the $y$ basis, one can introduce an infrared regulator $\epsilon$ and use it to match the norms and identify states in the two bases.  Then the reduced inner product, which is trivially evaluated in the collective coordinate basis, can be defined in the linearized $y$ basis.  Instead, we will take a faster approach.  We will directly apply these substitutions to Eq.~(\ref{amp}), which was itself derived using all of the steps above.

Let us first evaluate the following term from the left hand side of Eq.~(\ref{amp})
\bea
A|0,\psi\rangle_y^{0n}&=&\left(
\sqrt{Q_0}+ \ppin{k}\Delta_{kB}\left( B^\ddag_k+\frac{B_{-k}}{2\ok{}}\right)
\right)|0,\psi\rangle_y^{0n}\\
&=&\ppink{n}\gamma_\psi^{0n}(k_1\cdots k_n)
\left(
\sqrt{Q_0}+ \ppin{k\p}\Delta_{k\p B}\left( B^\ddag_{k\p}+\frac{B_{-k\p}}{2\okp{}}\right)\right)
|0,k_1\cdots k_n\rangle_y
\nonumber\\
&=&\sqrt{Q_0}|0,\psi\rangle_y^{0n}+\ppink{n+1}\gamma_\psi^{0n}(k_1\cdots k_n)\Delta_{k_{n+1},B}|0,k_1\cdots k_{n+1}\rangle_y\nonumber\\
&&+n\ppink{n}\gamma_\psi^{0n}(k_1\cdots k_n)\frac{\Delta_{-k_n,B}}{2\ok n}|0,k_1\cdots k_{n-1}\rangle_y
\nonumber
\eea
where, in the last line, we have assumed that $\gamma_\psi^{0n}$ is symmetrized over its arguments $k_i$.  Summing over $n$ one arrives at
\bea\label{A0psi}
A\sum_n|0,\psi\rangle_y^{0n}&=&\sum_n \ppink{n}
\left[ \sqrt{Q_0}\gamma_\psi^{0n}(k_1\cdots k_n)+
\gamma_\psi^{0,n-1}(k_1\cdots k_{n-1})\Delta_{k_n,B}
\right.\nonumber\\
&&\left.
+(n+1)\ppin{k_{n+1}}\gamma_\psi^{0,n+1}(k_1\cdots k_{n+1})\frac{\Delta_{-k_{n+1}, B}}{2\ok{n+1}}
\right]
|0,k_1\cdots k_{n}\rangle_y.
\eea

Using the oscillator algebra satisfied by $B$ and $B^\ddag$ and ${}_y\langle y_1|y_2\rangle_y=\delta(y_1-y_2)$ one finds the inner products of
\beq
|a_i,y_i\rangle=\ppink{n}a_i(k_1\cdots k_n)|y_i,k_1\cdots k_n\rangle_y
\eeq
where $a_i$ is symmetric in its arguments, to be
\beq
\langle a_1,y_1|a_2,y_2\rangle=n!\delta(y_1-y_2)\ppink{n}\frac{a_1^*(k_1\cdots k_n)a_2(k_1\cdots k_n)}{\prod_{i=1}^n(2\ok{i})}. \label{qfti}
\eeq

Finally we want to generalize the reduced inner product (\ref{qmpadr}) to quantum field theory.  To do this, we need to generalize the vector inner product of the $\hat v$ vectors.  Our definition is that this inner product is to be interpreted as the full inner product (\ref{qfti}) in quantum field theory, without the $\delta(y_1-y_2)$.  This statement is just the quantum field theory generalization of Eq.~(\ref{amp}).

We then obtain our master formula for the reduced inner product in quantum field theory
\bea
{}_{\rm{}}{}\langle \phi|\psi\rangle_{\rm{red}}&=&\sum_{n_1n_2}{}_{\ \ y}^{0n_1}\langle y_1,\phi|A|y_2,\psi\rangle_y^{0n_2}|_{{\rm Coefficient\ of}\ \delta(y_1-y_2){\rm \ at}\ y_1=0} \label{padqft}
\\
&=&
\sum_n n! \ppink{n}\frac{\gamma_\phi^{0n*}(k_1\cdots k_n)}{\prod_{i=1}^n(2\ok{i})}
\left[ \sqrt{Q_0}\gamma_\psi^{0n}(k_1\cdots k_n)+
\gamma_\psi^{0,n-1}(k_1\cdots k_{n-1})\Delta_{k_n,B}
\right.\nonumber\\
&&\left.
+(n+1)\ppin{k_{n+1}}\gamma_\psi^{0,n+1}(k_1\cdots k_{n+1})\frac{\Delta_{-k_{n+1},B}}{2\ok{n+1}}
\right].
\nonumber
\eea
This is our main result.  We remind the reader that all $\gamma$ must be symmetrized in their arguments before this formula applied, or else the $n!$ and $(n+1)!$ factors should be replaced with sums over $S_n$ and $S_{n+1}$ permutations.  The second and third terms in the square brackets are the Jacobian terms resulting from the off-diagonal part of $A$.   Both are proportional to $\Delta_{kB}$, which describes the mixing between the zero mode and the normal modes as the kink moves.

In the next section we will see that this formula satisfies some basic consistency checks.  For example, we will see that the reduced inner product of a 0-meson and 1-meson state vanishes, whereas one would obtain a nonzero result if one did not include the Jacobian terms in (\ref{padqft}). 

\section{Examples of Reduced Norms} \label{exsez}

In this section we will calculate the reduced norms of the kink ground state and also a kink with one excitation, which can be a continuum meson or a bound shape mode.  We will show that, up to corrections which are suppressed, with respect to the leading term, by a quantity of order $O(\lambda)$
\beq
\langle 0\vac_{\rm{red}}=\sqrt{Q_0}+O(\sl)\hsp
\langle\kt_1|\kt_2\rangle_{\rm{red}}=\frac{2\pi\delta(\kt_1-\kt_2)}{2\okt 1}\sqrt{Q_0}+O(\sl). \label{grred}
\eeq
Had there been corrections of order $O(\lambda^0)$, which would be suppressed with respect to the leading term by only one power of $\sl$, this would have invalidated calculations in Refs.~\cite{alberto} and \cite{memult}.

We will decompose each $\gamma^{mn}$ as
\beq
\gamma^{mn}=\sum_i Q_0^{-i/2}\gamma_i^{mn}.
\eeq

\subsection{The Reduced Norm of the Ground State}

The kink ground state, at subleading order, is characterized by the coefficients\footnote{These coefficients were calculated in Ref.~\cite{me2loop}.  As a result of a sign difference in the convention (\ref{gb}) for $\g_B(x)$, here the meson and kink momentum contributions to $P\p$ have a different relative sign.  This changes the signs of all $\Delta$ terms in all coefficients. Also, the convention for $\v3$ here differs by a factor of $\sqrt{\lambda}$.}
\bea
\gamma_0^{00}&=&1\hsp
\gamma_1^{12}(k_1,k_2)=\frac{\left(\ok 2-\ok 1\right)\Delta_{k_1k_2}}{2}\hsp
\gamma_1^{21}(k_1)=-\frac{\omega_{k_1}\Delta_{k_1B}}{2}\\
\gamma_1^{01}(k_1)&=&-\frac{\Delta_{k_1B}}{2}-\frac{\sqrt{\lambda Q_0}}{2}\frac{V_{\I k_1}}{\ok{1}}\hsp
\gamma_1^{03}(k_1,k_2,k_3)=-\frac{\sqrt{\lambda Q_0}}{6}\frac{V_{k_1k_2k_3}}{\ok{1}+\ok{2}+\ok{3}}\nonumber
\eea
where we have defined
\bea
V_{k_1\cdots k_n}&=&\int dx \V{n} \g_{k_1}(x)\cdots  \g_{k_n}(x)\\
V_{\I k_1\cdots k_n}&=&\int dx \V{n+2} \I(x) \g_{k_1}(x)\cdots\g_{k_n}(x)\nonumber\\
\I(x)&=&\pin{k}\frac{\left|{\g}_{k}(x)\right|^2-1}{2\omega_k}+\sum_S \frac{\left|{\g}_{S}(x)\right|^2}{2\omega_k}.
\nonumber
\eea
In the first two lines, the $k_i$ run over not just the continous momenta, but also the shape modes.



The reduced norm can be written as a sum of three terms corresponding to $n=0$,\ $1$\ and $3$\ in Eq.~(\ref{padqft})
\bea
|\vac|^2_{\rm{n,red}}&=&n! \ppink{n}\frac{\gamma^{0n*}(k_1\cdots k_n)}{\prod_{i=1}^n(2\ok{i})}
\left[ \sqrt{Q_0}\gamma^{0n}(k_1\cdots k_n)+
\gamma^{0,n-1}(k_1\cdots k_{n-1})\Delta_{k_n,B}
\right.\nonumber\\
&&\left.
+(n+1)\ppin{k_{n+1}}\gamma^{0,n+1}(k_1\cdots k_{n+1})\frac{\Delta_{-k_{n+1}, B}}{2\ok{n+1}}
\right].
\eea

These summands are
\bea
|\vac|^2_{\rm{0,red}}&=&
 \sqrt{Q_0}+\ppin{k_1}\gamma^{01}(k_1)\frac{\Delta_{-k_1 B}}{2\ok{1}}\label{0rednorm}\\
 &=&
 \sqrt{Q_0}-\frac{1}{4\sqrt{Q_0}}\ppin{k_1}\left[ 
 {\Delta_{k_1 B}}{}+{\sqrt{\lambda Q_0}}{}\frac{V_{\I k_1}}{\ok{1}}
 \right]\frac{\Delta_{-k_1 B}}{\ok{1}}\nonumber\\
|\vac|^2_{\rm{1,red}}&=& \ppin{k_1}\frac{\gamma^{01*}(k_1)}{2\ok{1}}
\left[ \sqrt{Q_0}\gamma^{01}(k_1)+
\Delta_{k_1B}
\right] \nonumber\\
&=& \frac{1}{8\sqrt{Q_0}}\ppin{k_1}\left[\frac{\lambda Q_0 |V_{\I k_1}|^2}{\ok{1}^3}-\frac{|\Delta_{k_1B}|^2}{\ok{1}}\right]\nonumber\\
|\vac|^2_{\rm{3,red}}&=&\frac{3\sqrt{Q_0}}{4} \ppink{3}\frac{\left|\gamma^{03}(k_1,k_2, k_3)\right|^2}{\prod_{i=1}^3\ok{i}}\nonumber\\
&=&\frac{\lambda\sqrt{Q_0}}{48} \ppink{3}\frac{\left|V_{k_1k_2 k_3}\right|^2}{\ok  1\ok 2\ok 3(\ok{1}+\ok 2+\ok 3)^2}.\nonumber
\eea

Altogether we find
\bea
|\vac|^2_{\rm{red}}&=&\sqrt{Q_0}+\frac{1}{8\sqrt{Q_0}}\ppin{k_1}\frac{1}{\ok 1}\left({\sqrt{\lambda Q_0}}{}\frac{V_{\I k_1}}{\ok{1}}+{\Delta_{k_1 B}}{}\right)\left({{\sqrt{\lambda Q_0}}{}\frac{V_{\I -k_1}}{\ok{1}}{}-3\Delta_{-k_1 B}}\right)\nonumber\\
&&+\frac{\lambda\sqrt{Q_0}}{48} \ppink{3}\frac{\left|V_{k_1k_2 k_3}\right|^2}{\ok  1\ok 2\ok 3(\ok{1}+\ok 2+\ok 3)^2}.
\eea
Note that we have adapted the convention $\gamma_2^{00}=0$.  Another convention for $\gamma_2^{00}$ would have resulted in a different norm.  This is the lowest order manifestation of the freedom in choosing the overall normalization, which is already present quantum mechanics.  Clearly there is such a freedom for every state.  Although the norms of all states are a matter of convention, the determination of the norm for a given convention is physically relevant, as the convention then fixes all of the reduced inner products.


\subsection{Inner Product of Zero and One-Meson State}

\subsubsection{The One-Meson States up to $O(\sqrt{\lambda})$}

Now let us consider a one-meson Hamiltonian eigenstate $|\kt\rangle$.  At leading order, it is $|\kt\rangle_0$, characterized by
\beq
\gamma_{0\kt}^{01}(k_1)=2\pi\delta(k_1-\kt). \label{g0}
\eeq
The next order corrections\footnote{They were calculated in Ref.~\cite{menormal}, again with the sign flip for all $\Delta$ symbols and $\sqrt{\lambda}$ for $V$ symbols resulting from the convention (\ref{gb}).} are summarized by the corresponding symbols $\gamma_{1\kt}^{mn}$
\bea
\gamma_{1\kt}^{11}(k_1)&=&\frac{1}{2}\Delta_{-\kt k_1}\left(1+\frac{\ok1}{\omega_{\kt}}\right)\hsp
\gamma_{1\kt}^{13}(k_1,k_2,k_3)=\ok3\Delta_{k_2k_3}2\pi\delta(k_1-\kt)\label{gammakt}\\
\gamma_{1\kt}^{22}(k_1,k_2)&=&-\frac{\ok2}{2}\Delta_{k_2 B}2\pi\delta(k_1-\kt)\hsp
\gamma_{1\kt}^{00}= \frac{\sqrt{Q_0\lambda}V_{\I, -\kt}}{4\omega^2_{\kt}}-\frac{\Delta_{-\kt B}}{4\omega_{\kt}}\nonumber\\
\gamma_{1\kt}^{02}(k_1,k_2)&=& -\frac{2\pi\delta(k_2-\kt)}{4}\left(\Delta_{k_1 B}+\sqrt{Q_0\lambda}\frac{V_{\I  k_1}}{\ok1}\right)+\frac{\sqrt{Q_0\lambda}V_{-\kt k_1 k_2}}{4\omega_{\kt}\left(\omega_{\kt}-\ok1-\ok2\right)}\nonumber\\
&&-\frac{2\pi\delta(k_1-\kt)}{4}\left(\Delta_{k_2 B}+\sqrt{Q_0\lambda}\frac{V_{\I  k_2}}{\ok 2}\right)\nonumber\\
\gamma_{1\kt}^{04}(k_1\cdots k_4)&=& -\frac{\sqrt{Q_0\lambda}V_{k_1 k_2 k_3}}{6\sum_{j=1}^3 \ok{j}}2\pi\delta(k_4-\kt)\hsp
\gamma_{1\kt}^{20}=\frac{1}{4}\Delta_{-\kt B}.\nonumber
\eea

\subsubsection{The Inner Product}

Let us calculate the reduced inner product of the kink ground state $\vac$ and a one-kink one-meson state $|\kt\rangle$ up to order $O(\lambda^0)$.  There are two contributions
\bea
\langle 0|\kt\rangle_{\rm{n,red}}&=&n! \ppink{n}\frac{\gamma^{0n*}(k_1\cdots k_n)}{\prod_{i=1}^n(2\ok{i})}
\left[ \sqrt{Q_0}\gamma^{0n}_{\kt}(k_1\cdots k_n)+
\gamma^{0,n-1}_{\kt}(k_1\cdots k_{n-1})\Delta_{k_n,B}
\right.\nonumber\\
&&\left.
+(n+1)\ppin{k\p}\gamma_{\kt}^{0,n+1}(k_1\cdots k_n,k\p)\frac{\Delta_{-k\p B}}{2\okp{}}
\right]
\eea
at this order.  These are
\bea
\langle 0|\kt\rangle_{\rm{0,red}}&=& \sqrt{Q_0}\gamma^{00}_{\kt}+\ppin{k\p}\gamma_{\kt}^{01}(k\p)\frac{\Delta_{-k\p B}}{2\okp{}}=\frac{\sqrt{Q_0\lambda}V_{\I -\kt}}{4\omega^2_{\kt}}+\frac{\Delta_{-\kt B}}{4\omega_{\kt}}\\
\langle 0|\kt\rangle_{\rm{1,red}}&=& \ppin{k_1}\frac{\gamma^{01*}(k_1)}{2\ok{1}}
\sqrt{Q_0}\gamma^{01}_{\kt}(k_1)=\frac{\gamma_1^{01*}(\kt)}{2\okt{}}=
-\frac{\Delta_{-\kt B}}{4\okt{}}-\frac{\sqrt{\lambda Q_0}}{2}\frac{V_{\I -\kt}}{2\okt{}^2}.\nonumber
\eea
These cancel precisely, leaving
\beq
\langle 0|\kt\rangle_{\rm{red}}=0
\eeq
at order $O(\lambda^0)$.  This is to be expected, as $\vac$ and $|\kt\rangle$ are eigenstates of  $H\p$ with distinct eigenvalues.  Note that the off-diagonal terms in $A$ contributed to $\langle 0|\kt\rangle_{\rm{0,red}}$ and were necessary for the reduced inner product to respect this orthogonality.

\subsection{Inner Product of Two One-Meson States}

We next turn our attention to the reduced inner product of two one-meson, one-kink states, $|\kt_1\rangle$ and $|\kt_2\rangle$, at $O(\sl)$.

\subsubsection{A Coefficient at $O(\lambda)$}

In addition to the $O(\lambda^0)$ coefficient given in Eq.~(\ref{g0}) and the $O(\sl)$ coefficients given in Eq.~(\ref{gammakt}), we will also need the $O(\lambda)$ coefficient $\gamma_{2\kt}^{01}(k_1)$ at $k_1\neq\kt$.  To calculate this, we use the eigenvalue equation
\beq
(H\p-E)|\kt\rangle=0.
\eeq
At order $O(\lambda)$ this consists of five terms
\beq
0=H\p_4|\kt\rangle_0+H\p_3|\kt\rangle_1+H\p_2|\kt\rangle_2-E_2|\kt\rangle_0-E_1|\kt\rangle_2. \label{schrod}
\eeq
Here $E_n$ is the $O(\lambda^{n-1})$ term in the energy of the 1-kink, 1-meson state $|\kt\rangle$.  In particular
\beq
E_1=\okt{}\hsp 
E_2=\sigma_{\kt}\hsp
H\p_2=\frac{\pi_0^2}{2}+\ppin{k}\ok{}\Bd{}B_k
\eeq
where $\sigma_{\kt}$ was calculated in Ref.~\cite{menormal}.  Note that, since $H\p_0=E_0=Q_0$ is a scalar, the $H\p_0$ and $E_0$ terms that one may be tempted to include would cancel.

We will impose Eq.~(\ref{schrod}) on the terms which are independent of $\phi_0$ and contain one meson, in other words the $m=0$, $n=1$ terms.  These can only result from the terms
\beq\label{m0n1}
|\kt\rangle_2\supset \frac{1}{Q_0}\ppin{k_1}\left[ 
\gamma_{2\kt}^{01}(k_1)+\phi_0^2\gamma_{2\kt}^{21}(k_1)
\right]|k_1\rangle_0
\eeq
in $|\kt\rangle_2$.  The last three terms in Eq.~(\ref{schrod}) are then easily written as
\beq
H\p_2|\kt\rangle_2-E_2|\kt\rangle_0-E_1|\kt\rangle_2=\frac{1}{Q_0}\ppin{k_1}\left[ 
(\ok{1}-\okt{})\gamma_{2\kt}^{01}(k_1)-\gamma_{2\kt}^{21}(k_1)
\right]|k_1\rangle_0-\sigma_{\kt} |\kt\rangle_0.
\eeq
Our strategy will be to determine $\gamma_{2\kt}^{01}(k_1)$ by matching the coefficient of $|k_1\rangle_0$ to
\beq
H\p_4|\kt\rangle_0+H\p_3|\kt\rangle_1=\frac{1}{Q_0}\ppin{k_1}\rho_{\kt}(k_1)
|k_1\rangle_0+\hat  \sigma_{\kt} |\kt\rangle_0
\eeq
where $\rho_{\kt}$ will be calculated below.  Here we have separated out of $\sigma_{\kt}$ the contribution $\hat\sigma_{\kt}$ from $\gamma_{2\kt}^{21}$ by decomposing
\beq
\gamma_{2\kt}^{21}(k_1)=\hat\gamma_{2\kt}^{21}(k_1)+2\pi\delta(k_1-\kt)Q_0\left(\hat\sigma_{\kt}-\sigma_{\kt}
\right)
\eeq
where $\hat\gamma_{\kt}(k_1)$ is continuous at $k_1=\kt$.

Matching the coefficients yields
\beq
\gamma_{2\kt}^{01}(k_1)=\frac{-\hat \gamma_{2\kt}^{21}(k_1)+\rho_{\kt}(k_1)}{\okt{}-\ok{1}}.
\eeq
This is undefined at the two poles, located at $k_1=\pm\kt$.  The ambiguity at $k_1=\kt$ reflects the choice of normalization of the state $|\kt\rangle$.  

The ambiguity at $k_1=-\kt$ results from the fact that the states $|\kt\rangle$ and $|-\kt\rangle$ have the same energy, and both have zero momentum as measured by $P\p$.  We have defined $|\kt\rangle$ as the $H\p$ eigenstate which is $|\kt\rangle_0$ at leading order, however this definition does not fix the mixing with $|-\kt\rangle$ at subleading orders.  We will see below, when we discuss meson multiplication, that a choice of definition of the pole corresponds to a choice of initial condition in meson-kink scattering.  In the future we intend to study elastic kink-meson scattering, with intermediate states consisting of two continuum modes, a continuum mode and a shape mode, or the two shape mode resonance.  We expect that a choice of $i\epsilon$ shift of the pole will be necessary for an initial condition for that process, to ensure that the initial meson is always moving towards the kink.


\subsubsection{Calculating $\rho_{\kt}$}

Let us begin with $H\p_4|\kt\rangle_0$.  Only one term which appears in Wick's theorem \cite{mewick} will contribute
\bea
H\p_4&=&\frac{\lambda}{24}\int dx \V4 :\phi^4(x):_a\supset \frac{\lambda}{4}\int dx \V4 \left(\I(x) :\phi^2(x):_b+\frac{\I^2(x)}{2}\right)\nonumber\\
&\supset&\frac{\lambda}{2}\ppink{2} V_{\I k_1 -k_2} \Bd 1 \frac{B_{k_2}}{2\ok 2}+\frac{\lambda V_{\I\I}}{8}\hsp
V_{\I\I}=\int dx \V{4} \I^2(x)
\eea
where we have defined the normal ordering $::_b$ which places $B^\ddag$ before $B$.  We then find the contribution
\beq
H\p_4|\kt\rangle_0\supset \frac{\lambda V_{\I\I}}{8}|\kt\rangle_0+\frac{\lambda}{4\okt{}}\ppin{k_1}V_{\I k_1 -\kt} |k_1\rangle_0.
\eeq
The first term contributes to $\hat \sigma_{\kt}$ and the second to $\rho_{\kt}$.

Three contributions arise from $H\p_3\ks_1$.  Following Ref.~\cite{menormal}
\bea
H_3\p\ks_1^{00}&=&\frac{\lambda}{8}\left(\frac{V_{\I  -\kt}}{\omega^2_{\kt}}-\frac{\Delta_{-\kt B}}{\omega_{\kt}\sqrt{\lambda Q_0}}\right)\ppin{k_1}V_{\I k_1}|k_1\rangle_0.\nonumber
\eea
The second is
\bea
H_3\p\ks_1^{02}&=&\frac{\sl}{\sqrt{Q_0}}\ppin{k_1}\left[\ppinkp{2}\frac{\sqrt{\lambda Q_0}V_{-\kt k\p_1 k\p_2}V_{-k\p_1-k\p_2k_1}}{16\omega_{\kt}\okp1\okp2\left(\omega_{\kt}-\okp1-\okp2\right)}\right.\nonumber\\
&&\left.+\ppin{k\p}\left(\frac{\left(-\okp{}\Delta_{k\p B}-\sqrt{\lambda Q_0}V_{\I  k\p}\right)
V_{-k\p-\kt k_1}}{8\okp{}^2\omega_{\kt}}+\frac{\sqrt{\lambda Q_0}V_{-\kt k\p k_1}V_{\I -k\p}}{8\omega_{\kt}\okp{}\left(\omega_{\kt}-\okp{}-\ok1\right)}\right)\right.\nonumber\\
&&+\left.
\frac{ \left(-\ok1\Delta_{k_1 B}-\sqrt{\lambda Q_0}V_{\I  k_1}\right)V_{\I -\kt}}{8\omega_{\kt}\ok1}\right]|k_1\rangle_0\\
&&
+\frac{\sl}{\sqrt{Q_0}}\left[\ppin{k\p}\frac{\left(-\okp{}\Delta_{k\p B}-\sqrt{\lambda Q_0}V_{\I  k\p}\right)
V_{\I -k\p}}{8\okp{}^2}\right]
|\kt\rangle_0.\nonumber
\eea
The third contribution is
\bea
H_3\p\ks_1^{04}&&=-\frac{\lambda}{16}\ppin{k_1}\left[\ppinkp{2}\frac{V_{k_1k\p_1k\p_2}V_{-\kt-k\p_1-k\p_2}}{\omega_{\kt}\okp1\okp2\left(\ok1+\okp1+\okp2\right)}\right]|k_1\rangle_0\nonumber\\
&&-\frac{\lambda}{48}\left[\ppinkp{3}\frac{V_{k\p_1k\p_2k\p_3}V_{-k\p_1-k\p_2-k\p_3}}{\okp1\okp2\okp3\left(\okp1+\okp2+\okp3\right)}
\right]|\kt\rangle_0.\nonumber
\eea

Adding these together, we may read off
\bea
\hat\sigma_k
&=&\frac{\lambda  V_{\I\I}}{8}+\frac{\sl}{\sqrt{Q_0}}\left[\ppin{k\p}\frac{\left(-\okp{}\Delta_{k\p B}-\sqrt{\lambda Q_0}V_{\I  k\p}\right)
V_{\I -k\p}}{8\okp{}^2}\right]\nonumber\\
&&-\frac{\lambda}{48}\left[\ppinkp{3}\frac{V_{k\p_1k\p_2k\p_3}V_{-k\p_1-k\p_2-k\p_3}}{\okp1\okp2\okp3\left(\okp1+\okp2+\okp3\right)}
\right]
\eea
and
\bea
\rho_{\kt}(k_1)&=&
\frac{\lambda Q_0}{4\okt{}}V_{\I k_1 -\kt}
+\frac{\lambda Q_0}{8}\left(\frac{V_{\I  -\kt}}{\omega^2_{\kt}}-\frac{\Delta_{-\kt B}}{\omega_{\kt}\sqrt{\lambda Q_0}}\right)V_{\I k_1}\\
&&+\sqrt{\lambda Q_0}\left[\ppinkp{2}\frac{\sqrt{\lambda Q_0}V_{-\kt k\p_1 k\p_2}V_{-k\p_1-k\p_2k_1}}{16\omega_{\kt}\okp1\okp2\left(\omega_{\kt}-\okp1-\okp2\right)}\right.\nonumber\\
&&\left.+\ppin{k\p}\left(\frac{\left(-\okp1\Delta_{k\p B}-\sqrt{\lambda Q_0}V_{\I  k\p}\right)
V_{-k\p-\kt k_1}}{8\okp{}^2\omega_{\kt}}+\frac{\sqrt{\lambda Q_0}V_{-\kt k\p k_1}V_{\I -k\p}}{8\omega_{\kt}\okp{}\left(\omega_{\kt}-\okp{}-\ok1\right)}\right)\right.\nonumber\\
&&+\left.
\frac{ \left(-\ok1\Delta_{k_1 B}-\sqrt{\lambda Q_0}V_{\I  k_1}\right)V_{\I -\kt}}{8\omega_{\kt}\ok1}\right]
-\frac{\lambda Q_0}{16}\ppinkp{2}\frac{V_{k_1k\p_1k\p_2}V_{-\kt-k\p_1-k\p_2}}{\omega_{\kt}\okp1\okp2\left(\ok1+\okp1+\okp2\right)}.
\nonumber
\eea

From Ref.~\cite{menormal}
\bea
\gamma_{2\kt}^{21}(k_1)&=&
2\pi\delta(k_1-\kt)\left[\ppin{k\p}\frac{\Delta_{-k\p B}}{8}\left(\Delta_{k\p B}-\frac{\sqrt{\lambda Q_0}V_{\I  k\p}}{\okp{}}\right)\right.\\
&&\left.
-\frac{1}{16}\ppinkp{2}\frac{\left(\okp1-\okp2\right)^2}{\okp1\okp2}\Delta_{k\p_1k\p_2}\Delta_{-k\p_1,-k\p_2}\right]\nonumber\\
&&+\frac{3}{8}\left(-1+\frac{\ok1}{\omega_{\kt}}\right)\Delta_{k_1 B}\Delta_{-\kt B}-\frac{1}{4}\ppin{k\p}\left(\frac{\ok1}{\okp{}}+\frac{\okp{}}{\omega_{\kt}}
\right)\Delta_{-\kt,-k\p}\Delta_{k_1k\p}\nonumber\\
&&-\frac{\sqrt{\lambda Q_0}}{8\omega_{\kt}}\left(\omega_{k_1}\Delta_{k_1 B}\frac{V_{\I -\kt}}{\omega_{\kt}}+\omega_{\kt}\Delta_{-\kt B}\frac{V_{\I  k_1}}{\ok1}\right)+\frac{1}{8}\ppin{k\p}
\frac{\sqrt{\lambda Q_0}\Delta_{-k\p B}V_{-\kt k_1 k\p}}{\omega_{\kt}\left(\omega_{\kt}-\ok1-\okp{}\right)}.\nonumber
\eea
Decomposing, this is
\bea
\hat\gamma_{2\kt}^{21}(k_1)&=&\frac{3}{8}\left(-1+\frac{\ok1}{\omega_{\kt}}\right)\Delta_{k_1 B}\Delta_{-\kt B}-\frac{1}{4}\ppin{k\p}\left(\frac{\ok1}{\okp{}}+\frac{\okp{}}{\omega_{\kt}}
\right)\Delta_{-\kt,-k\p}\Delta_{k_1k\p}\nonumber\\
&&-\frac{\sqrt{\lambda Q_0}}{8\omega_{\kt}}\left(\omega_{k_1}\Delta_{k_1 B}\frac{V_{\I -\kt}}{\omega_{\kt}}+\omega_{\kt}\Delta_{-\kt B}\frac{V_{\I  k_1}}{\ok1}\right)+\frac{1}{8}\ppin{k\p}
\frac{\sqrt{\lambda Q_0}\Delta_{-k\p B}V_{-\kt k_1 k\p}}{\omega_{\kt}\left(\omega_{\kt}-\ok1-\okp{}\right)}\nonumber\\
\hat\sigma_{\kt}-\sigma_{\kt}&=&\frac{1}{Q_0}\left[\ppin{k\p}\frac{\Delta_{-k\p B}}{8}\left(\Delta_{k\p B}-\frac{\sqrt{\lambda Q_0}V_{\I  k\p}}{\okp{}}\right)\right.\nonumber\\
&&\left.
-\frac{1}{16}\ppinkp{2}\frac{\left(\okp1-\okp2\right)^2}{\okp1\okp2}\Delta_{k\p_1k\p_2}\Delta_{-k\p_1,-k\p_2}\right].
\eea

In particular this implies $\sigma_{\kt}=Q_2$, where $Q_2$ is the two-loop correction to the kink ground state mass, found in Ref.~\cite{me2loop}.

\subsubsection{The Reduced Inner Products}

The inner product of the $O(\lambda)$ correction $|\kt\rangle_2$ with the leading term $|\kt\rangle_0$ yields
\bea
\langle\kt_1|\kt_2\rangle_{\rm{red}}&\supset&
\frac{1}{\sqrt{Q_0}}\frac{\gamma^{01}_{2\kt_2}(\kt_1)}{2\okt{1}}+\frac{1}{\sqrt{Q_0}}\frac{\gamma^{01*}_{2\kt_1}(\kt_2)}{2\okt{2}}\\
&=&\frac{-\okt2 \hat \gamma_{2\kt_2}^{21}(\kt_1)+\okt1 \hat \gamma_{2\kt_1}^{21 *}(\kt_2)}{2\sqrt{Q_0}\okt 1\okt 2(\okt 2-\okt 1)}
+\frac{\okt 2\rho_{\kt_2}(\kt_1)-\okt 1\rho^*_{\kt_1}(\kt_2)}{2\sqrt{Q_0}\okt 1\okt 2(\okt 2-\okt 1)}.
\nonumber
\eea
Due to the antisymmetry, there are many cancellations in the second numerator
\bea
&&\okt 2\rho_{\kt_2}(\kt_1)=
\frac{\lambda Q_0}{4}V_{\I \kt_1 -\kt_2}
+\frac{\lambda Q_0}{8}\left(\frac{V_{\I  -\kt_2}}{\omega_{\kt_2}}-\frac{\Delta_{-\kt_2 B}}{\sqrt{\lambda Q_0}}\right)V_{\I \kt_1}\nonumber\\
&&+\frac{\lambda Q_0}{16}\ppinkp{2}\frac{V_{-\kt_2 k\p_1 k\p_2}V_{-k\p_1-k\p_2\kt_1}}{\okp1\okp2\left(\okt2-\okp1-\okp2\right)}\nonumber\\
&&+\frac{\sqrt{\lambda Q_0}}{8}\ppin{k\p}\left(\frac{\left(-\okp1\Delta_{k\p B}-\sqrt{\lambda Q_0}V_{\I  k\p}\right)
V_{-k\p-\kt_2 \kt_1}}{\okp{}^2}+\frac{\sqrt{\lambda Q_0}V_{-\kt_2 k\p \kt_1}V_{\I -k\p}}{\okp{}\left(\okt2-\okt1-\okp{}\right)}\right)\nonumber\\
&&+
\frac{\lambda Q_0}{8} \left(-\frac{\Delta_{k_1 B}}{\sqrt{\lambda Q_0}}-\frac{V_{\I  \kt_1}}{\okt1}\right)V_{\I -\kt_2}
-\frac{\lambda Q_0}{16}\ppinkp{2}\frac{V_{\kt_1k\p_1k\p_2}V_{-\kt_2-k\p_1-k\p_2}}{\okp1\okp2\left(\okt{1}+\okp1+\okp2\right)}
\eea
and so
\bea
\frac{\okt 2\rho_{\kt_2}(\kt_1)-\okt 1\rho^*_{\kt_1}(\kt_2)}{2\sqrt{Q_0}\okt1\okt2(\okt2-\okt1)} \label{diff}
&&=-\frac{\lambda \sqrt{Q_0}}{8}\frac{V_{\I-\kt_2}V_{\I\kt_1}}{\okt1^2\okt2^2}\\
&&-\frac{\lambda \sqrt{Q_0}}{32\okt1\okt2}\ppinkp{2}\frac{V_{-\kt_2 k\p_1 k\p_2}V_{-k\p_1-k\p_2\kt_1}}{\okp1\okp2\left(\okt2-\okp1-\okp2\right)\left(\okt1-\okp1-\okp2\right)}
\nonumber\\
&&+\frac{\lambda \sqrt{Q_0}}{8\okt1\okt2}\ppin{k\p}\frac{V_{-\kt_2 k\p \kt_1}V_{\I -k\p}}{\okp{}\left[(\okt2-\okt1)^2-\okp{}^2\right]}\nonumber\\
&&-\frac{\lambda \sqrt{Q_0}}{32\okt1\okt2}\ppinkp{2}\frac{V_{\kt_1k\p_1k\p_2}V_{-\kt_2-k\p_1-k\p_2}}{\okp1\okp2\left(\okt{1}+\okp1+\okp2\right)\left(\okt{2}+\okp1+\okp2\right)}.\nonumber
\eea
Similarly
\bea
\okt2 \hat \gamma_{2\kt_2}^{21}(\kt_1)&=&
\frac{3}{8}\left({\okt1}-\okt2\right)\Delta_{\kt_1 B}\Delta_{-\kt_2 B}-\frac{1}{4}\ppin{k\p}\left(\frac{\okt1\okt2}{\okp{}}+{\okp{}}{}
\right)\Delta_{-\kt_2,-k\p}\Delta_{\kt_1k\p}\nonumber\\
&&-\frac{\sqrt{\lambda Q_0}}{8}\left(\okt1\Delta_{\kt_1 B}\frac{V_{\I -\kt_2}}{\okt2}+\okt2\Delta_{-\kt_2 B}\frac{V_{\I  \kt_1}}{\okt1}\right)\nonumber\\
&&+\frac{1}{8}\ppin{k\p}
\frac{\sqrt{\lambda Q_0}\Delta_{-k\p B}V_{-\kt_2 \kt_1 k\p}}{\left(\okt2-\okt1-\okp{}\right)} \label{somma}
\eea
leading to
\beq
\frac{-\okt2 \hat \gamma_{2\kt_2}^{21}(\kt_1)+\okt1 \hat \gamma_{2\kt_1}^{21 *}(\kt_2)}{2\sqrt{Q_0}\okt1\okt2\left({\okt2}-\okt1\right)} = \frac{3\Delta_{\kt_1 B}\Delta_{-\kt_2 B}}{8\sqrt{Q_0}\okt1\okt2}-\frac{1}{8\sqrt{Q_0}\okt1\okt2}\ppin{k\p}
\frac{\sqrt{\lambda Q_0}\Delta_{-k\p B}V_{-\kt_2 \kt_1 k\p}}{\left[(\okt2-\okt1)^2-\okp{}^2\right]} .\label{gh}
\eeq
This concludes our calculation of both contributions (\ref{diff}) and (\ref{gh}) to the reduced inner product $\langle\kt_1|\kt_2\rangle_{\rm{red}}$ involving the $O(\lambda)$ corrections to the states, encoded in $\gamma_2$.  

We will now calculate contributions to the inner product involving only terms of $O(\lambda^0)$ and $O(\sl)$, encoded in $\gamma_0$ and $\gamma_1$ respectively.  We will define $\langle\kt_1|\kt_2\rangle_{n,\rm{red}}$ to consist of all such terms in Eq.~(\ref{padqft}) at the corresponding value of $n$.  Then, using the coefficients in Eqs.~(\ref{g0}) and (\ref{gammakt}), one finds
\bea
\langle\kt_1|\kt_2\rangle_{\rm{1,red}}&=&
\ppin{k_1} \frac{\gamma_{\kt_1}^{01*}(k_1)}{ (2\ok{1})}\left[\sqrt{Q_0}\gamma_{\kt_2}^{01}(k_1)+\Delta_{k_1 B}\gamma_{\kt_2}^{00}+2\ppin{k\p}\frac{\Delta_{k\p B}}{2\okp{}}{\gamma_{\kt_2}^{02}(-k\p,k_1)}
\right]\nonumber\\
&=& \frac{1}{ 2\okt{1}}\left[\sqrt{Q_0}\gamma_{\kt_2}^{01}(\kt_1)+\Delta_{\kt_1 B}\gamma_{\kt_2}^{00}+2\ppin{k\p}\frac{\Delta_{k\p B}}{2\okp{}}{\gamma_{\kt_2}^{02}(-k\p,\kt_1)}
\right]\nonumber\\
&=&\frac{\sqrt{Q_0}2\pi\delta(\kt_1-\kt_2)}{ 2\okt{1}}+\frac{1}{ 2\okt{1}\sqrt{Q_0}}\left[ 
\Delta_{\kt_1 B}\left(
 \frac{\sqrt{Q_0\lambda}V_{\I -\kt_2}}{4\okt{2}^2}-\frac{\Delta_{-\kt_2 B}}{4\okt 2}
\right)
\right.\nonumber\\
&&\left.+\ppin{k\p}\frac{\Delta_{k\p B}}{\okp{}}\left(
 -\frac{2\pi\delta(\kt_1-\kt_2)}{4}\left(\Delta_{-k\p B}+\sqrt{Q_0\lambda}\frac{V_{\I  -k\p}}{\okp{}}\right)\right.\right.\nonumber\\
 &&\left.\left.+\frac{\sqrt{Q_0\lambda}V_{-\kt_2 -k\p \kt_1}}{4\okt 2\left(\okt 2-\okp{}-\okt 1\right)}-\frac{2\pi\delta(k\p+\kt_2)}{4}\left(\Delta_{\kt_1 B}+\sqrt{Q_0\lambda}\frac{V_{\I  \kt_1}}{\okt 1}\right)
\right)
\right]\nonumber\\
&=&\frac{2\pi\delta(\kt_1-\kt_2)}{2\okt 1}\left[ 
{\sqrt{Q_0}}
-\frac{1}{\sqrt{Q_0}}\ppin{k\p}\frac{\Delta_{k\p B}}{4\okp{}}\left(\Delta_{-k\p B}+\sqrt{Q_0\lambda}\frac{V_{\I  -k\p}}{\okp{}}\right)
\right]
\nonumber\\
&&
+\frac{\Delta_{\kt_1 B}}{ 8\okt{1}\okt 2\sqrt{Q_0}}
\left(
 \frac{\sqrt{Q_0\lambda}V_{\I -\kt_2}}{\okt{2}}-{\Delta_{-\kt_2 B}}{}
\right)
-\frac{\Delta_{-\kt_2 B}}{8\okt 1\okt{2}\sqrt{Q_0}}\left({\Delta_{\kt_1 B}}{}+\sqrt{Q_0\lambda}\frac{V_{\I  \kt_1}}{\okt 1}\right)
\nonumber\\
&&+\frac{\sqrt{\lambda}}{8\okt1\okt2}\ppin{k\p}\frac{\Delta_{k\p B}V_{-\kt_2 -k\p \kt_1}}{\okp{}(\okt 2-\okp{}-\okt 1)}
\label{eq1}
\eea
and
\bea
\langle\kt_1|\kt_2\rangle_{\rm{0,red}}&=&\frac{1}{16\sqrt{Q_0}\omega_{\kt_1}\omega_{\kt_2}}\left(\frac{\sqrt{Q_0\lambda}V_{\I \kt_1}}{\omega_{\kt_1}}-\Delta_{\kt _1B}\right)\left(\frac{\sqrt{Q_0\lambda}V_{\I -\kt_2}}{\omega_{\kt_2}}+\Delta_{-\kt_2B}\right)
\label{eq0}
\eea
and
\bea
\langle\kt_1|\kt_2\rangle_{\rm{2,red}}&=&\ppink{2} \frac{\gamma_{\kt_1}^{02*}(k_1,k_2)}{4\ok 1\ok 2}\left[\sqrt{Q_0}\gamma_{\kt_2}^{02}(k_1,k_2)+\Delta_{k_1 B}\gamma_{\kt_2}^{01}(k_2)+(k_1\leftrightarrow k_2)\right]\nonumber\\
&=&\ppink{2} \frac{1}{16\ok 1\ok 2\sqrt{Q_0}}\left[-{2\pi\delta(k_2-\kt_1)}\left(\Delta_{-k_1 B}+\sqrt{Q_0\lambda}\frac{V_{\I  -k_1}}{\ok1}\right)\right.\nonumber\\
&&\left.+\frac{\sqrt{Q_0\lambda}V^*_{-\kt_1 k_1 k_2}}{\omega_{\kt_1}\left(\omega_{\kt_1}-\ok1-\ok2\right)}-{2\pi\delta(k_1-\kt_1)}{}\left(\Delta_{-k_2 B}+\sqrt{Q_0\lambda}\frac{V_{\I  -k_2}}{\ok 2}\right)\right]\nonumber\\
&&\times\left[ {2\pi\delta(k_2-\kt_2)}{}\left(\Delta_{k_1 B}-\sqrt{Q_0\lambda}\frac{V_{\I  k_1}}{\ok1}\right)+\frac{\sqrt{Q_0\lambda}V_{-\kt_2 k_1 k_2}}{2\omega_{\kt_2}\left(\omega_{\kt_2}-\ok1-\ok2\right)}\right]\nonumber\\
&=&\frac{2\pi\delta(\kt_1-\kt_2)}{16\omega_{\kt_1}\sqrt{Q_0}}\pin{k_1}\frac{1}{\ok 1}\left[Q_0\lambda\frac{|V_{\I k_1}|^2}{\ok{1}^2}-|\Delta_{k_1B}|^2  
\right]\nonumber\\
&&+\frac{1}{16\omega_{\kt_1}\omega_{\kt_2}\sqrt{Q_0}}
\left(\sqrt{Q_0\lambda}\frac{V_{\I  -\kt_2}}{\omega_{\kt_2}}+\Delta_{-\kt_2 B}\right)\left(\sqrt{Q_0\lambda}\frac{V_{\I  \kt_1}}{\omega_{\kt_1}}-\Delta_{\kt_1 B}\right)
\nonumber\\
&&+\frac{\sqrt{\lambda}}{16\omega_{\kt_1}\omega_{\kt_2}}\ppin{k\p}\frac{V_{\kt_1-\kt_2 -k\p }}{\okp{}\left(\omega_{\kt_1}-\omega_{\kt_2}-\okp{}\right)}\left(\Delta_{k\p B}-\sqrt{Q_0\lambda}\frac{V_{\I  k\p}}{\okp{}}\right)\nonumber\\
&&+\frac{\sqrt{\lambda}}{16\omega_{\kt_1}\omega_{\kt_2}}\ppin{k\p}\frac{V_{\kt_1-\kt_2 -k\p }}{\okp{}\left(\omega_{\kt_1}-\omega_{\kt_2}+\okp{}\right)}\left(\Delta_{k\p B}+\sqrt{Q_0\lambda}\frac{V_{\I  k\p}}{\okp{}}\right)
\nonumber\\
&&+\frac{\sqrt{Q_0}\lambda}{32\omega_{\kt_1}\omega_{\kt_2}}\ppinkp{2}\frac{V_{\kt_1 -k\p_1 -k\p_2}V_{-\kt_2 k\p_1 k\p_2}}{\okp1\okp2\left(\omega_{\kt_1}-\okp1-\okp2\right)\left(\omega_{\kt_2}-\okp1-\okp2\right)}.
\eea
The $V_{\I -\kt_2}V_{\I\kt_1}$ term on the second line, plus that in Eq.~(\ref{eq0}) cancel that in the first line of Eq.~(\ref{diff}).  The $\Delta_{-\kt_2 B}\Delta_{\kt_1 B}$ term on the second line, added to the contribution in Eq.~(\ref{eq0}) and the two contributions in Eq.~(\ref{eq1}) exactly cancels the first term in Eq.~(\ref{gh}).  Adding the $V_{\kt_1-\kt_2-k\p}\Delta_{k\p B}$ terms on the third and fourth lines to the last line of Eq.~(\ref{eq1}) leads to a total which precisely cancels the other term in Eq.~(\ref{gh}).

Adding the third and forth lines, the $V_{\kt_1-\kt_2 k\p}V_{\I k\p}$ term cancels that on the third line of Eq.~(\ref{diff}).  The last line cancels the second line of Eq.~(\ref{diff}).  Finally, the $\Delta_{\kt_1}V_{\I\kt_2}$ terms in the second line, added to that in Eq.~(\ref{eq0}) and in the second line of the last expression of Eq.~(\ref{eq1}) vanishes, as does its conjugate $(\kt_1\leftrightarrow \kt_2)$.  

Finally, the $n=4$ contribution is
\bea
\langle\kt_1|\kt_2\rangle_{\rm{4,red}}&=&2\pi\delta(\kt_1-\kt_2)\frac{\lambda\sqrt{Q_0}}{96\omega_{\kt_1}}\ppink{3}
\frac{|V_{k_1k_2k_3}|^2}{\ok{1}\ok{2}\ok{3}(\ok{1}+\ok{2}+\ok{3})^2}\nonumber\\
&&+\frac{\lambda\sqrt{Q_0}}{32
\omega_{\kt_1}\omega_{\kt_2}}\ppink{2}
\frac{V^*_{\kt_2 k_1 k_2}V_{\kt_1 k_1 k_2}}{\ok{1}\ok{2}(\okt 1+\ok{1}+\ok{2})(\okt 2+\ok{1}+\ok{2})}.
\eea
The second line, which is the only one which survives at $\kt_1\neq\kt_2$, cancels the last line of Eq.~(\ref{diff}), completing the cancellation of the terms in Eq.~(\ref{diff}).

\subsubsection{Remarks}

Thus we conclude that at $\kt_1\neq\kt_2$ the reduced inner product vanishes.  This is as it must be, as these represent distinct eigenstates of $H\p$.  It is thus a consistency test of our main result (\ref{padqft}).

Our derivation does not apply to $O(\sl)$ corrections at $\kt_1=-\kt_2$ as there have been terms with $(\okt1-\okt2)$ in both the numerator and denominator, which we have canceled.  Indeed the states $|\kt_1\rangle$ and $|-\kt_1\rangle$ have the same energy and so may mix.  

The problem is not simply that we were not careful, indeed there is a degenerate eigenspace  and so one is free to define $|\kt_1\rangle$ to have any overlap with $|-\kt_1\rangle$.  However, for a given physical problem, there may be a more useful prescription for the pole at $\okt1=\okt2$.  When we turn to meson multiplication below, we will see how such a physical principle fixes a related pole.  In future work, we intend to use elastic meson-kink scattering to fix the prescription for defining the pole at $\kt_1=-\kt_2.$

Similarly, our derivation is not reliable at $\kt_1=\kt_2$ as the same manipulation is ill-defined.  This is simply a reflection of our freedom to choose the normalization of $|\kt_1\rangle$.   One may, for example, fix $\gamma_{i\kt}^{01}(\kt)=0$ for all $i>0$, analogously to the condition $\gamma_2^{00}=0$ that we imposed when computing the reduced norm of the 1-kink, 0-meson state.

\section{Initial and Final State Corrections} \label{multsez}

\subsection{Motivation}

In an experiment, any initial condition is allowed.  The choice of initial condition is at the discretion of the experimenter, as it depends on how the experiment is set up.  Similarly, the choice of final states in each detection channel is determined by the experimenter, as it depends on the design of the detector.  In Ref.~\cite{memult} we considered initial state wave packets constructed as a superposition of $H\p_2$ eigenstates $|k_1\rangle_0$, each corresponding to the leading semiclassical approximation to the desired state.  In other words, the initial one-meson state was constructed exclusively using the one-meson Fock space of the free kink Hamiltonian $H\p_2$.  Similarly, the probability calculated used a projector onto the two-meson Fock space of the free Hamiltonian, which is generated by the states $|k_2k_3\rangle_0$.  This procedure is well-defined and corresponds to the result of some experiment.

However, there was a choice.  One could, instead, have used eigenstates $|k_1\rangle$ of the full Hamiltonian $H\p$ to build the wave packet.  Each element of the one-meson Fock space of the full Hamiltonian $H\p$ contains a superposition of the various $n$-meson eigenstates $|k_1\cdots k_n\rangle_0$ of the free Hamiltonian $H\p_2$.  This choice is somewhat arbitrary as the wave packet itself will not be the eigenstate of either Hamiltonian.  However, one may ask whether the resulting probability depends on this choice.  This is an important point experimentally because, if the probability depends on the choice, then one needs to determine just to which choice a given preparation method and detector corresponds.  Theoretically it is also important because, if the results differ, one choice may be compatible with an LSZ reduction theorem while the other may not.

\subsection{The Initial and Final Conditions}

In Ref.~\cite{memult} we calculated the amplitude for an initial state with one kink and one meson to evolve to a final state with one kink and two mesons.  We called this process meson multiplication.  While the initial state and final state involved no powers of $\lambda$, the interaction contained a $\sqrt{\lambda}$ and so the amplitude was of order $O(\sqrt{\lambda})$.  However, if the initial state contained a quantum correction of $O(\sqrt{\lambda})$ which could evolve via the $\lambda$-free $H\p_2$ to the final state, this would contribute at the same order.  Similarly, if the admissible final states contained an $O(\sqrt{\lambda})$  correction which has an $O(1)$ inner product with the $H\p_2$-evolved initial state, it will also contribute at the same order.  Just such corrections arise if our initial state or projector is constructed as superpositions of eigenstates of the full Hamiltonian.

Thus we are motivated to consider a reflectionless kink, so that far from the kink the normal modes become plane waves, whose form we will review shortly in Eq.~(\ref{gk}). Letting the initial meson wave packet have the same superposition coefficients as in Ref.~\cite{memult}
\beq
\alpha_{k_1}=2\sigma\sqrt{\pi}\mb_{k_1}e^{-\sigma^2\left(k_1-k_0\right)^2}e^{i(k_0-k_1)x_0}
\eeq
but this time, as a superposition of the 1-meson states $|k_1\rangle$ which are eigenstates of the full kink Hamiltonian $H\p$.  Our initial state is
\beq
\left|\Phi\right\rangle=\int \frac{d k_1}{2 \pi} \alpha_{k_1}\left|k_1\right\rangle \label{is}
\eeq
unlike Ref.~\cite{memult} where the $H\p$-eigenstate $|k_1\rangle$ was replaced with, in the notation of the present paper, the $H\p_2$-eigenstate $|k_1\rangle_0$
\beq
\left|\Phi\right\rangle_0=\int \frac{d k_1}{2 \pi} \alpha_{k_1}\left|k_1\right\rangle_0.
\eeq
Note that in both cases, one integrates over continuum modes $k_1$ with no sum over bound modes, as these vanish exponentially far from the kink, and we have assumed that $|x_0|\gg 1/m$.

Now, instead of the matrix element ${}_0\langle k_2 k_3|e^{-it(H\p_2+H\p_3)}|\Phi\rangle_0$ computed in Ref.~\cite{memult}, we will be interested in a matrix element which we write as
\beq
{}_\rv\langle k_2 k_3|e^{-itH\p}|\Phi\rangle.
\eeq
Here $|k_2k_3\rangle_\rv$ is not the kink Hamiltonian eigenstate $|k_2k_3\rangle$.  If it were, then the $H\p$ in the evolution operator would just multiply it by a phase and the matrix element would evolve by a simple phase rotation and the probability that the state contains two mesons would be time-independent.  However we are interested in a probability which begins at zero, as the initial condition contains one meson, and evolves to a nonzero value as meson multiplication occurs.  Therefore  $|k_2k_3\rangle_\rv$ is instead the translation-invariant eigenstate of $H\p$ far to the left or right of the kink, which is defined by replacing $f(x)$ with $f(-\infty)$ or $f(+\infty)$ in its definition (\ref{dfd},\ref{df}).  We refer to these limits of $H\p$ as the left and right vacuum Hamiltonians. 

In the case of a nonreflective kink, at leading order, the only relevant quantum correction in $|k_2k_3\rangle_\rv$ is
\beq
|k_2k_3\rangle_\rv=|k_2k_3\rangle_0+\frac{\sl V^{(3)}(\sl f(-\infty))\mb_{-k_2}\mb_{-k_3}\mb_{k_2+k_3}}{4\ok2\ok3(\ok2+\ok3-\omega_{k_2+k_3})}|k_2+k_3\rangle_0\label{sf}
\eeq
for an inner product with a wave packet localized at $x\ll 0$ and
\beq
|k_2k_3\rangle_\rv=|k_2k_3\rangle_0+\frac{\sl V^{(3)}(\sl f(+\infty))\md_{-k_2}\md_{-k_3}\md_{k_2+k_3}}{4\ok2\ok3(\ok2+\ok3-\omega_{k_2+k_3})}|k_2+k_3\rangle_0\label{sf2}
\eeq
for an inner product with a wave packet localized at $x\gg 0$.  The projector $\mathcal{P}$ is assembled from an integral of wave packets of $|k_2k_3\rangle_{\rm{vac}}$ localized at $x\ll 0$ and $x\gg 0$ consisting of superpositions of (\ref{sf}) and (\ref{sf2}) respectively.  Note that only the first term in  (\ref{sf}) and in (\ref{sf2}) will be relevant to initial state corrections, and the second to final state corrections.

\subsection{Calculating Initial and Final State Corrections} \label{isc}

The state at time $t$ is
\beq
|t\rangle=\pin{k_1}\alpha_{k_1}
  e^{-itH\p}|k_1\rangle=\pin{k_1}\alpha_{k_1}
  e^{-it\tilde{\omega}_{k_1}}|k_1\rangle
\eeq
where $\tilde{\omega}_{k_1}$ is the quantum corrected energy of $|k_1\rangle$.  It is equal to $\ok{1}$ plus corrections of order $O(\lambda)$ \cite{menormal}.  As we are only considering corrections of order $O(\sqrt{\lambda})$ here, we can ignore these corrections and set it to $\ok{1}$.  Next, as $\alpha_{k_1}$ is localized near $k_1=k_0$, we may expand
\beq
\tilde{\omega}_{k_1}=\ok{1}=\ok{0}+\frac{k_0}{\ok{0}}(k_1-k_0).
\eeq
We then find
\bea
|t\rangle&=&\pin{k_1}2\sigma\sqrt{\pi}\mb_{k_1}e^{-\sigma^2\left(k_1-k_0\right)^2}e^{i(k_0-k_1)x_0}
  e^{-it\left(\ok{0}+\frac{k_0}{\ok{0}}(k_1-k_0)\right)}|k_1\rangle\\
&=&2\sigma\sqrt{\pi}\mb_{k_0}e^{-i\ok{0}t}
\pin{k_1}e^{-\sigma^2\left(k_1-k_0\right)^2}e^{-i(k_1-k_0)\left(x_0+\frac{k_0}{\ok{0}}t\right)}|k_1\rangle.
\nonumber
\eea

Now, let us a consider a specific contribution to $|k_1\rangle$ at $O(\sqrt{\lambda})$
\bea
|k_1\rangle&\supset&\frac{1}{\sqrt{Q_0}}\pin{k_2}\pin{k_3}\gamma_{1k_1}^{02}(k_2,k_3) |k_2k_3\rangle_0 \label{spec}\\
&\supset&
\frac{\sqrt{\lambda}}{4\ok 1}\pin{k_2}\pin{k_3}\frac{V_{-k_1 k_2 k_3}}{\left(\ok1-\ok2-\ok3\right)} |k_2k_3\rangle_0
\nonumber
\eea
where we have used the coefficients $\gamma_{1k_1}^{02}$ reviewed in Eq.~ (\ref{gammakt}).  The case in which $k_2$ or $k_3$ is a bound mode is interesting and will be the subject of a separate study on (anti-)Stokes scattering, and so here we will consider only continuum modes $k_2$ and $k_3$.  There is a pole at $\ok{1}=\ok{2}+\ok{3}$.  This pole of course is important, as meson multiplication occurs on the pole.  But let us first consider $k_2$ and $k_3$ far from this pole, as compared with $1/\sigma$, returning to the pole in Subsec.~\ref{polesez}.  Then we can set $k_1$ to $k_0$ in the denominator and (\ref{spec}) contributes
\bea
|t\rangle&\supset&\frac{\sqrt{\lambda}\mb_{k_0}e^{-i\ok{0}t}}{4\ok 0}\pin{k_2}\pin{k_3}\frac{1}{\ok 0-\ok 2-\ok 3}\int dx \V3 \g_{k_2}(x)\g_{k_3}(x)\nonumber\\
&&\times  \left[2\sigma\sqrt{\pi}\pin{k_1} e^{-\sigma^2\left(k_1-k_0\right)^2}e^{-i(k_1-k_0)\left(x_0+\frac{k_0}{\ok{0}}t\right)}\g_{-k_1}(x)\right]|k_2 k_3\rangle_0. \label{speccon}
\eea
Let us try to evaluate the integral in square brackets at $x\gg 0$ and $x\ll 0$, where
\bea
\g_k(x)&=&\left\{\begin{tabular}{lll}
$\mb_ke^{-ikx}$&\rm{if} & $x\ll  -1/m$\\
$\md_ke^{-ikx}$&\rm{if} & $x\gg 1/m$\\
\end{tabular}
\right. \label{gk}\\
|\mb_k|^2&=&|\md_k|^2=1\hsp
\mb^*_k=\mb_{-k}\hsp
\md^*_k=\md_{-k}.\nonumber
\eea
It is
\bea
&&2\sigma\sqrt{\pi}\pin{k_1} e^{-\sigma^2\left(k_1-k_0\right)^2}e^{-i(k_1-k_0)\left(x_0+\frac{k_0}{\ok{0}}t\right)}\g_{-k_1}(x)\\
&&\hspace{3cm}=e^{ik_0 x}{\rm{Exp}}\left[-\frac{\left(-x+x_0+\frac{k_0}{\ok 0}t\right)^2}{4\sigma^2}\right]\left\{\begin{tabular}{lll}
$\mb_{-k_0}$&\rm{if} & $x\ll  -1/m$\\
$\md_{-k_0}$&\rm{if} & $x\gg 1/m$.\\
\end{tabular}
\right. \nonumber
\eea

We see that $x$  is peaked near $x_t$ where
\beq
x_t=x_0+\frac{k_0}{\ok 0}t.
\eeq
When $|x_t|\gg 0$, the Gaussian is supported at $|x|\gg 0$.  Here $f(x)$ tends to a constant, and so $\V3$ also tends to a constant, corresponding to the third derivative of the potential in one of the two vacua of the theory.  The value of the constant depends on the sign of $x_t$.  Now let us turn to the $x$ integration.  For concreteness, let us consider $t$ much smaller than the time when the meson wave packet strikes the kink, so that $x\ll 0$, then
\bea
&&\int dx \V3 \g_{k_2}(x)\g_{k_3}(x) e^{ik_0 x}{\rm{Exp}}\left[-\frac{\left(-x+x_0+\frac{k_0}{\ok 0}t\right)^2}{4\sigma^2}\right]\mb_{-k_0}\\
&&\hspace{2cm}=2\sigma\sqrt{\pi}V^{(3)}(\sqrt{\lambda}f(-\infty))\mb_{-k_0}\mb_{k_2}\mb_{k_3}e^{-\sigma^2(k_0-k_2-k_3)^2}e^{ix_t(k_0-k_2-k_3)}.
\nonumber
\eea
When $t$ is large, so that $x_t\gg 0$, one simply changes the phases $\mb$ into $\md$ and $V^{(3)}$ is evaluated at the vacuum on the right of the kink.  

Summarizing, after the collision
\bea
|t\rangle&\supset&\frac{2\sigma\sqrt{\pi}V^{(3)}(\sqrt{\lambda}f(+\infty))\sqrt{\lambda}\mb_{k_0}\md_{-k_0}e^{-i\ok{0}t}}{4\ok 0}\label{prima}\\
&&\times\pin{k_2}\pin{k_3}\frac{\md_{k_2}\md_{k_3}}{\ok 0-\ok 2-\ok 3}e^{-\sigma^2(k_0-k_2-k_3)^2}e^{ix_t(k_0-k_2-k_3)} |k_2 k_3\rangle_0\nonumber\\
&=&\frac{2\sigma\sqrt{\pi}V^{(3)}(\sqrt{\lambda}f(+\infty))\sqrt{\lambda}\mb_{k_0}\md_{-k_0}e^{-i\ok{0}t+ik_0x_t}}{4\ok 0}\nonumber\\
&&\times\pin{k_2}\pin{k_3}\frac{\g_{k_2}(x_t)\g_{k_3}(x_t)}{\ok 0-\ok 2-\ok 3}e^{-\sigma^2(k_0-k_2-k_3)^2} |k_2 k_3\rangle_0\nonumber\\
&=&\frac{V^{(3)}(\sqrt{\lambda}f(+\infty))\sqrt{\lambda}\mb_{k_0}\md_{-k_0}e^{-i\ok{0}t}}{4\ok 0}\nonumber\\
&&\times\pin{k_2}\pin{k_3}\frac{\md_{k_2}\md_{k_3}}{\ok 0-\ok 2-\ok 3}2\pi\delta(k_0-k_2-k_3) |k_2 k_3\rangle_0.\nonumber
\eea
In the last equality we considered the limit $\sigma\rightarrow\infty$.  Before the collision
\bea
|t\rangle&\supset&\frac{2\sigma\sqrt{\pi}V^{(3)}(\sqrt{\lambda}f(-\infty))\sqrt{\lambda}\mb_{k_0}\mb_{-k_0}e^{-i\ok{0}t}}{4\ok 0}\label{dopo}\\
&&\times\pin{k_2}\pin{k_3}\frac{\mb_{k_2}\mb_{k_3}}{\ok 0-\ok 2-\ok 3}e^{-\sigma^2(k_0-k_2-k_3)^2}e^{ix_t(k_0-k_2-k_3)} |k_2 k_3\rangle_0.\nonumber\\
&=&\frac{2\sigma\sqrt{\pi}V^{(3)}(\sqrt{\lambda}f(-\infty))\sqrt{\lambda}e^{-i\ok{0}t+ik_0x_t}}{4\ok 0}\nonumber\\
&&\times\pin{k_2}\pin{k_3}\frac{\g_{k_2}(x_t)\g_{k_3}(x_t)}{\ok 0-\ok 2-\ok 3}e^{-\sigma^2(k_0-k_2-k_3)^2} |k_2 k_3\rangle_0\nonumber\\
&=&\frac{V^{(3)}(\sqrt{\lambda}f(-\infty))\sqrt{\lambda}e^{-i\ok{0}t}}{4\ok 0}\nonumber\\
&&\times\pin{k_2}\pin{k_3}\frac{\mb_{k_2}\mb_{k_3}}{\ok 0-\ok 2-\ok 3}2\pi\delta(k_0-k_2-k_3) |k_2 k_3\rangle_0.\nonumber
\eea
Notice that in either case, $k_0$ and $k_2+k_3$ differ by of order $1/\sigma$, which is by assumption much less than $m$.  Therefore $\ok{0}$ is quite far from $\ok{2}+\ok{3}$, and any creation of mesons of energies $\ok{2}$ and $\ok{3}$ from the initial wave packet will be far off-shell.  Thus we expect that such terms do not contribute to the meson multiplication probability.  Do they?

To answer this question, we need only calculate the reduced inner product of $|t\rangle$ with $|k_2k_3\rangle_\rv$ in Eq.~(\ref{sf}).

After the collision and to the order of $O(\sqrt{\lambda})$, $|k_2k_3\rangle_\rv$ is given in Eq.~(\ref{sf2}).
We can easily read the coefficients $\gamma$'s off from the states. We will always take the limit $\sigma\rightarrow\infty$ and first, let's consider the inner product after the collision.
\bea \label{gamma0k2k3}
\gamma_t^{01}(k)&=&\mb_{k}e^{-i\ok{} t}2\pi\delta(k-k_0)\\
\gamma_t^{02}(k\p_2k\p_3)&=&\frac{\sqrt{\lambda}V^{(3)}(\sqrt{\lambda}f(+\infty))\mb_{k_0}\md_{-k_0}\md_{k\p_2}\md_{k\p_3}e^{-i\ok{0}t}2\pi\delta(k_0-k\p_2-k\p_3)}{4\ok 0\left(\ok 0-\okp 2-\okp 3\right)}\nonumber\\
\gamma_{k_2k_3,\rv}^{02}(k\p_2k\p_3)&=&2\pi \delta(k\p_2-k_2)2\pi \delta(k\p_3-k_3)\nonumber\\
\gamma_{k_2k_3,\rv}^{01}(k)&=&\frac{\sl V^{(3)}(\sl f(+\infty))\md_{-k_2}\md_{-k_3}\md_{k}2\pi \delta(k-k_2-k_3)}{4\ok2 \ok3 (\ok2+\ok3-\ok{})}.\nonumber
\eea
Now we again use our master formula Eq.~(\ref{padqft}) to calculate the reduced inner product to $O(\sqrt{\lambda})$
\bea
{}_{\rv}\langle k_2k_3|t\rangle_{\rm{red}}&=&\ppin{k}\frac{\gamma_{k_2k_3,\rv}^{01*}(k)}{2 \ok{}}\sqrt{Q_0}\gamma_t^{01}(k)+2\int\hspace{-17pt}\sum \frac{dk\p_2dk\p_3}{(2\pi)^2}\frac{\gamma_{k_2k_3,\rv}^{02*}(k\p_2k\p_3)}{4\okp2\okp3}\sqrt{Q_0}\gamma_t^{02}(k\p_2k\p_3)\nonumber\\
&=&\ppin{k}\frac{\sqrt{Q_0}}{2 \ok{}}\frac{\sl V^{(3)}(\sl f(+\infty))\md_{k_2}\md_{k_3}\md_{-k}2\pi \delta(k-k_2-k_3)}{4\ok2 \ok3 (\ok2+\ok3-\ok{})}\mb_{k}e^{-i\ok{} t}2\pi\delta(k-k_0)\nonumber\\
&&+2\int\hspace{-17pt}\sum \frac{dk\p_2dk\p_3}{(2\pi)^2}\frac{\sqrt{Q_0}}{4\okp2\okp3}2\pi \delta(k\p_2-k_2)2\pi \delta(k\p_3-k_3)\nonumber\\
&&\times\frac{\sqrt{\lambda}V^{(3)}(\sqrt{\lambda}f(+\infty))\mb_{k_0}\md_{-k_0}\md_{k\p_2}\md_{k\p_3}e^{-i\ok{0}t}2\pi\delta(k_0-k\p_2-k\p_3)}{4\ok 0\left(\ok 0-\okp 2-\okp 3\right)}\nonumber\\
&=&\frac{\sqrt{\lambda Q_0} V^{(3)}(\sl f(+\infty))\mb_{k_2+k_3}\md_{k_2}\md_{k_3}\md_{-k_2-k_3}e^{-i\omega_{k_2+k_3} t}2\pi\delta(k_0-k_2-k_3)}{8\ok2 \ok3 \omega_{k_2+k_3}(\ok2+\ok3-\omega_{k_2+k_3})}\nonumber\\
&&+\frac{\sqrt{\lambda Q_0} V^{(3)}(\sl f(+\infty))\mb_{k_2+k_3}\md_{k_2}\md_{k_3}\md_{-k_2-k_3}e^{-i\omega_{k_2+k_3} t}2\pi\delta(k_0-k_2-k_3)}{8\ok2 \ok3 \omega_{k_2+k_3}(\omega_{k_2+k_3}-\ok2-\ok3)}\nonumber\\
&=&0.
\eea

The calculation of the inner product before the collision is similar. Here the $|k_2k_3\rangle_{\rm{vac}}$ that appear in the projector, and so in the matrix element, are given in Eq.~(\ref{sf}).  Therefore two of the $\gamma's$ in Eq.~(\ref{gamma0k2k3}) become
\bea
\gamma_t^{02}(k\p_2k\p_3)&=&\frac{\sqrt{\lambda}V^{(3)}(\sqrt{\lambda}f(-\infty))\mb_{k\p_2}\mb_{k\p_3}e^{-i\ok{0}t}2\pi\delta(k_0-k\p_2-k\p_3)}{4\ok 0\left(\ok 0-\okp 2-\okp 3\right)}\nonumber\\
\gamma_{k_2k_3,\rv}^{01}(k)&=&\frac{\sl V^{(3)}(\sl f(-\infty))\mb_{-k_2}\mb_{-k_3}\mb_{k}2\pi \delta(k-k_2-k_3)}{4\ok2 \ok3 (\ok2+\ok3-\ok{})}.\nonumber
\eea
Again to $O(\sl)$
\bea
{}_{\rv}\langle k_2k_3|t\rangle_{\rm{red}}&=&\ppin{k}\frac{\gamma_{k_2k_3,\rv}^{01*}(k)}{2 \ok{}}\sqrt{Q_0}\gamma_t^{01}(k)+2\int\hspace{-17pt}\sum \frac{dk\p_2dk\p_3}{(2\pi)^2}\frac{\gamma_{k_2k_3,\rv}^{02*}(k\p_2k\p_3)}{4\ok2\ok3}\sqrt{Q_0}\gamma_t^{02}(k\p_2k\p_3)\nonumber\\
&=&\ppin{k}\frac{\sqrt{Q_0}}{2 \ok{}}\frac{\sl V^{(3)}(\sl f(-\infty))\mb_{k_2}\mb_{k_3}\mb_{-k}2\pi \delta(k-k_2-k_3)}{4\ok2 \ok3 (\ok2+\ok3-\ok{})}\mb_{k}e^{-i\ok{} t}2\pi\delta(k-k_0)\nonumber\\
&&+2\int\hspace{-17pt}\sum \frac{dk\p_2dk\p_3}{(2\pi)^2}\frac{\sqrt{Q_0}}{4\ok2\ok3}2\pi \delta(k\p_2-k_2)2\pi \delta(k\p_3-k_3)\nonumber\\
&&\times\frac{\sqrt{\lambda}V^{(3)}(\sqrt{\lambda}f(-\infty))\mb_{k\p_2}\mb_{k\p_3}e^{-i\ok{0}t}2\pi\delta(k_0-k\p_2-k\p_3)}{4\ok 0\left(\ok 0-\okp 2-\okp 3\right)}\nonumber\\
&=&\frac{\sqrt{\lambda Q_0} V^{(3)}(\sl f(+\infty))\mb_{k_2}\mb_{k_3}e^{-i\omega_{k_2+k_3} t}2\pi\delta(k_0-k_2-k_3)}{8\ok2 \ok3 \omega_{k_2+k_3}(\ok2+\ok3-\omega_{k_2+k_3})}\nonumber\\
&&+\frac{\sqrt{\lambda Q_0} V^{(3)}(\sl f(-\infty))\mb_{k_2}\mb_{k_3}e^{-i\omega_{k_2+k_3} t}2\pi\delta(k_0-k_2-k_3)}{8\ok2 \ok3 \omega_{k_2+k_3}(\omega_{k_2+k_3}-\ok2-\ok3)}=0.
\eea
We see that the inner product also vanishes.  Thus initial and final state corrections do not contribute at this order away from the pole.  Of course this is to be expected, as the process only conserves energy at the pole.

\subsection{The Pole Contribution} \label{polesez}

In Eq.~(\ref{speccon}) we calculated the contribution of the term (\ref{spec}) to the meson multiplication amplitude, and found that it vanished.  However we ignored the contribution from the pole at $\ok 1=\ok 2+\ok 3$.  More precisely, we set $k_1$ to $k_0$ in the denominator, although in general they differ by of order $O(1/\sigma)$.  This approximation is reasonable except in a neighborhood of size $1/\sigma$ of the pole.  In the limit $\sigma\rightarrow\infty$ this becomes valid except in an infinitesimal neighborhood of the pole.  One thus expects that the error introduced depends only on the integrand in that infinitesimal neighborhood, and in particular only on the residue of the pole.  

At this pole, energy is conserved, and so this contribution to the multiplication would be on-shell.  Let us rewrite the contribution (\ref{speccon}) to the amplitude, now keeping the contribution from the pole
\bea
|t\rangle&\supset&\frac{\sqrt{\lambda}\mb_{k_0}e^{-i\ok{0}t}}{4}\pin{k_2}\pin{k_3}\int dx \V3 \g_{k_2}(x)\g_{k_3}(x)\nonumber\\
&&\times  2\sigma\sqrt{\pi}\left[\pin{k_1} \frac{e^{-\sigma^2\left(k_1-k_0\right)^2}e^{-i(k_1-k_0)x_t}}{\ok 1-\ok 2-\ok 3}\frac{\g_{-k_1}(x)}{\ok 1}\right]|k_2 k_3\rangle_0. \label{dint}
\eea
As is written, the integral is not defined at the pole.  

The problem is as follows.  Let us define the location of the pole by $k_1=k_I$ such that
\beq
\ok I=\ok 2+\ok 3\hsp k_I>0. \label{oki}
\eeq
There is another pole at $k_1=-k_I$.  The 2-meson contribution to the 1-meson Hamiltonian eigenstate $|k_1\rangle$ is summarized by the coefficient $\gamma_{1k_1}^{02}(k_2,k_3)$, which has simple poles at $k_1=\pm k_I$, where meson multiplication is on-shell.  At the locations of the poles the integrand is, of course, infinite and so the integral is ill-defined.  

The origin of this ambiguity can be seen in its derivation.  This coefficient was derived in Ref.~\cite{menormal} from the fact that $|k_1\rangle$ is an eigenstate of the kink Hamiltonian $H\p$
\beq
(H\p-E)|k_1\rangle=0. \label{se}
\eeq
Let us consider the $O(\sl)$ part of this equation, projected onto the 2-meson part of the free Fock space $|k_2k_3\rangle_0$.  There is no 2-meson contribution at $O(\lambda^0)$, so the state itself is already at $O(\sl)$, meaning that the $H\p-E$ can only contribute at $O(\lambda^0)$.  The only such contributions are
\beq
H\p_2|k_2k_3\rangle_0= \left(Q_1+\ok 2+\ok 3\right)|k_2k_3\rangle_0\hsp E|k_2k_3\rangle_0=(Q_1+\ok 1)|k_2 k_3\rangle_0.
\eeq
Using Eq.~(\ref{oki}) one sees that the two contributions to the left hand side of (\ref{se}) cancel if $k_1=\pm k_I$.  

Therefore, the eigenvalue equation (\ref{se}) is always satisfied when $k_1=\pm k_I$, whatever $\gamma_{1k_1}^{02}(k_2,k_3)$ is chosen.  This function is, as a result, undetermined for on-shell values of $k_2$ and $k_3$, in other words, at the pole.  One is free to add to $\gamma_{1k_1}^{02}(k_2,k_3)$ any function of $k_2$ multiplied by $\delta(k_1\pm k_I)$, which, by the Sokhotski–Plemelj theorem, roughly corresponds to adding a small imaginary function to the denominator of the pole.

Physically, this means that there are many kink Hamiltonian eigenstates which we would equally well call $|k_1\rangle$, each corresponding to a different mix of 2-meson states with the same energy.  These mixtures correspond to different prescriptions for evaluating the integral over the pole.  The question is, which of these choices of eigenstate $|k_1\rangle$ provides an appropriate initial condition, and therefore should be used to define the initial state in Eq.~(\ref{is})?   We are interested in calculating the probability of conversion of a 1-meson state into a 2-meson state upon a collision of the meson with a kink.  Therefore, we will impose that for times long before the collision, the probability that the initial state contains two mesons is equal to zero.  This additional physical criterion will allow us to fix our definition of $|k_1\rangle$, or equivalently the prescription for evaluating the integral at the pole.

At this point, one could add arbitrary functions to $\gamma_{1k_1}^{02}(k_2,k_3)$ at each pole, calculate the probability for each at small times and use the result to identify the correct function.  We will opt for a simpler, but let us direct approach.

Let us, for now, simply guess that the pole is defined using a principal value prescription. We can evaluate the contributions to the term in brackets in (\ref{dint}) at the poles using the Sokhotski–Plemelj theorem.  We have already argued that the contributions away from the poles do not contribute to the amplitude, so we only need to consider $\pm i\pi$ times the residue at each pole.  At the pole $k_1=-k_I$ the residue contains a factor of $e^{-\sigma^2(k_I+k_0)^2}$ which vanishes in the limit $\sigma\rightarrow\infty$ and so we will not consider that pole further.


If $x_t>x$ then the contour should be closed below yielding $-\pi i$ times the residue, otherwise it should be closed above yielding $\pi i$ times the residue.  Note that naively the Gaussian term diverges on such a contour.  However, it only contributes a constant factor to the integral in a $1/\sigma$-neighborhood of the pole in the $\sigma\rightarrow\infty$ limit, and so one can simply set the $k_1$ in the Gaussian to its value at the pole before performing the integration.  This affects the value of the integral over the real line, but as we have argued, only the integral in a neighborhood of the pole can contribute to the amplitude.  The term in brackets in (\ref{dint}) thus becomes
\beq
-\sign{x_t-x}\frac{i}{2k_I}  e^{-\sigma^2\left(k_I-k_0\right)^2} e^{-i(k_I-k_0)x_t}\g_{-k_I}(x). \label{fint}
\eeq

An alternate derivation, without use of contour integrals, is as follows.  At large $|x|$ the $\g_{-k_1}(x)$ in the square bracket, up to a constant phase $\mb_{-k_1}$ or $\md_{-k_1}$, is just $e^{i k_1 x}$.  Combining this with the $e^{-i(k_1-k_0)x_t}$ yields
\beq
e^{-i(k_1-k_0)x_t}e^{i k_1 x}
=e^{-i(k_1-k_I)(x_t-x)}e^{-i(k_I-k_0)x_t}e^{ik_Ix}. \label{fasa}
\eeq
The third term on the right hand side, together with the phase $\mb_{-k_1}$ or $\md_{-k_1}$, becomes the $g_{-k_I}(x)$ in (\ref{fint}).   The second also appears in (\ref{fint}).  At large $\sigma$, we may expand the denominator $(\ok1-\ok I)\ok 1$ of the term in brackets to linear order in $(k_1-k_I)$, yielding $(k_1-k_I)k_1$.  This denominator is odd in $(k_1-k_I)$, and so only the odd term in the first term on the right hand side of (\ref{fasa}) contributes.  Dividing this by $(k_1-k_I)k_1$ one identifies the nascent delta function
\beq
\lim{|x_t-x|\rightarrow\infty}-\frac{{\rm sin}\left[(k_1-k_I)(x_t-x)  \right]}{(k_1-k_I)k_1}=-\pi{\rm sign}(x_t-x)\frac{\delta(k_1-k_I)}{k_1}
\eeq
which can then be used to perform the $k_1$ integral in the square brackets in Eq.~(\ref{dint}), leading again to Eq.~(\ref{fint}).

This does not satisfy our physical criterion that the probability for the state to contain two mesons should be zero at early times and nonzero at late times.  On the contrary, the amplitude is, up to a sign, symmetric in time and so the probability of observing two mesons will be the same in the far past and the far future.

Let us break the time reversal symmetry with another choice of prescription for interpreting the pole in $\gamma_{1 k_1}^{02}$.  Instead of the principal value prescription, let us try
\beq
\gamma_{1 k_1}^{02}(k_2,k_3)= \frac{2\pi\delta(k_3-k_1)}{2}\left(-\Delta_{k_2 B}-\sqrt{Q_0\lambda}\frac{V_{\I  k_2}}{\ok2}\right)+\frac{\sqrt{Q_0\lambda}V_{-k_1 k_2 k_3}}{4\omega_{k_1}\left(\omega_{k_1}-\ok2-\ok3+i\epsilon\right)}. \label{newinit}
\eeq
In the next subsection we will explain why such a shift leads to another Hamiltonian eigenstate with the same energy and so is allowed.

Now the pole on the complex $k_1$ plane is at an infinitesimal negative imaginary value.  As a result, if $x_t<x$ then the pole is not included in the contour.  Now the term in brackets becomes
\beq
-\Theta(x_t-x)\frac{i}{k_I}  e^{-\sigma^2\left(k_I-k_0\right)^2} e^{-i(k_I-k_0)x_t}\g_{-k_I}(x)
\eeq
where $\Theta$ is the Heaviside step function.  The corresponding contribution to $|t\rangle$ is
\bea
|t\rangle&\supset&-\frac{\sqrt{\lambda}\mb_{k_0}e^{-i\ok{0}t}}{4}\pin{k_2}\pin{k_3}\int_{-\infty}^{x_t} dx \V3 \g_{k_2}(x)\g_{k_3}(x)\nonumber\\
&&\times  2\sigma\sqrt{\pi}\left[\frac{i }{k_I}  e^{-\sigma^2\left(k_I-k_0\right)^2} e^{-i(k_I-k_0)x_t}\g_{-k_I}(x)\right]|k_2 k_3\rangle_0.
\eea
Now if $x_t\ll 0$, so that the meson wave packet has not reached the kink, then the $x$-integral will only cover the asymptotic region where $\g_{k_2}(x)\g_{k_3}(x)\g_{-k_I}(x)\sim e^{ix(k_I-k_2-k_3)}$ oscillates rapidly, exponentially suppressing the amplitude.  On the other hand, after the collision $x_t\gg 0$ and so the integral is, up to an exponentially suppressed correction, equal to $V_{-k_Ik_2k_3}$.  The state is then
\beq
|t\rangle\supset-\Theta(x_t)\frac{i\sigma\sqrt{\pi\lambda}\mb_{k_0}e^{-i\ok{0}t}}{2 }\pin{k_2}\pin{k_3} e^{-\sigma^2\left(k_I-k_0\right)^2} e^{-i(k_I-k_0)x_t}\frac{V_{-k_I k_2 k_3}}{k_I}|k_2 k_3\rangle_0.
\eeq


Using (\ref{grred}) to evaluate the denominator, this leads to the reduced matrix-element
\beq
\frac{{{}_{\rm{vac}}\langle k_2 k_3|t\rangle_{\rm{red}}}}{{}_{\rm{}}\langle 0|0\rangle_{\rm{red}}}\supset-\Theta(x_t)\frac{i\sigma\sqrt{\pi\lambda}\mb_{k_0}e^{-i\ok{0}t}}{4\ok 2\ok 3k_I }e^{-\sigma^2\left(k_I-k_0\right)^2} e^{-i(k_I-k_0)x_t}V_{-k_I k_2 k_3}
\eeq
plus corrections of order $O(\lambda^{3/2})$, in agreement with the amplitude reported in Ref.~\cite{memult}.  Here we used the fact that in the large $\sigma$ limit, the Gaussian term is supported at final momenta such that $|k_I-k_0|\sim 1/\sigma$.  Thus the only final momenta which can contribute to the matrix element are those such that the $e^{-i(k_I-k_0)x_t}$ term tends to $1$.  

However, here we have considered a translation-invariant initial and final state.  Thus we conclude that the higher order corrections to the initial and final state which lead to translation-invariance do not affect the meson multiplication amplitude at order $O(\sqrt{\lambda}).$



\subsection{Degenerate Eigenstates}


We defined $|k_1\rangle$ to be the $H\p$ eigenstate which is annihilated by $P\p$ and whose leading order term is $|k_1\rangle_0$.  This does not completely characterize the state, because there are other translation-invariant states with the same energy.  Consider any $k_2$ and $k_3$ such that $\tilde{\omega}_{k_2}+\tilde{\omega}_{k_3}=\tilde{\omega}_{k_1}$.  Recall that, up to corrections of order $O(\lambda)$, which we do not consider, this condition is $\ok{2}+\ok{3}=\ok{1}$.  Then the state $|k_2 k_3\rangle$ has the same energy as $|k_1\rangle$ and it is, by construction, also translation invariant.  

Let us shift the definition of $|k_1\rangle$ by
\beq
|k_1\rangle\longrightarrow |k_1\rangle+c_{k_1k_2k_3}\sqrt{\lambda}|k_2 k_3\rangle
\eeq
where $c$ is of order $O(\lambda^0)$ and is nonvanishing only when $\tilde{\omega}_{k_2}+\tilde{\omega}_{k_3}=\tilde{\omega}_{k_1}$.  Now the energy eigenvalues match in the new term, so the argument used above, to argue that the contributions to $|k_1\rangle$ in $\gamma_{1k_1}^{m2}$ do not contribute to the amplitude, cannot be applied.  

This new choice of $|k_1\rangle$ also satisfies our definition.   However it differs from the old choice by a change in $\gamma_1^{20}(k_2,k_3)$.  In fact, any value of $\gamma_1^{20}(k_2,k_3)$ corresponds to some choice of $c_{k_1k_2k_3}$ so long as it agrees with the old value in Eq.~(\ref{gammakt}) when $\ok 1\neq \ok 2+\ok 3$.  Intuitively, one may only add something proportional to $\delta(\ok 1-\ok 2-\ok 3)$.  The infinitesimal shift in the pole in (\ref{newinit}) is exactly of this form.



We thus claim that the correct initial condition in Ref.~\cite{memult} corresponds to Eq.~(\ref{gammakt}) with $\gamma_1^{21}$ replaced by Eq.~(\ref{newinit}).   What if the meson wave packet scatters with the kink from the other side?  Then $x_0>0$ and $k_0<0$.  In this case, we want the integral to vanish when $x_t<x$, so that the $k_1$ contour is closed on the bottom of the complex plane.  This requires the pole to be shifted by $+i\epsilon$.  However, as $k_0<0$, this still corresponds to a negative imaginary part for $\ok 1$, and so still corresponds to the modification (\ref{newinit}).   We remind the reader that this state has the same energy, momentum and $O(\lambda^0)$ term as the state defined by Ref.~\cite{memult}, but does not lead to a two-meson component in the initial wave packet $|\Phi\rangle$.

\section{Remarks}

Given a stationary kink solution, one may compute its normal modes and even their interactions \cite{shapeinter}.  Every year, this is done for new classes of models \cite{wshifman,takyi1,takyi2}, including recently even gravitating kinks \cite{yuan1,yuan2}.  With these normal modes in hand, one can construct the quantum states corresponding to kinks.   Recently there has even been progress towards to a quantum treatment of nontopological solitons \cite{quantosc,kovbreather}.  However every such treatment needs to deal with the fact that translation-invariant kink states are non-normalizable, as a result of the infinite volume of the translation group.

There are many proposed solutions to this problem, each useful in some settings.  Many have the drawback that they destroy translation-invariance, they do not preserve local quantities or they cause finite shifts.  In the present note, we have proposed another method of dealing with this problem, replacing inner products by reduced inner products where we have quotiented by the translation group.  This is, in our opinion, a reasonable approach as the volume of the translation group appears in the numerator and denominator of observable quantities and so is canceled.  We have found that this greatly simplifies many calculations, as we are able to fix the translation symmetry so that all terms with zero modes $\phi_0$ vanish.  

The problem was complicated by our choice of coordinates $y$, defined to be the eigenvalue of $\phi_0$.  Intuitively, this can be understood as follows.  Let $f(x)$ be a classical kink solution.  A shift in the collective coordinate transforms $f(x)$ to $f(x-x_0)$, and so acts linearly on the position.  On the other hand, a shift in $y$ changes $f(x)$ to $f(x)-y_0 f\p(x_0)/\sqrt{Q_0}$.  This does not correspond to a shifted kink solution, unless one simultaneously compensates by shifting the normal modes.  Thus the nondiagonal part of the Jacobian factor arising from a quotient by this translation symmetry is proportional to $\Delta_{Bk}$, which is the mixing between the zero mode $\g_B(x)$ and the other normal modes, when one shifts $x$.  However, at small $y_0$ these two transformations are related by a simple proportionality factor of $\sqrt{Q_0}$, which allows us to easily define a matching condition and evaluate the necessary Jacobian.

In Ref.~\cite{menormal}, the leading corrections to a 1-meson state $|\kt\rangle$ were found, summarized here in Eq.~(\ref{gammakt}).  However, if $\omega_{\kt}\geq 2m$ then this state has the same energy and momenta as some 2-meson states.  The state always contains a cloud of off-shell 2-meson states.  In Eq.~(\ref{gammakt}), the degenerate states are on-shell.  The physically correct initial condition to study 2-meson production is to begin with a state that does not contain a component with two on-shell mesons.   In Eq.~(\ref{newinit}) we present this state.  It is equal to that of Ref.~\cite{menormal}, with a subleading contribution from a degenerate 2-meson state.  This subleading contribution is added by including an infinitesimal, imaginary shift of a pole.   We suspect more generally that such poles in states represent on-shell contributions from degenerate states, which can and often should be removed via such imaginary shifts.  Said differently, we suspect that the prescription for evaluating such poles corresponds in general to a physical choice of Hamiltonian eigenstate in a degenerate eigenspace.





\section* {Acknowledgement}

\noindent
JE is supported by NSFC MianShang grants 11875296 and 11675223. HL acknowledges the support from CAS-DAAD Joint Fellowship Programme for Doctoral students of UCAS.

\end{document}

\section{Stokes Scattering}

\bea
H_I&=&\frac{\sqrt{\lambda}}{2} \int \frac{d k_1}{2 \pi} \frac{d k_2}{2 \pi}  \frac{V_{-k_1 k_2 S}}{\omega_{k_1}} B_{k_2}^{\ddagger} B_{S}^{\ddagger} B_{k_1} \\
V_{-k_1 k_2 S}&=&\int d x V^{(3)}(\sqrt{\lambda} f(x)) \mathfrak{g}_{-k_1}(x) \mathfrak{g}_{k_2}(x) \mathfrak{g}_{S}(x).\nonumber
\eea

\beq
\Phi(x)=\operatorname{Exp}\left[-\frac{\left(x-x_0\right)^2}{4 \sigma^2}+i x k_0\right], \quad x_0 \ll-\frac{1}{ m}, \quad  \frac{1}{k_0},\frac{1}{m}\ll\sigma \ll\left|x_0\right| .
\eeq

\begin{equation}
\alpha_k=\int d x \Phi(x) \mathfrak{g}_k^*(x).
\end{equation}

\beq
|S k\rangle=B_s^\ddag B^\ddag_k\vac_0\hsp |k\rangle=B^\ddag_k\vac_0.
\eeq

\begin{equation}
\left|\Phi\right\rangle=\pin{k_1} \alpha_{k_1}\left|k_1\right\rangle
\end{equation}

\beq
e^{-it(H\p_2+H_I)}=e^{-itH\p_2}-i\int _0^t dt_1 e^{-i(t-t_1)H\p_2}H_I e^{-i t_1 H\p_2}+O(\lambda)
\eeq

\beq
\ok{I}=\ok{2}+\os\hsp k_I>0
\eeq

\beq
e^{-iH t}|k_1\rangle=\frac{-i\sqrt{\lambda}}{2\omega_{k_1}} \pin{k_2}V_{S,k_2,-k_1}e^{-\frac{it}{2}(\omega_{k_1}+\os+\ok{2})}\frac{\rm{sin}\left[\left(\frac{\os+\ok{2}-\ok{1}}{2}\right)t\right]}{(\os+\ok{2}-\ok{1})/2}|S k_2\rangle
\eeq

\beq
\frac{\rm{sin}\left[\left(\frac{\os+\ok{2}-\ok{1}}{2}\right)t\right]}{(\os+\ok{2}-\ok{1})/2}=2\pi\delta(\os+\ok{2}-\ok{1})=\left(\frac{\ok{I}}{k_I}\right)\left(2\pi\delta(k_1-k_I)+2\pi\delta(k_1+k_I)\right)
\eeq

\bea
e^{-iH t}|\Phi\rangle&=&-i\sqrt{\lambda}\pin{k_1}\frac{\alpha_{k_1}}{2\ok{1}} \pin{k_2}V_{S,k_2,-k_1}e^{-\frac{it}{2}(\ok{1}+\os+\ok{2})}\frac{\rm{sin}\left[\left(\frac{\os+\ok{2}-\ok{1}}{2}\right)t\right]}{(\os+\ok{2}-\ok{1})/2}|S k_2\rangle\nonumber\\
&=&\frac{-i\sqrt{\lambda}}{2} \pin{k_2}e^{-i\ok{I}t}\left(\frac{1}{k_I}\right)\left(\alpha_{k_I}V_{S,k_2,-k_I}+\alpha_{-k_I}V_{S,k_2,k_I}
\right)|S k_2\rangle
\eea

\bea
\g_k(x)&=&\left\{\begin{tabular}{lll}
$\mb_ke^{ikx}+\mc_ke^{-ikx}$&\rm{if} & $x\ll  -1/m$\\
$\md_ke^{ikx}+\me_k e^{-ikx}$&\rm{if} & $x\gg 1/m$\\
\end{tabular}
\right. \label{gk}\\
|\mb_k|^2+|\mc_k|^2&=&|\md_k|^2+|\me_k|^2=1\hsp
\mb^*_k=\mb_{-k}\hsp
\mc^*_k=\mc_{-k}\hsp
\md^*_k=\md_{-k}\hsp
\me^*_k=\me_{-k}.\nonumber
\eea

\bea
\alpha_{k_I}&=&2\sigma\sqrt{\pi}\left[\mb^*_{k_I}e^{ix_0(k_0-k_I)}e^{-\sigma^2(k_0-k_I)^2}+\mc^*_{k_I}e^{ix_0(k_0+k_I)}e^{-\sigma^2(k_0+k_I)^2}
\right]\\
&=&2\sigma\sqrt{\pi}\mb^*_{k_I}e^{ix_0(k_0-k_I)}e^{-\sigma^2(k_0-k_I)^2}\nonumber
\eea

\bea
\alpha_{-k_I}&=&2\sigma\sqrt{\pi}\left[\mb_{k_I}e^{ix_0(k_0+k_I)}e^{-\sigma^2(k_0+k_I)^2}+\mc_{k_I}e^{ix_0(k_0-k_I)}e^{-\sigma^2(k_0-k_I)^2}
\right]\\
&=&2\sigma\sqrt{\pi}\mc_{k_I}e^{ix_0(k_0-k_I)}e^{-\sigma^2(k_0-k_I)^2}\nonumber
\eea

\bea
e^{-iH t}|\Phi\rangle&=&-i\sigma\sqrt{\pi\lambda} \pin{k_2}e^{ix_0(k_0-k_I)}e^{-\sigma^2(k_0-k_I)^2}e^{-i\ok{I}t}\left(\frac{\tilde{V}_{S,k_2,-k_I}}{k_I}\right)|S k\rangle\\
\tilde{V}_{S,k_2,-k_I}&=&\mb^*_{k_I}V_{S,k_2,-k_I}+\mc_{k_I}V_{S,k_2,k_I}
\nonumber
\eea

\beq
\langle S k_1|S k_2\rangle=\frac{2\pi\delta(k_1-k_2)}{4\os\ok{1}}{}_0\langle 0\vac_0.
\eeq

\beq
\frac{\langle S k_2|e^{-iH t}|\Phi\rangle}{{}_0\langle 0\vac_0}=\frac{-i\sigma\sqrt{\pi\lambda}}{4\os\ok{2}k_I} e^{ix_0(k_0-k_I)}e^{-\sigma^2(k_0-k_I)^2}e^{-i\ok{I}t}\tilde{V}_{S,k_2,-k_I}
\eeq

\bea
\left|\frac{\langle S k_2|e^{-iH t}|\Phi\rangle}{{}_0\langle 0\vac_0}\right|^2&=&\frac{\sigma^2\pi\lambda}{16\os^2\ok{2}^2k^2_I}\left|\tilde{V}_{S,k_2,-k_I}\right|^2e^{-2\sigma^2(k_0-k_I)^2}\\
&=&\frac{\sigma\pi^{3/2}\lambda}{16\sqrt{2}\os^2\ok{2}^2k^2_I}\left|\tilde{V}_{S,k_2,-k_I}\right|^2\delta(k_I-k_0)\nonumber
\eea

\beq
\mathcal{P}=\pin{k_2} \frac{4\os\ok{2}}{{}_0\langle 0\vac_0} |Sk_2\rangle\langle Sk_2|
\eeq

\beq
\frac{\langle k_1|k_2\rangle}{{}_0\langle 0\vac_0}=\frac{2\pi\delta(k_1-k_2)}{2\ok{1}}
\eeq

\bea
\frac{\langle\Phi|\Phi\rangle}{{}_0\langle 0\vac_0}&=&\pink{2}\alpha_{k_1}\alpha^*_{k_2}\frac{\langle k_2|k_1\rangle}{{}_0\langle 0\vac_0}=\pin{k}\frac{|\alpha_k|^2}{2\ok{}}=\frac{1}{2\omega_{k_0}}\pin{k}|\alpha_k|^2\\
&=&\frac{1}{2\omega_{k_0}}\pin{k}\int dx\int dy g_k^*(x)g_k(y)\Phi(x)\Phi^*(y)
\nonumber\\
&=&\frac{1}{2\omega_{k_0}}\int dx |\Phi(x)|^2=\frac{\sigma\sqrt{\pi}}{\sqrt{2}\omega_{k_0}}\nonumber
\eea

\bea
P&=&\frac{\langle\Phi|e^{iHt}\mathcal{P}e^{-iHt}|\Phi\rangle}{\langle\Phi|\Phi\rangle}=\pin{k_2} \frac{4\os\ok{2}}{{}_0\langle 0\vac_0}\frac{\left|\langle Sk_2|e^{-iHt}|\Phi\rangle\right|^2}{\langle\Phi|\Phi\rangle/{}_0\langle 0\vac_0}\frac{1}{{}_0\langle 0\vac_0}\\
&=&\pin{k_2}4\os\ok{2}\frac{\frac{\sigma\pi^{3/2}\lambda}{16\sqrt{2}\os^2\ok{2}^2k^2_I}\left|\tilde{V}_{S,k_2,-k_I}\right|^2\delta(k_I-k_0)}{\left(\frac{\sigma\sqrt{\pi}}{\sqrt{2}\omega_{k_0}}\right)}
\nonumber\\
&=&\frac{\pi\lambda\ok{0}}{4\os (\ok{0}-\os)k_0^2}\pin{k_2}\left|\tilde{V}_{S,k_2,-k_I}\right|^2\delta(k_I-k_0)\nonumber\\
&=&\lambda\frac{\left|\tilde{V}_{S,\sqrt{(\ok{0}-\ok{S})^2-m^2},-k_0}\right|^2+\left|\tilde{V}_{S,-\sqrt{(\ok{0}-\ok{S})^2-m^2},-k_0}\right|^2
}{8\os k_0\sqrt{(\ok{0}-\ok{S})^2-m^2}}\nonumber
\eea

\section{Anti-Stokes Scattering}

\bea
H_I&=&\frac{\sqrt{\lambda}}{4\os} \int \frac{d k_1}{2 \pi} \frac{d k_2}{2 \pi}  \frac{V_{-k_1 k_2 S}}{\omega_{k_1}} B_{k_2}^{\ddagger} B_{S} B_{k_1} 
\eea

\beq
\left|\Phi\right\rangle=\pin{k_1} \alpha_{k_1}\left|S k_1\right\rangle
\eeq

\section{Example: $\phi^4$ Double-Well Model}

{\blu{Below I took the complex conjugate of the $\phi^4$ paper ref, is that the right way to change the convention?  $k\rightarrow -k$ wouldn't do anything}}
\beq
V_{k_1k_2S}=i\pi \frac{3\sqrt{3\lambda}}{8}\frac{\left(17\b^4-(\ok1^2-\ok2^2)^2\right)(\b^2+k_1^2+k_2^2)+8\b^2k_1^2k_2^2}{\b^{3/2}\ok1\ok2\sqrt{\b^2+k_1^2}\sqrt{\b^2+k_2^2}}\sech\left(\frac{\pi(k_1+k_2)}{2\b}\right).
\eeq

\section{Remarks}

\end{document}

Two-dimensional scalar models provide an ideal sandbox for developing tools to treat real-world solitons.  If a scalar field is subjected to a potential with degenerate minima, then the theory will enjoy kink and antikink solutions.  In general, at weak coupling, one can decompose a given configuration into kinks and also perturbative, elementary quanta of the scalar field, called mesons.  An understanding of these theories at weak coupling is then reduced to understanding the interactions of mesons with one another, of kinks with (anti)kinks and of kinks with mesons.

The interactions of mesons with one another is largely as in the perturbative theory with no kinks, and so is well understood.  Interactions of kinks with (anti)kinks in classical field theory is a rich field and has been a subject of intense investigation since the discovery of resonance windows \cite{csw} and related phenomena \cite{osc,osc3d}.  It was once thought that these phenomena can be understood in terms of the internal excitations of the kink, but it has been found in Ref.~\cite{doreyf6} that resonances persist in the $\phi^6$ theory, whose kink has no internal excitations.  Instead, although certainly the internal excitations do affect the scattering phenomenology \cite{multex22a,multex22b}, it is now widely believed \cite{sfal21,col22} that a decisive role is played by the interactions of kinks with bulk excitations, which are not localized to a single kink and in this sense are related to mesons.

Kink-meson interactions have received relatively little attention, despite being the simplest nonperturbative scattering processes in such models.  In classical field theory, the mesons correspond to radiation.  Using the perturbative approach to the classical equations of motion for radiation introduced in Ref.~\cite{mm}, incident radiation upon a kink was studied in Refs.~\cite{tomrad1,tomrad2}.  It was found that if the kink is reflectionless, and the radiation is monochromatic with frequency $\omega$, then some of the transmitted radiation will have a frequency of $2\omega$ and this frequency doubling will exert a negative pressure on the kink.  In a quantized model this is easy to understand, it represents the process kink$+2$mesons$ \rightarrow $kink$+$meson.  One can show that energy conservation among the mesons, which is exact at leading order, implies that the final state meson has more momentum than the two merged mesons, with the difference causing a negative recoil of the kink.  This, including higher-order meson merging, is the only processes admitted in the case of classical reflectionless kinks.  In the case of reflective kinks, Ref.~\cite{tomrad3} found that there is also meson reflection, yielding a positive contribution to the pressure.

In the present note we consider a new process, meson multiplication, in which a meson incident on a kink splits into two mesons.  This process appears to have no classical counterpart, in the sense that the perturbative approach of Ref.~\cite{mm} is able to solve any initial value problem which begins with frequency $\omega$ monochromatic radiation perturbatively, and it only yields radiation components whose frequencies are integer multiples of $\omega$. 

We will thus show that meson-kink interactions have a very different character in the quantum regime as compared with the classical regime, with the former leading to positive pressure and the second negative pressure.  To some extent this is not surprising, as an initial state consisting of $N$ mesons will yield a number of meson multiplication events proportional to $N$, while the probability of meson fusion will be of order $O(N^2)$.  Thus one expects meson fusion to dominate for sufficiently intense meson sources.

We begin in Sec.~\ref{revsez} with a review of the linearized kink perturbation theory of Refs.~\cite{mekink, me2loop}.  This quantum field theoretic approach is much more economical than the traditional collective coordinate approach of Refs.~\cite{gjscc,gj76}, in particular in the one-kink sector.  Next in Sec.~\ref{moltsez} we calculate the probability of meson multiplication in a general (1+1)d scalar field theory.  In Sec.~\ref{exsez} we apply this formula to two reflectionless kinks: the sine-Gordon soliton and the $\phi^4$ kink.  As a result of integrability, of course, this process does not occur in the sine-Gordon case.  Finally, in Sec.~\ref{numsez}, we numerically evaluate various probabilities associated with meson multiplication in the $\phi^4$ model, such as probability densities and recoil probabilities.

\section{Review} \label{revsez}

We will consider a 1+1d quantum field theory of a Schrodinger picture scalar field $\phi(x)$ and its conjugate $\pi(x)$, defined by the Hamiltonian
\begin{equation}
H=\int d x: \mathcal{H}(x):_a, \quad \mathcal{H}(x)=\frac{\pi^2(x)}{2}+\frac{\left(\partial_x \phi(x)\right)^2}{2}+\frac{V(\sqrt{\lambda} \phi(x))}{\lambda}.
\end{equation}
Here $\lambda$ is a coupling constant.  We consider a potential $V$ with degenerate minima, so that the classical equations of motion have a kink solution $\phi(x,t)=f(x)$.  Here $::_a$ is the usual normal ordering at the mass scale $m$, defined by
\beq
m^2=V^{(2)}(\sqrt{\lambda} f(\pm \infty))\hsp
V^{(n)}(\sqrt{\lambda} \phi(x))=\frac{\partial^n V(\sqrt{\lambda} \phi(x))}{(\partial \sqrt{\lambda} \phi(x))^n}.
\eeq
We assume that the two values of the mass, as defined at $x=\infty$ and $x=-\infty$, are equal, as otherwise the vacuum on one side of the kink will be a false vacuum \cite{wstabile}.

As usual, creation operators can be constructed via a plane wave decomposition of the fields.  These create elementary mesons.  Acting them on the vacuum state creates the Fock space of mesons, which we will call the vacuum sector.  Similarly, we will construct creation operators which create mesons in the one-kink sector.  Configurations consisting of a single kink plus any number of mesons will be called the one-kink sector.

Consider the unitary displacement operator
\beq
\df={{\rm Exp}}\left[-i\int dx f(x)\pi(x)\right]. \label{dfd}
\eeq
Acting $\df$ on the vacuum, one arrives at a state in the one-kink sector.  As always, this state can be time-translated using the Hamiltonian $H$.  

Instead of this active transformation point of view, we wish to view $\df$ as a passive transformation of the Hilbert space which preserves the states but transforms the operators.  Let us explain this more precisely.  We refer to the usual representation of the Hilbert space as the {\it{defining frame}}, in which $H$ is the Hamiltonian which generates time translations and whose eigenvalues are energies.  We define the {\it{kink frame}} as follows.  The Dirac ket $|\psi\rangle$ in the kink frame is defined to represent the state $\df|\psi\rangle$ in the defining frame.


Let us try to understand the properties of the kink frame.  First, consider a state represented by the ket $|K\rangle$ in the defining frame.  Then in the kink frame, this state will be represented by the ket $\df^\dag|K\rangle$.  These are two representations of the same state and so clearly they the have the same number of kinks.   Now, if we used the same operator to measure the number of kinks in both frames, then $\df^\dag|K\rangle$ would have one less kink than $|K\rangle$, which is not the case.  Therefore the kink number operator is different in the two frames, in fact the two realizations of the kink number operator are related by conjugation with $\df$, as is the case with all operators.  For example, the Hamiltonian in the kink frame is the kink Hamiltonian~$H\p$
\beq
H\p=\df^\dag H\df. \label{df}
\eeq
To see this, note that if $|K\rangle$ has energy $E_K$, so that
\beq
H|K\rangle=E_K|K\rangle \label{schrodvec}
\eeq
then
\beq
H\p\df^\dag|K\rangle=\df^\dag H|K\rangle=E\df^\dag|K\rangle \label{schrod}
\eeq
and so its eigenvalues yield the correct spectrum.  Similarly, $e^{-iH\p t}$ is the time evolution operator in the kink frame.

The reason that we introduce the kink frame is that, while the defining-frame eigenvalue equation (\ref{schrodvec}) is nonperturbative if $|K\rangle$ is in the one-kink sector, the corresponding kink-frame equation (\ref{schrod}) is perturbative.  Thus, one can solve for kink states $\df^\dag|K\rangle$ using perturbation theory in the kink frame, and then transform the answer back to the defining frame if needed using $\df$.  This has been done to obtain quantum corrections to kink states and masses in Refs.~\cite{mekink,me2loop}.

What is the kink Hamiltonian $H\p$?  Let $Q_n$ be the $n$-loop quantum correction to the kink mass.  Then we may expand $H\p$ into terms $H\p_n$ which have $n$ factors of $\phi(x)$ and $\pi(x)$ when normal-ordered.  One easily finds
\beq
H\p_0=Q_0\hsp H\p_1=0\hsp
H\p_{n>2}=\lambda^{\frac{n}{2}-1}\int dx \frac{V^{(n)}(\sqrt{\lambda} f(x))}{n !}: \phi^n(x):_a.\label{hn}
\eeq

What about $H\p_2$?  This is the most important term, as its eigenstates are the starting points of the perturbative expansion of the entire one-kink sector.  To write it simply, we will need a short digression.

The kink's normal modes $\g(x)$ are the constant frequency solutions of the classical equations of motion corresponding to $H\p_2$
\beq
\V{2}{\g}(x)=\omega^2{\g}(x)+{\g}^{\prime\prime}(x)\hsp \phi(x,t)=e^{-i\omega t}\g(x). \label{sl}
\eeq
There are three kinds of normal mode.  The first is the real zero-mode $\g_B(x)$ which has zero frequency $\omega_B=0$.  Next, there are complex continuum modes $\g_k(x)$ with frequencies $\ok{}=\sqrt{m^2+k^2}$.  Finally, some kinks enjoy discrete, real shape modes $\g_S(x)$ with $0<\omega_S<m$.
We will fix their normalization via the conditions
 $\g^*_k=\g_{-k}$ and 
\beq
\int dx |{\g}_{B}(x)|^2=1,\
\int dx {\g}_{k_1} (x) {\g}^*_{k_2}(x)=2\pi \delta(k_1-k_2),\ 
\int dx {\g}_{S_1}(x){\g}^*_{S_2}(x)=\delta_{S_1S_2}. \label{comp}
\eeq

As $\g(x)$ satisfy a Sturm-Liouville equation (\ref{sl}), they are a complete basis of the space of bounded functions and so can be used to decompose the Schrodinger picture field \cite{cahill76}
\bea
\phi(x) &=&\phi_0 \mathfrak{g}_B(x)+\ppin{k} \left(B_k^{\ddag}+\frac{B_{-k}}{2 \omega_k}\right) \mathfrak{g}_k(x) \label{dec}\\
\pi(x) &=&\pi_0 \mathfrak{g}_B(x)+i \ppin{k}\left(\omega_k B_k^{\ddag}-\frac{B_{-k}}{2}\right) \mathfrak{g}_k(x) \nonumber
\eea
where $B_k^{\ddagger}=B_k^{\dagger} /\left(2 \omega_k\right)$ and $B_{-S}=B_S$.  The symbol $\dint$ is an integral over continuum modes $k$ plus a sum over shape modes $S$.  We have decomposed $\phi(x)$ and $\pi(x)$ into operators $\phi_0,\ \pi_0,\ B$\ and $B^\ddag$ which satisfy the algebra
\beq
\left[\phi_0, \pi_0\right]=i, \quad\left[B_{S_1}, B_{S_2}^{\ddagger}\right]=\delta_{S_1 S_2}, \quad\left[B_{k_1}, B_{k_2}^{\ddagger}\right]=2 \pi \delta\left(k_1-k_2\right).
\eeq

Using this basis, we can write $H\p_2$ as
\begin{equation}
H\p_2=Q_1+H_{\text {free }}, \quad H_{\text {free }}=\frac{\pi_0^2}{2}+\omega_S B_S^{\ddag} B_S+\int \frac{d k}{2 \pi} \omega_k B_k^{\ddag} B_k. \label{h2}
\end{equation}
Now we can interpret the operators.  $\phi_0$ and $\pi_0$ are the position and momentum of a free quantum mechanical particle representing the center of mass of the kink plus mesons.  The operators $B_S^\ddag$ and $B_k^\ddag$ create bound and continuum normal modes respectively.  The ground state $\vac_0$ of $H\p_2$, which is the kink frame first approximation to the kink ground state $\vac$, is the simultaneous ground state of each of the quantum mechanics terms in Eq.~(\ref{h2}).  Therefore it is the solution of the conditions
\beq
\pi_0\vac_0=B_k\vac_0=B_S\vac_0=0. \label{v0}
\eeq
A general one-meson, one-kink state is, at this leading order, $|k\rangle=B^\ddag_k\vac_0$ while acting on this with $B^\ddag_{k\p}$ yields a two-meson, one-kink state 
\beq
|kk\p\rangle=B^\ddag_{k\p}B^\ddag_k\vac_0. \label{2m}
\eeq

\section{Meson Multiplication} \label{moltsez}

\subsection{Gaussian Wave Packets}
Our initial condition will be a meson wave packet centered at $x_0$
\begin{equation}
\Phi(x)=\operatorname{Exp}\left[-\frac{\left(x-x_0\right)^2}{4 \sigma^2}+i x k_0\right], \quad x_0 \ll-\frac{1}{ m}, \quad  \frac{1}{k_0},\frac{1}{m}\ll\sigma \ll\left|x_0\right| .
\end{equation}
The bounds on $x_0$ and $|x_0|$ ensure that the initial wave packet, which starts at $x=x_0$, does not overlap with the kink, which is centered at $x=0$.  The lower bounds on $\sigma$ ensure that the meson momentum is sufficiently strongly peaked that all components move towards the kink and also we can approximate, as described below, the wave packet to be monochromatic.

The evolution of the wave packet will be simpler after a kind of Fourier transform 
\begin{equation}
\Phi(x)=\int \frac{d k}{2 \pi} \alpha_k \mathfrak{g}_k(x), \quad \alpha_k=\int d x \Phi(x) \mathfrak{g}_k^*(x).
\end{equation}
This transform is not with respect to the plane waves, which are solutions of the free equations of motion in the vacuum sector, but rather with respect to the normal modes, which are solutions in the one-kink sector.  The shape modes and zero mode need not be included in the transform, as they have support at $|x|$ of order $O(1/m)$, where $\Phi(x)$ is negligibly small.

The initial one-kink, one-meson state $\left|\Phi\right\rangle$ can be constructed, in the kink frame, in terms of the free kink ground state $\vac_0$ as
\begin{equation}
\left|\Phi\right\rangle=\int d x \Phi(x)\left|x\right\rangle=\int \frac{d k}{2 \pi} \alpha_k\left|k\right\rangle, \quad\left|k\right\rangle=B_k^{\ddagger}|0\rangle_0, \quad|x\rangle=\int \frac{d k}{2 \pi} \mathfrak{g}_{k}^*(x)\left|k\right\rangle.
\end{equation}

\subsection{Time Evolution}
The interactions in the kink frame are summarized by the Hamiltonian terms in Eq.~(\ref{hn}).  These are organized into a power series in $\sqrt{\lambda}$.  At the leading order, $O(\sqrt{\lambda})$, the only term which contributes to meson multiplication is\footnote{Here we have exchanged the order of the $k$ and $x$ integrals with respect to the definition in Eqs.~(\ref{hn}) and (\ref{dec}).  These integrals do not actually commute, and as a result $V_{-k_1k_2k_3}$ appears to be the integral of a nonintegrable function.  It should therefore be remembered that to make sense of this integral, one needs to perform the $k$ integration first.  It turns out that this is equivalent to first performing the $x$ integration using a principal value prescription which will be defined in Eq.~(\ref{iden}).\label{foot}}
\bea
H_I&=&\frac{\sqrt{\lambda}}{4} \int \frac{d k_1}{2 \pi} \frac{d k_2}{2 \pi} \frac{d k_3}{2 \pi} V_{-k_1 k_2 k_3} \frac{1}{\omega_{k_1}} B_{k_2}^{\ddagger} B_{k_3}^{\ddagger} B_{k_1} \\
V_{-k_1 k_2 k_3}&=&\int d x V^{(3)}(\sqrt{\lambda} f(x)) \mathfrak{g}_{-k_1}(x) \mathfrak{g}_{k_2}(x) \mathfrak{g}_{k_3}(x).\nonumber
\eea
$H_I$ converts a one-meson state into a two-meson state
\begin{equation}
H_I |k_1\rangle=\frac{\sqrt{\lambda}}{4 \omega_{k_1}} \int \frac{d k_2}{2 \pi} \frac{d k_3}{2 \pi} V_{-k_1 k_2 k_3}\left|k_2 k_3\right\rangle.
\end{equation}

At time $t$,  at order $O(\sqrt{\lambda})$, the wave packet evolves to
\begin{equation}
\begin{aligned}
|\Phi(t)\rangle&=e^{-i\left(H_{\text {free }}+H_I\right) t}|_{O(\sqrt{\lambda})}\left|\Phi\right\rangle \\
&=\sum_{n=1}^{\infty} \frac{(-i t)^n}{n !}\left(H_{\text {free }}+H_I\right)^n|_{O(\sqrt{\lambda})}\left|\Phi\right\rangle =\sum_{n=1}^{\infty} \frac{(-i t)^n}{n !} \sum_{m=0}^{n-1} H_{\text {free }}^m H_I H_{\text {free }}^{n-m-1}\left|\Phi\right\rangle \\
&=\int \frac{d k_1}{2\pi} \frac{d k_2}{2\pi} \frac{d k_3}{2 \pi} \frac{\sqrt{\lambda}}{4} \alpha_{k_1} V_{-k_1 k_2 k_3} \sum_{n=1}^{\infty} \frac{(-i t)^n}{n !} \sum_{m=0}^{n-1}\left(\omega_{k_2}+\omega_{k_3}\right)^m \omega_{k_1}^{n-m-2}\left|k_2 k_3\right\rangle \\
&=-\frac{i \sqrt{\lambda}}{4} \int \frac{d k_1}{2 \pi} \frac{d k_2}{2 \pi} \frac{d k_3}{2 \pi} \frac{\alpha_{k_1} }{\omega_{k_1}} V_{-k_1 k_2 k_3} {\rm Exp}\left[-i \frac{\omega_{k_1}+\omega_{k_2}+\omega_{k_3}}{2} t\right] \frac{\sin \left(\frac{\omega_{k_2}+\omega_{k_3}-\omega_{k_1}}{2} t \right)}{\left(\omega_{k_2}+\omega_{k_3}-\omega_{k_1}\right)/2}  \left|k_2 k_3\right\rangle.
\end{aligned}
\end{equation}
Here we dropped the $O(\lambda^0)$ term which will not contribute to the matrix elements below.  

One may define the Dirac bra corresponding to a one-kink, two-meson state (\ref{2m}) by
\begin{equation}
\langle k_2 k_3|= \left(B_{k_2}^{\ddagger} B_{k_3}^{\ddagger}|0\rangle_0\right)^\dag={}_0\langle 0|\frac{B_{k_2}}{2\ok{2}}\frac{B_{k_3}}{2\ok{3}}.
\end{equation}
This leads to the normalization
\begin{equation}
\left\langle k_2 k_3|k_2^{\prime} k_3^{\prime}\right\rangle=\frac{{_0}{\langle 0}|0\rangle_0}{4 \omega_{k_2} \omega_{k_3}} \left(2 \pi \delta\left(k_2^{\prime}-k_2\right) 2 \pi \delta\left(k_3^{\prime}-k_3\right)+2 \pi \delta\left(k_2^{\prime}-k_3\right) 2 \pi \delta\left(k_3^{\prime}-k_2\right)\right).
\end{equation}
Our master formula for the unnormalized meson multiplication amplitude is then
\begin{equation}
\langle k_2 k_3 | \Phi(t)\rangle=-\frac{i \sqrt{\lambda}}{8 \omega_{k_2} \omega_{k_3}} \int \frac{d k_1}{2 \pi}\frac{ \alpha_{k_1} }{\omega_{k_1}} V_{-k_1 k_2 k_3} {\rm Exp}\left[-i \frac{\omega_{k_1}+\omega_{k_2}+\omega_{k_3}}{2} t\right] \frac{\sin \left(\frac{\omega_{k_2}+\omega_{k_3}-\omega_{k_1}}{2} t \right)}{\left(\omega_{k_2}+\omega_{k_3}-\omega_{k_1}\right)/2}  {_0}\langle 0| 0\rangle_0. \label{elt}
\end{equation}

\subsection{Amplitude at Finite Times}

Writing the amplitude as
\beq
\langle k_2 k_3 | \Phi(t)\rangle=\frac{ \sqrt{\lambda}}{8 \omega_{k_2} \omega_{k_3}}  \int \frac{d k_1}{2 \pi}\frac{ \alpha_{k_1} }{\omega_{k_1}} V_{-k_1 k_2 k_3} \frac{e^{-i(\ok{2}+\ok{3}) t }-e^{-i\ok{1}t }}{\left(\omega_{k_2}+\omega_{k_3}-\omega_{k_1}\right)}  {_0}\langle 0| 0\rangle_0 \label{amp}
\eeq
we may factor out an overall phase and constant
\beq
A_{k_2k_3}(t)=\frac{e^{i(\ok{2}+\ok{3}) t }}{{_0}\langle 0| 0\rangle_0}\langle k_2 k_3 | \Phi(t)\rangle = \frac{ \sqrt{\lambda}}{8 \omega_{k_2} \omega_{k_3}}  \int \frac{d k_1}{2 \pi}\frac{ \alpha_{k_1} }{\omega_{k_1}} V_{-k_1 k_2 k_3} \frac{1-e^{i(\ok{2}+\ok{3}-\ok{1}) t }}{\left(\omega_{k_2}+\omega_{k_3}-\omega_{k_1}\right)}.
\eeq
At $t=0$, the matrix element vanishes as the sine in the numerator of Eq.~(\ref{elt}) vanishes.  Taking the time derivative one finds
\bea
\dot{A}_{k_2k_3}(t)&=& -i\frac{ \sqrt{\lambda}}{8 \omega_{k_2} \omega_{k_3}}  \int \frac{d k_1}{2 \pi}\frac{ \alpha_{k_1} }{\omega_{k_1}} V_{-k_1 k_2 k_3} e^{i(\ok{2}+\ok{3}-\ok{1}) t }.\label{aeq}
\eea
This can be simplified with a few good approximations.  

\subsubsection{Reflectionless Kinks}

First of all, $|x_0|\gg\sigma$ and $|x_0|\gg1/m$ and so the Gaussian factor in $\alpha_{k_1}$ has support in the large $|x|$ region, where $\g^*_{k_1}$ is a sum of plane waves.  Let us first consider the case of a reflectionless kink, in which case
\bea
\g_k(x)&=&\left\{\begin{tabular}{lll}
$\mb_ke^{ikx}$&\rm{if} & $x\ll  -1/m$\\
$\md_ke^{ikx}$&\rm{if} & $x\gg 1/m$\\
\end{tabular}
\right. \label{gk}\\
|\mb_k|^2&=&|\md_k|^2=1\hsp
\mb^*_k=\mb_{-k}\hsp
\md^*_k=\md_{-k}\nonumber
\eea
where the phases $\mb_k$ and $\md_k$ vary on scales of order $O(m)$ in $k$-space
\beq
\frac{\partial_k\mb_k}{\mb_k}\sim\frac{\partial_k\md_k}{\md_k}\sim O\left(\frac{1}{m}\right).
\eeq
As $x_0\ll -1/m$, this approximation yields
\beq \label{ak1}
\alpha_{k_1}=2\sigma\sqrt{\pi}\mb_{-k_1}e^{-\sigma^2\left(k_1-k_0\right)^2}e^{i(k_0-k_1)x_0}.
\eeq

Next, let us consider $t\gg1/m$.  We will not assume that the time is big enough for the meson to arrive at the kink.  So with this approximation, the process will be roughly on-shell, and so $\ok{1}$ can be replaced with $\ok{2}+\ok{3}$.  This needs to be done delicately, as terms of order $\ok{2}+\ok{3}-\ok{1}$ have appeared in various places.  Each expression should be treated as an expansion in powers of $\ok{2}+\ok{3}-\ok{1}$.  However, this replacement can safely by done on the $\ok{1}$ in the denominator of Eq.~(\ref{aeq}), as this term is of zeroth order in $\ok{2}+\ok{3}-\ok{1}$.  

With these two approximations we find
\bea \label{adot}
\dot{A}_{k_2k_3}(t)&=& -i2\sigma\sqrt{\pi}\frac{ \sqrt{\lambda}}{8 \omega_{k_2} \omega_{k_3}(\ok{2}+\ok{3})}  \pin{k_1}\mb_{-k_1}
e^{-\sigma^2\left(k_1-k_0\right)^2} e^{i(k_0-k_1)x_0}\nonumber\\
&&\times\left[ \int d y V^{(3)}(\sqrt{\lambda} f(y)) \mathfrak{g}_{-k_1}(y) \mathfrak{g}_{k_2}(y) \mathfrak{g}_{k_3}(y) \right]e^{i(\ok{2}+\ok{3}-\ok{1}) t }.
\eea
$k_1$ is always close to $k_0$, as $\sigma\gg 1/m$, and so we may expand
\begin{equation}\label{om}
\omega_{k_1}=\omega_{k_0}+\left(k_1-k_0\right) \frac{k_0}{\omega_{k_0}}\hsp \mb_{-k_1}=\mb_{-k_0}\hsp \g_{-k_1}=\g_{-k_0}.
\end{equation}
Inserting Eq.~(\ref{om}) into Eq.~(\ref{adot}),
\bea
\dot{A}_{k_2k_3}(t)&=& -i2\sigma\sqrt{\pi}\mb_{-k_0}\frac{ \sqrt{\lambda}e^{i(\ok{2}+\ok{3}-\ok{0}) t }}{8 \omega_{k_2} \omega_{k_3}(\ok{2}+\ok{3})}  \left[ \int d y V^{(3)}(\sqrt{\lambda} f(y)) \mathfrak{g}_{-k_0}(y) \mathfrak{g}_{k_2}(y) \mathfrak{g}_{k_3}(y) \right]\nonumber\\
&&\times\int \frac{d k_1}{2 \pi}
e^{-\sigma^2\left(k_1-k_0\right)^2} e^{i(k_0-k_1)(x_0+\frac{k_0}{\ok{0}}t)}\nonumber\\
&=&-i\mb_{-k_0}\frac{ \sqrt{\lambda}e^{i(\ok{2}+\ok{3}-\ok{0}) t }}{8 \omega_{k_2} \omega_{k_3}(\ok{2}+\ok{3})} {\rm Exp}\left[-\frac{(x_0+\frac{k_0}{\ok{0}}t)^2}{4\sigma^2}\right] V_{-k_0 k_2 k_3}.
\eea

\subsubsection{Reflective Kinks}

So far we have only considered reflectionless kinks, such as those of the sine-Gordon and $\phi^4$ models.  However, in general kinks are reflective, and so asymptotically the normal modes are of the form
\bea
\g_k(x)&=&\left\{\begin{tabular}{lll}
$\mb_ke^{ikx}+\mc_ke^{-ikx}$&\rm{if} & $x\ll  -1/m$\\
$\md_ke^{ikx}+\me_k e^{-ikx}$&\rm{if} & $x\gg 1/m$\\
\end{tabular}
\right. \label{gk}\\
|\mb_k|^2+|\mc_k|^2&=&|\md_k|^2+|\me_k|^2=1\hsp
\mb^*_k=\mb_{-k}\hsp
\mc^*_k=\mc_{-k}\hsp
\md^*_k=\md_{-k}\hsp
\me^*_k=\me_{-k}.\nonumber
\eea
Again, our initial wave packet is supported near $x_0\ll-1/m$ and so this approximation allows us to simplify the coefficients $\alpha_{k_1}$
\beq \label{ak1}
\alpha_{k_1}=2\sigma\sqrt{\pi}\left[\mb_{-k_1}e^{-\sigma^2\left(k_1-k_0\right)^2}e^{i(k_0-k_1)x_0}+\mc_{-k_1}e^{-\sigma^2\left(k_1+k_0\right)^2}e^{i(k_0+k_1)x_0}\right].
\eeq

Substituting this into Eq.~(\ref{aeq}) one finds
\bea
\dot{A}_{k_2k_3}(t)&=& -i2\sigma\sqrt{\pi}\frac{ \sqrt{\lambda}}{8 \omega_{k_2} \omega_{k_3}(\ok{2}+\ok{3})} \int \frac{d k_1}{2 \pi}V_{-k_1 k_2 k_3}e^{i(\ok{2}+\ok{3}-\ok{1}) t }\nonumber\\
&&\times \left[
\mb_{k_1}^* e^{-\sigma^2\left(k_1-k_0\right)^2} e^{i(k_0-k_1)x_0}+\mc_{k_1}^* e^{-\sigma^2\left(k_1+k_0\right)^2} e^{i(k_0+k_1)x_0}\right].\label{aref}
\eea

Recall that we have fixed $k_0>0$ so that the wave packet moves to the right, towards the kink.  In the reflectionless case this implied that $k_1>0$.  Now we see that there are two Gaussian factors, the first is supported at $k_1\sim k_0$ but the second is instead supported at $k_1\sim -k_0.$  Thus, while the initial motion of the meson is always to the right, in the reflective case this corresponds to two distinct regions in the one-meson Fock space.

As a result, we will need to consider the expansion of $k_1$ about both $k_0$ and also $-k_0$, which leads to the corresponding expansion for the frequencies
\begin{equation}
\omega_{k_1}=\omega_{k_0}+\left(\pm k_1-k_0\right) \frac{k_0}{\omega_{k_0}}. \label{svil}
\end{equation}

Inserting these two expansions into Eq.~(\ref{aref}), we obtain
\bea
\dot{A}_{k_2k_3}(t)&=& -i2\sigma\sqrt{\pi}\frac{ \sqrt{\lambda}e^{i(\ok{2}+\ok{3}-\ok{0}) t }}{8 \omega_{k_2} \omega_{k_3}(\ok{2}+\ok{3})}
 \int \frac{d k_1}{2 \pi}V_{-k_1 k_2 k_3}
\label{adr}\\
&&\times  \left[\mb_{k_1}^*
e^{-\sigma^2\left(k_1-k_0\right)^2} e^{i(k_0-k_1)(x_0+\frac{k_0}{\ok{0}}t)}+\mc_{k_1}^*
e^{-\sigma^2\left(k_1+k_0\right)^2} e^{i(k_1+k_0)(x_0+\frac{k_0}{\ok{0}}t)}\right]\nonumber\\
&=&-i\frac{ \sqrt{\lambda}e^{i(\ok{2}+\ok{3}-\ok{0}) t }}{8 \omega_{k_2} \omega_{k_3}(\ok{2}+\ok{3})} {\rm Exp}\left[-\frac{(x_0+\frac{k_0}{\ok{0}}t)^2}{4\sigma^2}\right]\tilde{V}_{-k_0 k_2 k_3}\nonumber
\eea
where we have defined the shorthand
\beq \label{tildv}
\tilde{V}_{-k_0 k_2 k_3}=\mb_{-k_0} V_{-k_0 k_2 k_3}+\mc_{k_0} V_{k_0 k_2 k_3}.
\eeq

\subsubsection{Remarks}

As a result of the Gaussian factor, this time derivative of the amplitude is only appreciable when the exponent
\beq
x_t=x_0+\frac{k_0}{\ok{0}}t
\eeq
is small, which occurs at time
\beq
t\sim t_1=  -\frac{\ok{0}}{k_0}x_0
\eeq
when the meson strikes the kink.  

In particular, since $t\geq 0$, we see that this requires $k_0$ and $x_0$ to have opposite signs, which of course is necessary for the meson to move towards the kink.  As $A(0)=0$, we learn that the amplitude $A(t)$ vanishes at $t\ll t_1$, before the collision.

\subsection{Amplitude in the Asymptotic Future}

\subsubsection{The Large Time Limit}

We are interested in the large time limit, when the meson has already scattered with the kink.  At large times $t$ we may integrate Eq.~(\ref{adr}) to obtain
\bea
\stackrel{\rm{lim}}{{}_{t\rightarrow\infty}}A_{k_2k_3}(t)&=&
-i\frac{ \sqrt{\lambda} \tilde{V}_{-k_0 k_2 k_3}}{8 \omega_{k_2} \omega_{k_3}(\ok{2}+\ok{3})}\int_{-\infty}^{\infty} dt  {\rm Exp}\left[-\frac{(x_0+\frac{k_0}{\ok{0}}t)^2}{4\sigma^2}\right]e^{i(\ok{2}+\ok{3}-\ok{0}) t }\nonumber\\
&=&-i\frac{ \sqrt{\lambda} \tilde{V}_{-k_0 k_2 k_3}}{4\omega_{k_2} \omega_{k_3}(\ok{2}+\ok{3})}\sigma\sqrt{\pi}\frac{\ok{0}}{k_0}\nonumber\\
&&\times{\rm{Exp}}
\left[-\sigma^2\frac{\ok{0}^2}{k^2_0}\left(\ok{2}+\ok{3}-\ok{0}\right)^2-i\left(\ok{2}+\ok{3}-\ok{0}\right)\frac{\ok{0}}{k_0}x_0
\right].
\eea
Therefore
\beq
\stackrel{\rm{lim}}{{}_{t\rightarrow\infty}}\frac{\left| \langle k_2 k_3 | \Phi(t)\rangle\right|^2}{|{}_0\langle 0\vac_0|^2}=
\frac{ \pi\lambda\sigma^2 \left|\tilde{V}_{-k_0 k_2 k_3}\right|^2}{16\omega^2_{k_2} \omega^2_{k_3}(\ok{2}+\ok{3})^2}\left(\frac{\ok{0}}{k_0}
\right)^2{\rm{Exp}}
\left[-2\sigma^2\frac{\ok{0}^2}{k^2_0}\left(\ok{2}+\ok{3}-\ok{0}\right)^2
\right]. \label{lim}
\eeq

Let us define the on-shell initial momentum $k_I$ by
\beq\label{I23}
 k_I \equiv  \sqrt{\left(\ok{2}+\ok{3}\right)^2-m^2}
\eeq
so that $\ok{I}=\ok{2}+\ok{3}.$  The Gaussian factor in Eq.~(\ref{lim}) has support at $\ok{0}\sim\ok{I}$.  Therefore, as $k_0$ and $k_I$ are both defined to be positive, in the region in $k_2-k_3$-space with the largest contribution to the probability, $k_0\sim k_I$.  We thus expand
\beq
k_0=k_I+(k_0-k_I)
\eeq
and keep only the leading nonvanishing term in each expression.  This yields
\beq
\stackrel{\rm{lim}}{{}_{t\rightarrow\infty}}\frac{\left| \langle k_2 k_3 | \Phi(t)\rangle\right|^2}{|{}_0\langle 0\vac_0|^2}=
\frac{ \pi\lambda\sigma^2 \left|\tilde{V}_{-k_I k_2 k_3}\right|^2}{16\omega^2_{k_2} \omega^2_{k_3}k_I^2}{\rm{Exp}}
\left[-2\sigma^2\frac{\ok{I}^2}{k^2_I}\left(\ok{I}-\ok{0}\right)^2
\right].
\eeq
Using the same expansion as in Eq.~(\ref{svil}) this simplifies further to 
\beq
\stackrel{\rm{lim}}{{}_{t\rightarrow\infty}}\frac{\left| \langle k_2 k_3 | \Phi(t)\rangle\right|^2}{|{}_0\langle 0\vac_0|^2}=
\frac{ \pi\lambda\sigma^2 \left|\tilde{V}_{-k_I k_2 k_3}\right|^2}{16\omega^2_{k_2} \omega^2_{k_3}k_I^2}e^{
-2\sigma^2\left(k_{I}-k_{0}\right)^2
}.
\eeq

\subsubsection{A Faster Derivation}

A faster approach, which however sheds no light on the evolution at intermediate times, is to directly take the $t\rightarrow\infty$ limit of Eq.~(\ref{elt}).  Using the identity
\beq
\stackrel{\rm{lim}}{{}_{t\rightarrow\infty}}
\frac{\sin \left(\frac{\omega_{k_2}+\omega_{k_3}-\omega_{k_1}}{2} t \right)}{\left(\omega_{k_2}+\omega_{k_3}-\omega_{k_1}\right)/2} 
=2 \pi \delta\left(\omega_{k_2}+\omega_{k_3}-\omega_{k_1}\right)=\frac{\omega_{k_I}}{k_I}\left(2 \pi \delta\left(k_1-k_I\right)+2 \pi \delta\left(k_1+k_I\right)\right)
\eeq
the amplitude can be simplified to 
\begin{equation}
\stackrel{\rm{lim}}{{}_{t\rightarrow\infty}}
\frac{\langle k_2 k_3 | \Phi(t)\rangle}{{_0}\langle 0| 0\rangle_0}=-\frac{i \sqrt{\lambda}}{8 \omega_{k_2} \omega_{k_3} k_I}  e^{-i \omega_{k_I} t}\left(\alpha_{k_I} V_{-k_I k_2 k_3}+\alpha_{-k_I} V_{k_I k_2 k_3}\right).
\end{equation}
As $k_I$ and $k_0$ are both large and positive, the Gaussians in Eq.~(\ref{ak1}) with $(k_I+k_0)$ are exponentially suppressed, leaving only the $\mb_{-k_I}$ term in $\alpha_{k_I}$ and the $\mc_{k_I}$ term in $\alpha_{-k_I}$.  Altogether we find
\beq
\stackrel{\rm{lim}}{{}_{t\rightarrow\infty}}
\frac{\langle k_2 k_3 | \Phi(t)\rangle}{{_0}\langle 0| 0\rangle_0}=-\frac{i\sigma \sqrt{\pi\lambda}}{4 \omega_{k_2} \omega_{k_3} k_I}  e^{-i \omega_{k_I} t}e^{-\sigma^2(k_0-k_I)^2}\tilde{V}_{-k_I k_2 k_3}
\eeq
in agreement with the longer derivation above.

\subsection{The Probability}

The probability $P$ that $|\Phi(t)\rangle$, the state at time $t$, is in a given subspace of the Hilbert space is given by
\begin{equation}
P=\frac{\langle \Phi(t)|\mathcal{P}|  \Phi(t)\rangle}{\langle \Phi(t) |  \Phi(t)\rangle}\label{pdef}
\end{equation}
where $\mathcal{P}$ is a projector onto that subspace.

We are interested in the probability $P_{\rm{tot}}$ that the final state has two mesons, corresponding to the projector 
\begin{equation}
\mathcal{P}_{\rm{tot}}|k_2 k_3\rangle=|k_2 k_3\rangle\hsp
k_2,\ k_3\in \R.
\end{equation}
We are also interested in the corresponding probability density $P_{\rm{diff}}(k_2,k_3)$ that the final mesons have momenta $k_2$ and $k_3$.  This is related to the total probability by
\beq
P_{\rm{tot}}=\frac{1}{2}\int dk_2 dk_3 P_\text{diff}(k_2,k_3)
\eeq
where the factor of $1/2$ results from the fact that $|k_2k_3\rangle$ and $|k_3k_2\rangle$ represent the same state.  $P_{\rm{diff}}$ is defined by a formula similar to (\ref{pdef}) in which the operator $\mathcal{P}_{\rm{diff}}$ annihilates all states with $k$ not equal to $k_2$ and $k_3$.  It is not a projector, as it has an infinite eigenvalue.  These two equations are easily solved, yielding the operators
\beq
\mathcal{P}_\text{diff}(k_2,k_3)=\frac{\omega_{k_2} \omega_{k_3}}{\pi^2{_0}\langle 0 |0\rangle_0}|k_2 k_3\rangle\langle k_2 k_3|\hsp\mathcal{P}_\text{tot}=\frac{1}{2}\int d k_2 d k_3\mathcal{P}_\text{diff}(k_2,k_3).
\eeq

Consider a general reflective kink with $\alpha_{k_1}$ of the form of Eq.~(\ref{ak1})
\begin{equation}
\langle \Phi(t) |  \Phi(t)\rangle=\langle \Phi |  \Phi \rangle =\pin{k_1} \alpha_{k_1} \alpha_{k_1}^{*} \frac{{_0}\langle 0 |0\rangle_0}{2 \omega_{k_1}} =\sqrt{2\pi}\sigma\frac{{_0}\langle 0 |0\rangle_0}{2 \omega_{k_0}}
\end{equation}
where we used $\ok{1}\sim\ok{0}$.

The probability density at a large time $t$ is
\bea \label{pdiffeq}
P_{\rm{diff}}(k_2,k_3)&=&\stackrel{\rm{lim}}{{}_{t\rightarrow\infty}}\frac{\langle \Phi(t)|\mathcal{P}_\text{diff}(k_2,k_3) | \Phi(t)\rangle}{\langle \Phi(t) |  \Phi(t)\rangle} =\stackrel{\rm{lim}}{{}_{t\rightarrow\infty}}\frac{\sqrt{2} \ok{0}\ok{2}\ok{3}}{\pi^{5/2}\sigma} \frac{\left| \langle k_2 k_3 | \Phi(t)\rangle\right|^2}{|{}_0\langle 0\vac_0|^2}\\
&=&\frac{\lambda\sigma\ok{0} \left|\tilde{V}_{-k_I k_2 k_3}\right|^2}{8\sqrt{2}\pi^{3/2}\omega_{k_2} \omega_{k_3}k_I^2}e^{
-2\sigma^2\left(k_{I}-k_{0}\right)^2
}. \nonumber
\eea
Integrating this yields total probability for meson multiplication at a large time $t$ 
\begin{equation} 
P_{\rm{tot}}=\frac{1}{2}\int dk_2 dk_3 P_{\rm{diff}}(k_2,k_3)=
\frac{ \lambda\sigma\ok{0} }{16\sqrt{2}\pi^{3/2} }
\int  dk_2 dk_3
\frac{  \left|\tilde{V}_{-k_I k_2 k_3}\right|^2}{\omega_{k_2} \omega_{k_3}k_I^2}e^{-2\sigma^2\left(k_{I}-k_{0}\right)^2}.
\end{equation}
As $\sigma\gg 1/m$ we may approximate the Gaussian to be a Dirac delta function, yielding
\bea 
P_{\rm{diff}}(k_2,k_3)&=&\frac{\lambda\ok{I} \left|\tilde{V}_{-k_I k_2 k_3}\right|^2}{16\pi\omega_{k_2} \omega_{k_3}k_I^2}\delta(k_I-k_0)
\label{ptoteq}\\
P_{\rm{tot}}&=&\frac{\lambda\ok{0} }{32\pi k_0^2}
\int dk_2 dk_3
\frac{  \left|\tilde{V}_{-k_I k_2 k_3}\right|^2}{\omega_{k_2} \omega_{k_3}}\delta(k_I-k_0)\nonumber\\
&=&\frac{ \lambda }{32\pi k_0}
\int dk_2
\frac{  \left|\tilde{V}_{-k_0, k_2, \sqrt{(\ok{0}-\ok{2})^2-m^2}}\right|^2+\left|\tilde{V}_{-k_0, k_2, -\sqrt{(\ok{0}-\ok{2})^2-m^2}}\right|^2}{\omega_{k_2} \sqrt{(\ok{0}-\ok{2})^2-m^2}}\nonumber
\eea
where we used
\beq
\frac{\partial k_I}{\partial k_3}=\frac{\ok{0}k_3}{k_0\ok{3}}=\frac{\ok{0}\sqrt{(\ok{0}-\ok{2})^2-m^2}}{k_0(\ok{0}-\ok{2})}.
\eeq


\section{Examples: The Sine-Gordon Soliton and $\phi^4$ Kink} \label{exsez}

\subsection{The Sine-Gordon Soliton}
In the sine-Gordon theory, defined by
\beq
V(\sqrt{\lambda}\phi(x))=m^2\left(1-{\rm{cos}}(\sqrt{\lambda}\phi(x)\right)
\eeq
the symbol $V_{k_1k_2k_3}$ is given\footnote{We have taken $k\rightarrow -k$ with respect to Ref.~\cite{me2loop} so that at large $k$, $k$ approaches the momentum.} in Ref.~\cite{me2loop}
\bea
V_{k_1k_2k_3}&=&-\frac{\pi i\sqrt{\lambda}}{4}{\rm{sign}}(k_1k_2k_3){\rm{sech}}\left(\frac{\pi(k_1+k_2+k_3)}{2m}\right)\\
&&\times\frac{(\ok{1}+\ok{2}+\ok{3})(\ok{1}+\ok{2}-\ok{3})(\ok{1}+\ok{3}-\ok{2})(\ok{2}+\ok{3}-\ok{1})}{\ok{1}\ok{2}\ok{3}}.\nonumber
\eea
As a result
\beq
V_{\pm k_Ik_2k_3}=0
\eeq
because it is proportional to $\ok{2}+\ok{3}-\ok{I}=0$.  This in turn implies that
\beq
\tilde{V}_{- k_Ik_2k_3}=0
\eeq
as it is a linear combination (\ref{tildv}) of $V_{\pm k_Ik_2k_3}$.  Eq.~(\ref{pdiffeq}) then implies that the differential probability vanishes for all $k_2$ and $k_3$.

This is to be expected, the integrability of the sine-Gordon model implies that the number of mesons is conserved and so meson multiplication does not appear in the $S$-matrix.

\subsection{The $\phi^4$ Kink}

\subsubsection{Review}

We will need an expression for $\tilde{V}_{-k_1k_2k_3}$ in the case of the $\phi^4$ double-well model, with potential
\beq
V(\sqrt{\lambda}\phi(x))=\frac{\lambda\phi^2(x)}{4}\left(\sqrt{\lambda}\phi(x)-\sqrt{2}m\right)^2
.
\eeq
This requires a knowledge of $\mb_k,\ \mc_k$\ and $V_{k_1k_2k_3}$.  The first two are easily read off of the normal modes
\beq
\g_k(x)=\frac{e^{ikx}}{\ok{} \sqrt{k^2+\b^2}}\left[k^2-2\b^2+3\b^2\sech^2(\b x)+3i\b k\tanh(\b x)\right]\hsp\b=\frac{m}{2}. \label{norm}
\eeq
At $x\ll-1/\beta$ this becomes a plane wave with phase
\beq \label{coeffbc}
\mb_k=\frac{k^2-2\beta^2-3i\beta k}{\ok{}\sqrt{k^2+\beta^2}}\hsp \mc_k=0.
\eeq
Our convention for normal modes is the complex conjugate of that in Ref.~\cite{phi42loop}, so that $k$ becomes approximately the meson momentum at high $k$.  As a result $\mc_k$ vanishes, as opposed to $\mb_k$ in that reference.  As the $\phi^4$ kink is reflectionless, the product $\mb_k\mc_k$ vanishes in any convention \cite{merif}.  

Using Eq.~(\ref{tildv}) and $|\mb_k|=1$, the reflectionless condition thus leads to the simplification
\beq 
\left|\tilde{V}_{-k_0 k_2 k_3}\right|=\left|V_{-k_0 k_2 k_3}\right|.
\eeq
We then need only calculate $V_{k_1k_2k_3}$.  In Ref.~\cite{phi42loop} this is calculated in terms of a sum of integrals over $x$, however those integrals are not evaluated because that paper was concerned with infrared divergences which required a delicate treatment of the integrand.  We will see a similar infrared divergence here, arising from the fact that the 3-point interaction responsible for meson multiplication has a nonzero constant norm even far from the kink.  Meson multiplication far from the kink is suppressed only because the corresponding matrix element oscillates quickly, leading to destructive interference when the initial momentum is integrated over even a very small interval.

Let us begin by reviewing the expression for $V_{k_1k_2k_3}$ in Ref.~\cite{phi42loop}.  First, the third derivative of the potential is 
\beq
V^{(3)}(\sqrt{\lambda}f(x))=6\sqrt{2}\b \tanh(\b x).
\eeq
Note that it is of order $O(\sqrt{\lambda})$, and so that will be the order of our amplitude.  Also notice that it tends to a nonzero constant at large $x$ and $-x$.

We will perform the $x$-integrals using the identities
\bea
\int dx e^{ikx}\sech^{2n}(\b x)&=&\left\{
\begin{array}{cl}
2\pi\delta(k) &  {\rm{\ \ \ if}}\  n=0 \\ \frac{\pi}{(2n-1)!k}\left[\prod_{j=0}^{n-1}\left(\frac{k^2}{\b^2}+(2j)^2\right)\right]\ck   & {\rm{\ \ \ if}}\ n>0
\end{array}
\right.\nonumber\\
\int dx e^{ikx}\sech^{2n}(\b x)\tanh(\b x)&=&i\frac{\pi}{(2n)!\b}\left[\prod_{j=0}^{n-1}\left(\frac{k^2}{\b^2}+(2j)^2\right)\right]\ck \label{iden}.
\eea
Note that in the $n=0$ cases of the two integrals, the integrand does not become small at large $|x|$.  These formulas correspond to a kind of principal value prescription for evaluating the integrals.  We have checked that this principal value prescription is indeed the right one, as it yields the same answer as would be achieved by integrating over a small region in $k_1$ with a smooth weight function.  Such a coherent integral was indeed present in our master formula (\ref{elt}) for the amplitude, it is the integral over the momentum in the initial wave packet.  The fact that the $k$ integral should be performed before the $x$ integral was explained in Footnote~\ref{foot}.

$V_{k_1k_2k_3}$ consists of a sum of terms which are each integrals over $x$ of $\sech^{2I}(\beta x)\tanh^J(\beta x)$ where $I\in\{0,1,2,3\}$ and $J\in\{0,1\}$.  The case $I=J=0$ yields a $\delta(k_1+k_2+k_3)$ which will vanish in our case, as $\ok{I}=\ok{2}+\ok{3}$.  We will keep it, as our expression for $V_{k_1k_2k_3}$ may be useful for future problems, however we will separate it as it will not contribute to meson multiplication at tree level.  Thus we decompose
\beq
V_{k_1k_2k_3}=V^{00}_{k_1k_2k_3}+\hat{V}_{k_1k_2k_3}\hsp
V^{00}_{k_1k_2k_3}=\frac{9\sqrt{2}i\beta^2 k_1k_2k_3\left(6\b^2+k_{1}^2+k_2^2+k_{3}^2\right)2\pi\delta(k)}{\ok1\ok2\ok3\sqrt{\b^2+k_1^2}\sqrt{\b^2+k_2^2}\sqrt{\b^2+k_3^2}}
\eeq
where $V^{00}$ contains all of the $\delta(k)$ terms and only $\hat{V}$ will be relevant below.

Let us define the symbols $u$ by
\beq
\hat{V}_{k_1k_2k_3}=\frac{6\sqrt{2}\pi\b\ck}{\ok1\ok2\ok3\sqrt{\b^2+k_1^2}\sqrt{\b^2+k_2^2}\sqrt{\b^2+k_3^2}}\sum_{J=0}^1\sum_{I=1-J}^3 u_{k_1k_2k_3}^{IJ}
\eeq
where the sum does not include $I=J=0$, as that term is in $V^{00}$.  

Each $u^{IJ}$ is defined to be the term in $V_{k_1k_2k_3}$ with an $x$ integral of $e^{ixk}\sech^{2I}(\b x)\tanh^J(\b x)$.  Let us define the symbol $\Phi$ to summarize the coefficients
\beq
u_{k_1k_2k_3}^{IJ}=\frac{\sinh\left(\frac{\pi k}{2\beta}\right)}{\pi}\Phi_{k_1k_2k_3}^{IJ}\int dxe^{ixk}\sech^{2I}(\b x)\tanh^J(\b x).
\eeq
Ref.~\cite{phi42loop} provided the components of $\Phi$ 
\bea
\Phi_{k_1k_2k_3}^{10}&=&3i\b\left[-16\b^4S_1^1+\b^2\left(5S_2^{21}+18S_3^1\right)-S_3^1S_2^1\right]\\
\Phi_{k_1k_2k_3}^{20}&=&9i\b^3\left[7\b^2S^1_1-S_2^{21}-3S_3^1\right]\hsp \Phi_{k_1k_2k_3}^{30}=-27i\b^5S_1^1\nonumber\\
\Phi_{k_1k_2k_3}^{01}&=&-8\b^6+\b^4(18S_2^1+4S_1^2)+\b^2(-2S_2^2-9S_3^1S_1^1)+S_3^2
\nonumber\\
\Phi_{k_1k_2k_3}^{11}&=&3\b^2\left[12\b^4+\b^2(-15S_2^1-4S_1^2)+(S_2^2+3S_3^1S_1^1)\right]
\nonumber\\
\Phi_{k_1k_2k_3}^{21}&=&9\b^4\left[-6\b^2+(3S_2^1+S_1^2)\right]
\hsp
\Phi_{k_1k_2k_3}^{31}=27\b^6
\nonumber
\eea
in terms of symmetric combinations of the $k$'s
\bea
S_1^n&=&k_1^n+k_2^n+k_3^n\hsp 
S_2^n=(k_1k_2)^n+(k_1k_3)^n+(k_2k_3)^n\hsp
S_3^n=(k_1k_2k_3)^n\nonumber\\
S_2^{mn}&=&k_1^mk_2^n+k_1^mk_3^n+k_2^mk_3^n+k_1^nk_2^m+k_1^nk_3^m+k_2^nk_3^m.
\eea

\subsubsection{The Calculation}

We may now perform the $x$ integrals using Eq.~(\ref{iden}) 
\bea
u_{k_1k_2k_3}^{I0}&=&\Phi_{k_1k_2k_3}^{I0}\frac{1}{(2I-1)!k}\left[\prod_{j=0}^{I-1}\left(\frac{k^2}{\b^2}+(2j)^2\right)\right]\\
u_{k_1k_2k_3}^{I1}&=&\Phi_{k_1k_2k_3}^{I1}\frac{i}{(2I)!\b}\left[\prod_{j=0}^{I-1}\left(\frac{k^2}{\b^2}+(2j)^2\right)\right].\nonumber
\eea
In particular, we find
\bea
u_{k_1k_2k_3}^{10}&=&3ik\left[-16\b^3S_1^1+\b\left(5S_2^{21}+18S_3^1\right)-\frac{1}{\beta}S_3^1S_2^1\right]\\
u_{k_1k_2k_3}^{20}&=&\frac{3ik}{2}\left(\frac{k^2}{\beta^2}+4\right)\left[7\b^3 S^1_1-\b S_2^{21}-3\b S_3^1\right]\nonumber\\
u_{k_1k_2k_3}^{30}&=&-\frac{9i k}{40}\left(\frac{k^4}{\beta^4}+20\frac{k^2}{\beta^2}+64\right)\left[\beta^3S_1^1\right]\nonumber\\
u_{k_1k_2k_3}^{01}&=&i\left[-8\b^5+\b^3(18S_2^1+4S_1^2)+\b^1(-2S_2^2-9S_3^1S_1^1)+\frac{S_3^2}{\b}\right]
\nonumber\\
u_{k_1k_2k_3}^{11}&=&\frac{3ik^2}{2}\left[12\b^3+\b(-15S_2^1-4S_1^2)+\frac{1}{\b}(S_2^2+3S_3^1S_1^1)\right]\nonumber\\
u_{k_1k_2k_3}^{21}&=&\frac{3ik^2}{8}\left(\frac{k^2}{\beta^2}+4\right)\left[-6\b^3+\b(3S_2^1+S_1^2)\right]\nonumber\\
u_{k_1k_2k_3}^{31}&=&\frac{3ik^2}{80}\left(\frac{k^4}{\beta^4}+20\frac{k^2}{\beta^2}+64\right)\left[\b^3\right].\nonumber
\eea

Reassembling these components, we finally arrive at
\bea \label{vphi4}
\hat{V}_{k_1k_2k_3}
&=&\frac{6\sqrt{2} \pi \csch\left(\frac{\pi (k_1+k_2+k_3)}{2 \b}\right)}{\ok1\ok2\ok3\sqrt{\b^2+k_1^2}\sqrt{\b^2+k_2^2}\sqrt{\b^2+k_3^2}}\nonumber\\
&&\times \Bigg\{-8i\b^6  - 5i \b^4 (k_1^2+k_2^2+k_3^2)-2i \b^2  (k_1^2 k_2^2+k_1^2 k_3^2+k_2^2 k_3^2)\nonumber\\
&&\quad -i\left[\frac{3}{16}(-k_1^6-k_2^6-k_3^6+k_1^4 k_2^2+k_1^4 k_3^2+k_2^4 k_3^2\right.\nonumber\\
&&\left.\quad\qquad+k_2^4 k_1^2+k_3^4 k_1^2+k_3^4 k_2^2)+\frac{1}{8}k_1^2k_2^2k_3^2\right]\Bigg\}.
\eea
Recall that the meson multiplication probability density (\ref{pdiffeq}) and total probability (\ref{ptoteq}) only require the special case $k_1=-k_I$.  In this case the coefficients simplify to
\bea \label{vphi4I23}
V_{-k_I k_2 k_3}&=&\frac{48\sqrt{2}\pi i \ok2\ok3\ok{I}\csch\left(\frac{\pi \left(k_2+k_3-k_I\right)}{m}\right)}{\sqrt{4k_2^2+m^2}\sqrt{4k_3^2+m^2}\sqrt{4k_I^2+m^2 }}\\
&=&\frac{48\sqrt{2}\pi i \ok2\ok3\left(\ok2+\ok3\right)\csch\left(\frac{\pi \left(k_2+k_3-\sqrt{k_2^2+k_3^2+m^2+2 \ok2\ok3}\right)}{m}\right)}{\sqrt{4k_2^2+m^2}\sqrt{4k_3^2+m^2}\sqrt{4k_2^2+4k_3^2+5m^2+8\ok2 \ok3 }}.\nonumber
\eea



For completeness we provide $\tilde{V}$
\bea \label{vtildephi4}
\tilde{V}_{-k_I k_2 k_3}&=&\mb_{-k_I} V_{-k_I k_2 k_3}+\mc_{k_I} V_{k_I k_2 k_3}
=\frac{k_I^2-2\beta^2+3i\beta k_I}{\ok{I}\sqrt{k_I^2+\beta^2}}V_{-k_I k_2 k_3}\nonumber\\
&=&\frac{48\sqrt{2}\pi\ok2 \ok3 \left(i \left(2 k_2^2+2k_3^2+m^2+4\ok2\ok3)\right)-3m\sqrt{k_2^2+k_3^2+m^2+2\ok2\ok3}\right)}{\sqrt{4k_2^2+m^2}\sqrt{4k_3^2+m^2}\left(4k_2^2+4k_3^2+5m^2+8\ok2 \ok3 \right)}\nonumber\\
&&\times \csch\left(\frac{\pi \left(k_2+k_3-\sqrt{k_2^2+k_3^2+m^2+2 \ok2\ok3}\right)}{m}\right)
\eea
where we used Eq.~(\ref{coeffbc}) and Eq.~(\ref{I23}).  However, as a result of $(\ref{tildv})$, at tree level we only need the absolute value $|\tilde{V}|$ which is equal to $|\hat{V}|$ for a reflectionless kink and to $|V|$ at $k_1\sim - k_I$.

Substituting Eq.~(\ref{vtildephi4}) into  Eq.~(\ref{ptoteq}), we find the probability density and total probability for meson multiplication. Our main result is the following analytic expression for the probability density
\bea 
P_{\rm{diff}}(k_2,k_3)&=&\frac{\lambda\ok{I} \left|\tilde{V}_{-k_I k_2 k_3}\right|^2}{16\pi\omega_{k_2} \omega_{k_3}k_I^2}\delta(k_I-k_0)\label{princ}\\
&=&\frac{288\pi \lambda \ok2\ok3\ok{I}^3\csch^2\left(\frac{\pi \left(k_2+k_3-k_I\right)}{m}\right)}{k_I^2(4k_2^2+m^2)(4k_3^2+m^2)(4k_I^2+m^2 )}\delta(k_I-k_0). \nonumber
\eea
As expected, it is order $O(\lambda)$.  The Dirac $\delta$ function imposes exact energy conservation.  On the other hand, momentum conservation among mesons is imposed by the csch.  This is not a $\delta$ function, and so the momentum can be transferred between the mesons and the kink.  

In the ultrarelativistic limit $k_0\gg m$, Eq.~(\ref{princ}) becomes
\bea
P_{\rm{diff}}(k_2,k_3)
&=&\frac{9\pi \lambda  \csch^2\left(\frac{\pi m}{2k_2k_3k_I}\left(k_I^2-k_2k_3 \right)\right)}{2  k_2 k_3 k_I}\delta(k_I-k_0)\\
&=&\frac{18 \lambda k_2k_3k_0}{\pi m^2\left(k_0^2-k_2k_3 \right)^2}\delta(k_2+k_3-k_0).\nonumber
\eea
This is supported when $k_2,\ k_3$\ and $k_I$ are all of order $k_0$, and so it is proportional to $1/k_0$.  To obtain the total probability, one integrates over the $k_2-k_3$ plane, or more precisely the line $k_2+k_3=k_0$ with $k_2,\ k_3>0$.  The length of this line is of order $O(k_0)$, and so the total probability asymptotes to a constant at large $k_0$.   Letting $k_2=k_0 x$ we find that in the ultrarelativistic limit
\beq
P_{\rm{tot}}
=\frac{9\lambda}{\pi m^2} \int_0^{1} dx \frac{  x (1-x)}{\left(1-x+x^2 \right)^2}
=\frac{\lambda}{m^2} \left(\frac{6}{\pi}-\frac{2}{\sqrt{3}}  \right)\sim 0.755 \frac{\lambda}{m^2}. \label{asy}
\eeq

\section{Numerical Results for the $\phi^4$ Kink} \label{numsez}
In this section we will numerically evaluate some of the probabilities just calculated for the $\phi^4$ double-well model.

At order $O(\lambda)$ the probability density $P_{\rm{diff}}$ and the total probability $P_{\rm{tot}}$ are proportional to $\lambda$, so in the plots we will divide them by $\lambda$. We use the parameters $m=1$, $\sigma=20$. We have numerically checked that as long as the value of $\sigma$ satisfies $1/m\ll\sigma$
, the value of $\sigma$ will not affect the numerical results.

We begin in Fig.~\ref{pdiff} by plotting the probability density
\beq
P_{\rm{diff}}(k_2)=\int dk_3 P_{\rm{diff}}(k_2,k_3)
\eeq
that one of the two final mesons will have momentum $k_2$.  The shoulder on the right of each curve is not a numerical artifact.  It results from the fact that, with fixed $k_0$, the Jacobian factor in the $k_3$ integral diverges at threshold for the production of the corresponding meson.
\begin{figure}[htbp]
\centering
\includegraphics[width = 0.6\textwidth]{pdiff.pdf}
\caption{The probability density, $P_{\rm{diff}}(k_2)$, that one of the final mesons has momentum $k_2$, plotted for various values of $k_0$.  The factor of $\lambda$ has been divided out.}\label{pdiff}
\end{figure}

Next, in Fig.~\ref{ptot}, we plot the total probability for meson multiplication, as a function of the initial meson momentum $k_0$.  Note that, at high $k_0$, the probability asymptotes to the value found in Eq.~(\ref{asy}).

\begin{figure}[htbp]
\centering
\includegraphics[width = 0.6\textwidth]{ptot.pdf}
\caption{The total meson multiplication probability $P_{\rm{tot}}$ as a function of $k_0$, rescaled by $1/\lambda$.  The dashed line is the asymptotic value derived in Eq.~(\ref{asy}).}\label{ptot}
\end{figure}

Finally in Fig.~\ref{p0p1p2} we plot the probability, $P_n$, that precisely $n$ of the final mesons have $k<0$, so that they travel backwards from the kink.  This plot shows that, at order $O(\lambda)$, even reflectionless kinks lead to some reflection.  However, as might be expected, this is very rare when the momentum $k_0$ of the initial meson is much greater than the meson mass $m$.
\begin{figure}[htbp]
\centering
\includegraphics[width = 0.6\textwidth]{p0p1p2.pdf}
\caption{The probability $P_n$ that $n$ of the momenta of the outgoing mesons are negative. These are all rescaled by $1/\lambda$ and also by other factors, given in the legend, to make them visible in the plot.  The dashed line is again the asymptotic value in Eq.~(\ref{asy}).}\label{p0p1p2}
\end{figure}

\section{Remarks}
Expanding the potential of the $\phi^4$ double-well model about one of its minima, one finds a cubic interaction.  This interaction, in principle, allows a meson to split into two mesons.  However, this process is forbidden in the vacuum because it is not possible to simultaneously conserve energy and momentum.

On the other hand, in the presence of a kink the situation changes.  At leading order in perturbation theory, the mesons still cannot transfer energy to the kink.  However the momentum can be transferred if the meson splits sufficiently close to a kink.  This transfer appears in the probability density (\ref{princ}) as a csch${}^2$ term which enforces approximate momentum conservation among the mesons.

The momentum transfer at a distance nonetheless complicates our calculations, as the meson splitting can occur at any position and all of these positions need to be integrated over, naively leading to these divergences.  We have found three ways of treating these divergences.  First, the coherent integral over the momentum of the initial meson wave packet causes the rapidly oscillating amplitude at large $|x|$ to be suppressed.  Next, adding an exponential damping term to the amplitude and then taking the limit as the damping vanishes also removes the divergence.  Finally, the principal value prescription for the $x$ integral of tanh, used above, renders it finite.  We have checked that all three methods of removing the divergence yield the same results.  Only the first is justified, as it results from the intrinsic spread of the wave packet and not an {\it{ad hoc}} modification.  However the later two methods are much more easily implemented in our calculations.

There are only two inelastic processes that may occur in the scattering of a kink with a single meson at order $O(\lambda)$.  One is meson splitting, treated here.  The second is the (de)excitation of a shape mode while the meson is transmitted or reflected.  We intend to turn to this process in the near future.

\section* {Acknowledgement}

\noindent
JE is supported by NSFC MianShang grants 11875296 and 11675223. HL acknowledges the support from CAS-DAAD Joint Fellowship Programme for Doctoral students of UCAS.

\end{document}